\documentclass[]{raa}            
\usepackage{graphicx,times}
\usepackage{natbib}
\usepackage{lscape}

\providecommand{\bm}[1]{\mbox{\boldmath$#1$\unboldmath}}

\begin{document}
\title{Spectral indices for radio emission of 228 pulsars}

 \volnopage{ {\bf 2016} Vol.\ {\bf ?} No. {\bf ??}, 000--000}
   \setcounter{page}{1}

   \author{Jun Han
      \inst{1,2}
   \and Chen Wang
   \inst{1}
   \and Jun Xu
   \inst{1}
   \and J. L. Han
   \inst{1}
   }

\institute{
National Astronomical Observatories, Chinese Academy of
  Sciences, Beijing 100012, China; \\
\and
School of Astronomy, University of Chinese Academy of Sciences, Beijing 100049, China \\
\vs \no
{\small Received [year] [month] [day]; accepted [year] [month] [day]}}

\abstract{ We determine spectral indices of 228 pulsars by using
  Parkes pulsar data observed at 1.4~GHz, among which 200 spectra are
  newly determined. The indices are distributed in the range from $-4.84$ to
  $-0.46$. Together with known pulsar spectra from literature, we tried
  to find clues to pulsar emission process. The weak correlations
  between the spectral index and the spin-down energy loss rate
  $\dot{E}$ and the potential drop in the polar gap $\Delta\Psi$ 
  hint that emission properties are related to particle
  acceleration process in a pulsar's magnetosphere.
}

\authorrunning{Jun Han et al.}    
\titlerunning{Spectral indices for radio emission of 228 pulsars}
   \maketitle


%

\section{Introduction}
\label{sect:intro}

Pulsar emission mechanism is a long standing problem since the
discovery of the first pulsar in 1968 \citep{hbp+68}. There has been
much effort to understand emission properties on both observational
and theorectical aspects.  Based on observations, cone and core
components, which have different spectral indices and polarization
properties, have been identified from mean pulsar profiles
\citep{r83}. Theoretically, it is not clear yet whether the emission
is generated above the polar cap or near the light cylinder though
radio polarization angle curves hint for the region near the polar cap
\citep{rc69,k70}, while multi-band observations of the Crab pulsar
show that the emission comes from the very outer region near the light
cylinder in pulsar magnetosphere \citep{mh96}.
Very steep spectra and very high temperature of pulsar emission are
very intriguing, which implies that the emission is generated by a
bunch of particles in a coherent mode \citep{m06} when they drift out
in pulsar magnetosphere.  Because of scintillation due to motion of
irregular clumps of interstellar medium, the radio flux densities of
point sources such as nearby pulsars are difficult to measure
especially at low frequencies.  Previously flux densities of pulsars
have to be measured at many epochs at several frequencies, and then
are averaged and fitted for spectral indexes to overcome different
scintillation at different frequencies.

\begin{table}
\caption{Pulsar spectral analyses and frequency ranges}
\label{tab1}
\begin{center}
  \begin{tabular}{lcc}
\hline
Reference  &Number    &Frequency (GHz)      \\
\hline
Sieber (1973)           & $27$     &$0.1\sim10$     \\
Malofeev et al. (1994)  & $20$     &$0.1\sim30$     \\
Lorimer et al. (1995)   &$280$     &$0.4\sim1.6$    \\
Malofeev et al. (1996)  & $284$    &$0.1\sim10$     \\
Toscano et al. (1998)   &$216$     &$0.4\sim1.6$    \\
Kijak et al. (1998)     &$144$     &$1.4\sim4.9$    \\
Maron et al. (2000)    &$266$     &$0.4\sim23$     \\[1mm]
This paper             &$228$     &$1.2-1.5$       \\
\hline
  \end{tabular}
\end{center}
\end{table}

The flux density measurements $S_{\nu}$ observed at a number of
frequencies $\nu$ are fitted by a power law, $S_{\nu} \sim
\nu^{\alpha}$ to get a spectral index, $\alpha$.  Pulsar spectral
indices have previously been determined from measurements in wide
frequency ranges (see Table~\ref{tab1}).
\citet{s73} made the first extensive analysis of spectra of 27
pulsars observed with the German 100~m and 25~m telescopes combined
with flux densities published in literature, and found the turn-over
behaviour at low frequencies and spectral break at high frequencies.
\citet{mgj+94} published spectra for 45 pulsars by using flux
densities over a wide frequency range from 0.1 GHz to 30 GHz.
A comprehensive database for pulsar flux densities of 280 pulsars from
0.4 GHz to 1.6 GHz was published by \citet{lyl+95}. They found a weak
inverse correlation between spectral index and characteristic age for
normal pulsars and a strong positive correlation between spectral
index and period for millisecond pulsars.
\citet{tbm+98} published flux densities for 216 southern pulsars and
19 millisecond pulsars between 436~MHz and 1660~MHz and showed that
normal pulsars have a spectral index between $-3.5$ and $+0.5$ with a
peak around $-1.7$, while millisecond pulsars have spectral indices in
a narrow range between $-3$ and $-1$.
\citet{kxl+98} also analyzed spectra of 23 millisecond pulsars and
found the indices distribute in the range from $-1.0$ to $-2.0$, which is much
narrower than the range for normal pulsars.
\citet{kkw+98} analyzed pulse shape at 4.85~GHz and determined
spectral indices from flux densities at 1.4 GHz and 4.85 GHz for 144
weak pulsars, and showed the spectral index distribution peaks around
$-1.8$ in the range of $-3.7$ $\sim$ $-0.3$.
\citet{mkk+00} collected flux densities between 0.4 GHz to 23 GHz in
literature and analyzed spectra for 281 pulsars, and found that some
pulsars have a spectral break near 1 GHz or lower frequencies, which
are probably due to some components disappearing at high frequencies
as noticed previously \citep{srw75, ikms79, kll+99}.

\begin{table}[htpb]\addtolength{\tabcolsep}{-4pt}
\center
\caption{The flux densities ($S_i$, in mJy) at 4 sub-bands (frequency
  in MHz) and spectral indices $\alpha$ we obtained for 228
  pulsars. Their dispersion measures (in $\mathrm{cm^{-3}pc}$) are 
  listed after pulsar names.}
\label{tab2}
\begin{footnotesize}
\renewcommand\arraystretch{1.2}
\begin{tabular}{llllllc}

\hline  
Pulsar Name   &DM        &Frequency1 / $S_{1}$          &Frequency2 / $S_{2}$          &Frequency3 / $S_{3}$            &Frequency4 / $S_{4}$       &$\alpha$  \\
\hline     
J0719-2545    &253.89    &1278.75 / 1.00$\pm$0.02       &1336.75 / 0.87$\pm$0.02       &1400.29 / 0.82$\pm$0.02       &1459.32 / 0.76$\pm$0.02    &$-1.99\pm0.25$  \\          
J0738-4042    &160.80    &1277.35 / 103.86$\pm$0.15     &1337.61 / 91.67$\pm$0.10      &1400.04 / 83.71$\pm$0.08      &1460.04 / 80.83$\pm$0.11   &$-1.89\pm0.01$  \\             
J0818-3232    &131.80    &1278.75 / 1.07$\pm$0.02       &1336.75 / 0.95$\pm$0.02       &1400.29 / 0.86$\pm$0.02       &1459.32 / 0.77$\pm$0.02    &$-2.41\pm0.20$  \\             
J0820-3921    &179.40    &1277.06 / 0.69$\pm$0.05       &1337.61 / 0.58$\pm$0.03       &1400.04 / 0.56$\pm$0.03       &1460.04 / 0.50$\pm$0.03    &$-2.08\pm0.58$  \\             
J0821-4221    &270.60    &1276.68 / 0.70$\pm$0.03       &1337.61 / 0.58$\pm$0.02       &1400.04 / 0.51$\pm$0.02       &1460.04 / 0.46$\pm$0.02    &$-3.15\pm0.49$  \\             
J0831-4406    &254.00    &1276.05 / 0.70$\pm$0.04       &1337.61 / 0.64$\pm$0.02       &1400.04 / 0.59$\pm$0.02       &1461.16 / 0.60$\pm$0.03    &$-1.10\pm0.47$  \\             
J0838-2621    &116.90    &1277.10 / 0.65$\pm$0.04       &1337.79 / 0.46$\pm$0.03       &1400.04 / 0.40$\pm$0.02       &1460.04 / 0.34$\pm$0.03    &$-4.84\pm0.74$  \\             
J0843-5022    &178.47    &1277.14 / 0.51$\pm$0.03       &1337.36 / 0.46$\pm$0.02       &1400.04 / 0.41$\pm$0.02       &1460.04 / 0.39$\pm$0.02    &$-1.99\pm0.61$  \\             
J0932-3217    &102.10    &1277.64 / 0.45$\pm$0.02       &1337.09 / 0.40$\pm$0.01       &1400.60 / 0.32$\pm$0.01       &1459.61 / 0.33$\pm$0.01    &$-2.74\pm0.43$  \\             
J0942-5552    &180.20    &1277.30 / 12.38$\pm$0.07      &1336.75 / 10.65$\pm$0.05      &1400.45 / 9.57$\pm$0.04       &1459.61 / 8.64$\pm$0.05    &$-2.64\pm0.05$  \\             
J0945-4833    &101.10    &1278.75 / 0.47$\pm$0.02       &1336.75 / 0.38$\pm$0.02       &1400.29 / 0.35$\pm$0.02       &1459.32 / 0.33$\pm$0.02    &$-2.73\pm0.53$  \\             
J0952-3839    &167.00    &1277.15 / 0.54$\pm$0.03       &1337.61 / 0.40$\pm$0.02       &1400.20 / 0.34$\pm$0.02       &1460.04 / 0.31$\pm$0.02    &$-4.27\pm0.54$  \\             
J1017-5621    &438.70    &1278.75 / 2.56$\pm$0.03       &1336.26 / 2.23$\pm$0.02       &1400.29 / 2.01$\pm$0.02       &1459.32 / 1.82$\pm$0.02    &$-2.56\pm0.11$  \\             
J1032-5911    &418.20    &1277.39 / 1.16$\pm$0.04       &1337.61 / 0.93$\pm$0.03       &1400.04 / 0.79$\pm$0.02       &1460.04 / 0.77$\pm$0.03    &$-3.24\pm0.37$  \\             
J1058-5957    &334.00    &1277.39 / 0.75$\pm$0.02       &1337.61 / 0.67$\pm$0.02       &1400.04 / 0.60$\pm$0.02       &1460.04 / 0.53$\pm$0.02    &$-2.57\pm0.33$  \\             
J1107-5947    &158.40    &1278.17 / 1.01$\pm$0.04       &1336.91 / 1.00$\pm$0.03       &1400.04 / 0.87$\pm$0.02       &1460.04 / 0.85$\pm$0.03    &$-1.59\pm0.36$  \\             
J1107-6143    &406.00    &1342.75 / 0.74$\pm$0.01       &1399.99 / 0.71$\pm$0.01       &1465.60 / 0.65$\pm$0.01       &1522.10 / 0.60$\pm$0.02    &$-1.58\pm0.23$  \\             
J1115-6052    &228.20    &1342.75 / 0.52$\pm$0.03       &1399.88 / 0.46$\pm$0.02       &1465.27 / 0.48$\pm$0.02       &1521.94 / 0.44$\pm$0.03    &$-1.06\pm0.61$  \\             
J1123-6651    &111.20    &1277.30 / 0.53$\pm$0.06       &1336.75 / 0.46$\pm$0.04       &1400.60 / 0.42$\pm$0.03       &1459.61 / 0.39$\pm$0.05    &$-2.34\pm1.08$  \\             
J1130-6807    &148.73    &1277.10 / 0.80$\pm$0.07       &1336.91 / 0.64$\pm$0.05       &1400.20 / 0.55$\pm$0.05       &1460.04 / 0.49$\pm$0.06    &$-3.62\pm0.96$  \\             
J1156-5707    &243.50    &1276.14 / 0.39$\pm$0.04       &1337.61 / 0.35$\pm$0.02       &1400.20 / 0.31$\pm$0.02       &1460.04 / 0.29$\pm$0.03    &$-2.24\pm0.86$  \\             
J1220-6318    &347.00    &1277.39 / 1.03$\pm$0.03       &1337.44 / 0.93$\pm$0.02       &1400.29 / 0.78$\pm$0.02       &1459.52 / 0.72$\pm$0.03    &$-2.87\pm0.33$  \\             
J1225-6035    &176.10    &1278.75 / 0.36$\pm$0.02       &1337.08 / 0.34$\pm$0.02       &1400.29 / 0.31$\pm$0.02       &1459.32 / 0.29$\pm$0.02    &$-1.69\pm0.62$  \\             
J1237-6725    &176.50    &1277.19 / 1.02$\pm$0.03       &1336.75 / 0.75$\pm$0.02       &1400.29 / 0.68$\pm$0.01       &1459.32 / 0.53$\pm$0.02    &$-4.48\pm0.27$  \\             
J1303-6305    &343.00    &1277.16 / 0.38$\pm$0.03       &1336.75 / 0.29$\pm$0.02       &1400.29 / 0.28$\pm$0.02       &1459.32 / 0.23$\pm$0.02    &$-3.36\pm0.85$  \\             
J1314-6101    &309.00    &1277.10 / 0.78$\pm$0.03       &1336.91 / 0.64$\pm$0.02       &1400.04 / 0.57$\pm$0.02       &1460.04 / 0.51$\pm$0.02    &$-3.09\pm0.40$  \\             
J1319-6105    &442.20    &1278.50 / 1.48$\pm$0.04       &1336.75 / 1.38$\pm$0.03       &1400.45 / 1.28$\pm$0.03       &1459.32 / 1.26$\pm$0.03    &$-1.27\pm0.24$  \\             
J1326-6408    &502.70    &1278.75 / 2.85$\pm$0.03       &1336.75 / 2.40$\pm$0.02       &1400.29 / 2.16$\pm$0.02       &1459.32 / 2.00$\pm$0.03    &$-2.71\pm0.12$  \\             
J1348-6307    &597.00    &1277.70 / 0.71$\pm$0.04       &1337.09 / 0.60$\pm$0.02       &1400.45 / 0.49$\pm$0.02       &1459.61 / 0.49$\pm$0.03    &$-3.06\pm0.53$  \\             
J1352-6803    &214.60    &1278.75 / 1.90$\pm$0.04       &1336.75 / 1.66$\pm$0.03       &1400.29 / 1.42$\pm$0.03       &1459.52 / 1.30$\pm$0.03    &$-2.99\pm0.22$  \\             
J1355-5925    &354.80    &1342.75 / 0.75$\pm$0.02       &1399.59 / 0.70$\pm$0.02       &1465.35 / 0.64$\pm$0.02       &1522.34 / 0.63$\pm$0.03    &$-1.52\pm0.36$  \\             
J1412-6111    &311.80    &1275.57 / 0.62$\pm$0.05       &1337.61 / 0.57$\pm$0.03       &1400.04 / 0.50$\pm$0.03       &1460.04 / 0.46$\pm$0.03    &$-2.27\pm0.67$  \\             
J1415-6621    &260.17    &1278.75 / 0.59$\pm$0.03       &1336.91 / 0.51$\pm$0.02       &1400.29 / 0.45$\pm$0.02       &1459.32 / 0.45$\pm$0.02    &$-2.13\pm0.47$  \\             
J1424-5556    &198.70    &1277.64 / 0.55$\pm$0.03       &1336.75 / 0.51$\pm$0.02       &1400.60 / 0.45$\pm$0.02       &1459.11 / 0.45$\pm$0.03    &$-1.65\pm0.56$  \\             
J1444-5941    &177.10    &1277.49 / 0.83$\pm$0.05       &1336.75 / 0.68$\pm$0.03       &1400.29 / 0.62$\pm$0.03       &1459.61 / 0.56$\pm$0.03    &$-2.85\pm0.58$  \\             
J1502-6128    &256.50    &1277.39 / 0.87$\pm$0.03       &1337.61 / 0.69$\pm$0.02       &1400.04 / 0.61$\pm$0.02       &1460.04 / 0.60$\pm$0.03    &$-2.95\pm0.41$  \\             
J1507-6640    &129.80    &1277.10 / 1.41$\pm$0.02       &1336.91 / 1.23$\pm$0.02       &1400.20 / 1.05$\pm$0.02       &1460.04 / 0.97$\pm$0.02    &$-2.96\pm0.17$  \\             
J1511-5835    &332.00    &1277.18 / 0.91$\pm$0.05       &1336.91 / 0.81$\pm$0.03       &1399.87 / 0.74$\pm$0.03       &1460.04 / 0.57$\pm$0.03    &$-3.09\pm0.56$  \\             
J1524-5706    &833.00    &1342.50 / 0.47$\pm$0.02       &1399.88 / 0.41$\pm$0.02       &1465.39 / 0.40$\pm$0.02       &1523.09 / 0.38$\pm$0.02    &$-1.56\pm0.49$  \\             
J1534-5405    &190.82    &1278.50 / 1.99$\pm$0.04       &1336.75 / 1.63$\pm$0.03       &1400.29 / 1.39$\pm$0.03       &1459.11 / 1.26$\pm$0.04    &$-3.55\pm0.24$  \\             
J1537-5645    &707.00    &1277.28 / 1.41$\pm$0.08       &1337.61 / 1.31$\pm$0.06       &1400.04 / 1.26$\pm$0.05       &1460.04 / 1.22$\pm$0.06    &$-1.04\pm0.51$  \\             
J1548-4927    &141.20    &1278.75 / 0.75$\pm$0.03       &1336.75 / 0.66$\pm$0.02       &1400.44 / 0.62$\pm$0.02       &1459.32 / 0.52$\pm$0.02    &$-2.48\pm0.37$  \\             
J1553-5456    &210.00    &1278.75 / 1.36$\pm$0.04       &1336.75 / 1.16$\pm$0.03       &1400.29 / 0.93$\pm$0.03       &1459.32 / 0.87$\pm$0.03    &$-3.60\pm0.29$  \\             
J1614-3937    &152.44    &1277.28 / 0.55$\pm$0.04       &1336.91 / 0.51$\pm$0.03       &1400.20 / 0.42$\pm$0.02       &1460.04 / 0.43$\pm$0.03    &$-2.05\pm0.66$  \\             
J1615-5537    &124.48    &1277.11 / 1.11$\pm$0.03       &1336.91 / 0.98$\pm$0.02       &1400.20 / 0.91$\pm$0.02       &1460.04 / 0.82$\pm$0.03    &$-2.14\pm0.30$  \\             
J1622-4802    &364.30    &1277.28 / 0.91$\pm$0.04       &1337.61 / 0.86$\pm$0.02       &1400.04 / 0.82$\pm$0.02       &1460.04 / 0.82$\pm$0.03    &$-0.84\pm0.36$  \\             
J1623-4256    &295.00    &1277.38 / 2.93$\pm$0.08       &1336.91 / 2.64$\pm$0.06       &1400.20 / 2.40$\pm$0.05       &1460.04 / 2.41$\pm$0.07    &$-1.57\pm0.27$  \\             
J1624-4411    &139.40    &1277.17 / 0.57$\pm$0.04       &1336.91 / 0.50$\pm$0.03       &1400.20 / 0.48$\pm$0.03       &1460.04 / 0.46$\pm$0.03    &$-1.39\pm0.65$  \\             
J1626-4537    &237.00    &1277.45 / 1.08$\pm$0.03       &1337.09 / 0.98$\pm$0.02       &1400.60 / 0.91$\pm$0.02       &1459.61 / 0.89$\pm$0.02    &$-1.49\pm0.26$  \\             
J1627-4845    &557.80    &1277.56 / 0.68$\pm$0.07       &1336.75 / 0.63$\pm$0.04       &1400.44 / 0.61$\pm$0.04       &1459.61 / 0.59$\pm$0.05    &$-0.92\pm0.87$  \\             
J1627-5547    &166.20    &1342.75 / 0.67$\pm$0.02       &1399.76 / 0.58$\pm$0.02       &1465.60 / 0.56$\pm$0.02       &1522.34 / 0.50$\pm$0.03    &$-2.15\pm0.43$  \\     
\hline  
\end{tabular}
\end{footnotesize}
\end{table}

\begin{table}[htpb]\addtolength{\tabcolsep}{-4pt}
\center
\addtocounter{table}{-1}
\caption{--continue.}
\begin{footnotesize}
\renewcommand\arraystretch{1.2}
\begin{tabular}{llllllc}
\hline
Pulsar Name  &DM        &Frequency1  / $S_{1}$         &Frequency2  / $S_{2}$          &Frequency3  / $S_{3}$         &Frequency4  / $S_{4}$       &$\alpha$  \\
\hline     
J1628-4804   &952.00    &1274.90 / 1.38$\pm$0.08       &1337.07 / 1.29$\pm$0.06       &1399.87 / 1.12$\pm$0.05       &1460.04 / 1.13$\pm$0.06     &$-1.68\pm0.55$ \\             
J1632-4621   &562.90    &1278.75 / 1.08$\pm$0.03       &1336.75 / 0.95$\pm$0.02       &1400.12 / 0.84$\pm$0.02       &1459.96 / 0.78$\pm$0.03     &$-2.52\pm0.31$ \\             
J1635-5954   &134.90    &1277.31 / 2.15$\pm$0.04       &1336.91 / 1.84$\pm$0.02       &1400.20 / 1.56$\pm$0.02       &1460.04 / 1.39$\pm$0.03     &$-3.35\pm0.17$ \\             
J1636-4803   &503.00    &1278.75 / 2.27$\pm$0.08       &1336.75 / 1.95$\pm$0.06       &1400.12 / 1.80$\pm$0.06       &1459.32 / 1.67$\pm$0.06     &$-2.26\pm0.36$ \\             
J1636-4933   &542.70    &1277.31 / 0.69$\pm$0.06       &1336.91 / 0.67$\pm$0.04       &1400.03 / 0.47$\pm$0.03       &1460.04 / 0.44$\pm$0.04     &$-4.28\pm0.80$ \\             
J1637-4721   &448.00    &1277.19 / 0.87$\pm$0.06       &1336.91 / 0.76$\pm$0.04       &1400.03 / 0.72$\pm$0.04       &1460.04 / 0.61$\pm$0.04     &$-2.41\pm0.70$ \\             
J1638-3815   &238.00    &1277.32 / 0.85$\pm$0.04       &1337.13 / 0.70$\pm$0.02       &1400.20 / 0.67$\pm$0.02       &1460.04 / 0.61$\pm$0.03     &$-2.26\pm0.44$ \\             
J1638-5226   &170.10    &1277.39 / 0.75$\pm$0.03       &1337.61 / 0.66$\pm$0.02       &1400.20 / 0.59$\pm$0.02       &1460.04 / 0.56$\pm$0.03     &$-2.29\pm0.45$ \\             
J1644-4559   &478.80    &1276.39 / 405.85$\pm$0.22     &1337.07 / 342.68$\pm$0.13     &1399.87 / 296.43$\pm$0.10     &1460.04 / 272.71$\pm$0.10   &$-2.90\pm0.00$ \\             
J1648-4458   &925.00    &1276.91 / 0.64$\pm$0.08       &1337.61 / 0.57$\pm$0.05       &1399.87 / 0.54$\pm$0.04       &1460.04 / 0.47$\pm$0.05     &$-2.17\pm1.14$ \\             
J1649-3805   &213.80    &1277.14 / 1.54$\pm$0.05       &1337.61 / 1.23$\pm$0.03       &1400.20 / 1.16$\pm$0.03       &1460.04 / 1.15$\pm$0.04     &$-2.12\pm0.30$ \\             
J1649-4653   &332.00    &1277.32 / 0.38$\pm$0.05       &1336.91 / 0.36$\pm$0.03       &1400.03 / 0.34$\pm$0.03       &1460.04 / 0.28$\pm$0.03     &$-2.17\pm1.18$ \\             
J1651-4246   &482.00    &1277.69 / 23.34$\pm$0.17      &1336.75 / 21.23$\pm$0.12      &1400.29 / 18.61$\pm$0.10      &1459.61 / 16.99$\pm$0.12    &$-2.46\pm0.07$ \\             
J1651-5255   &164.00    &1277.04 / 3.96$\pm$0.05       &1336.75 / 3.41$\pm$0.03       &1400.60 / 2.89$\pm$0.03       &1459.61 / 2.65$\pm$0.04     &$-3.11\pm0.13$ \\             
J1658-4958   &193.40    &1278.75 / 2.11$\pm$0.04       &1336.75 / 1.89$\pm$0.03       &1400.44 / 1.63$\pm$0.03       &1459.32 / 1.53$\pm$0.04     &$-2.56\pm0.21$ \\             
J1659-4439   &535.00    &1277.25 / 0.50$\pm$0.04       &1336.91 / 0.44$\pm$0.03       &1400.03 / 0.44$\pm$0.02       &1460.04 / 0.42$\pm$0.03     &$-1.10\pm0.74$ \\             
J1700-3611   &232.70    &1277.13 / 1.11$\pm$0.05       &1337.07 / 0.99$\pm$0.03       &1400.20 / 0.92$\pm$0.03       &1460.04 / 0.79$\pm$0.03     &$-2.27\pm0.41$ \\             
J1702-4217   &629.00    &1276.88 / 1.36$\pm$0.16       &1337.07 / 1.30$\pm$0.10       &1399.87 / 1.18$\pm$0.08       &1460.26 / 1.18$\pm$0.10     &$-1.23\pm0.93$ \\             
J1705-3950   &207.10    &1278.75 / 2.01$\pm$0.06       &1336.75 / 1.92$\pm$0.05       &1400.12 / 1.77$\pm$0.05       &1459.32 / 1.83$\pm$0.05     &$-0.84\pm0.30$ \\             
J1707-4053   &360.00    &1277.26 / 12.24$\pm$0.24      &1336.91 / 10.21$\pm$0.17      &1400.15 / 8.84$\pm$0.15       &1460.04 / 8.31$\pm$0.18     &$-2.99\pm0.20$ \\             
J1707-4341   &398.20    &1342.75 / 0.60$\pm$0.03       &1399.86 / 0.53$\pm$0.02       &1465.60 / 0.48$\pm$0.02       &1522.10 / 0.49$\pm$0.03     &$-1.85\pm0.50$ \\             
J1707-4729   &268.30    &1278.75 / 2.87$\pm$0.10       &1336.75 / 2.71$\pm$0.08       &1400.44 / 2.66$\pm$0.08       &1459.32 / 2.46$\pm$0.09     &$-1.04\pm0.35$ \\             
J1708-3827   &788.00    &1276.78 / 0.52$\pm$0.05       &1336.75 / 0.54$\pm$0.03       &1400.29 / 0.44$\pm$0.03       &1459.61 / 0.40$\pm$0.03     &$-2.53\pm0.85$ \\             
J1709-3626   &393.60    &1276.22 / 0.63$\pm$0.05       &1337.07 / 0.61$\pm$0.03       &1400.04 / 0.57$\pm$0.03       &1460.04 / 0.57$\pm$0.04     &$-0.84\pm0.68$ \\             
J1715-4034   &254.00    &1278.75 / 2.40$\pm$0.04       &1337.08 / 2.07$\pm$0.03       &1400.12 / 1.86$\pm$0.03       &1459.52 / 1.74$\pm$0.03     &$-2.48\pm0.17$ \\             
J1716-3720   &682.70    &1276.91 / 0.57$\pm$0.04       &1336.91 / 0.49$\pm$0.02       &1400.03 / 0.38$\pm$0.02       &1460.04 / 0.37$\pm$0.03     &$-3.75\pm0.68$ \\             
J1717-3953   &466.00    &1277.70 / 1.77$\pm$0.07       &1337.09 / 1.58$\pm$0.06       &1400.29 / 1.28$\pm$0.05       &1459.61 / 1.17$\pm$0.06     &$-3.33\pm0.45$ \\             
J1718-3718   &371.10    &1276.72 / 3.03$\pm$0.26       &1336.89 / 2.55$\pm$0.18       &1399.87 / 2.16$\pm$0.17       &1460.21 / 1.84$\pm$0.18     &$-3.69\pm0.89$ \\             
J1719-4302   &297.70    &1276.31 / 0.51$\pm$0.04       &1336.75 / 0.45$\pm$0.03       &1400.44 / 0.40$\pm$0.02       &1459.32 / 0.31$\pm$0.03     &$-3.33\pm0.85$ \\             
J1724-3149   &409.00    &1278.09 / 0.49$\pm$0.05       &1336.91 / 0.44$\pm$0.03       &1400.20 / 0.43$\pm$0.03       &1460.04 / 0.39$\pm$0.03     &$-1.49\pm0.88$ \\             
J1725-3546   &744.00    &1276.62 / 1.01$\pm$0.06       &1337.07 / 0.93$\pm$0.04       &1399.87 / 0.71$\pm$0.03       &1460.04 / 0.61$\pm$0.04     &$-4.07\pm0.61$ \\             
J1727-2739   &147.00    &1278.75 / 4.04$\pm$0.05       &1336.75 / 3.59$\pm$0.04       &1400.44 / 3.28$\pm$0.04       &1459.52 / 3.09$\pm$0.05     &$-2.04\pm0.14$ \\             
J1728-4028   &231.00    &1277.69 / 1.12$\pm$0.05       &1336.75 / 1.02$\pm$0.03       &1400.60 / 0.90$\pm$0.03       &1460.07 / 0.77$\pm$0.04     &$-2.82\pm0.42$ \\             
J1730-3353   &256.00    &1278.02 / 0.51$\pm$0.04       &1336.91 / 0.46$\pm$0.03       &1400.03 / 0.39$\pm$0.03       &1460.04 / 0.28$\pm$0.03     &$-4.08\pm0.90$ \\             
J1732-4128   &195.30    &1276.57 / 1.37$\pm$0.04       &1337.61 / 1.09$\pm$0.02       &1400.44 / 1.01$\pm$0.02       &1459.52 / 0.93$\pm$0.03     &$-2.76\pm0.27$ \\             
J1733-3322   &524.00    &1342.75 / 1.23$\pm$0.04       &1399.98 / 1.14$\pm$0.03       &1465.75 / 0.98$\pm$0.03       &1521.58 / 0.83$\pm$0.04     &$-2.96\pm0.38$ \\             
J1733-4005   &317.80    &1276.84 / 0.81$\pm$0.03       &1336.91 / 0.68$\pm$0.02       &1400.20 / 0.57$\pm$0.02       &1460.04 / 0.53$\pm$0.02     &$-3.32\pm0.34$ \\             
J1736-2457   &170.00    &1276.90 / 1.17$\pm$0.04       &1336.91 / 0.99$\pm$0.02       &1400.20 / 0.84$\pm$0.02       &1460.04 / 0.76$\pm$0.02     &$-3.24\pm0.30$ \\             
J1736-2843   &331.00    &1277.32 / 0.49$\pm$0.03       &1336.91 / 0.43$\pm$0.02       &1400.20 / 0.36$\pm$0.02       &1460.78 / 0.32$\pm$0.02     &$-3.16\pm0.68$ \\             
J1737-3102   &280.00    &1275.83 / 0.77$\pm$0.04       &1337.32 / 0.65$\pm$0.03       &1399.70 / 0.57$\pm$0.02       &1460.04 / 0.57$\pm$0.03     &$-2.26\pm0.49$ \\             
J1737-3137   &488.20    &1278.50 / 1.20$\pm$0.07       &1336.75 / 1.06$\pm$0.06       &1400.12 / 0.98$\pm$0.05       &1459.11 / 1.01$\pm$0.06     &$-1.33\pm0.58$ \\             
J1738-2330   & 99.30    &1277.32 / 1.47$\pm$0.06       &1336.91 / 1.21$\pm$0.04       &1400.20 / 1.05$\pm$0.03       &1460.04 / 0.88$\pm$0.05     &$-3.69\pm0.42$ \\             
J1738-2647   &182.20    &1277.25 / 0.70$\pm$0.04       &1336.91 / 0.53$\pm$0.03       &1400.20 / 0.50$\pm$0.02       &1460.04 / 0.48$\pm$0.03     &$-2.74\pm0.58$ \\             
J1738-3316   &273.00    &1275.67 / 0.92$\pm$0.07       &1337.07 / 0.69$\pm$0.04       &1400.04 / 0.62$\pm$0.04       &1460.04 / 0.53$\pm$0.05     &$-3.83\pm0.77$ \\             
J1739-3159   &337.00    &1277.70 / 1.10$\pm$0.07       &1337.09 / 1.06$\pm$0.04       &1400.29 / 0.98$\pm$0.04       &1459.61 / 0.81$\pm$0.05     &$-2.21\pm0.58$ \\             
J1740-3052   &738.78    &1278.75 / 1.14$\pm$0.05       &1336.75 / 0.90$\pm$0.04       &1399.95 / 0.79$\pm$0.04       &1459.32 / 0.74$\pm$0.04     &$-3.39\pm0.54$ \\             
J1741-2733   &149.20    &1278.50 / 3.02$\pm$0.06       &1336.75 / 2.67$\pm$0.05       &1400.45 / 2.34$\pm$0.04       &1459.32 / 2.18$\pm$0.05     &$-2.52\pm0.21$ \\             
J1741-3016   &382.00    &1278.75 / 3.49$\pm$0.08       &1336.75 / 3.06$\pm$0.05       &1399.95 / 2.74$\pm$0.05       &1459.32 / 2.71$\pm$0.06     &$-1.98\pm0.21$ \\             
J1744-3130   &192.90    &1276.42 / 0.81$\pm$0.04       &1337.61 / 0.66$\pm$0.02       &1400.04 / 0.67$\pm$0.02       &1460.29 / 0.61$\pm$0.03     &$-1.66\pm0.41$ \\             
J1749-2629   &409.00    &1342.50 / 0.74$\pm$0.04       &1399.99 / 0.72$\pm$0.03       &1465.39 / 0.62$\pm$0.03       &1522.35 / 0.55$\pm$0.04     &$-2.38\pm0.59$ \\             
J1750-2438   &476.00    &1275.87 / 0.83$\pm$0.03       &1337.07 / 0.63$\pm$0.02       &1400.04 / 0.55$\pm$0.02       &1460.04 / 0.49$\pm$0.02     &$-3.82\pm0.38$ \\            
\hline
\end{tabular}
\end{footnotesize}
\end{table}

\begin{table}[htpb]\addtolength{\tabcolsep}{-4pt}
\center
\addtocounter{table}{-1}
\caption{--continue.}
\begin{footnotesize}
\renewcommand\arraystretch{1.2}
\begin{tabular}{llllllc}
\hline
Pulsar Name      &DM        &Frequency1  / $S_{1}$         &Frequency2  / $S_{2}$          &Frequency3  / $S_{3}$         &Frequency4  / $S_{4}$        &$\alpha$  \\
\hline     
J1751-3323       &296.70    &1278.50 / 2.19$\pm$0.05       &1336.92 / 1.95$\pm$0.04       &1400.60 / 1.91$\pm$0.04       &1459.52 / 1.74$\pm$0.05     &$-1.53\pm0.24$  \\             
J1754-3443       &187.70    &1276.70 / 0.77$\pm$0.04       &1336.91 / 0.63$\pm$0.02       &1400.20 / 0.61$\pm$0.02       &1460.04 / 0.58$\pm$0.02     &$-1.82\pm0.43$  \\             
J1755-2725       &115.00    &1277.56 / 1.16$\pm$0.11       &1336.75 / 1.01$\pm$0.07       &1400.60 / 0.94$\pm$0.06       &1459.61 / 0.91$\pm$0.07     &$-1.74\pm0.84$  \\             
J1756-2251       &121.20    &1276.34 / 0.94$\pm$0.05       &1336.75 / 0.94$\pm$0.03       &1400.60 / 0.81$\pm$0.03       &1459.61 / 0.80$\pm$0.03     &$-1.67\pm0.43$  \\             
J1757-2223       &239.30    &1278.75 / 1.36$\pm$0.07       &1336.91 / 1.32$\pm$0.05       &1400.29 / 1.35$\pm$0.05       &1459.52 / 1.25$\pm$0.05     &$-0.46\pm0.44$  \\             
J1758-2206       &678.00    &1276.46 / 0.63$\pm$0.06       &1336.91 / 0.54$\pm$0.04       &1400.20 / 0.43$\pm$0.03       &1460.04 / 0.42$\pm$0.04     &$-3.36\pm0.95$  \\             
J1758-2540       &218.20    &1276.69 / 1.06$\pm$0.06       &1336.91 / 0.86$\pm$0.04       &1399.89 / 0.72$\pm$0.03       &1460.04 / 0.75$\pm$0.04     &$-2.73\pm0.55$  \\             
J1758-2630       &328.00    &1276.17 / 0.23$\pm$0.03       &1336.91 / 0.24$\pm$0.02       &1400.20 / 0.22$\pm$0.02       &1460.04 / 0.20$\pm$0.02     &$-1.28\pm1.09$  \\             
J1759-1940       &302.70    &1276.34 / 1.49$\pm$0.09       &1337.07 / 1.32$\pm$0.06       &1400.04 / 1.11$\pm$0.05       &1460.26 / 1.10$\pm$0.07     &$-2.54\pm0.57$  \\             
J1759-3107       &128.60    &1342.75 / 1.66$\pm$0.03       &1400.15 / 1.52$\pm$0.02       &1465.39 / 1.43$\pm$0.02       &1521.98 / 1.31$\pm$0.03     &$-1.82\pm0.18$  \\             
J1801-1909       &264.00    &1277.08 / 0.77$\pm$0.04       &1336.91 / 0.63$\pm$0.02       &1400.20 / 0.51$\pm$0.02       &1460.04 / 0.45$\pm$0.03     &$-4.23\pm0.53$  \\             
J1802-2124       &149.63    &1342.50 / 1.01$\pm$0.03       &1400.15 / 0.89$\pm$0.03       &1465.15 / 0.75$\pm$0.03       &1521.66 / 0.63$\pm$0.03     &$-3.73\pm0.44$  \\             
J1802-2426       &711.00    &1276.92 / 0.48$\pm$0.04       &1337.07 / 0.46$\pm$0.03       &1399.89 / 0.42$\pm$0.03       &1460.04 / 0.39$\pm$0.03     &$-1.60\pm0.80$  \\             
J1803-1857       &392.00    &1277.96 / 0.45$\pm$0.02       &1336.91 / 0.39$\pm$0.01       &1400.20 / 0.37$\pm$0.01       &1460.04 / 0.36$\pm$0.02     &$-1.59\pm0.46$  \\             
J1804-0735       &186.32    &1277.01 / 1.02$\pm$0.07       &1336.75 / 0.99$\pm$0.05       &1400.76 / 0.84$\pm$0.05       &1459.83 / 0.77$\pm$0.05     &$-2.35\pm0.67$  \\             
J1805-1504       &225.00    &1277.10 / 5.40$\pm$0.09       &1337.07 / 4.79$\pm$0.05       &1400.04 / 4.32$\pm$0.05       &1460.78 / 4.23$\pm$0.06     &$-1.83\pm0.14$  \\             
J1806-1154       &122.41    &1278.75 / 3.48$\pm$0.04       &1336.75 / 3.00$\pm$0.03       &1400.29 / 2.50$\pm$0.03       &1459.52 / 2.36$\pm$0.03     &$-3.14\pm0.13$  \\             
J1809-1429       &411.30    &1278.75 / 1.07$\pm$0.04       &1336.75 / 0.91$\pm$0.03       &1400.29 / 0.83$\pm$0.03       &1459.32 / 0.80$\pm$0.03     &$-2.19\pm0.35$  \\             
J1810-1820       &452.20    &1275.64 / 0.76$\pm$0.07       &1337.07 / 0.72$\pm$0.05       &1399.89 / 0.67$\pm$0.04       &1460.26 / 0.63$\pm$0.05     &$-1.44\pm0.84$  \\             
J1812-1718       &255.10    &1278.50 / 1.52$\pm$0.04       &1336.75 / 1.33$\pm$0.03       &1400.45 / 1.22$\pm$0.03       &1459.11 / 1.16$\pm$0.03     &$-2.02\pm0.28$  \\             
J1812-1733       &518.00    &1278.50 / 5.68$\pm$0.10       &1336.75 / 5.15$\pm$0.08       &1400.45 / 4.87$\pm$0.07       &1459.11 / 4.64$\pm$0.07     &$-1.47\pm0.17$  \\             
J1812-2102       &547.20    &1278.50 / 1.66$\pm$0.06       &1336.75 / 1.69$\pm$0.04       &1400.45 / 1.58$\pm$0.04       &1459.11 / 1.48$\pm$0.05     &$-1.00\pm0.33$  \\             
J1813-2113       &462.30    &1275.34 / 0.65$\pm$0.05       &1336.91 / 0.54$\pm$0.03       &1400.20 / 0.50$\pm$0.03       &1460.04 / 0.43$\pm$0.03     &$-2.74\pm0.70$  \\             
J1814-1649       &782.00    &1278.75 / 1.54$\pm$0.07       &1337.08 / 1.38$\pm$0.05       &1400.29 / 1.32$\pm$0.05       &1459.32 / 1.25$\pm$0.05     &$-1.48\pm0.42$  \\             
J1814-1744       &792.00    &1276.87 / 0.83$\pm$0.06       &1336.91 / 0.71$\pm$0.04       &1400.04 / 0.61$\pm$0.03       &1460.04 / 0.53$\pm$0.04     &$-3.39\pm0.69$  \\             
J1816-1729       &525.50    &1277.38 / 1.66$\pm$0.05       &1336.91 / 1.45$\pm$0.03       &1400.20 / 1.31$\pm$0.03       &1460.04 / 1.27$\pm$0.03     &$-1.97\pm0.26$  \\             
J1818-1422       &622.00    &1277.75 / 10.69$\pm$0.16      &1336.91 / 9.69$\pm$0.10       &1400.04 / 8.41$\pm$0.09       &1460.04 / 7.85$\pm$0.12     &$-2.46\pm0.14$  \\             
J1819-0925       &378.00    &1278.75 / 1.04$\pm$0.03       &1336.75 / 0.89$\pm$0.03       &1400.29 / 0.77$\pm$0.02       &1459.52 / 0.72$\pm$0.03     &$-2.92\pm0.35$  \\             
J1819-1510       &421.70    &1275.39 / 1.04$\pm$0.06       &1337.07 / 0.87$\pm$0.04       &1399.89 / 0.76$\pm$0.03       &1460.04 / 0.65$\pm$0.04     &$-3.39\pm0.57$  \\             
J1820-1529       &772.00    &1278.50 / 1.17$\pm$0.07       &1336.75 / 0.99$\pm$0.06       &1400.45 / 0.91$\pm$0.05       &1459.32 / 0.88$\pm$0.06     &$-2.20\pm0.65$  \\             
J1822-1400       &651.10    &1342.75 / 1.26$\pm$0.04       &1399.82 / 1.13$\pm$0.03       &1466.05 / 1.00$\pm$0.03       &1521.60 / 0.98$\pm$0.05     &$-2.25\pm0.43$  \\             
J1823-1347       &1044.00   &1277.56 / 0.54$\pm$0.06       &1336.75 / 0.45$\pm$0.04       &1400.45 / 0.44$\pm$0.03       &1459.61 / 0.44$\pm$0.04     &$-1.31\pm0.94$  \\             
J1823-1526       &611.00    &1277.44 / 0.48$\pm$0.04       &1336.75 / 0.47$\pm$0.03       &1400.45 / 0.41$\pm$0.02       &1459.61 / 0.36$\pm$0.02     &$-2.38\pm0.70$  \\             
J1824-1423       &428.30    &1278.75 / 1.28$\pm$0.04       &1336.75 / 1.15$\pm$0.03       &1400.29 / 1.13$\pm$0.03       &1459.32 / 1.07$\pm$0.04     &$-1.23\pm0.31$  \\             
J1826-1526       &530.00    &1277.09 / 0.42$\pm$0.04       &1336.91 / 0.34$\pm$0.02       &1400.04 / 0.35$\pm$0.02       &1460.04 / 0.29$\pm$0.02     &$-2.09\pm0.81$  \\             
J1827-0750       &381.00    &1277.14 / 3.41$\pm$0.06       &1337.61 / 3.03$\pm$0.04       &1400.04 / 2.70$\pm$0.04       &1460.04 / 2.56$\pm$0.04     &$-2.19\pm0.16$  \\             
J1827-0958       &430.00    &1278.75 / 2.79$\pm$0.07       &1336.91 / 2.51$\pm$0.06       &1400.29 / 2.17$\pm$0.05       &1459.32 / 2.02$\pm$0.06     &$-2.56\pm0.28$  \\             
J1828-0611       &363.20    &1278.50 / 1.89$\pm$0.04       &1336.92 / 1.74$\pm$0.03       &1400.45 / 1.62$\pm$0.03       &1459.55 / 1.53$\pm$0.03     &$-1.58\pm0.21$  \\             
J1829+0000       &114.00    &1275.18 / 0.69$\pm$0.07       &1336.91 / 0.57$\pm$0.04       &1400.20 / 0.54$\pm$0.04       &1460.04 / 0.50$\pm$0.05     &$-2.14\pm0.88$  \\             
J1830-1135       &257.00    &1277.18 / 1.51$\pm$0.05       &1336.91 / 1.33$\pm$0.03       &1400.80 / 1.12$\pm$0.03       &1460.04 / 1.03$\pm$0.04     &$-2.98\pm0.34$  \\             
J1831-0823       &245.90    &1278.75 / 1.18$\pm$0.03       &1336.91 / 1.11$\pm$0.02       &1400.29 / 1.01$\pm$0.02       &1459.75 / 0.93$\pm$0.03     &$-1.83\pm0.28$  \\             
J1831-1223       &342.00    &1278.75 / 1.50$\pm$0.05       &1336.91 / 1.37$\pm$0.04       &1400.29 / 1.18$\pm$0.03       &1459.32 / 1.10$\pm$0.04     &$-2.51\pm0.33$  \\             
J1831-1329       &338.00    &1276.91 / 0.78$\pm$0.03       &1336.91 / 0.67$\pm$0.02       &1400.20 / 0.60$\pm$0.02       &1460.04 / 0.49$\pm$0.02     &$-3.31\pm0.37$  \\             
J1832-1021       &475.70    &1278.75 / 2.44$\pm$0.05       &1336.75 / 2.09$\pm$0.04       &1400.29 / 1.83$\pm$0.04       &1459.52 / 1.66$\pm$0.04     &$-2.93\pm0.22$  \\             
J1833-0559       &353.00    &1277.39 / 1.02$\pm$0.08       &1337.07 / 1.04$\pm$0.05       &1400.04 / 0.94$\pm$0.05       &1460.04 / 0.87$\pm$0.06     &$-1.44\pm0.69$  \\             
J1833-1055       &543.00    &1275.09 / 1.75$\pm$0.10       &1337.07 / 1.51$\pm$0.06       &1400.04 / 1.42$\pm$0.06       &1460.04 / 1.27$\pm$0.07     &$-2.15\pm0.53$  \\             
J1834-0602       &445.00    &1278.75 / 1.14$\pm$0.05       &1336.75 / 0.97$\pm$0.04       &1400.29 / 0.96$\pm$0.03       &1459.32 / 0.85$\pm$0.04     &$-1.98\pm0.41$  \\             
J1834-1202       &342.40    &1277.15 / 0.90$\pm$0.05       &1337.07 / 0.88$\pm$0.03       &1400.04 / 0.84$\pm$0.03       &1461.24 / 0.81$\pm$0.05     &$-0.79\pm0.55$  \\             
J1834-1710       &123.80    &1278.75 / 0.80$\pm$0.03       &1337.08 / 0.78$\pm$0.02       &1400.29 / 0.73$\pm$0.02       &1459.32 / 0.71$\pm$0.03     &$-0.99\pm0.39$  \\             
J1834-1855       &185.20    &1276.86 / 0.49$\pm$0.03       &1336.91 / 0.46$\pm$0.02       &1400.20 / 0.39$\pm$0.02       &1460.04 / 0.37$\pm$0.02     &$-2.50\pm0.64$  \\             
J1835-0643       &472.90    &1277.82 / 3.27$\pm$0.15       &1336.75 / 2.92$\pm$0.12       &1400.45 / 2.66$\pm$0.10       &1459.61 / 2.52$\pm$0.12     &$-2.01\pm0.47$  \\   
\hline
\end{tabular}
\end{footnotesize}
\end{table}

\begin{table}[hpbt]\addtolength{\tabcolsep}{-4pt}
\center
\addtocounter{table}{-1}
\caption{--continue.}
\begin{footnotesize}
\renewcommand\arraystretch{1.2}
\begin{tabular}{llllllc}
\hline
Pulsar Name  &DM        &Frequency1  / $S_{1}$         &Frequency2  / $S_{2}$          &Frequency3  / $S_{3}$         &Frequency4  / $S_{4}$       &$\alpha$  \\
\hline     
J1835-0924   &471.00    &1277.07 / 0.46$\pm$0.05       &1336.75 / 0.42$\pm$0.04       &1400.45 / 0.41$\pm$0.03       &1459.54 / 0.39$\pm$0.04     &$-1.07\pm1.06$  \\             
J1835-0944   &277.20    &1276.74 / 0.72$\pm$0.06       &1336.91 / 0.58$\pm$0.04       &1400.04 / 0.55$\pm$0.03       &1460.04 / 0.54$\pm$0.04     &$-1.95\pm0.75$  \\             
J1835-1020   &113.70    &1278.75 / 2.24$\pm$0.04       &1336.91 / 2.08$\pm$0.03       &1400.29 / 1.96$\pm$0.03       &1459.32 / 1.91$\pm$0.03     &$-1.22\pm0.17$  \\             
J1835-1106   &132.68    &1278.75 / 2.81$\pm$0.04       &1336.75 / 2.40$\pm$0.03       &1400.29 / 2.14$\pm$0.03       &1459.52 / 2.05$\pm$0.03     &$-2.41\pm0.15$  \\             
J1836-0436   &231.50    &1278.75 / 2.43$\pm$0.04       &1337.08 / 2.21$\pm$0.03       &1400.29 / 2.09$\pm$0.03       &1459.96 / 1.86$\pm$0.04     &$-1.85\pm0.18$  \\             
J1836-1008   &316.98    &1276.45 / 6.02$\pm$0.08       &1337.07 / 5.18$\pm$0.05       &1400.04 / 4.40$\pm$0.05       &1460.26 / 4.11$\pm$0.06     &$-2.98\pm0.14$  \\             
J1837-0559   &317.80    &1342.75 / 0.67$\pm$0.05       &1400.12 / 0.63$\pm$0.04       &1465.36 / 0.57$\pm$0.04       &1518.87 / 0.58$\pm$0.05     &$-1.34\pm0.83$  \\             
J1837-0653   &316.10    &1278.75 / 4.41$\pm$0.07       &1336.91 / 3.96$\pm$0.05       &1400.29 / 3.76$\pm$0.05       &1459.52 / 3.54$\pm$0.06     &$-1.58\pm0.16$  \\             
J1837-1837   &100.74    &1278.21 / 0.44$\pm$0.03       &1336.91 / 0.42$\pm$0.02       &1400.20 / 0.34$\pm$0.02       &1460.04 / 0.33$\pm$0.02     &$-2.69\pm0.59$  \\             
J1838-1046   &208.00    &1277.18 / 0.53$\pm$0.03       &1336.91 / 0.44$\pm$0.02       &1400.41 / 0.41$\pm$0.01       &1460.04 / 0.31$\pm$0.02     &$-3.49\pm0.51$  \\             
J1839-0643   &497.90    &1278.75 / 2.08$\pm$0.08       &1336.91 / 1.87$\pm$0.06       &1400.29 / 1.79$\pm$0.06       &1459.96 / 1.59$\pm$0.07     &$-1.83\pm0.40$  \\             
J1840-0809   &349.80    &1278.75 / 3.07$\pm$0.03       &1336.75 / 2.74$\pm$0.03       &1400.29 / 2.60$\pm$0.02       &1459.52 / 2.45$\pm$0.03     &$-1.63\pm0.11$  \\             
J1840-0815   &233.20    &1278.75 / 2.39$\pm$0.03       &1336.91 / 2.06$\pm$0.02       &1400.29 / 1.87$\pm$0.02       &1459.52 / 1.71$\pm$0.02     &$-2.51\pm0.13$  \\             
J1840-0840   &272.00    &1276.49 / 2.16$\pm$0.06       &1336.91 / 1.99$\pm$0.04       &1400.04 / 1.72$\pm$0.03       &1460.04 / 1.61$\pm$0.04     &$-2.34\pm0.25$  \\             
J1841-0157   &475.00    &1278.75 / 1.97$\pm$0.04       &1336.75 / 1.75$\pm$0.03       &1400.29 / 1.66$\pm$0.03       &1459.32 / 1.65$\pm$0.03     &$-1.30\pm0.21$  \\             
J1842-0153   &434.00    &1278.75 / 1.07$\pm$0.05       &1336.75 / 0.91$\pm$0.04       &1400.29 / 0.89$\pm$0.04       &1459.52 / 0.83$\pm$0.04     &$-1.77\pm0.46$  \\             
J1842-0415   &188.00    &1277.49 / 0.40$\pm$0.04       &1336.91 / 0.40$\pm$0.03       &1400.04 / 0.37$\pm$0.02       &1460.04 / 0.37$\pm$0.03     &$-0.80\pm0.89$  \\             
J1842-0905   &343.30    &1278.75 / 1.18$\pm$0.04       &1336.75 / 1.06$\pm$0.03       &1400.29 / 0.97$\pm$0.02       &1459.32 / 0.87$\pm$0.03     &$-2.22\pm0.30$  \\             
J1843-0000   &101.50    &1278.75 / 4.55$\pm$0.05       &1337.08 / 4.19$\pm$0.04       &1400.29 / 3.92$\pm$0.04       &1459.32 / 3.69$\pm$0.04     &$-1.57\pm0.11$  \\             
J1843-0211   &441.70    &1278.75 / 1.62$\pm$0.05       &1336.91 / 1.46$\pm$0.04       &1400.29 / 1.34$\pm$0.03       &1459.52 / 1.20$\pm$0.04     &$-2.18\pm0.30$  \\             
J1843-0459   &444.10    &1278.75 / 2.38$\pm$0.05       &1336.75 / 2.16$\pm$0.04       &1400.29 / 1.80$\pm$0.04       &1459.52 / 1.72$\pm$0.05     &$-2.71\pm0.24$  \\             
J1843-0806   &215.80    &1277.34 / 0.57$\pm$0.04       &1336.91 / 0.48$\pm$0.02       &1400.04 / 0.42$\pm$0.02       &1460.04 / 0.39$\pm$0.03     &$-2.75\pm0.65$  \\             
J1844-0030   &605.00    &1276.90 / 0.55$\pm$0.05       &1336.91 / 0.48$\pm$0.03       &1400.04 / 0.43$\pm$0.02       &1460.04 / 0.38$\pm$0.03     &$-2.81\pm0.75$  \\             
J1844-0244   &429.00    &1278.75 / 1.50$\pm$0.06       &1336.75 / 1.28$\pm$0.05       &1400.29 / 1.32$\pm$0.05       &1459.32 / 1.09$\pm$0.05     &$-1.89\pm0.44$  \\             
J1844-0310   &836.10    &1277.11 / 1.09$\pm$0.11       &1337.07 / 0.84$\pm$0.07       &1400.04 / 0.82$\pm$0.06       &1460.04 / 0.79$\pm$0.08     &$-2.14\pm0.98$  \\             
J1845-0316   &500.00    &1277.04 / 0.54$\pm$0.08       &1336.75 / 0.45$\pm$0.05       &1400.45 / 0.44$\pm$0.04       &1459.61 / 0.35$\pm$0.05     &$-2.66\pm1.33$  \\             
J1845-0434   &230.80    &1277.39 / 2.67$\pm$0.07       &1337.07 / 2.57$\pm$0.05       &1400.04 / 2.34$\pm$0.04       &1460.04 / 2.32$\pm$0.05     &$-1.18\pm0.24$  \\             
J1845-0743   &281.00    &1278.75 / 4.03$\pm$0.04       &1336.75 / 3.66$\pm$0.04       &1400.29 / 3.44$\pm$0.03       &1459.52 / 3.15$\pm$0.04     &$-1.79\pm0.11$  \\             
J1846+0051   &140.00    &1275.45 / 0.36$\pm$0.04       &1336.91 / 0.30$\pm$0.03       &1400.04 / 0.26$\pm$0.02       &1460.04 / 0.21$\pm$0.03     &$-3.75\pm1.12$  \\             
J1847-0438   &229.00    &1276.31 / 0.68$\pm$0.03       &1337.07 / 0.62$\pm$0.02       &1400.04 / 0.55$\pm$0.02       &1460.26 / 0.54$\pm$0.02     &$-1.87\pm0.37$  \\             
J1847-0605   &207.90    &1278.75 / 1.27$\pm$0.05       &1336.75 / 1.10$\pm$0.04       &1400.29 / 1.01$\pm$0.04       &1459.52 / 0.91$\pm$0.04     &$-2.50\pm0.40$  \\             
J1848-1414   &134.47    &1277.56 / 0.70$\pm$0.04       &1336.75 / 0.61$\pm$0.03       &1400.60 / 0.55$\pm$0.02       &1459.61 / 0.48$\pm$0.03     &$-2.66\pm0.57$  \\             
J1849-0614   &119.60    &1274.49 / 0.86$\pm$0.04       &1337.32 / 0.75$\pm$0.02       &1400.04 / 0.64$\pm$0.02       &1460.26 / 0.60$\pm$0.02     &$-2.76\pm0.39$  \\             
J1850+0026   &201.40    &1278.75 / 1.81$\pm$0.05       &1336.75 / 1.67$\pm$0.04       &1400.29 / 1.51$\pm$0.03       &1459.32 / 1.35$\pm$0.04     &$-2.21\pm0.26$  \\             
J1851+0418   &115.54    &1278.20 / 2.43$\pm$0.08       &1336.91 / 2.27$\pm$0.05       &1400.03 / 1.96$\pm$0.05       &1460.04 / 1.89$\pm$0.05     &$-2.09\pm0.29$  \\             
J1852+0031   &787.00    &1278.50 / 6.11$\pm$0.12       &1336.92 / 5.72$\pm$0.10       &1400.45 / 5.60$\pm$0.09       &1459.32 / 5.18$\pm$0.10     &$-1.12\pm0.19$  \\             
J1852+0305   &320.00    &1276.95 / 0.27$\pm$0.04       &1337.07 / 0.22$\pm$0.02       &1400.04 / 0.18$\pm$0.02       &1460.04 / 0.18$\pm$0.03     &$-3.27\pm1.41$  \\             
J1852-0635   &171.00    &1278.75 / 13.44$\pm$0.06      &1337.08 / 12.52$\pm$0.05      &1400.29 / 11.97$\pm$0.05      &1459.52 / 11.6$\pm$0.05     &$-1.09\pm0.05$  \\             
J1853-0004   &438.20    &1278.75 / 1.37$\pm$0.07       &1336.75 / 0.99$\pm$0.06       &1400.29 / 0.98$\pm$0.05       &1459.52 / 0.75$\pm$0.07     &$-4.01\pm0.65$  \\             
J1853+0545   &198.70    &1278.75 / 3.18$\pm$0.06       &1337.08 / 2.95$\pm$0.05       &1400.29 / 2.82$\pm$0.05       &1459.96 / 2.78$\pm$0.05     &$-1.02\pm0.19$  \\             
J1855+0307   &402.50    &1278.75 / 0.75$\pm$0.03       &1336.99 / 0.65$\pm$0.02       &1400.29 / 0.62$\pm$0.02       &1459.52 / 0.57$\pm$0.03     &$-1.93\pm0.42$  \\             
J1855-0941   &151.99    &1278.27 / 1.48$\pm$0.06       &1336.91 / 1.24$\pm$0.04       &1400.20 / 1.01$\pm$0.04       &1460.04 / 0.89$\pm$0.04     &$-3.94\pm0.41$  \\             
J1856+0404   &341.30    &1278.01 / 0.49$\pm$0.05       &1336.91 / 0.42$\pm$0.03       &1399.87 / 0.38$\pm$0.03       &1460.04 / 0.33$\pm$0.04     &$-2.61\pm1.04$  \\             
J1900-0051   &136.80    &1278.16 / 0.56$\pm$0.04       &1336.75 / 0.46$\pm$0.03       &1400.60 / 0.43$\pm$0.02       &1459.61 / 0.39$\pm$0.03     &$-2.63\pm0.70$  \\             
J1901+0254   &185.00    &1277.33 / 1.22$\pm$0.06       &1337.07 / 1.15$\pm$0.04       &1400.04 / 1.07$\pm$0.04       &1460.04 / 1.00$\pm$0.05     &$-1.50\pm0.45$  \\             
J1901+0413   &352.00    &1278.75 / 1.21$\pm$0.06       &1337.08 / 1.21$\pm$0.04       &1400.29 / 1.05$\pm$0.03       &1459.72 / 0.98$\pm$0.04     &$-1.94\pm0.44$  \\             
J1902+0615   &502.90    &1278.75 / 2.22$\pm$0.03       &1336.91 / 1.82$\pm$0.02       &1400.29 / 1.62$\pm$0.02       &1459.32 / 1.43$\pm$0.02     &$-3.27\pm0.15$  \\             
J1904+0800   &438.80    &1278.31 / 0.50$\pm$0.05       &1336.91 / 0.47$\pm$0.03       &1400.03 / 0.47$\pm$0.03       &1460.04 / 0.45$\pm$0.04     &$-0.61\pm0.83$  \\             
J1904+1011   &135.00    &1277.44 / 0.54$\pm$0.04       &1336.75 / 0.55$\pm$0.03       &1400.45 / 0.47$\pm$0.03       &1459.61 / 0.46$\pm$0.03     &$-1.65\pm0.69$  \\             
J1904-1224   &118.23    &1277.53 / 0.43$\pm$0.03       &1336.91 / 0.35$\pm$0.02       &1400.20 / 0.29$\pm$0.01       &1460.04 / 0.25$\pm$0.02     &$-4.09\pm0.63$  \\             
J1905-0056   &229.13    &1278.75 / 0.95$\pm$0.03       &1336.75 / 0.83$\pm$0.02       &1400.29 / 0.71$\pm$0.02       &1459.52 / 0.69$\pm$0.02     &$-2.64\pm0.30$  \\ 
\hline
\end{tabular}
\end{footnotesize}
\end{table}

\begin{table}\addtolength{\tabcolsep}{-4pt}
\center
\addtocounter{table}{-1}
\caption{--continue.}
\begin{footnotesize}
\renewcommand\arraystretch{1.2}
\begin{tabular}{llllllc}
\hline
Pulsar Name  &DM    &Frequency1 / $S_{1}$   &Frequency2  / $S_{2}$  &Frequency3  / $S_{3}$  &Frequency4  / $S_{4}$      &$\alpha$  \\
\hline    
J1905+0600   &730.10    &1276.98 / 0.64$\pm$0.06       &1336.91 / 0.53$\pm$0.04       &1400.04 / 0.42$\pm$0.03       &1460.04 / 0.43$\pm$0.04     &$-3.40\pm0.84$  \\             
J1905+0616   &256.05    &1275.37 / 0.61$\pm$0.04       &1337.07 / 0.52$\pm$0.03       &1400.04 / 0.42$\pm$0.02       &1460.04 / 0.42$\pm$0.03     &$-3.17\pm0.62$  \\             
J1905+0709   &245.34    &1278.50 / 2.44$\pm$0.05       &1336.75 / 2.04$\pm$0.04       &1400.45 / 1.84$\pm$0.03       &1459.11 / 1.75$\pm$0.04     &$-2.58\pm0.20$  \\             
J1906+0641   &472.80    &1278.75 / 3.63$\pm$0.06       &1336.75 / 3.35$\pm$0.05       &1400.29 / 3.26$\pm$0.04       &1459.32 / 3.13$\pm$0.05     &$-1.03\pm0.16$  \\             
J1907+0249   &261.00    &1276.88 / 0.56$\pm$0.05       &1336.91 / 0.46$\pm$0.03       &1400.04 / 0.44$\pm$0.03       &1460.04 / 0.41$\pm$0.04     &$-2.25\pm0.80$  \\             
J1907+0740   &332.00    &1276.80 / 0.77$\pm$0.04       &1336.75 / 0.62$\pm$0.03       &1400.45 / 0.63$\pm$0.02       &1459.61 / 0.57$\pm$0.03     &$-1.79\pm0.49$  \\             
J1908+0457   &360.00    &1278.75 / 1.33$\pm$0.04       &1336.75 / 1.15$\pm$0.03       &1400.29 / 1.06$\pm$0.02       &1459.32 / 0.89$\pm$0.03     &$-2.80\pm0.29$  \\             
J1908+0500   &201.42    &1278.75 / 1.63$\pm$0.04       &1336.75 / 1.37$\pm$0.03       &1400.29 / 1.29$\pm$0.03       &1459.32 / 1.17$\pm$0.03     &$-2.37\pm0.24$  \\             
J1909+0254   &171.73    &1278.75 / 1.20$\pm$0.03       &1336.75 / 1.07$\pm$0.02       &1400.29 / 1.00$\pm$0.02       &1459.32 / 0.86$\pm$0.02     &$-2.36\pm0.25$  \\             
J1910+0225   &209.00    &1277.39 / 1.14$\pm$0.05       &1337.07 / 1.03$\pm$0.03       &1400.04 / 0.85$\pm$0.03       &1460.04 / 0.79$\pm$0.03     &$-2.99\pm0.38$  \\             
J1910+0714   &124.06    &1277.40 / 0.53$\pm$0.04       &1336.91 / 0.46$\pm$0.02       &1399.87 / 0.42$\pm$0.02       &1460.04 / 0.37$\pm$0.03     &$-2.53\pm0.66$  \\             
J1913+0446   &109.10    &1342.75 / 0.99$\pm$0.03       &1400.04 / 0.91$\pm$0.03       &1465.60 / 0.86$\pm$0.03       &1518.49 / 0.76$\pm$0.04     &$-1.84\pm0.41$  \\             
J1913+1000   &422.00    &1275.66 / 0.71$\pm$0.04       &1337.07 / 0.65$\pm$0.02       &1400.04 / 0.64$\pm$0.02       &1460.04 / 0.56$\pm$0.03     &$-1.53\pm0.47$  \\             
J1913+1011   &178.80    &1342.75 / 0.59$\pm$0.04       &1400.18 / 0.52$\pm$0.04       &1465.60 / 0.50$\pm$0.03       &1516.38 / 0.42$\pm$0.04     &$-2.20\pm0.87$  \\             
J1913+1145   &637.00    &1275.55 / 0.63$\pm$0.05       &1336.91 / 0.55$\pm$0.03       &1399.87 / 0.53$\pm$0.03       &1460.04 / 0.48$\pm$0.03     &$-1.77\pm0.68$  \\             
J1914+0219   &233.80    &1276.45 / 1.74$\pm$0.04       &1337.61 / 1.57$\pm$0.03       &1400.04 / 1.53$\pm$0.03       &1460.04 / 1.47$\pm$0.03     &$-1.14\pm0.22$  \\             
J1915+0227   &192.60    &1277.24 / 0.58$\pm$0.04       &1337.07 / 0.40$\pm$0.03       &1400.04 / 0.40$\pm$0.02       &1460.04 / 0.36$\pm$0.03     &$-3.18\pm0.76$  \\             
J1915+1606   &168.77    &1277.11 / 1.26$\pm$0.05       &1337.07 / 1.00$\pm$0.03       &1400.04 / 0.85$\pm$0.03       &1460.04 / 0.83$\pm$0.03     &$-3.23\pm0.39$  \\             
J1916+0844   &339.40    &1342.75 / 0.63$\pm$0.03       &1399.76 / 0.61$\pm$0.02       &1465.60 / 0.57$\pm$0.02       &1513.92 / 0.54$\pm$0.03     &$-1.28\pm0.51$  \\             
J1916+1023   &329.80    &1277.71 / 0.93$\pm$0.09       &1336.91 / 0.80$\pm$0.06       &1399.87 / 0.76$\pm$0.05       &1460.04 / 0.74$\pm$0.06     &$-1.62\pm0.86$  \\             
J1917+2224   &134.93    &1277.25 / 0.45$\pm$0.04       &1337.24 / 0.42$\pm$0.03       &1400.03 / 0.36$\pm$0.03       &1460.04 / 0.35$\pm$0.03     &$-2.22\pm0.80$  \\             
J1920+1040   &304.00    &1277.23 / 1.28$\pm$0.10       &1336.91 / 1.12$\pm$0.06       &1400.04 / 0.96$\pm$0.06       &1460.04 / 0.82$\pm$0.07     &$-3.37\pm0.79$  \\             
J1927+1856   & 99.00    &1276.60 / 0.72$\pm$0.05       &1336.91 / 0.56$\pm$0.03       &1399.87 / 0.55$\pm$0.03       &1460.04 / 0.49$\pm$0.04     &$-2.52\pm0.70$  \\             
J1935+1745   &214.60    &1277.12 / 0.20$\pm$0.02       &1336.91 / 0.18$\pm$0.01       &1400.04 / 0.16$\pm$0.01       &1460.04 / 0.13$\pm$0.02     &$-3.04\pm1.12$  \\ 
\hline
\end{tabular}
\end{footnotesize}
\end{table}


In this paper, we present new measurements for flux densities of 228
pulsars at the 4 sub-bands within a total bandwidth of 256~MHz around
1.4~GHz observed with the Parkes 64~m telescope, from which spectral
indices are determined. In Section 2, we describe our observations and
data reduction procedures. In Section 3, our results are compared with
data in literatures, and the spectral indices are correlated with
pulsar characteristic parameters to seek any hint for emission
process. The conclusions are given in Section 4.

\section{Observations and data reduction}

About 500 pulsars have been observed with the Parkes telescope between
2006 August and 2008 February. Data were taken by using the central
beam of the multibeam receiver at a central frequency of 1369 MHz and
a bandwidth of 256 MHz, with one session in 2007 March exceptionally
in which the H-OH receiver was used to observe at a central frequency
of 1433 MHz.  These observations are all well-calibrated with the on
and off observations of the strong radio source Hydra A, which has a
flux density of 43 Jy at 1400 MHz. All data are on-line folded to
produce mean pulse profiles in every 1-minute sub-integrations for 512
frequency channels.

In the offline data reduction, we cleaned the radio frequency
interference in channal-time date-cube by removing some bad
channels and some strong term interference in the wide band. We then
obtained the mean pulsar profiles in four sub-bands by summing data of
all sub-intergrations and two polarizations, and measured the pulsar
mean flux densities at these four sub-bands, as listed in
Table~\ref{tab2}. The uncertainty of the mean flux density was
estimated from the standard deviation of off-pulse window and the
sampling number of the on-pulse window.

\begin{figure}[hpt!]
  \center
\includegraphics[bb = 110 72 518 685,clip,angle=-90,width=0.175\textwidth]{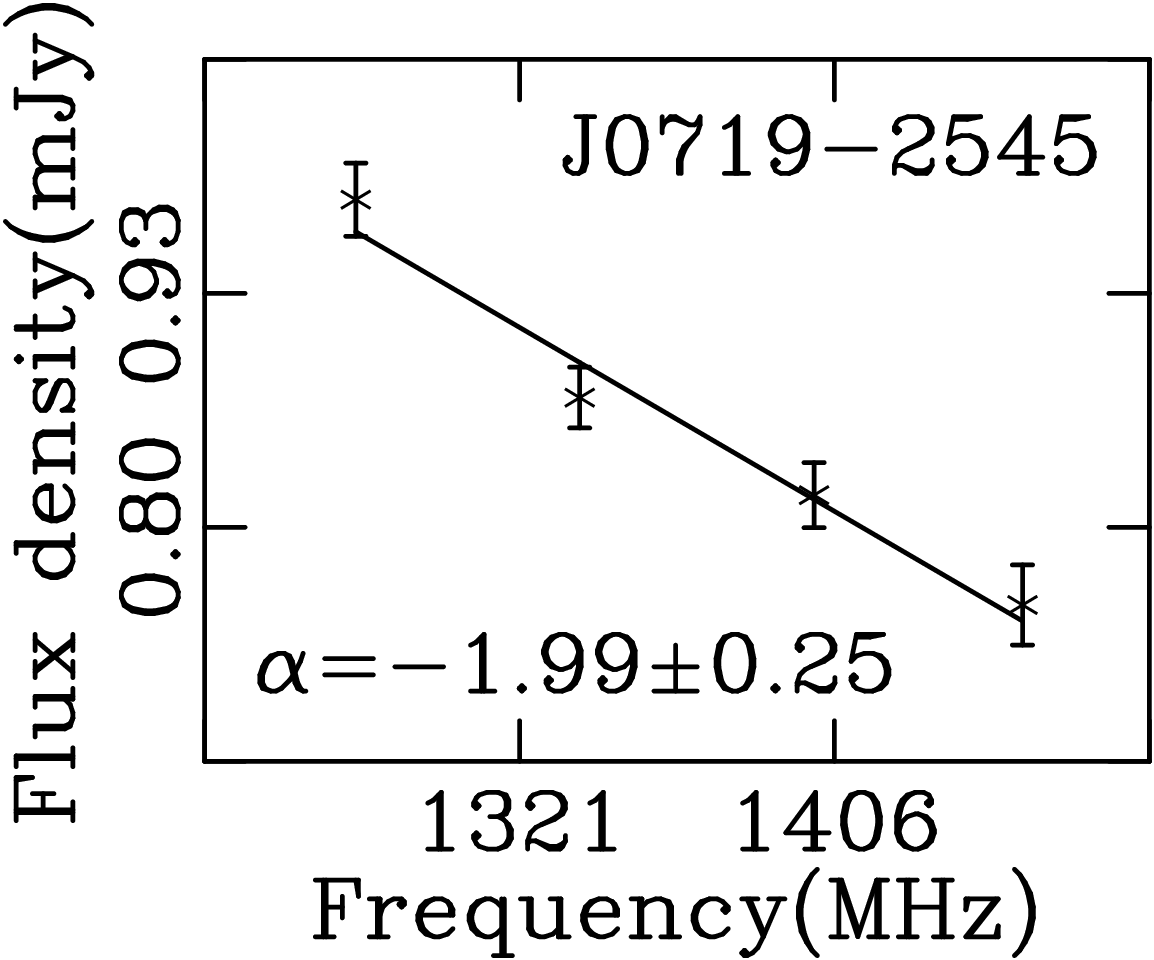}           
\includegraphics[bb = 110 120 518 685,clip,angle=-90,width=0.16\textwidth]{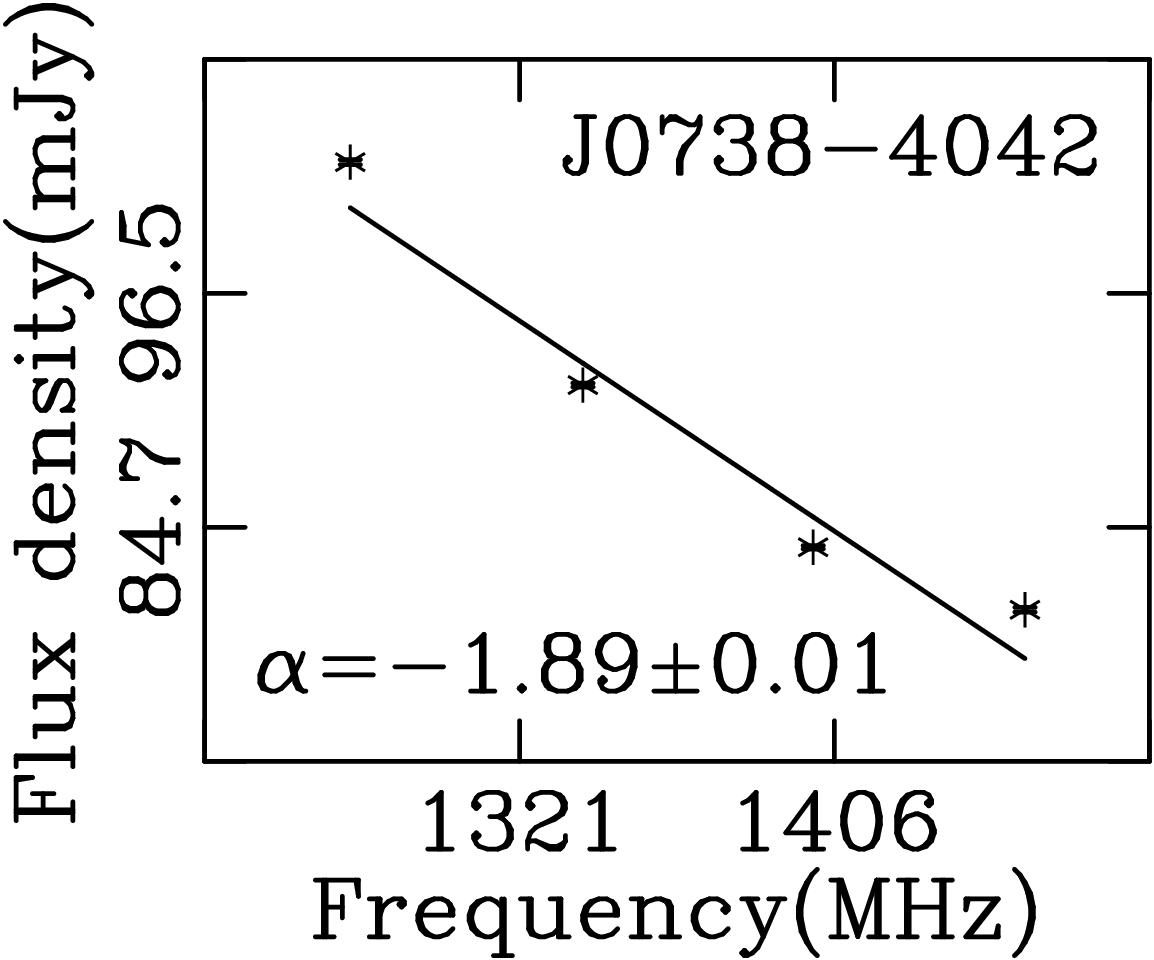}               
\includegraphics[bb = 110 120 518 685,clip,angle=-90,width=0.16\textwidth]{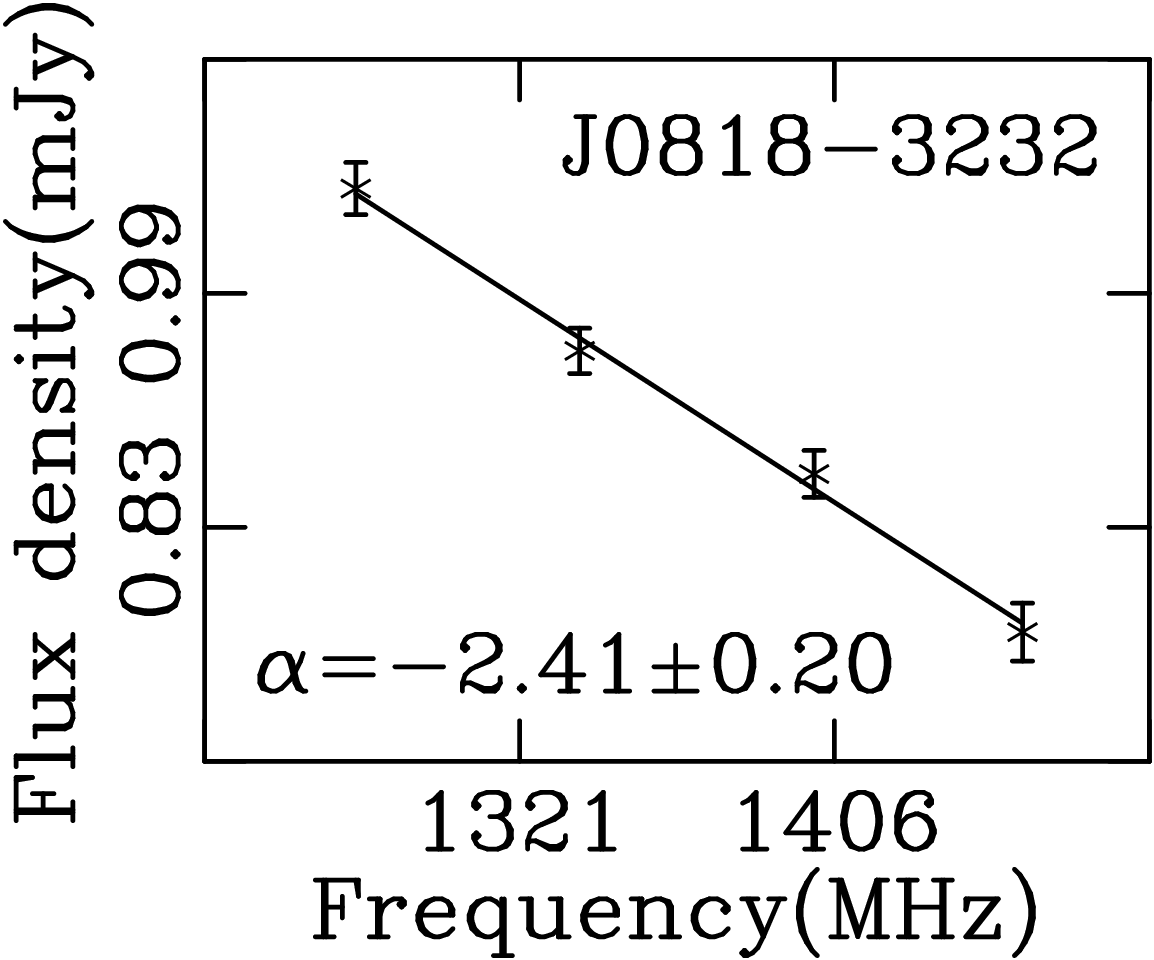}               
\includegraphics[bb = 110 120 518 685,clip,angle=-90,width=0.16\textwidth]{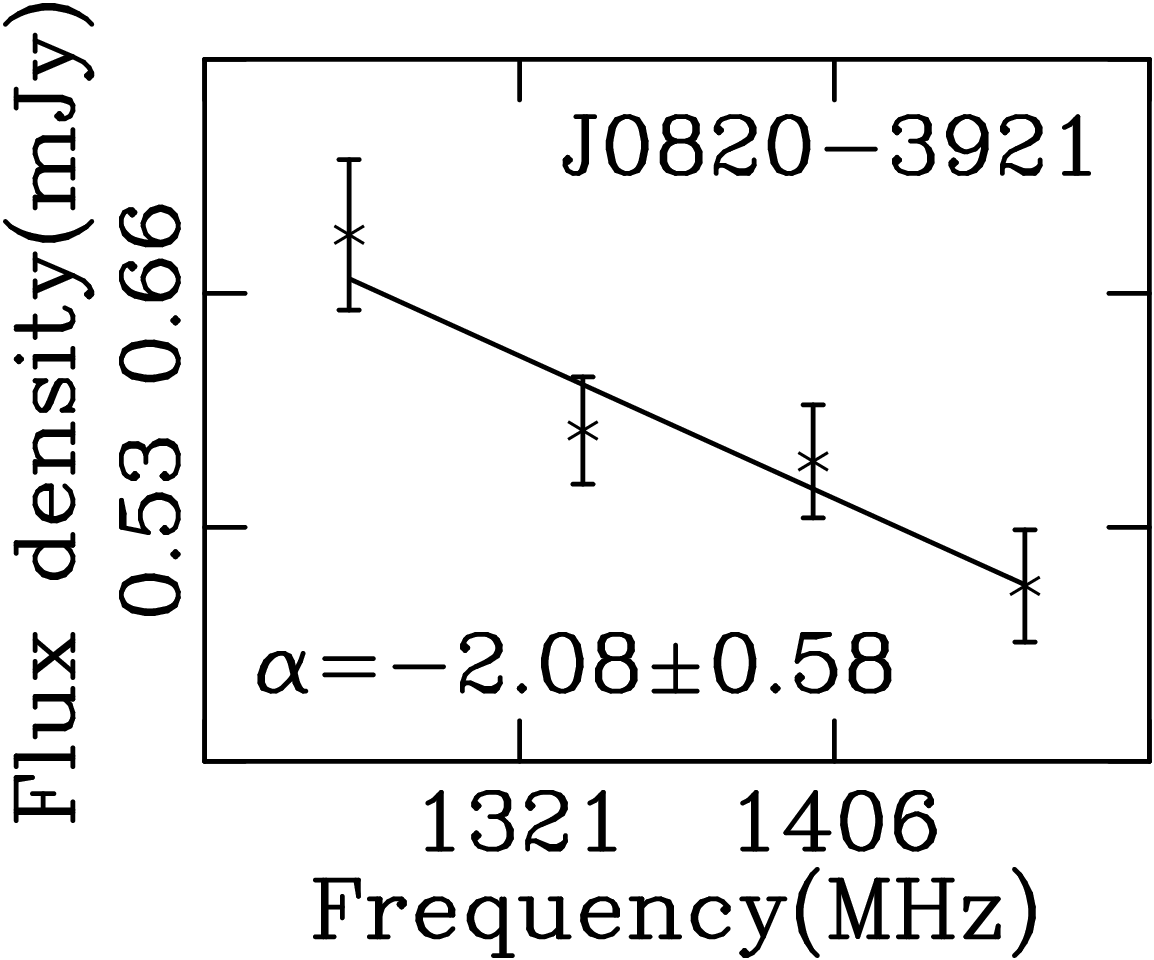}               
\includegraphics[bb = 110 120 518 685,clip,angle=-90,width=0.16\textwidth]{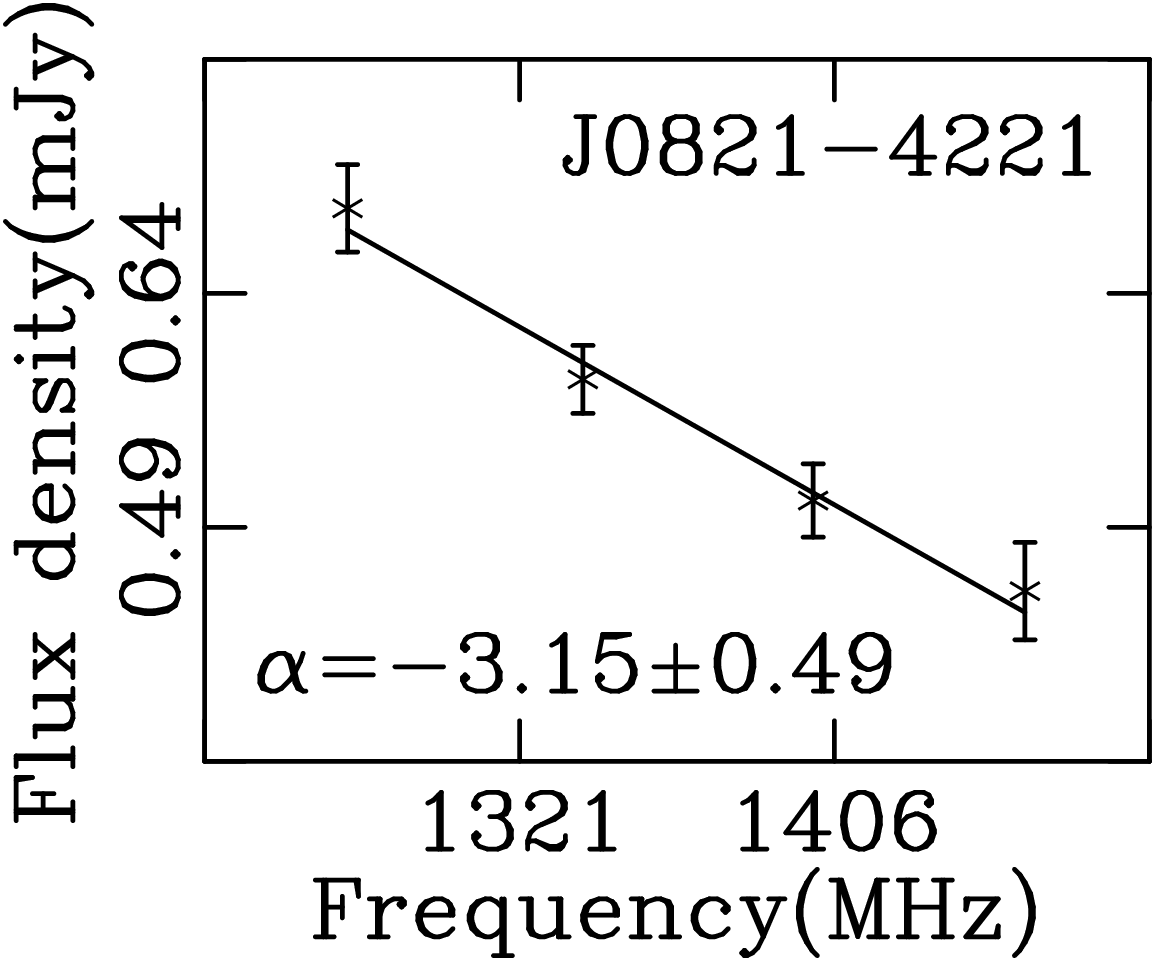}               
\includegraphics[bb = 110 120 518 685,clip,angle=-90,width=0.16\textwidth]{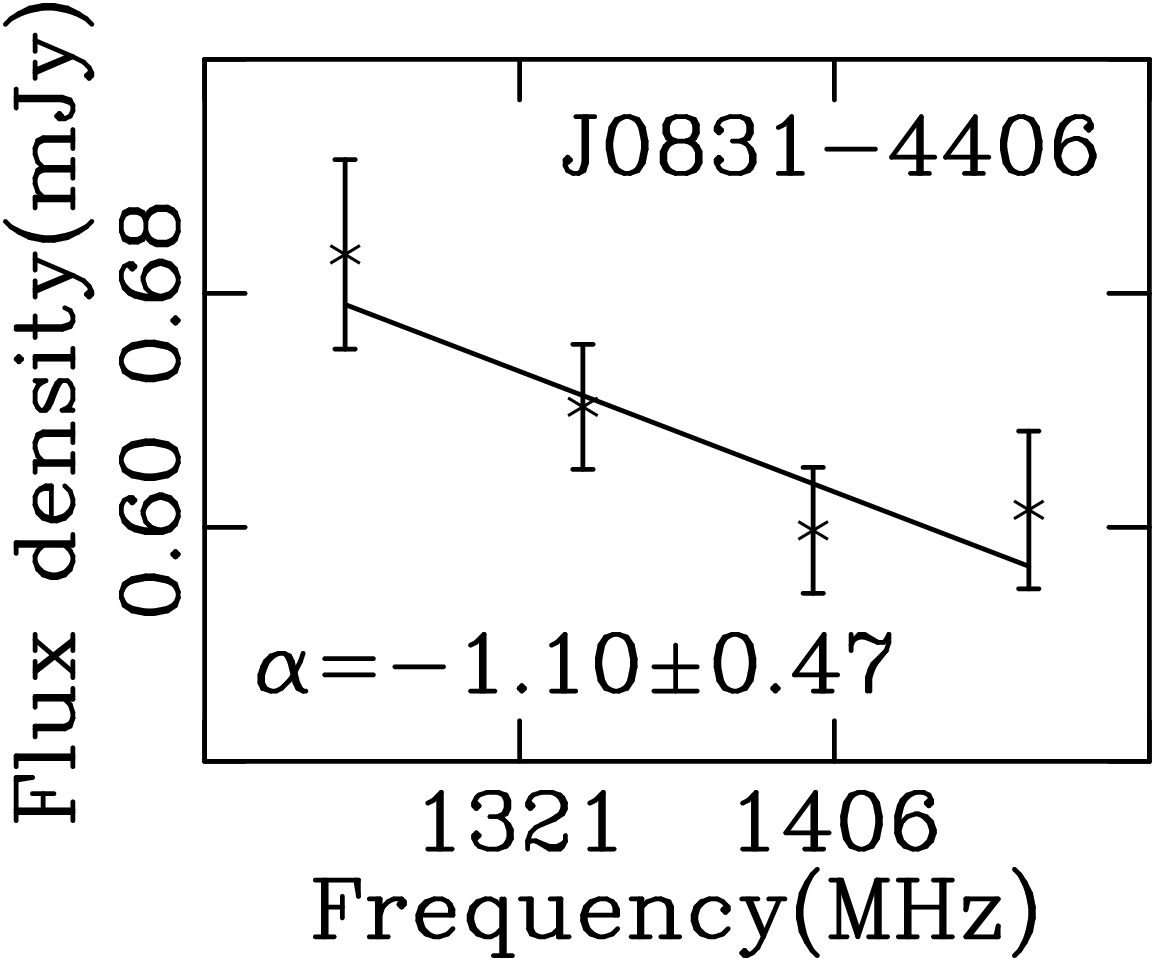}
 \\[2mm]              
\includegraphics[bb = 110 72 518 685,clip,angle=-90,width=0.175\textwidth]{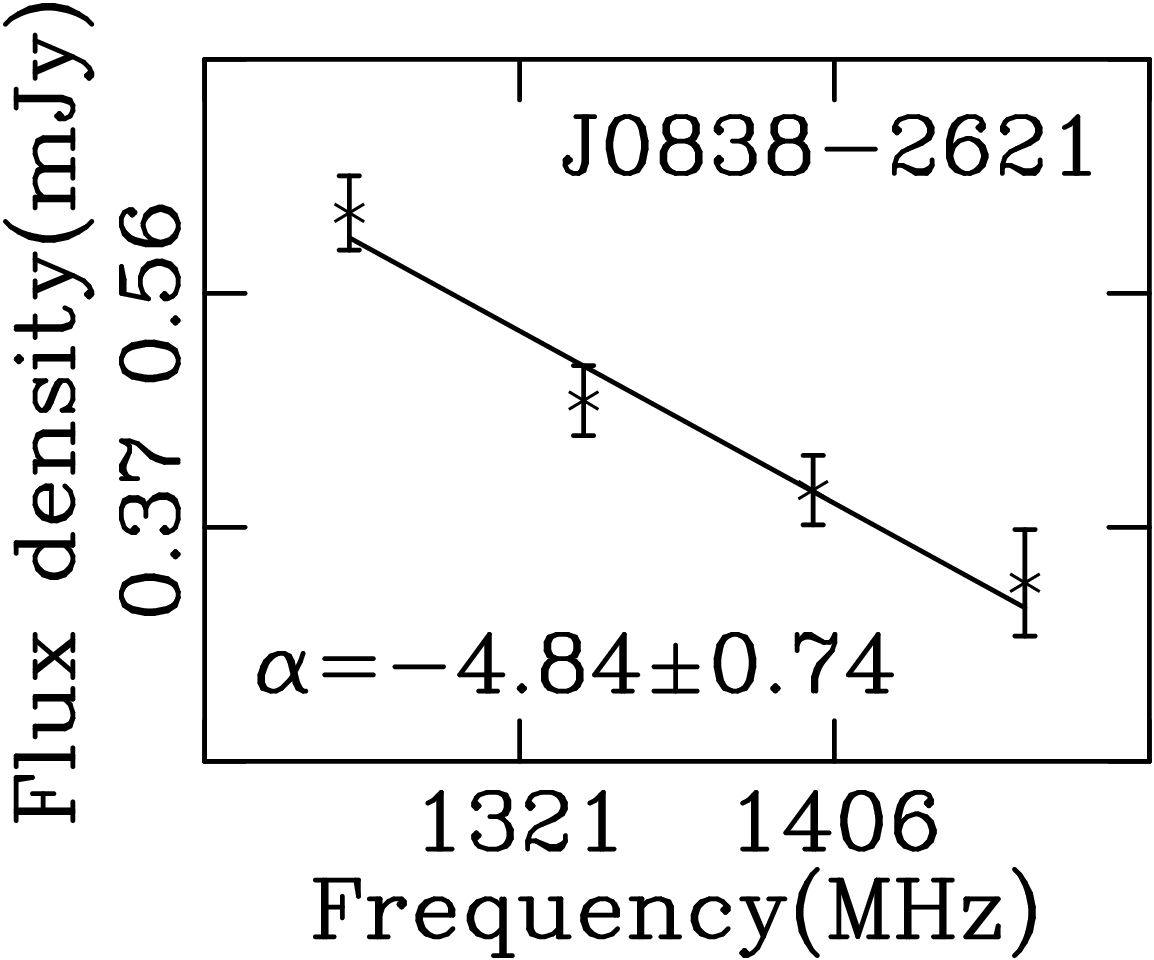}               
\includegraphics[bb = 110 120 518 685,clip,angle=-90,width=0.16\textwidth]{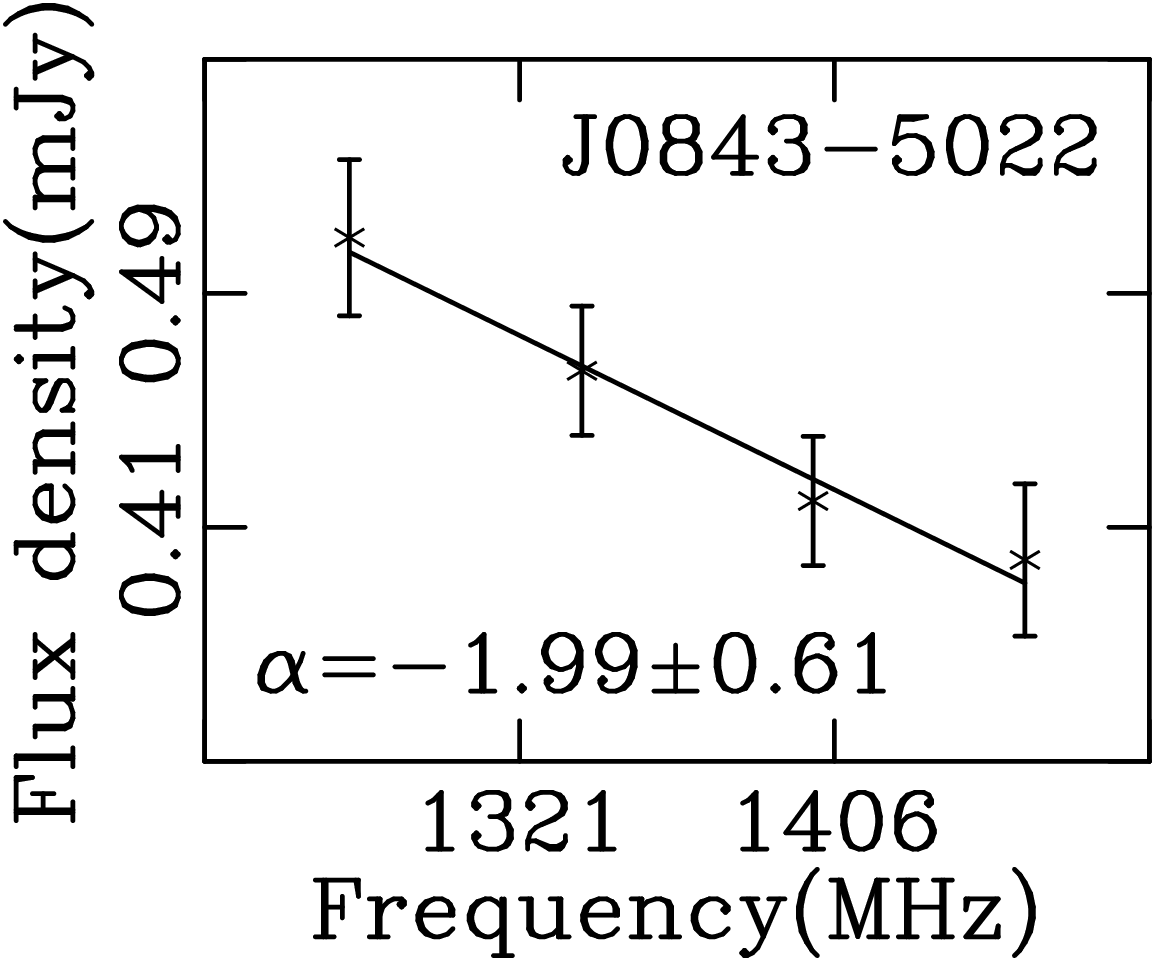}               
\includegraphics[bb = 110 120 518 685,clip,angle=-90,width=0.16\textwidth]{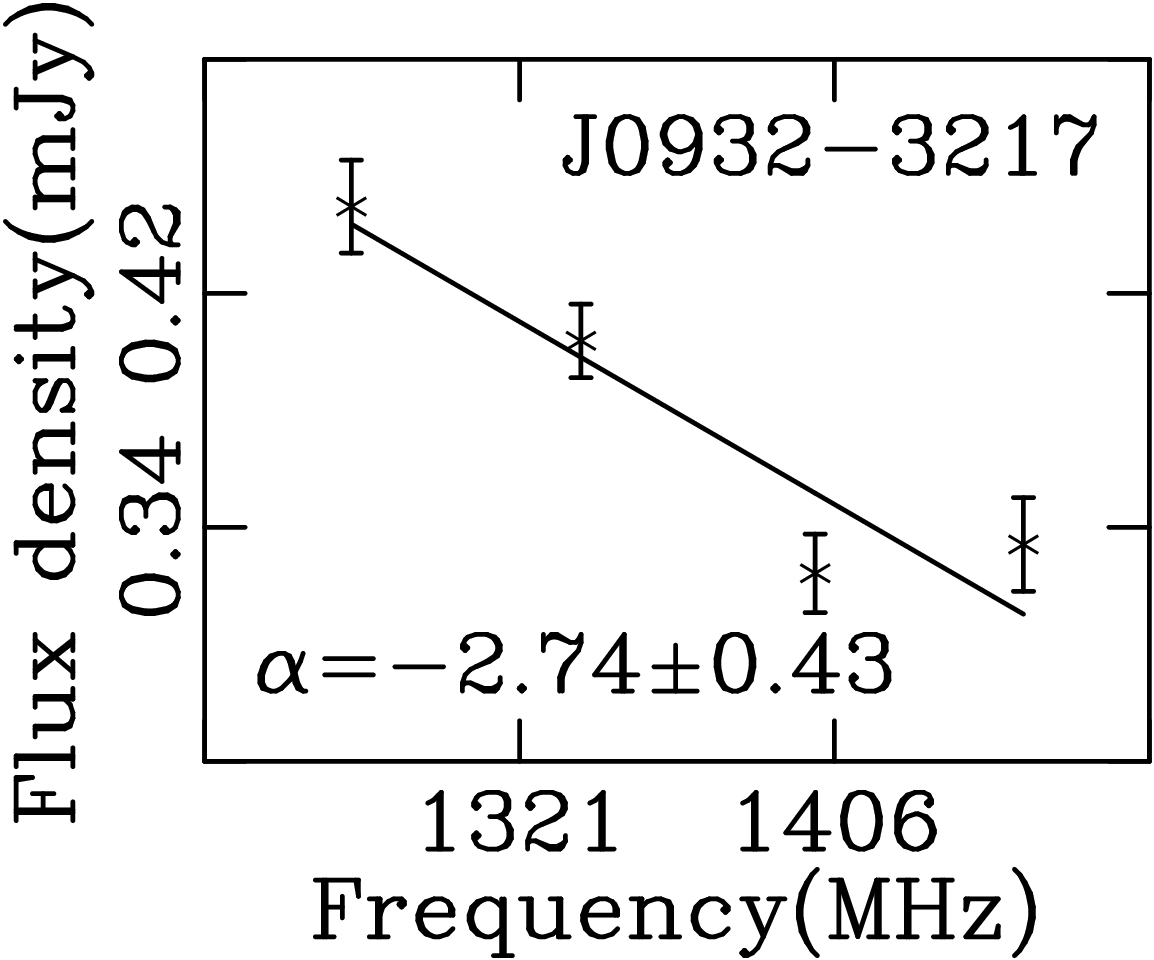}               
\includegraphics[bb = 110 120 518 685,clip,angle=-90,width=0.16\textwidth]{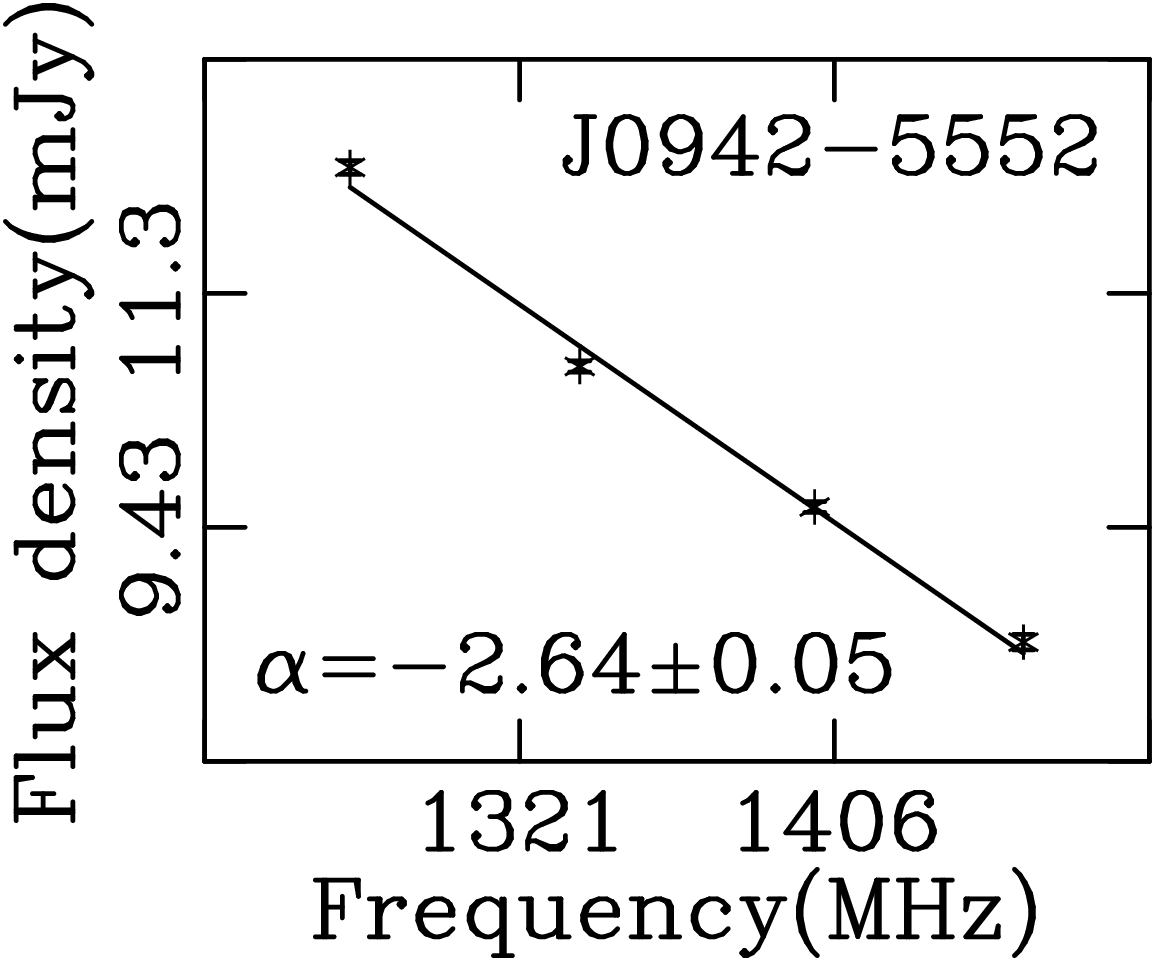}               
\includegraphics[bb = 110 120 518 685,clip,angle=-90,width=0.16\textwidth]{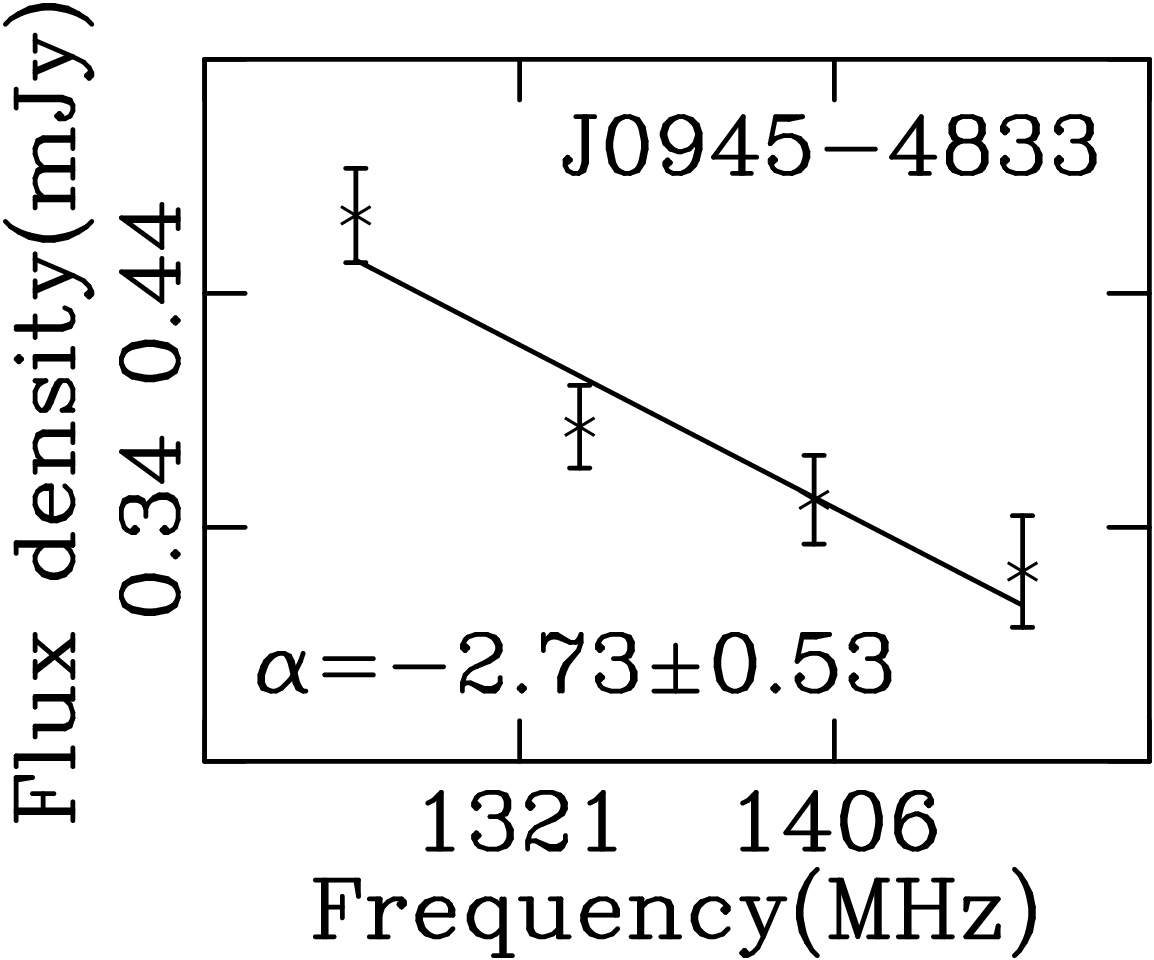}               
\includegraphics[bb = 110 120 518 685,clip,angle=-90,width=0.16\textwidth]{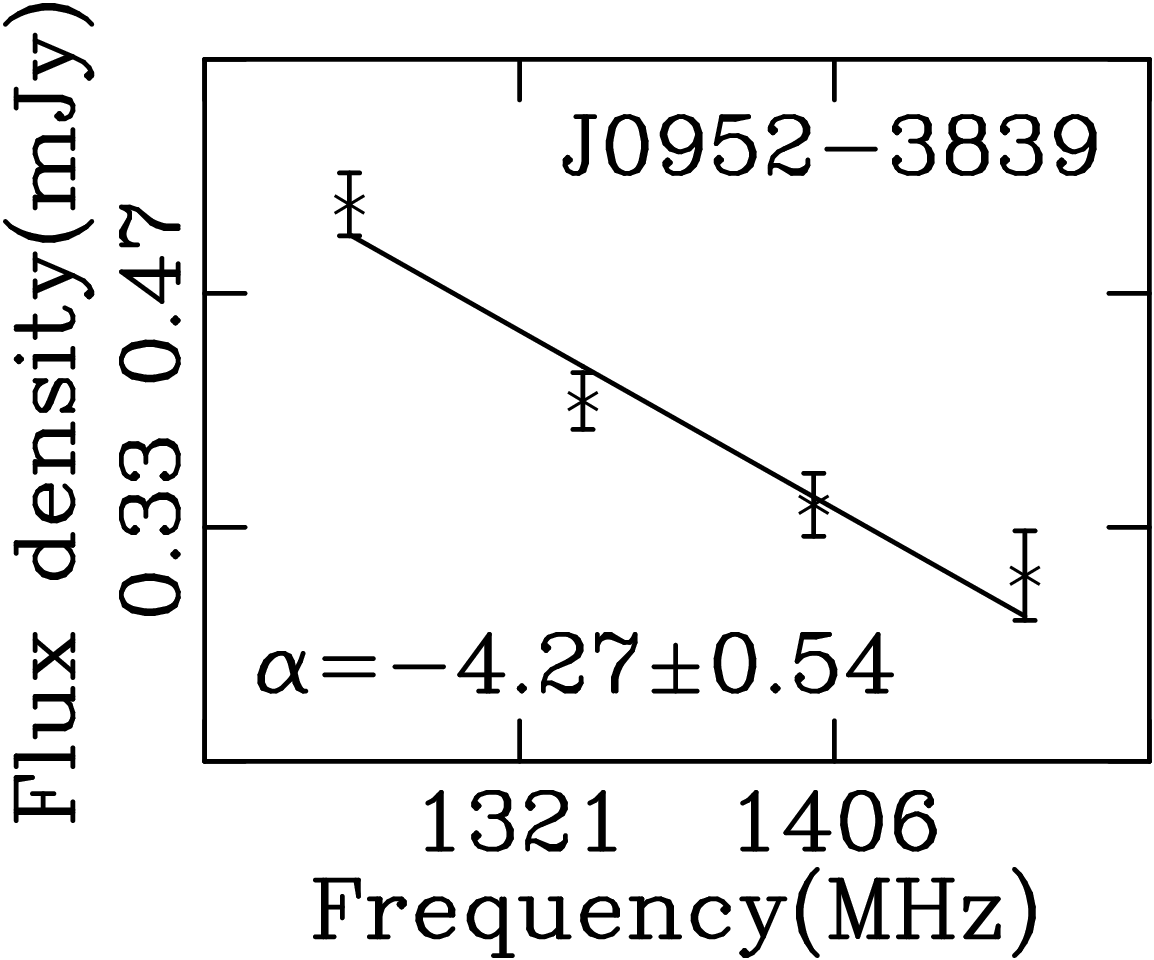}
\\[2mm]               
\includegraphics[bb = 110 72 518 685,clip,angle=-90,width=0.175\textwidth]{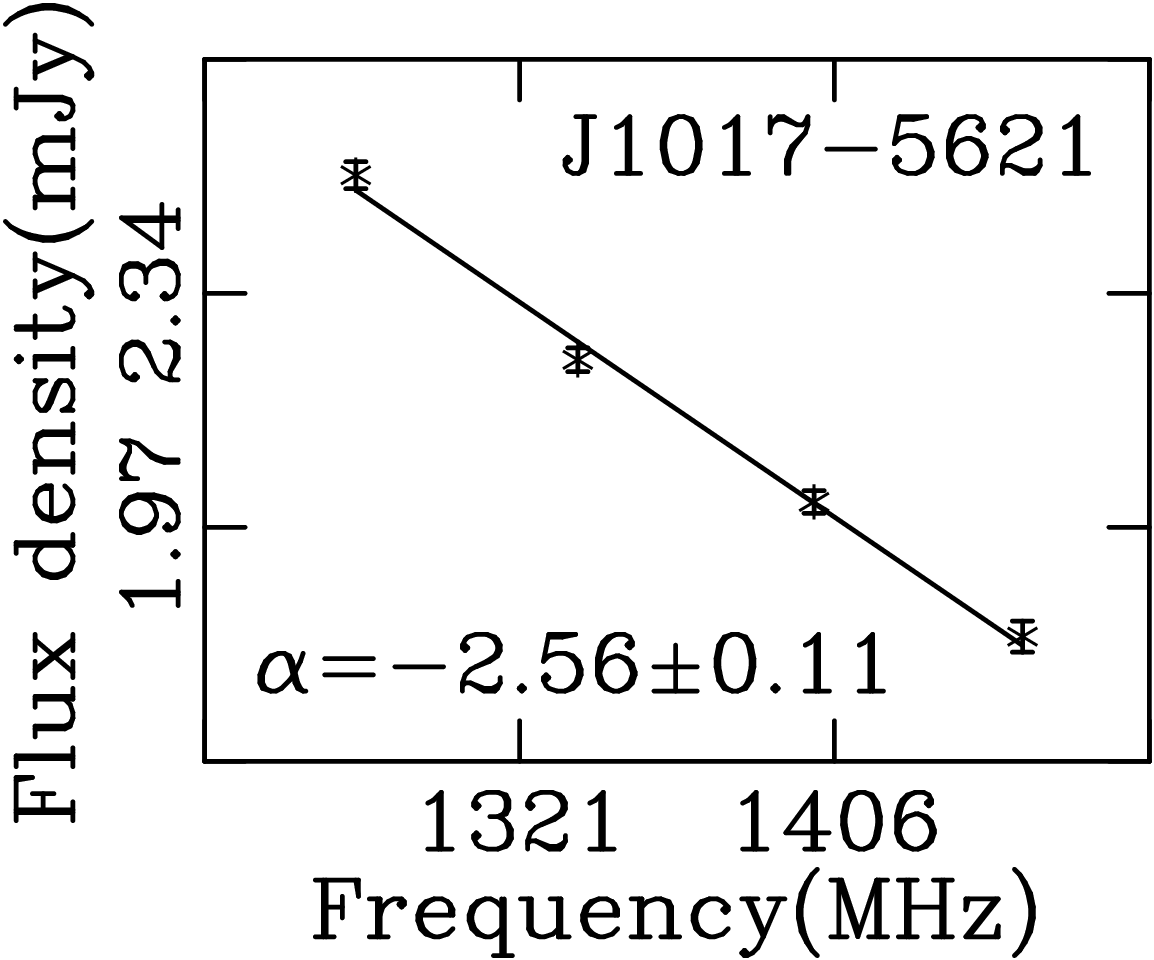}               
\includegraphics[bb = 110 120 518 685,clip,angle=-90,width=0.16\textwidth]{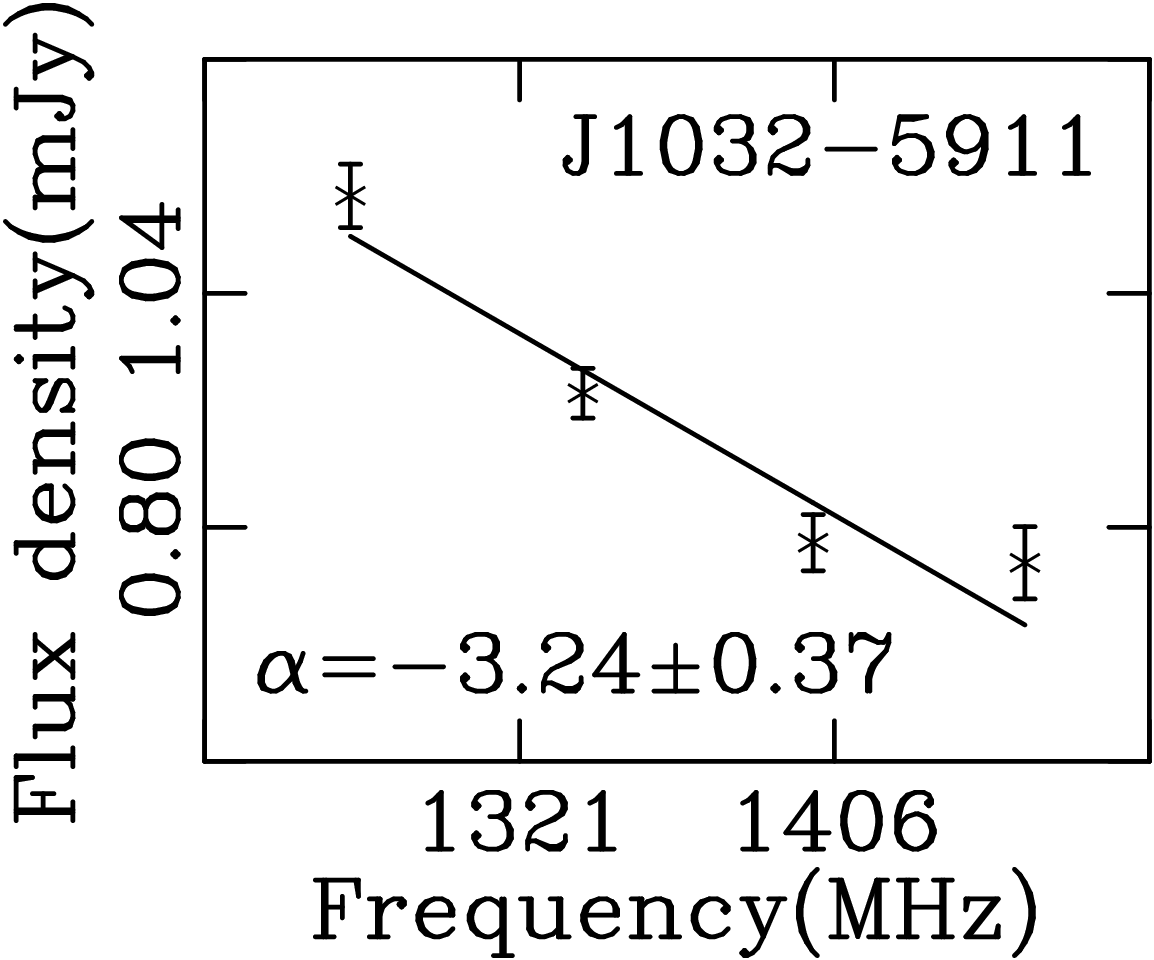}               
\includegraphics[bb = 110 120 518 685,clip,angle=-90,width=0.16\textwidth]{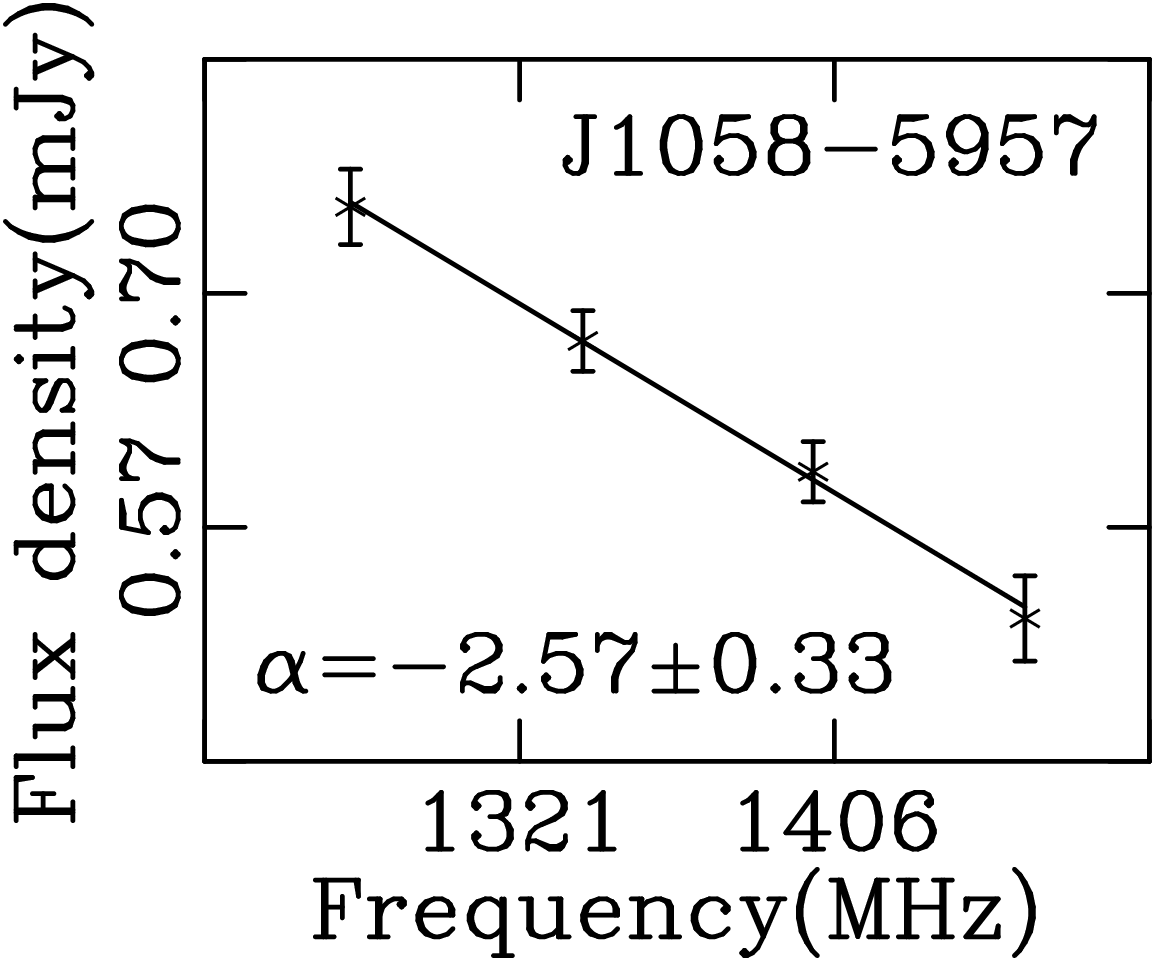}               
\includegraphics[bb = 110 120 518 685,clip,angle=-90,width=0.16\textwidth]{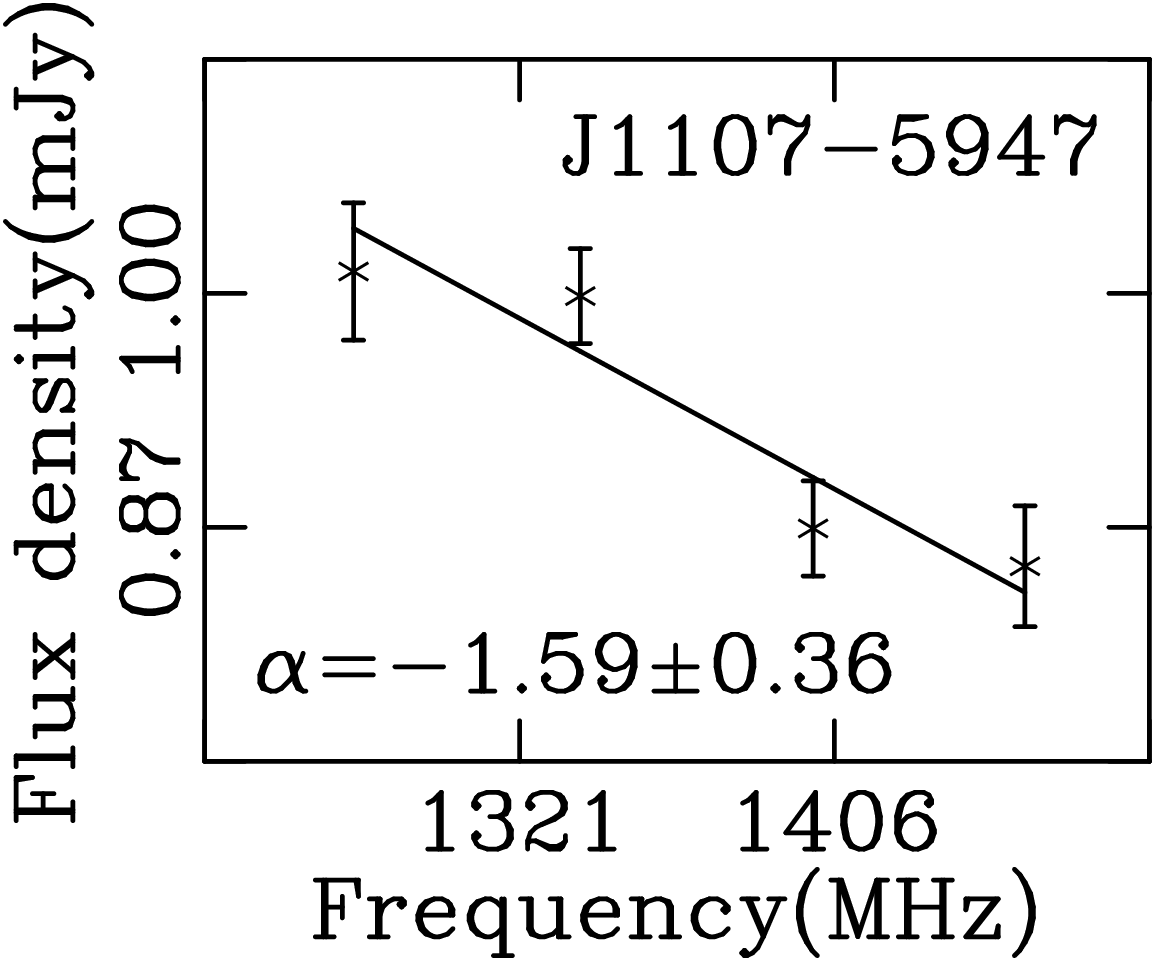}               
\includegraphics[bb = 110 120 518 685,clip,angle=-90,width=0.16\textwidth]{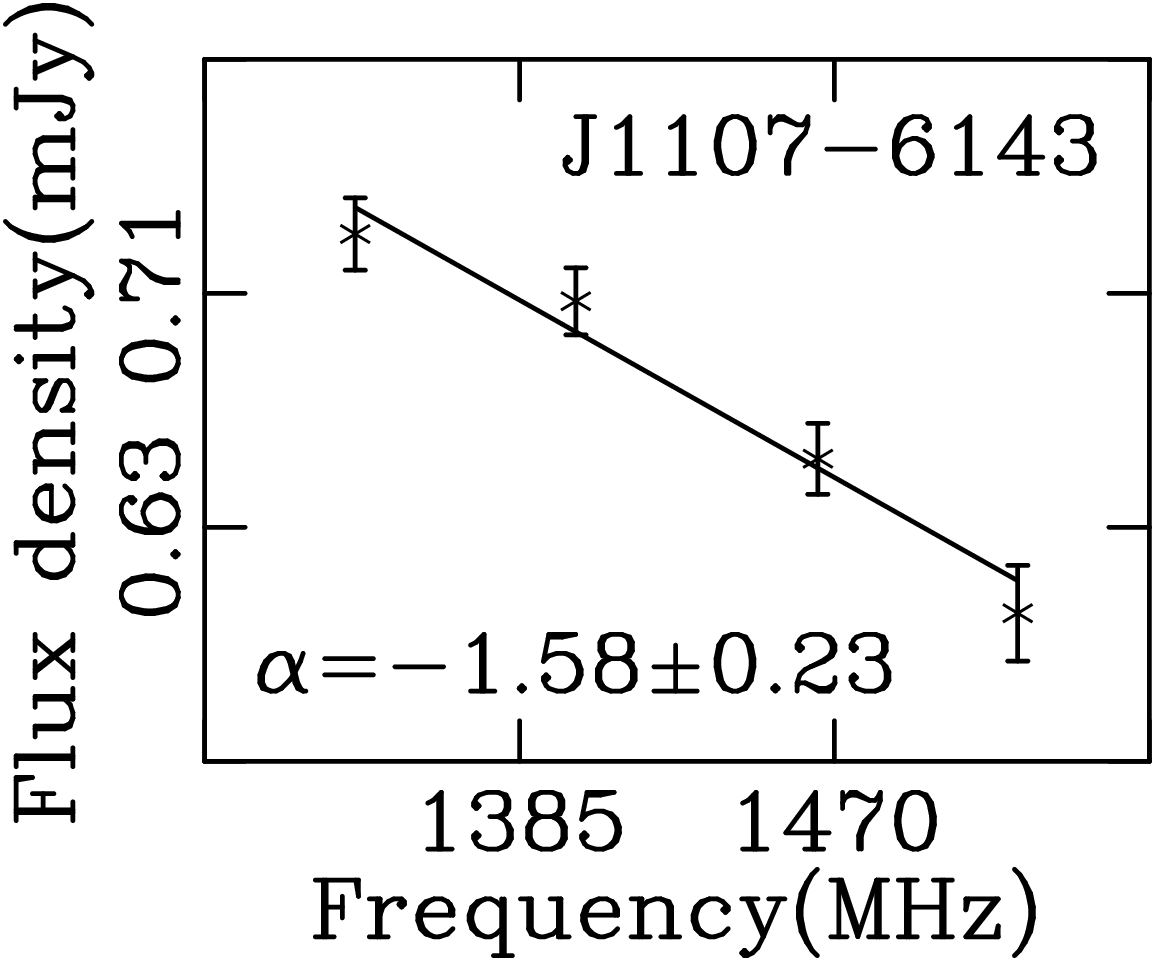}               
\includegraphics[bb = 110 120 518 685,clip,angle=-90,width=0.16\textwidth]{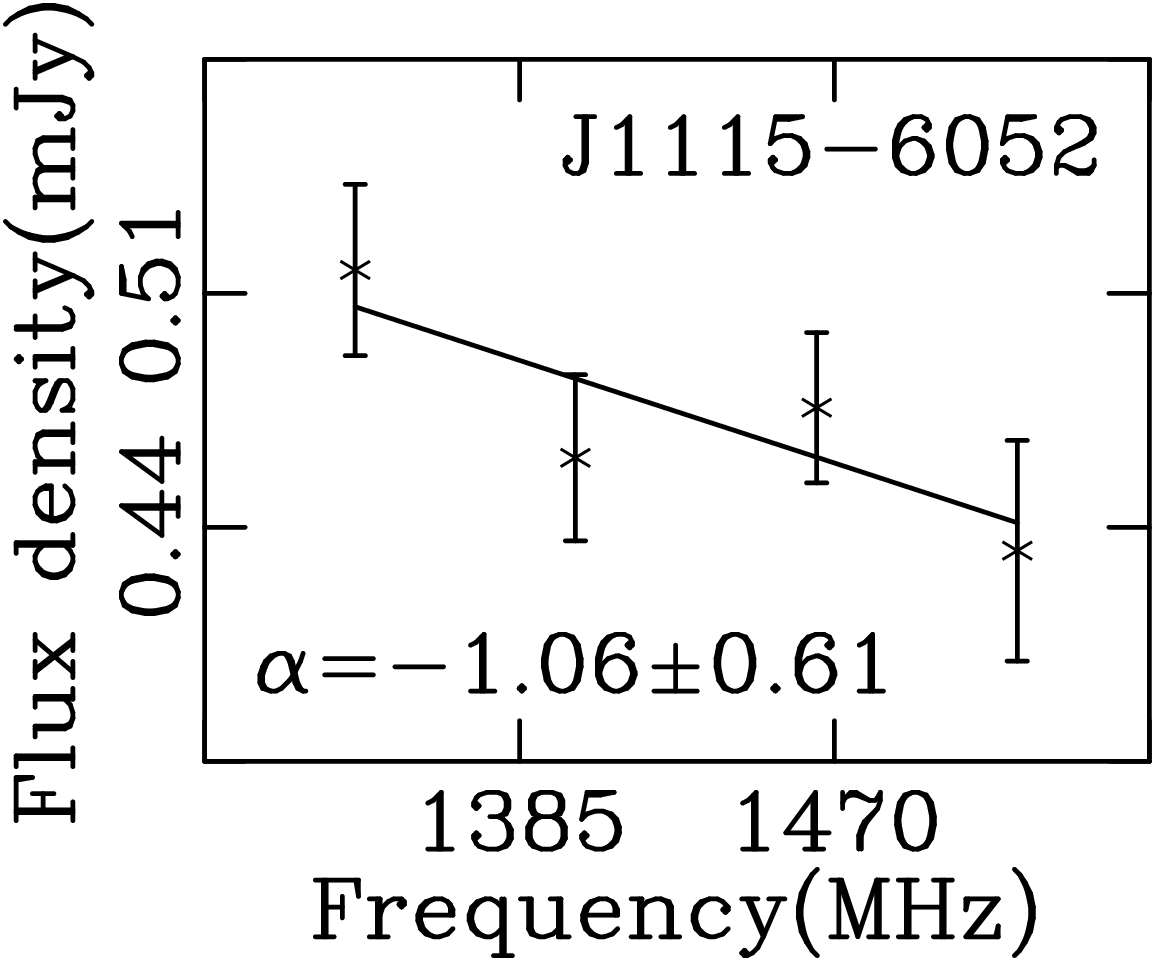}
\\[2mm]               
\includegraphics[bb = 110 72 518 685,clip,angle=-90,width=0.175\textwidth]{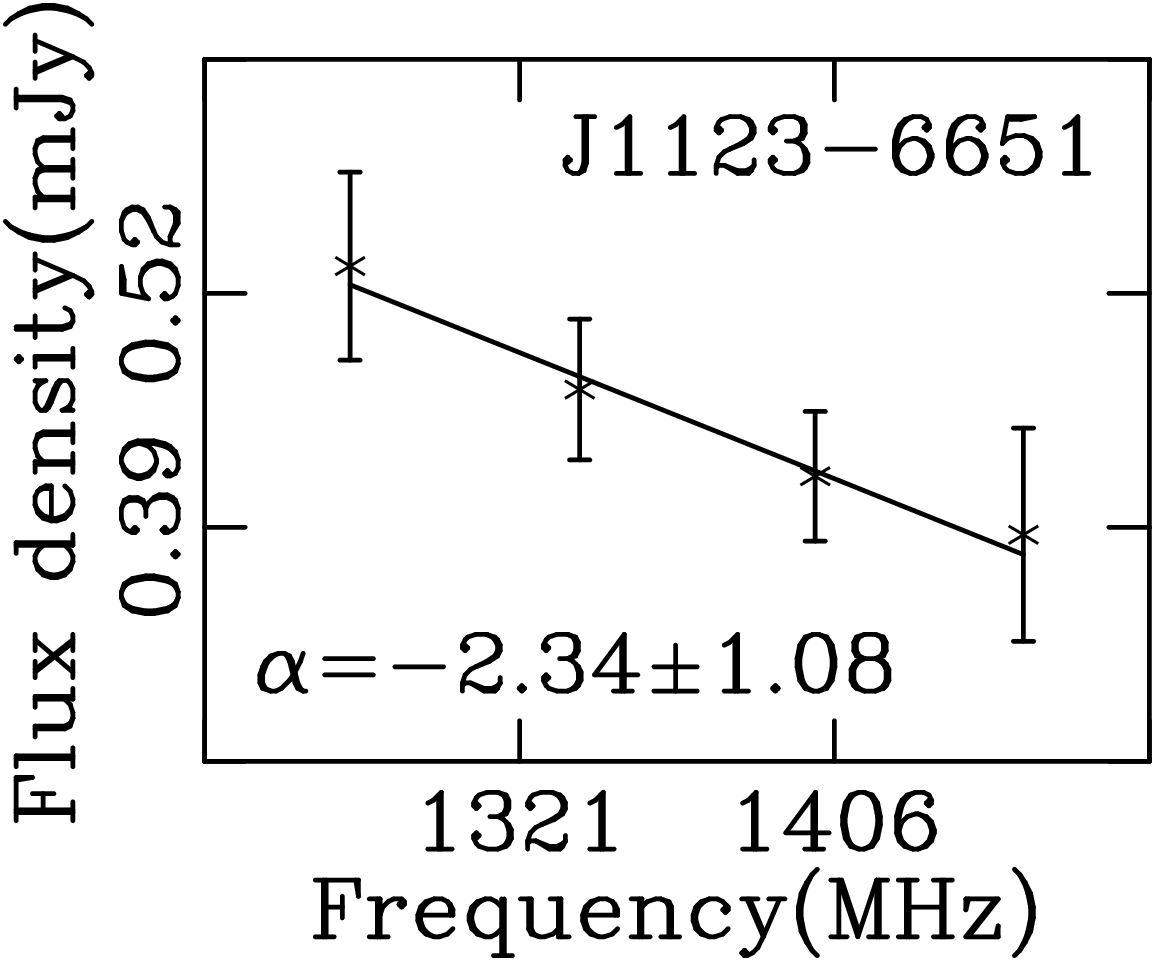}               
\includegraphics[bb = 110 120 518 685,clip,angle=-90,width=0.16\textwidth]{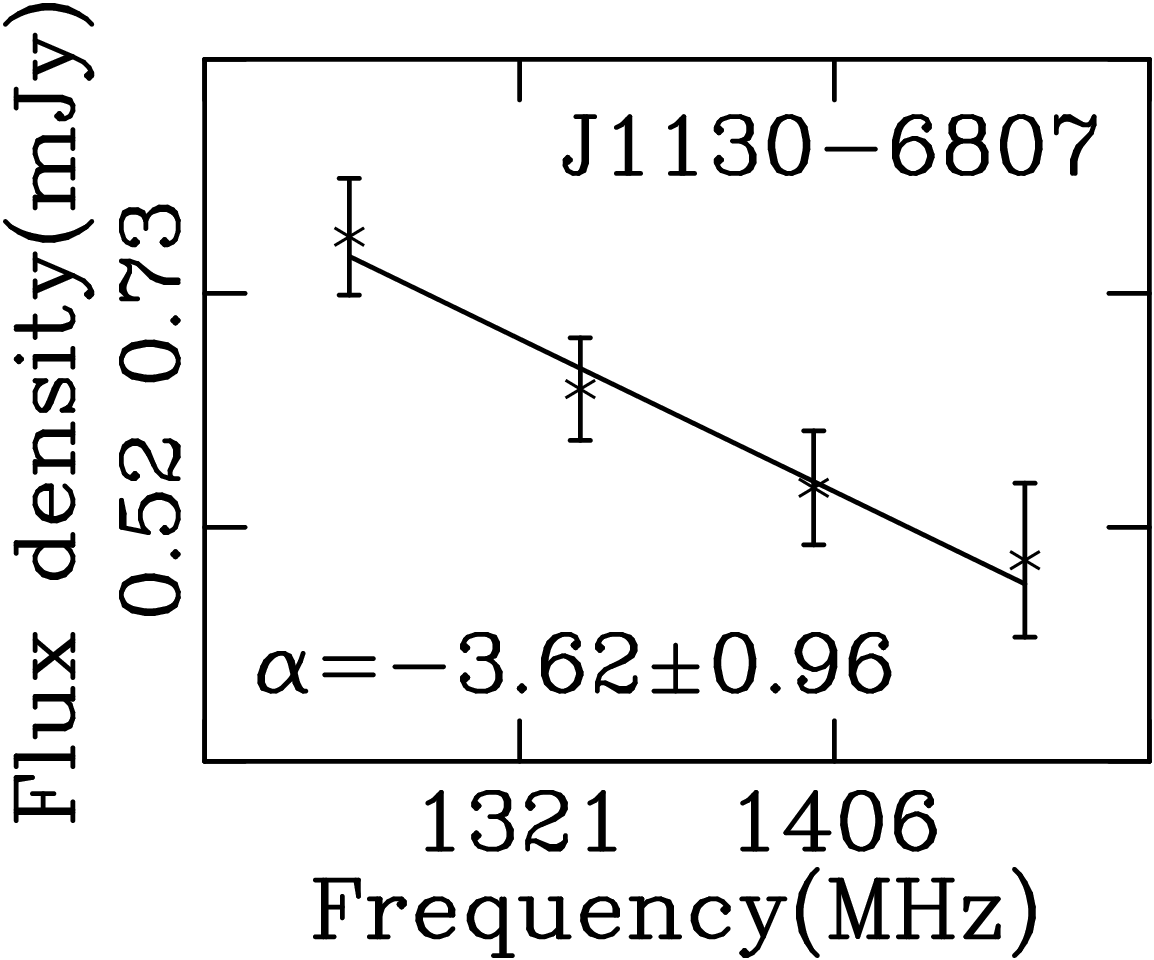}               
\includegraphics[bb = 110 120 518 685,clip,angle=-90,width=0.16\textwidth]{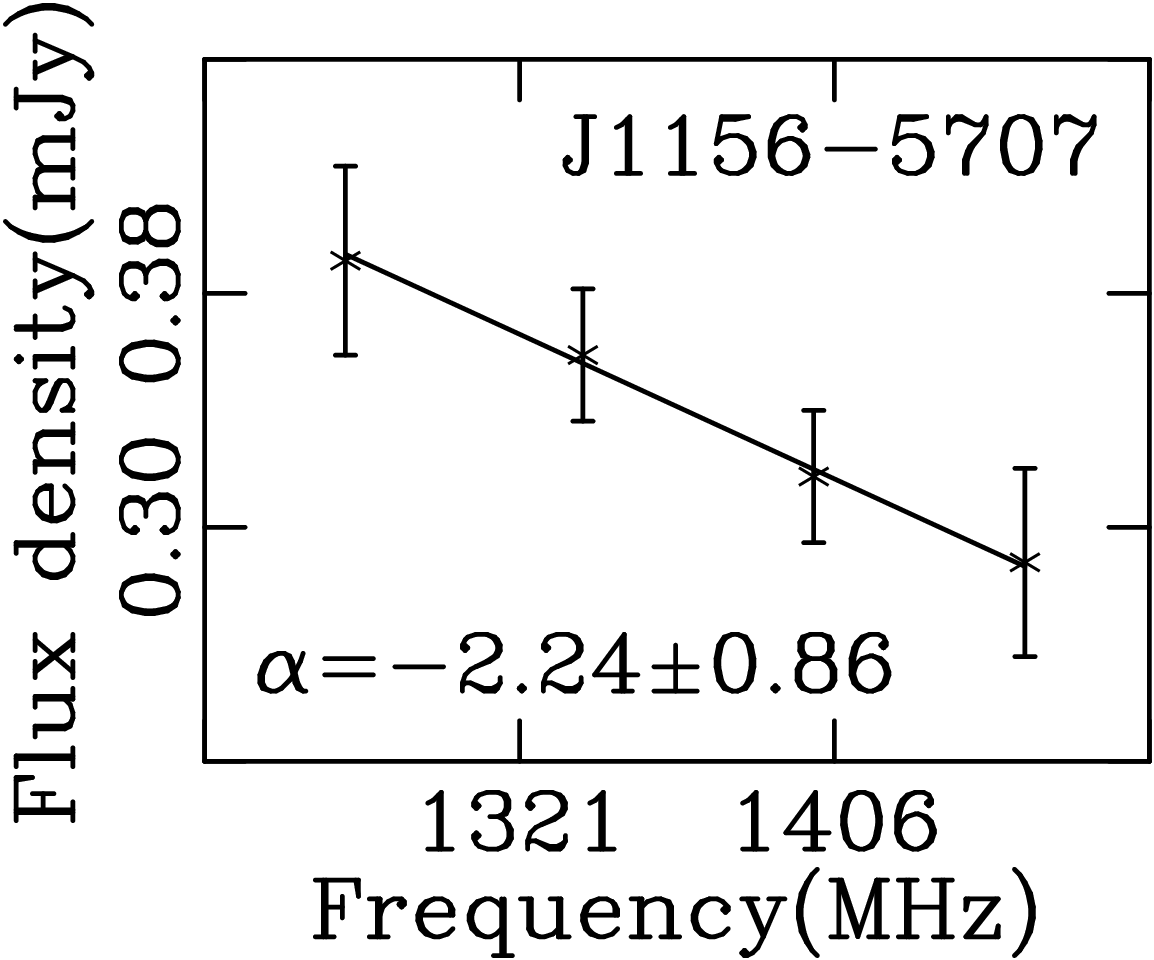}               
\includegraphics[bb = 110 120 518 685,clip,angle=-90,width=0.16\textwidth]{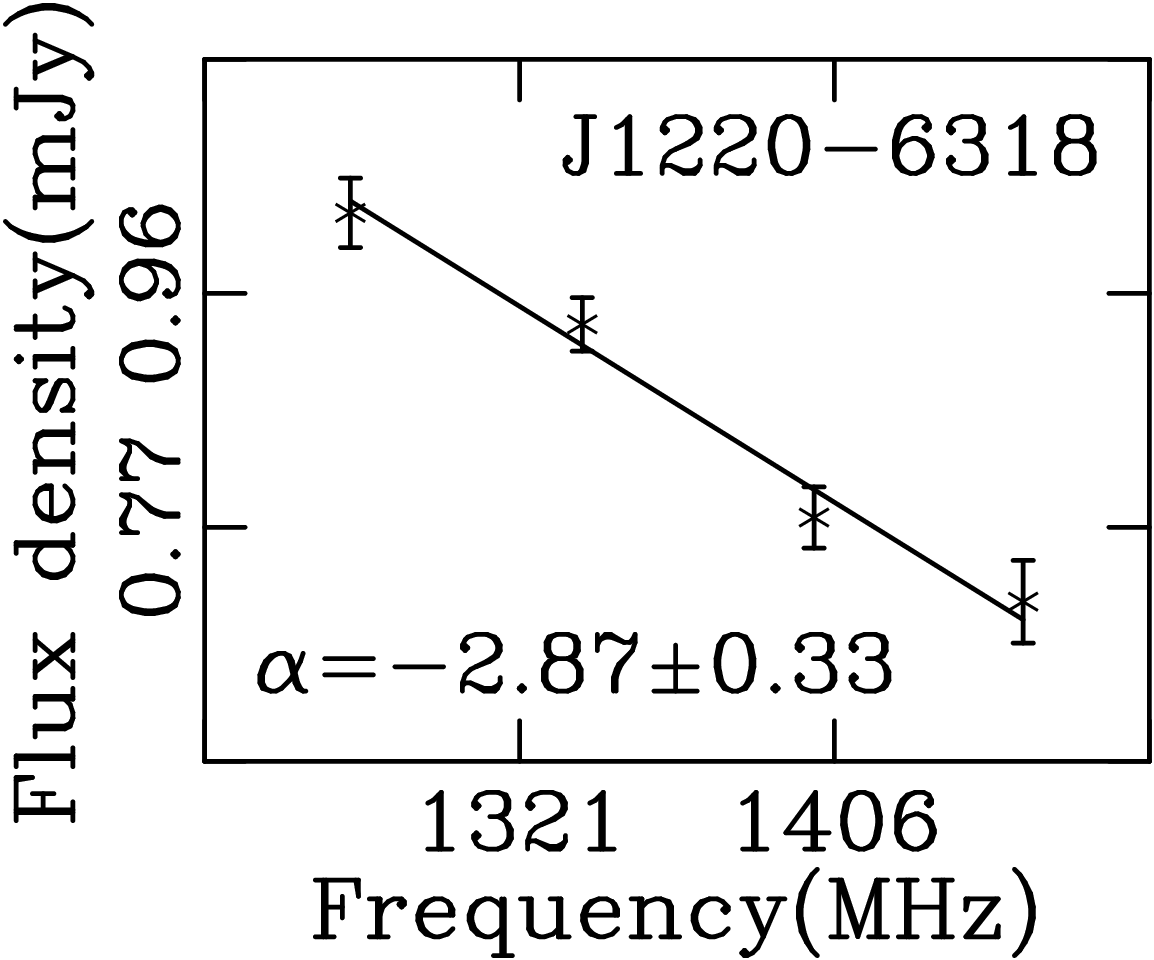}               
\includegraphics[bb = 110 120 518 685,clip,angle=-90,width=0.16\textwidth]{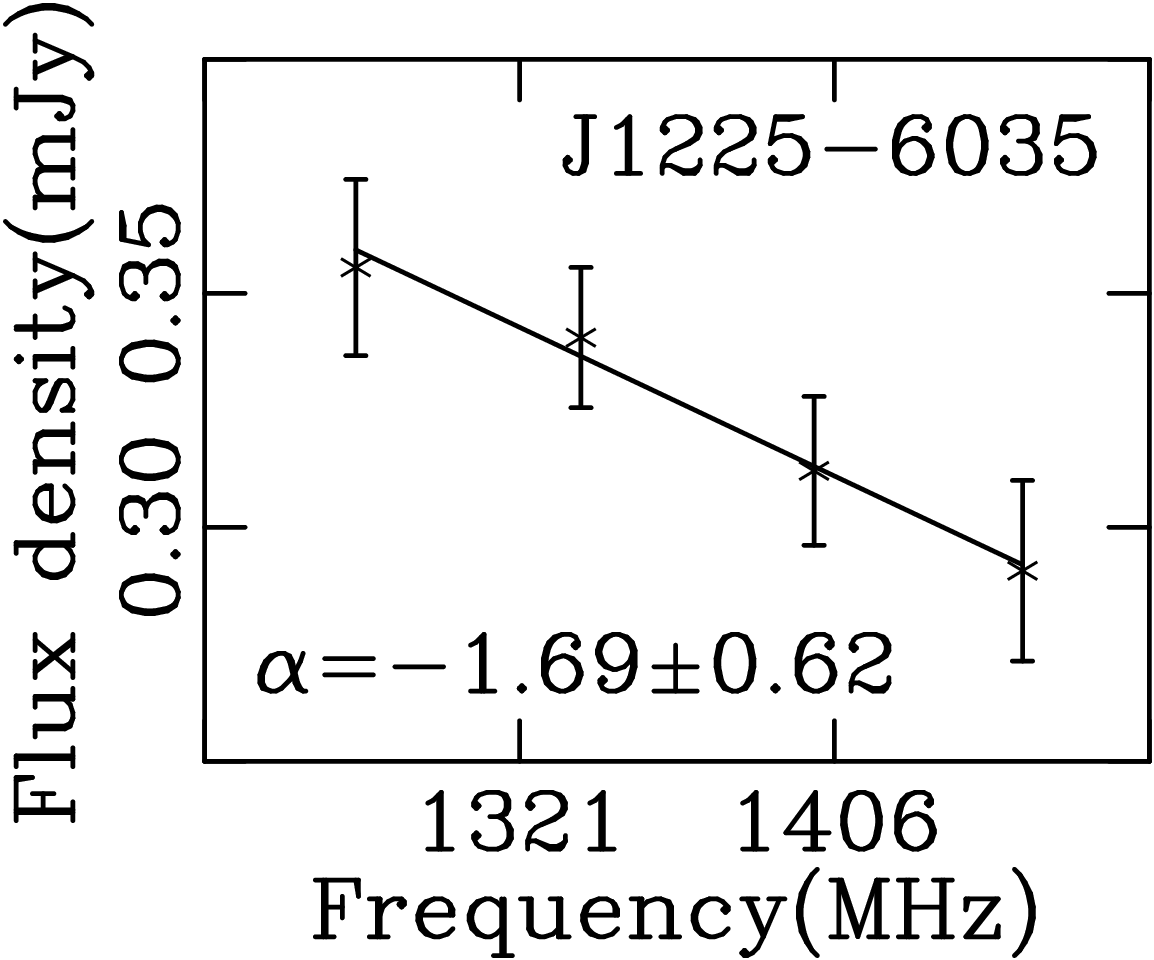}               
\includegraphics[bb = 110 120 518 685,clip,angle=-90,width=0.16\textwidth]{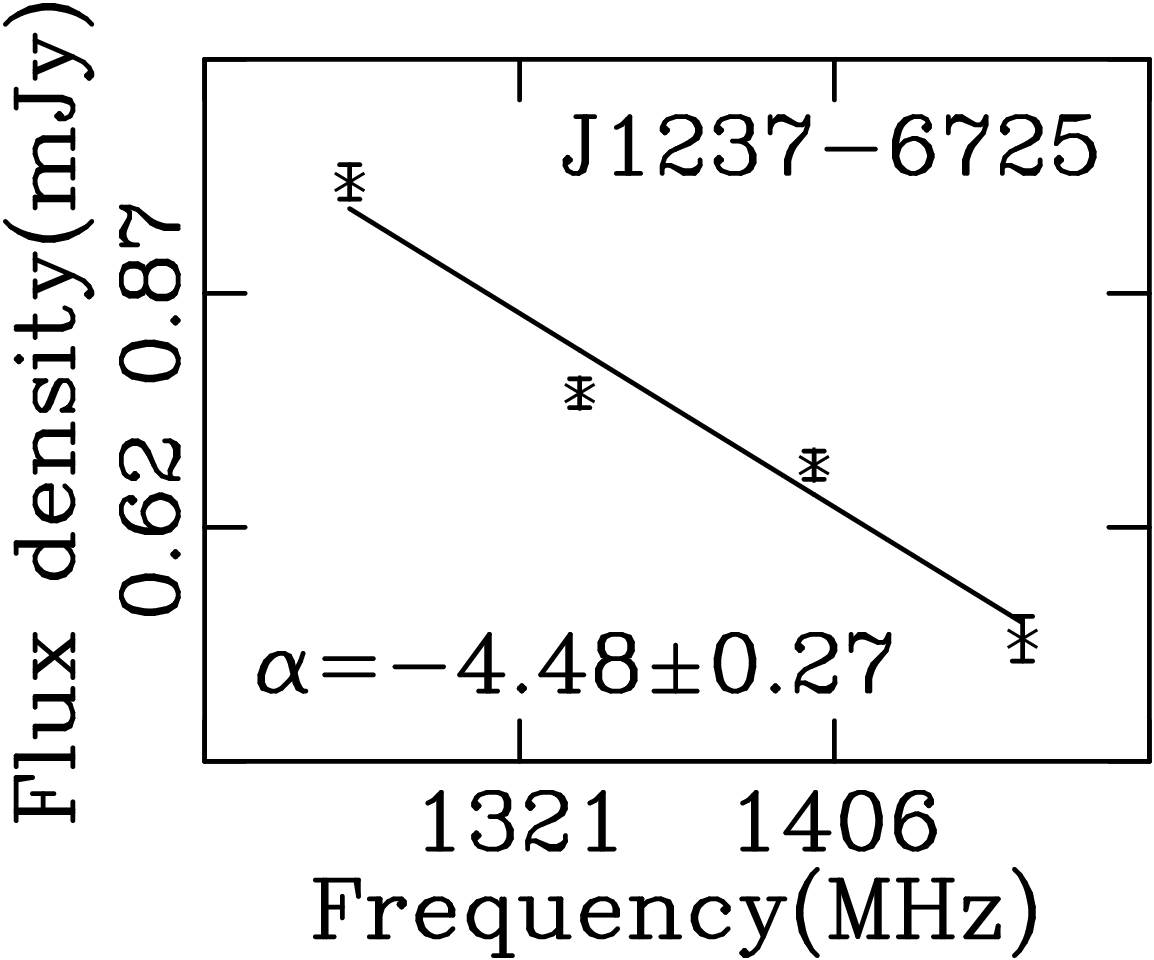}
 \\[2mm]              
\includegraphics[bb = 110 72 518 685,clip,angle=-90,width=0.175\textwidth]{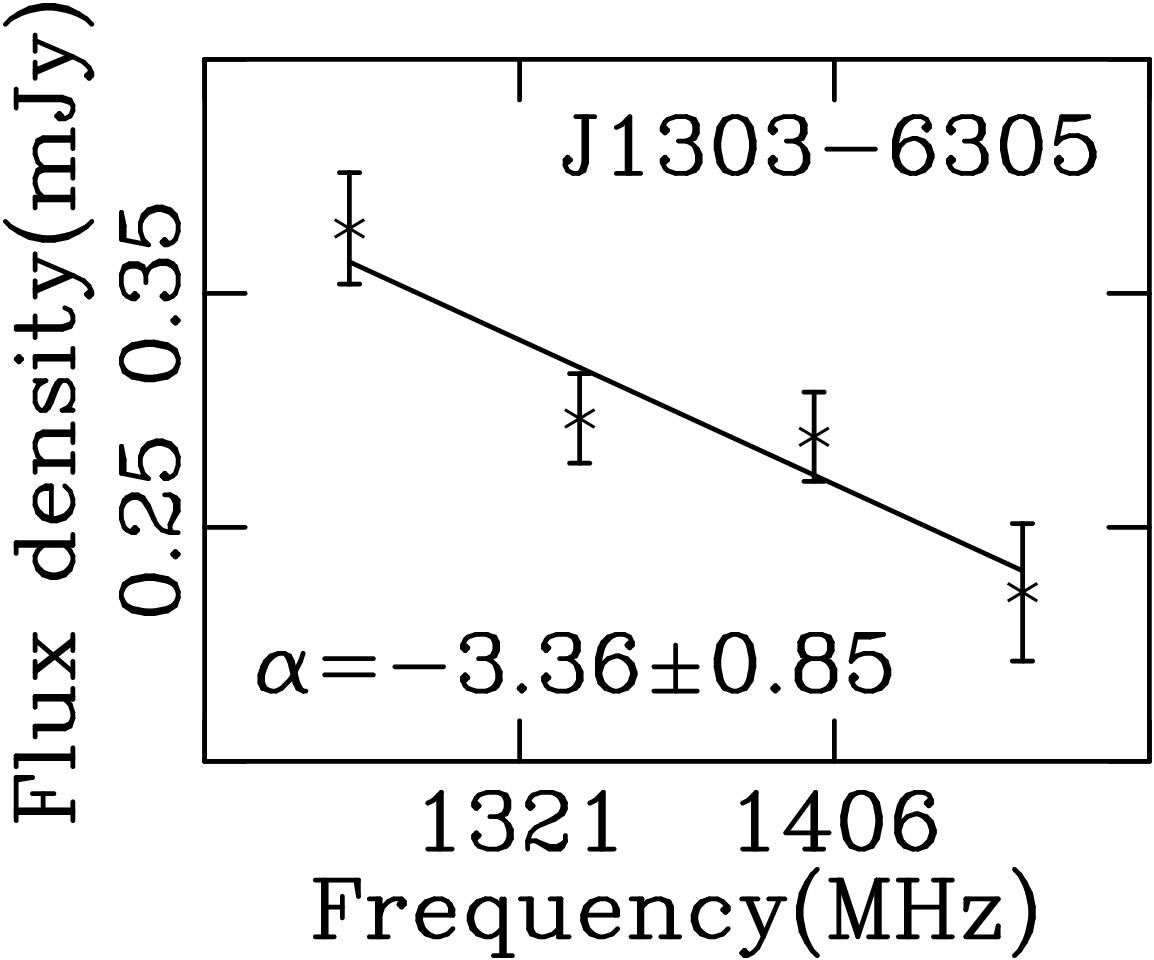}               
\includegraphics[bb = 110 120 518 685,clip,angle=-90,width=0.16\textwidth]{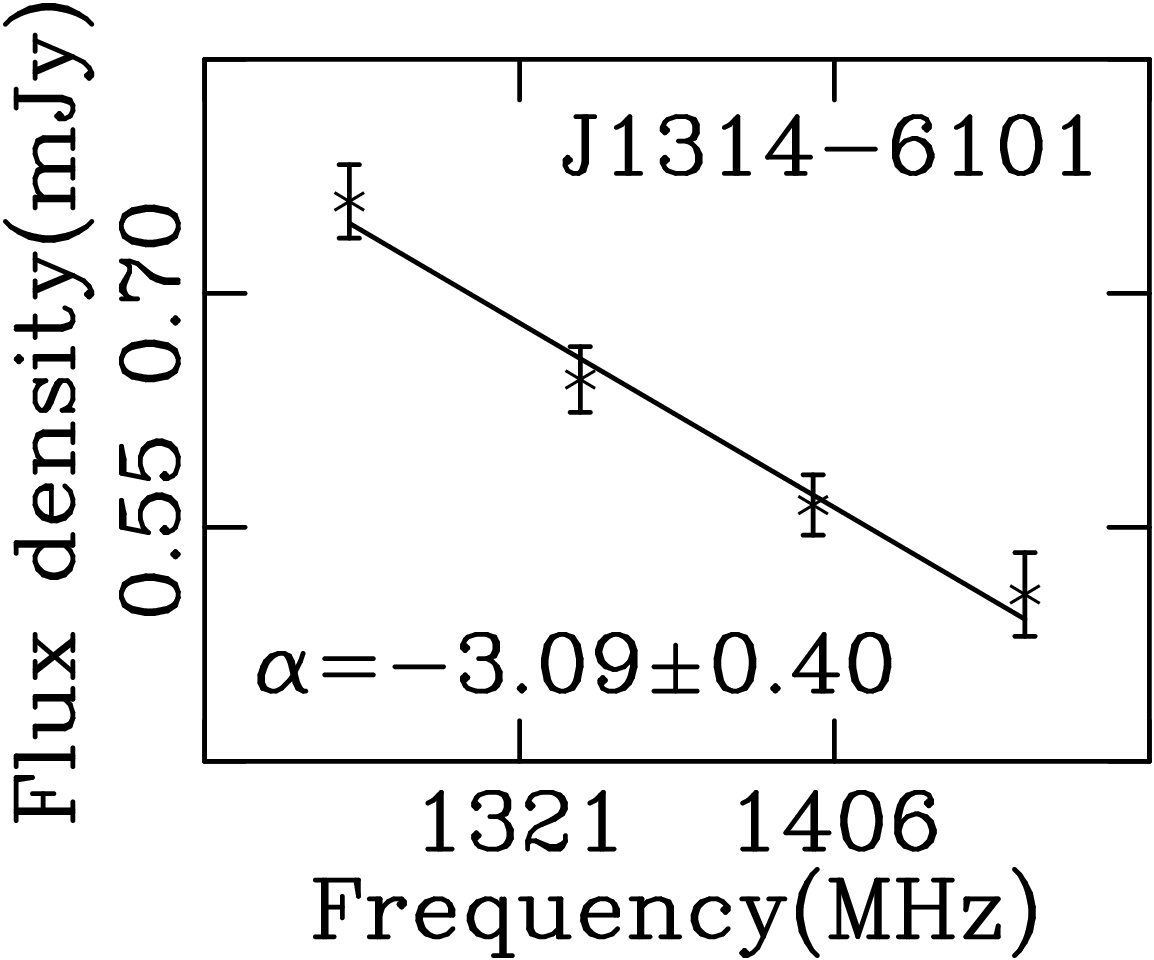}               
\includegraphics[bb = 110 120 518 685,clip,angle=-90,width=0.16\textwidth]{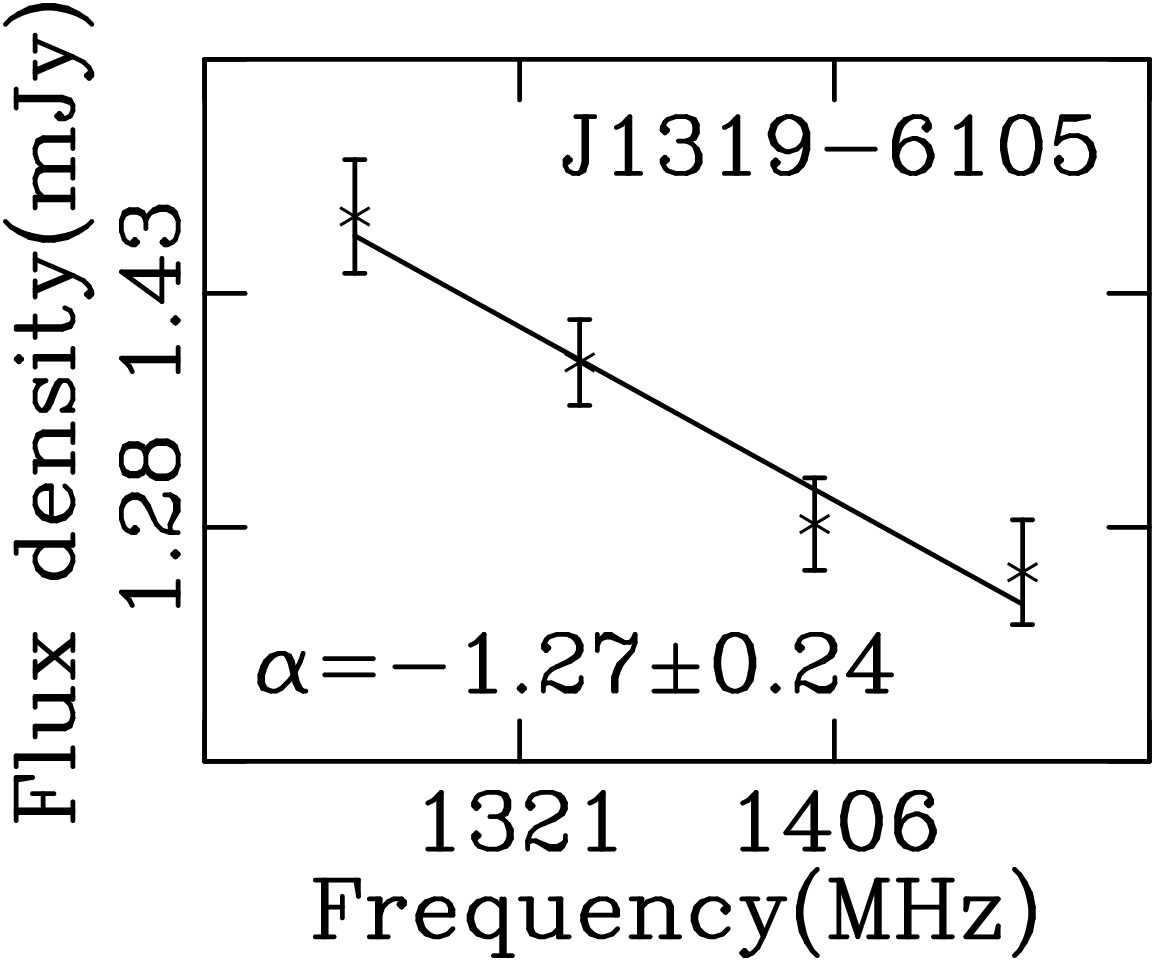}               
\includegraphics[bb = 110 120 518 685,clip,angle=-90,width=0.16\textwidth]{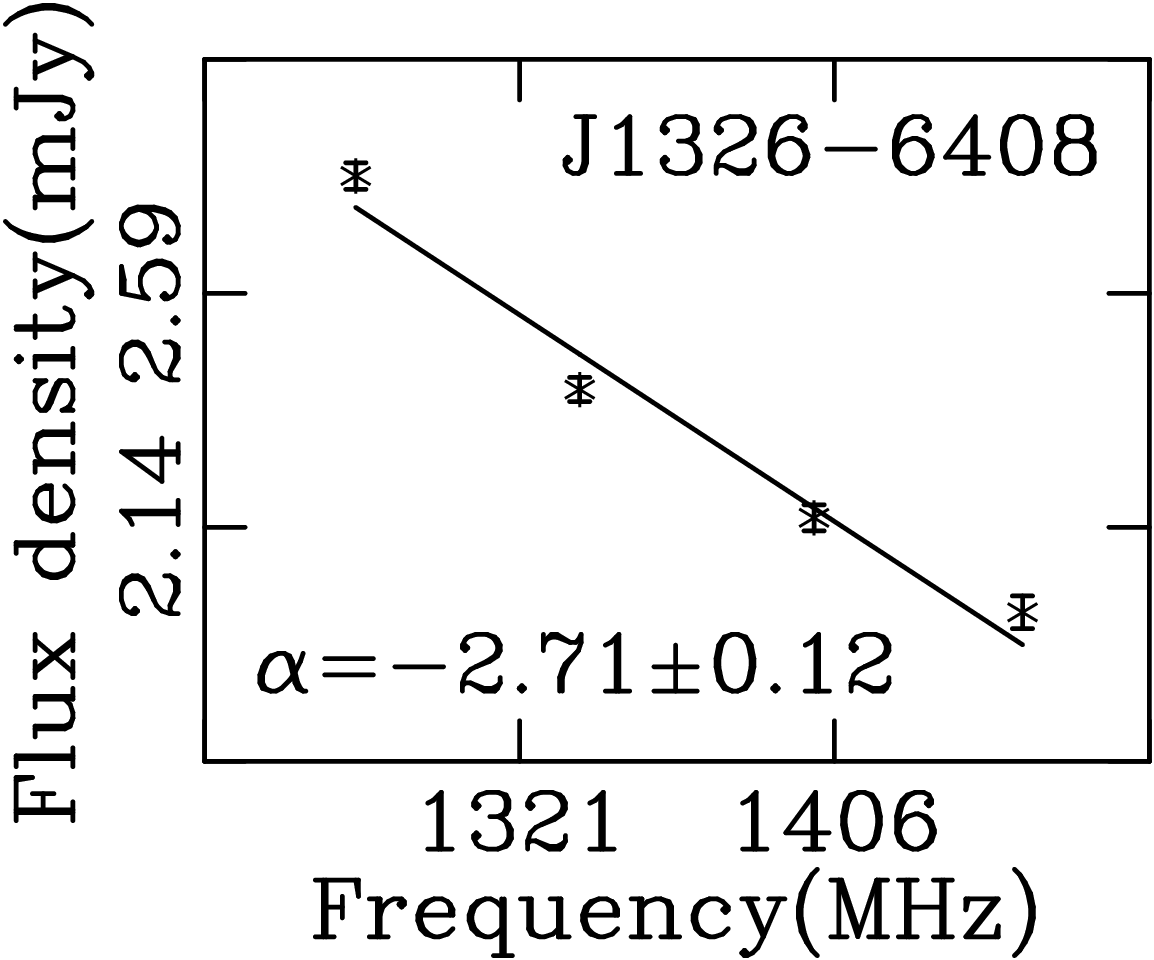}               
\includegraphics[bb = 110 120 518 685,clip,angle=-90,width=0.16\textwidth]{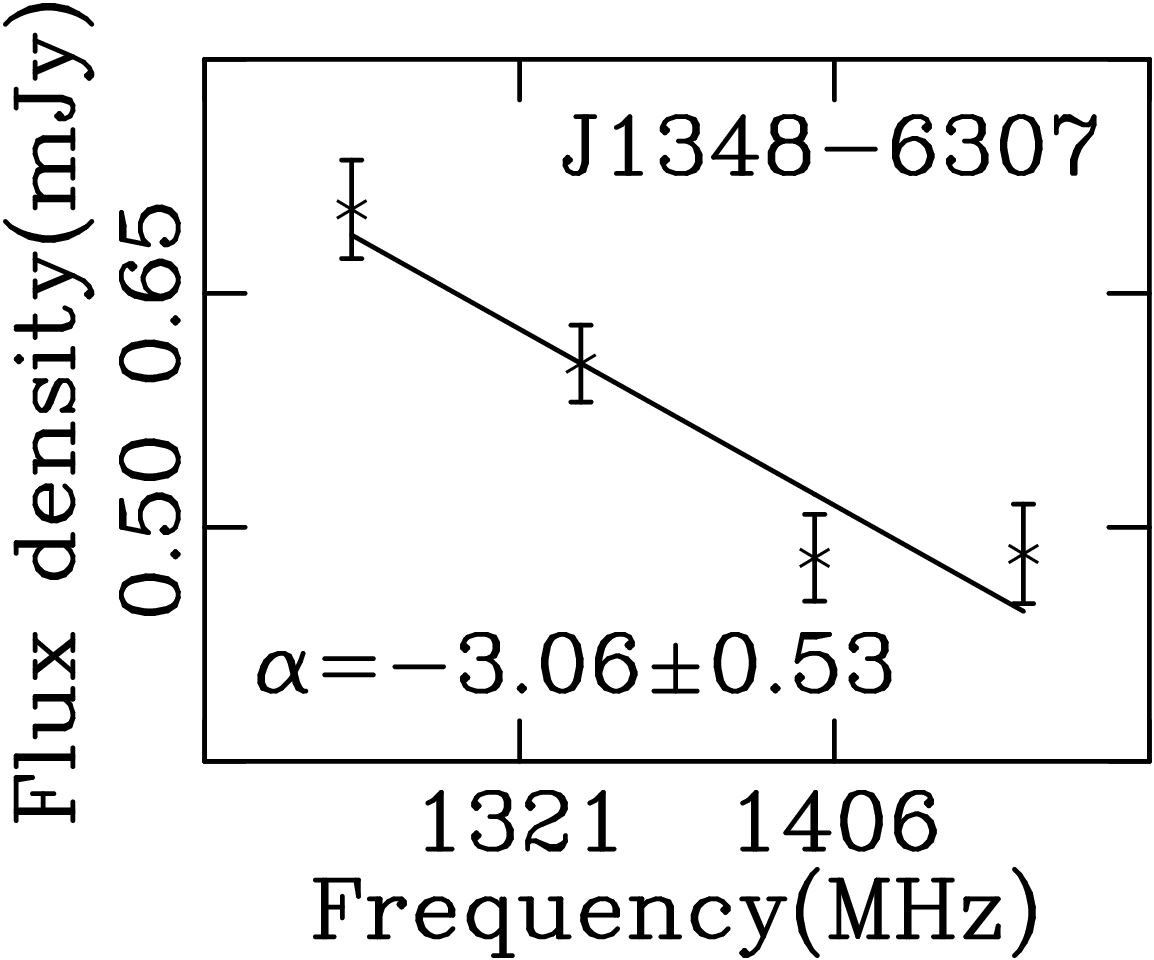}               
\includegraphics[bb = 110 120 518 685,clip,angle=-90,width=0.16\textwidth]{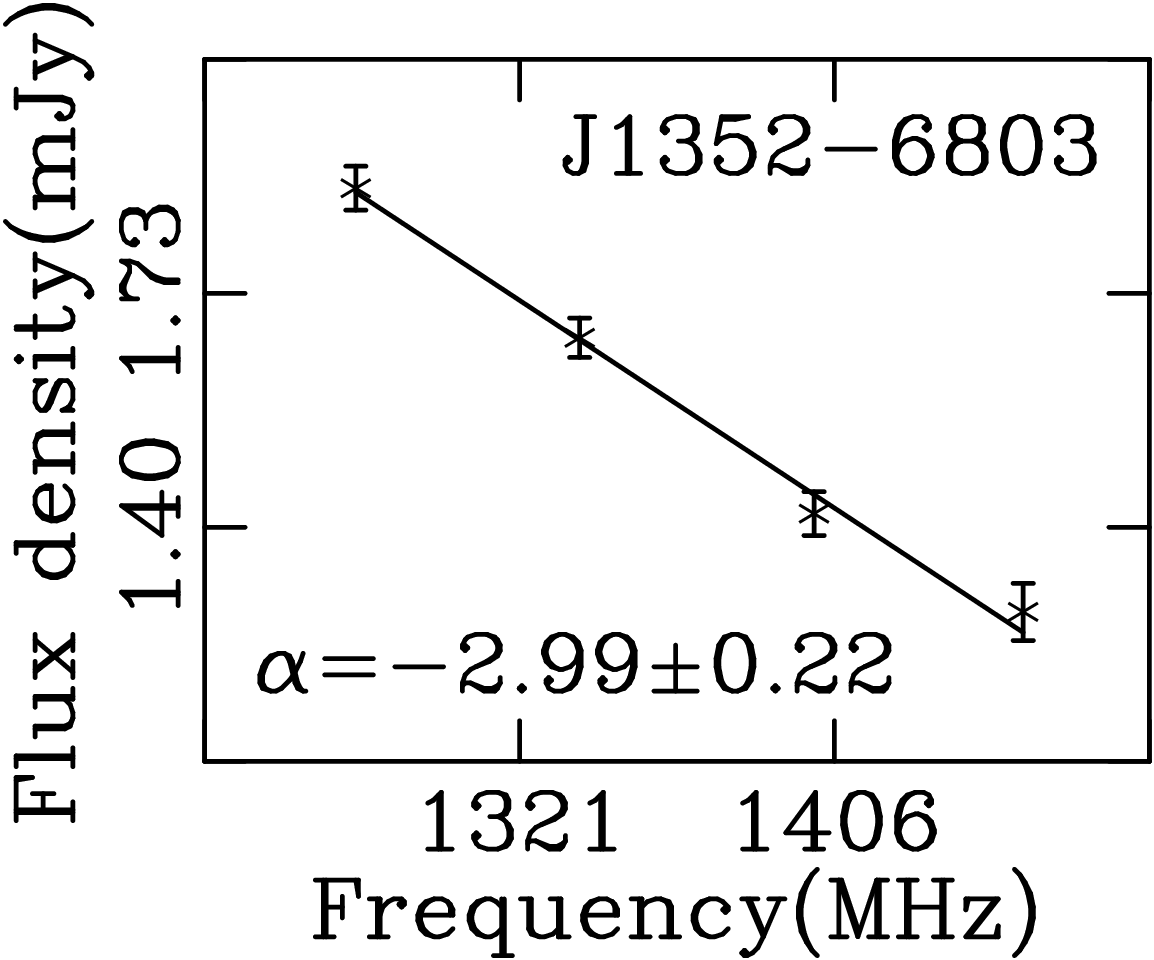}
\\[2mm]               
\includegraphics[bb = 110 72 518 685,clip,angle=-90,width=0.175\textwidth]{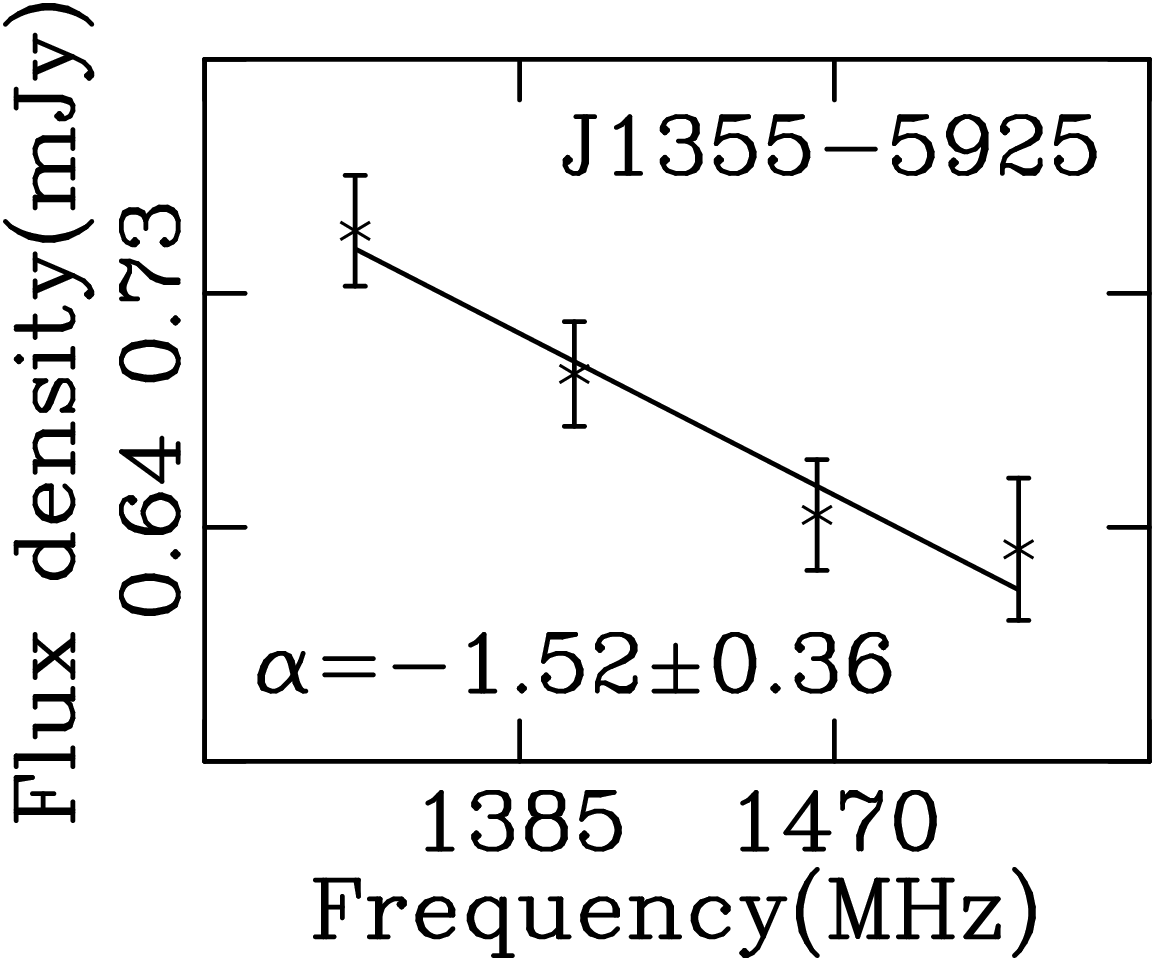}               
\includegraphics[bb = 110 120 518 685,clip,angle=-90,width=0.16\textwidth]{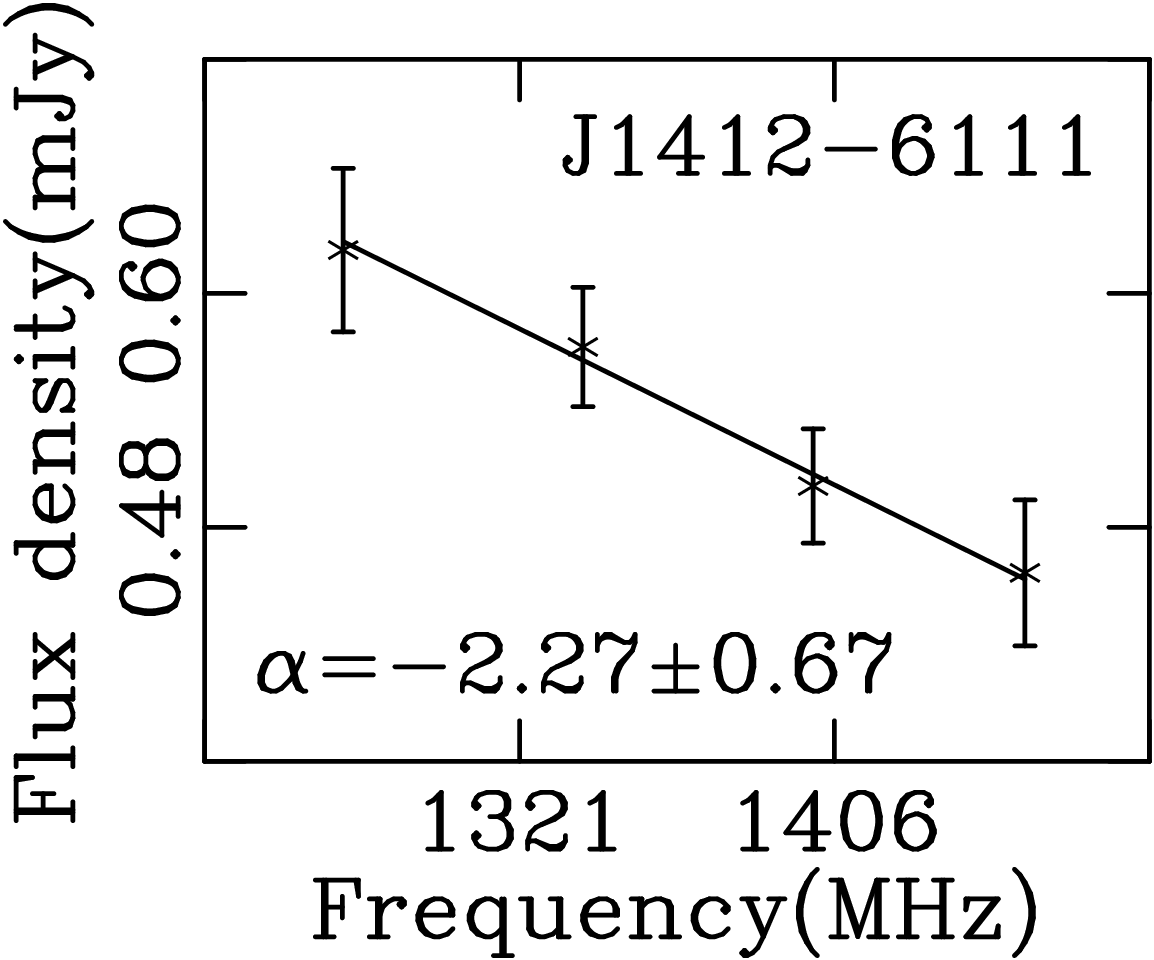}               
\includegraphics[bb = 110 120 518 685,clip,angle=-90,width=0.16\textwidth]{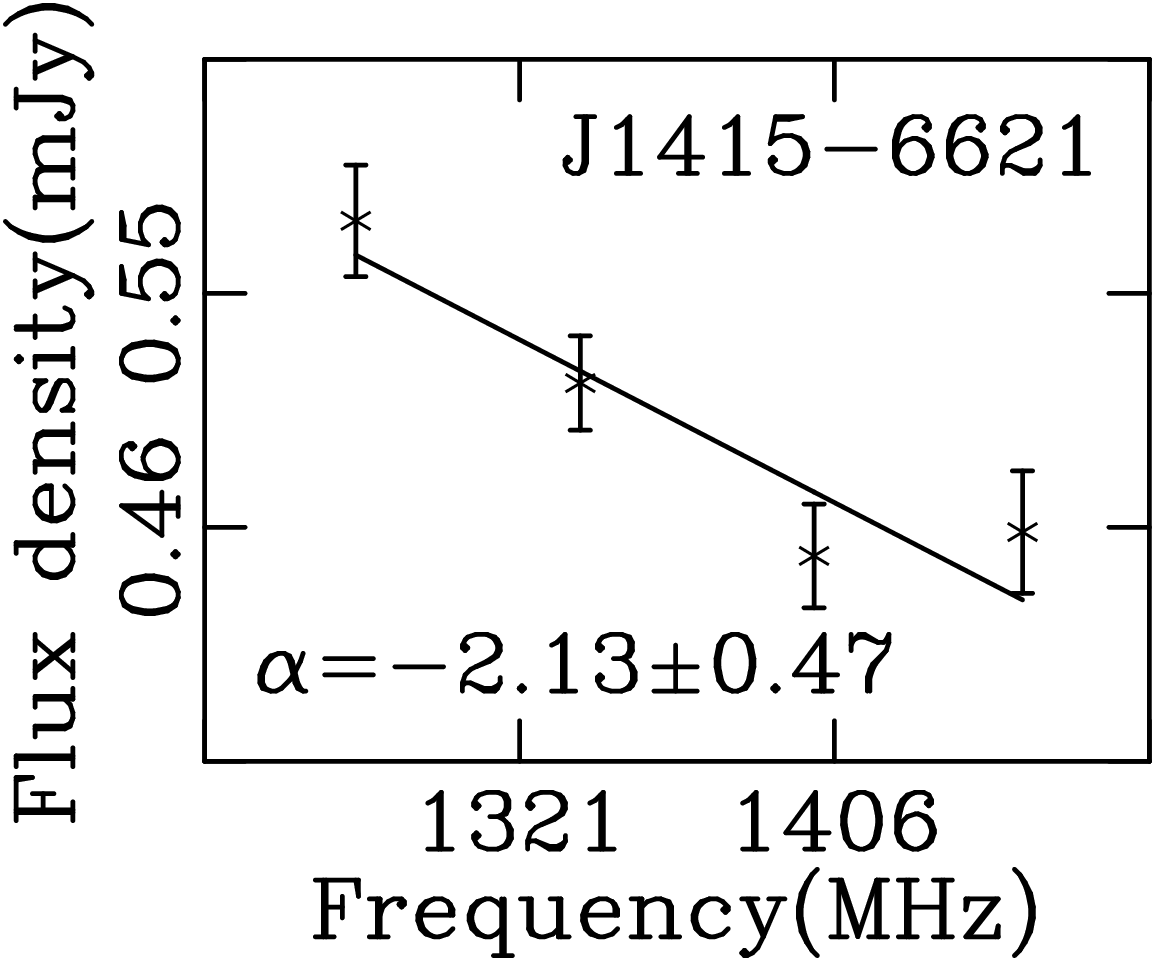}               
\includegraphics[bb = 110 120 518 685,clip,angle=-90,width=0.16\textwidth]{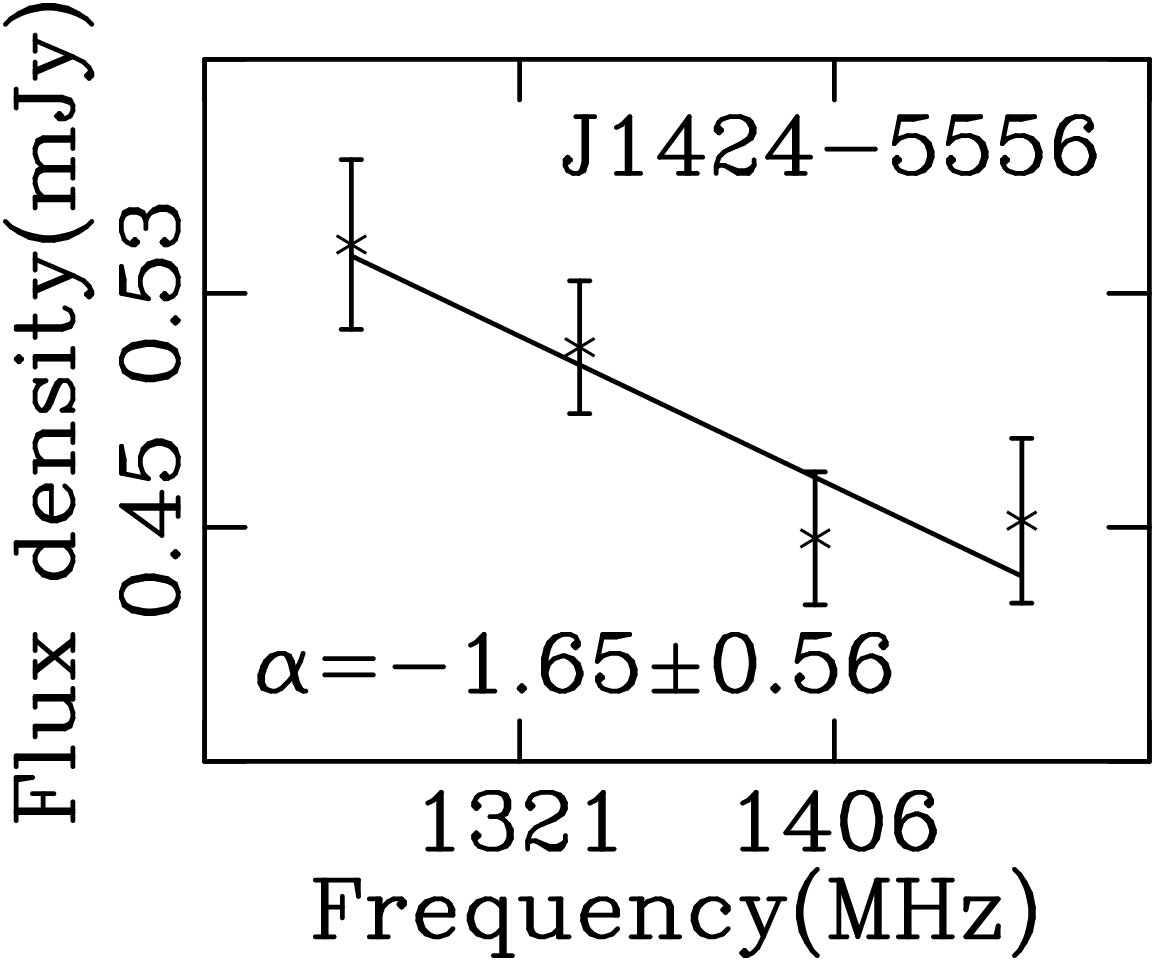}               
\includegraphics[bb = 110 120 518 685,clip,angle=-90,width=0.16\textwidth]{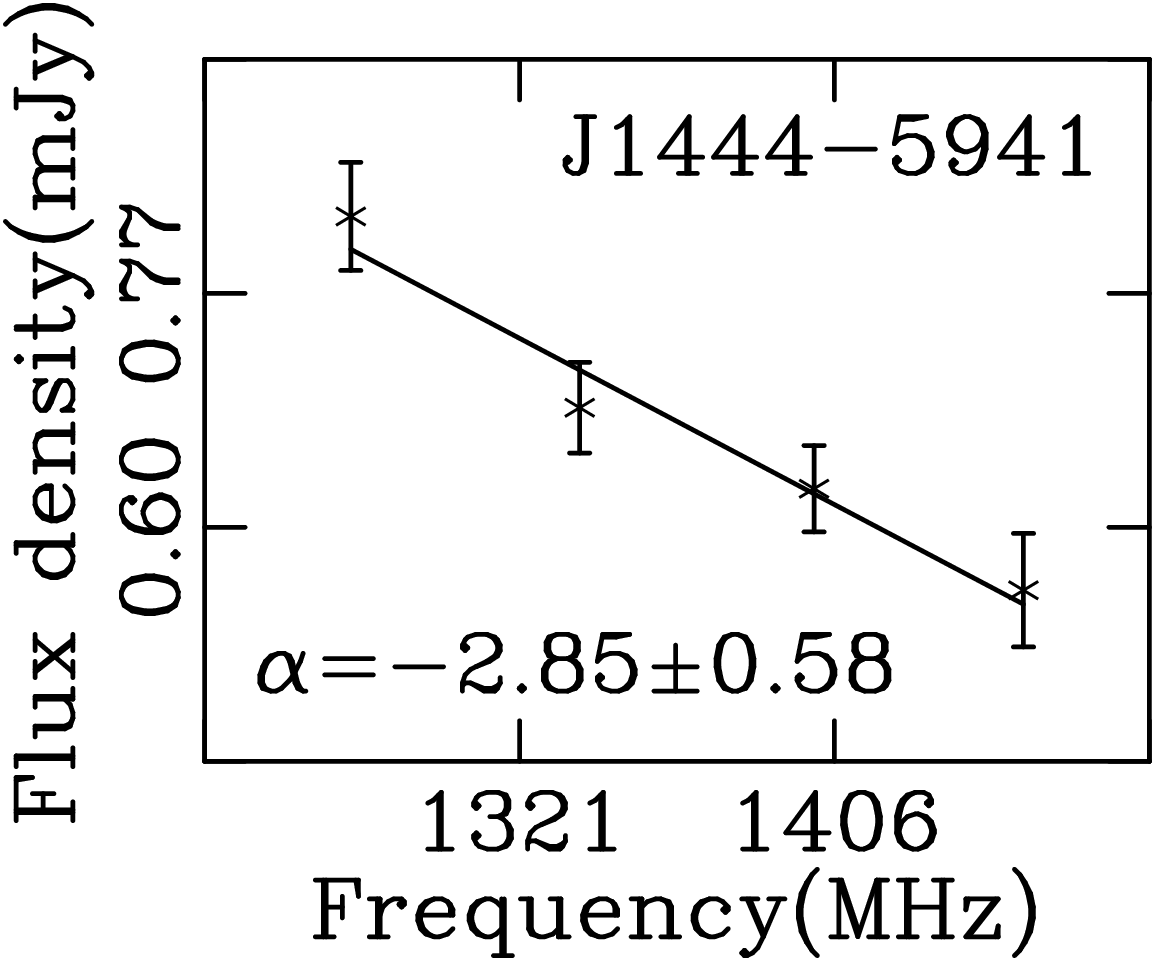}               
\includegraphics[bb = 110 120 518 685,clip,angle=-90,width=0.16\textwidth]{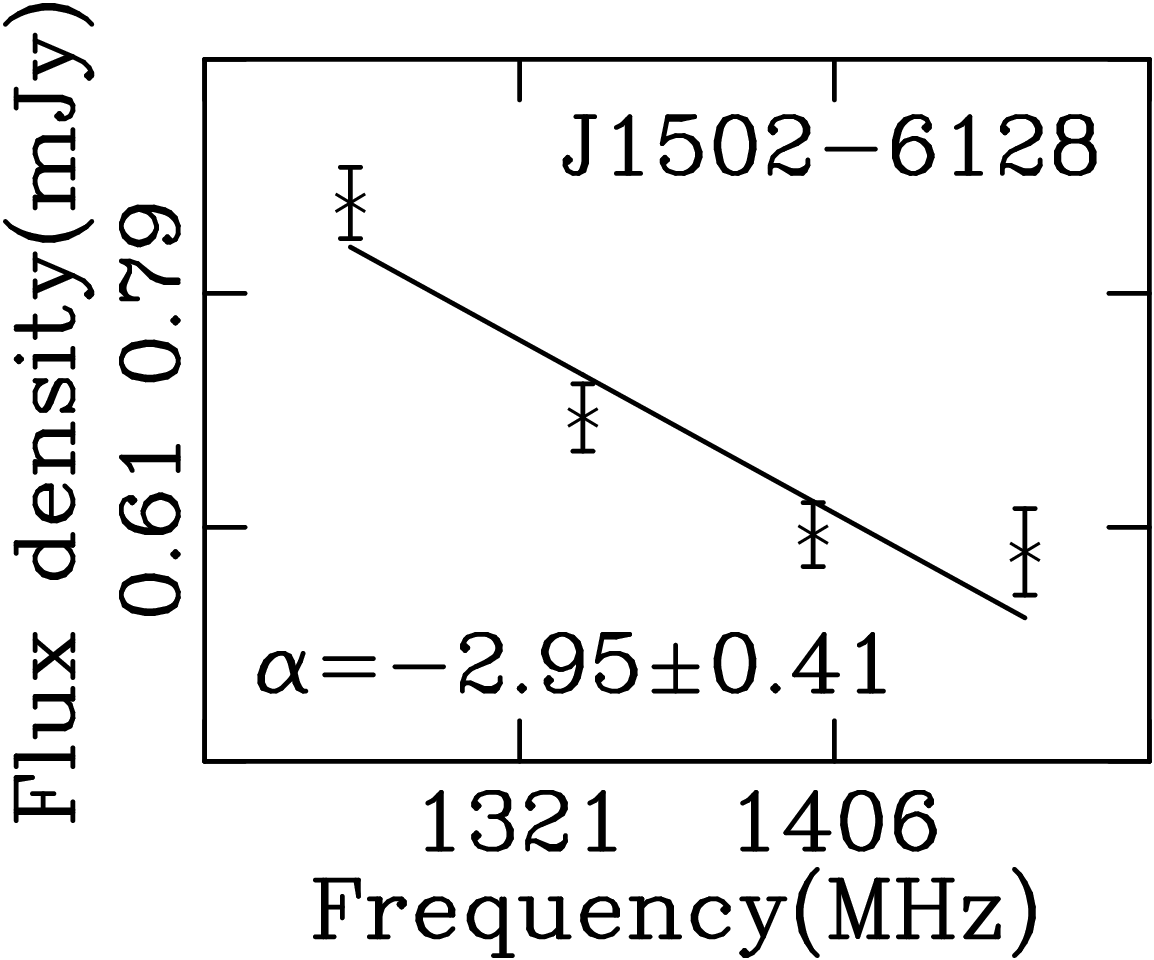} 
 \\[2mm]             
\includegraphics[bb = 110 72 518 685,clip,angle=-90,width=0.175\textwidth]{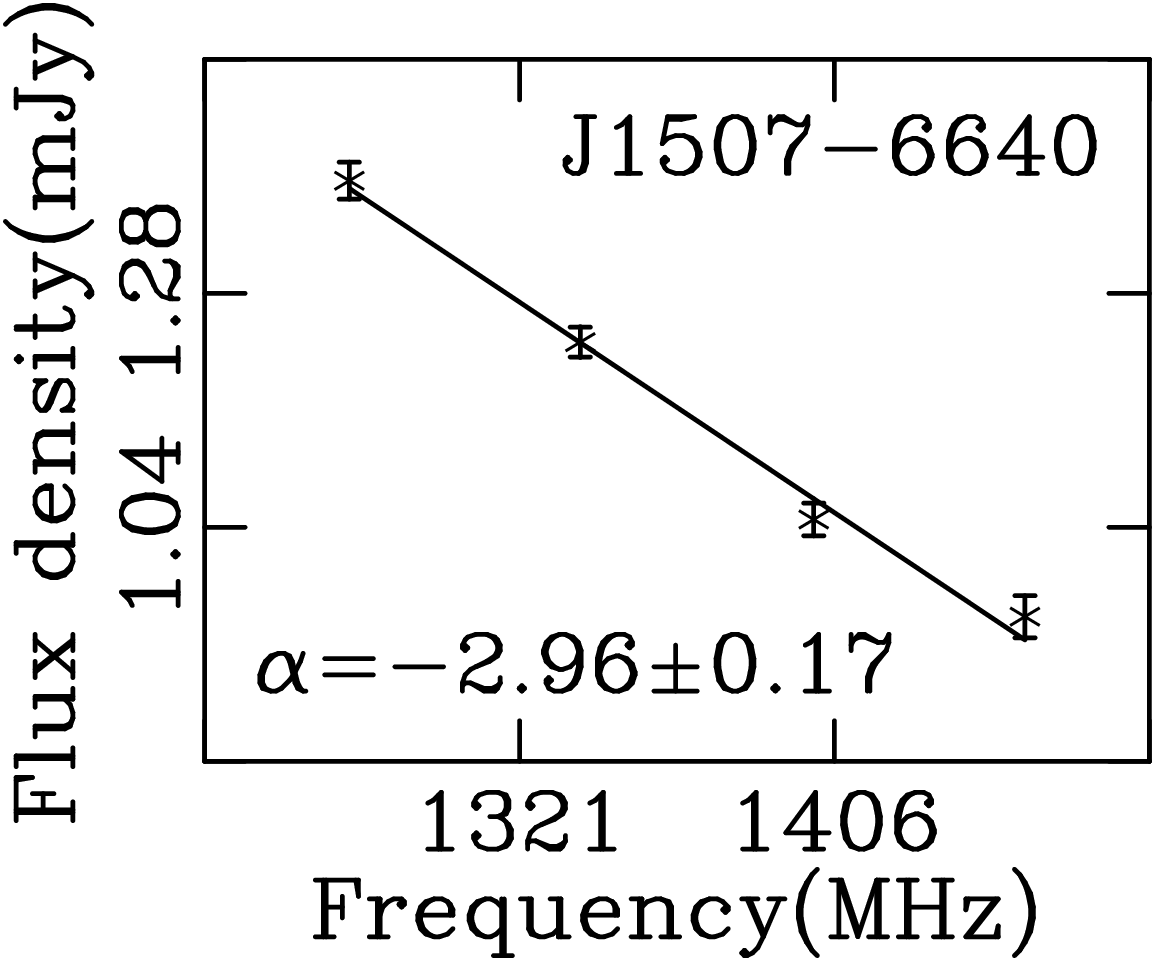}               
\includegraphics[bb = 110 120 518 685,clip,angle=-90,width=0.16\textwidth]{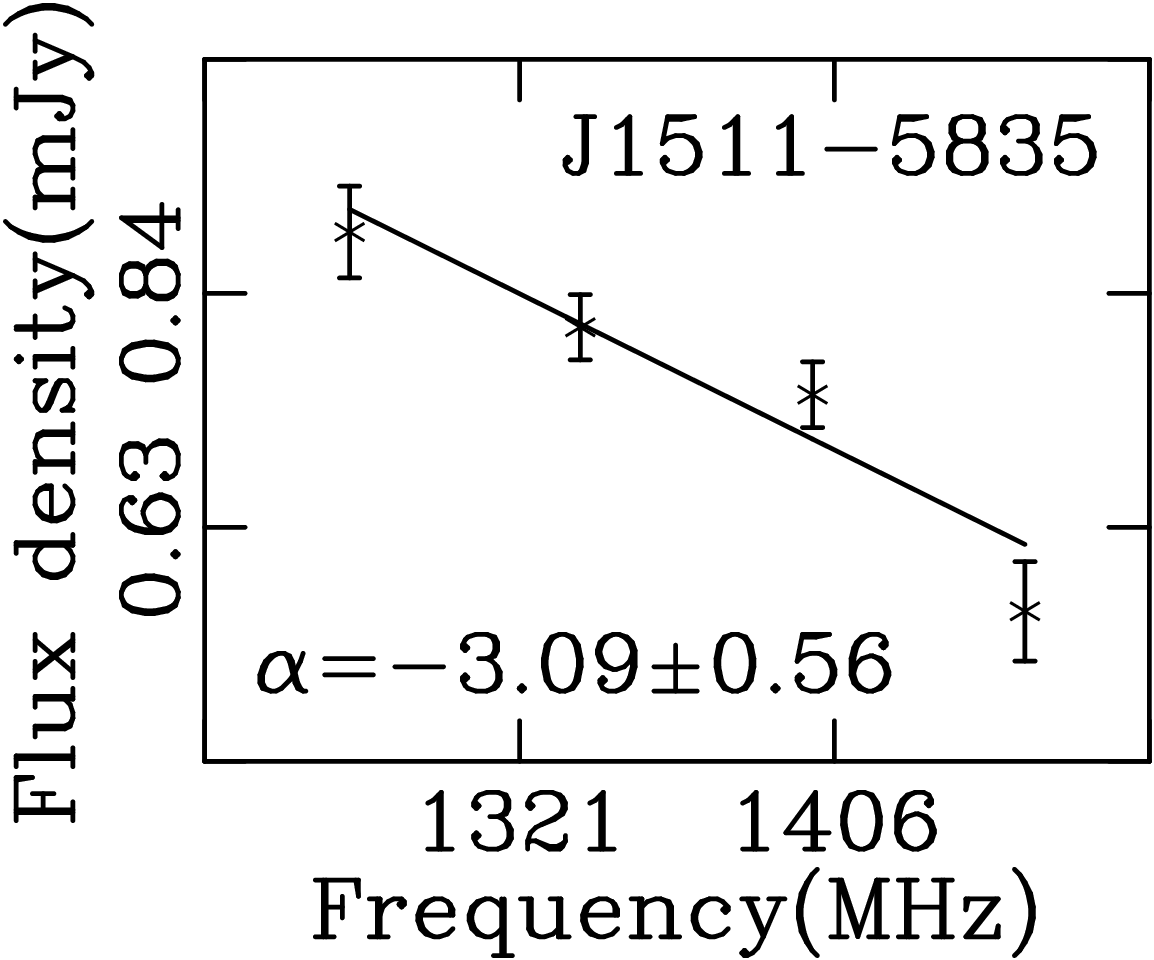}               
\includegraphics[bb = 110 120 518 685,clip,angle=-90,width=0.16\textwidth]{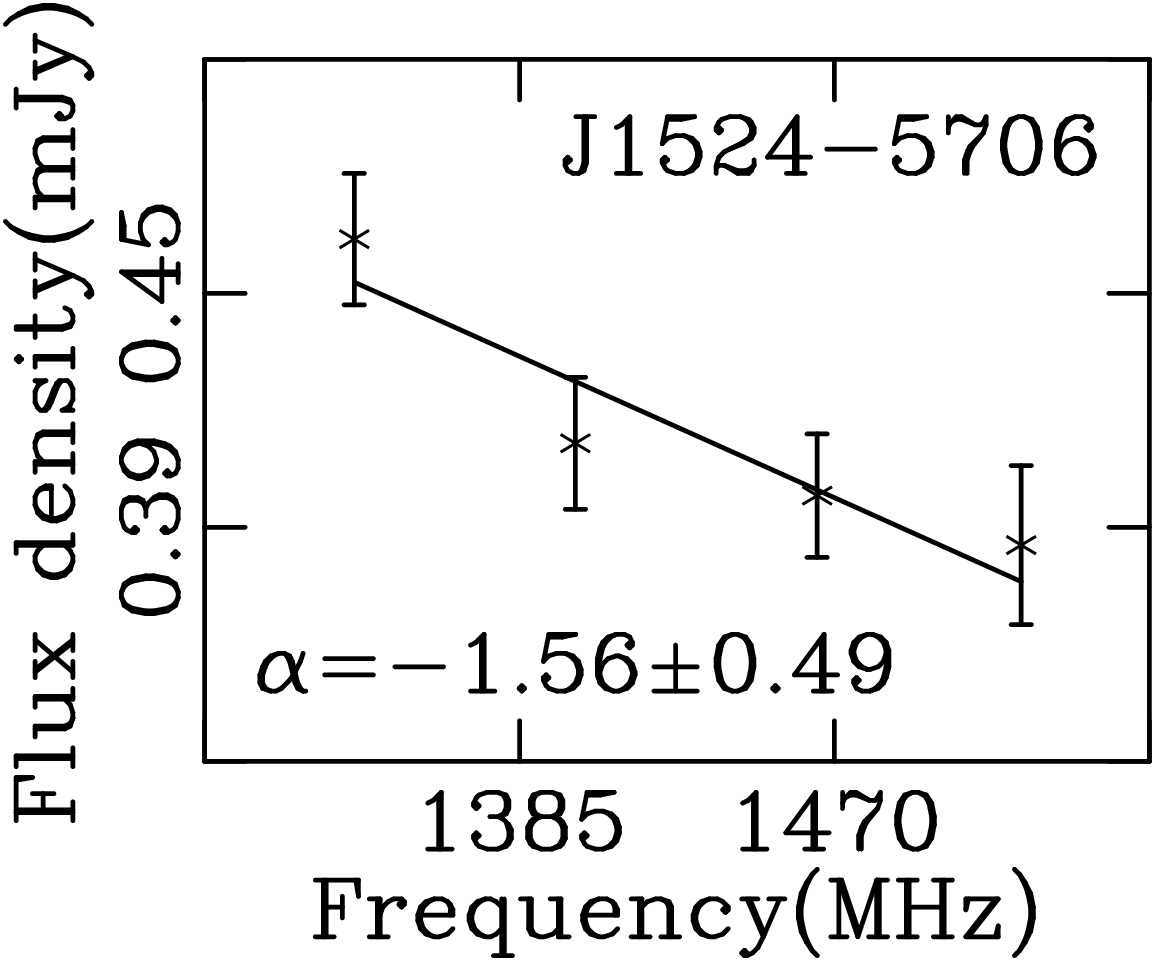}               
\includegraphics[bb = 110 120 518 685,clip,angle=-90,width=0.16\textwidth]{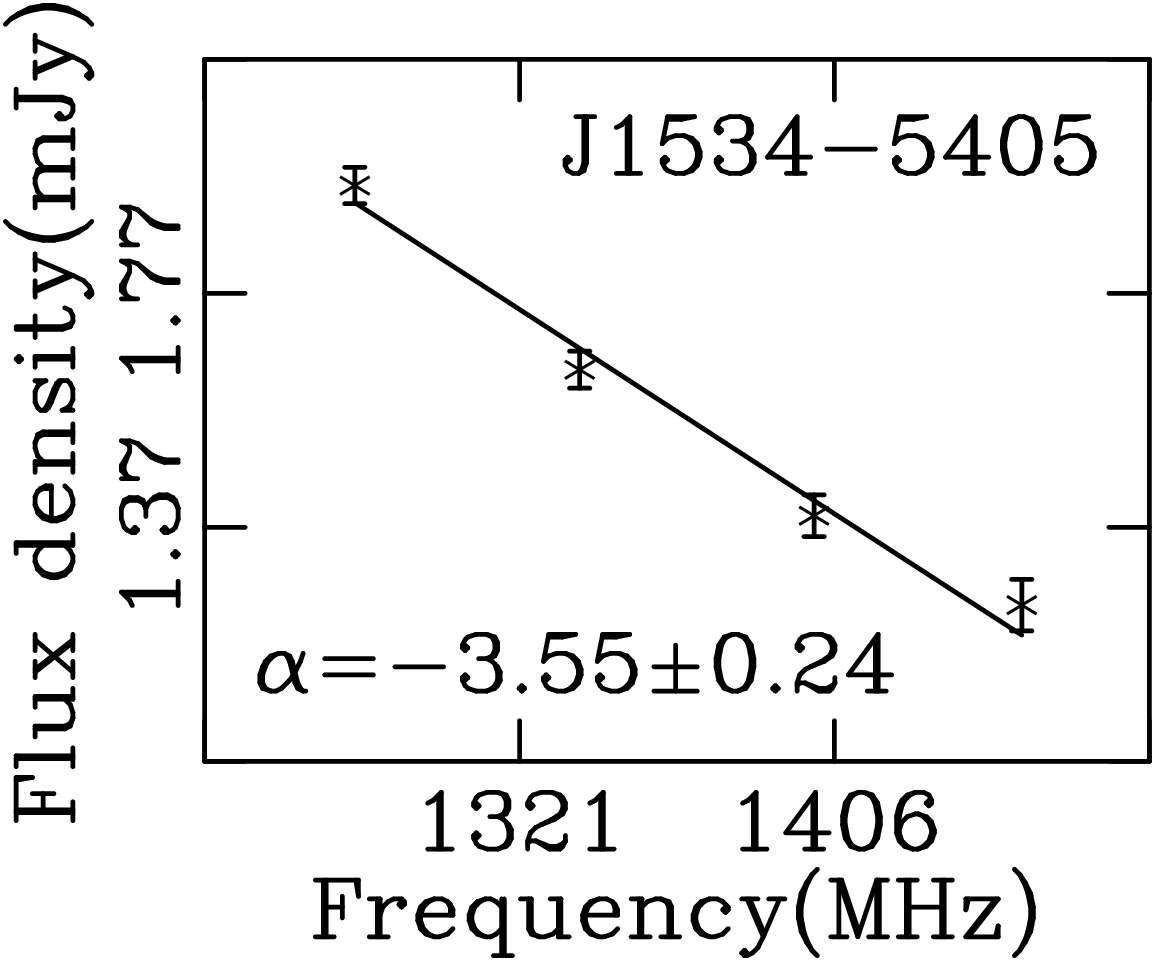}               
\includegraphics[bb = 110 120 518 685,clip,angle=-90,width=0.16\textwidth]{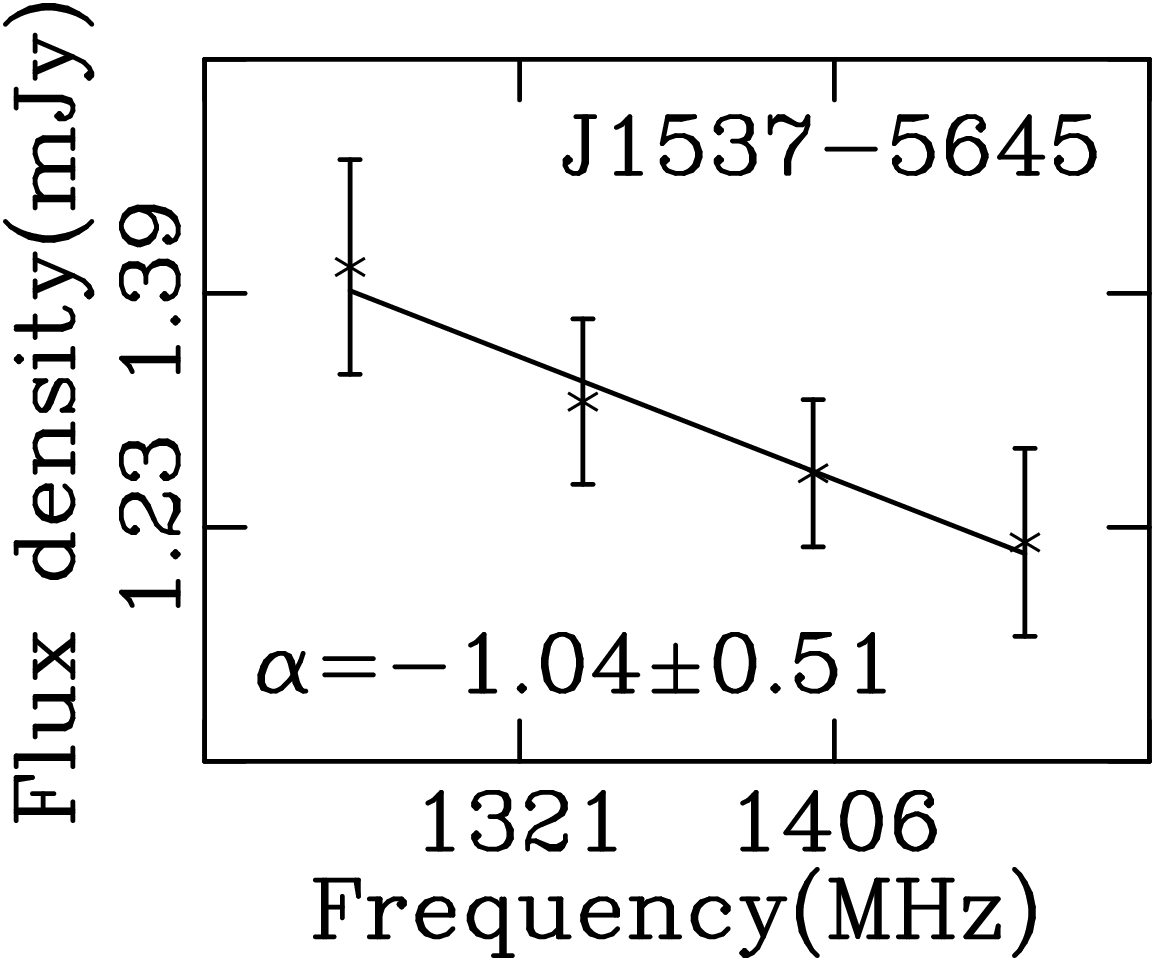}               
\includegraphics[bb = 110 120 518 685,clip,angle=-90,width=0.16\textwidth]{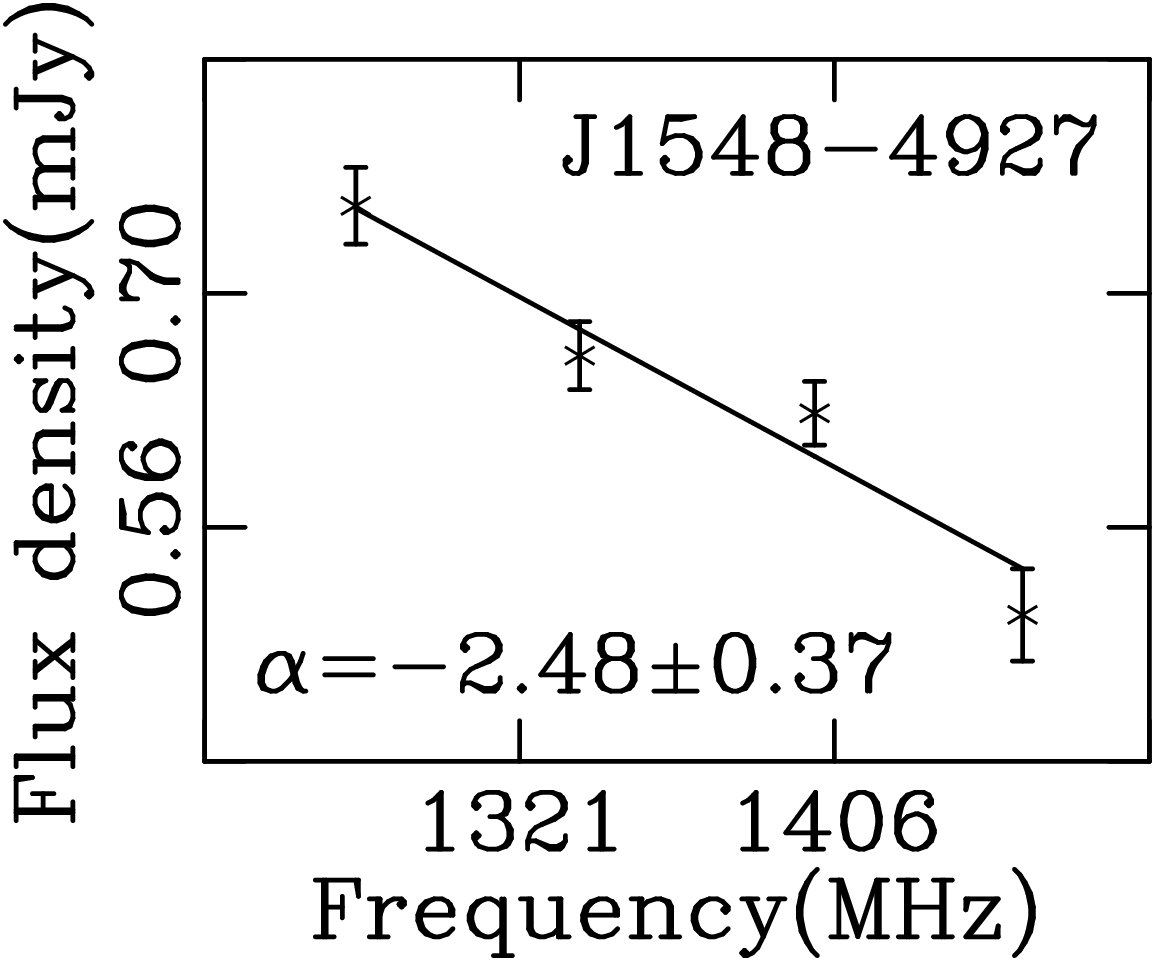}
 \\[2mm]              
\includegraphics[bb = 110 72 518 685,clip,angle=-90,width=0.175\textwidth]{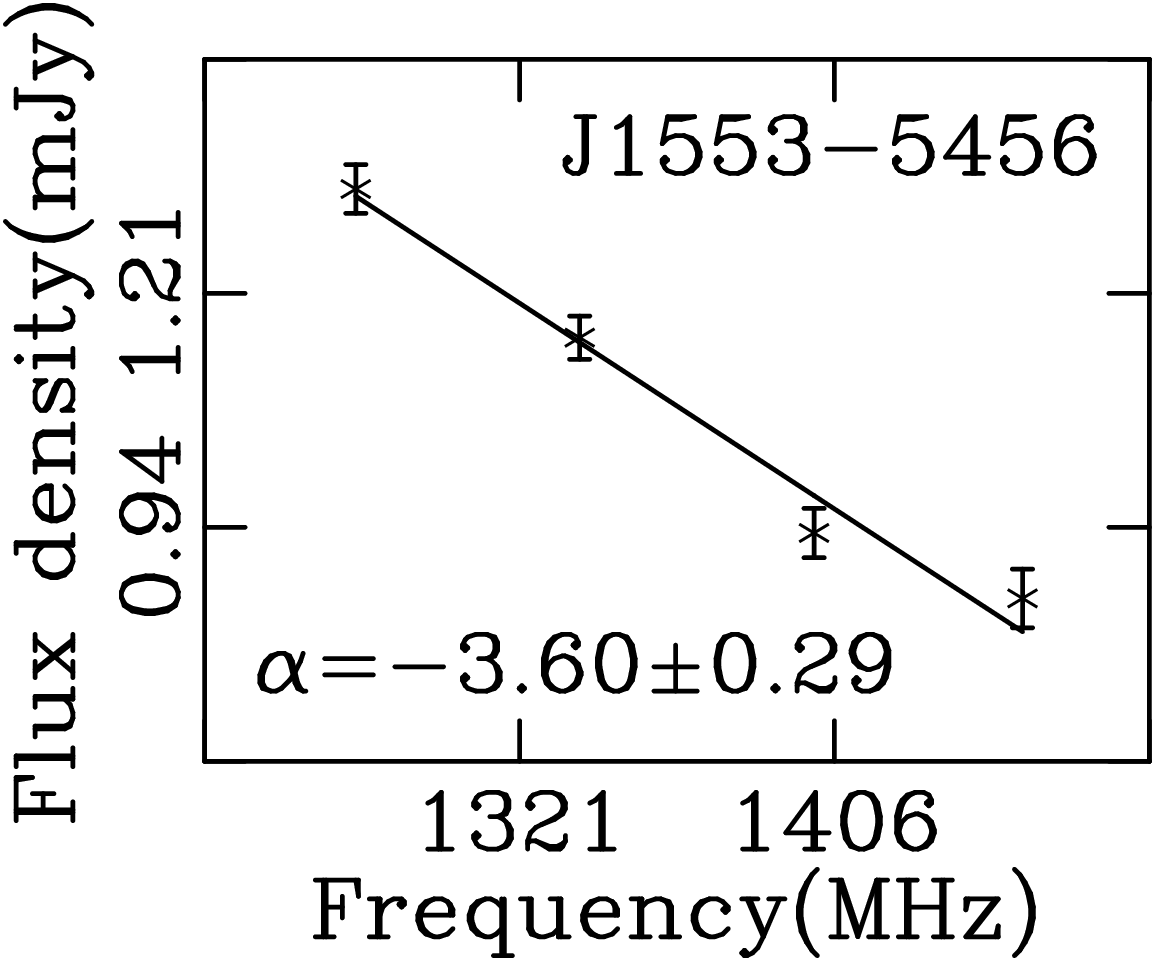}               
\includegraphics[bb = 110 120 518 685,clip,angle=-90,width=0.16\textwidth]{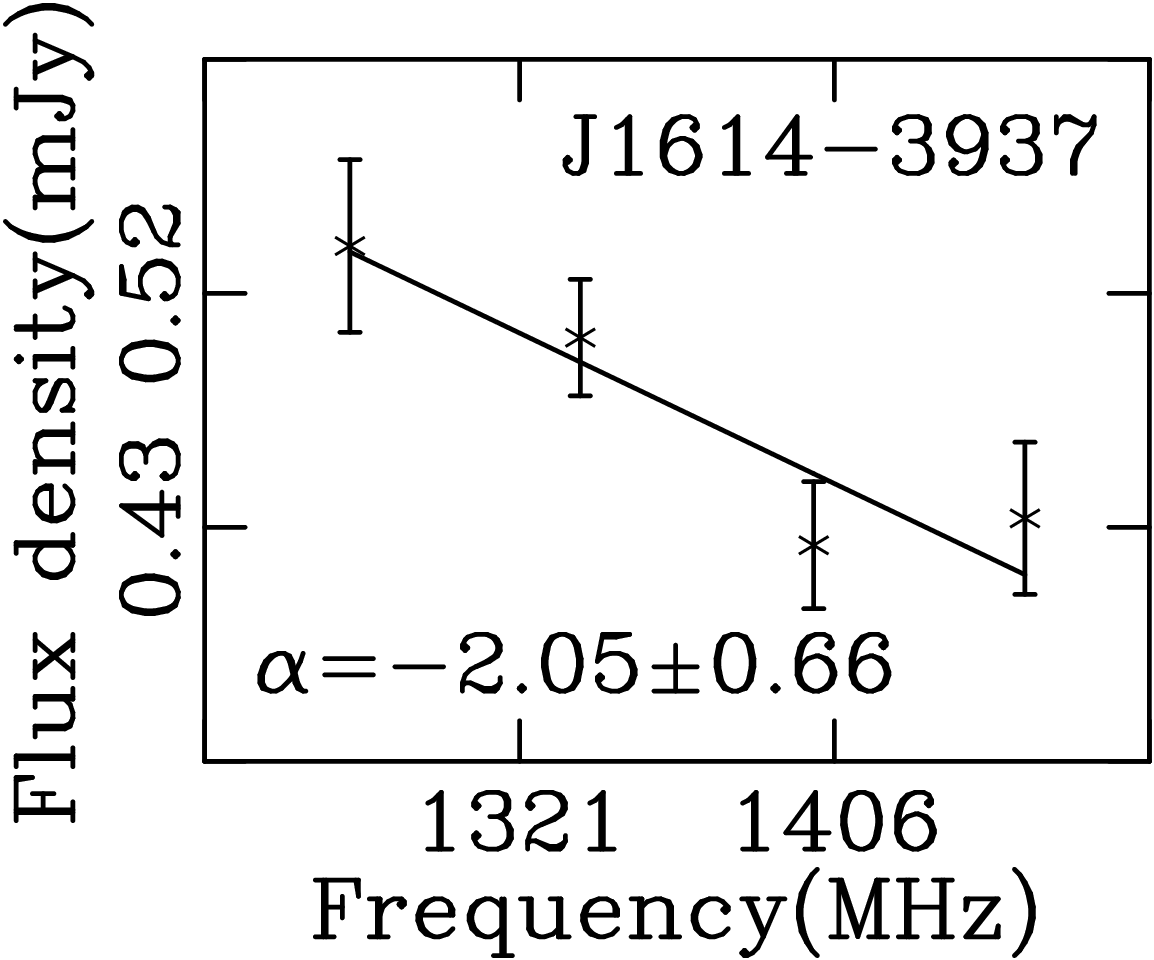}               
\includegraphics[bb = 110 120 518 685,clip,angle=-90,width=0.16\textwidth]{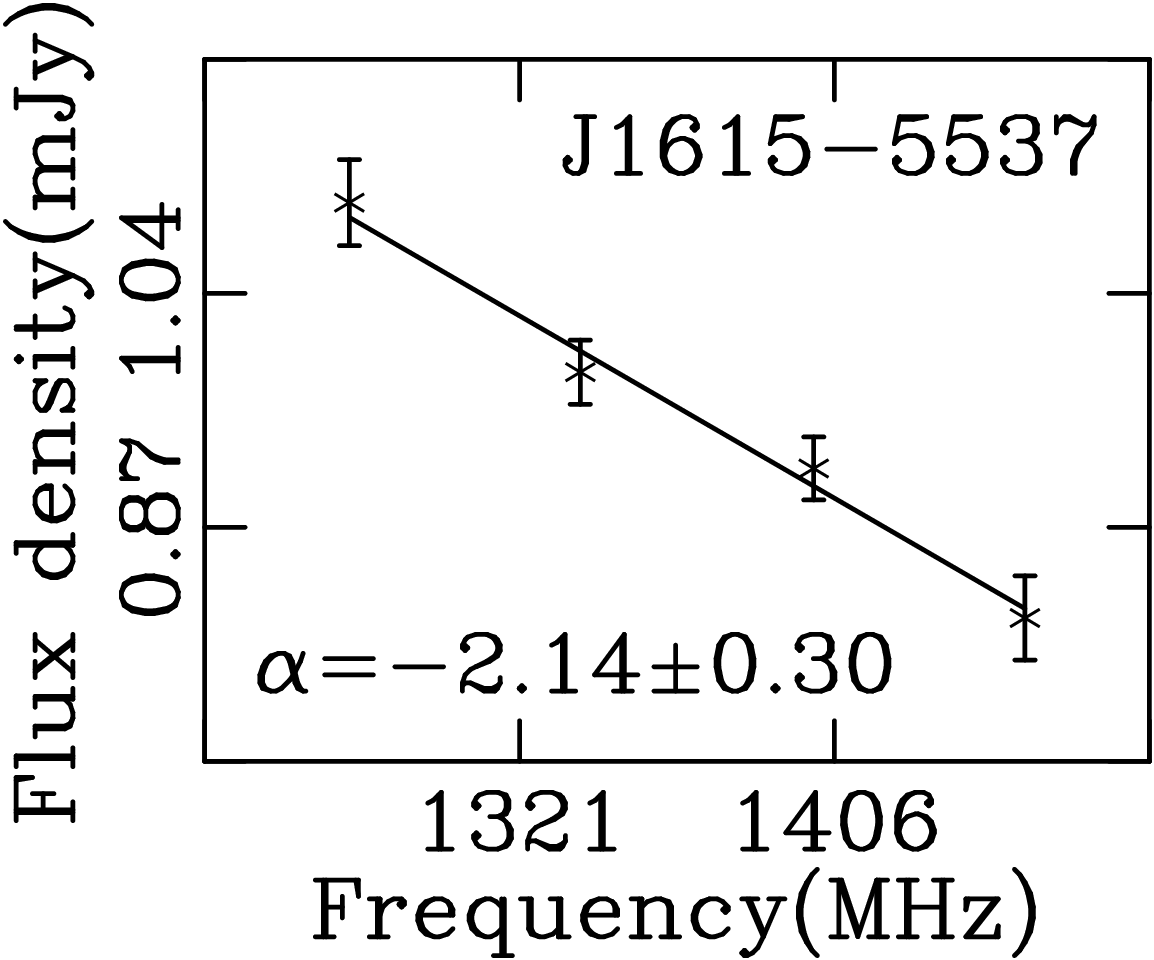}               
\includegraphics[bb = 110 120 518 685,clip,angle=-90,width=0.16\textwidth]{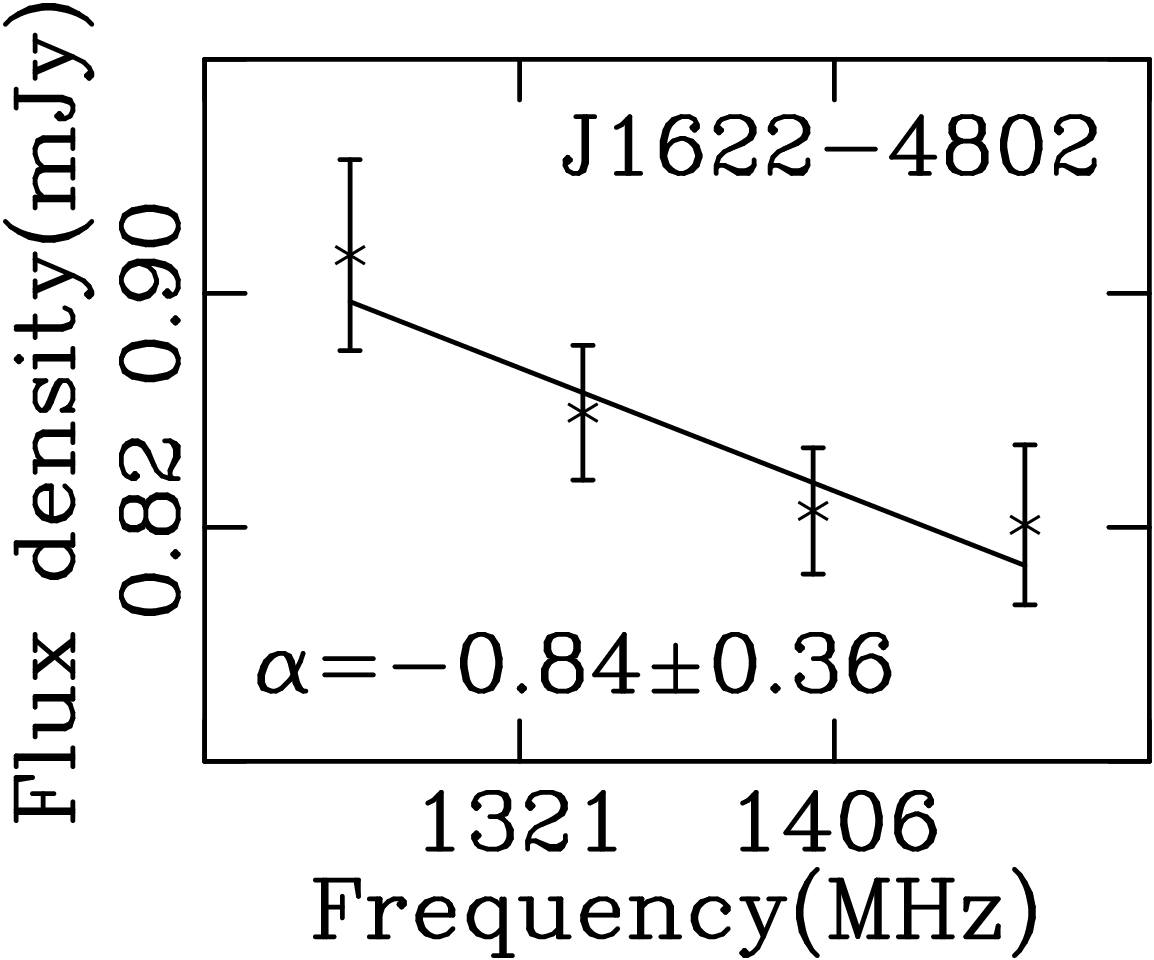}               
\includegraphics[bb = 110 120 518 685,clip,angle=-90,width=0.16\textwidth]{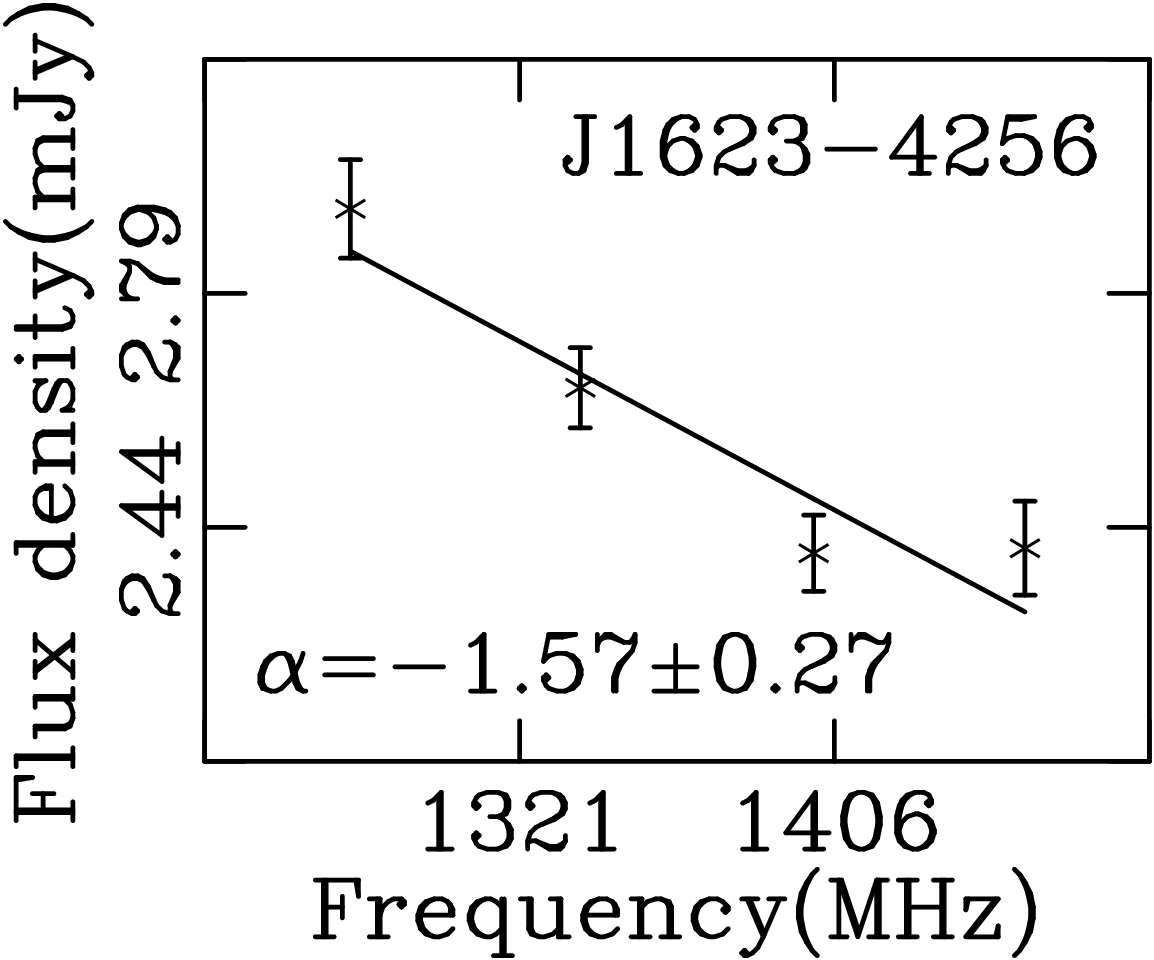}               
\includegraphics[bb = 110 120 518 685,clip,angle=-90,width=0.16\textwidth]{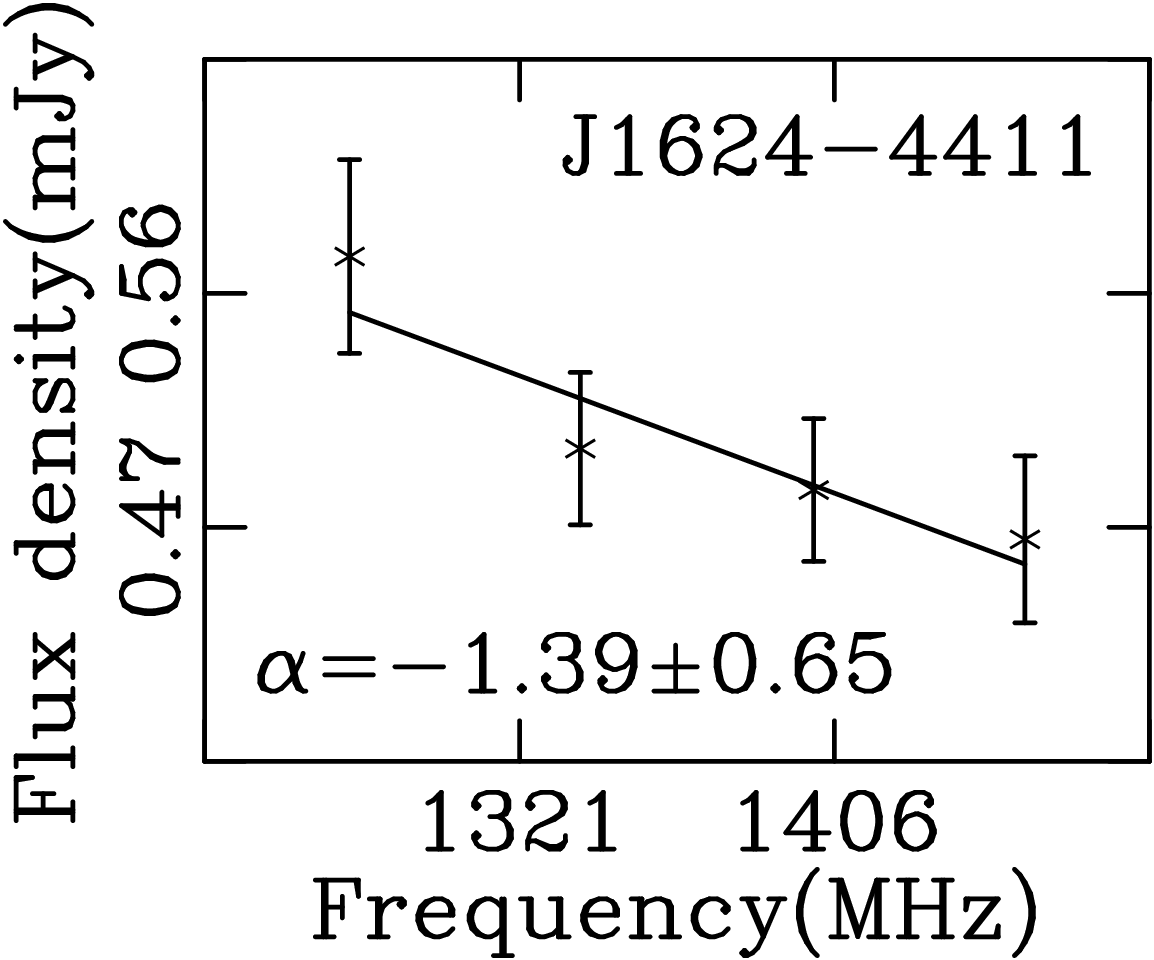}
 \\[2mm]              
\includegraphics[bb = 110 72 518 685,clip,angle=-90,width=0.175\textwidth]{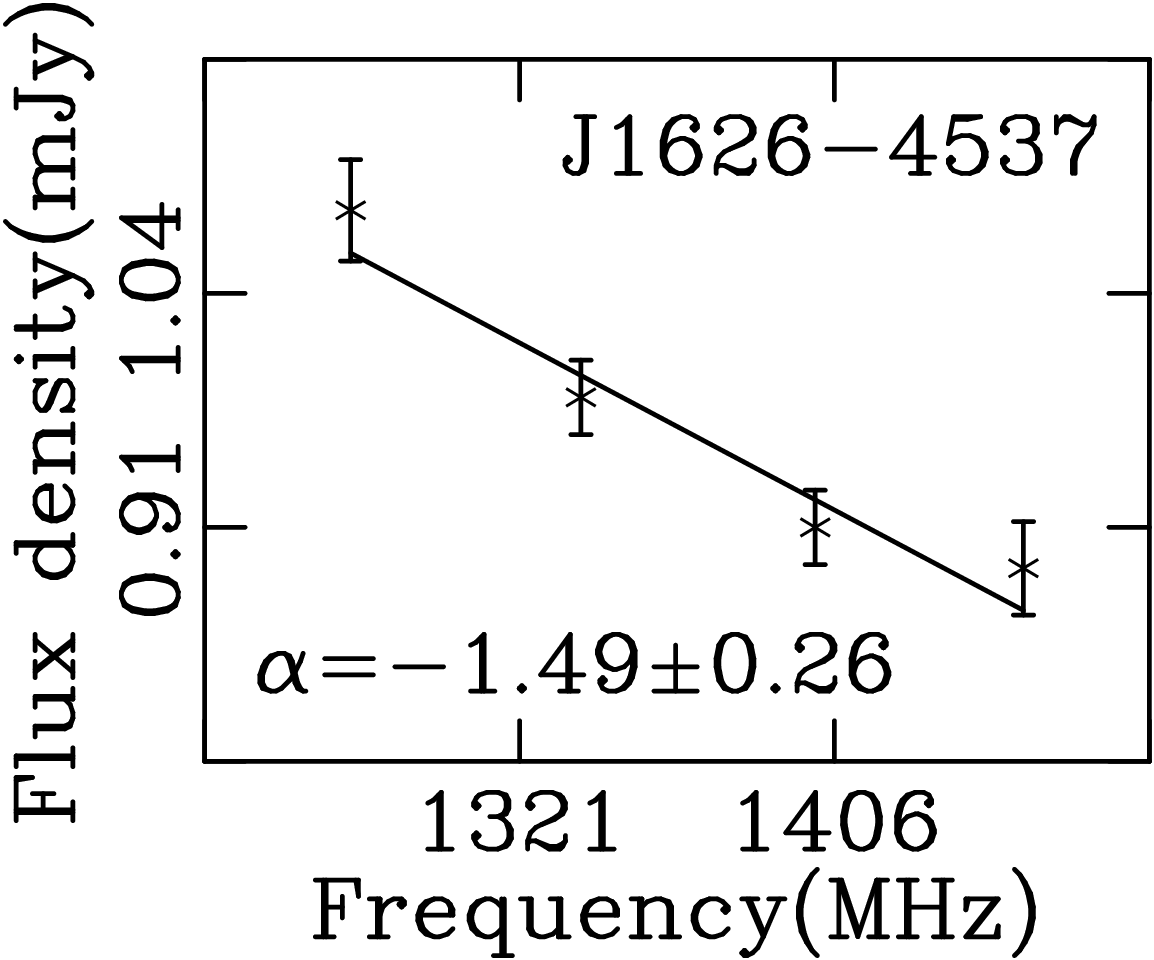}               
\includegraphics[bb = 110 120 518 685,clip,angle=-90,width=0.16\textwidth]{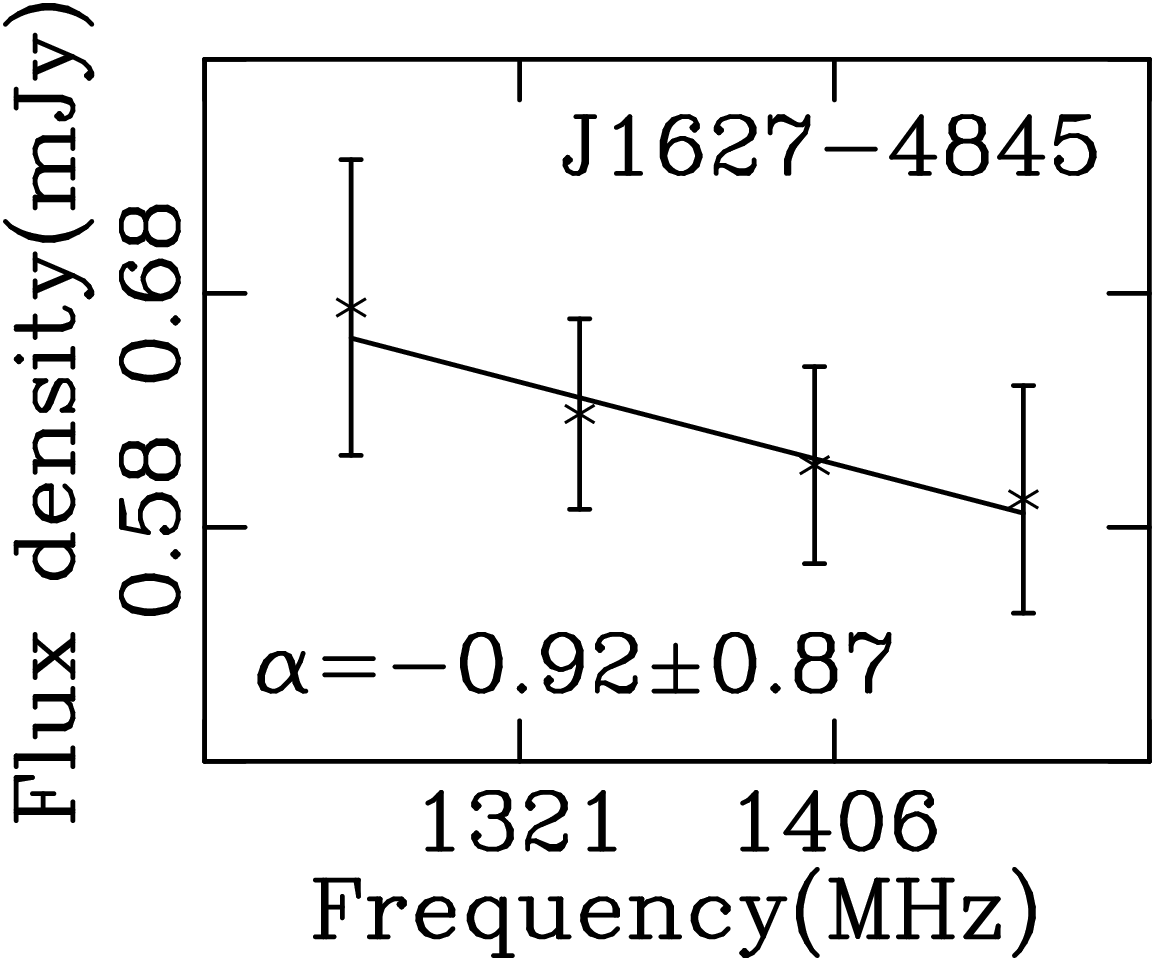}               
\includegraphics[bb = 110 120 518 685,clip,angle=-90,width=0.16\textwidth]{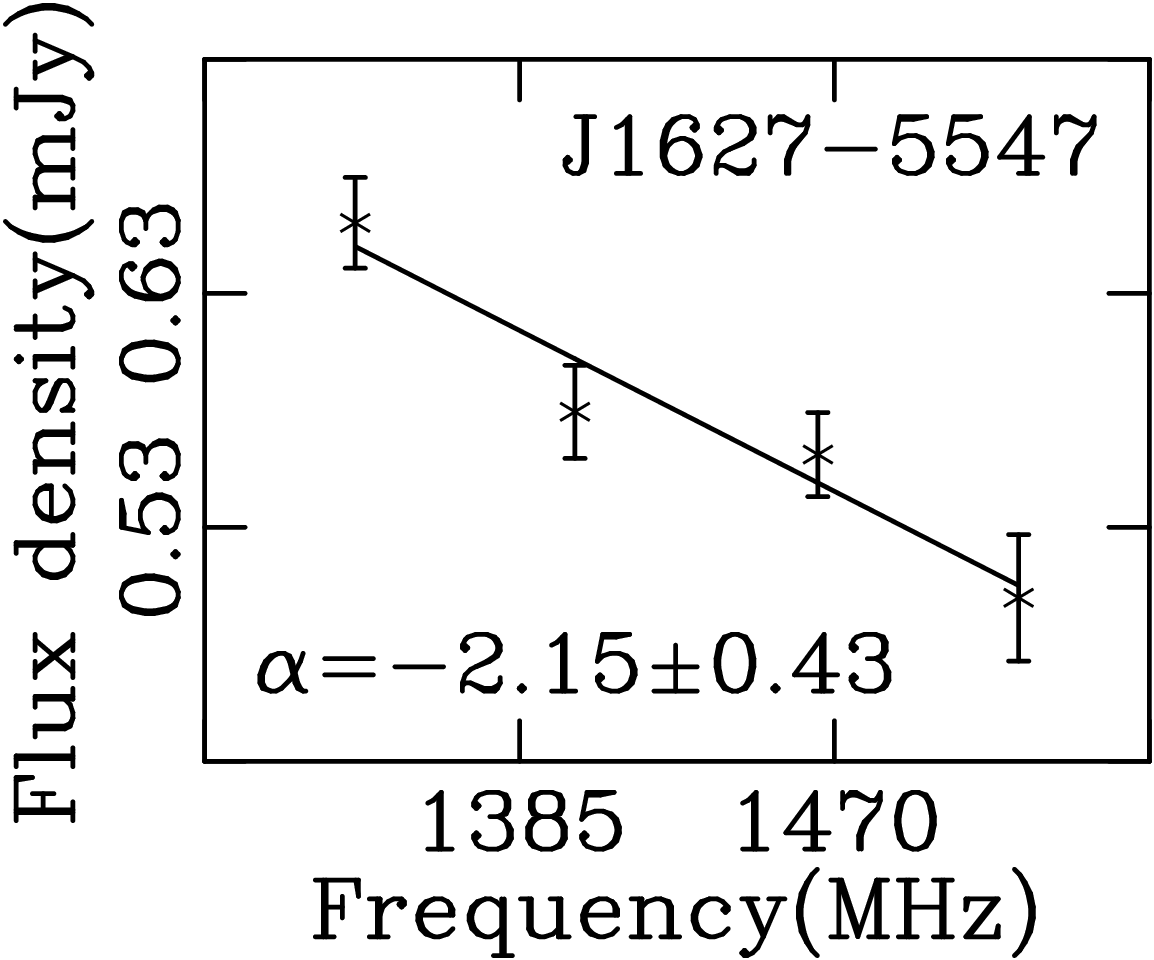}                   
\includegraphics[bb = 110 120 518 685,clip,angle=-90,width=0.16\textwidth]{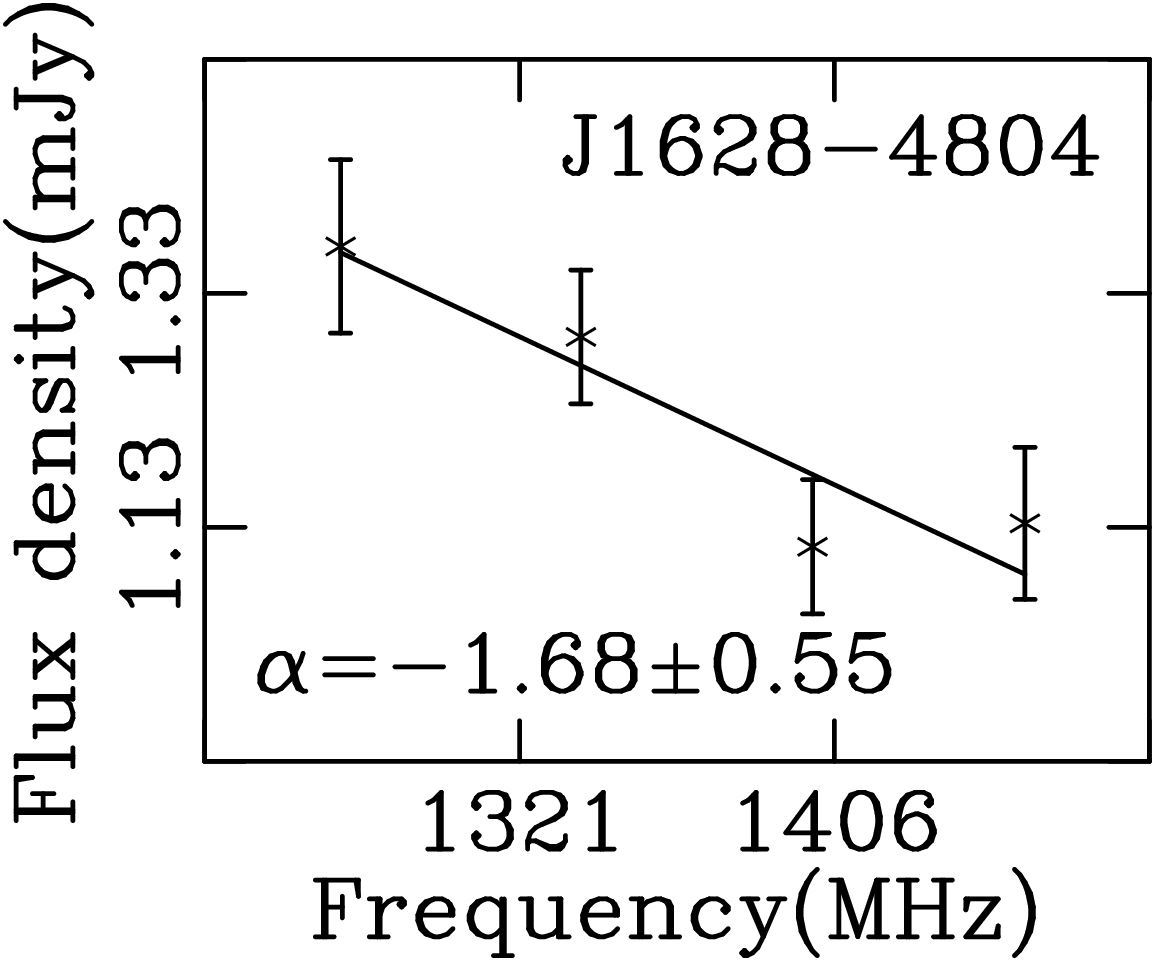}              
\includegraphics[bb = 110 120 518 685,clip,angle=-90,width=0.16\textwidth]{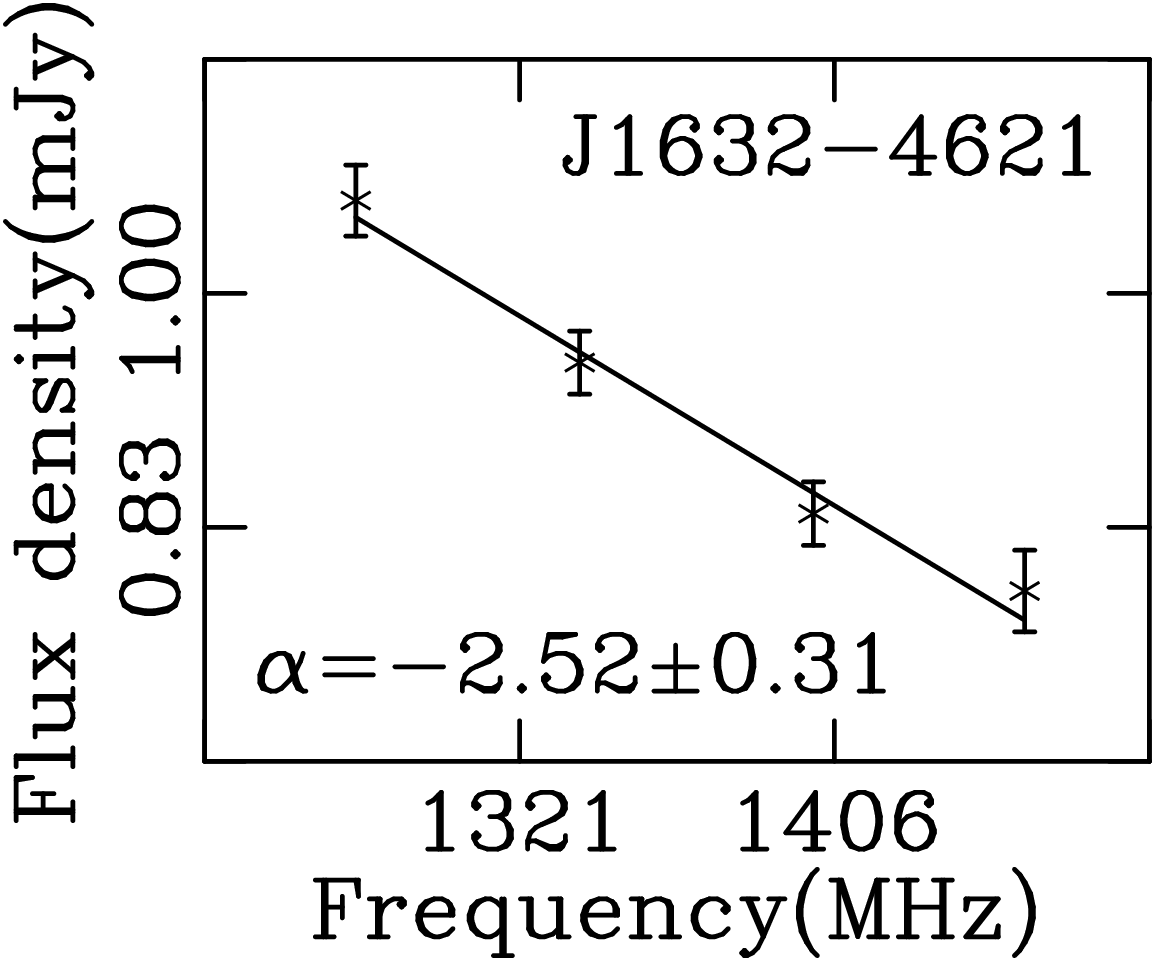}              
\includegraphics[bb = 110 120 518 685,clip,angle=-90,width=0.16\textwidth]{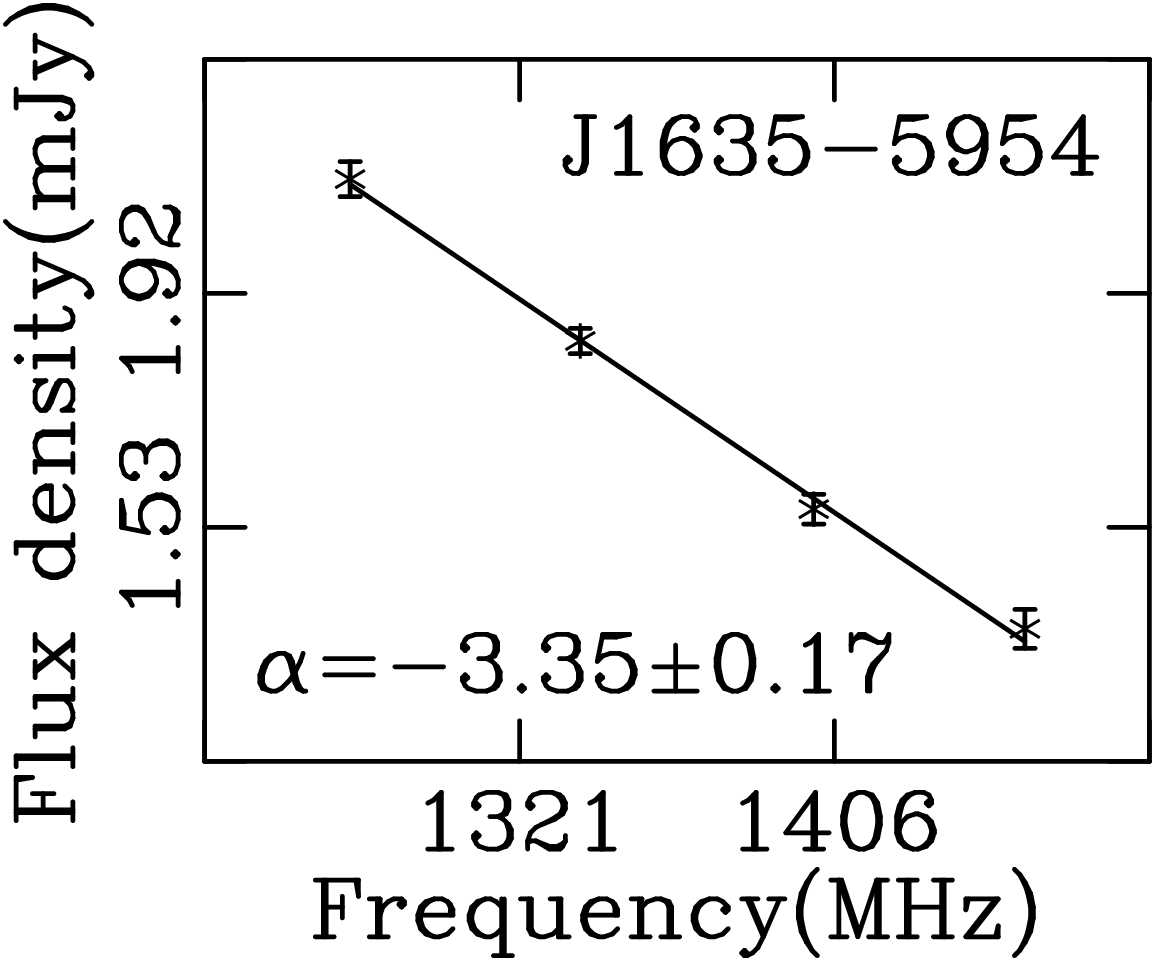}
 \\[2mm]             
\includegraphics[bb = 110 72 518 685,clip,angle=-90,width=0.175\textwidth]{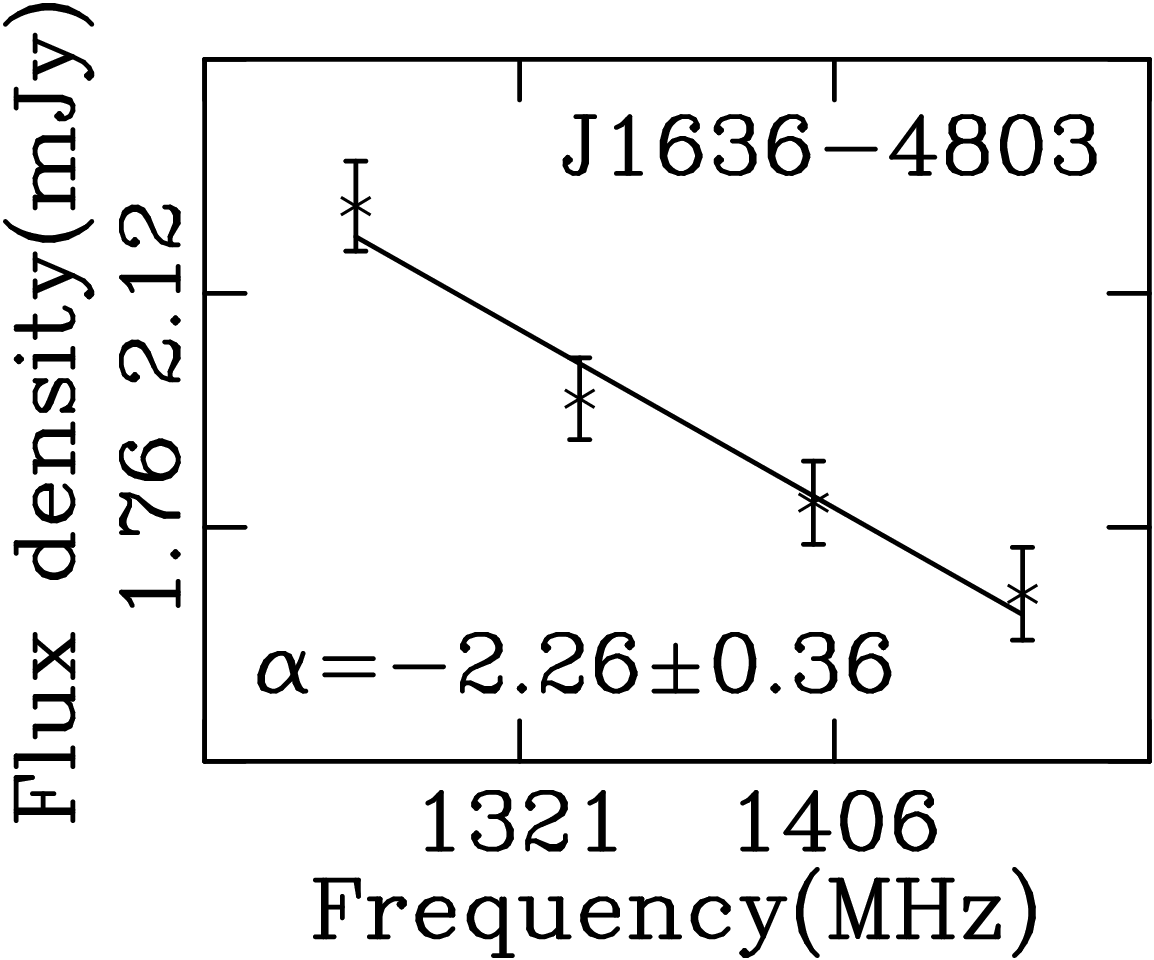}              
\includegraphics[bb = 110 120 518 685,clip,angle=-90,width=0.16\textwidth]{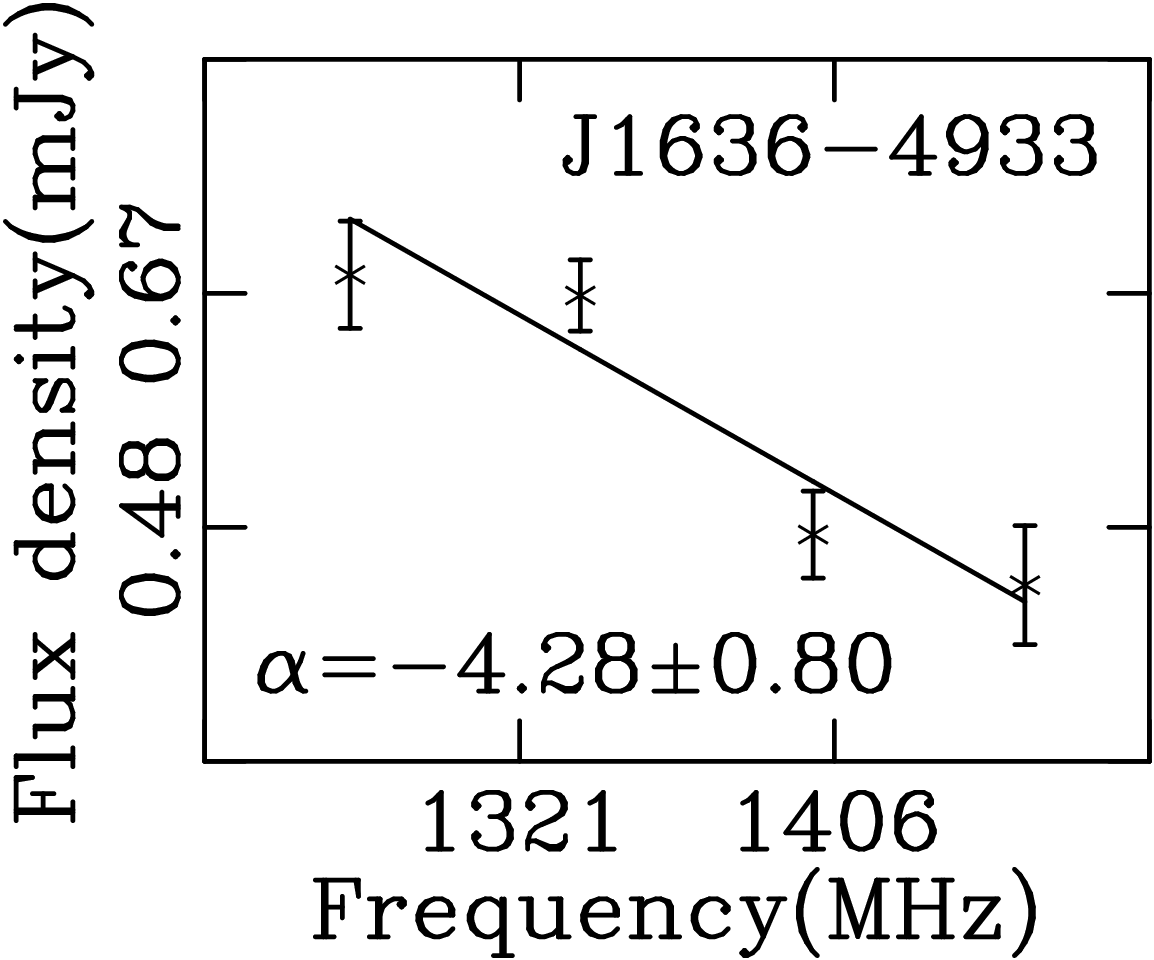}              
\includegraphics[bb = 110 120 518 685,clip,angle=-90,width=0.16\textwidth]{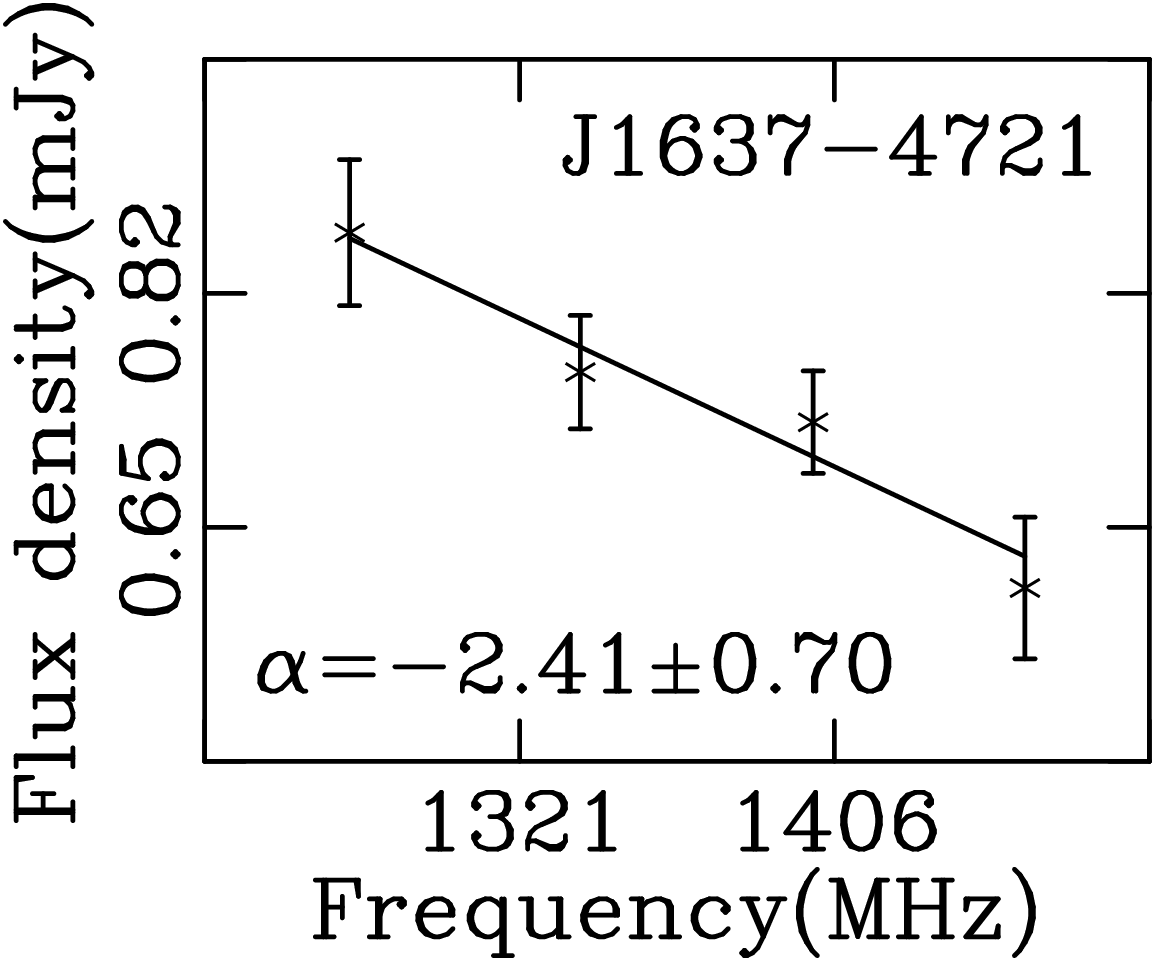}              
\includegraphics[bb = 110 120 518 685,clip,angle=-90,width=0.16\textwidth]{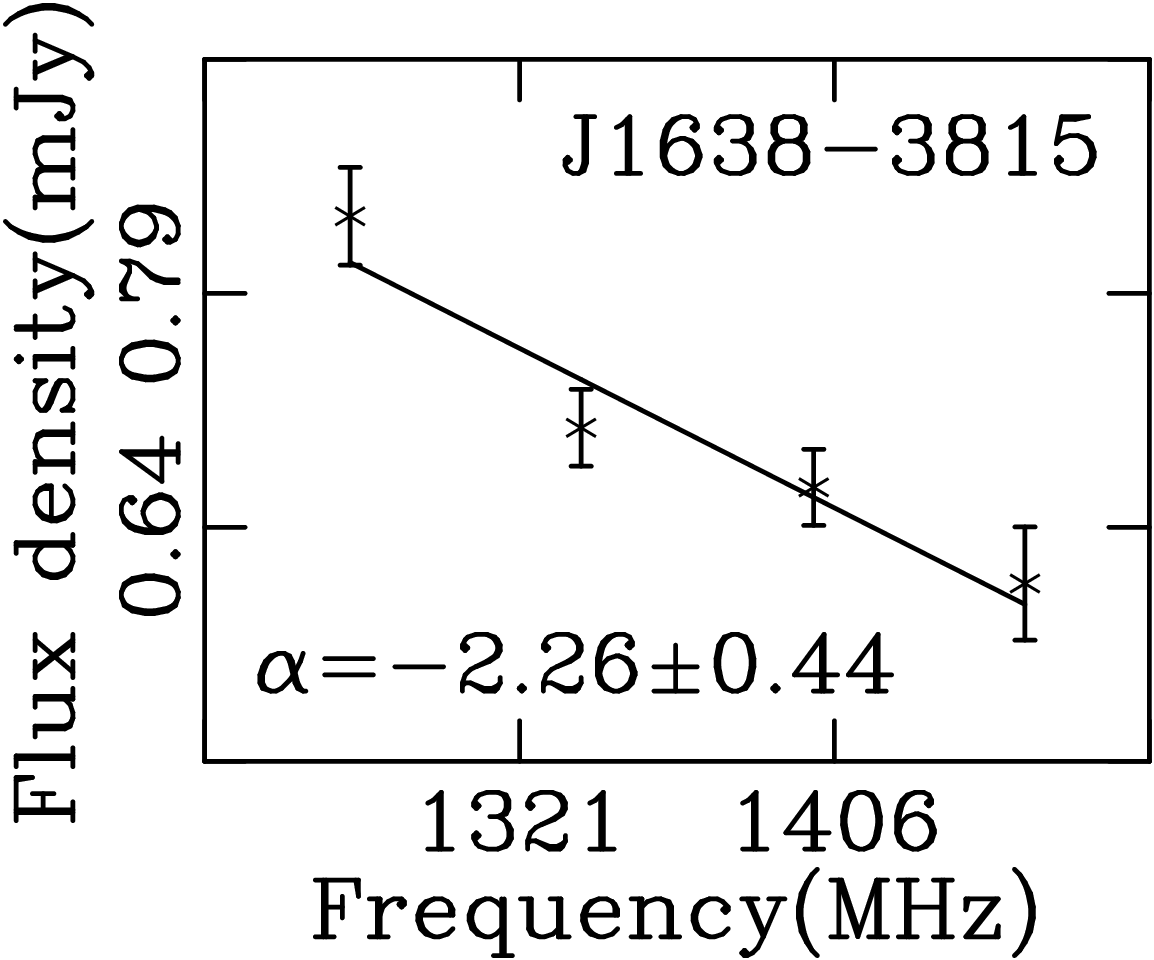}              
\includegraphics[bb = 110 120 518 685,clip,angle=-90,width=0.16\textwidth]{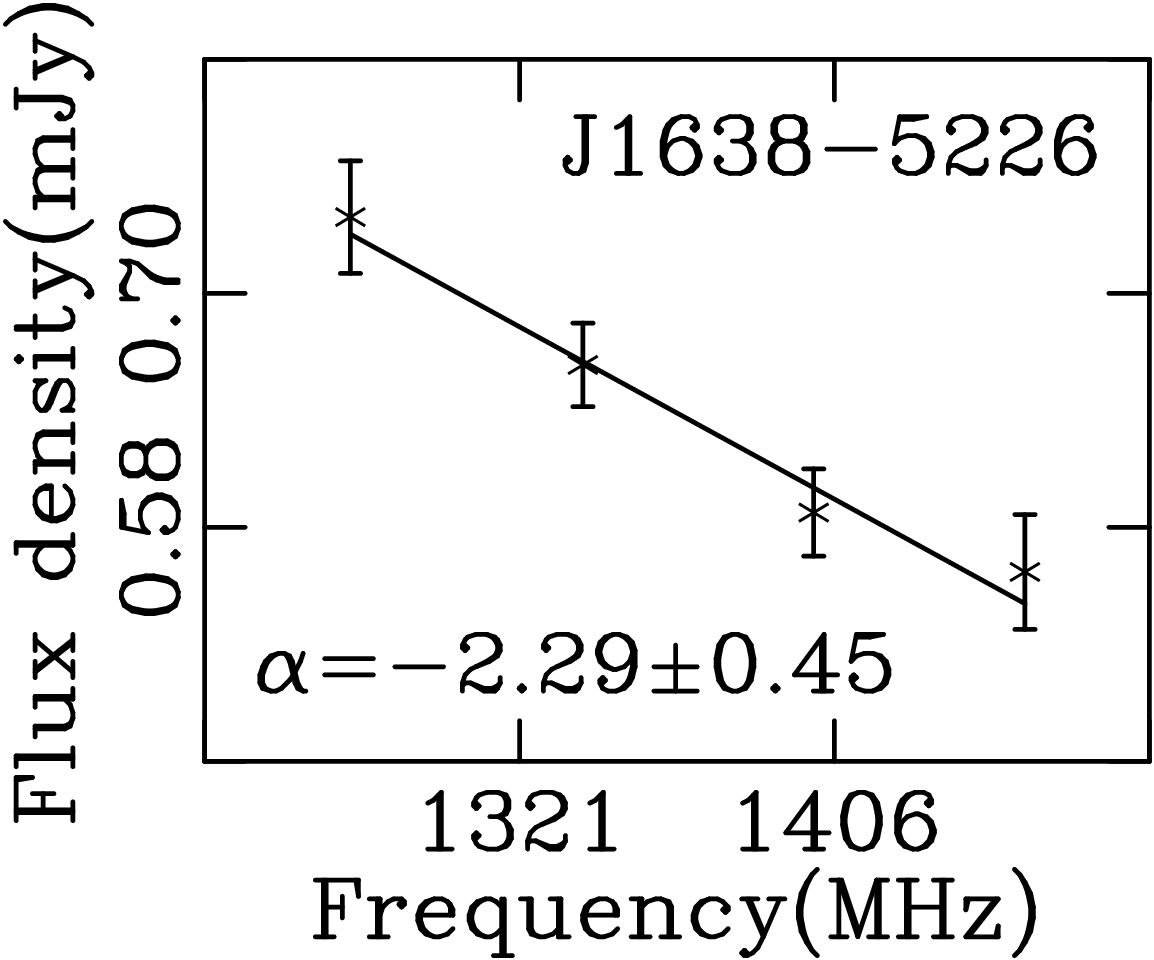}              
\includegraphics[bb = 110 120 518 685,clip,angle=-90,width=0.16\textwidth]{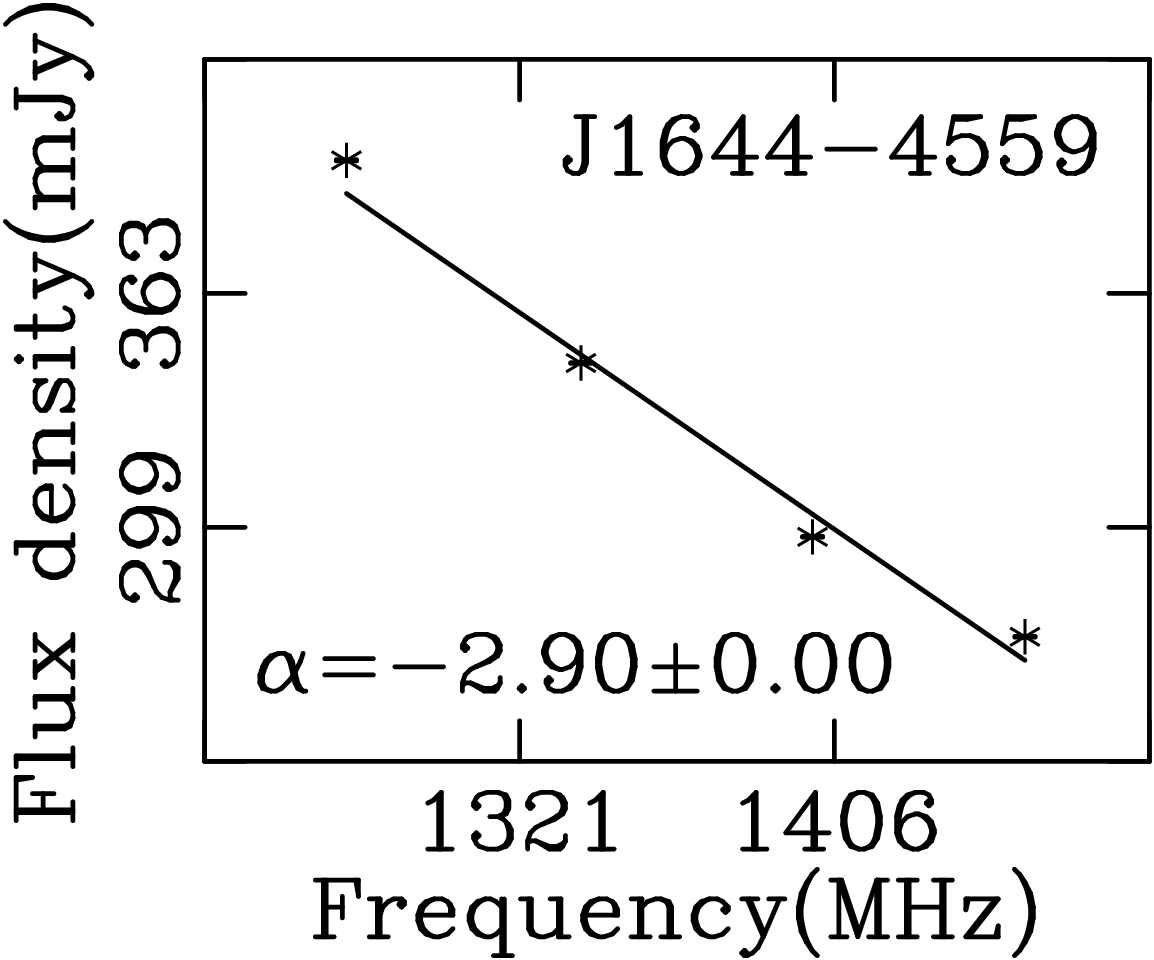}
 \\[2mm]             
\includegraphics[bb = 110 72 570 685,clip,angle=-90,width=0.175\textwidth]{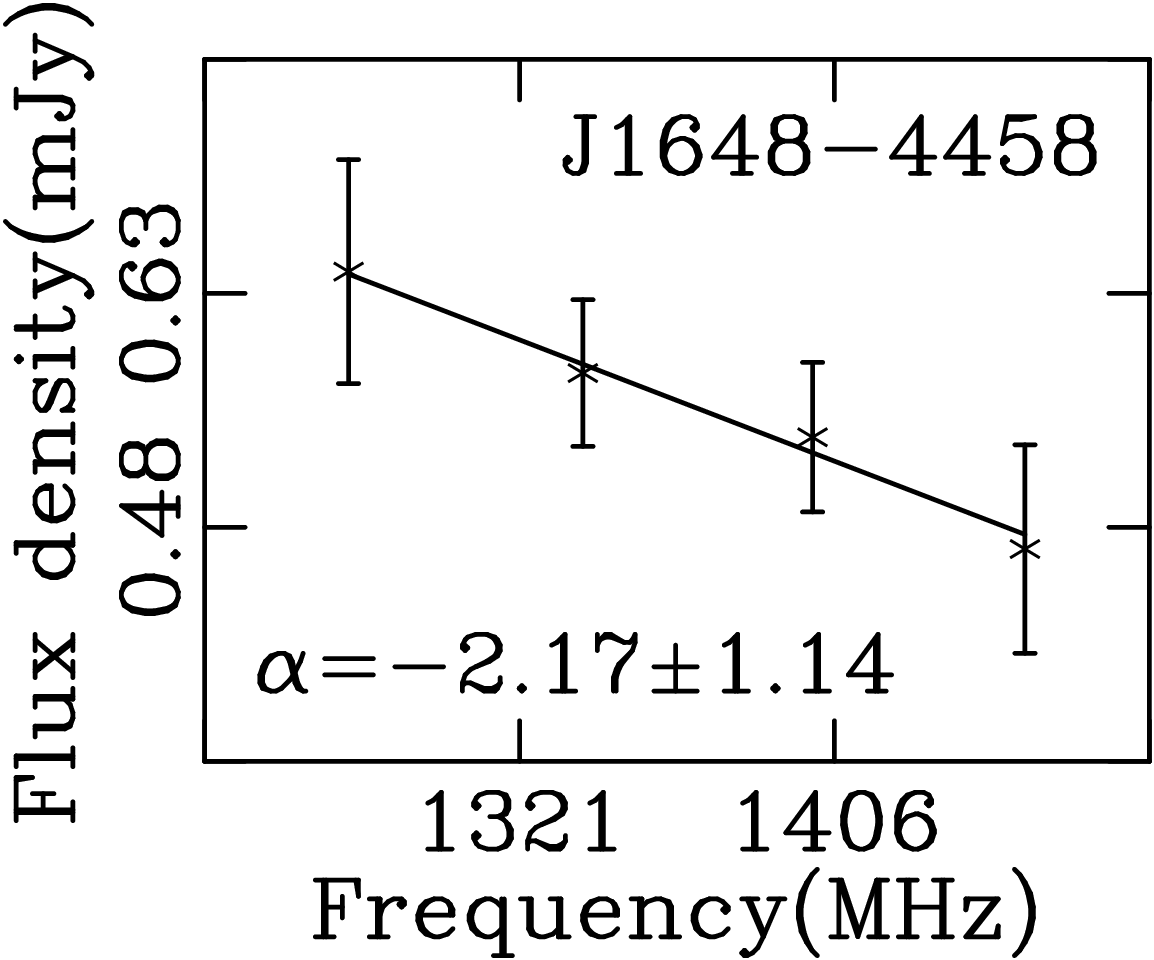}              
\includegraphics[bb = 110 120 570 685,clip,angle=-90,width=0.16\textwidth]{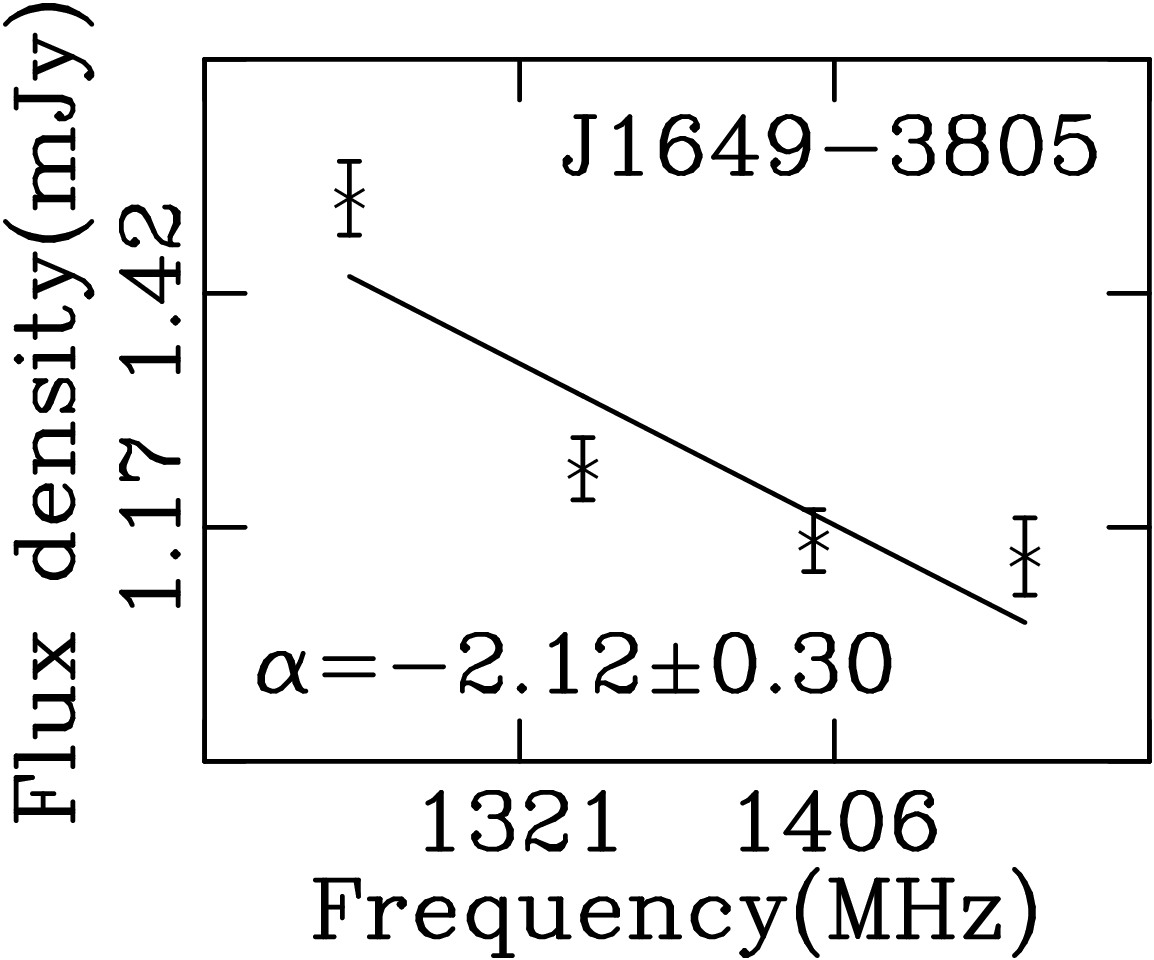}              
\includegraphics[bb = 110 120 570 685,clip,angle=-90,width=0.16\textwidth]{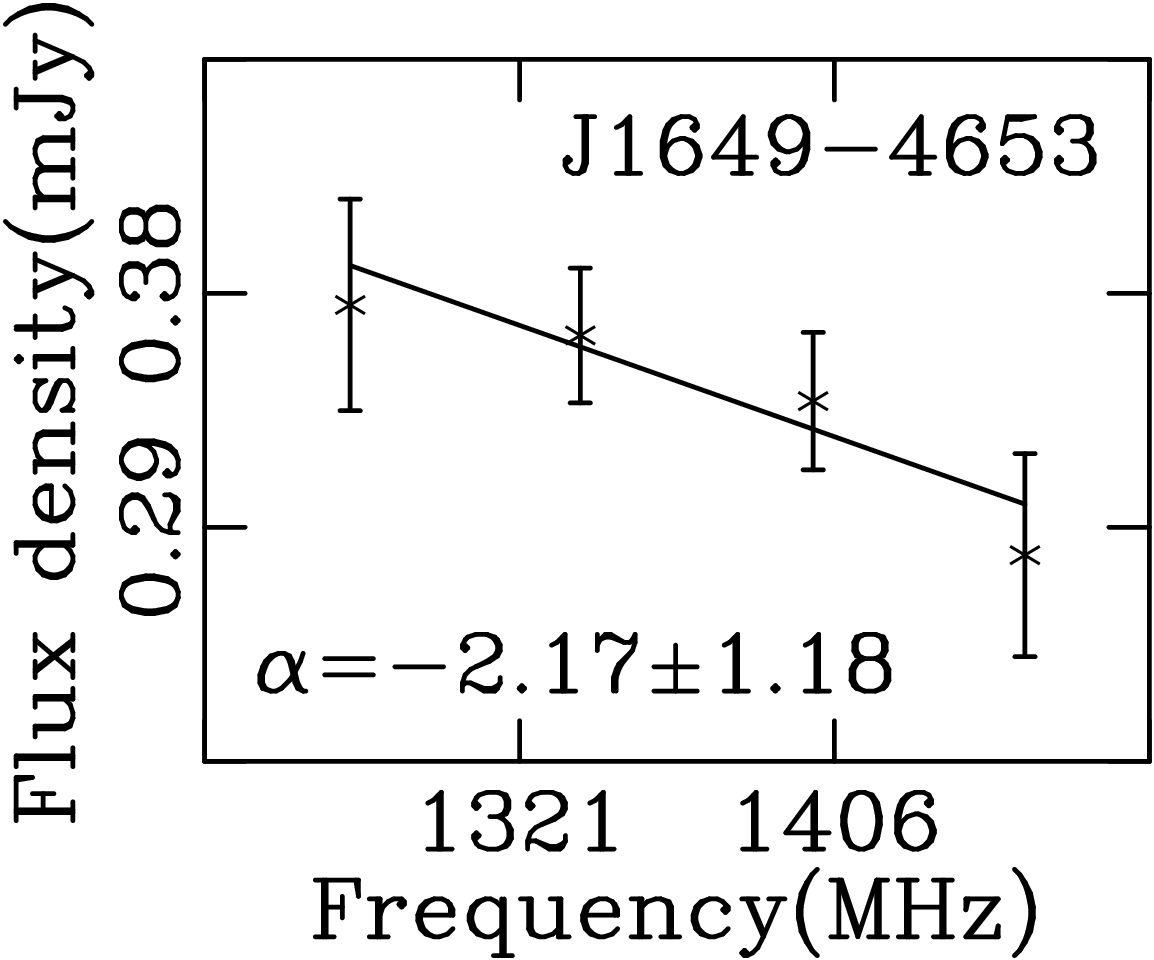}              
\includegraphics[bb = 110 120 570 685,clip,angle=-90,width=0.16\textwidth]{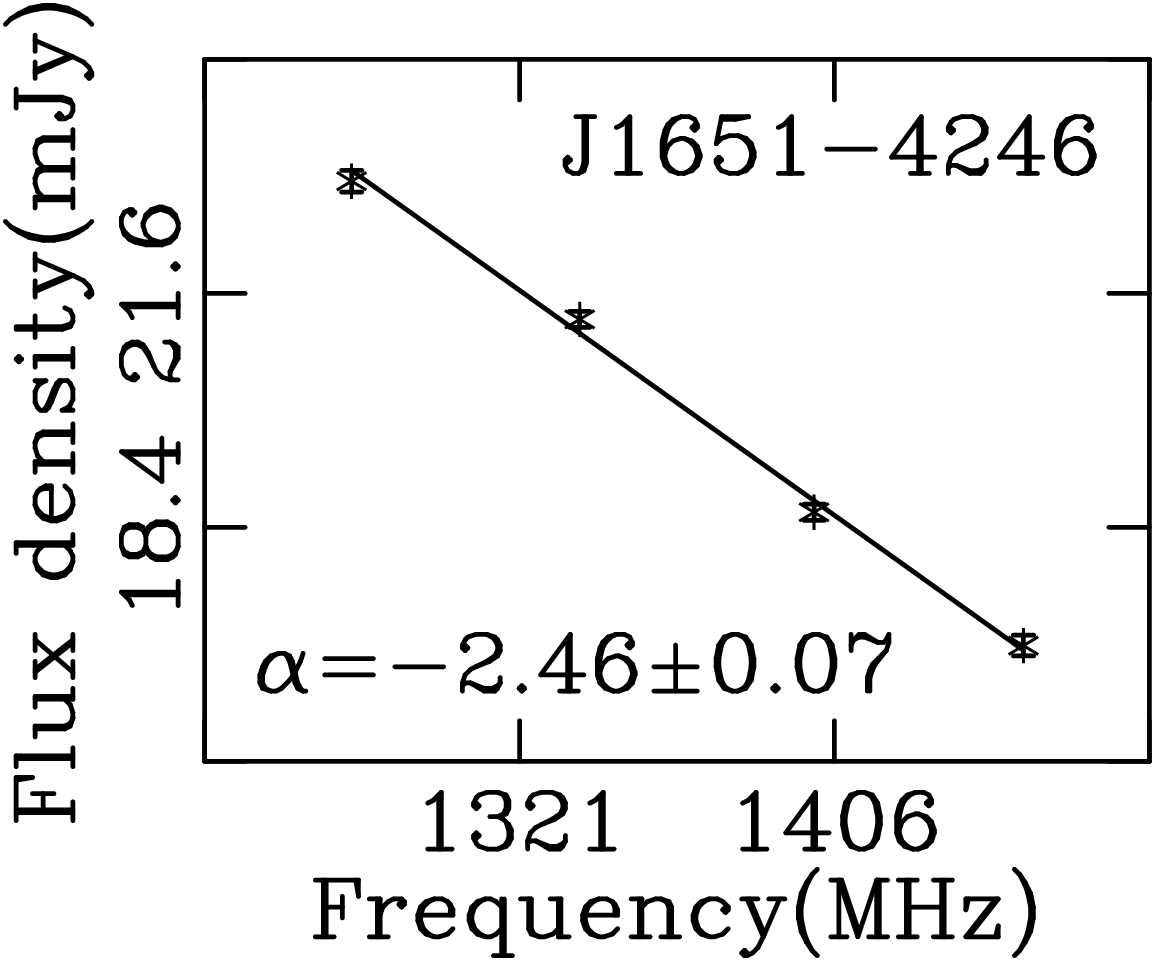}              
\includegraphics[bb = 110 120 570 685,clip,angle=-90,width=0.16\textwidth]{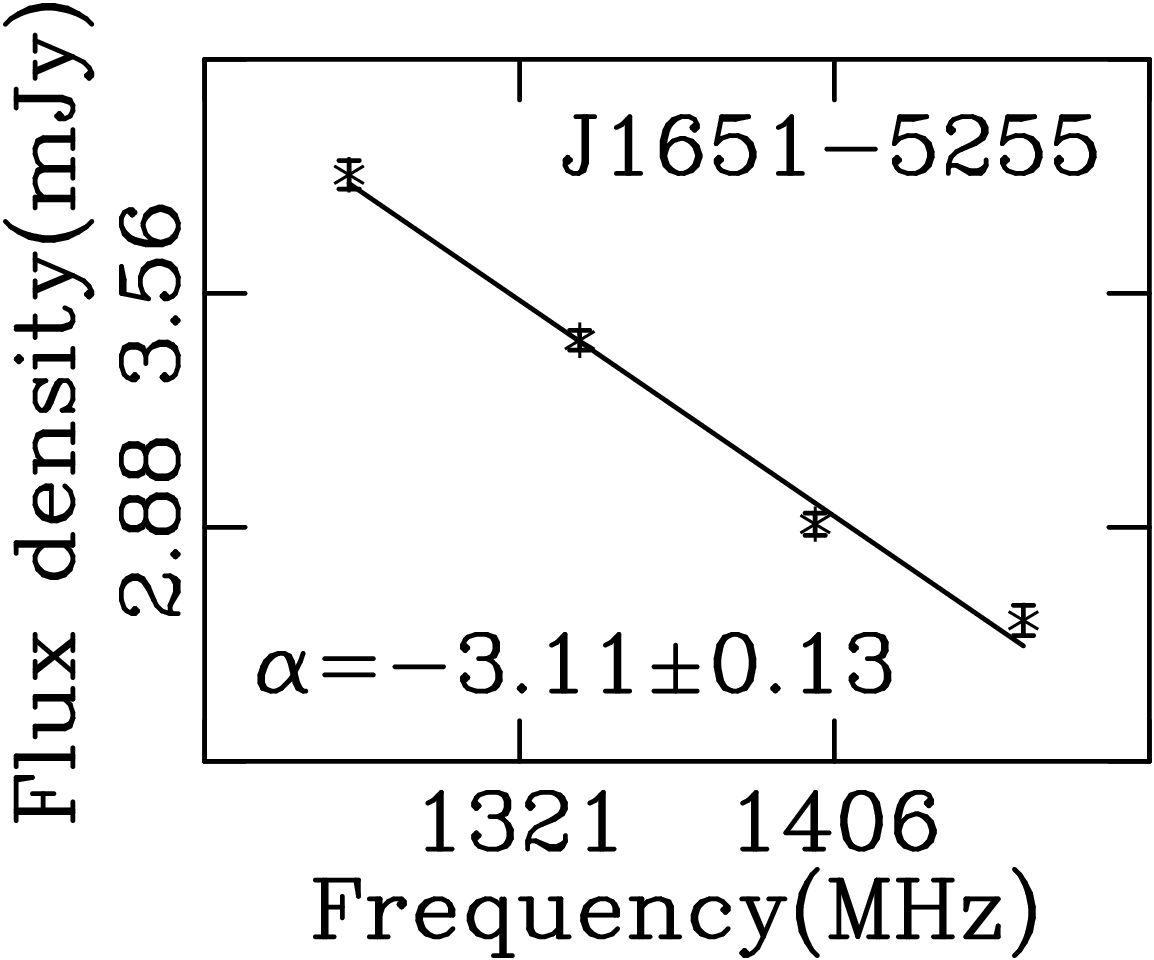}              
\includegraphics[bb = 110 120 570 685,clip,angle=-90,width=0.16\textwidth]{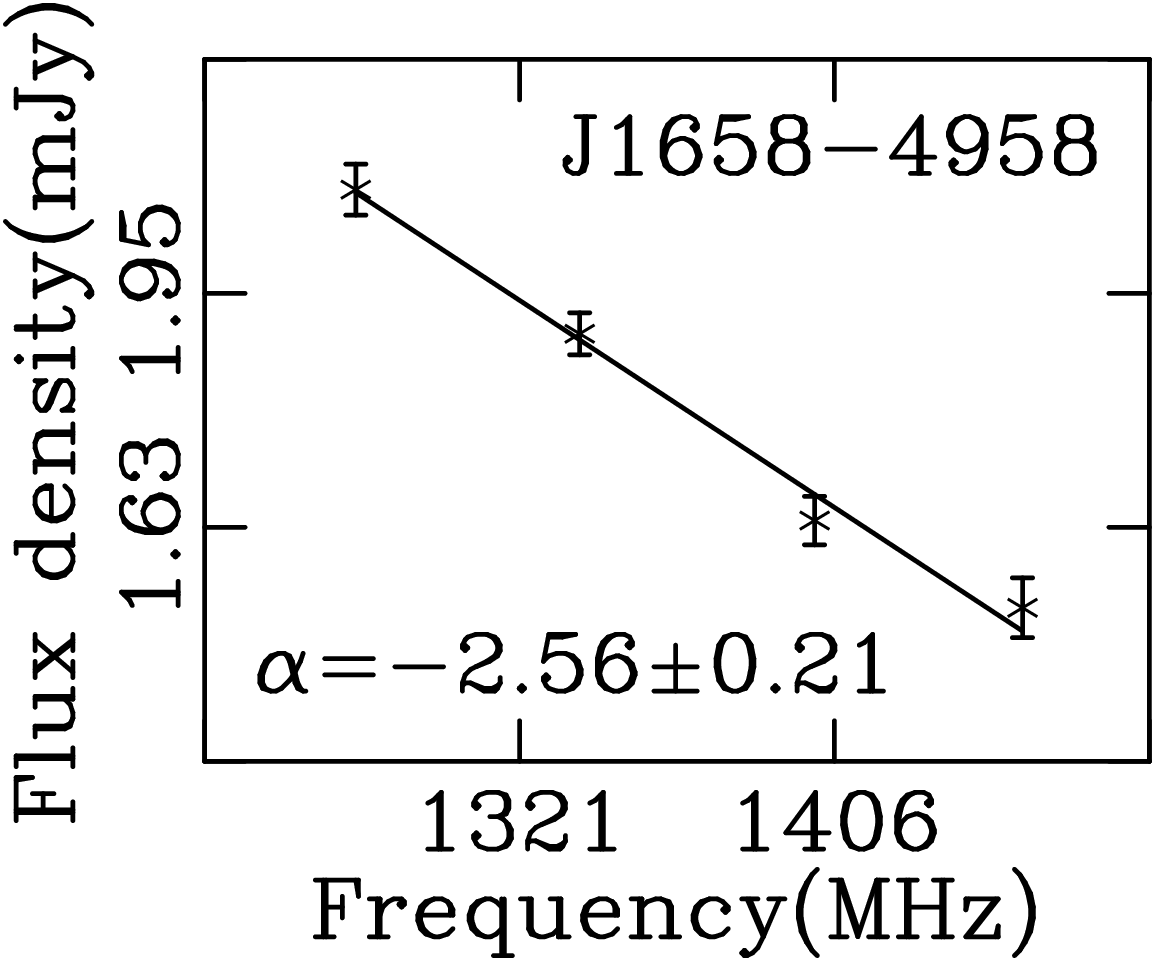}
 \center
\caption{The flux densities at 4 subbands are fitted for spectral
  indices of 228 pulsars. Both the X-axis for the frequency and Y-axis
  for the flux density are plotted in the logarithm scale. To be
  continued in the next pages.}
\end{figure}

\begin{figure}
  \center            
\includegraphics[bb = 110 72 518 685,clip,angle=-90,width=0.175\textwidth]{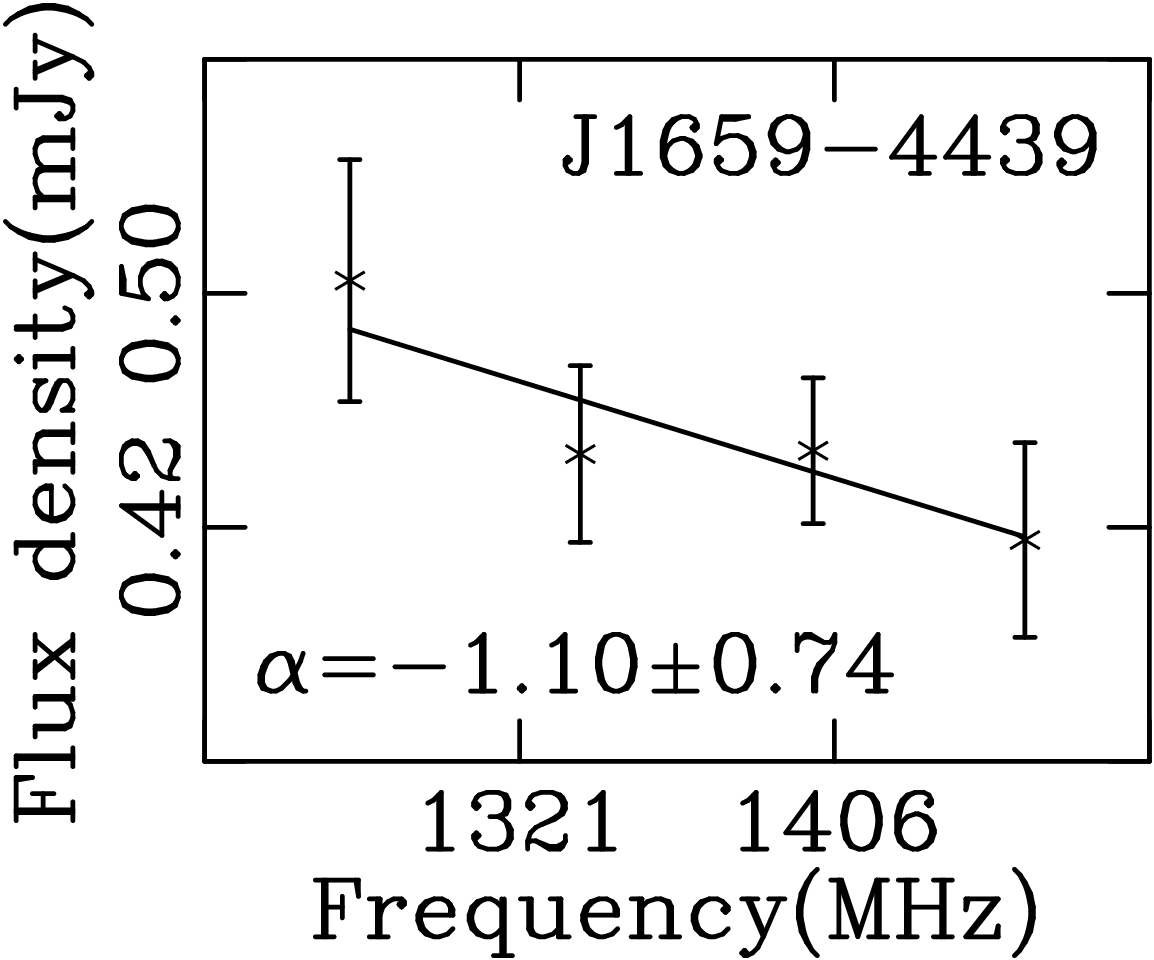}              
\includegraphics[bb = 110 120 518 685,clip,angle=-90,width=0.16\textwidth]{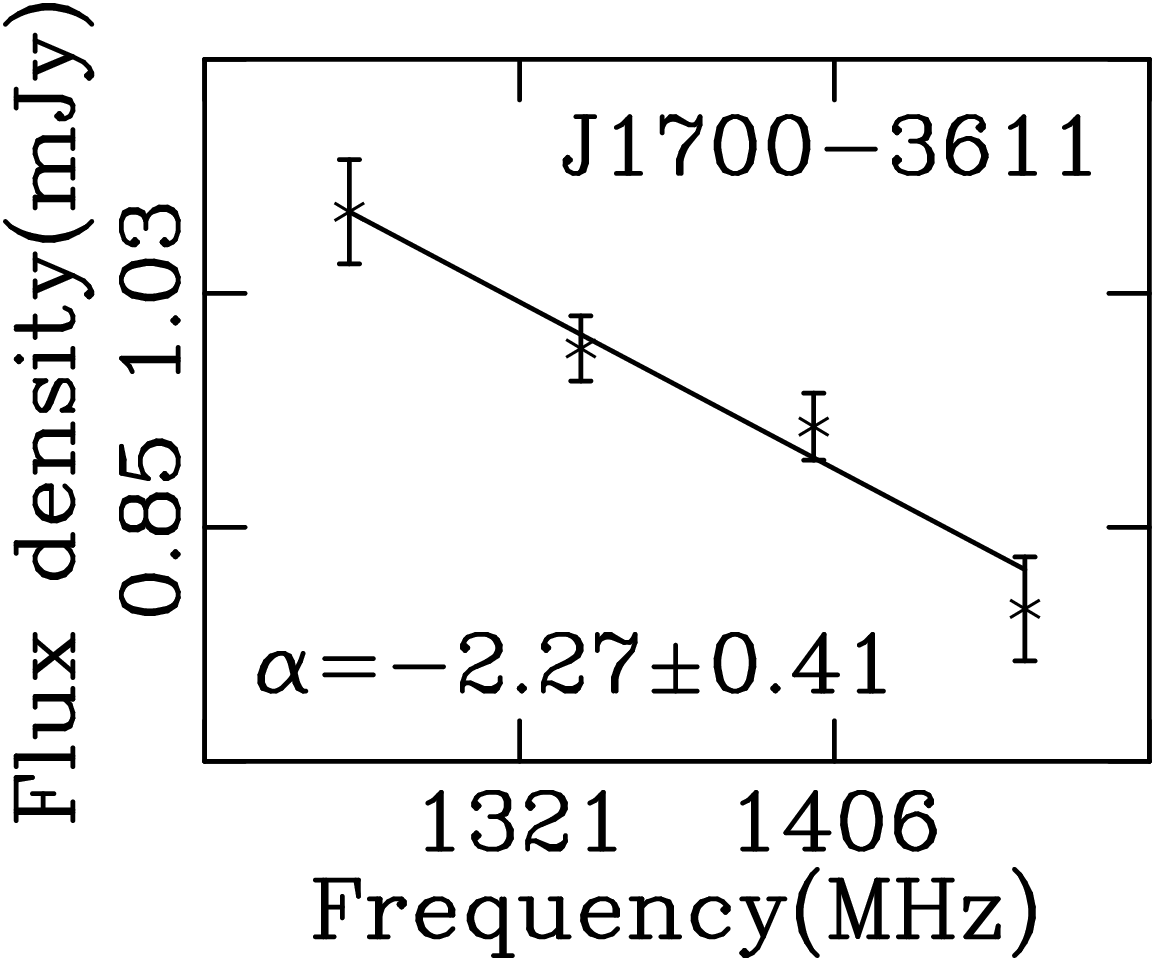}              
\includegraphics[bb = 110 120 518 685,clip,angle=-90,width=0.16\textwidth]{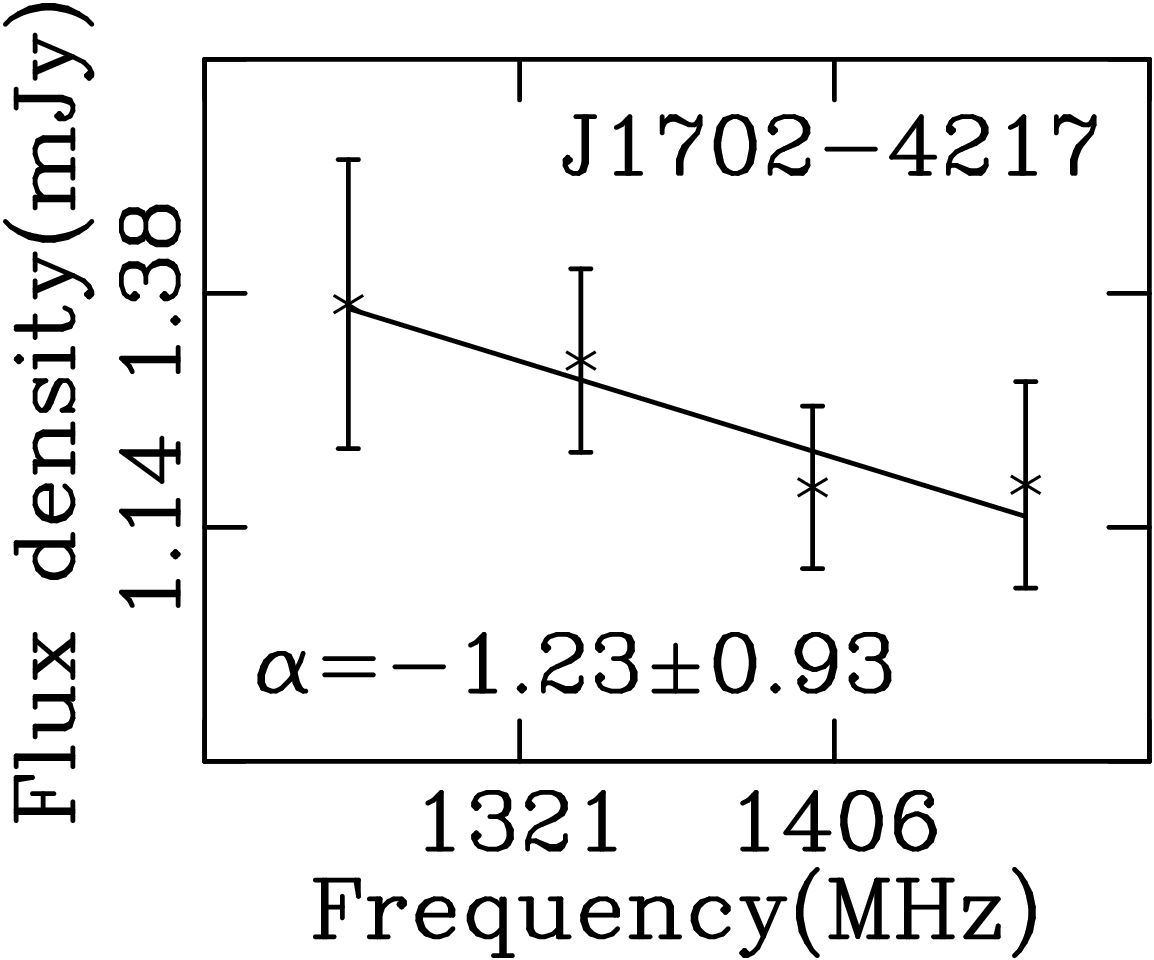}              
\includegraphics[bb = 110 120 518 685,clip,angle=-90,width=0.16\textwidth]{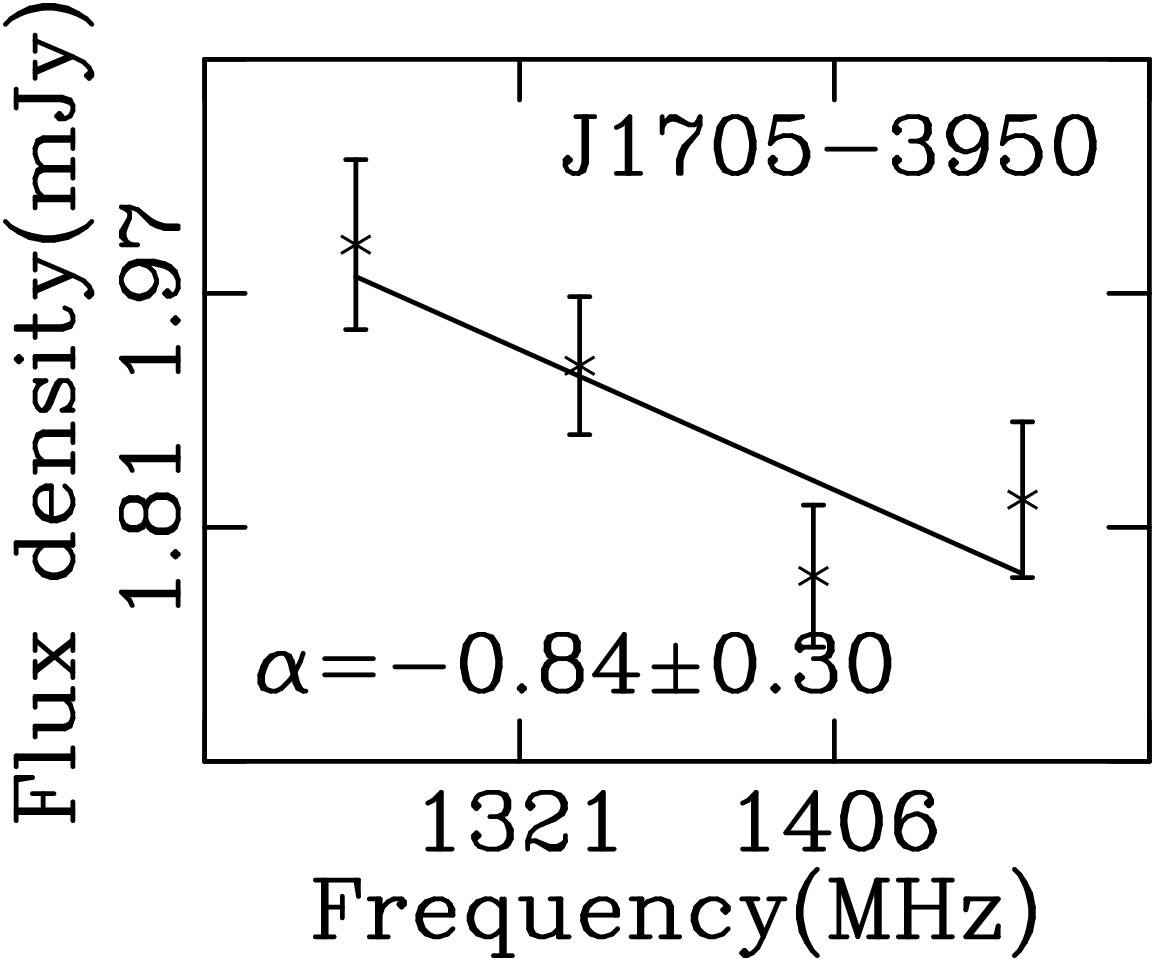}              
\includegraphics[bb = 110 120 518 685,clip,angle=-90,width=0.16\textwidth]{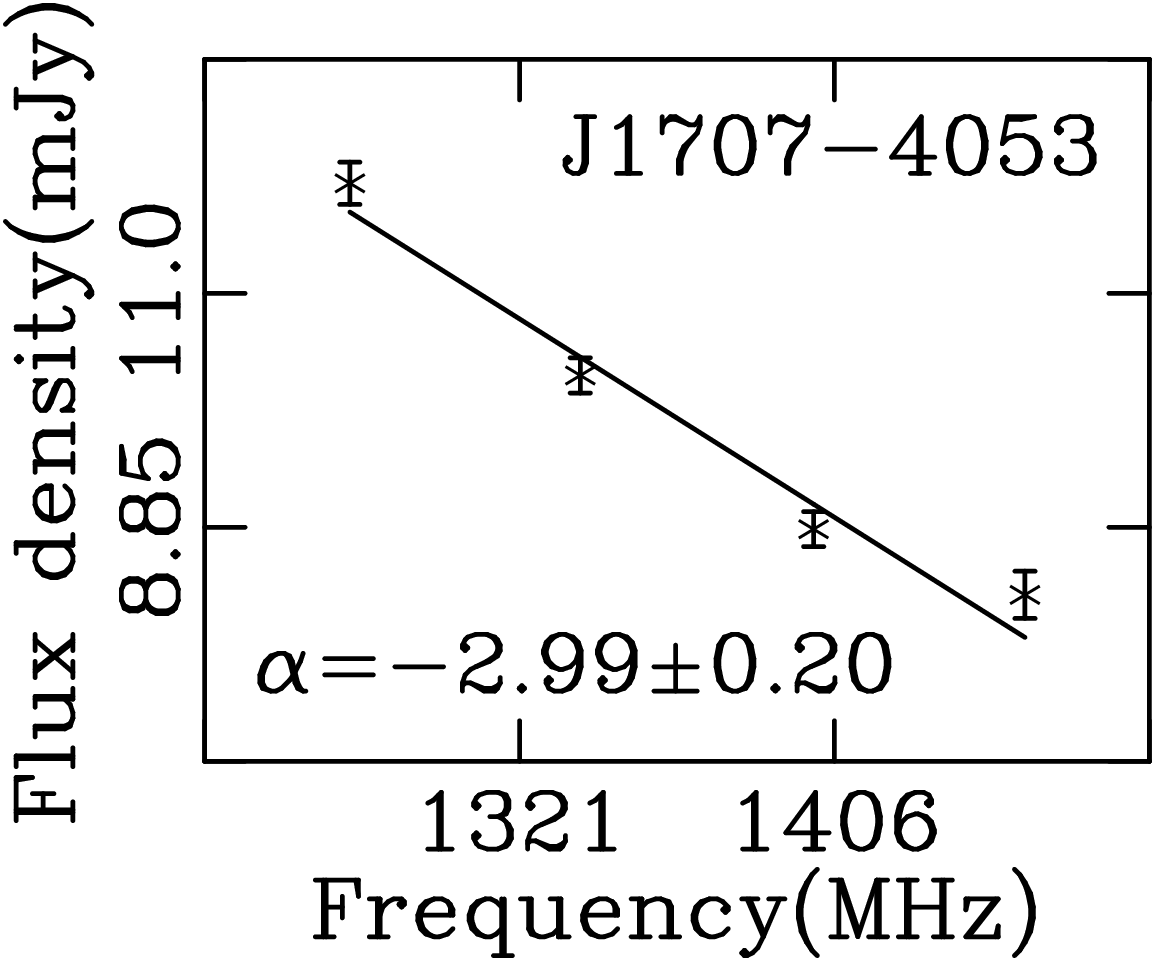}              
\includegraphics[bb = 110 120 518 685,clip,angle=-90,width=0.16\textwidth]{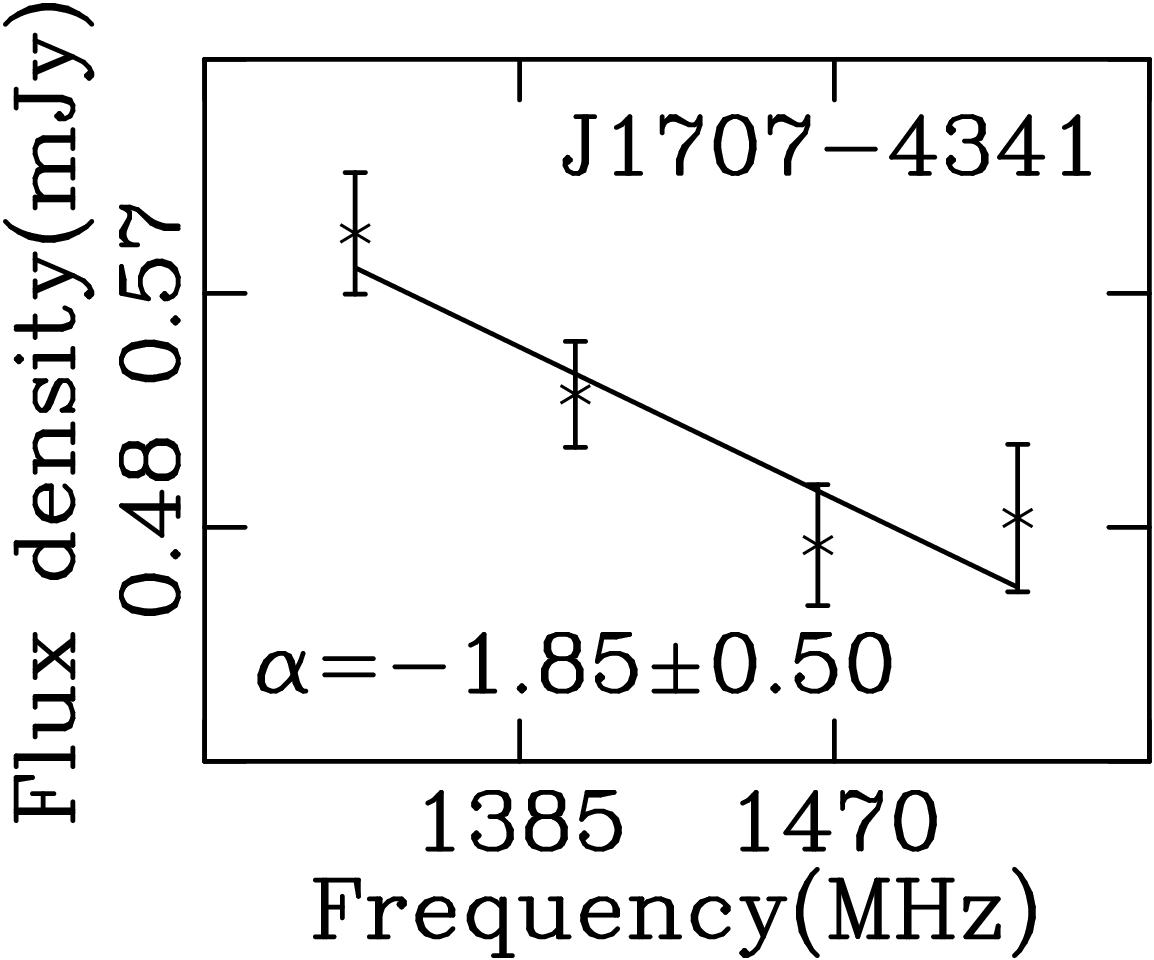}
\\[2mm]               
\includegraphics[bb = 110 72 518 685,clip,angle=-90,width=0.175\textwidth]{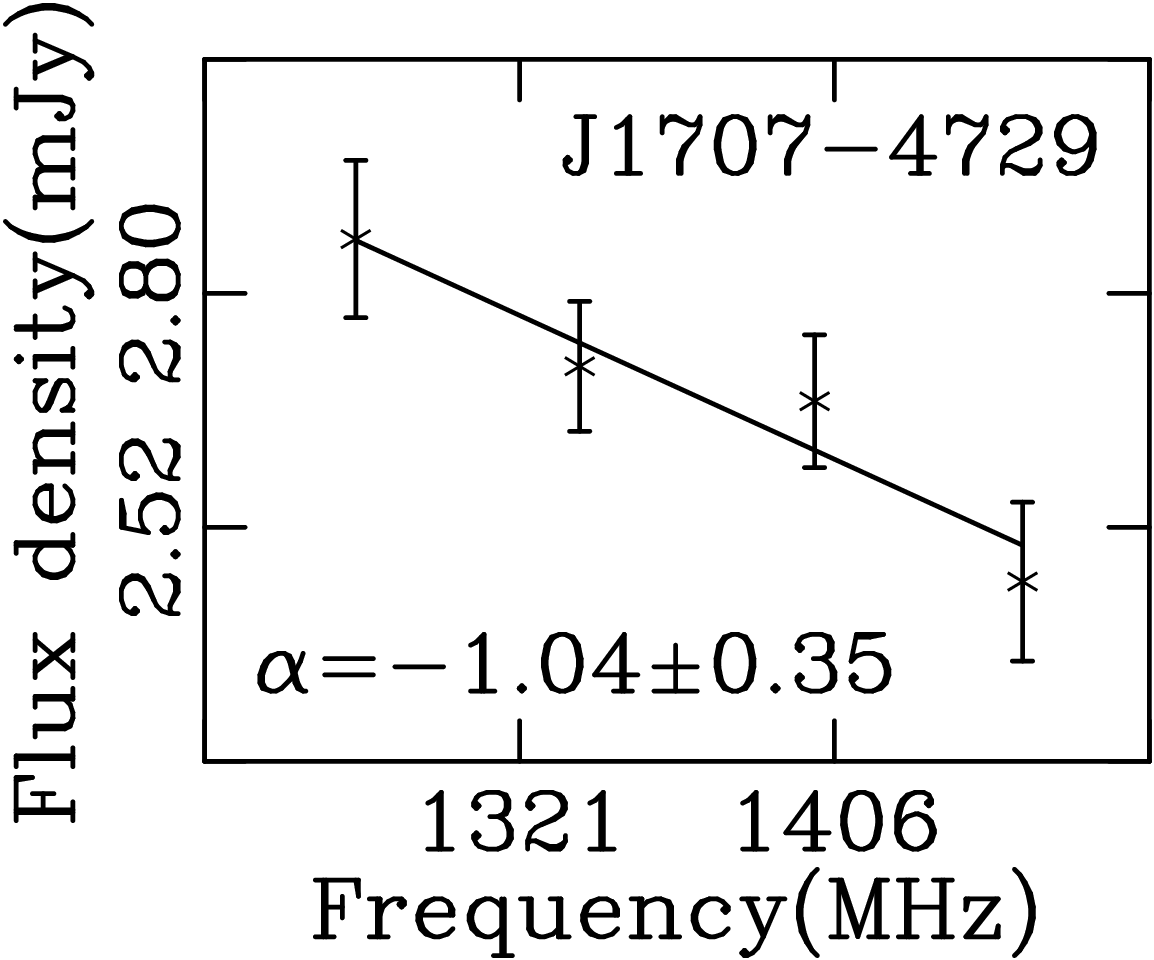}              
\includegraphics[bb = 110 120 518 685,clip,angle=-90,width=0.16\textwidth]{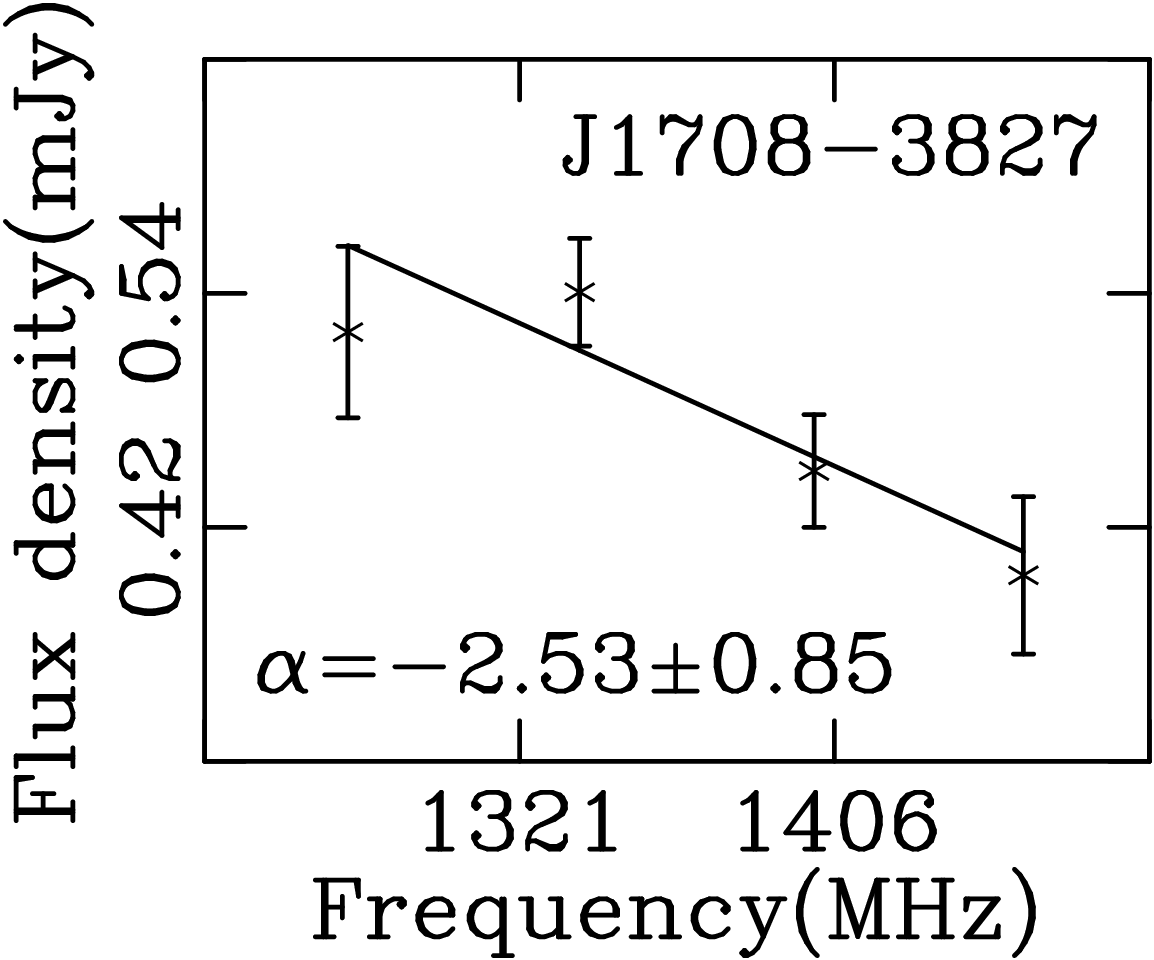}              
\includegraphics[bb = 110 120 518 685,clip,angle=-90,width=0.16\textwidth]{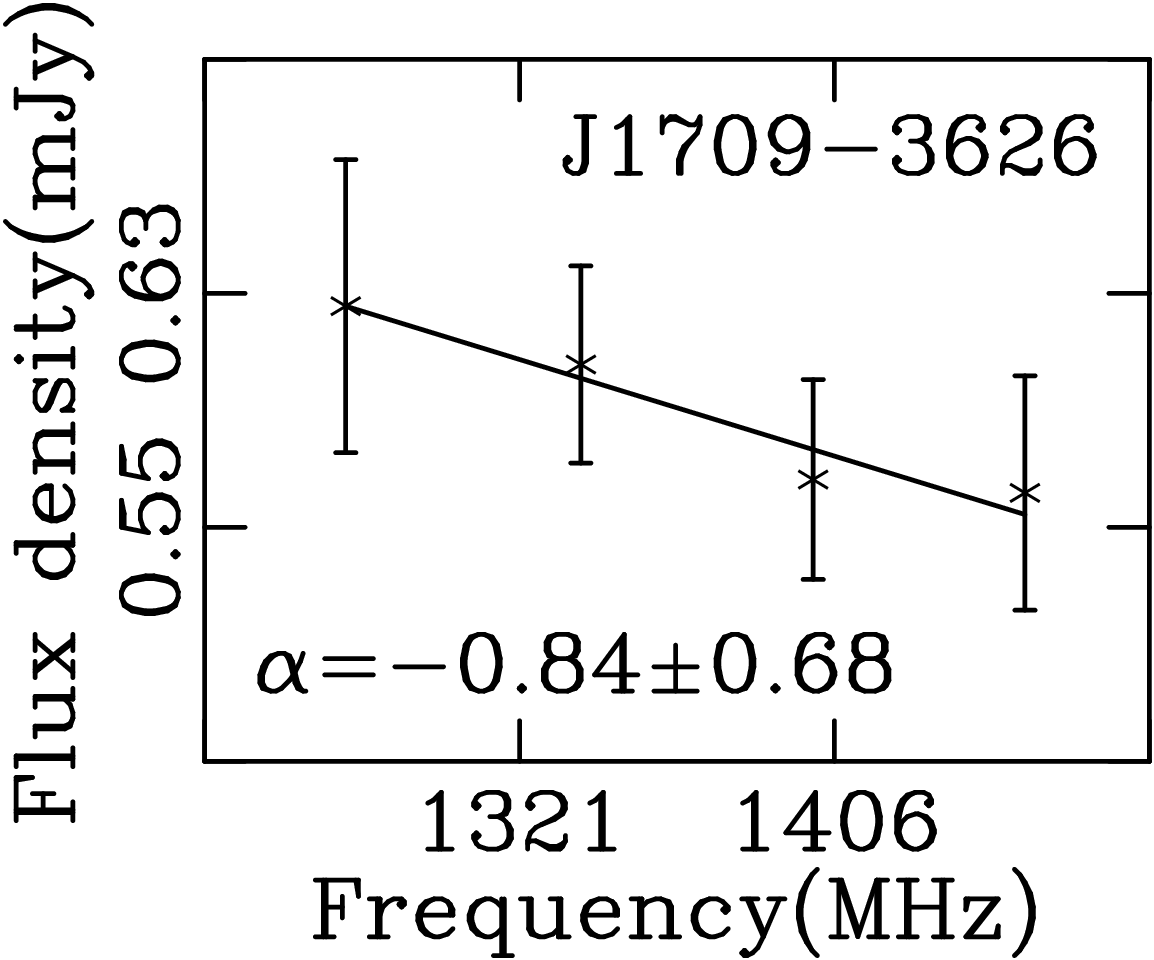}              
\includegraphics[bb = 110 120 518 685,clip,angle=-90,width=0.16\textwidth]{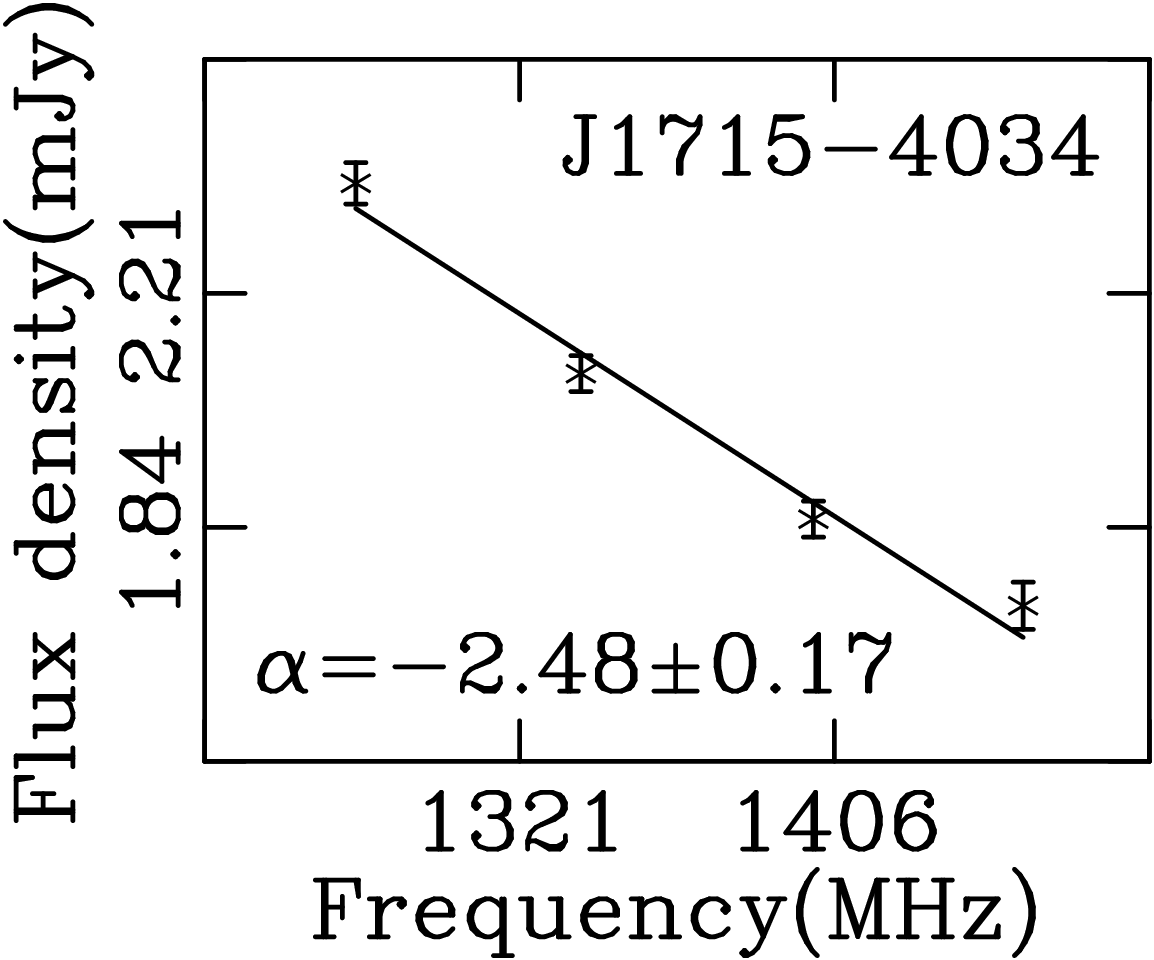}              
\includegraphics[bb = 110 120 518 685,clip,angle=-90,width=0.16\textwidth]{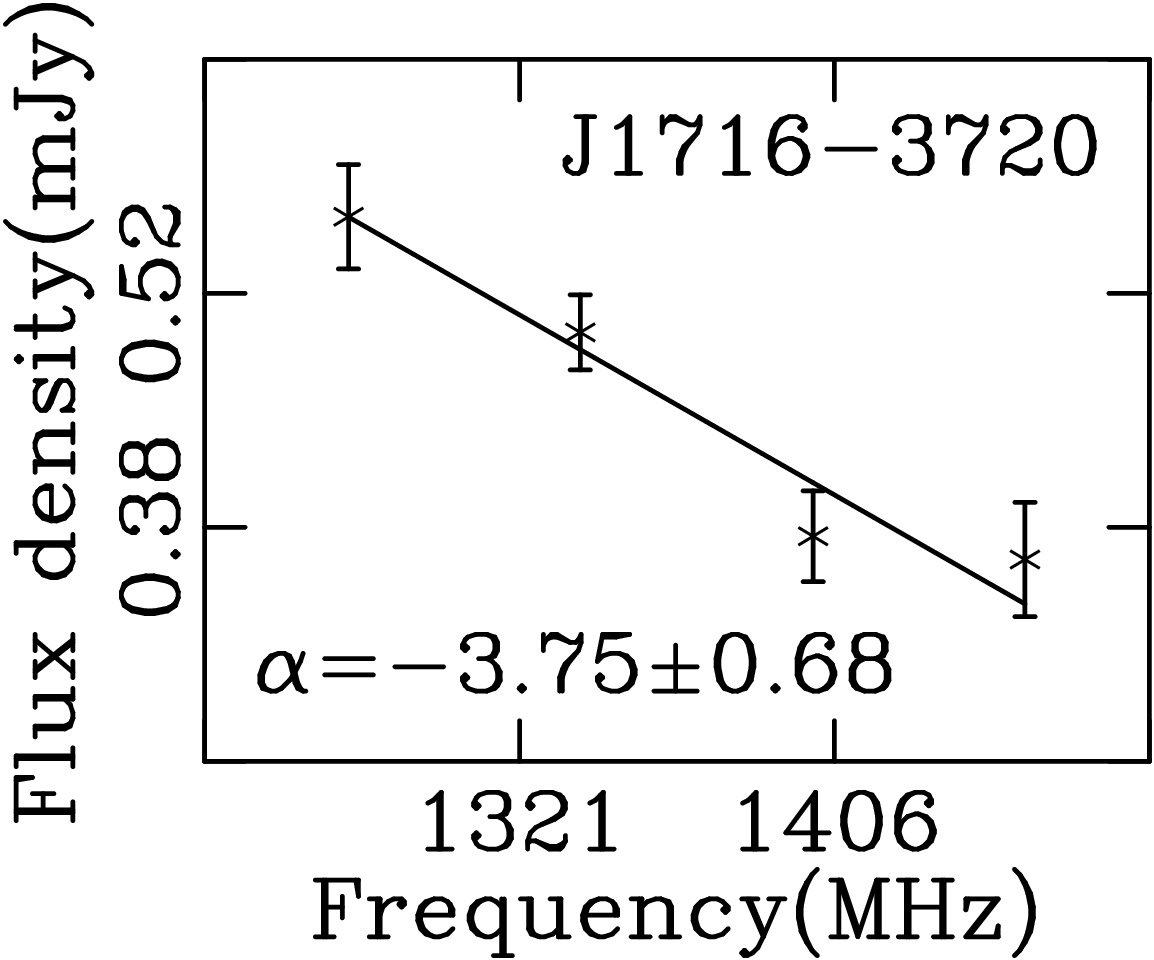}              
\includegraphics[bb = 110 120 518 685,clip,angle=-90,width=0.16\textwidth]{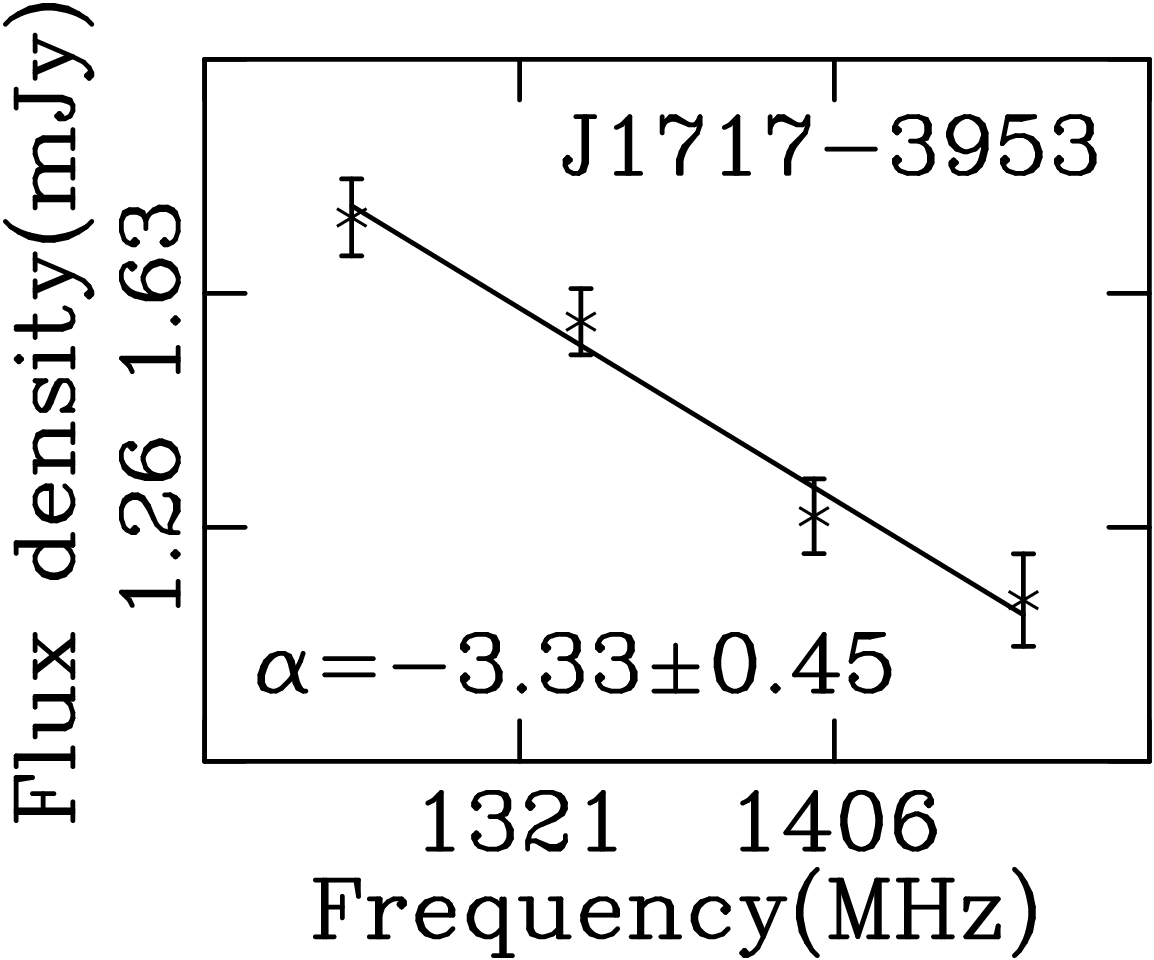}
\\[2mm]               
\includegraphics[bb = 110 72 518 685,clip,angle=-90,width=0.175\textwidth]{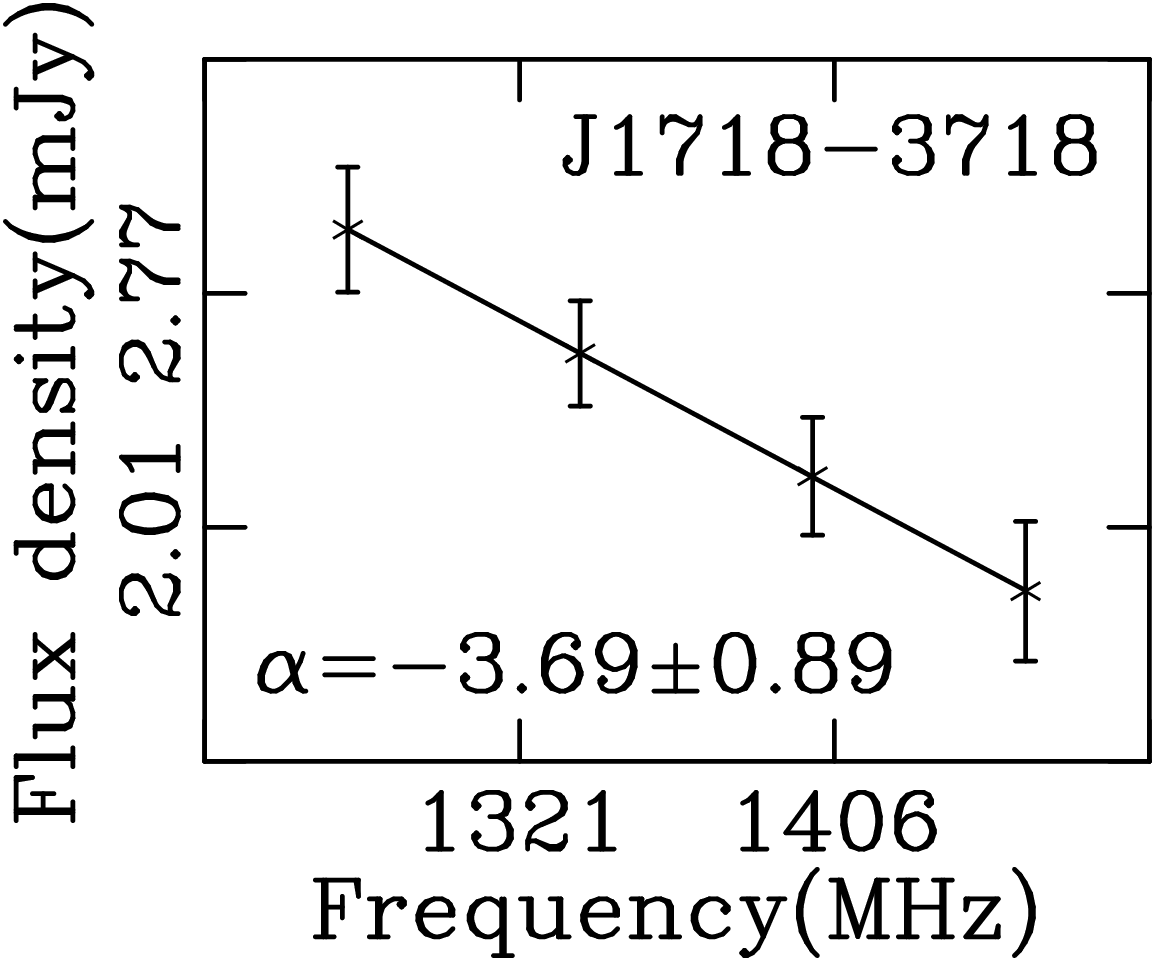}              
\includegraphics[bb = 110 120 518 685,clip,angle=-90,width=0.16\textwidth]{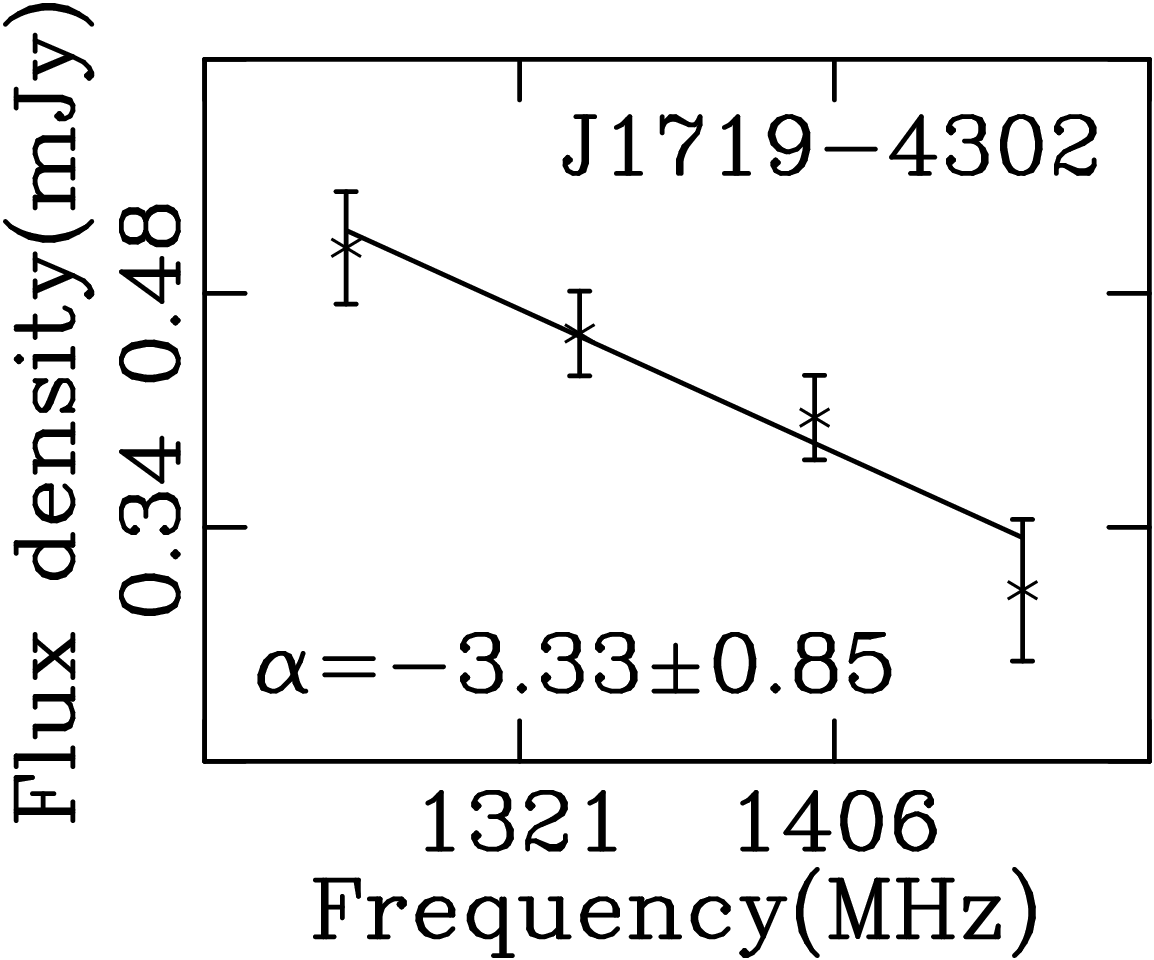}              
\includegraphics[bb = 110 120 518 685,clip,angle=-90,width=0.16\textwidth]{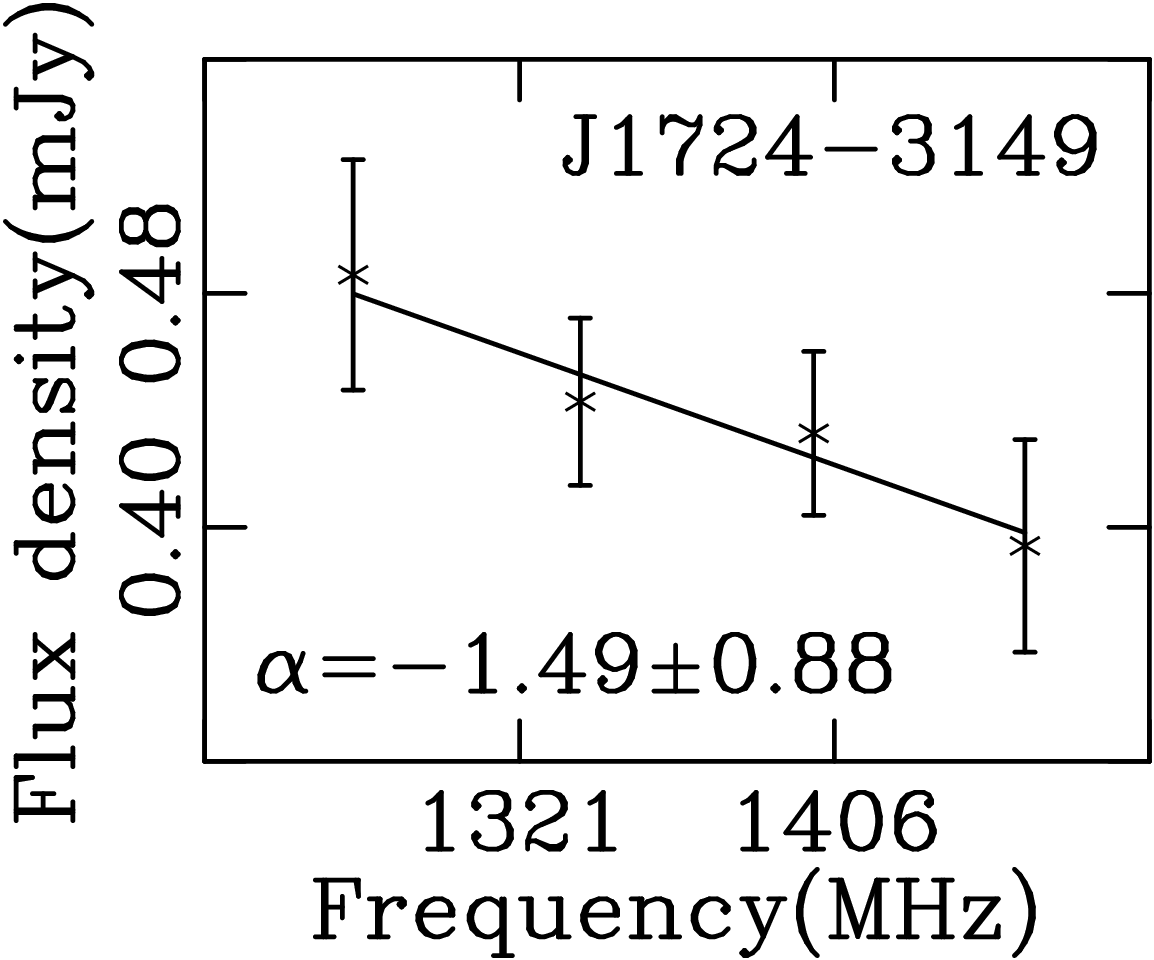}              
\includegraphics[bb = 110 120 518 685,clip,angle=-90,width=0.16\textwidth]{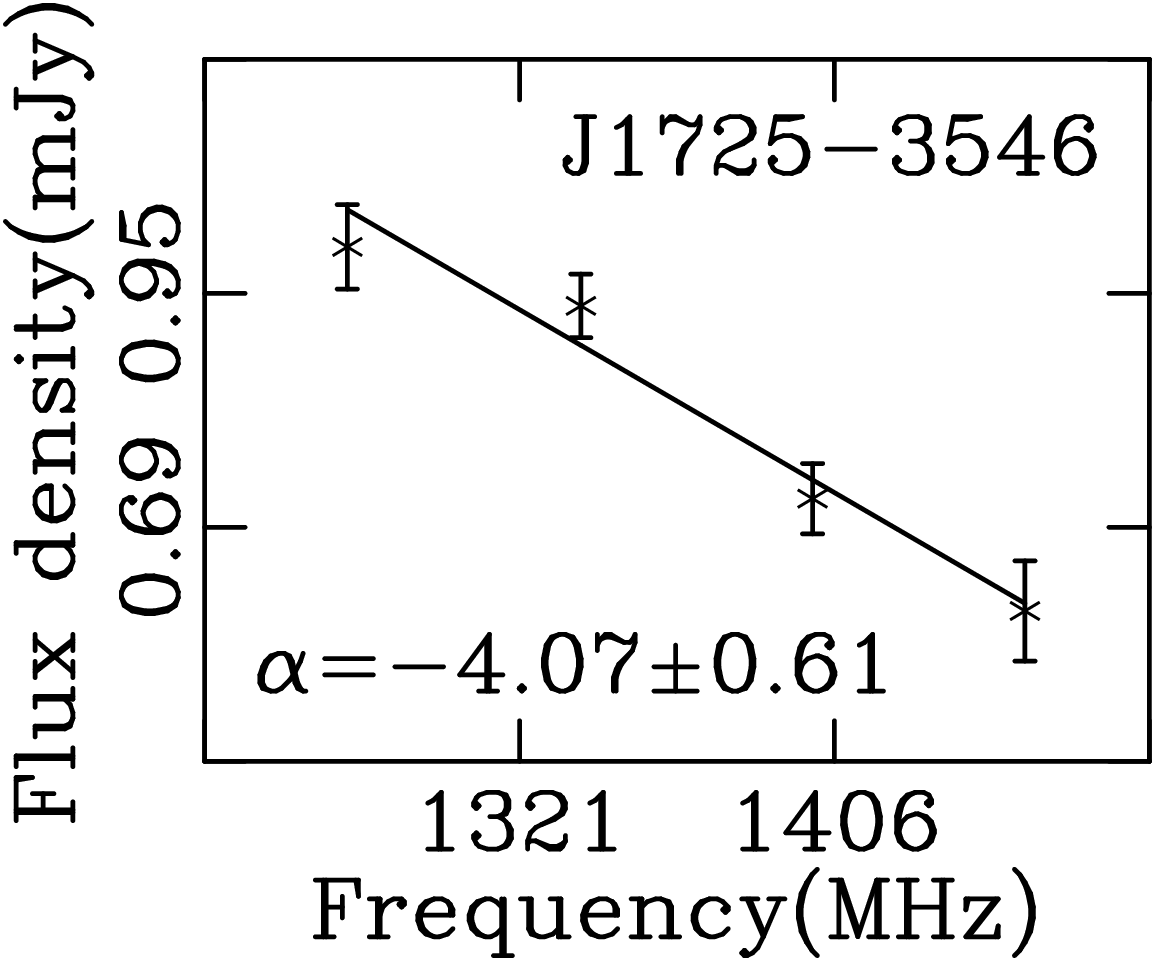}              
\includegraphics[bb = 110 120 518 685,clip,angle=-90,width=0.16\textwidth]{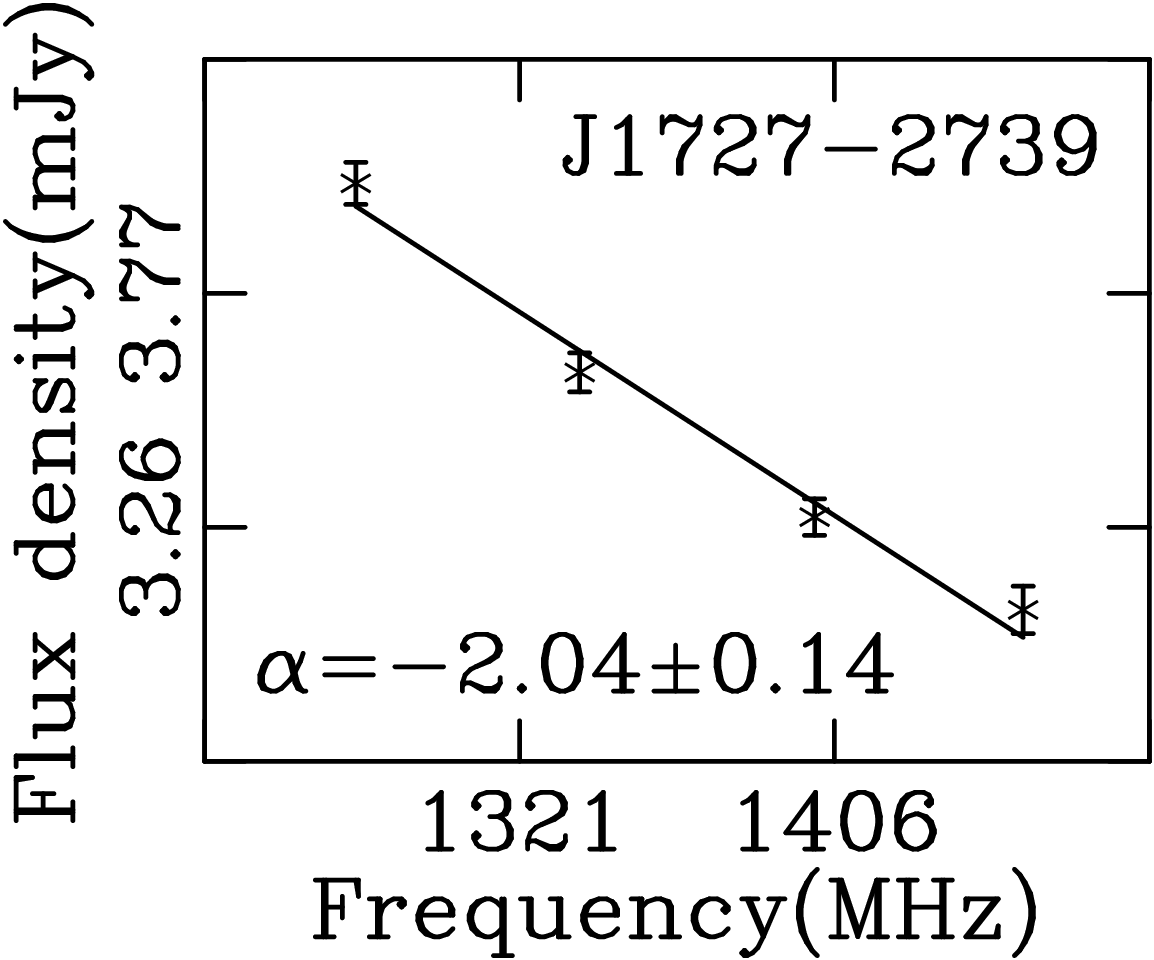}              
\includegraphics[bb = 110 120 518 685,clip,angle=-90,width=0.16\textwidth]{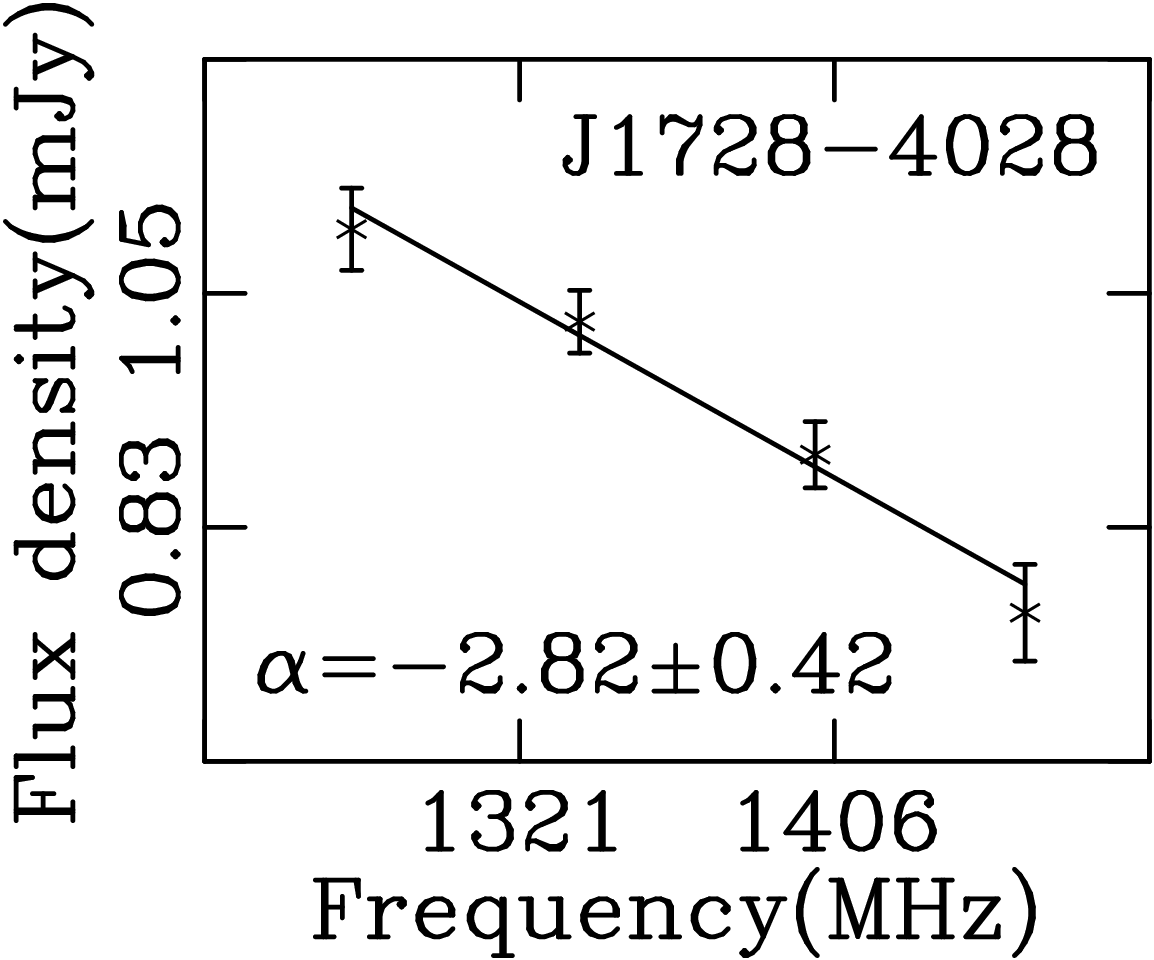} 
\\[2mm]             
\includegraphics[bb = 110 72 518 685,clip,angle=-90,width=0.175\textwidth]{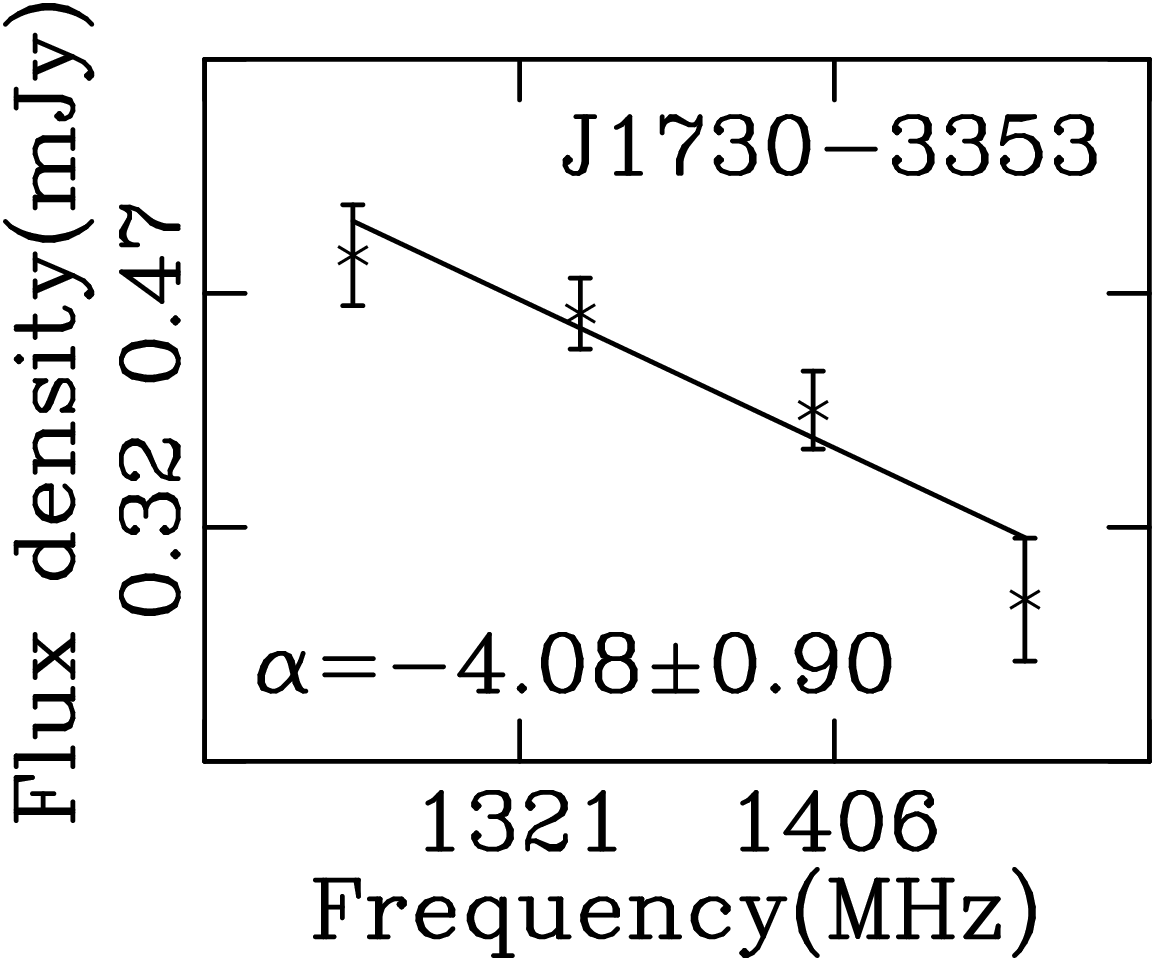}              
\includegraphics[bb = 110 120 518 685,clip,angle=-90,width=0.16\textwidth]{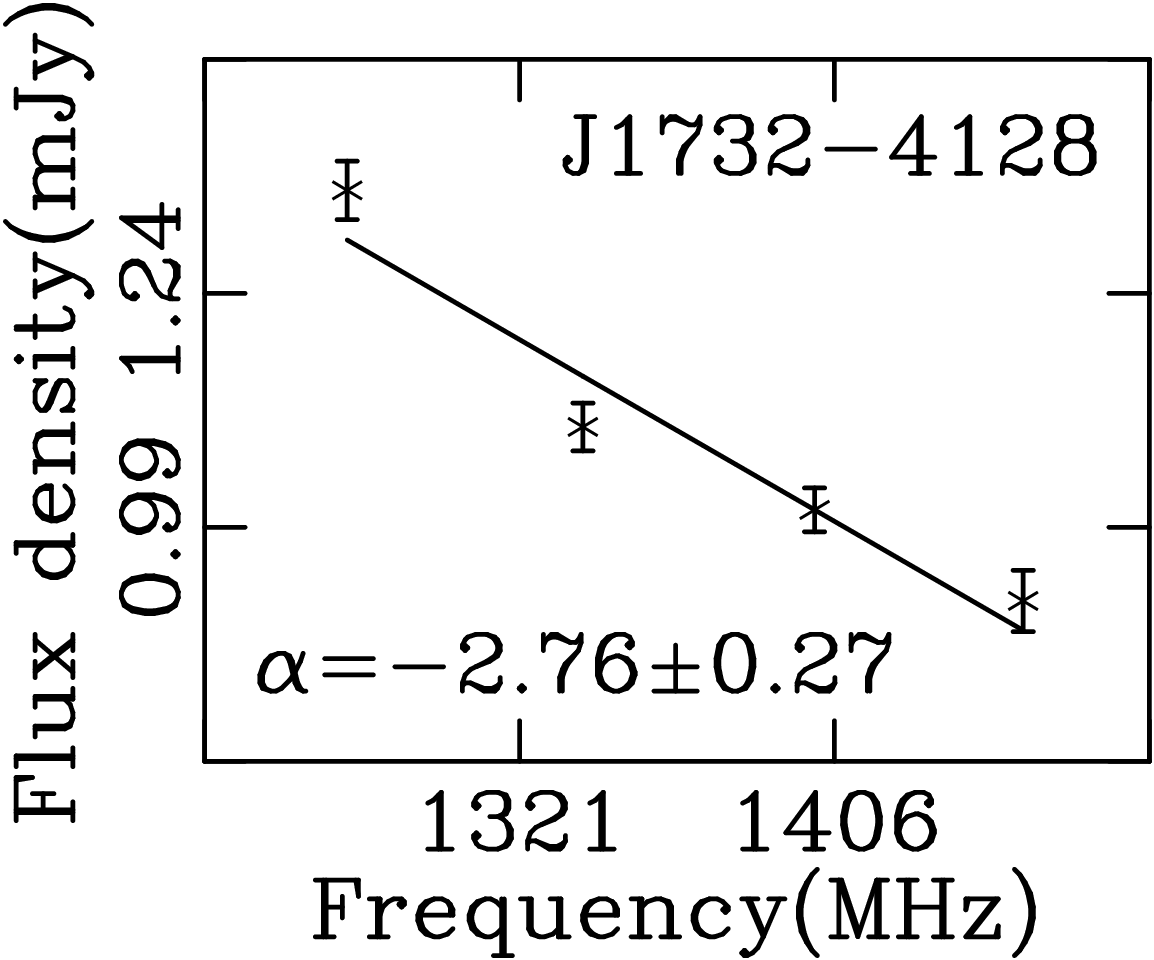}              
\includegraphics[bb = 110 120 518 685,clip,angle=-90,width=0.16\textwidth]{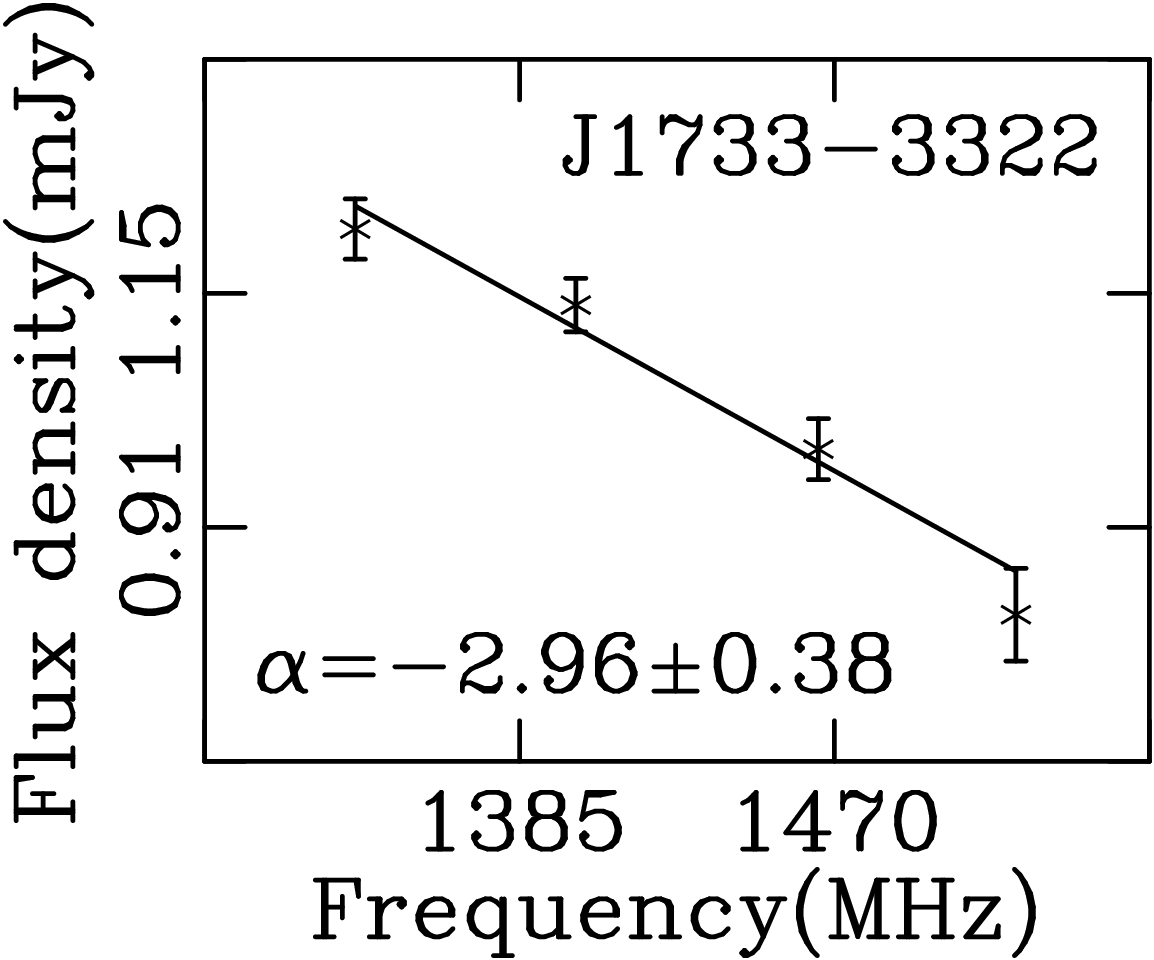}              
\includegraphics[bb = 110 120 518 685,clip,angle=-90,width=0.16\textwidth]{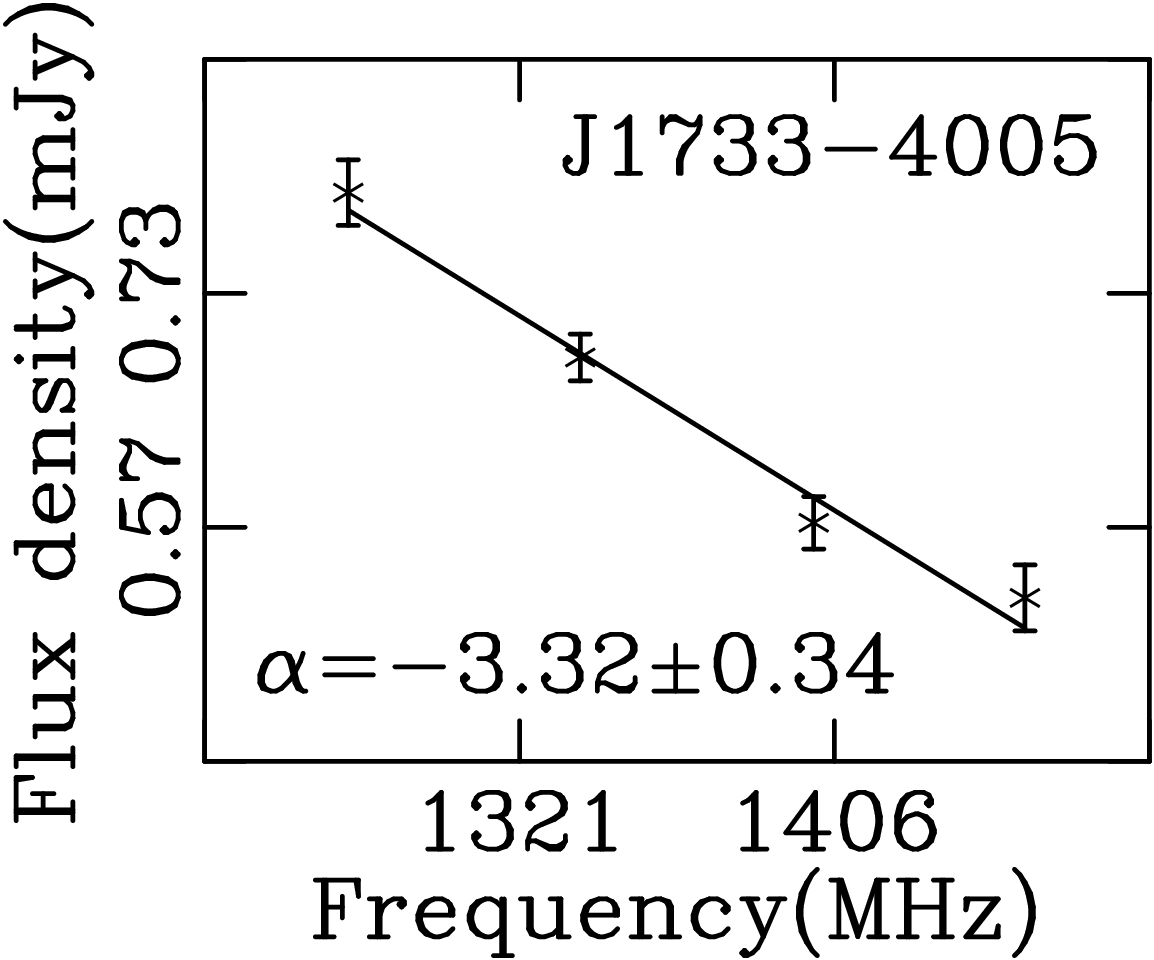}              
\includegraphics[bb = 110 120 518 685,clip,angle=-90,width=0.16\textwidth]{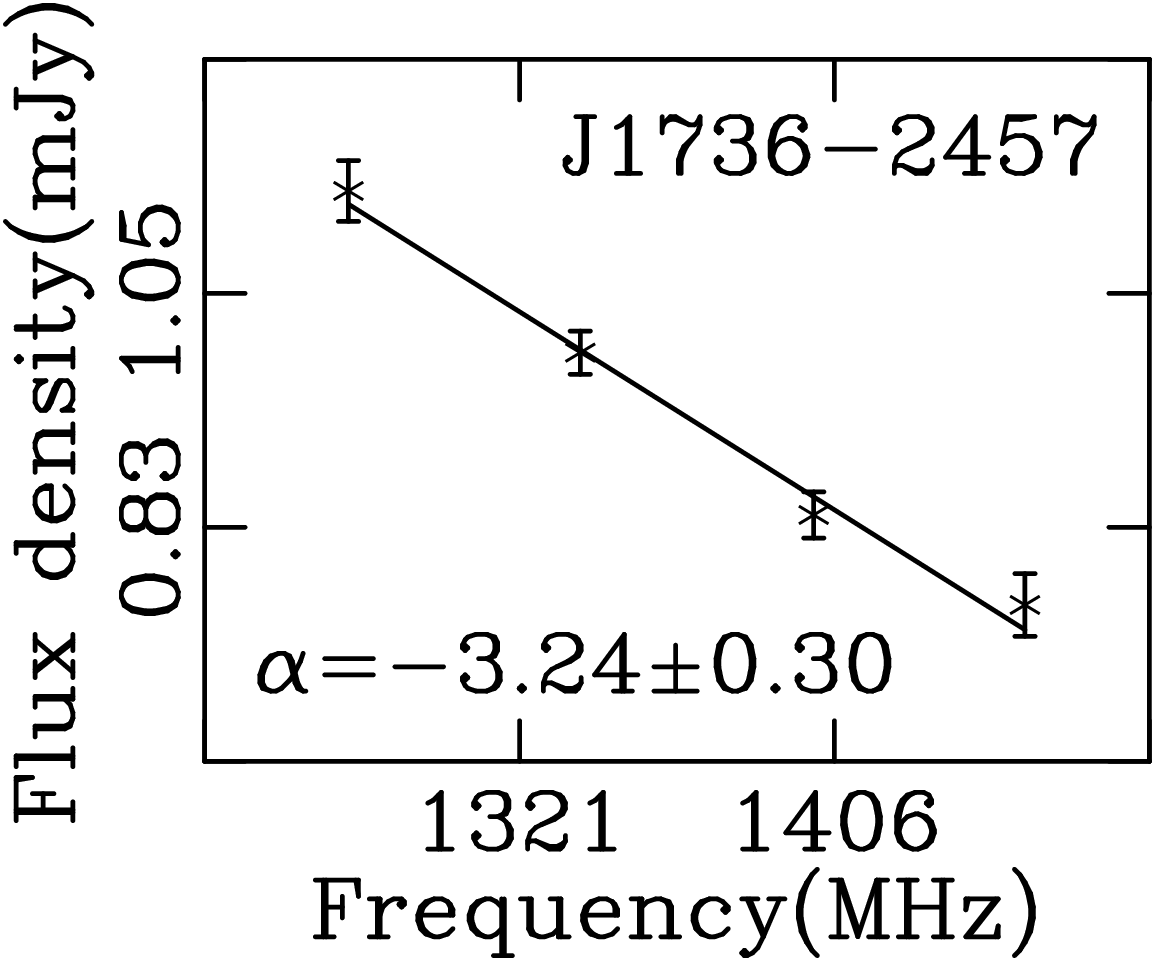}              
\includegraphics[bb = 110 120 518 685,clip,angle=-90,width=0.16\textwidth]{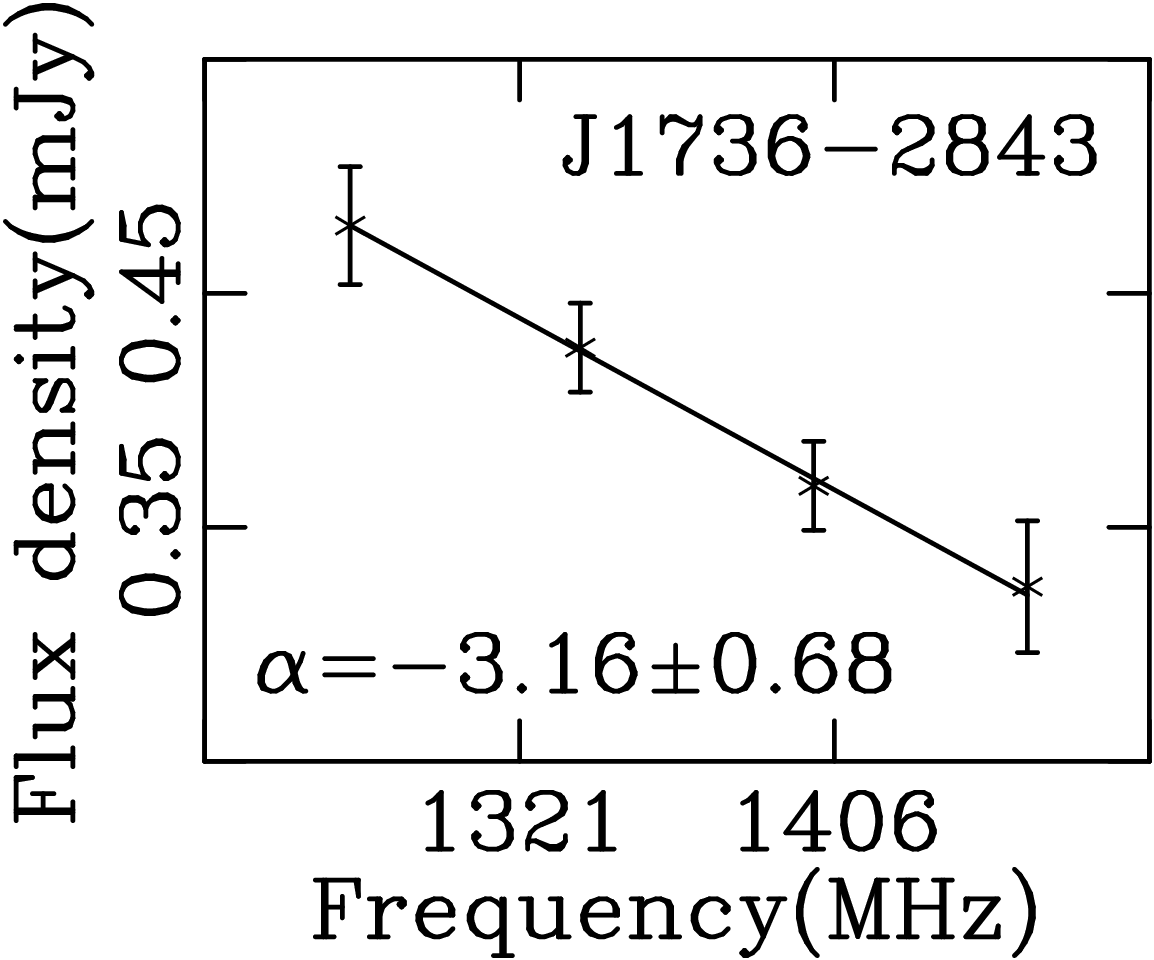}
\\[2mm]              
\includegraphics[bb = 110 72 518 685,clip,angle=-90,width=0.175\textwidth]{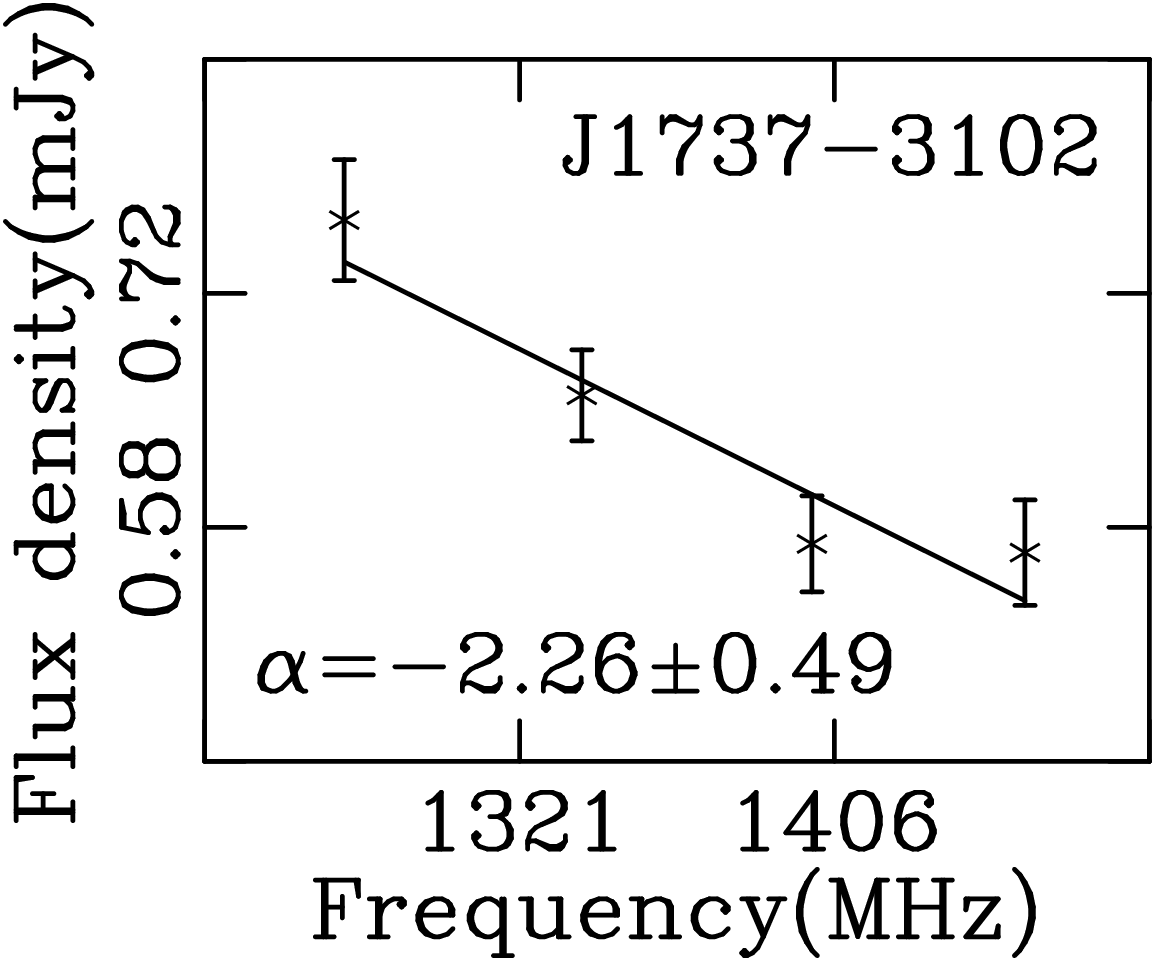}              
\includegraphics[bb = 110 120 518 685,clip,angle=-90,width=0.16\textwidth]{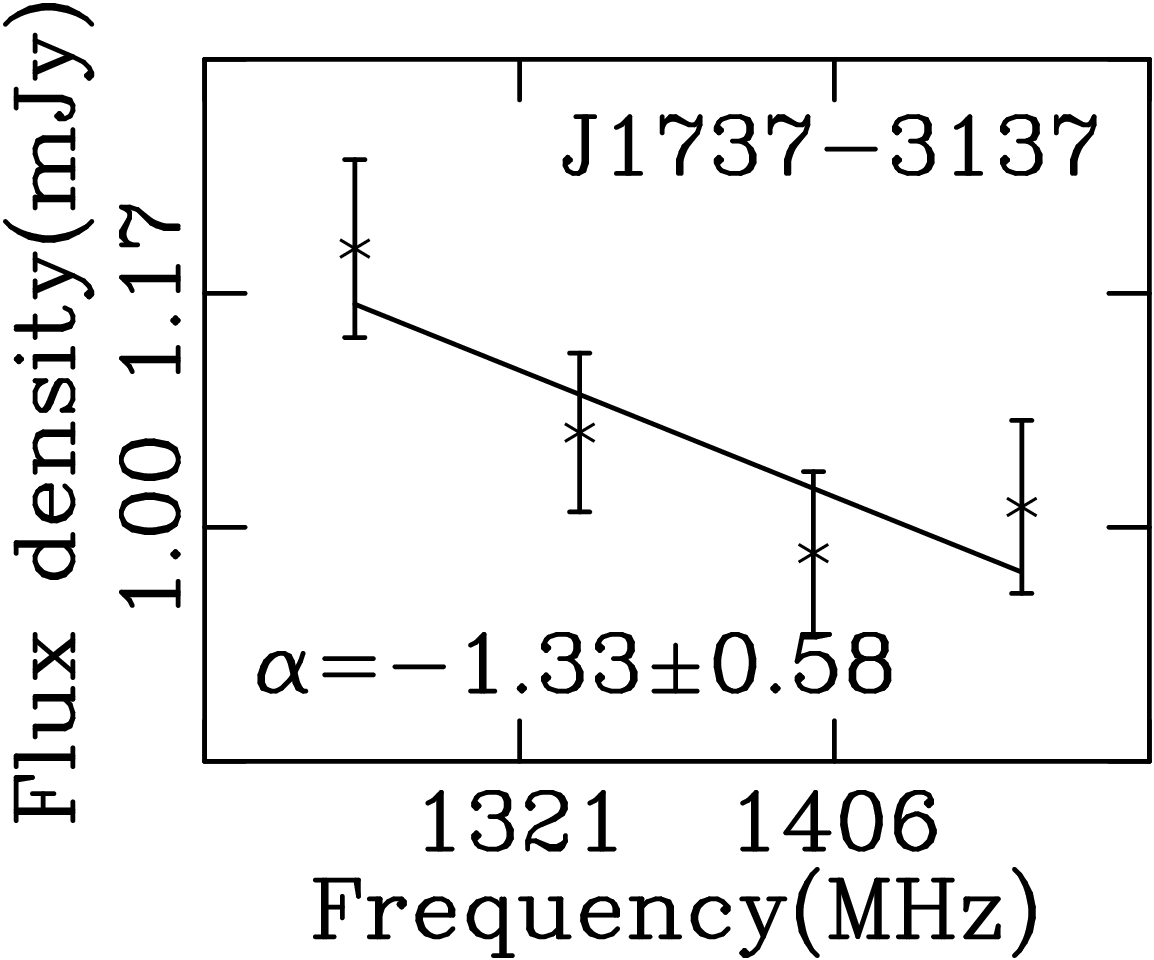}              
\includegraphics[bb = 110 120 518 685,clip,angle=-90,width=0.16\textwidth]{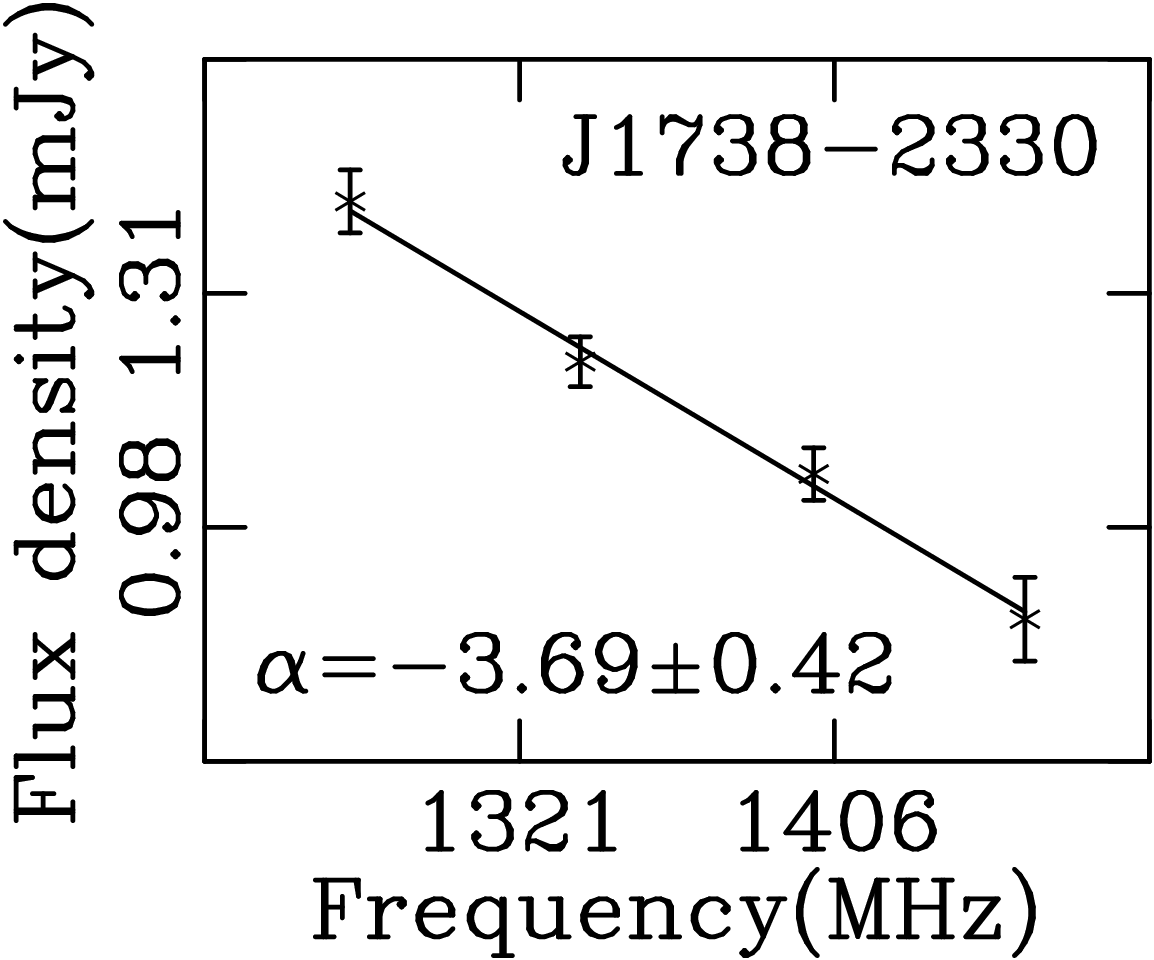}              
\includegraphics[bb = 110 120 518 685,clip,angle=-90,width=0.16\textwidth]{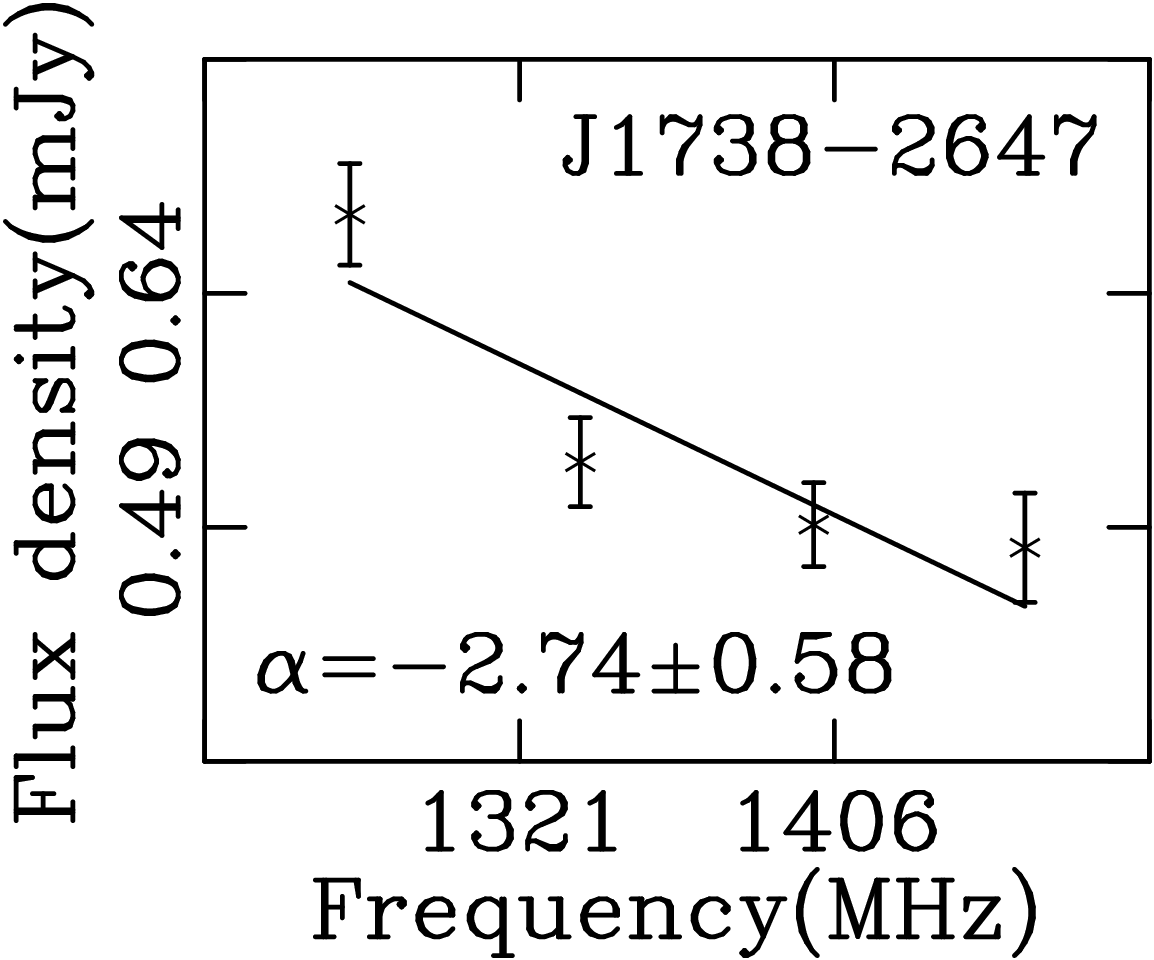}              
\includegraphics[bb = 110 120 518 685,clip,angle=-90,width=0.16\textwidth]{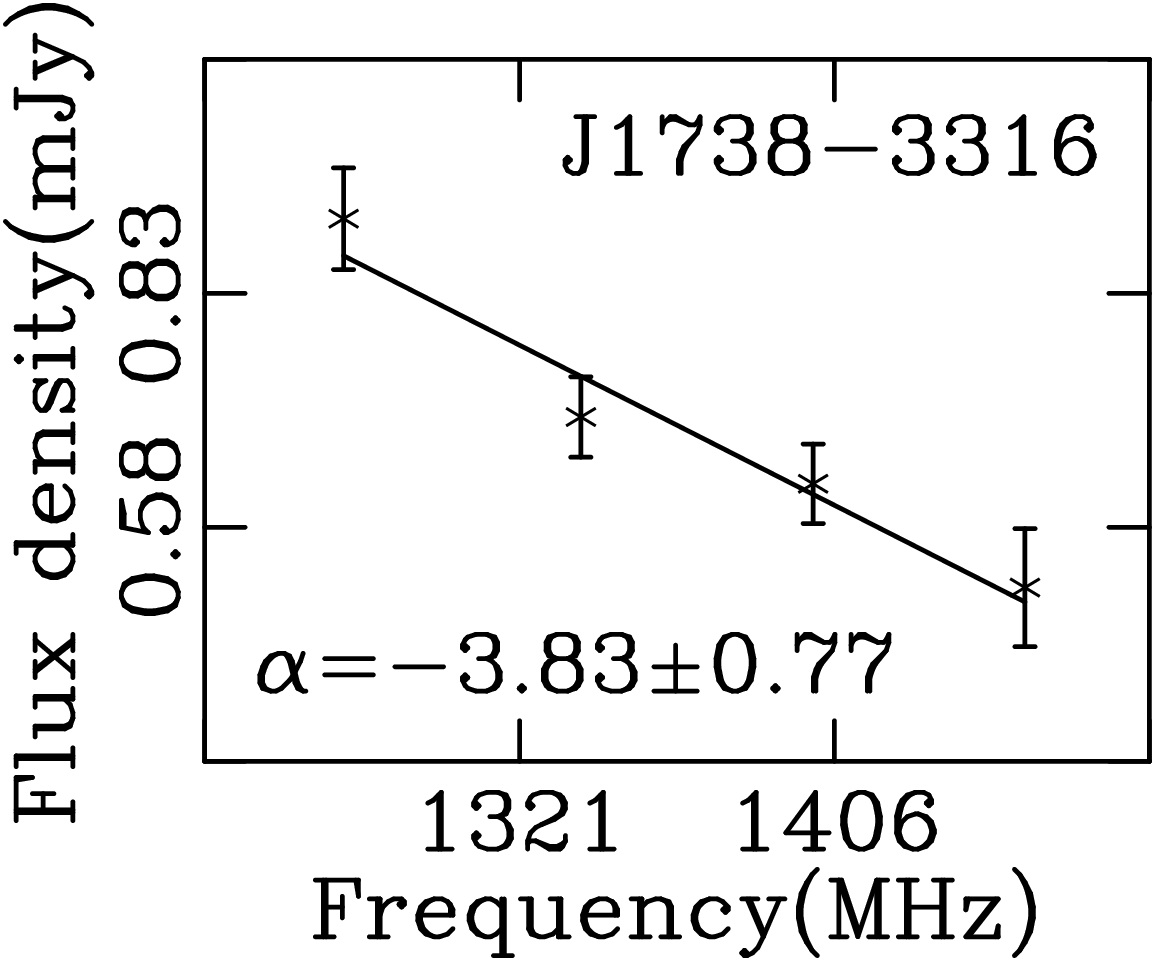}              
\includegraphics[bb = 110 120 518 685,clip,angle=-90,width=0.16\textwidth]{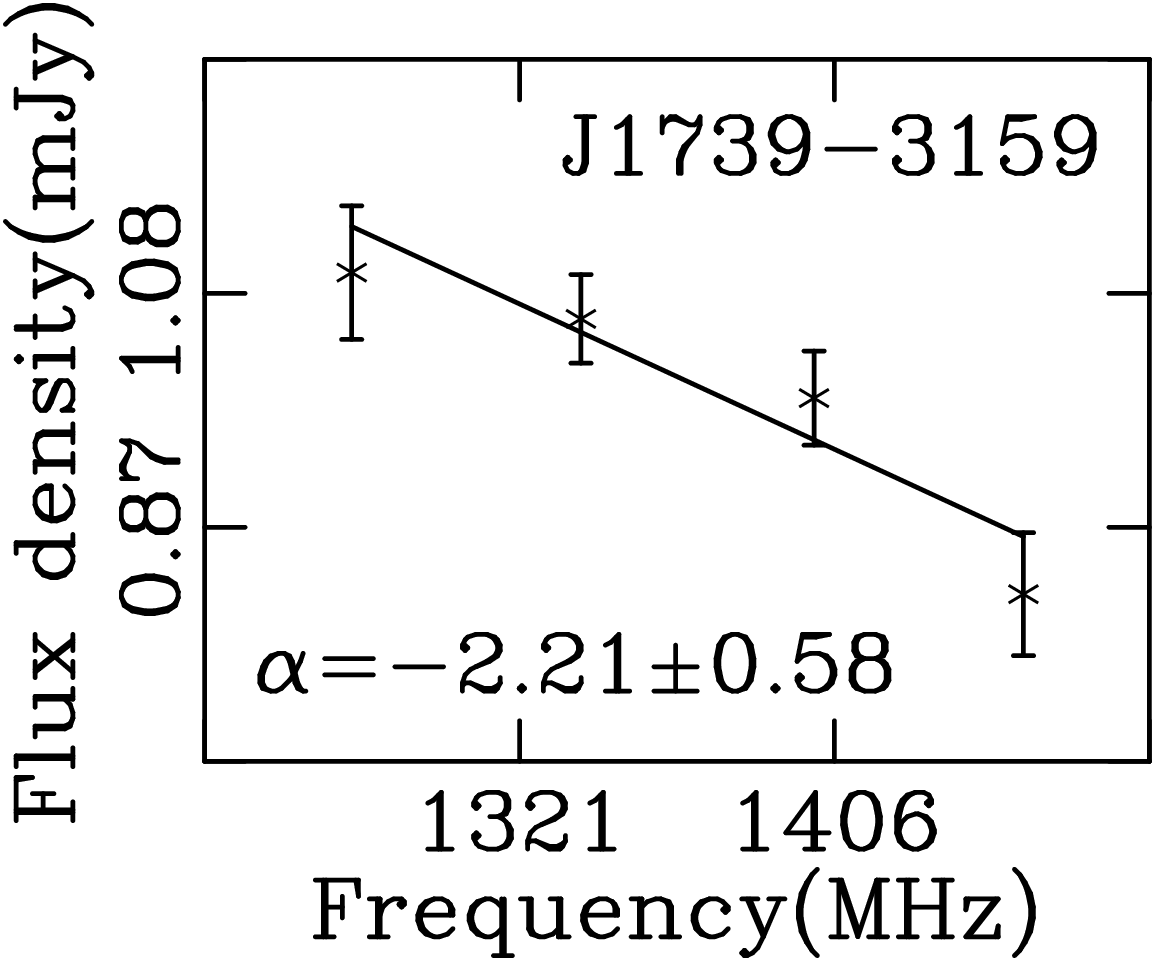}
 \\[2mm]             
\includegraphics[bb = 110 72 518 685,clip,angle=-90,width=0.175\textwidth]{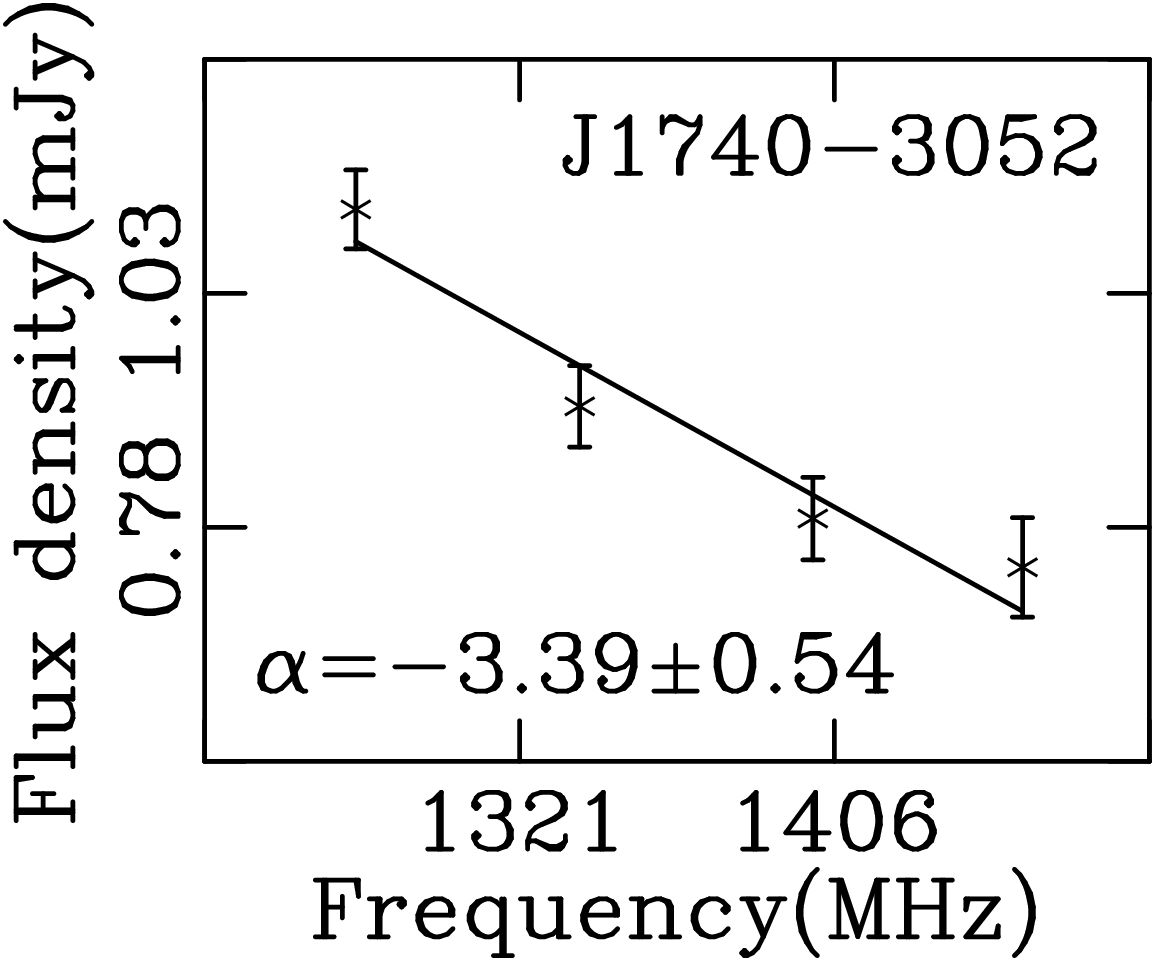}              
\includegraphics[bb = 110 120 518 685,clip,angle=-90,width=0.16\textwidth]{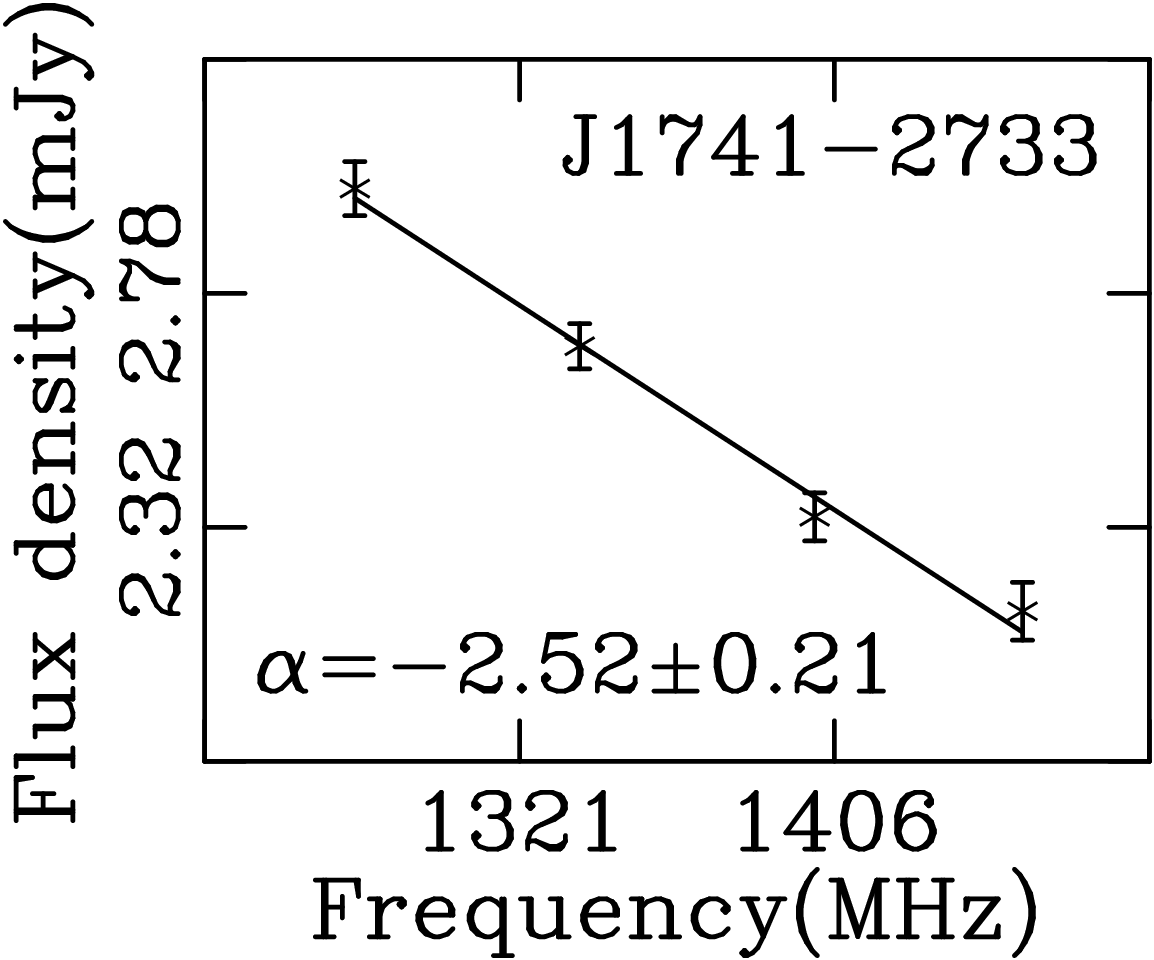}              
\includegraphics[bb = 110 120 518 685,clip,angle=-90,width=0.16\textwidth]{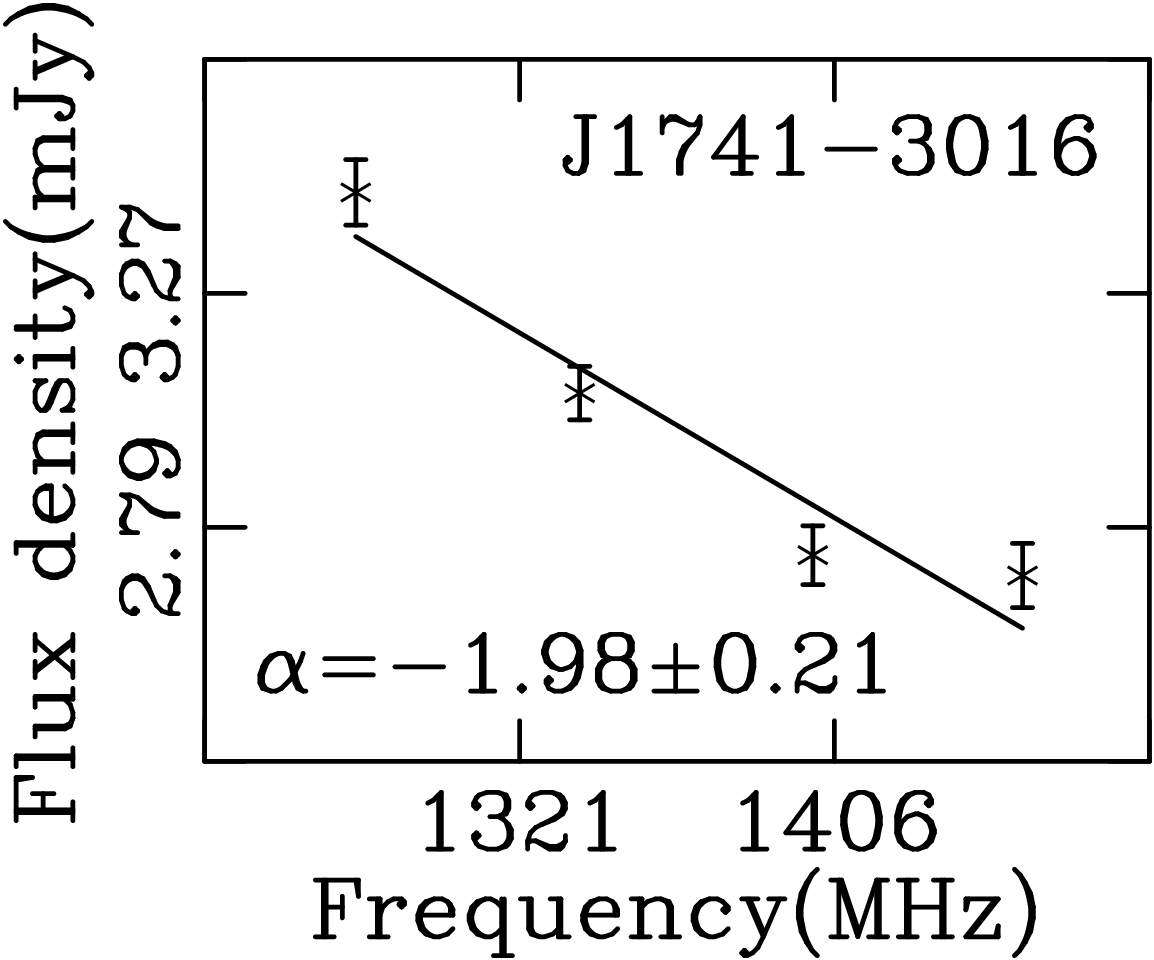}              
\includegraphics[bb = 110 120 518 685,clip,angle=-90,width=0.16\textwidth]{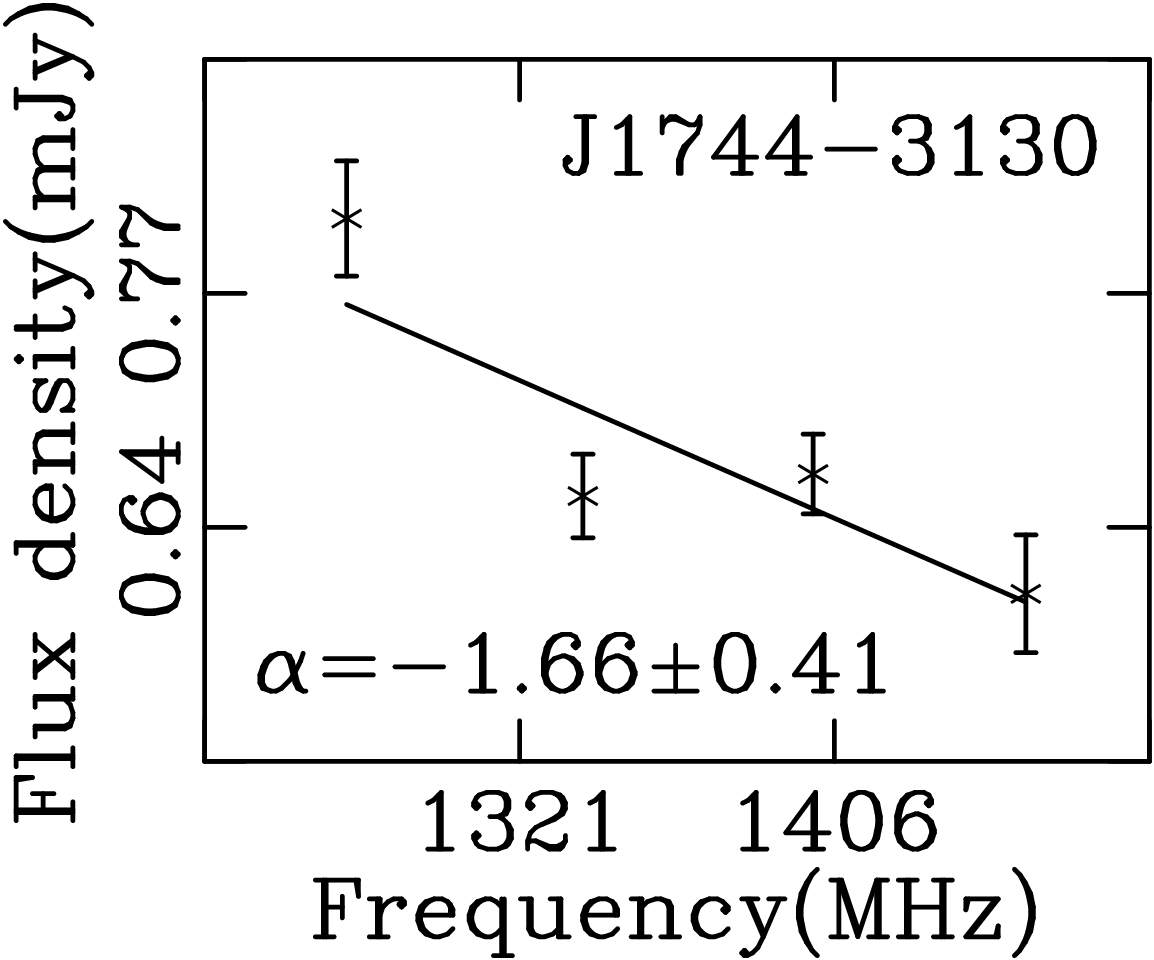}              
\includegraphics[bb = 110 120 518 685,clip,angle=-90,width=0.16\textwidth]{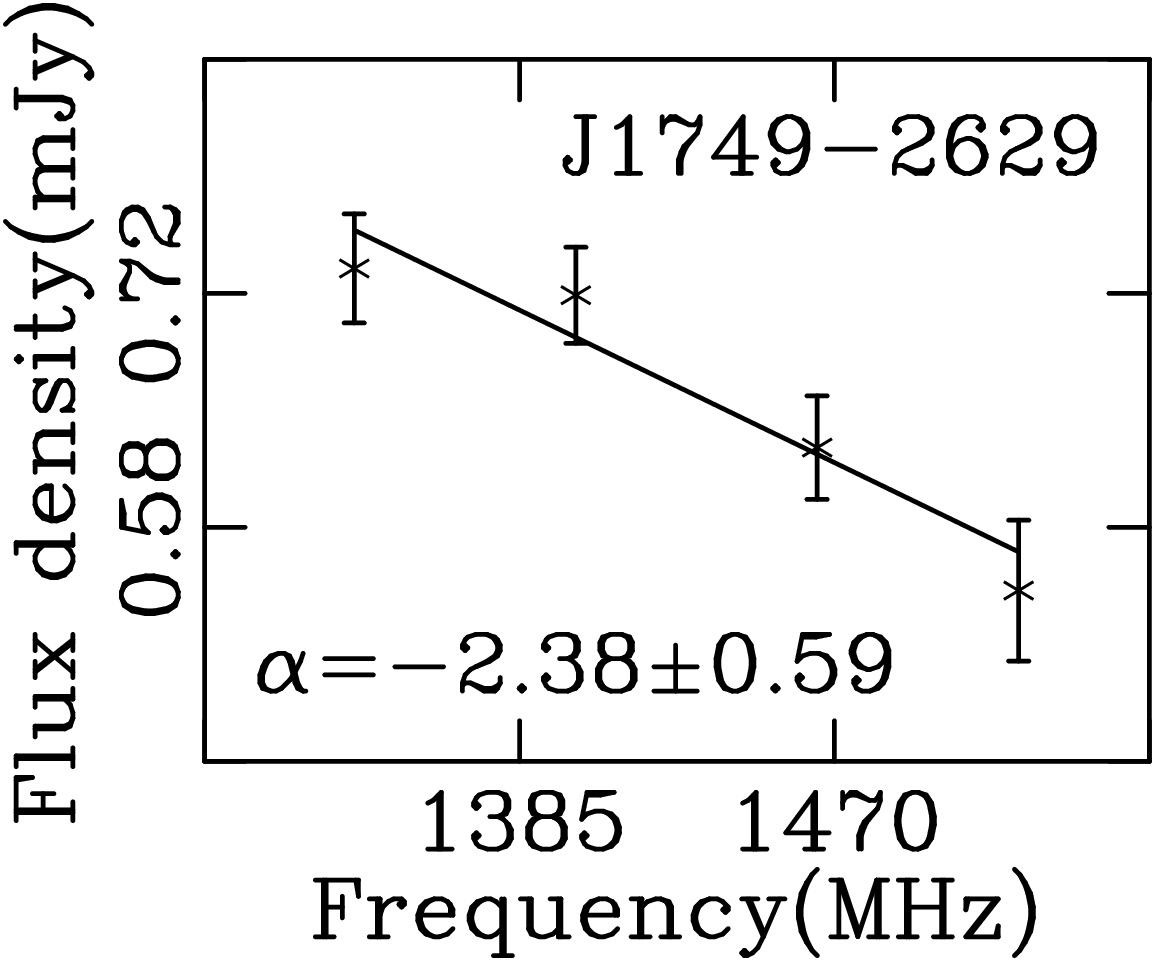}              
\includegraphics[bb = 110 120 518 685,clip,angle=-90,width=0.16\textwidth]{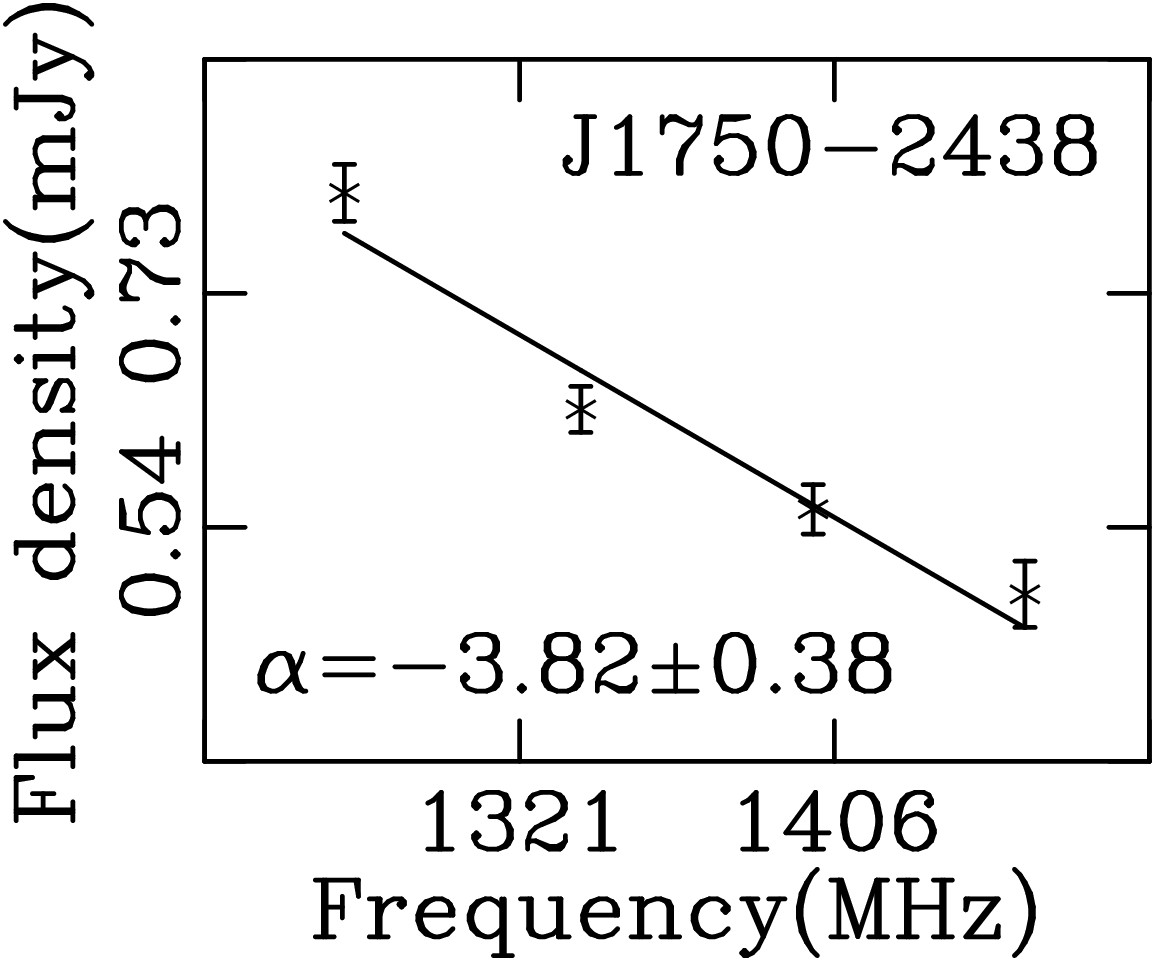}
\\[2mm]                   
\includegraphics[bb = 110 72 518 685,clip,angle=-90,width=0.175\textwidth]{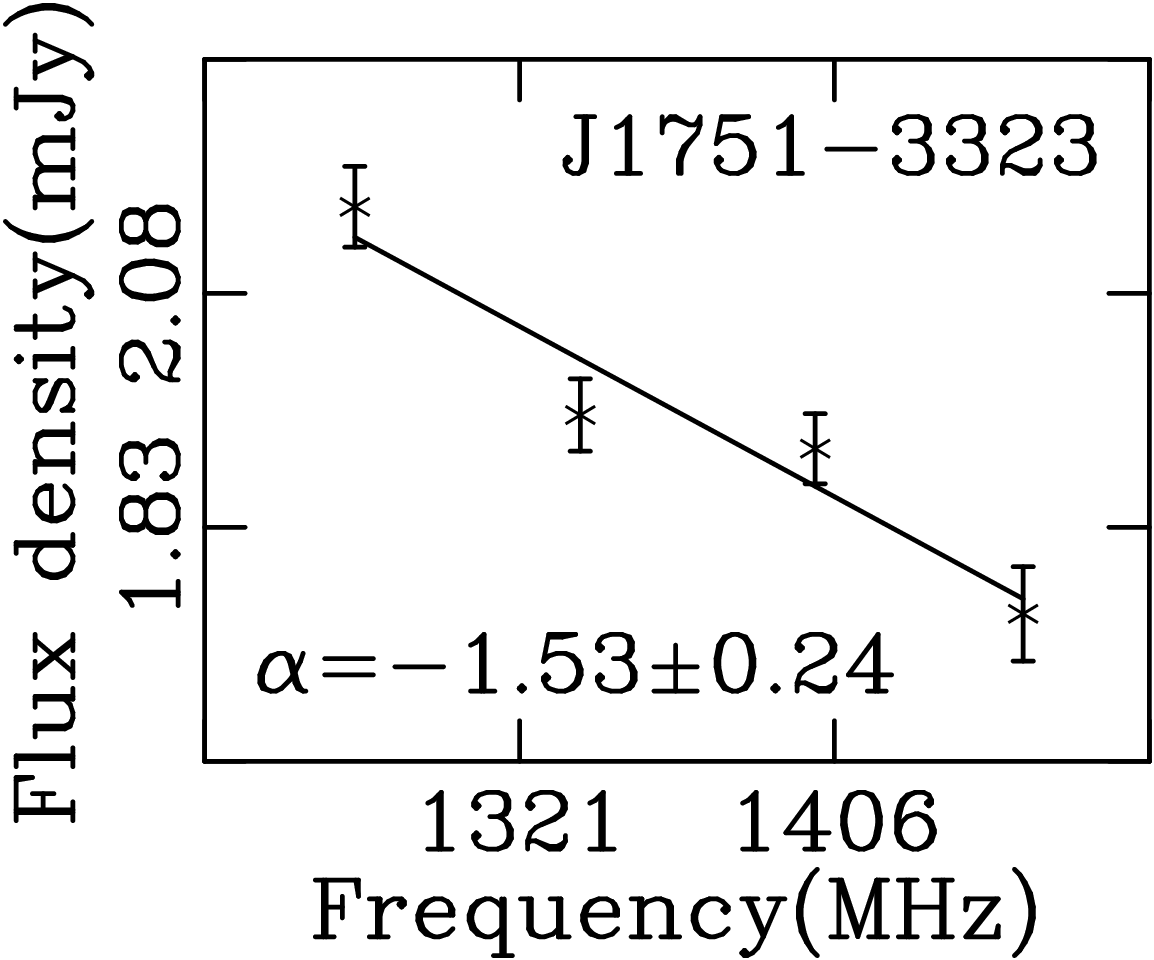}               
\includegraphics[bb = 110 120 518 685,clip,angle=-90,width=0.16\textwidth]{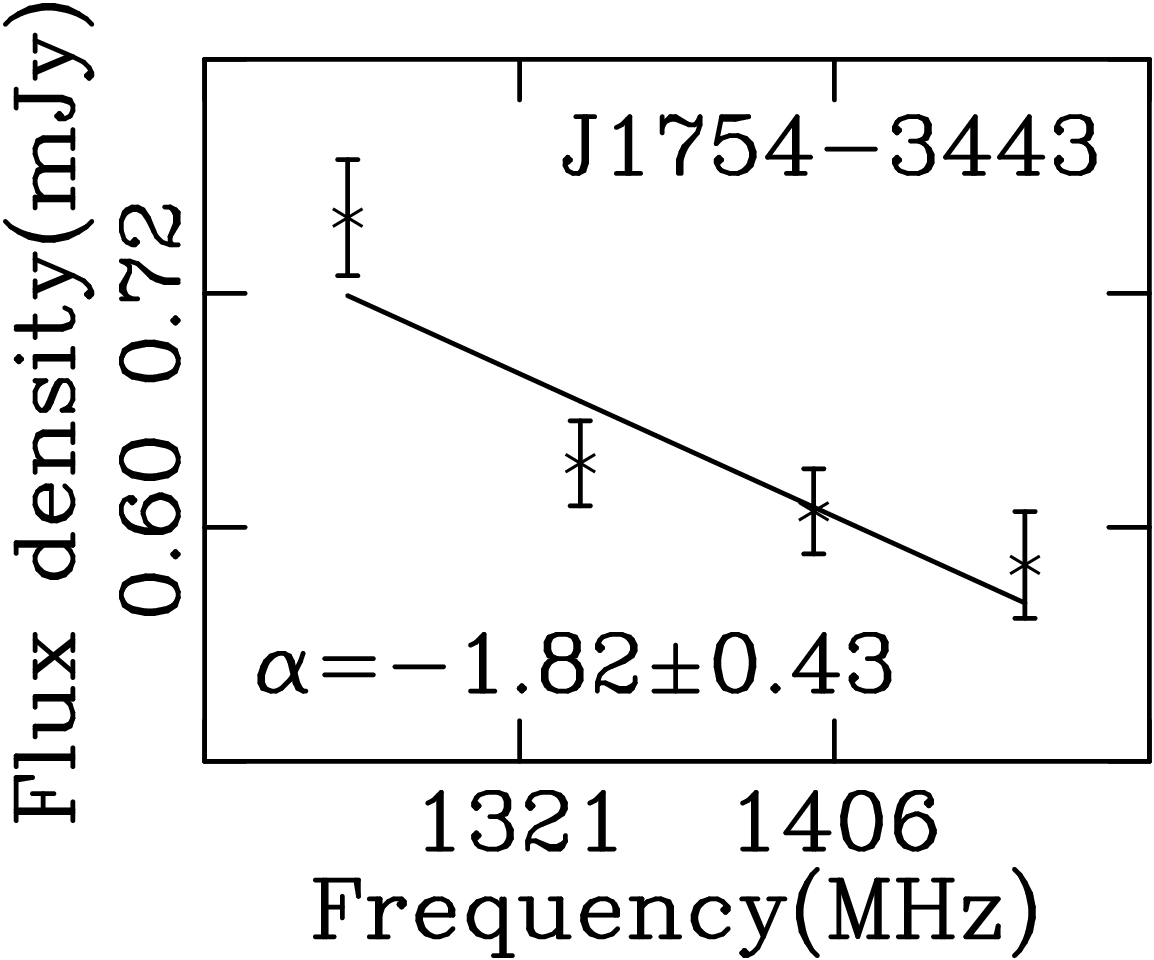}               
\includegraphics[bb = 110 120 518 685,clip,angle=-90,width=0.16\textwidth]{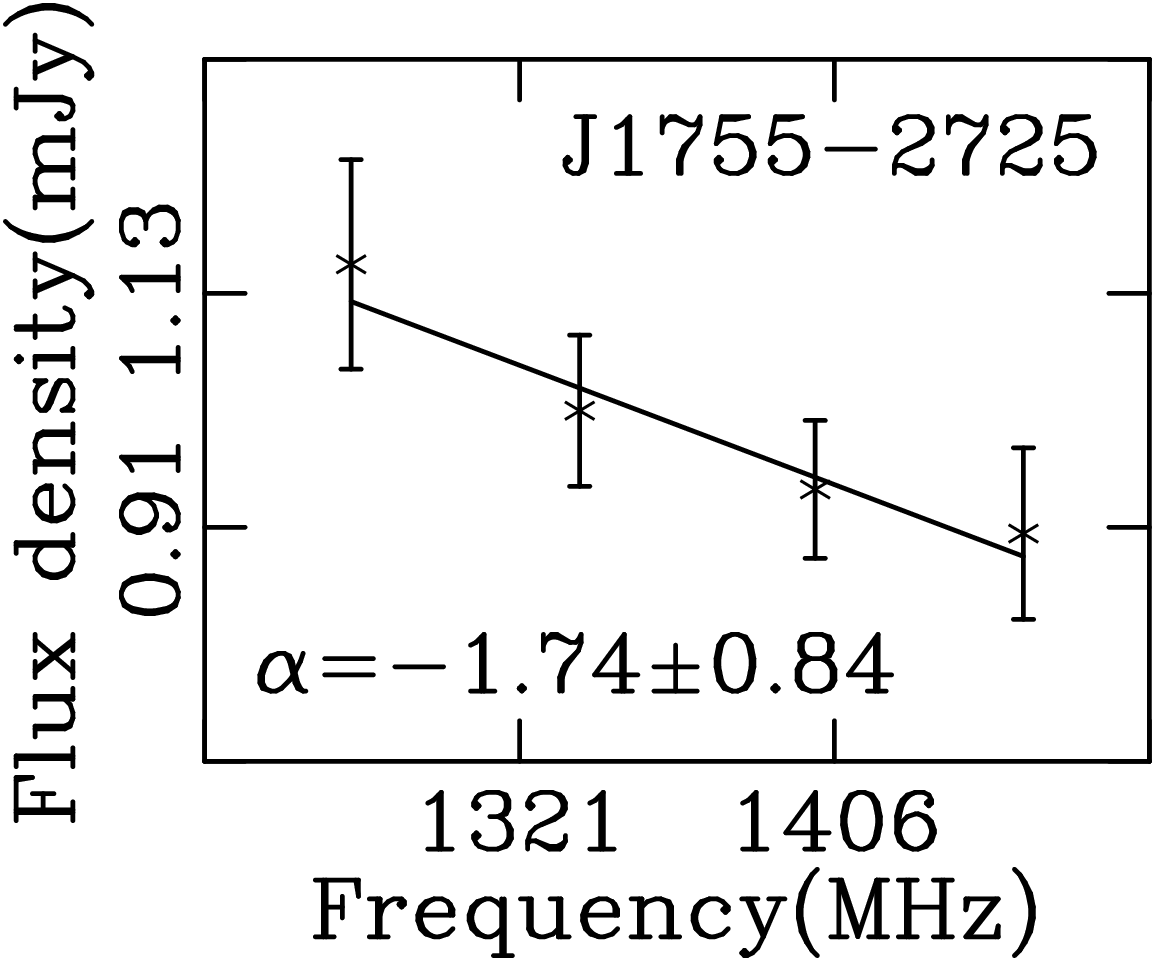}               
\includegraphics[bb = 110 120 518 685,clip,angle=-90,width=0.16\textwidth]{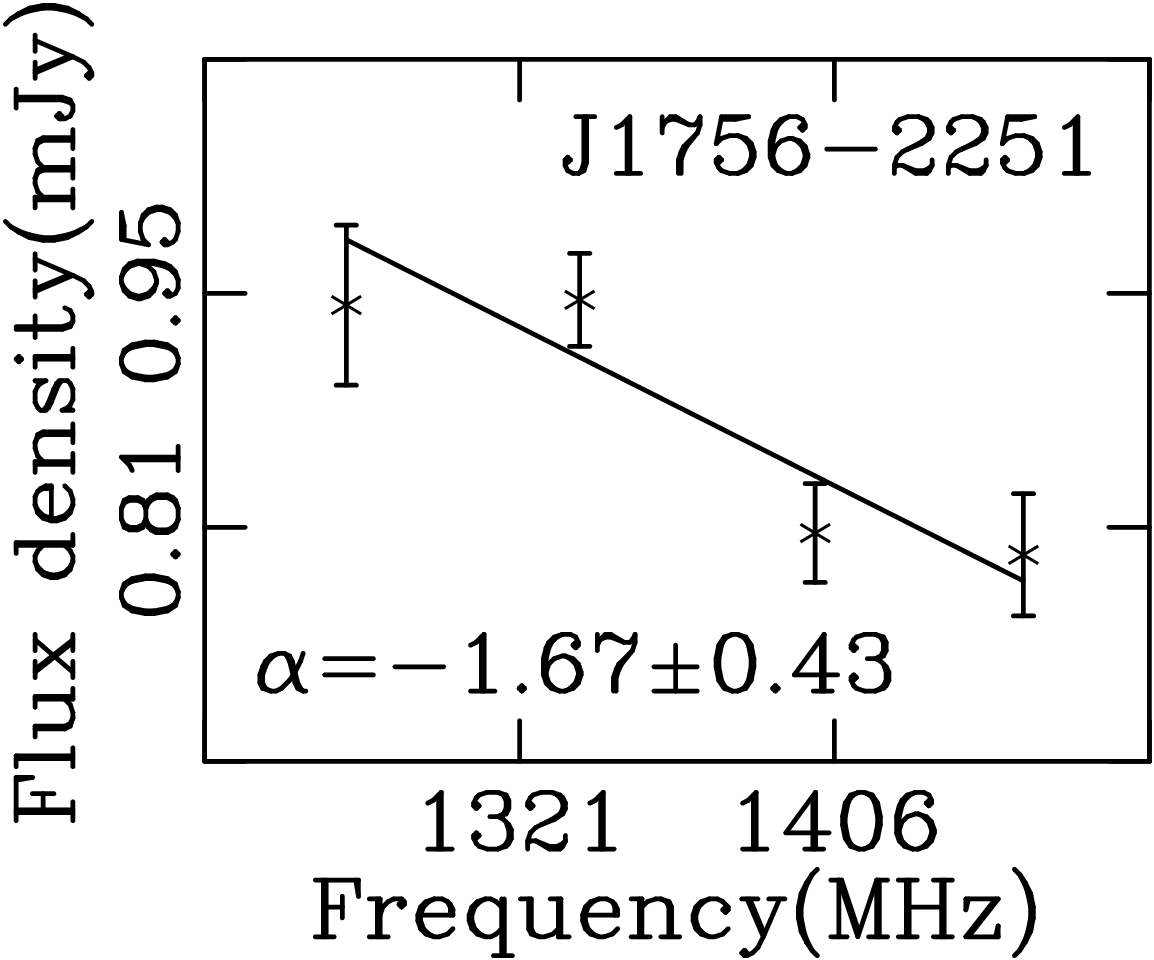}               
\includegraphics[bb = 110 120 518 685,clip,angle=-90,width=0.16\textwidth]{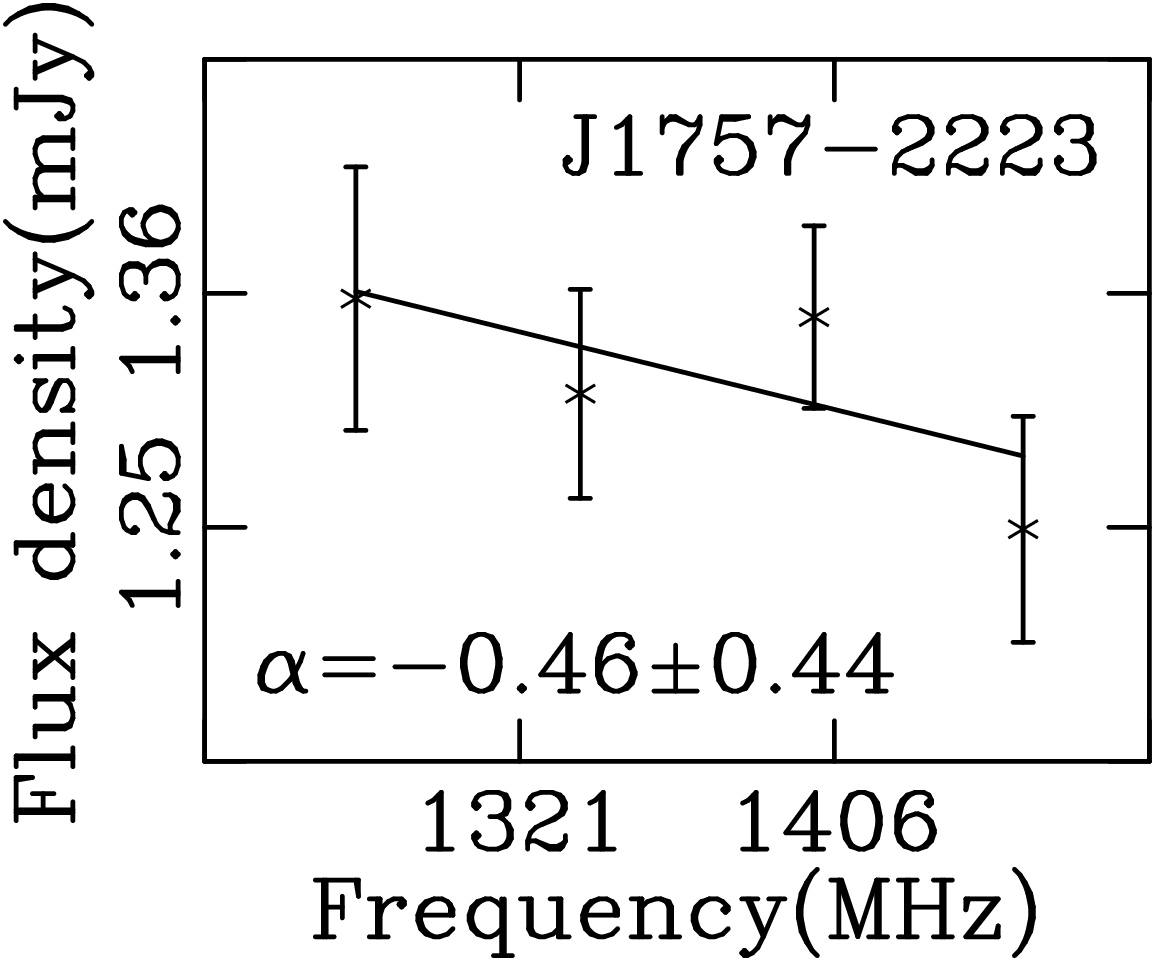}               
\includegraphics[bb = 110 120 518 685,clip,angle=-90,width=0.16\textwidth]{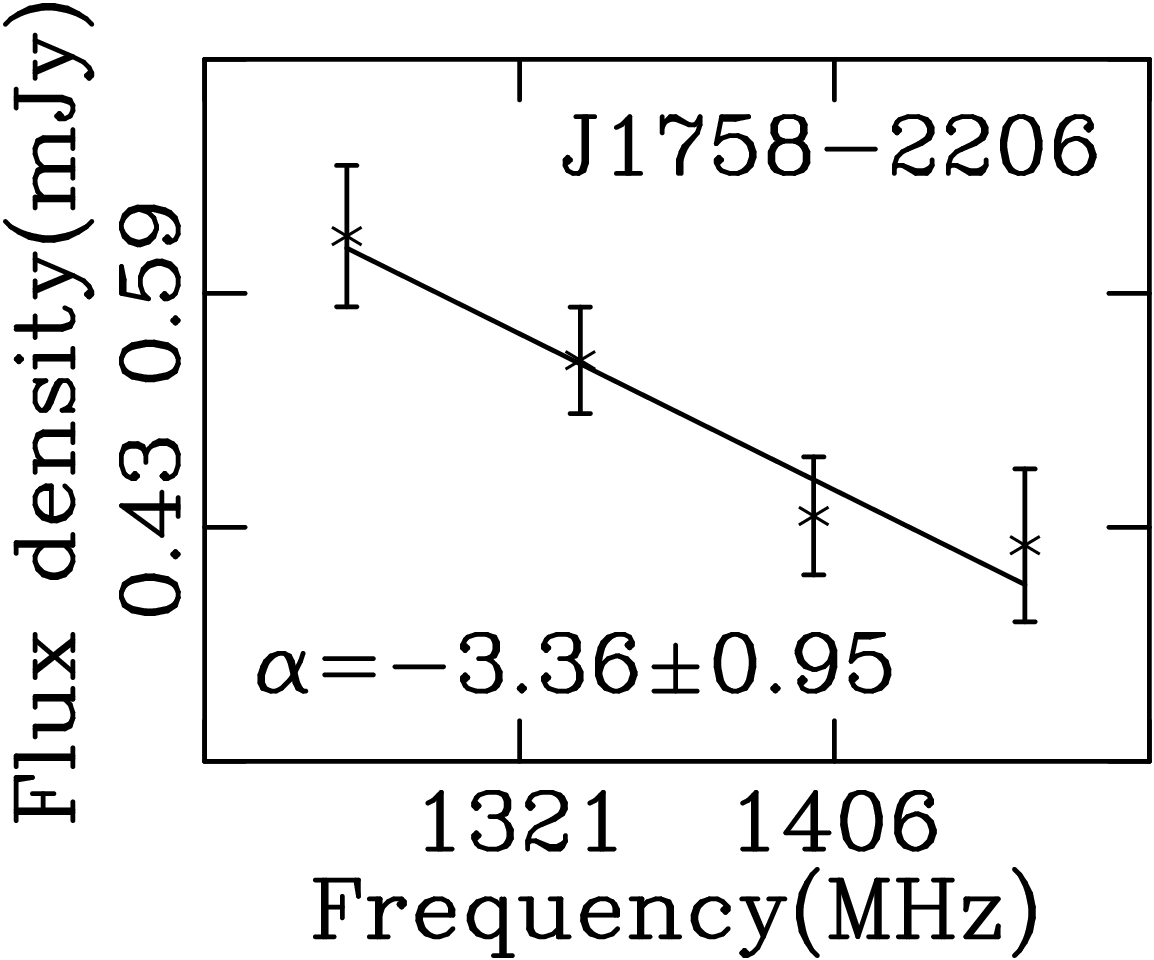}
\\[2mm]               
\includegraphics[bb = 110 72 518 685,clip,angle=-90,width=0.175\textwidth]{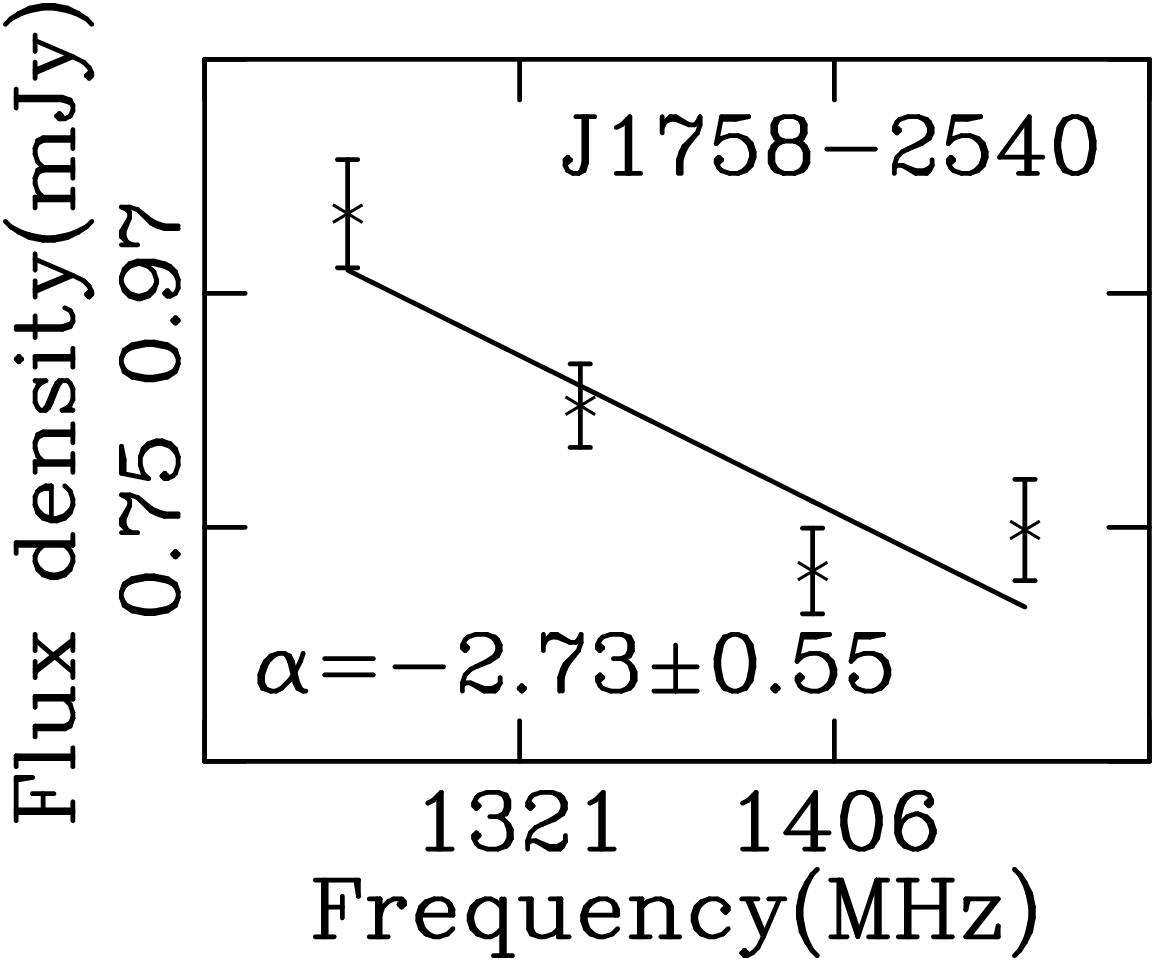}               
\includegraphics[bb = 110 120 518 685,clip,angle=-90,width=0.16\textwidth]{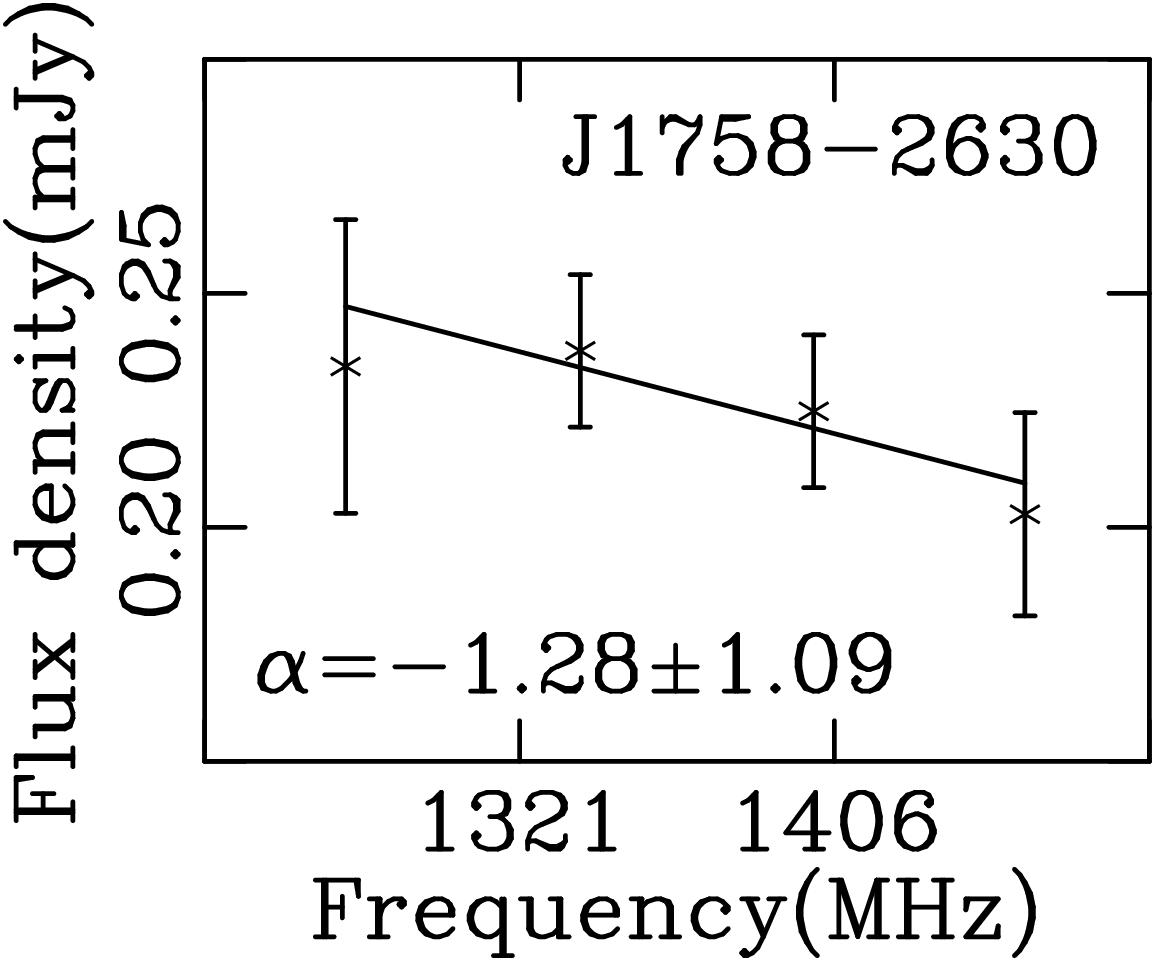}               
\includegraphics[bb = 110 120 518 685,clip,angle=-90,width=0.16\textwidth]{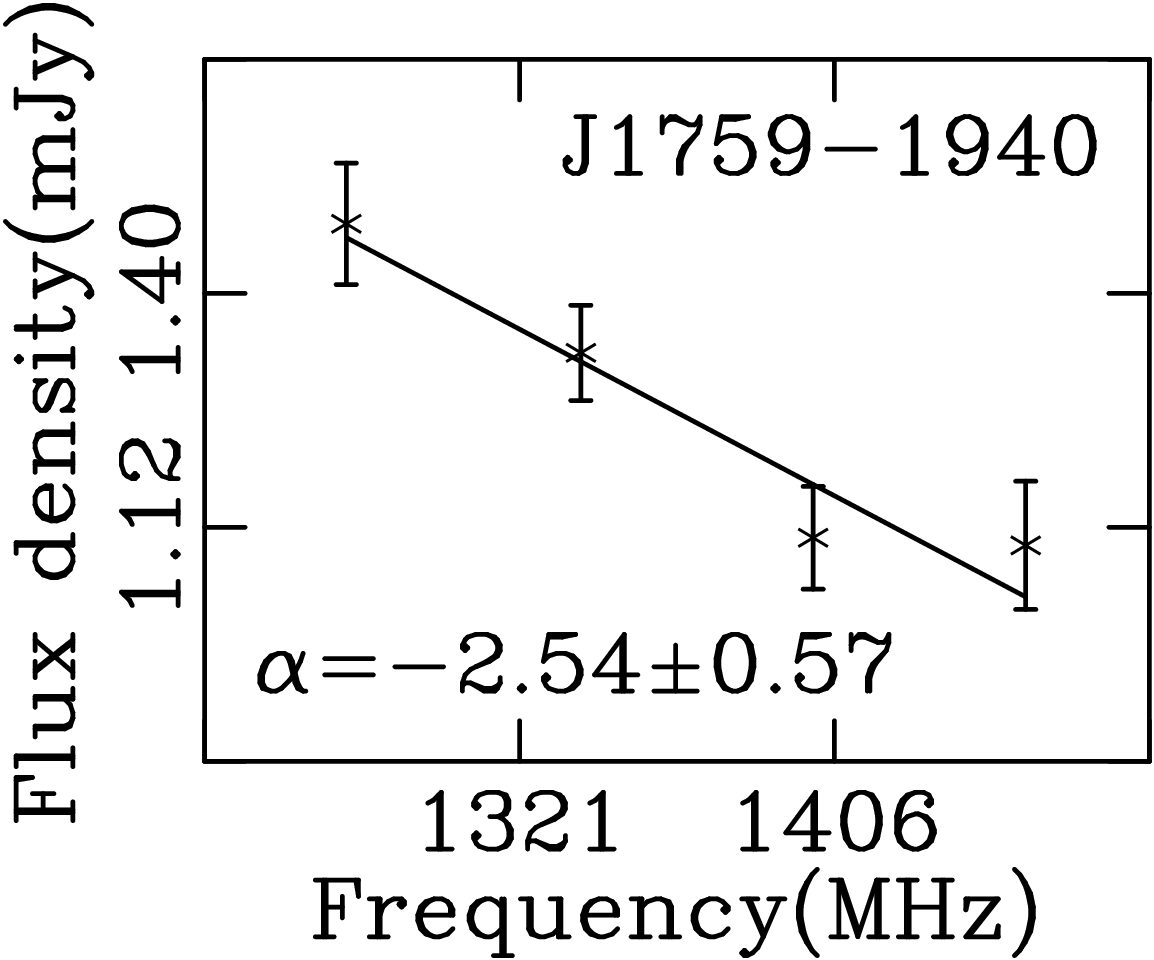}               
\includegraphics[bb = 110 120 518 685,clip,angle=-90,width=0.16\textwidth]{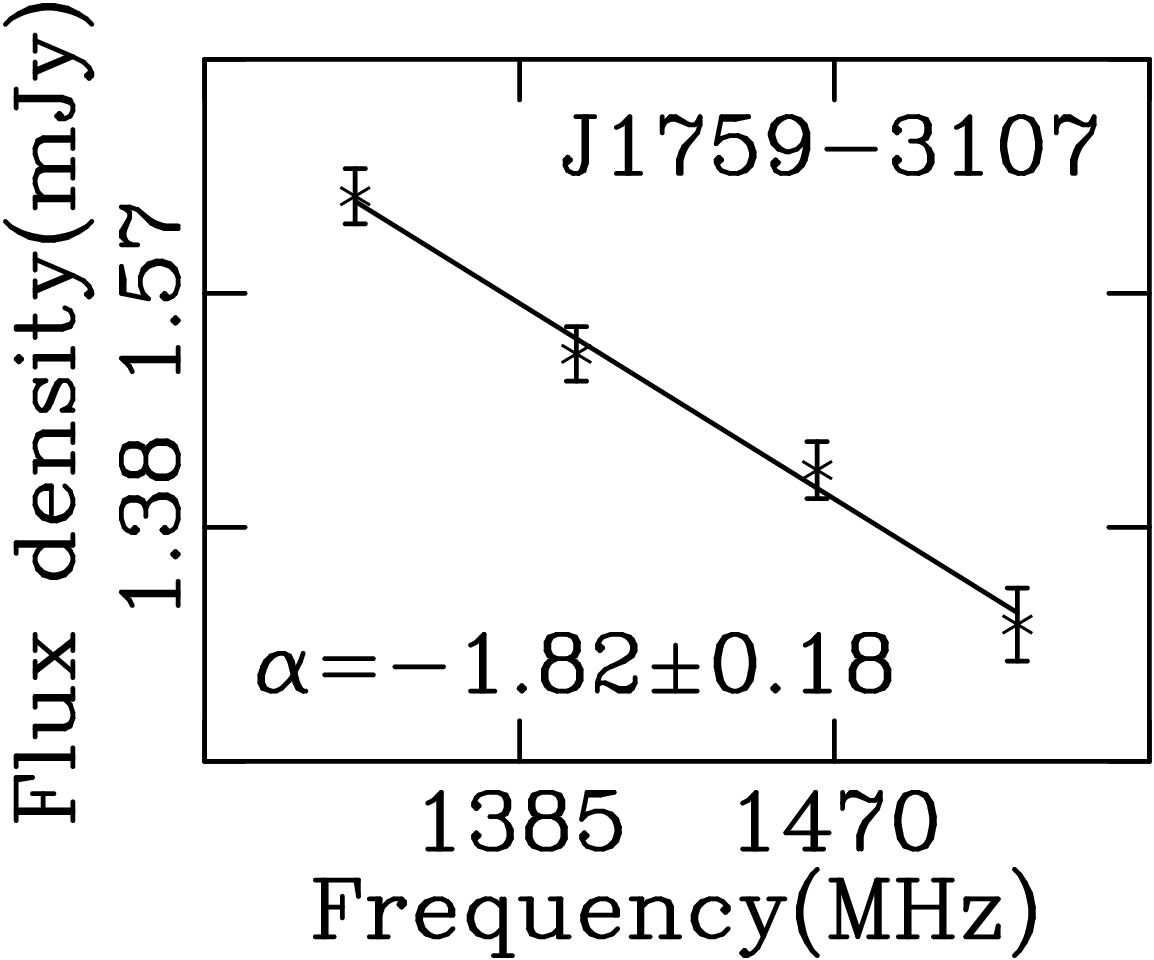}               
\includegraphics[bb = 110 120 518 685,clip,angle=-90,width=0.16\textwidth]{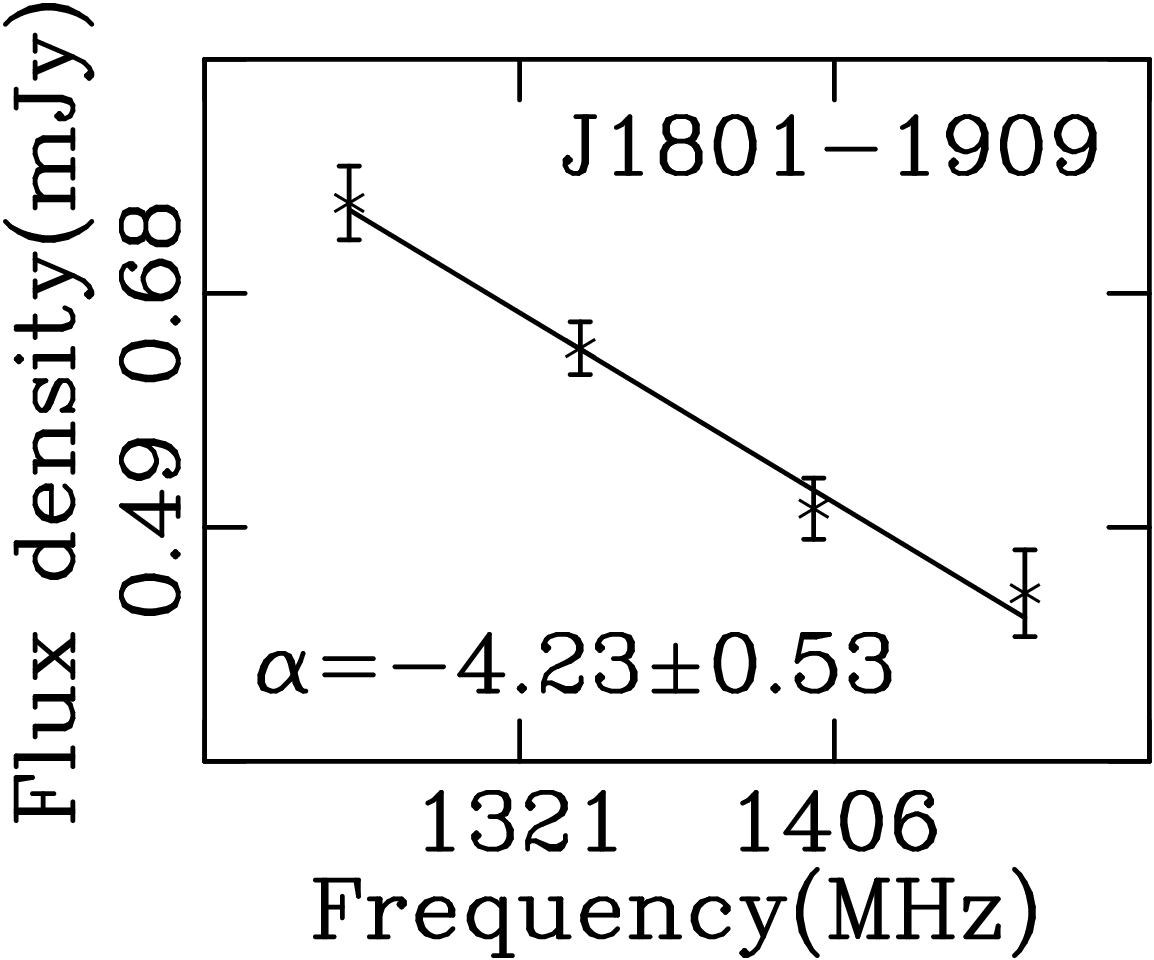}               
\includegraphics[bb = 110 120 518 685,clip,angle=-90,width=0.16\textwidth]{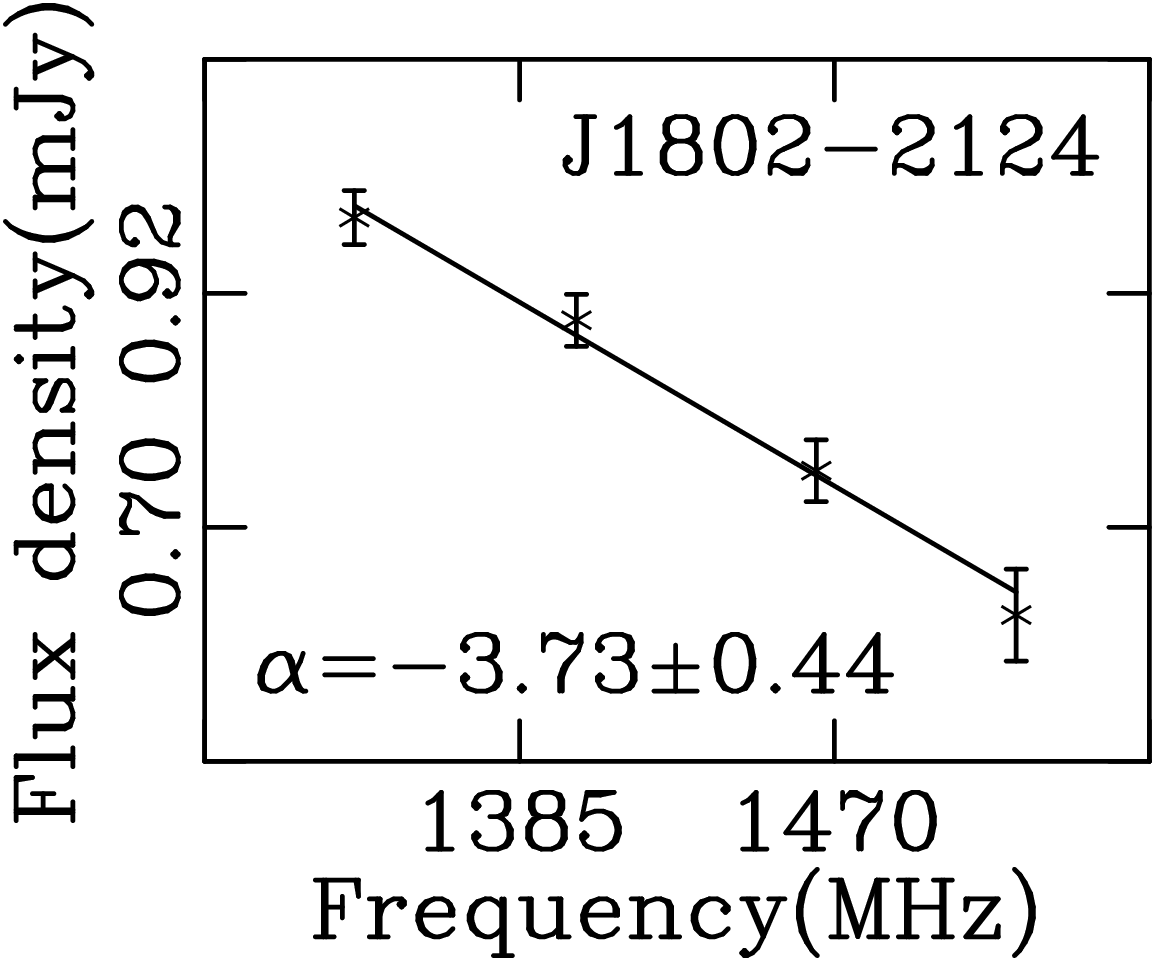} 
\\[2mm]              
\includegraphics[bb = 110 72 518 685,clip,angle=-90,width=0.175\textwidth]{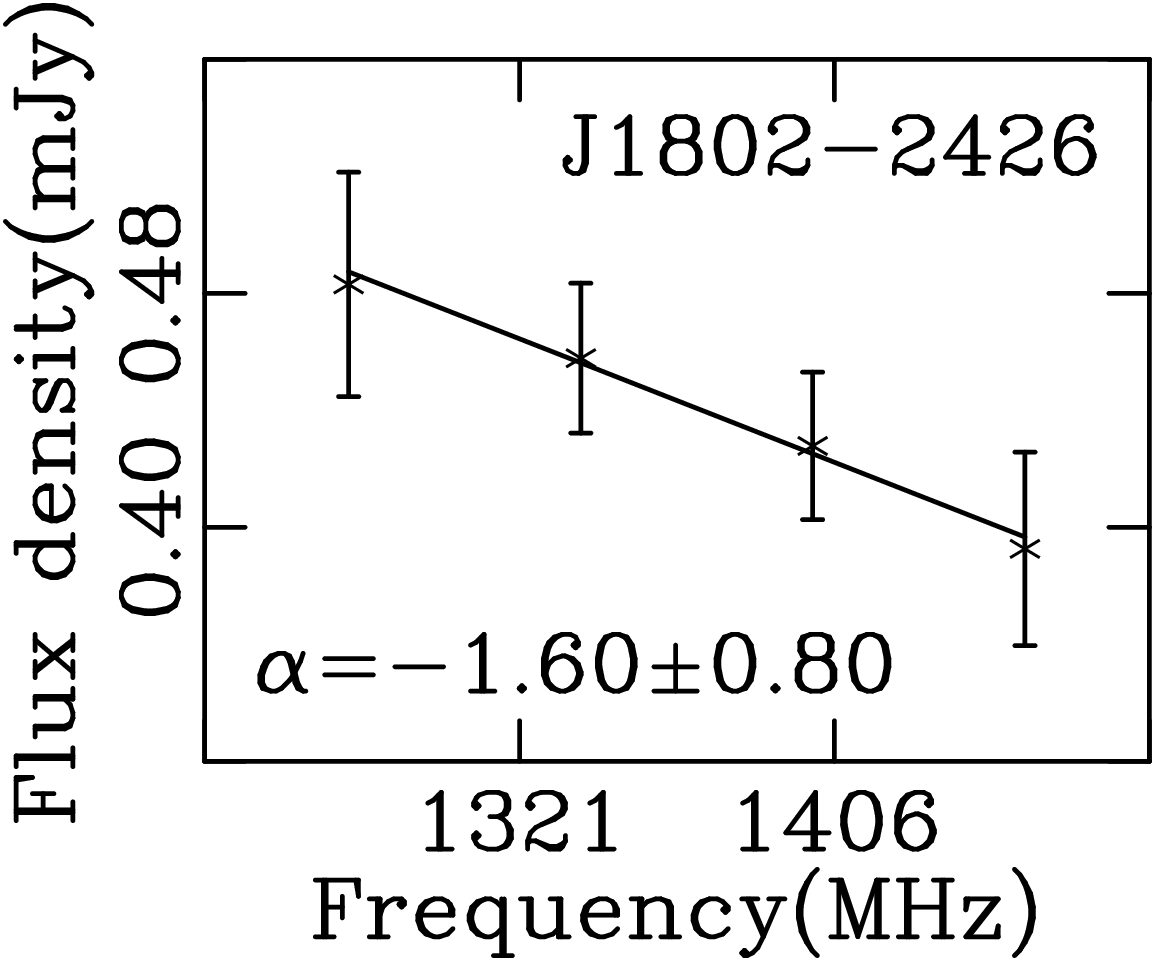}               
\includegraphics[bb = 110 120 518 685,clip,angle=-90,width=0.16\textwidth]{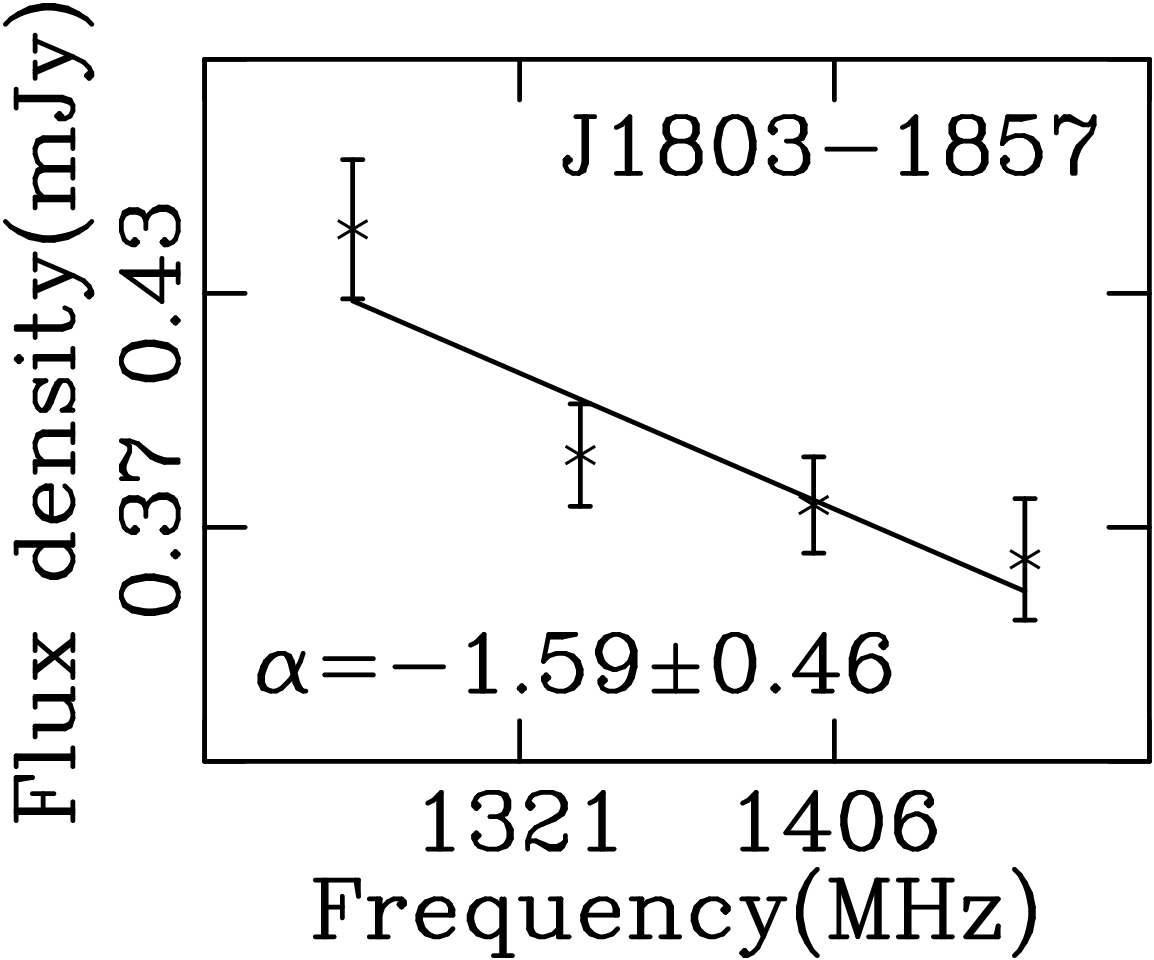}               
\includegraphics[bb = 110 120 518 685,clip,angle=-90,width=0.16\textwidth]{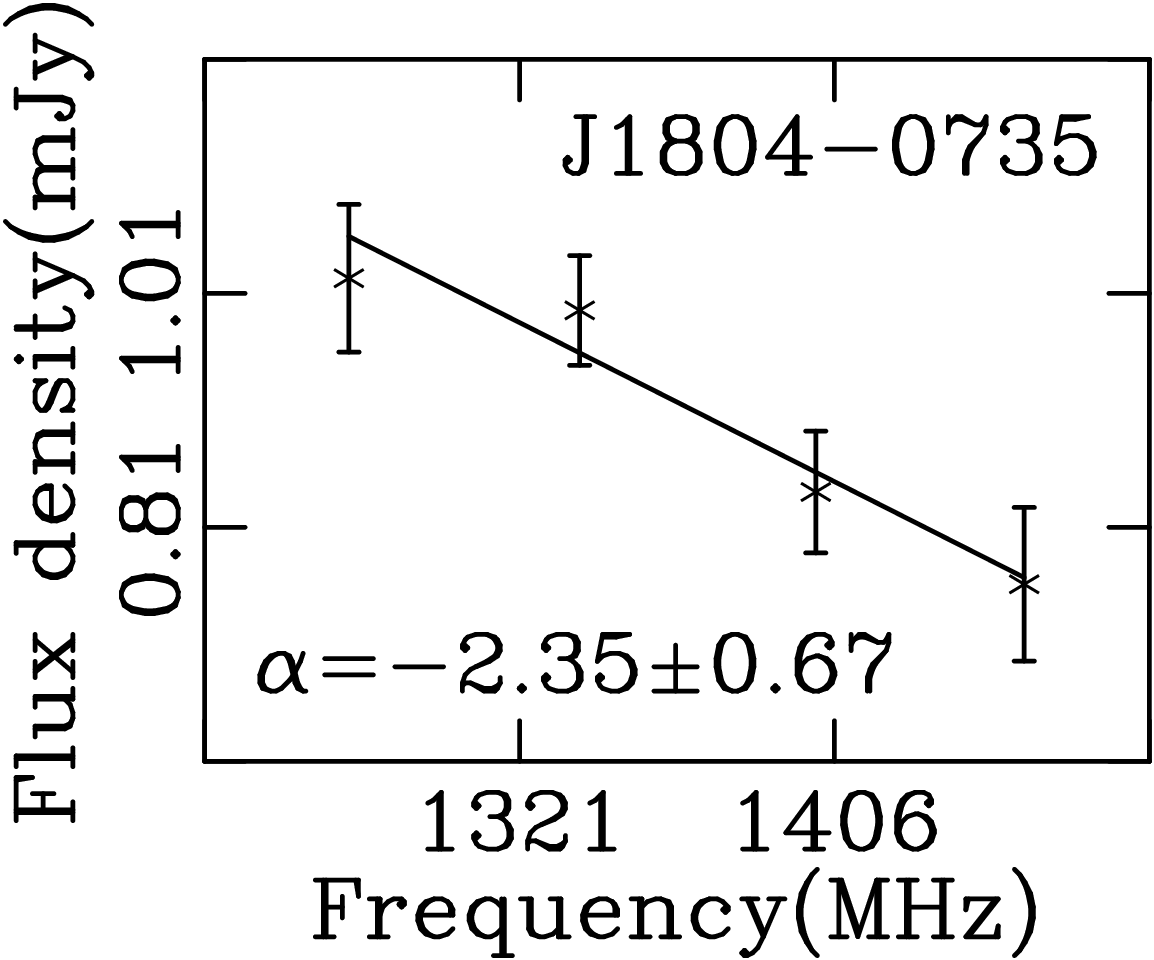}               
\includegraphics[bb = 110 120 518 685,clip,angle=-90,width=0.16\textwidth]{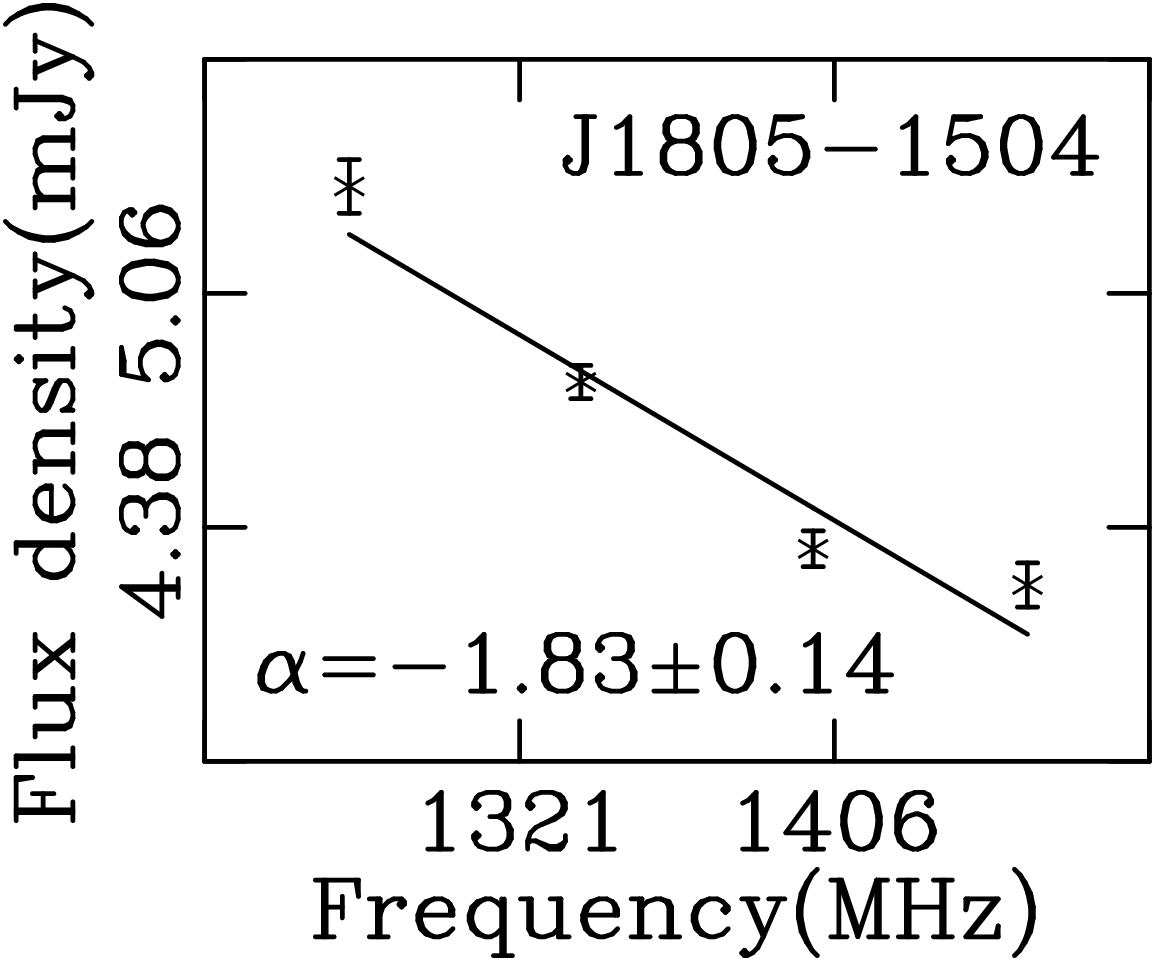}               
\includegraphics[bb = 110 120 518 685,clip,angle=-90,width=0.16\textwidth]{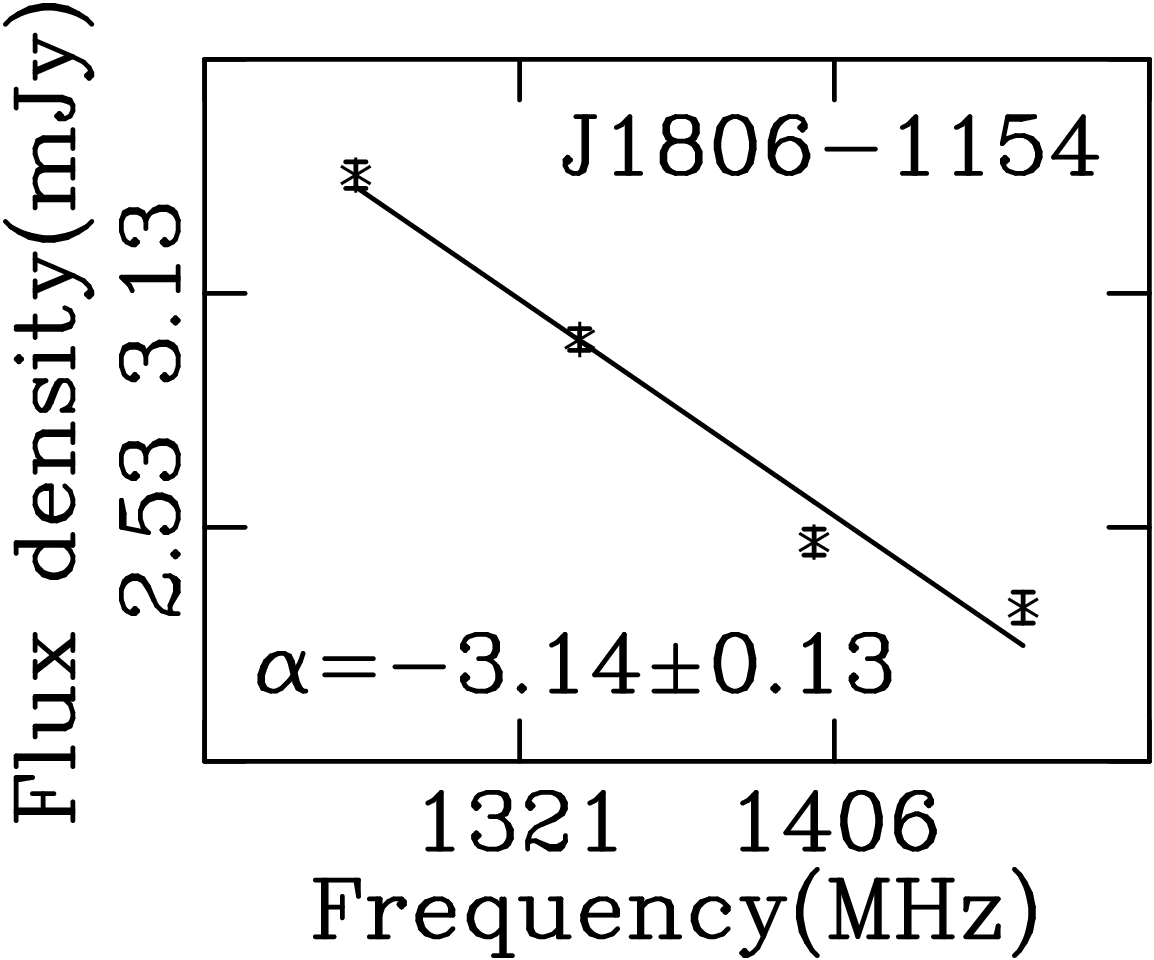}               
\includegraphics[bb = 110 120 518 685,clip,angle=-90,width=0.16\textwidth]{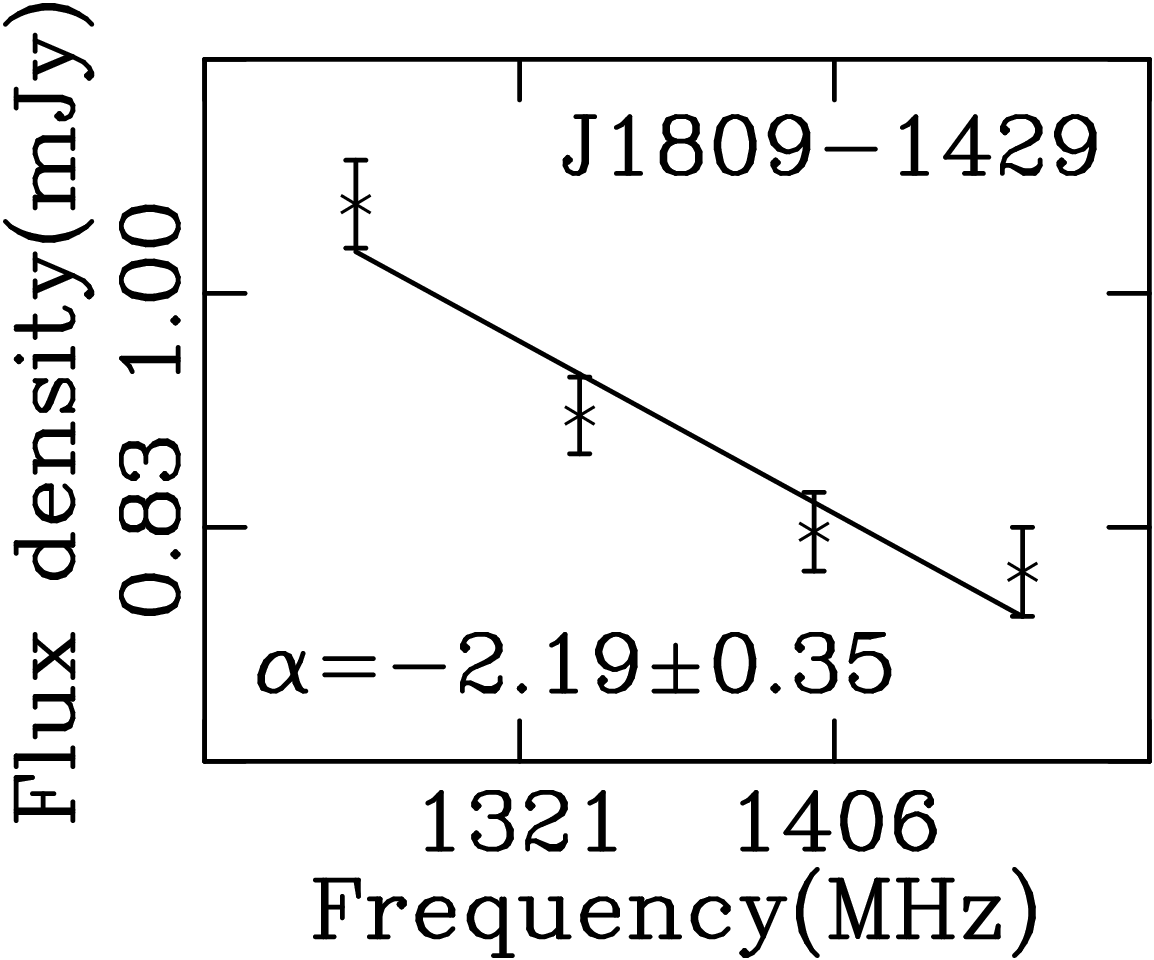} 
\\[2mm]              
\includegraphics[bb = 110 72 518 685,clip,angle=-90,width=0.175\textwidth]{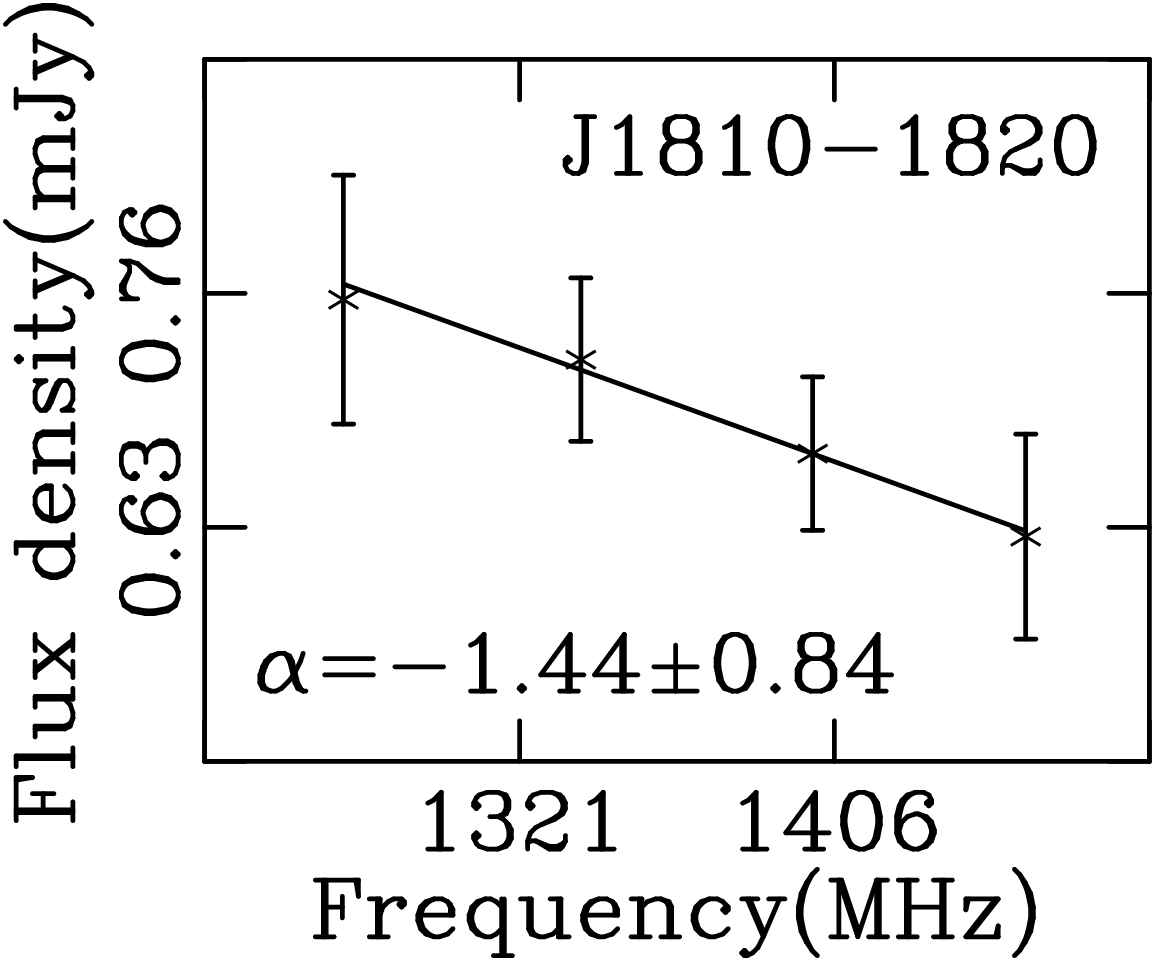}               
\includegraphics[bb = 110 120 518 685,clip,angle=-90,width=0.16\textwidth]{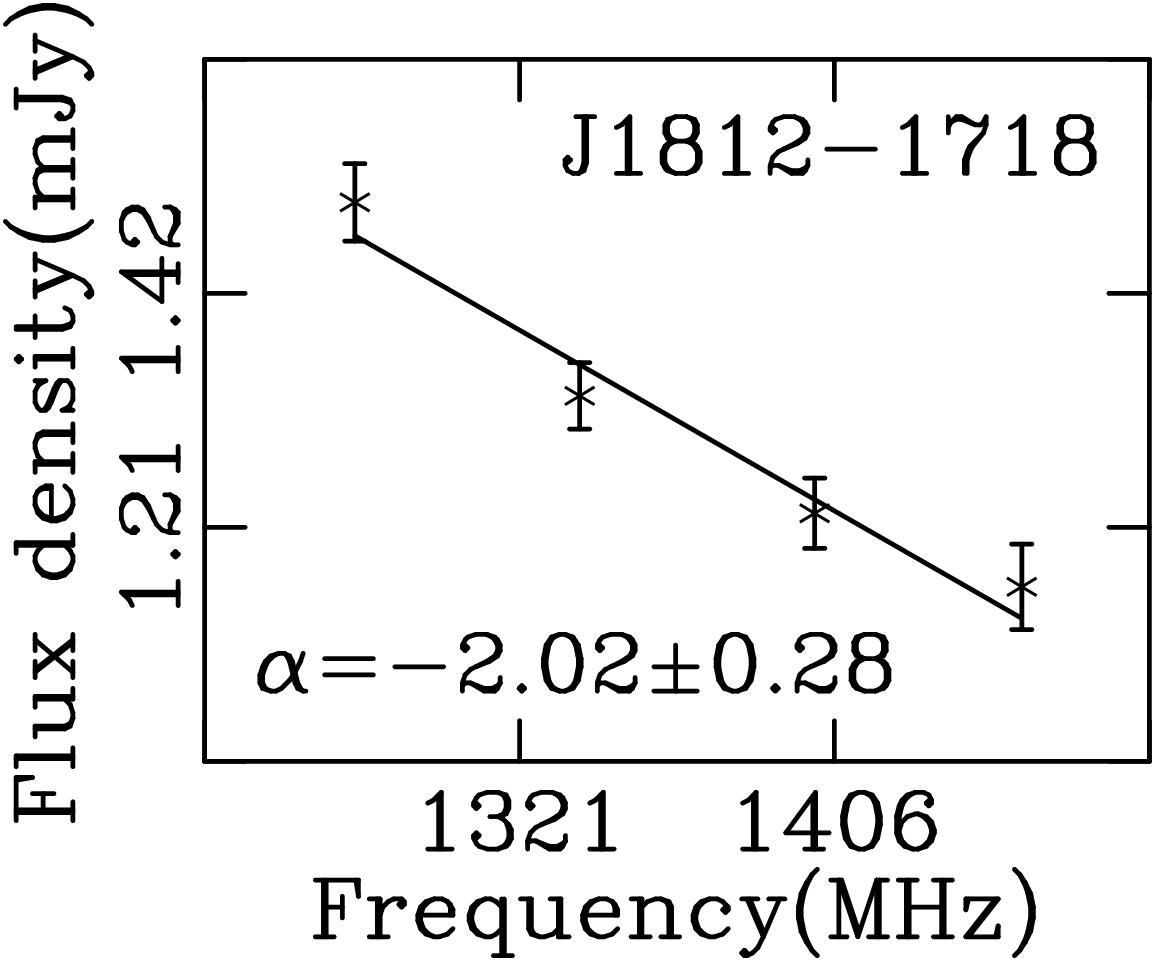}               
\includegraphics[bb = 110 120 518 685,clip,angle=-90,width=0.16\textwidth]{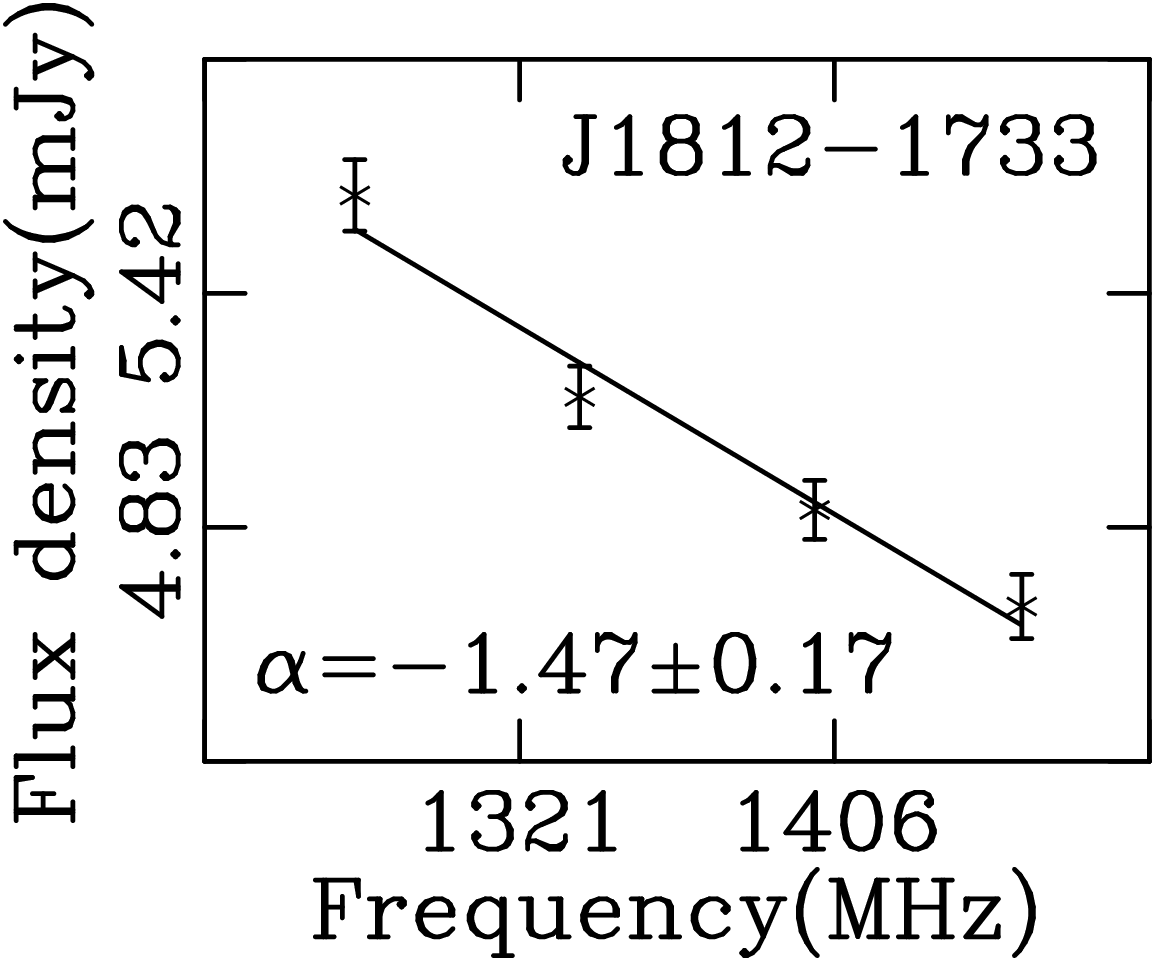}               
\includegraphics[bb = 110 120 518 685,clip,angle=-90,width=0.16\textwidth]{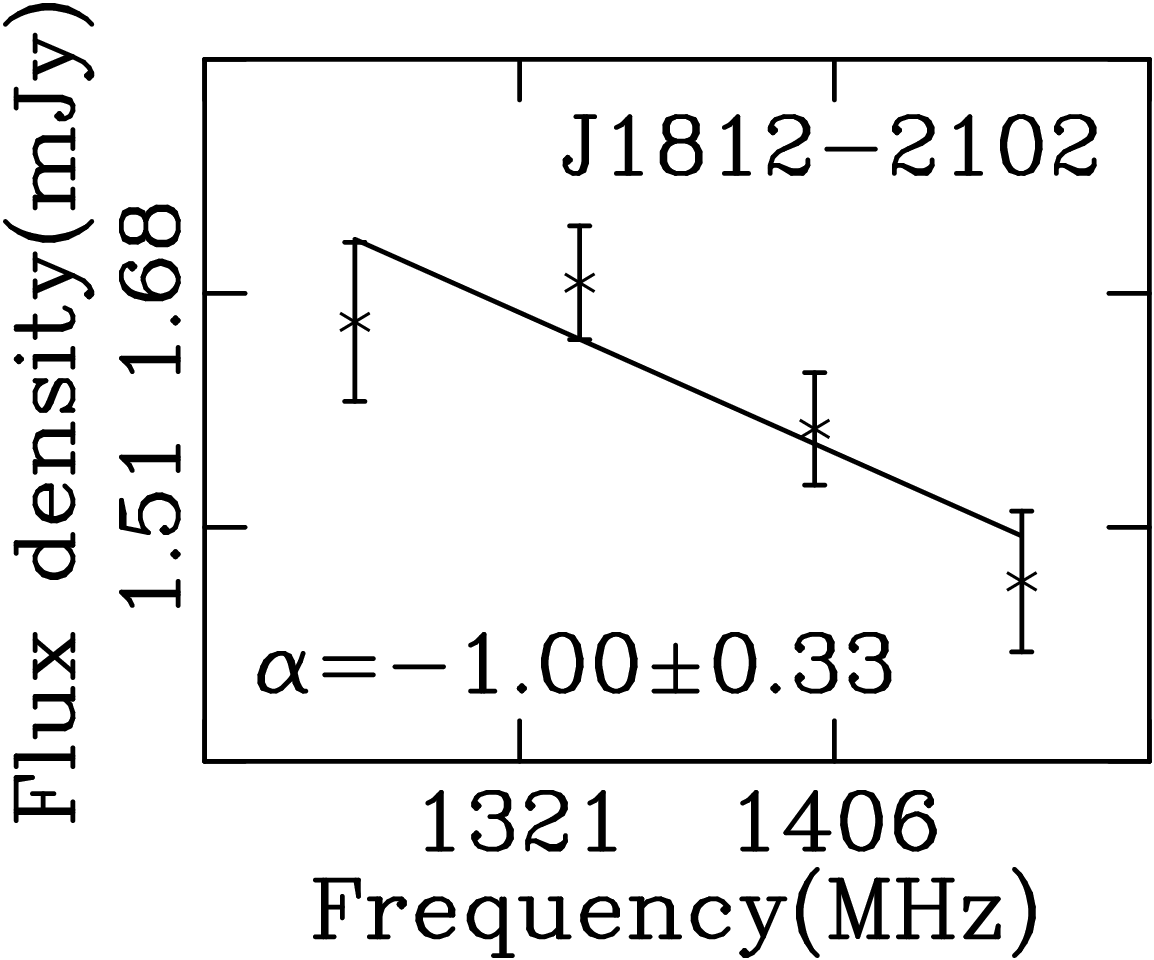}               
\includegraphics[bb = 110 120 518 685,clip,angle=-90,width=0.16\textwidth]{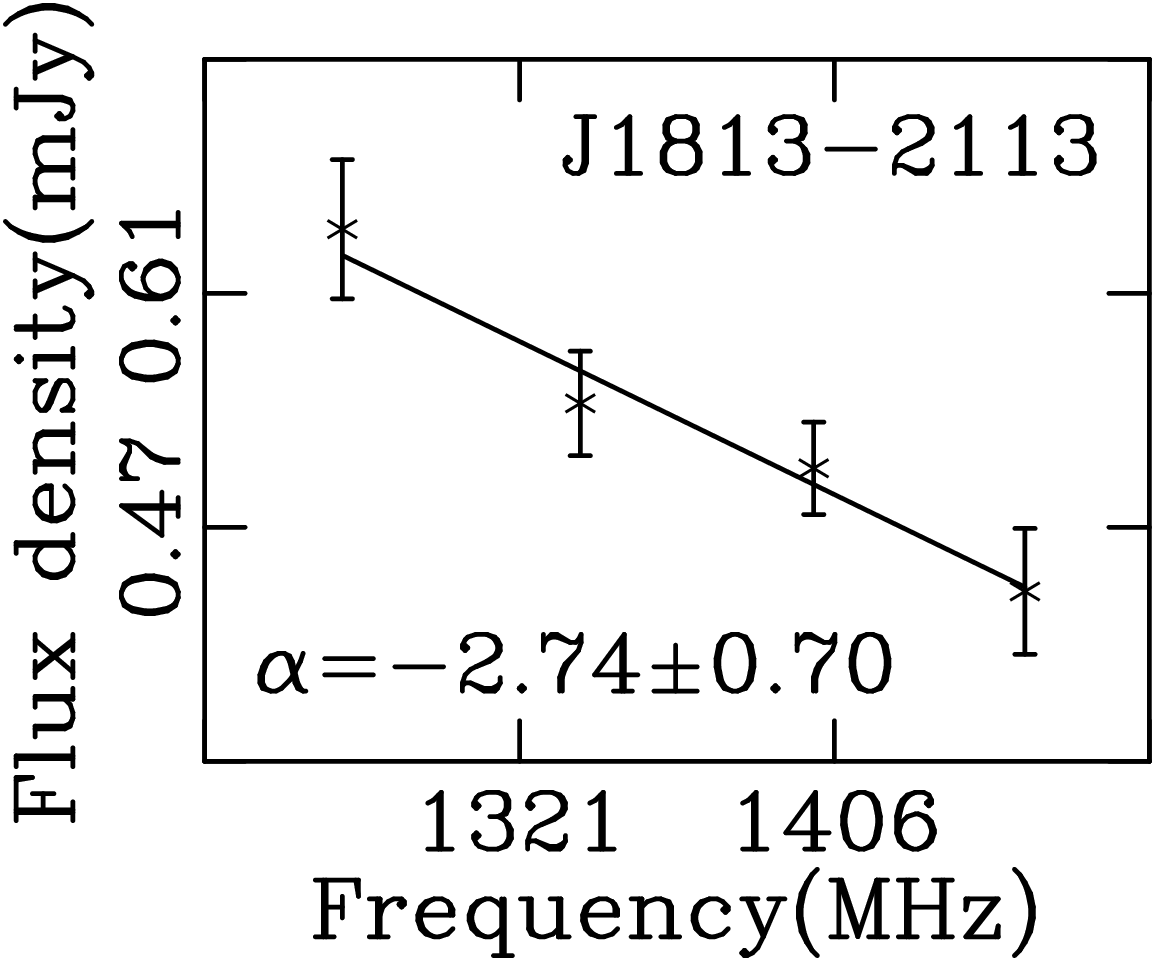}               
\includegraphics[bb = 110 120 518 685,clip,angle=-90,width=0.16\textwidth]{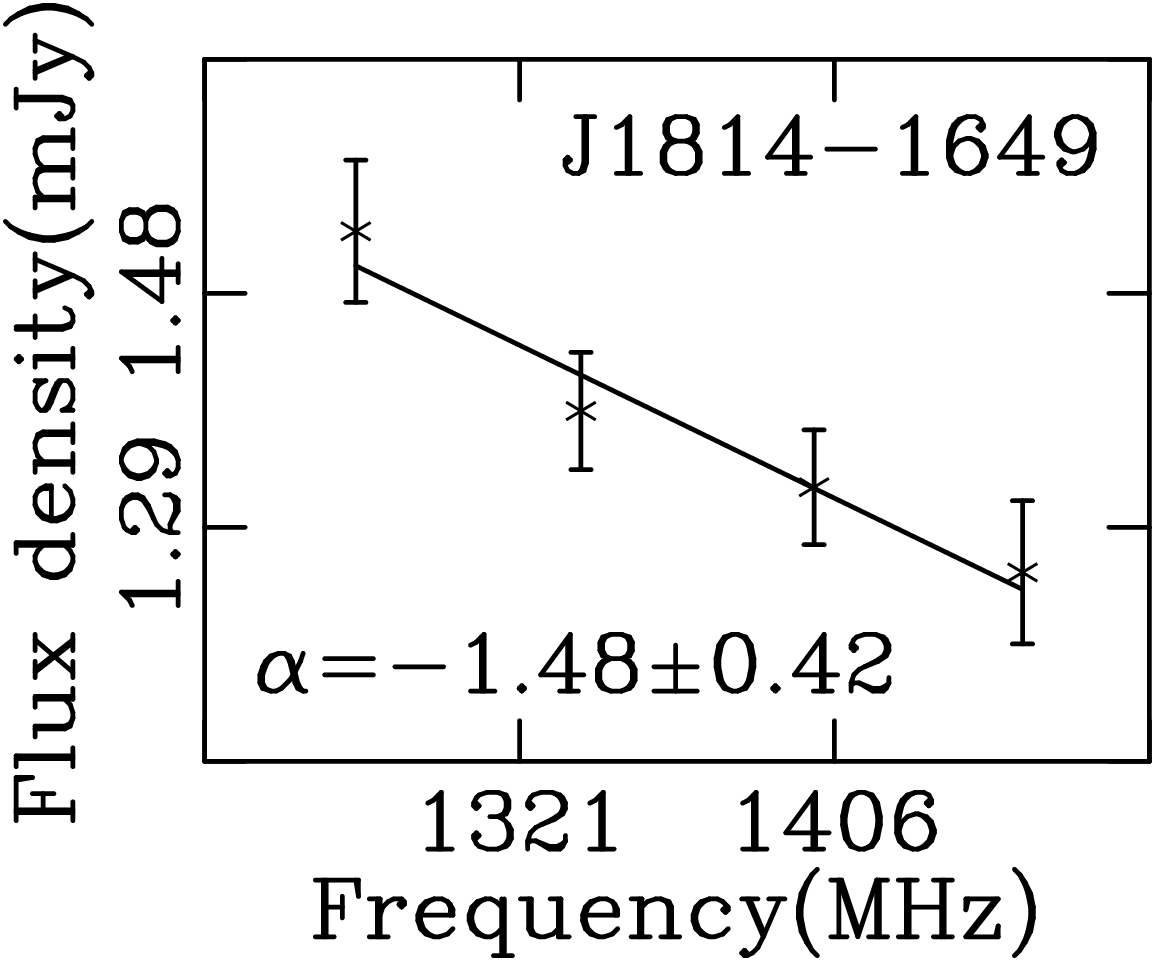}
 \\[2mm]              
\includegraphics[bb = 110 72 570 685,clip,angle=-90,width=0.175\textwidth]{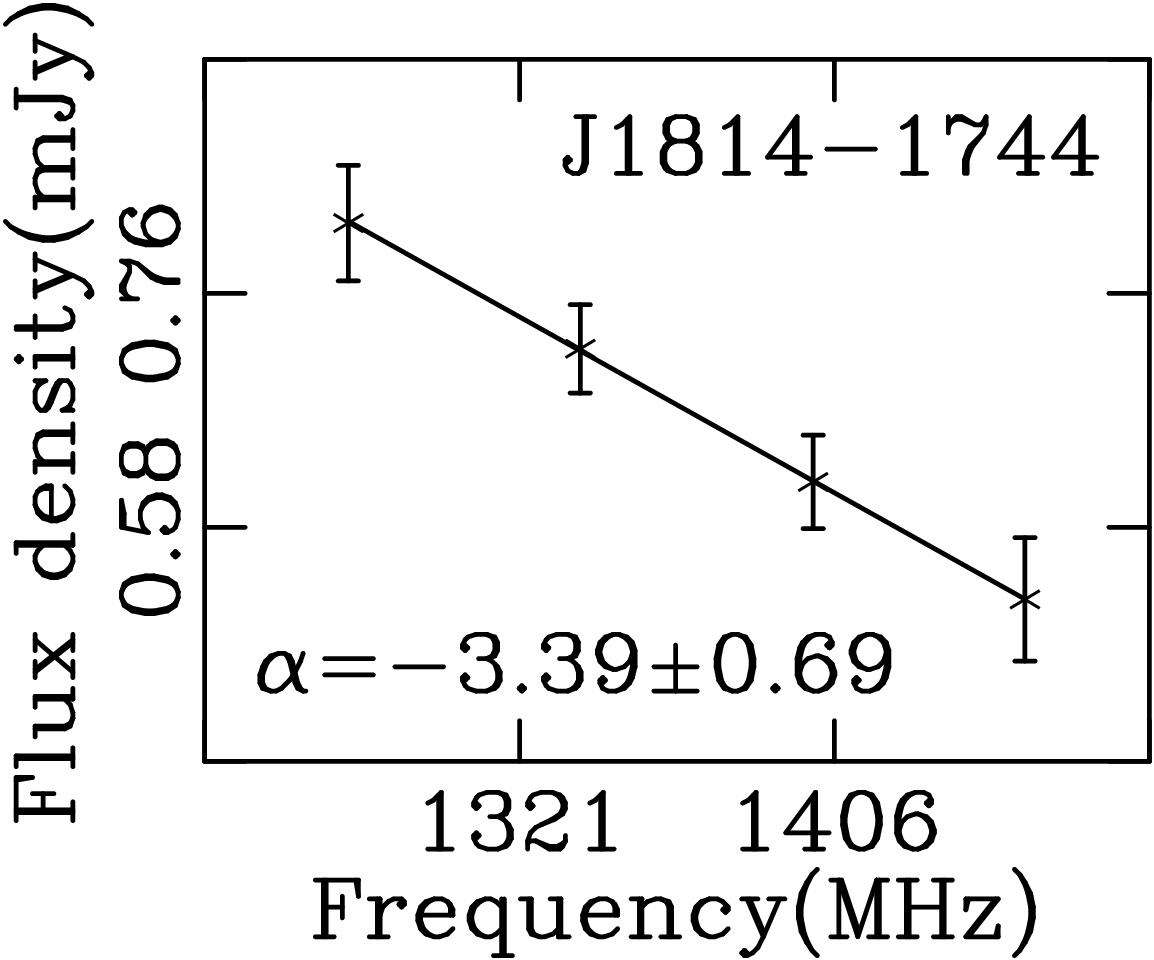}               
\includegraphics[bb = 110 120 570 685,clip,angle=-90,width=0.16\textwidth]{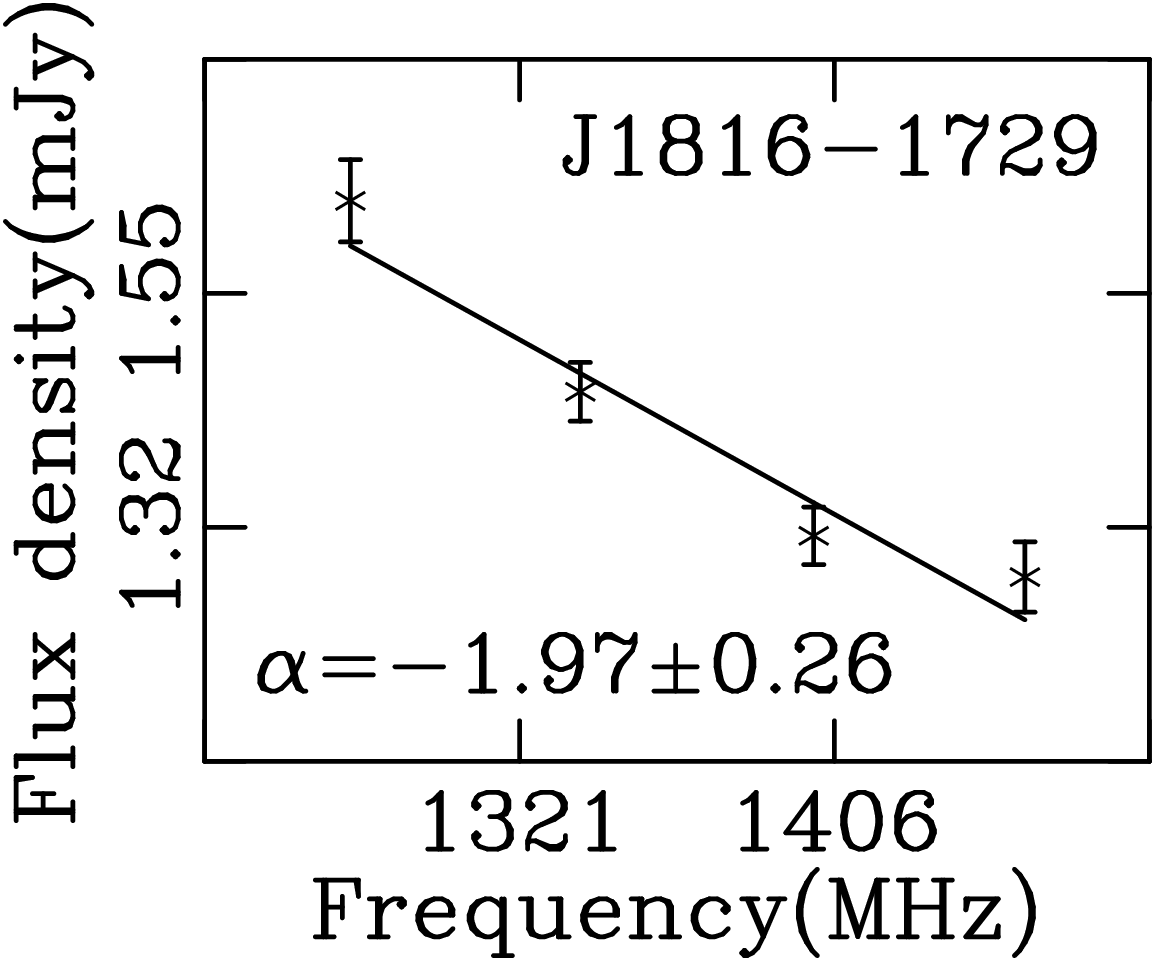}               
\includegraphics[bb = 110 120 570 685,clip,angle=-90,width=0.16\textwidth]{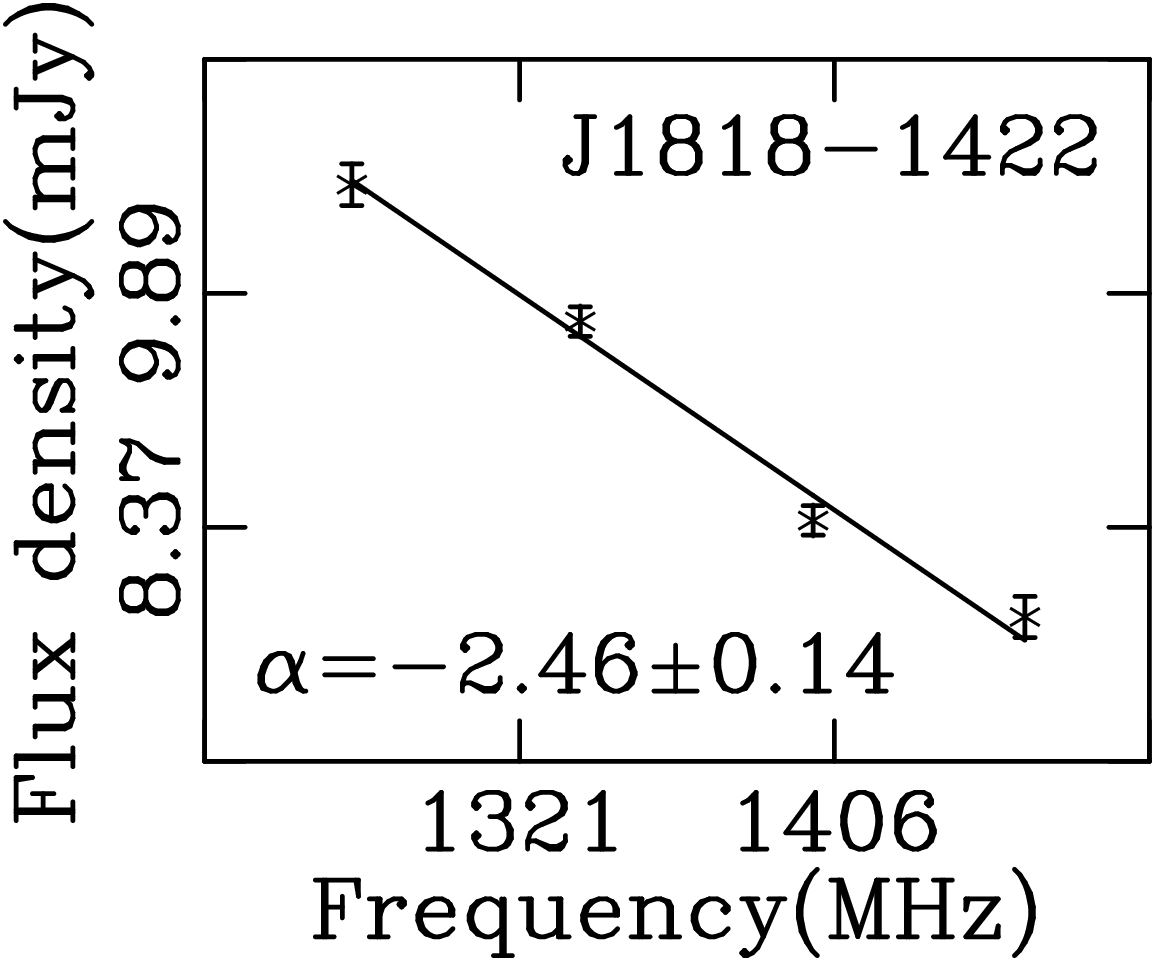}               
\includegraphics[bb = 110 120 570 685,clip,angle=-90,width=0.16\textwidth]{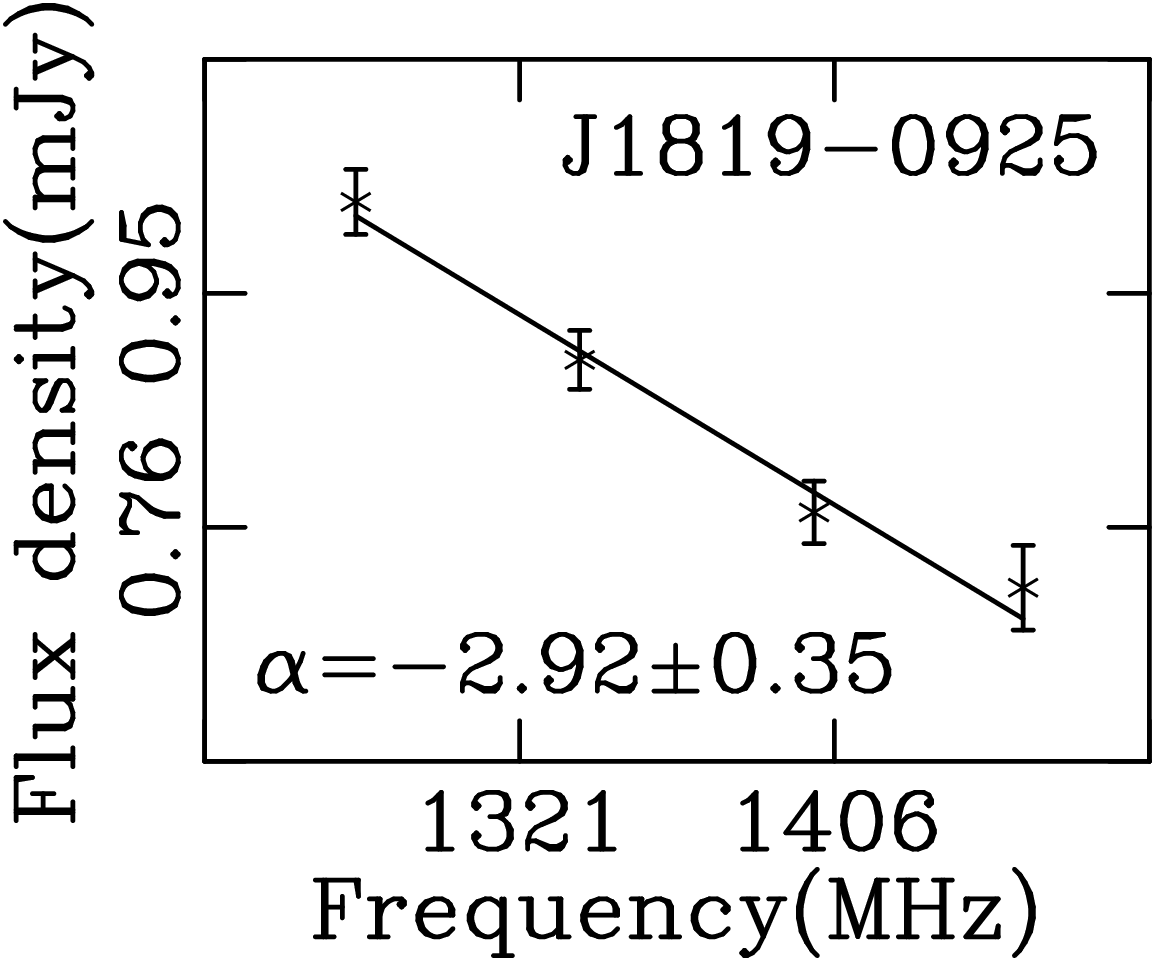}               
\includegraphics[bb = 110 120 570 685,clip,angle=-90,width=0.16\textwidth]{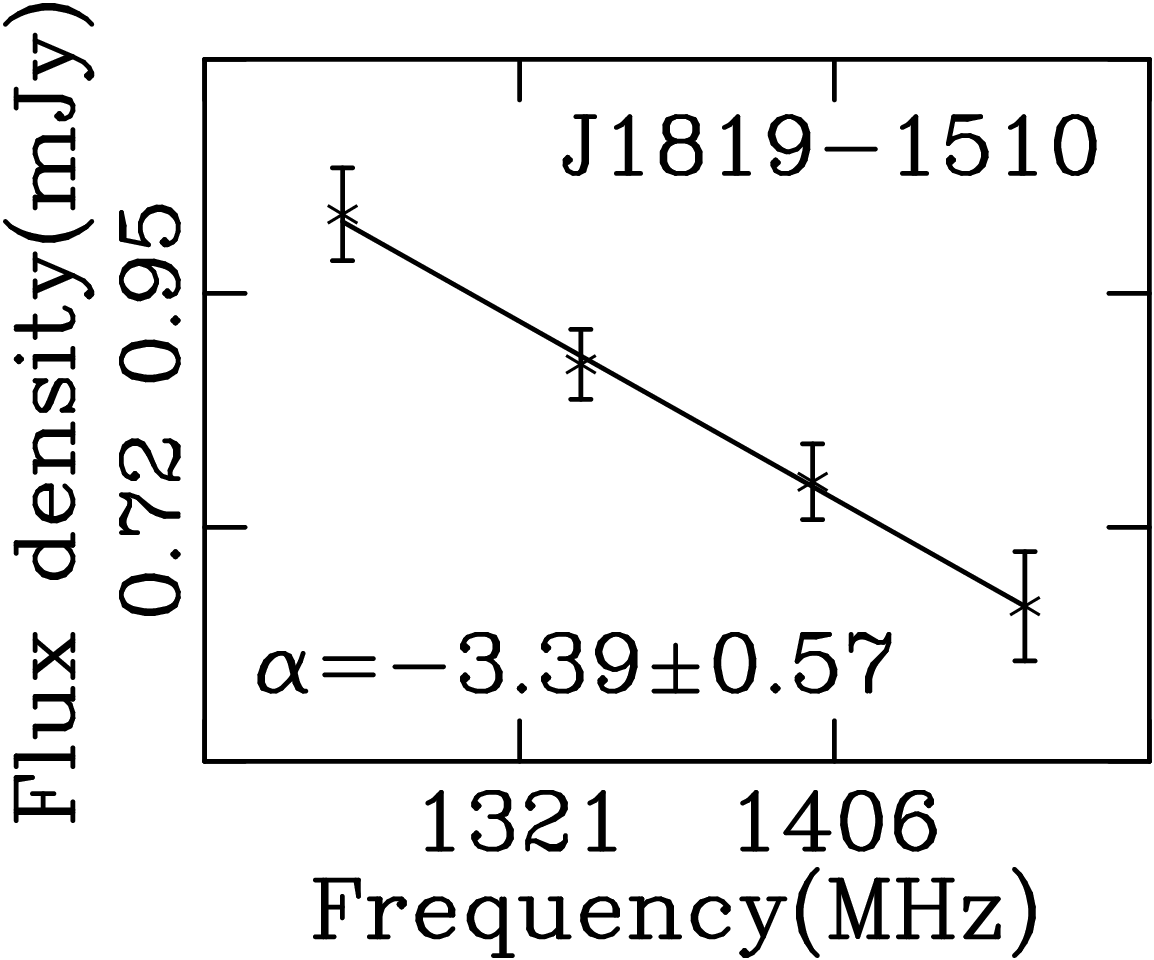}               
\includegraphics[bb = 110 120 570 685,clip,angle=-90,width=0.16\textwidth]{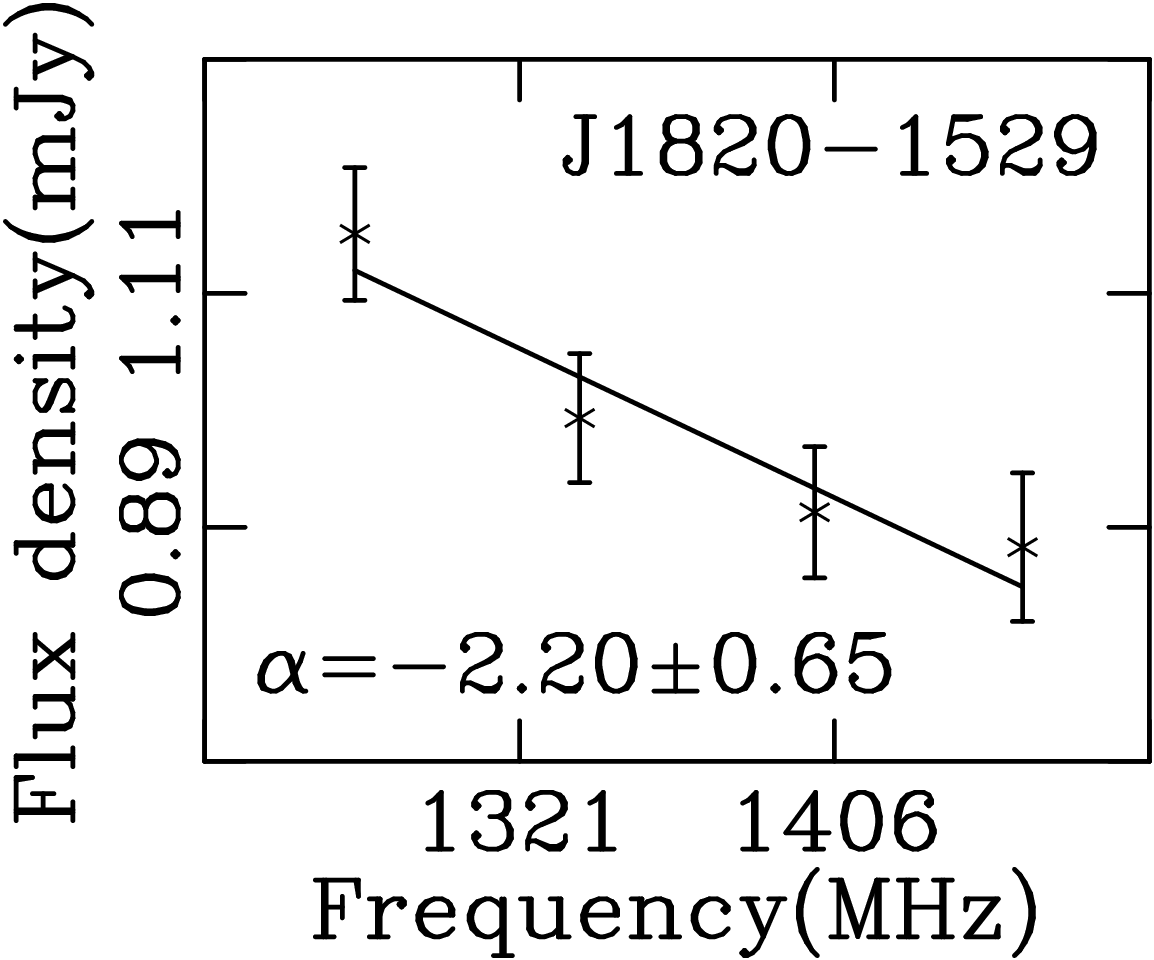}
\center
\addtocounter{figure}{-1}
\caption{--- continued.}
\end{figure}

\begin{figure}
  \center             
\includegraphics[bb = 110 72 518 685,clip,angle=-90,width=0.175\textwidth]{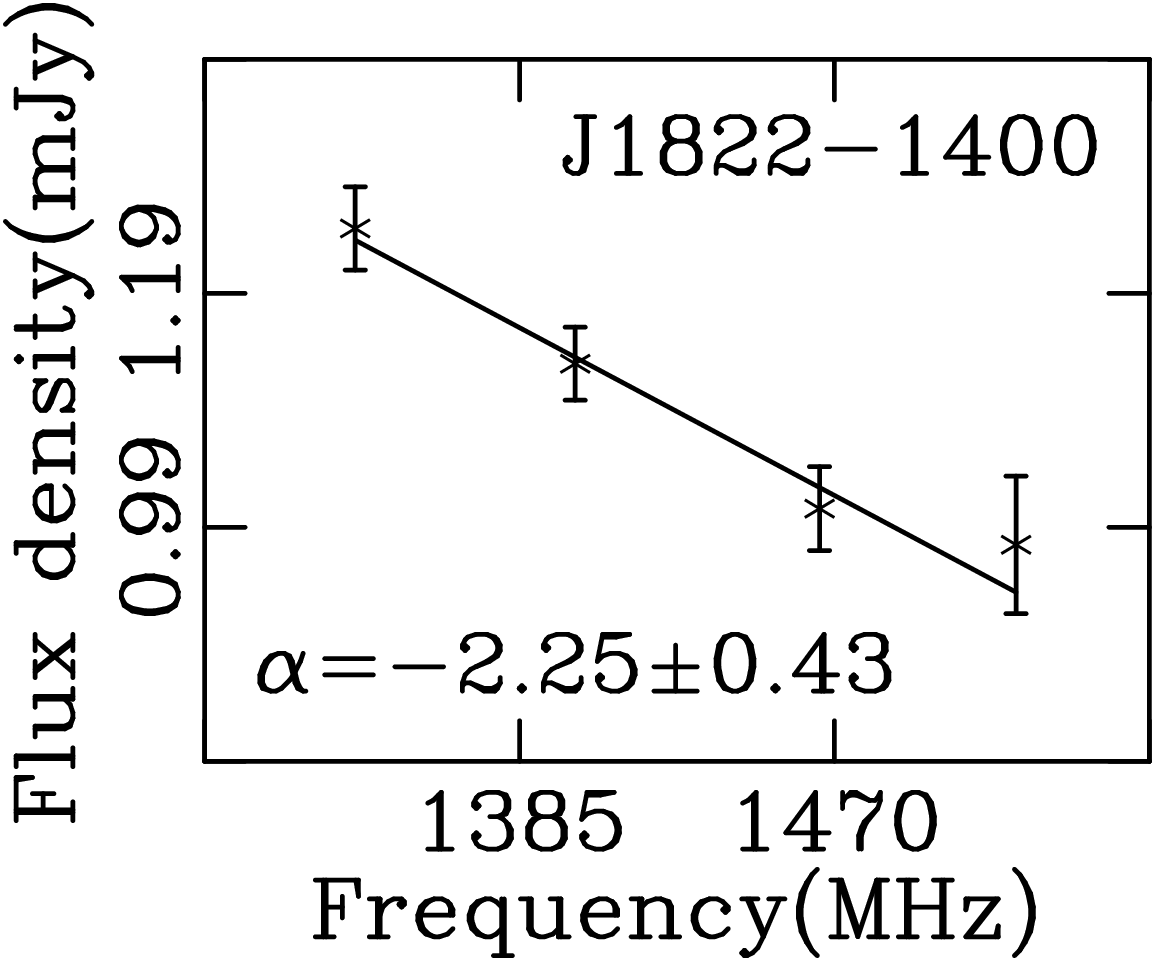}               
\includegraphics[bb = 110 120 518 685,clip,angle=-90,width=0.16\textwidth]{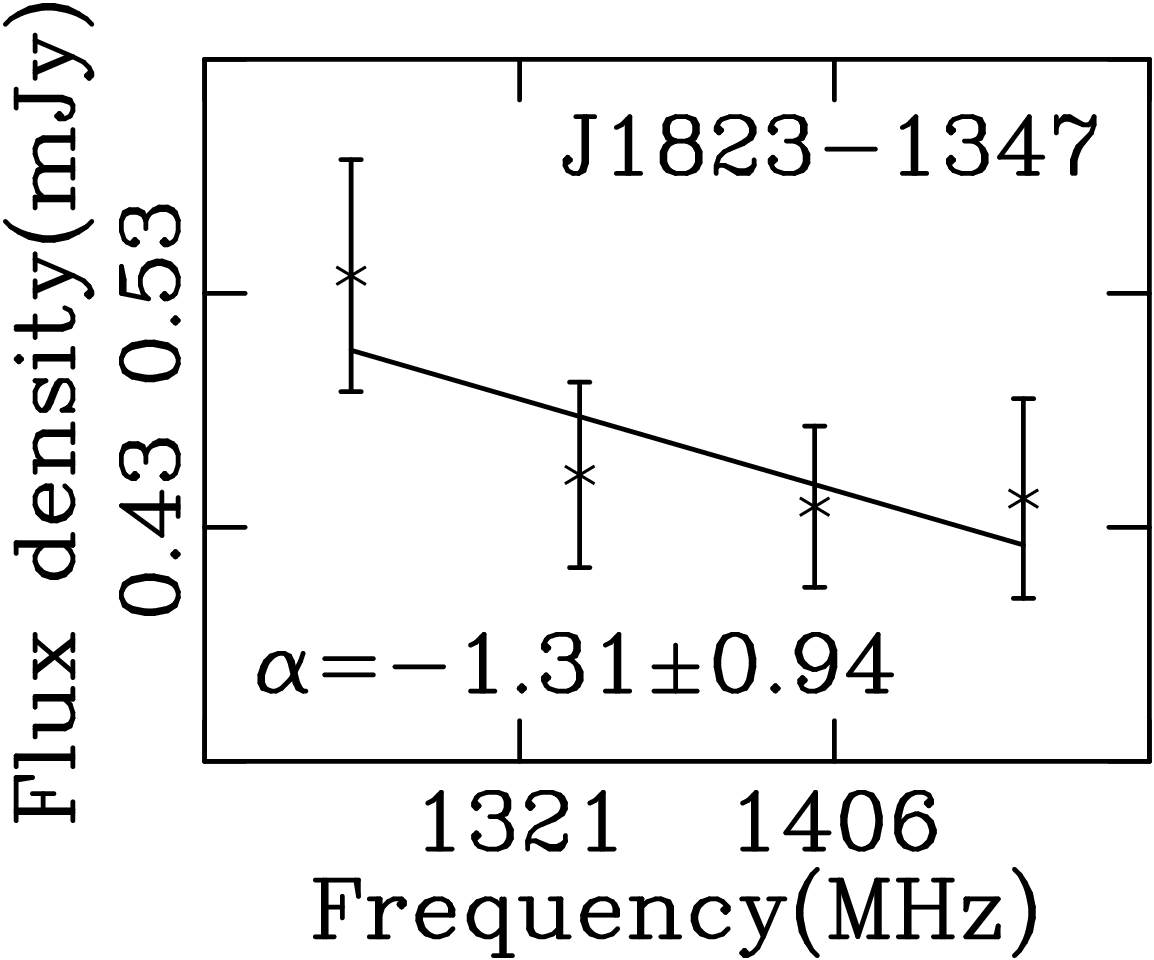}               
\includegraphics[bb = 110 120 518 685,clip,angle=-90,width=0.16\textwidth]{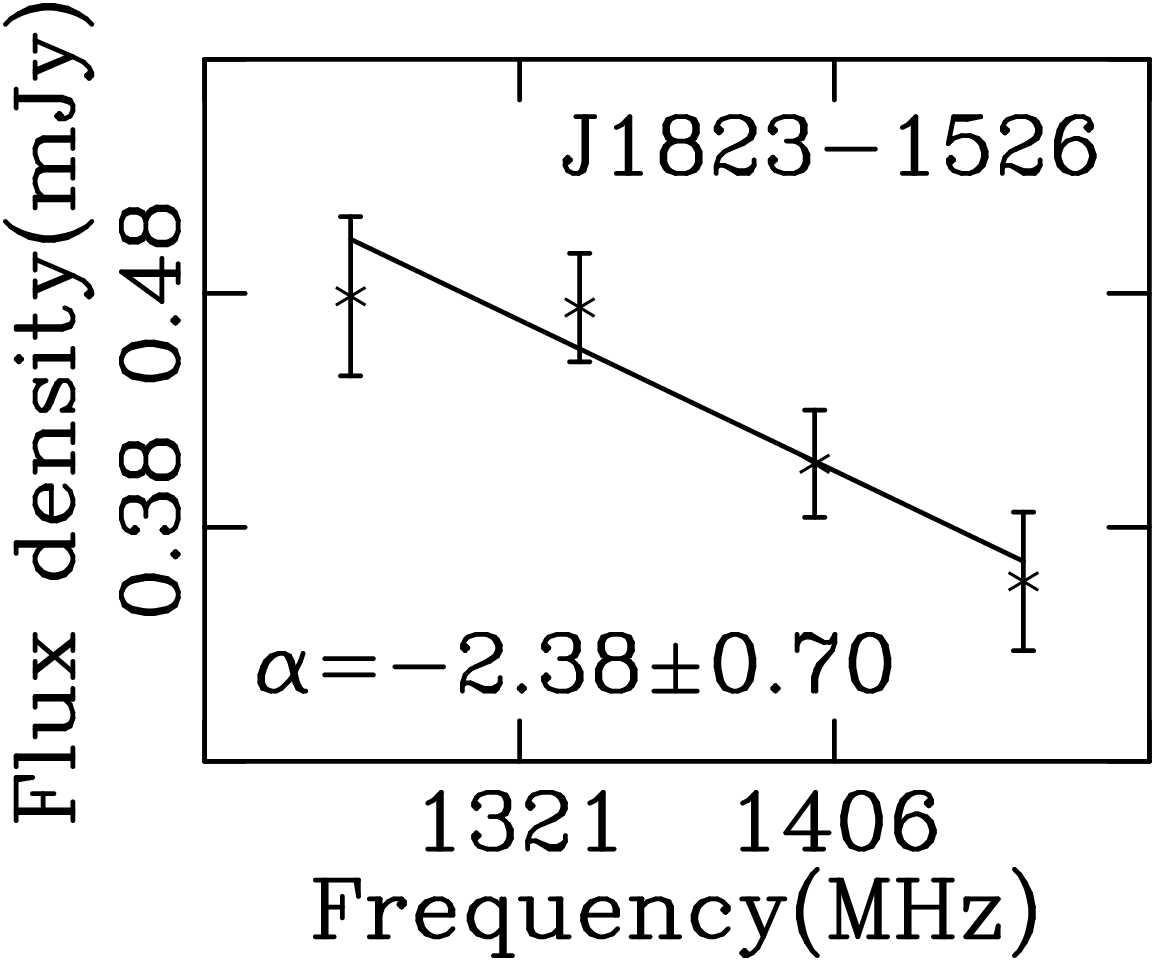}               
\includegraphics[bb = 110 120 518 685,clip,angle=-90,width=0.16\textwidth]{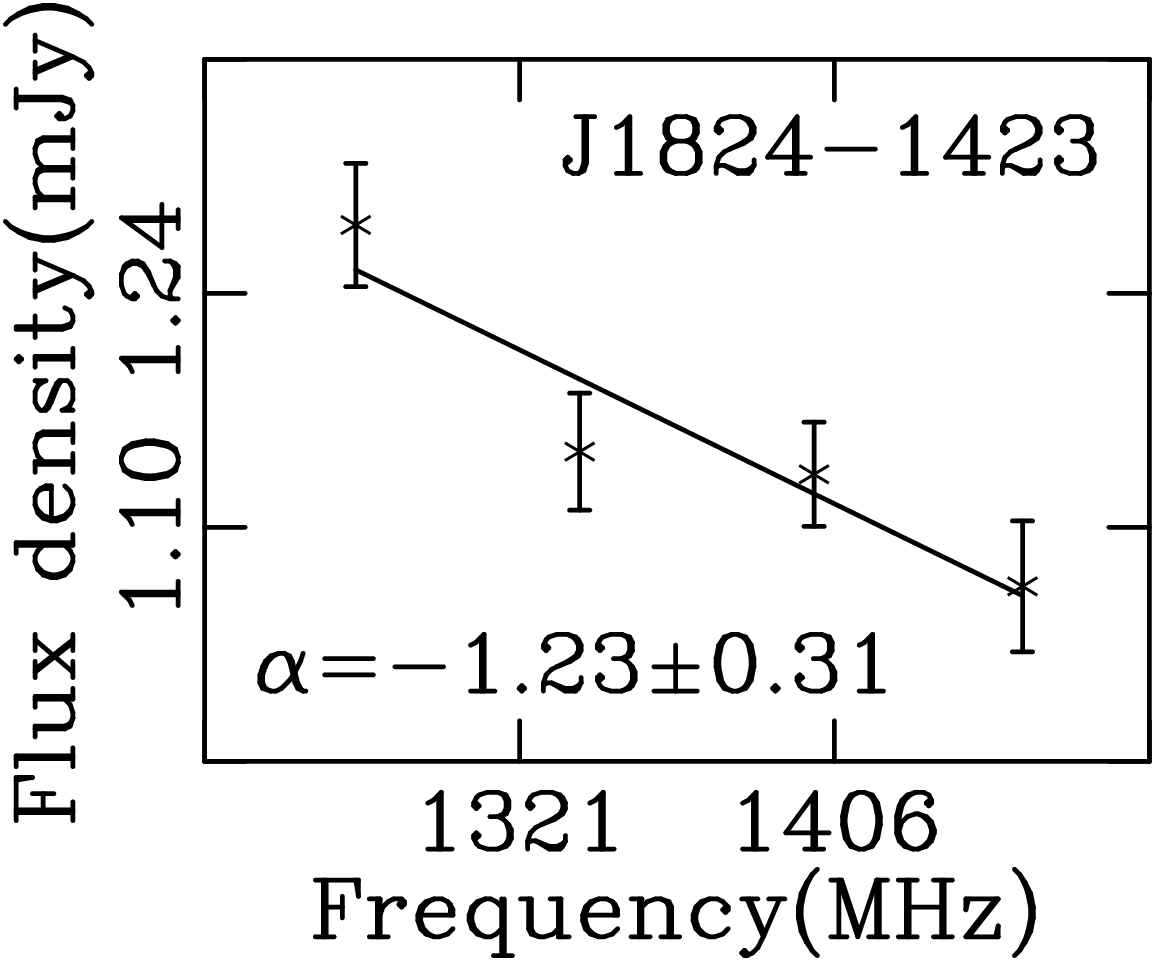}               
\includegraphics[bb = 110 120 518 685,clip,angle=-90,width=0.16\textwidth]{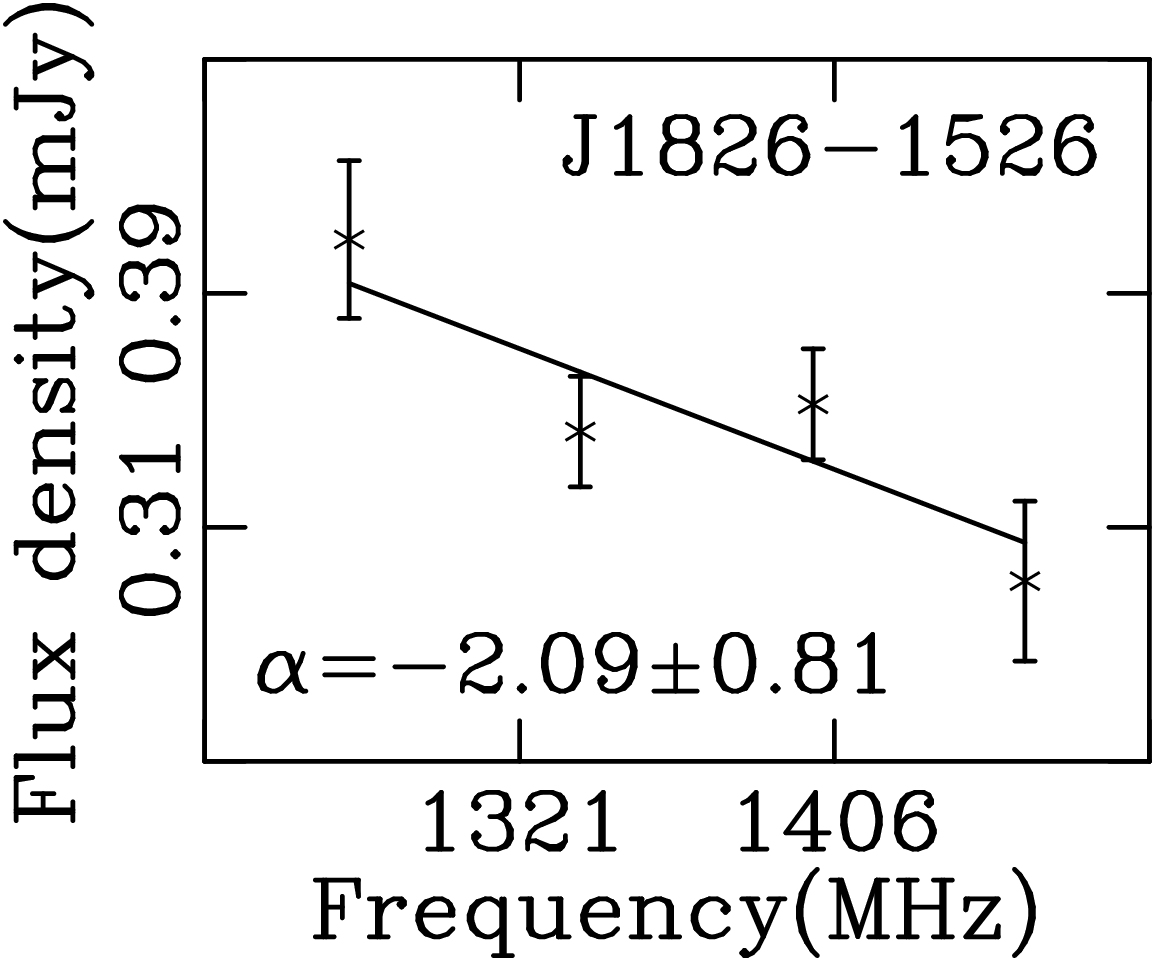}               
\includegraphics[bb = 110 120 518 685,clip,angle=-90,width=0.16\textwidth]{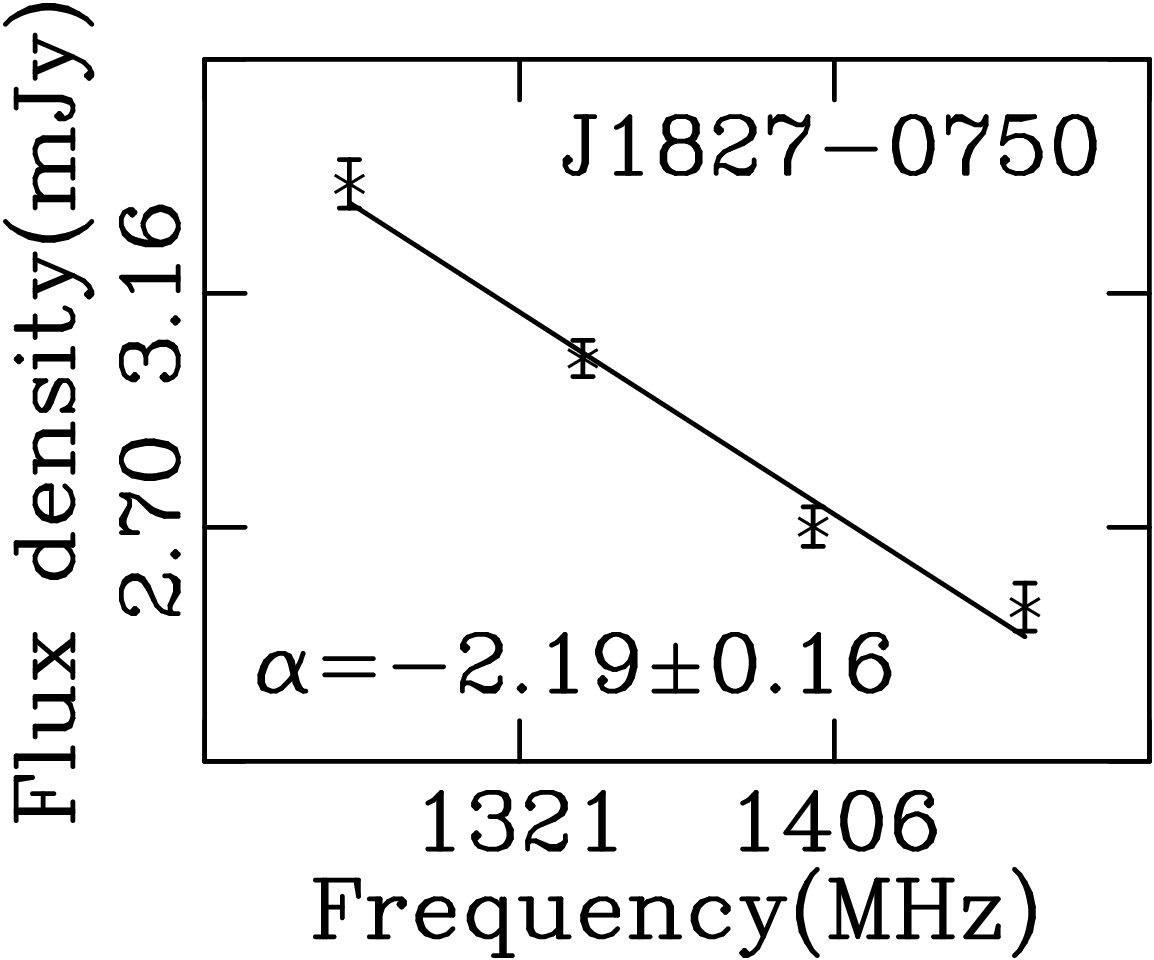}
\\[2mm]                
\includegraphics[bb = 110 72 518 685,clip,angle=-90,width=0.175\textwidth]{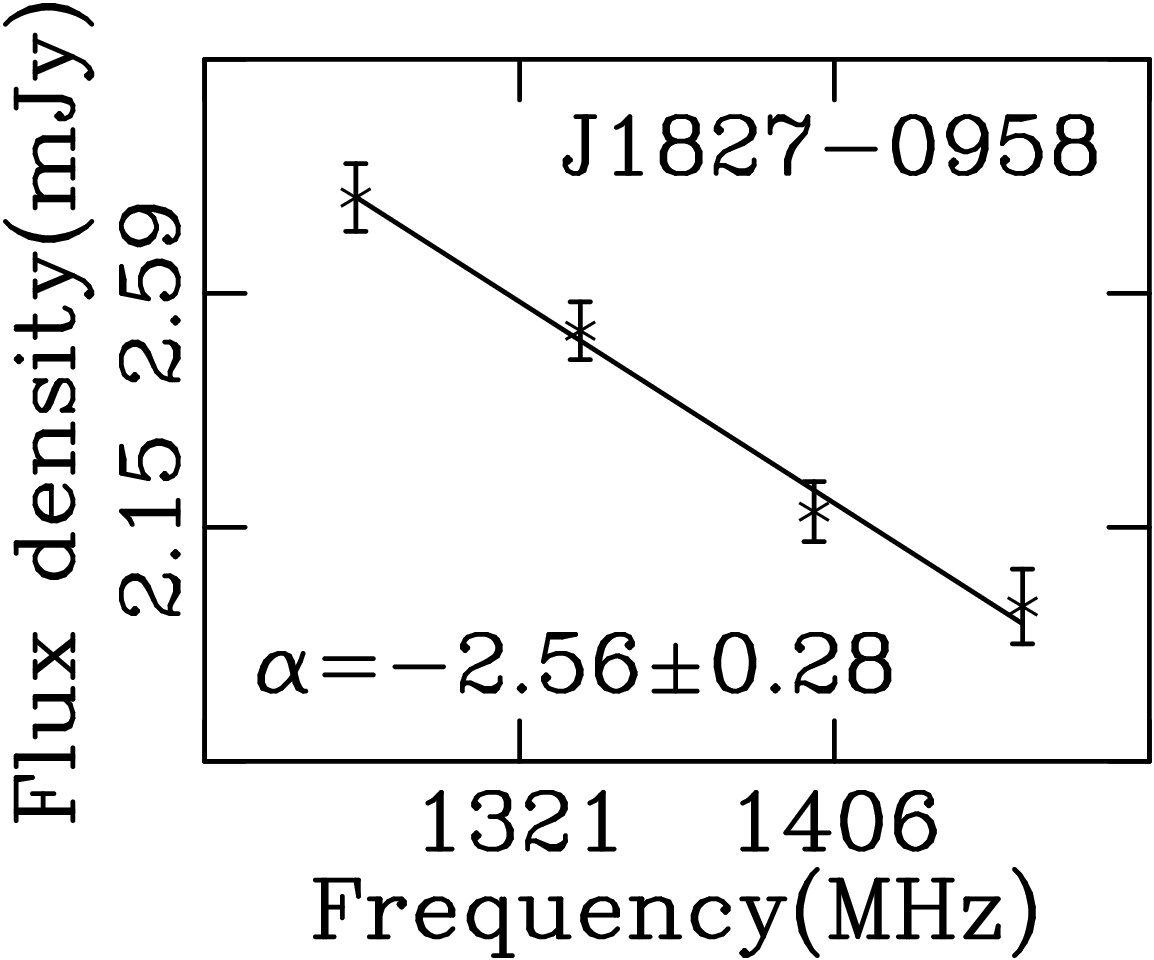}               
\includegraphics[bb = 110 120 518 685,clip,angle=-90,width=0.16\textwidth]{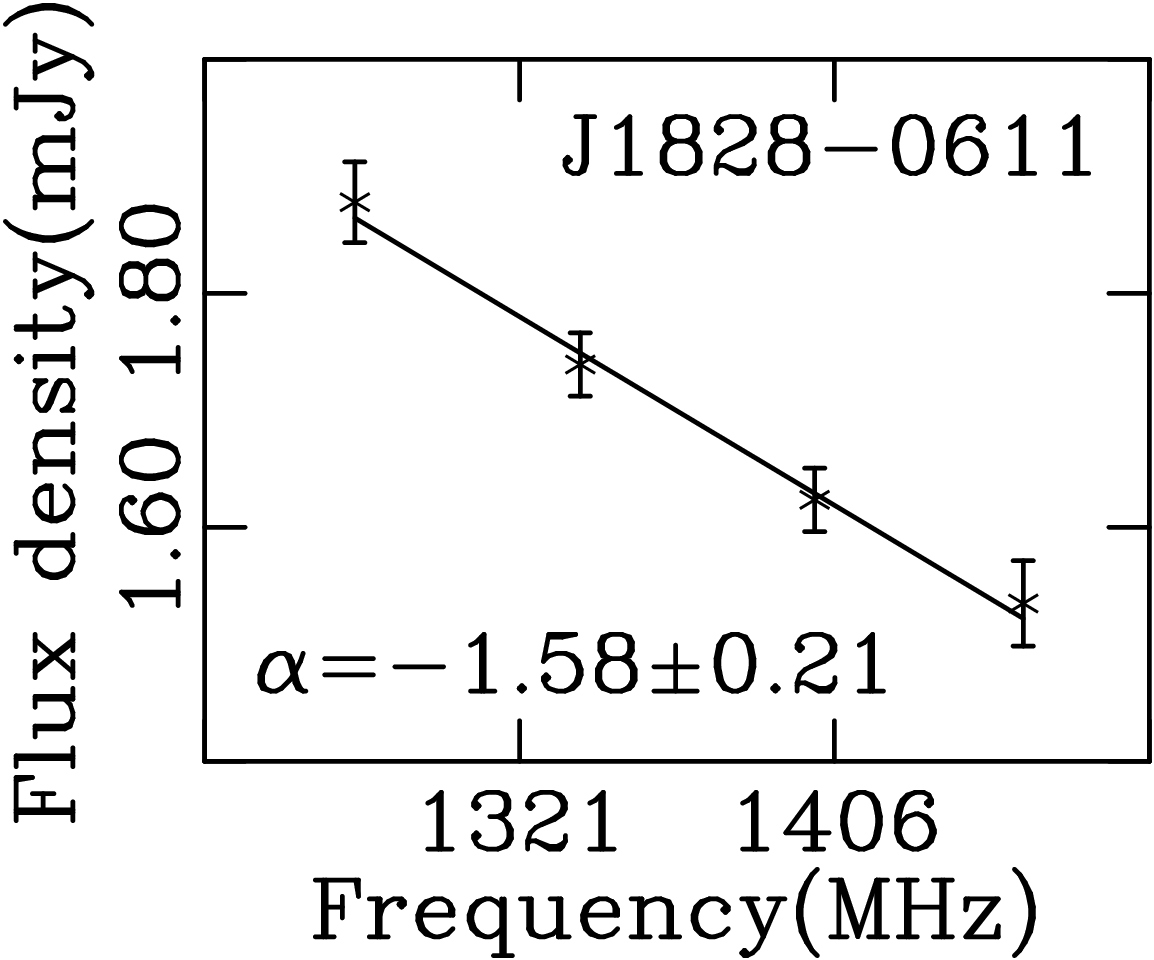}               
\includegraphics[bb = 110 120 518 685,clip,angle=-90,width=0.16\textwidth]{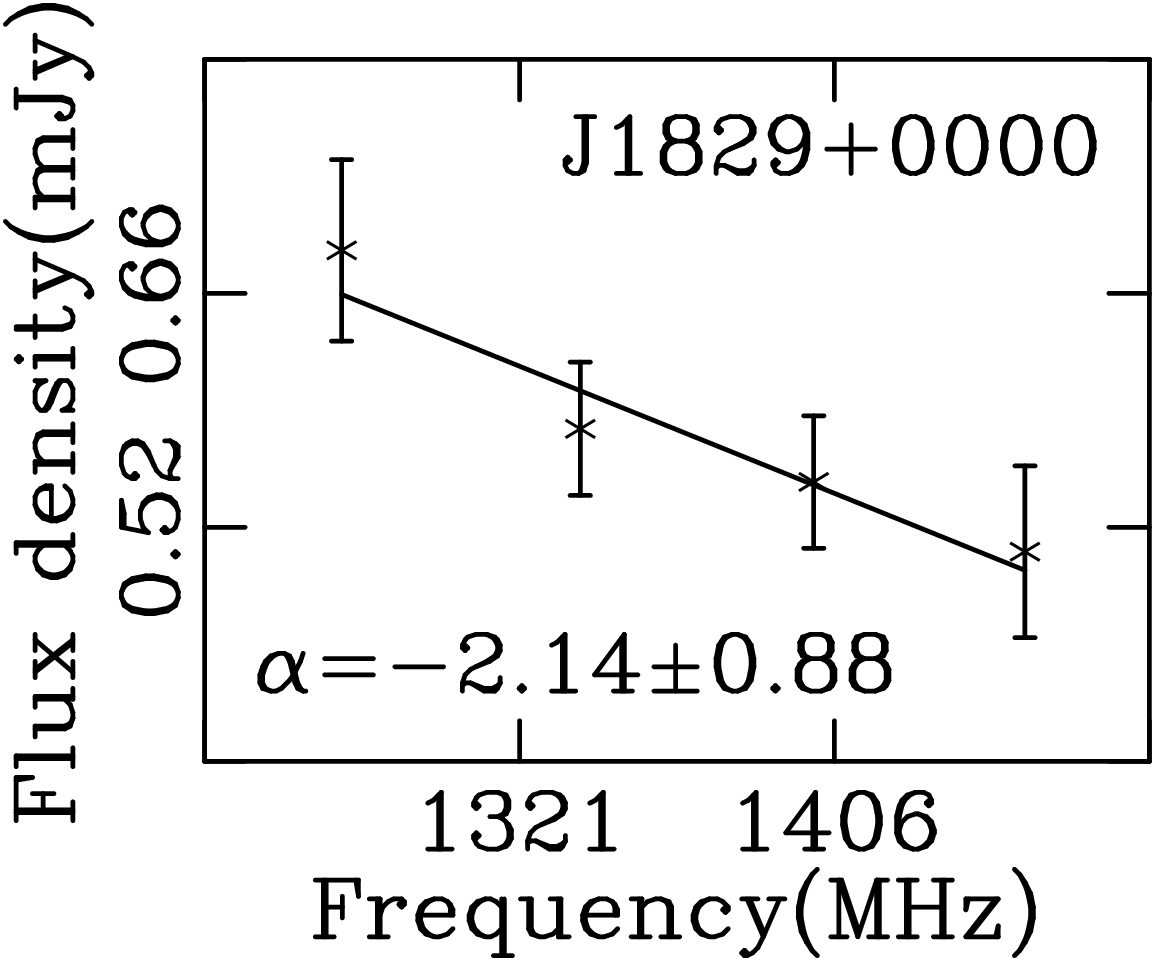}               
\includegraphics[bb = 110 120 518 685,clip,angle=-90,width=0.16\textwidth]{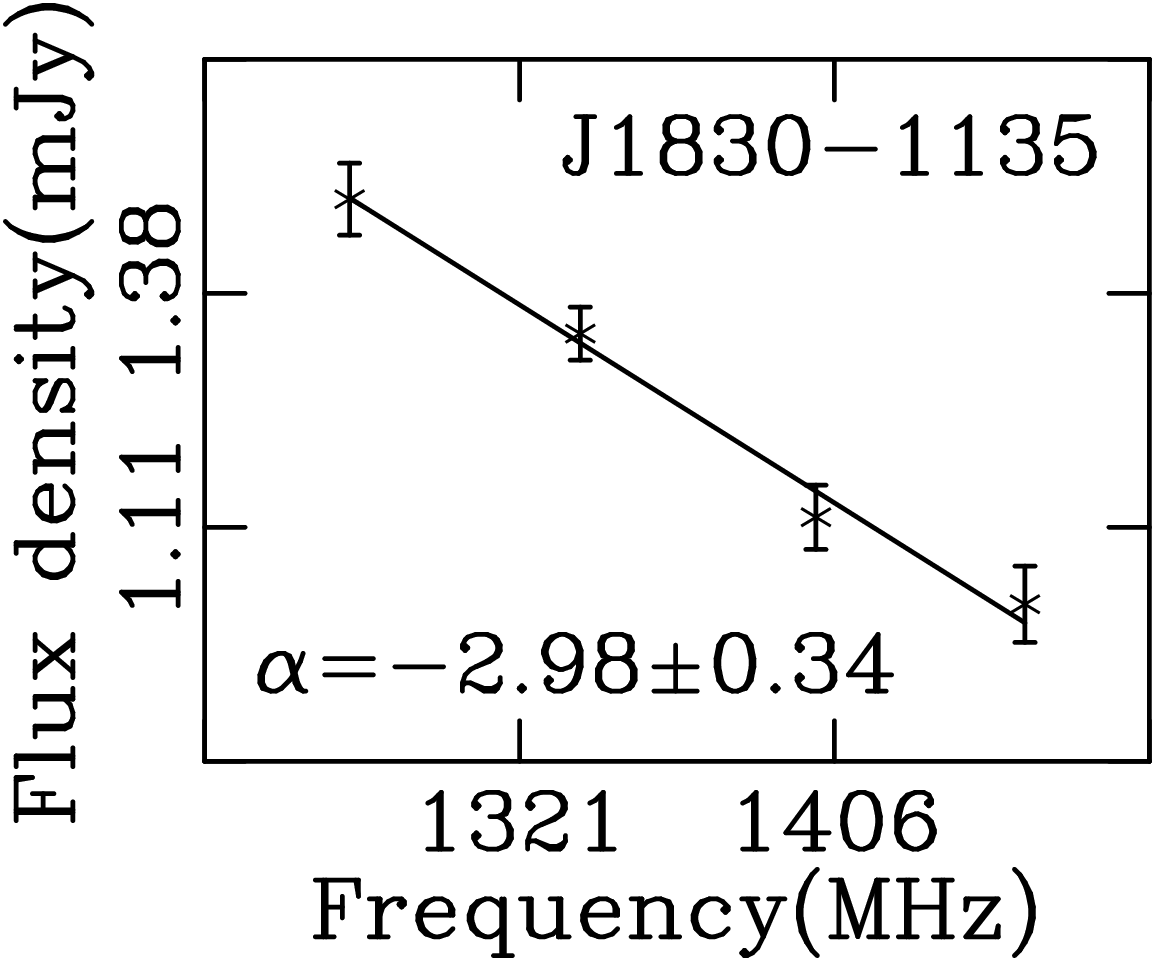}               
\includegraphics[bb = 110 120 518 685,clip,angle=-90,width=0.16\textwidth]{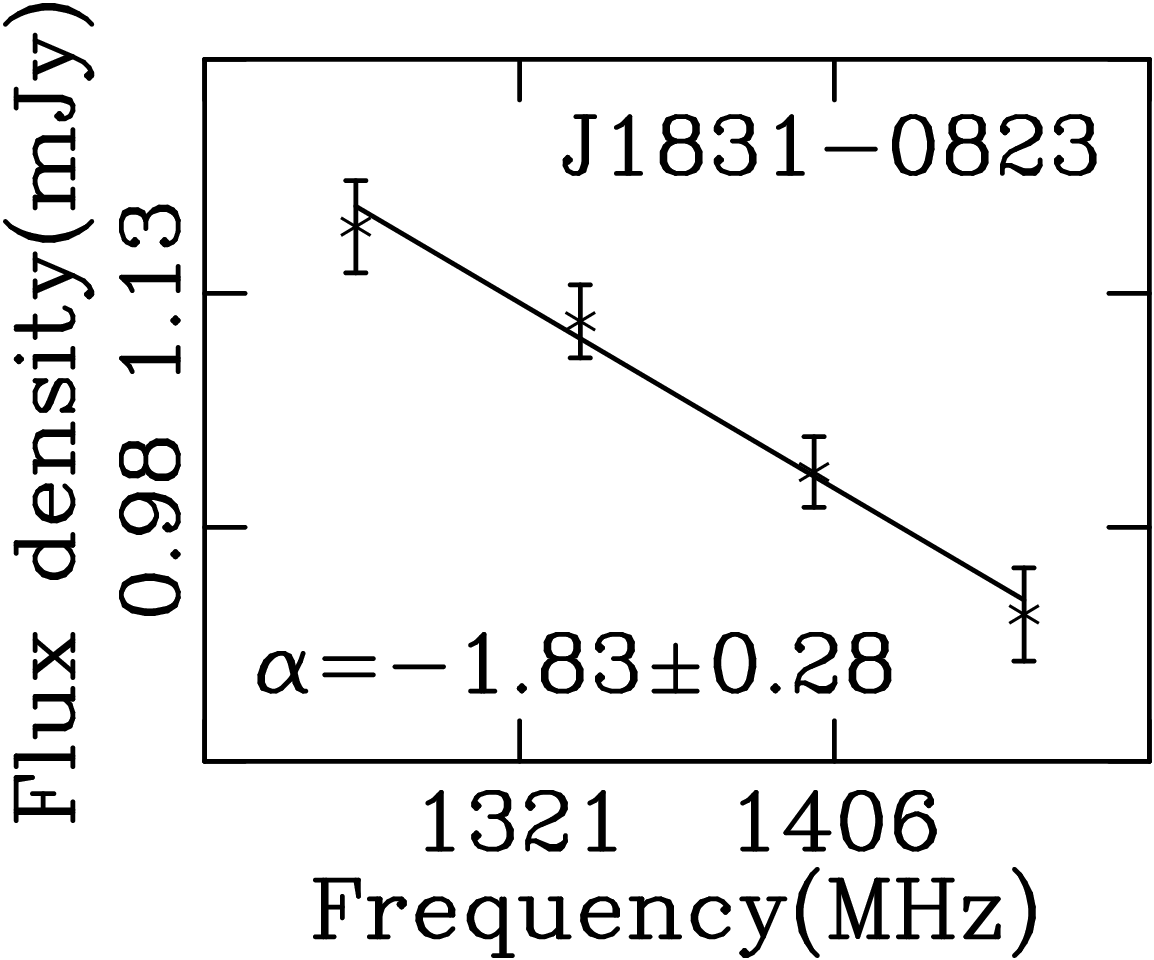}               
\includegraphics[bb = 110 120 518 685,clip,angle=-90,width=0.16\textwidth]{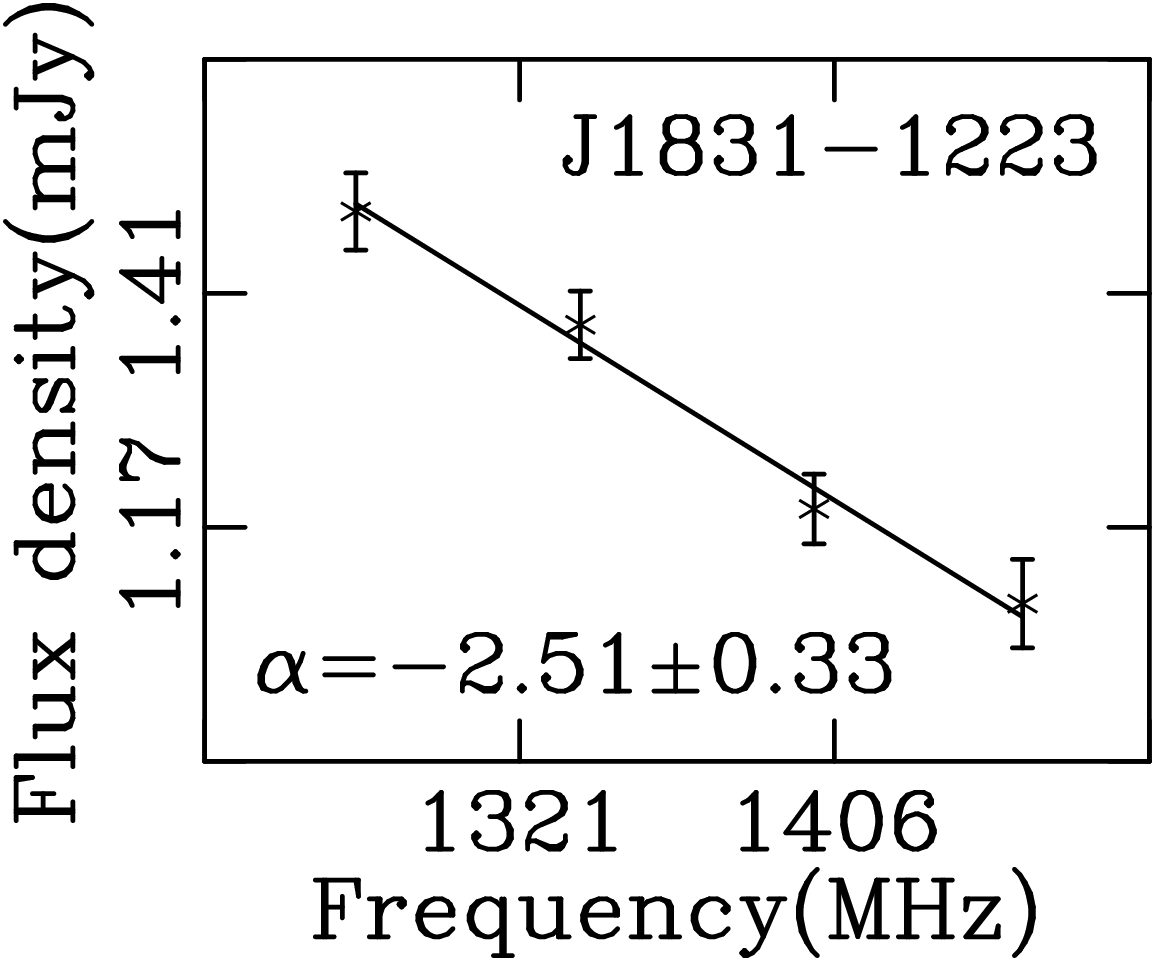}
\\[2mm]                
\includegraphics[bb = 110 72 518 685,clip,angle=-90,width=0.175\textwidth]{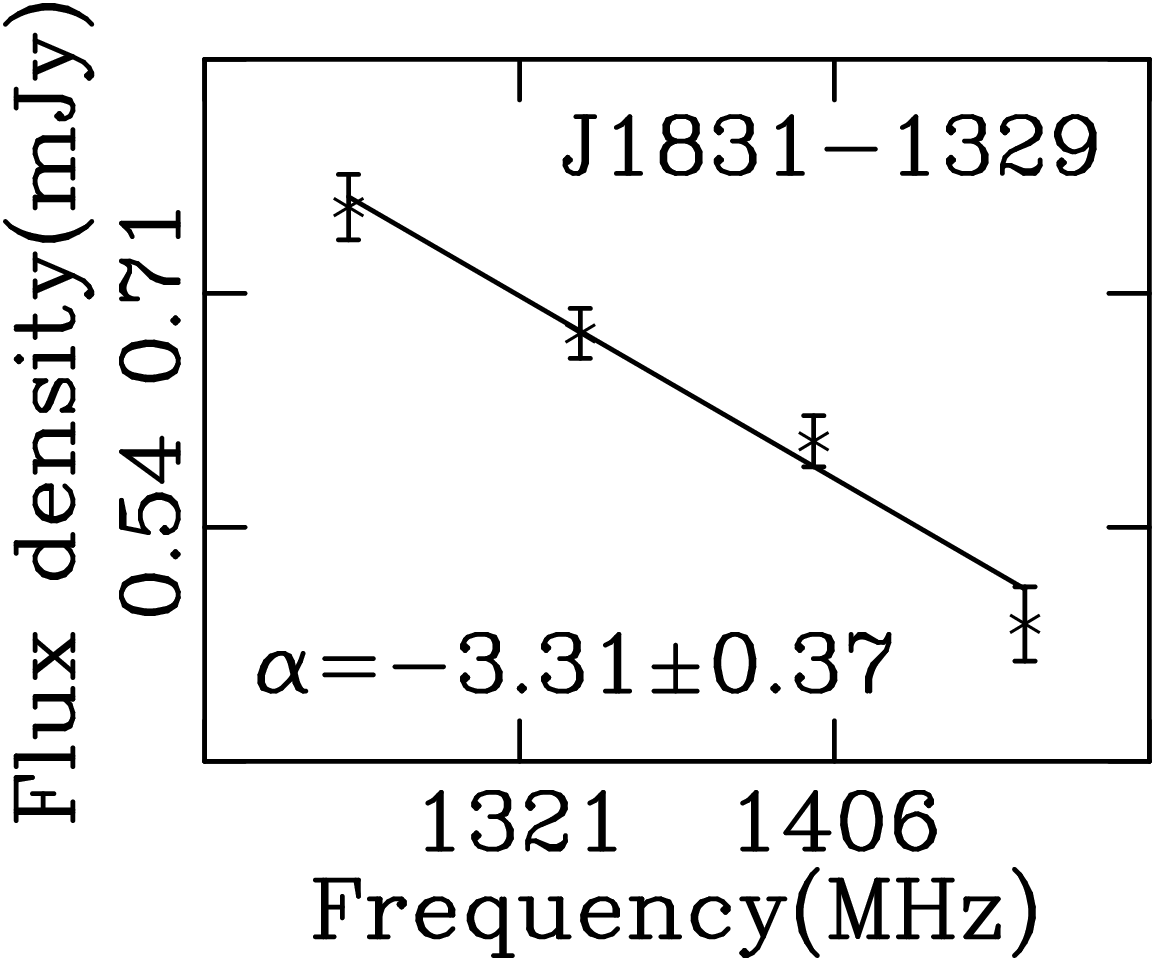}               
\includegraphics[bb = 110 120 518 685,clip,angle=-90,width=0.16\textwidth]{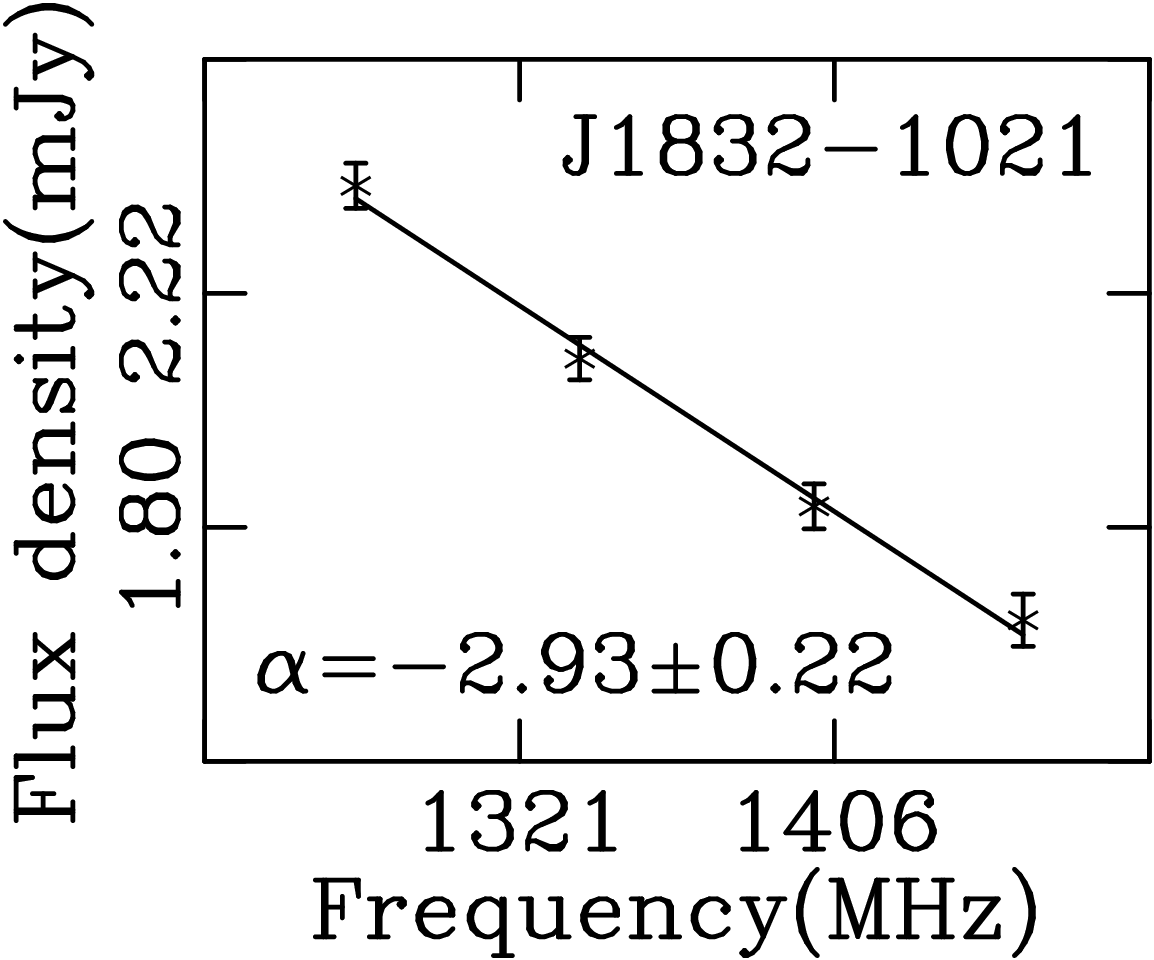}               
\includegraphics[bb = 110 120 518 685,clip,angle=-90,width=0.16\textwidth]{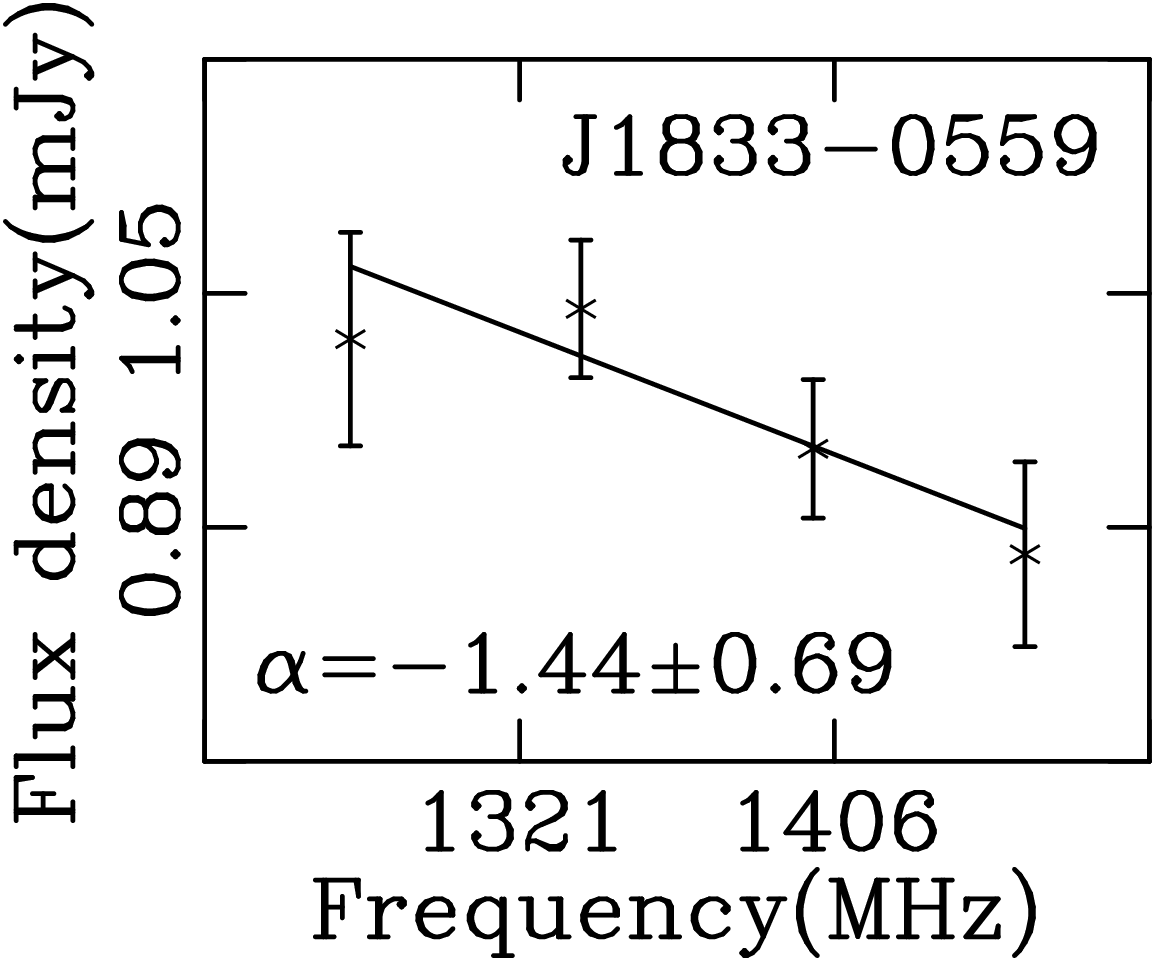}               
\includegraphics[bb = 110 120 518 685,clip,angle=-90,width=0.16\textwidth]{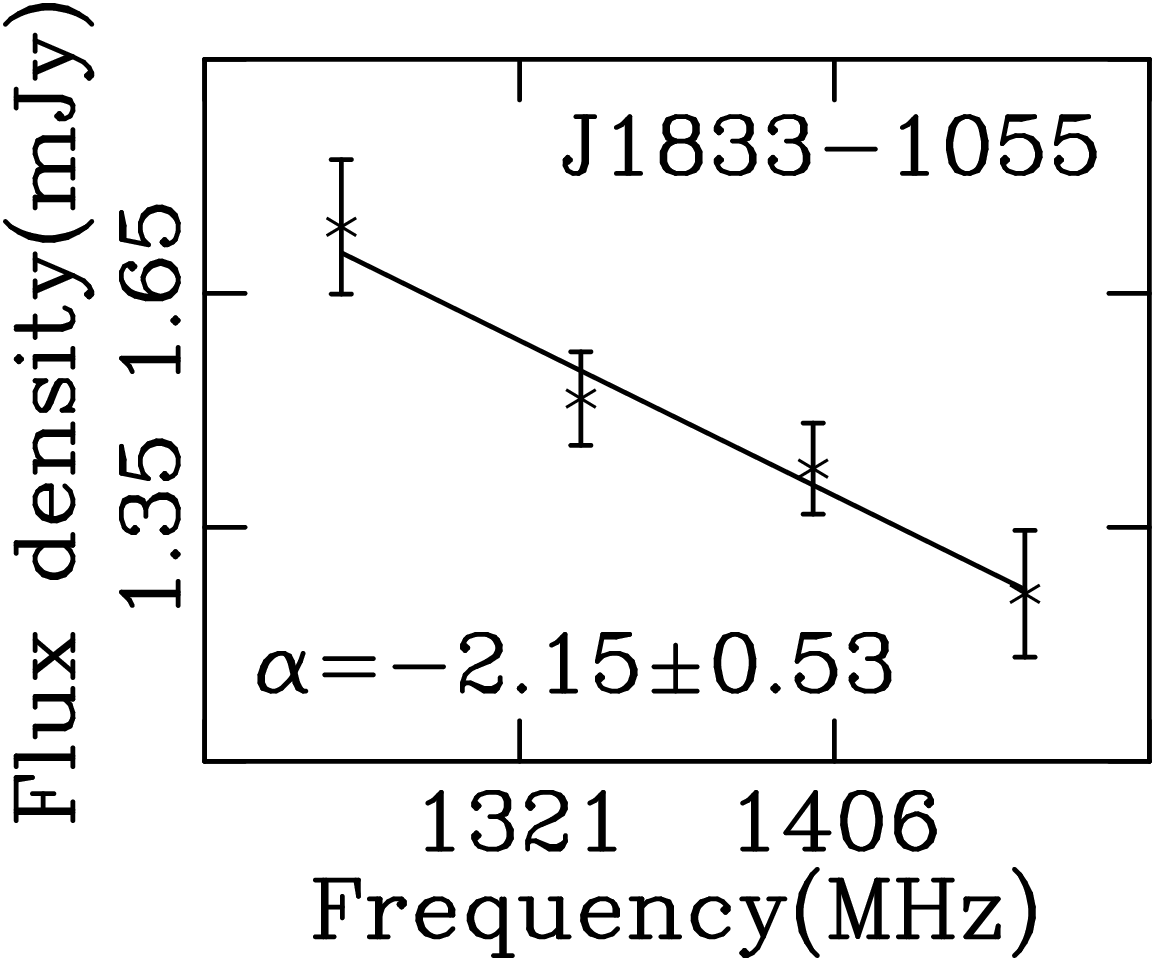}               
\includegraphics[bb = 110 120 518 685,clip,angle=-90,width=0.16\textwidth]{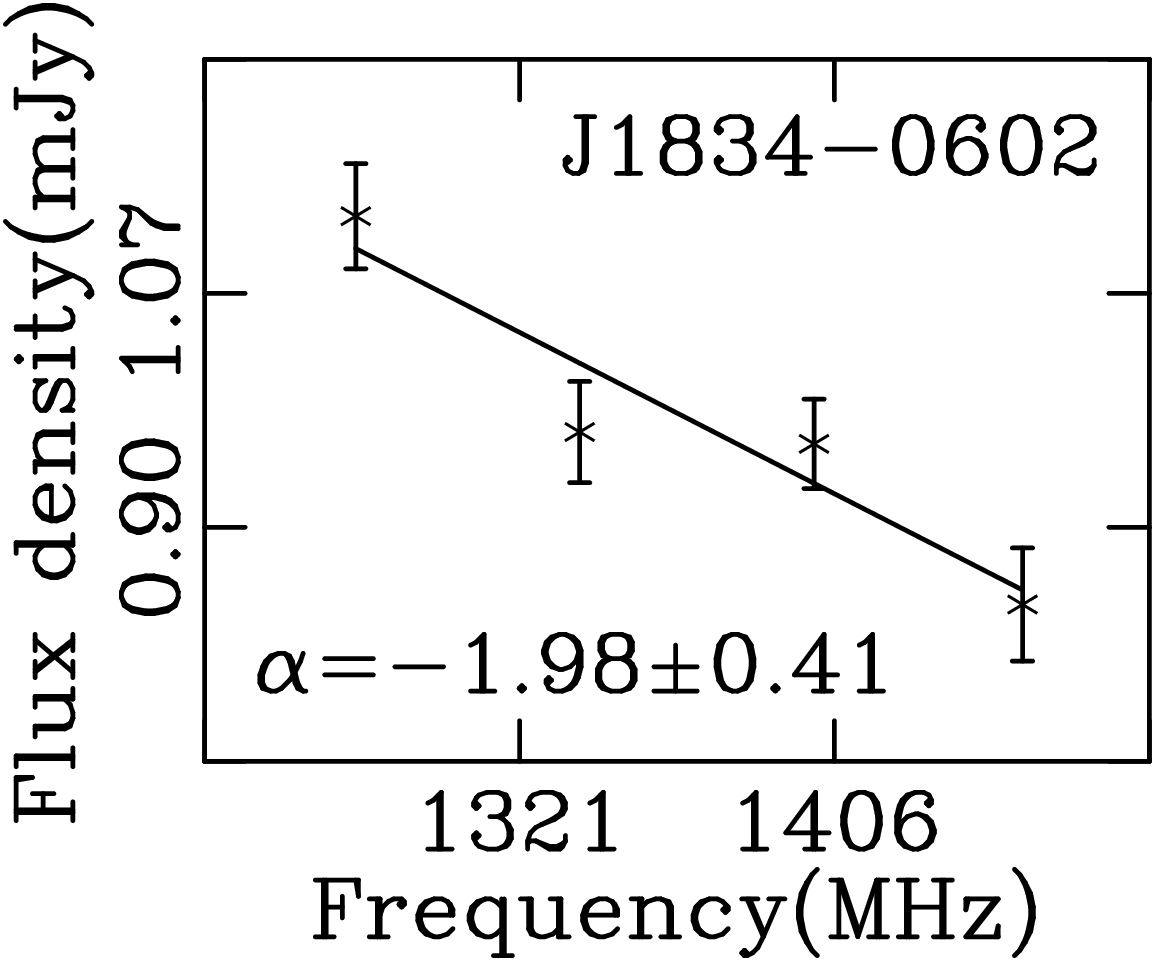}               
\includegraphics[bb = 110 120 518 685,clip,angle=-90,width=0.16\textwidth]{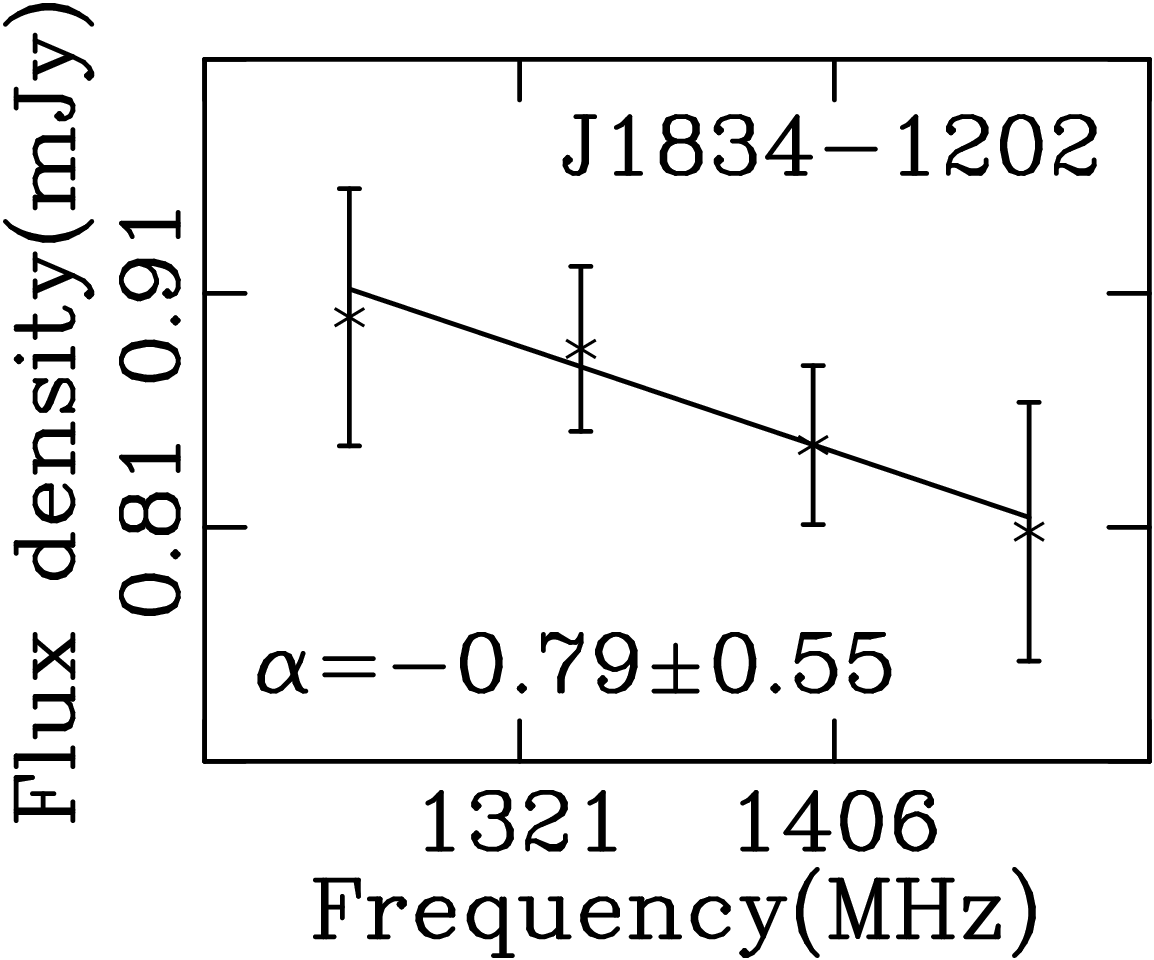}
\\[2mm]                
\includegraphics[bb = 110 72 518 685,clip,angle=-90,width=0.175\textwidth]{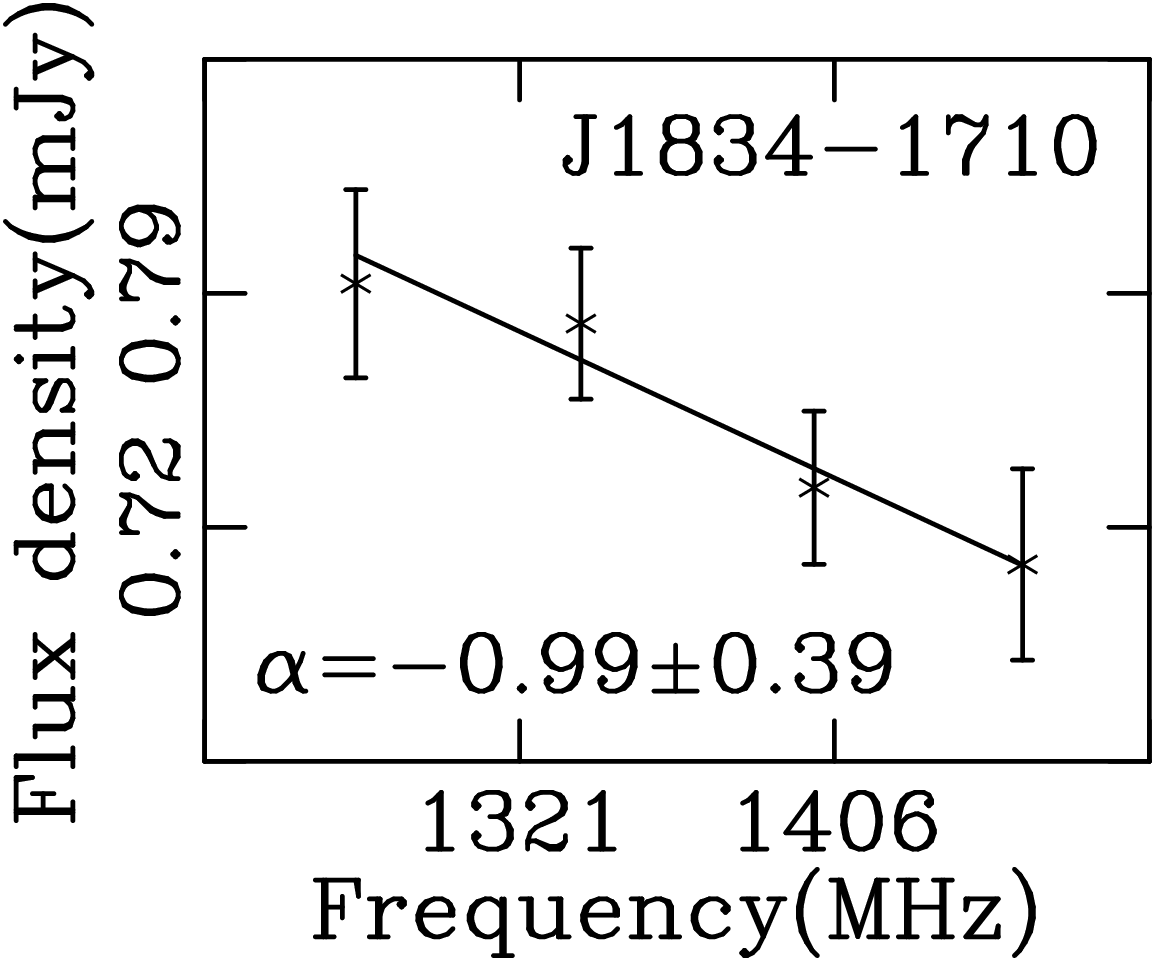}               
\includegraphics[bb = 110 120 518 685,clip,angle=-90,width=0.16\textwidth]{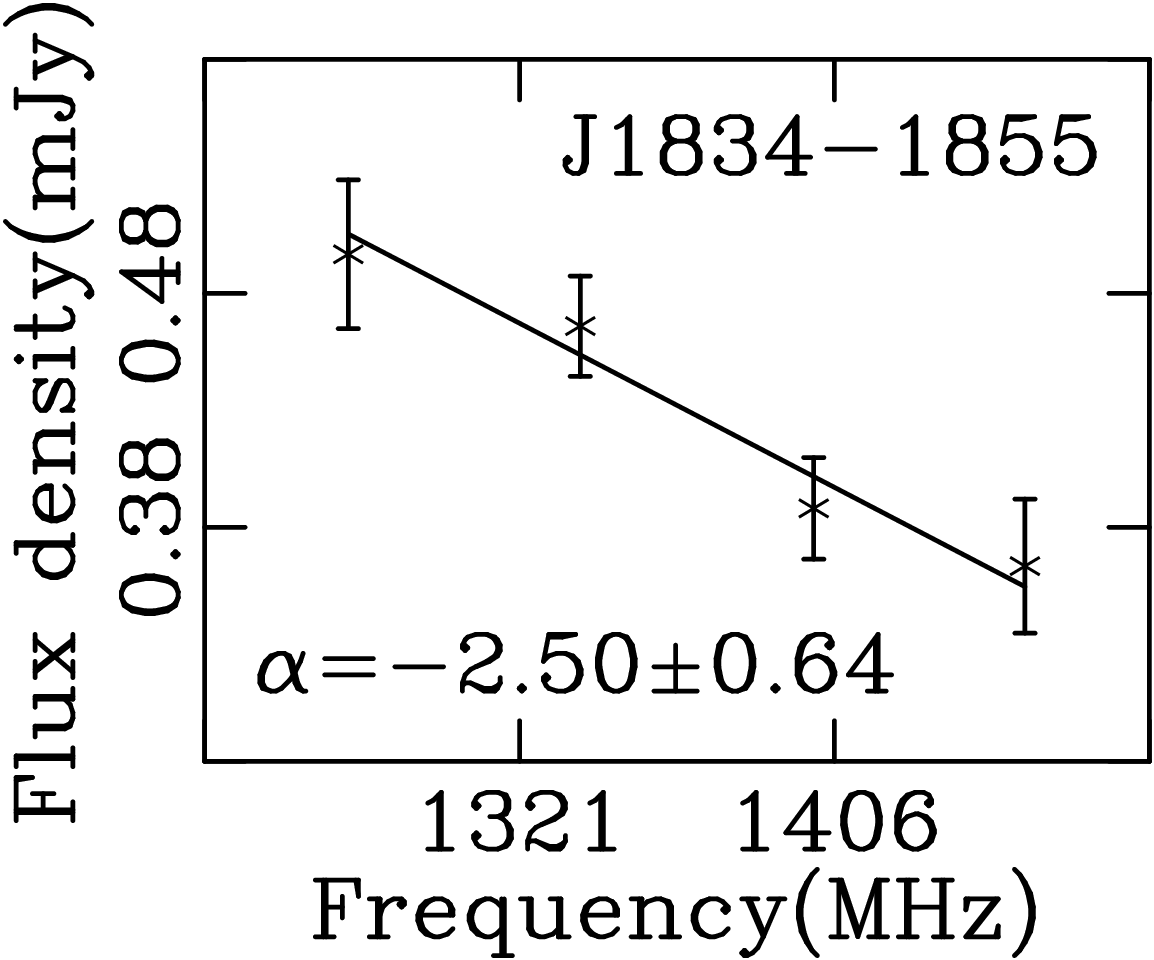}               
\includegraphics[bb = 110 120 518 685,clip,angle=-90,width=0.16\textwidth]{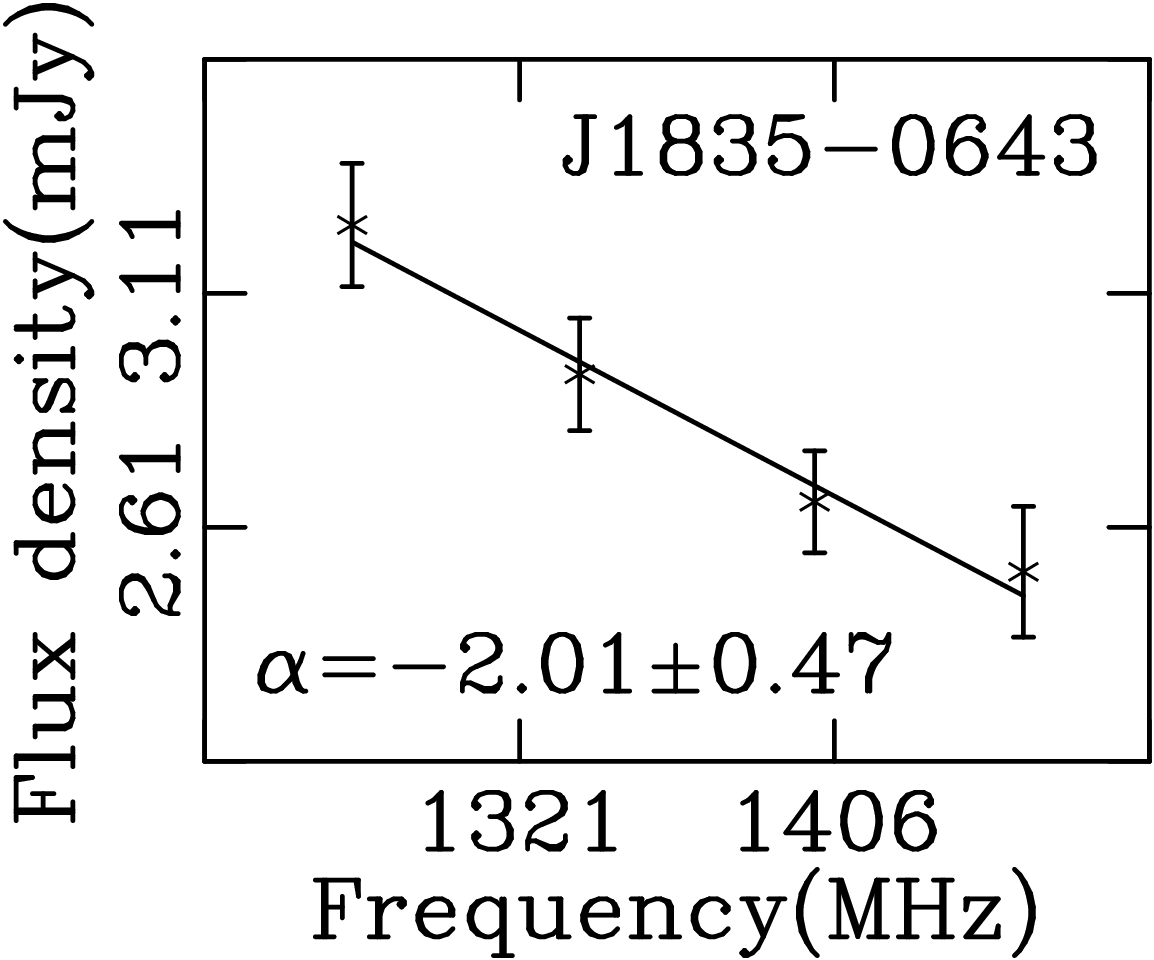}                             
\includegraphics[bb = 110 120 518 685,clip,angle=-90,width=0.16\textwidth]{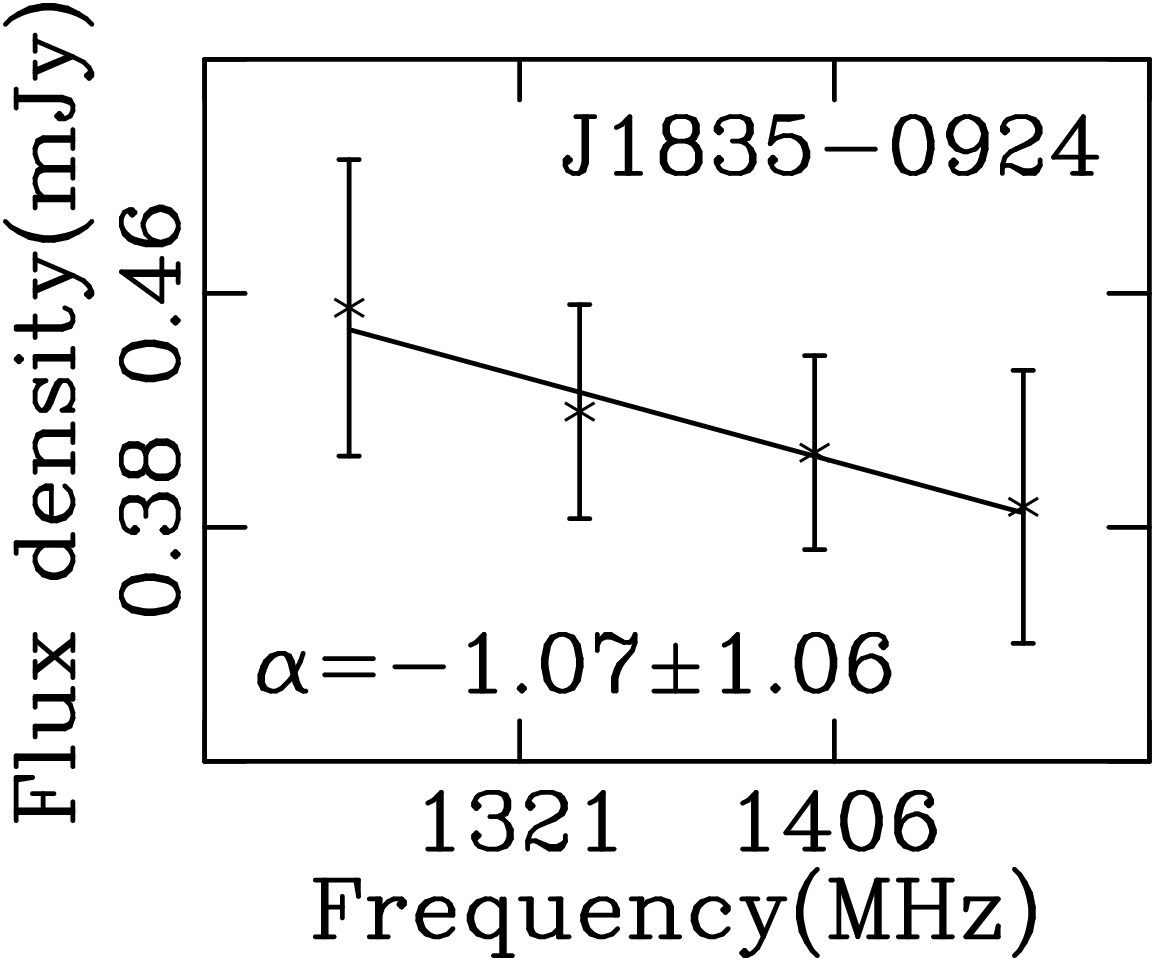}               
\includegraphics[bb = 110 120 518 685,clip,angle=-90,width=0.16\textwidth]{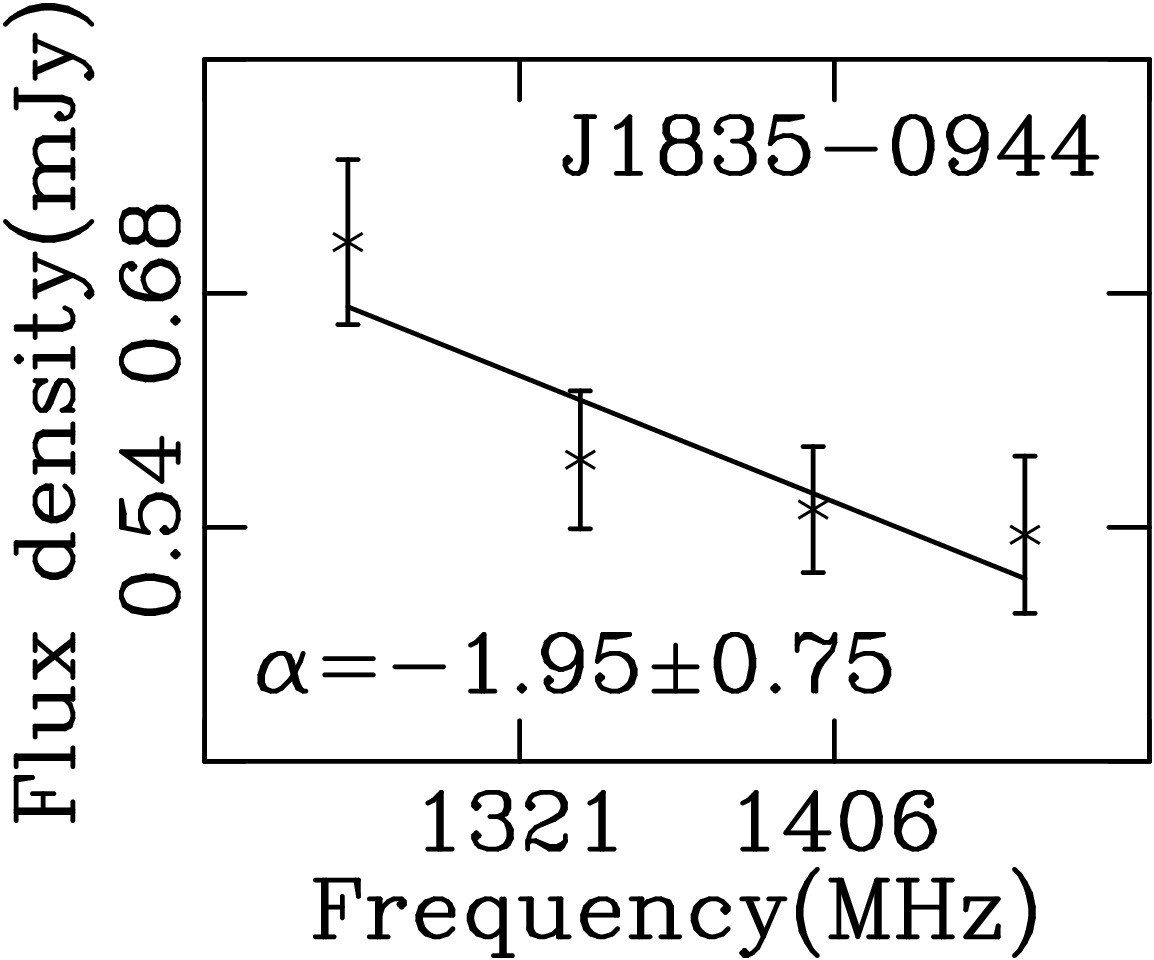}               
\includegraphics[bb = 110 120 518 685,clip,angle=-90,width=0.16\textwidth]{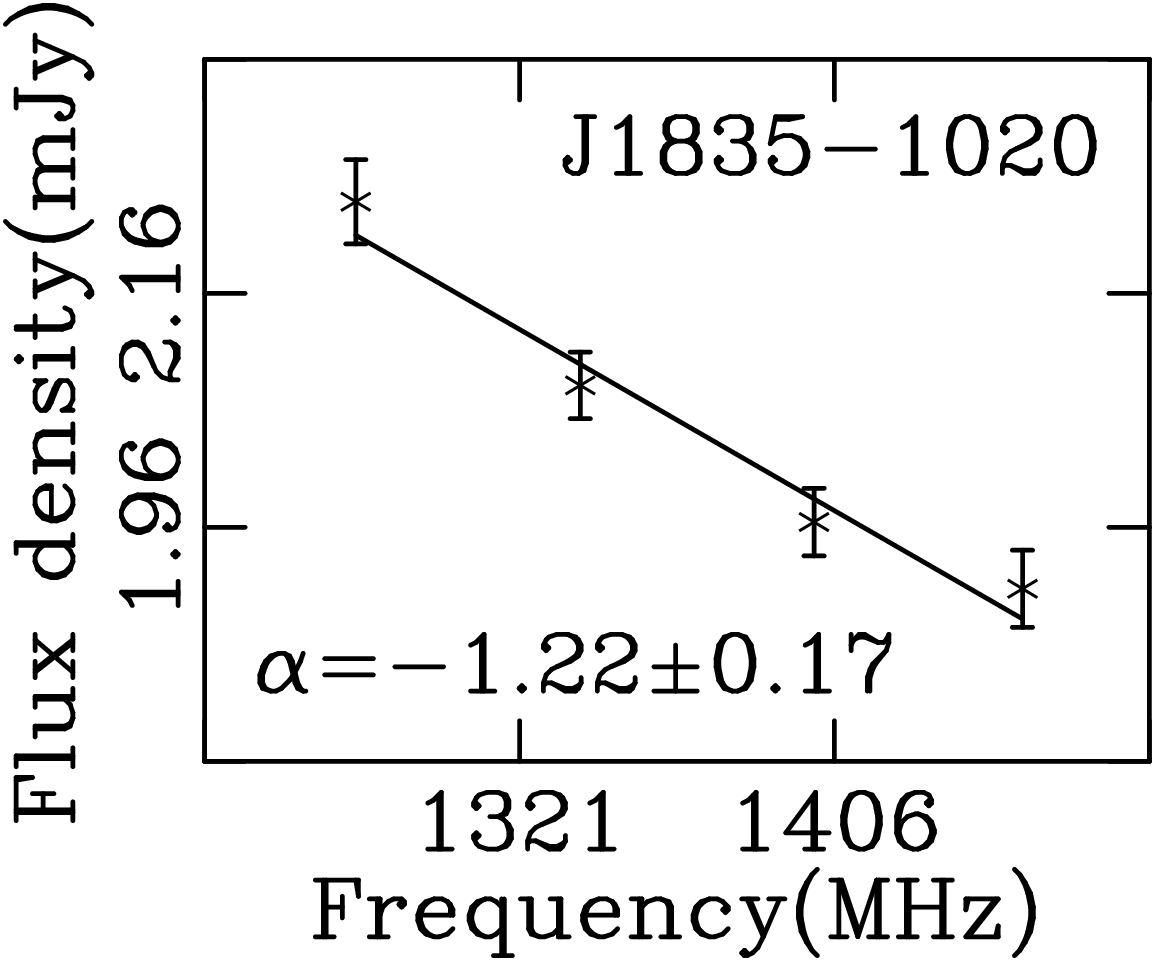}
\\[2mm]                
\includegraphics[bb = 110 72 518 685,clip,angle=-90,width=0.175\textwidth]{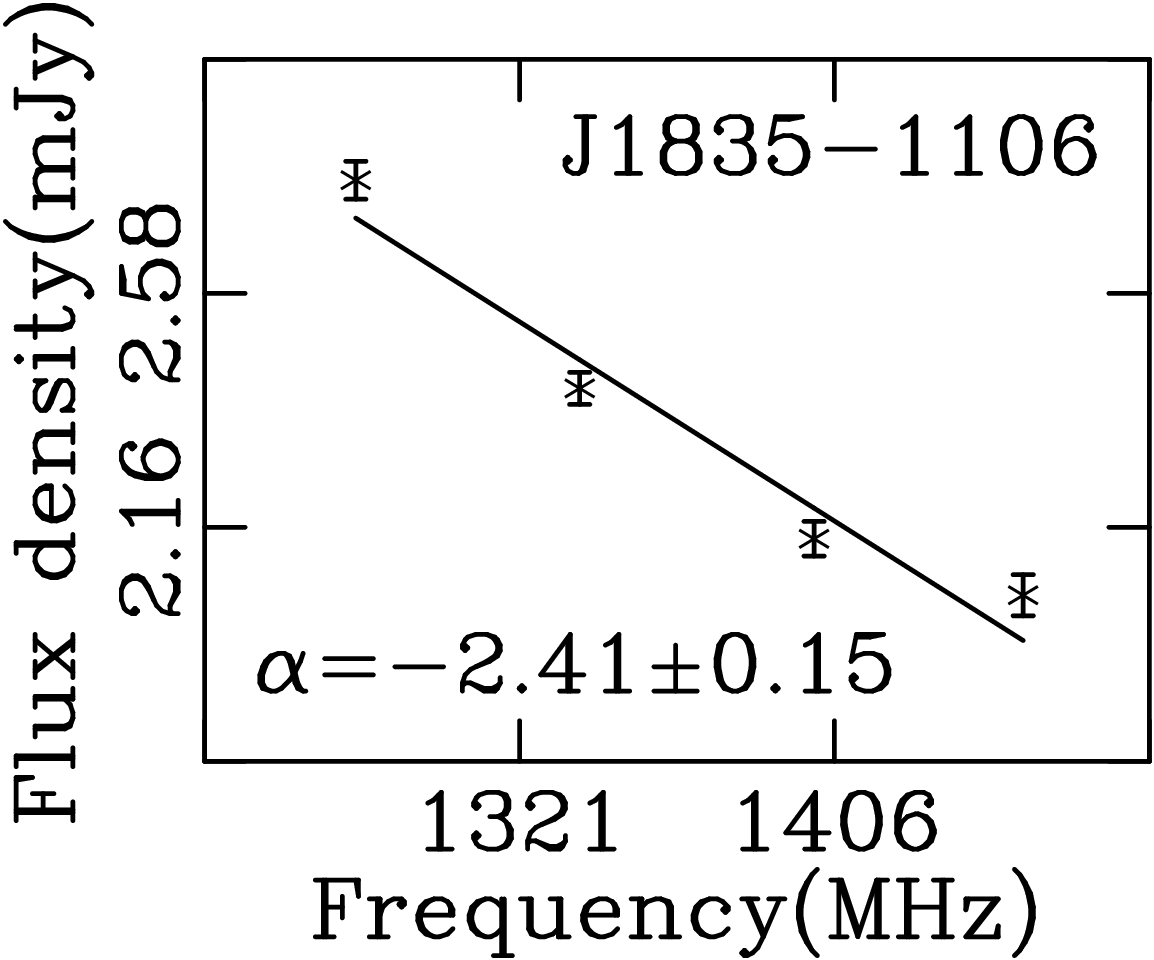}               
\includegraphics[bb = 110 120 518 685,clip,angle=-90,width=0.16\textwidth]{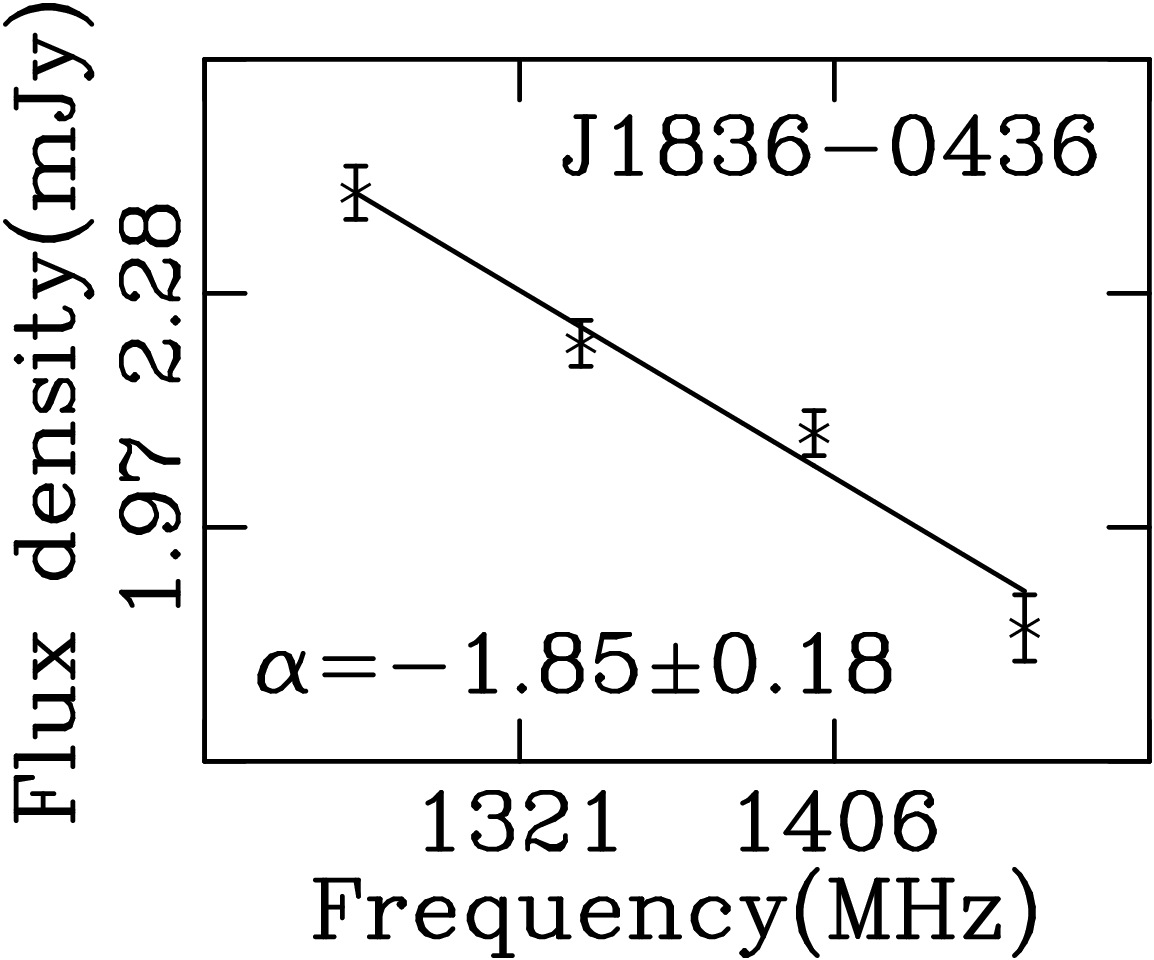}               
\includegraphics[bb = 110 120 518 685,clip,angle=-90,width=0.16\textwidth]{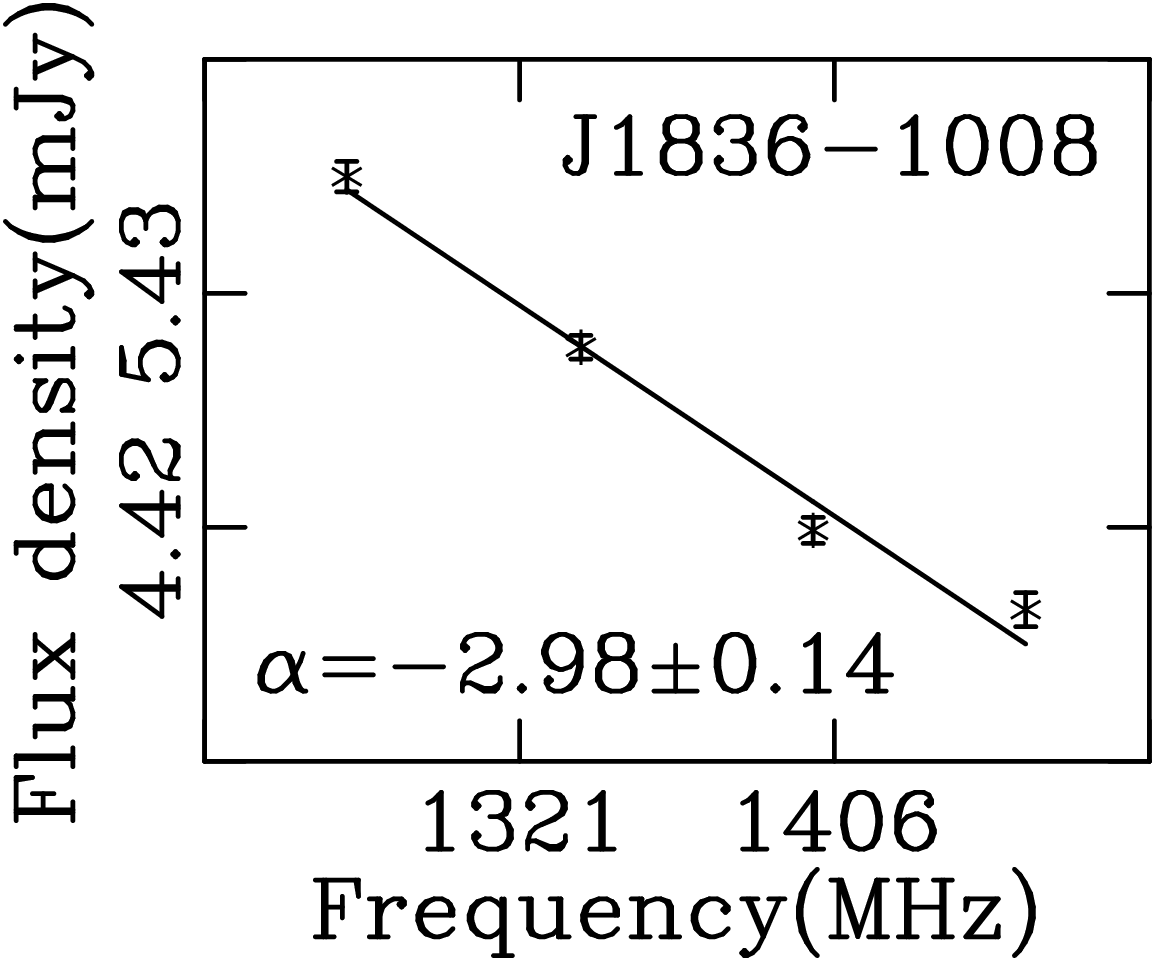}               
\includegraphics[bb = 110 120 518 685,clip,angle=-90,width=0.16\textwidth]{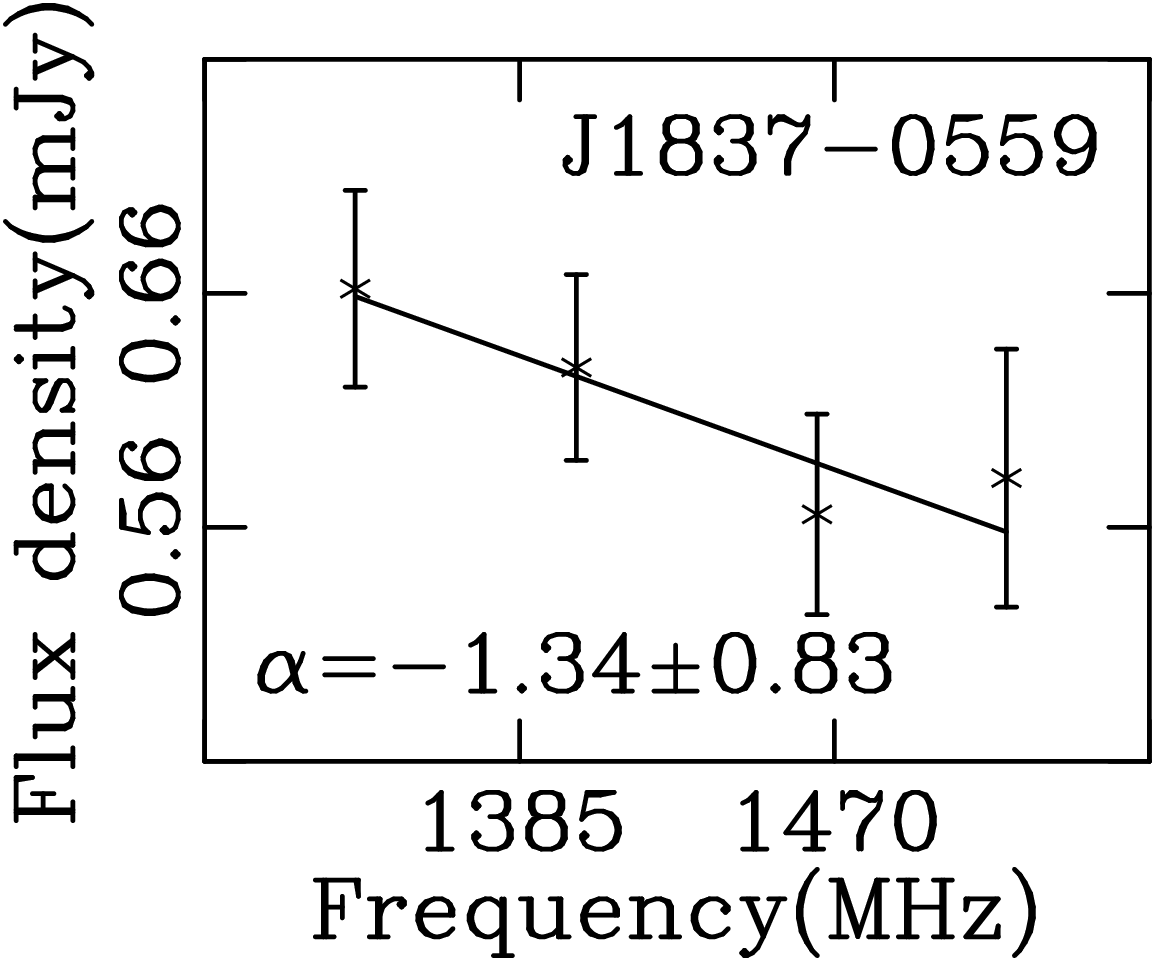}               
\includegraphics[bb = 110 120 518 685,clip,angle=-90,width=0.16\textwidth]{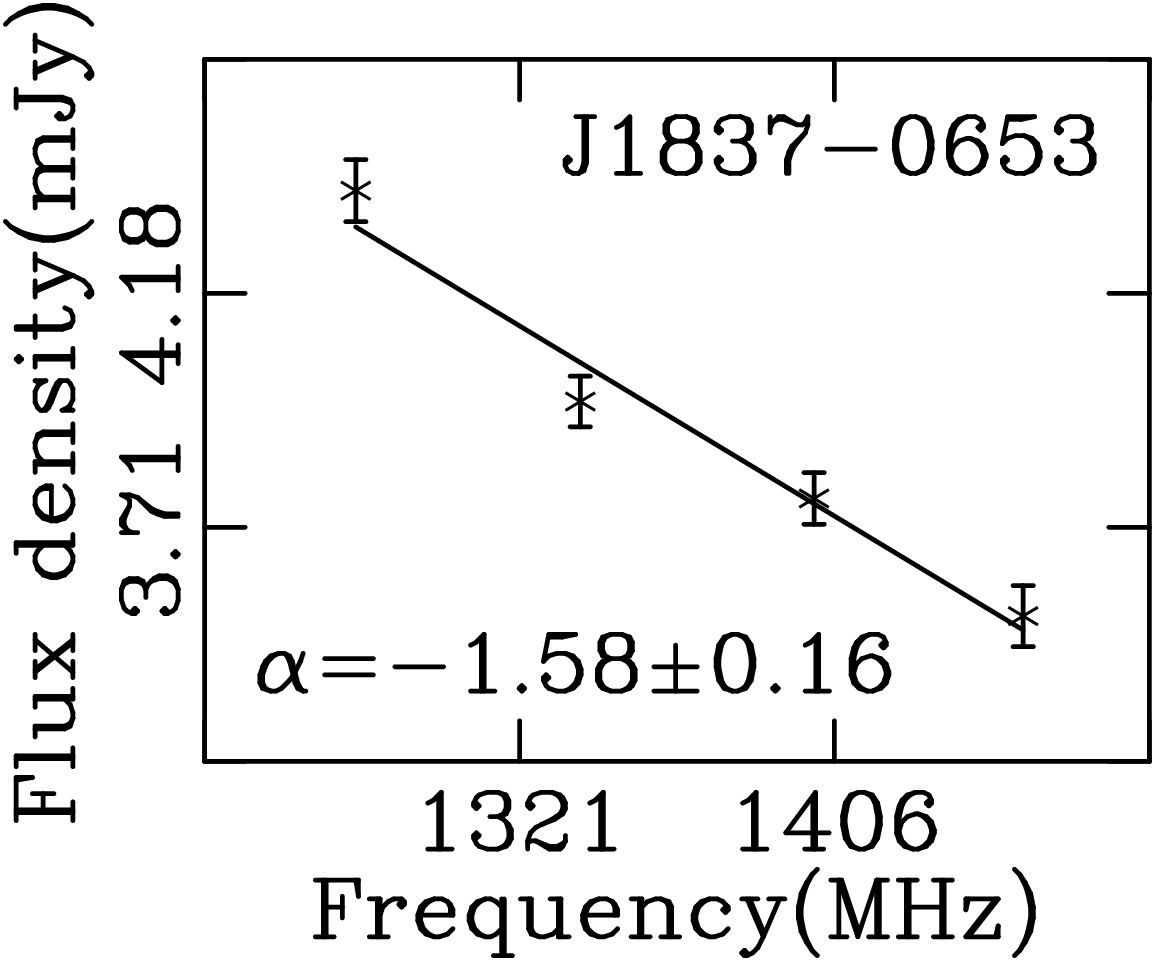}               
\includegraphics[bb = 110 120 518 685,clip,angle=-90,width=0.16\textwidth]{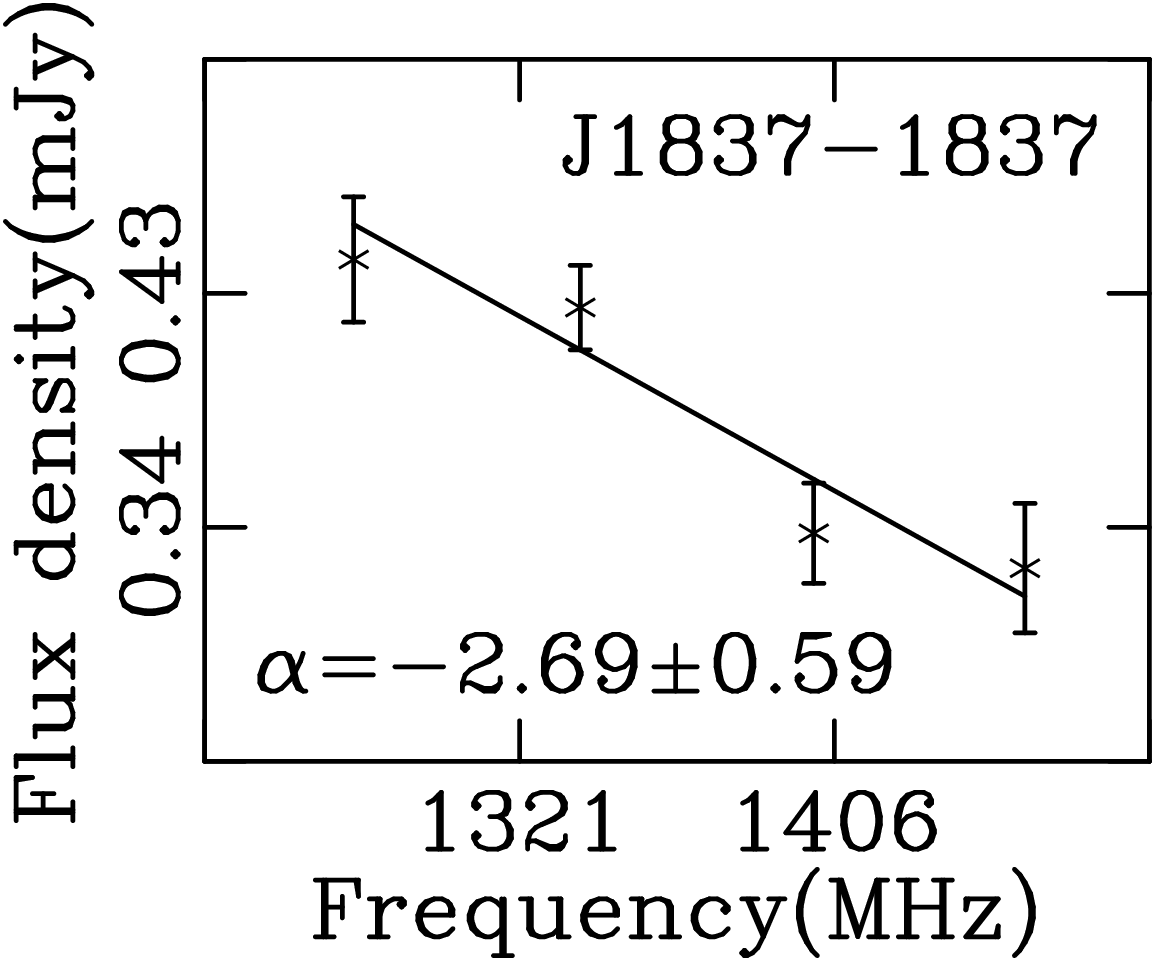}
 \\[2mm]               
\includegraphics[bb = 110 72 518 685,clip,angle=-90,width=0.175\textwidth]{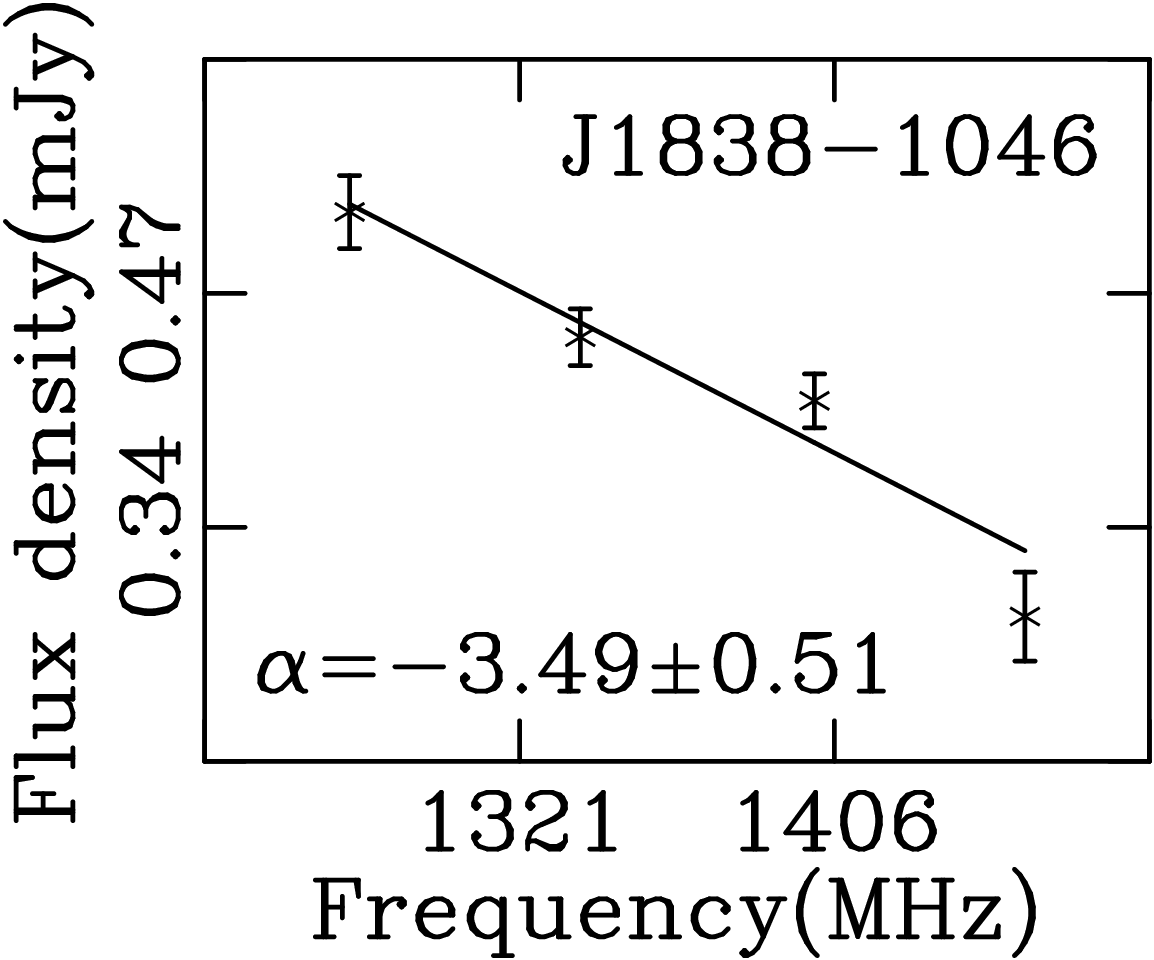}               
\includegraphics[bb = 110 120 518 685,clip,angle=-90,width=0.16\textwidth]{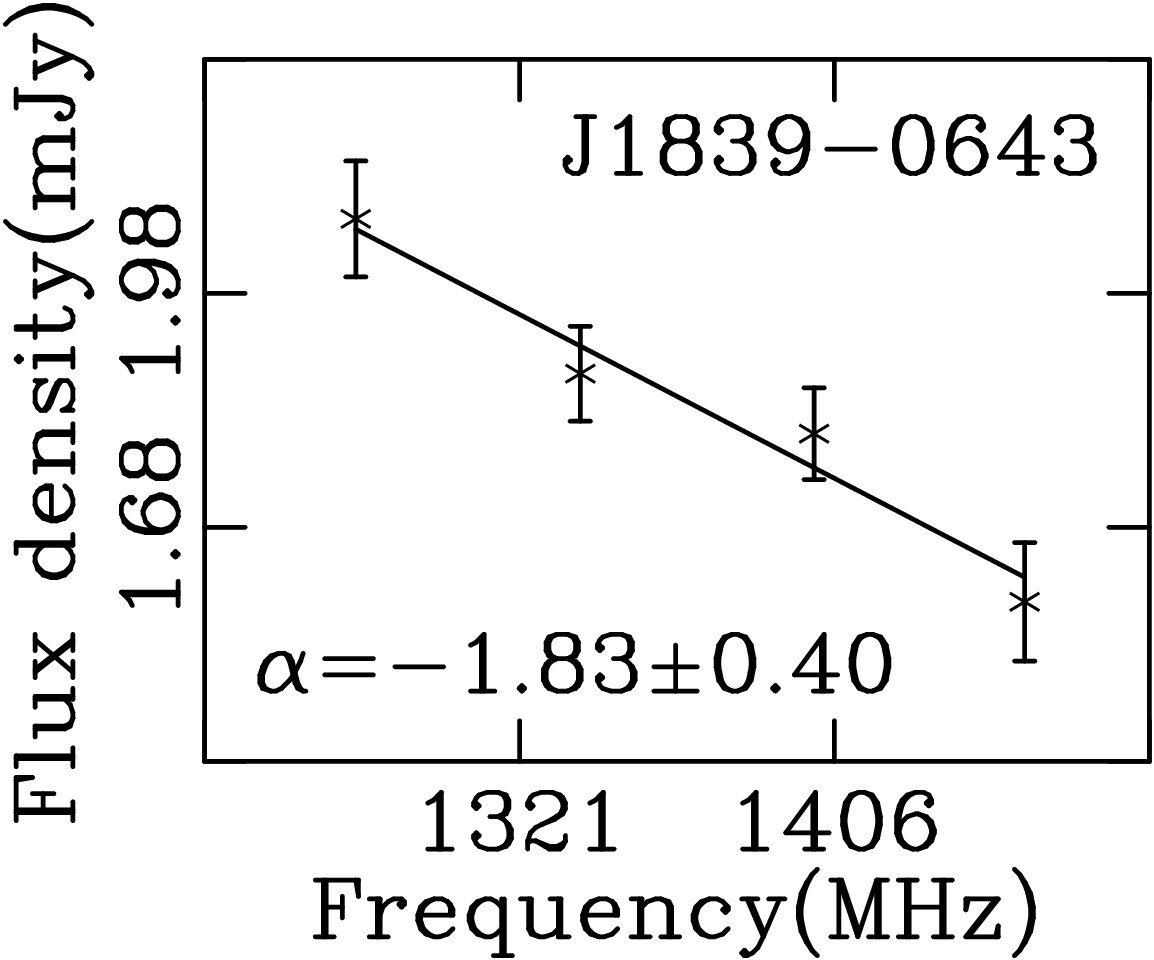}               
\includegraphics[bb = 110 120 518 685,clip,angle=-90,width=0.16\textwidth]{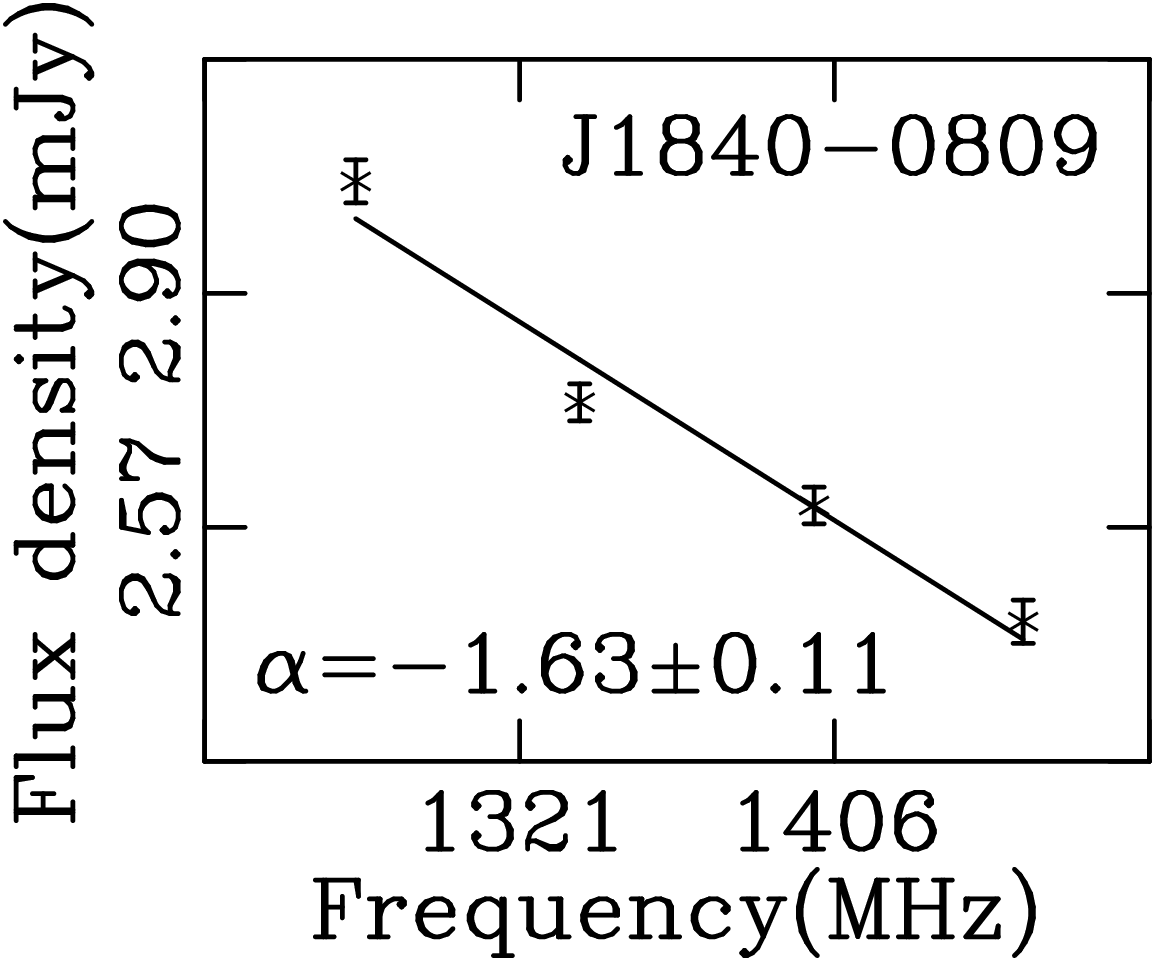}               
\includegraphics[bb = 110 120 518 685,clip,angle=-90,width=0.16\textwidth]{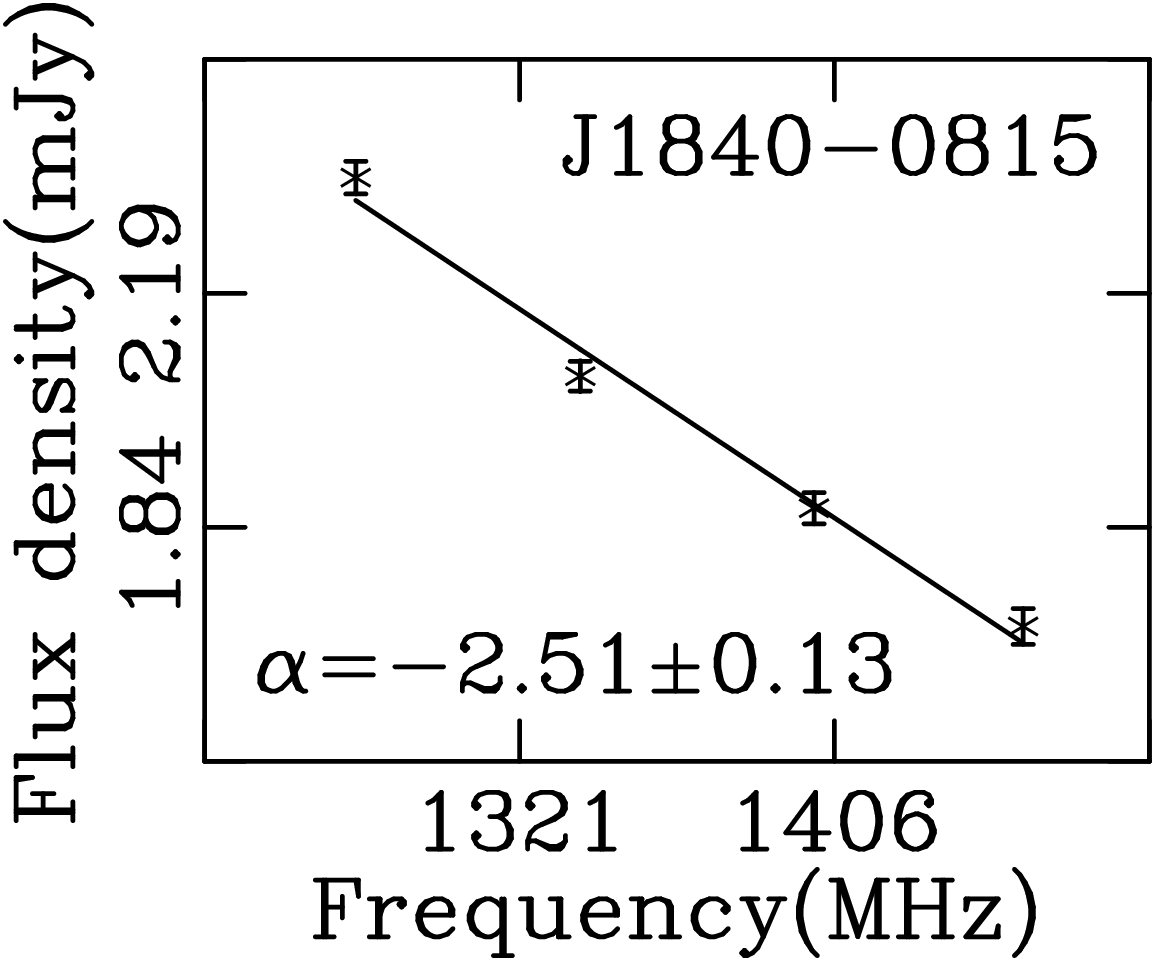}               
\includegraphics[bb = 110 120 518 685,clip,angle=-90,width=0.16\textwidth]{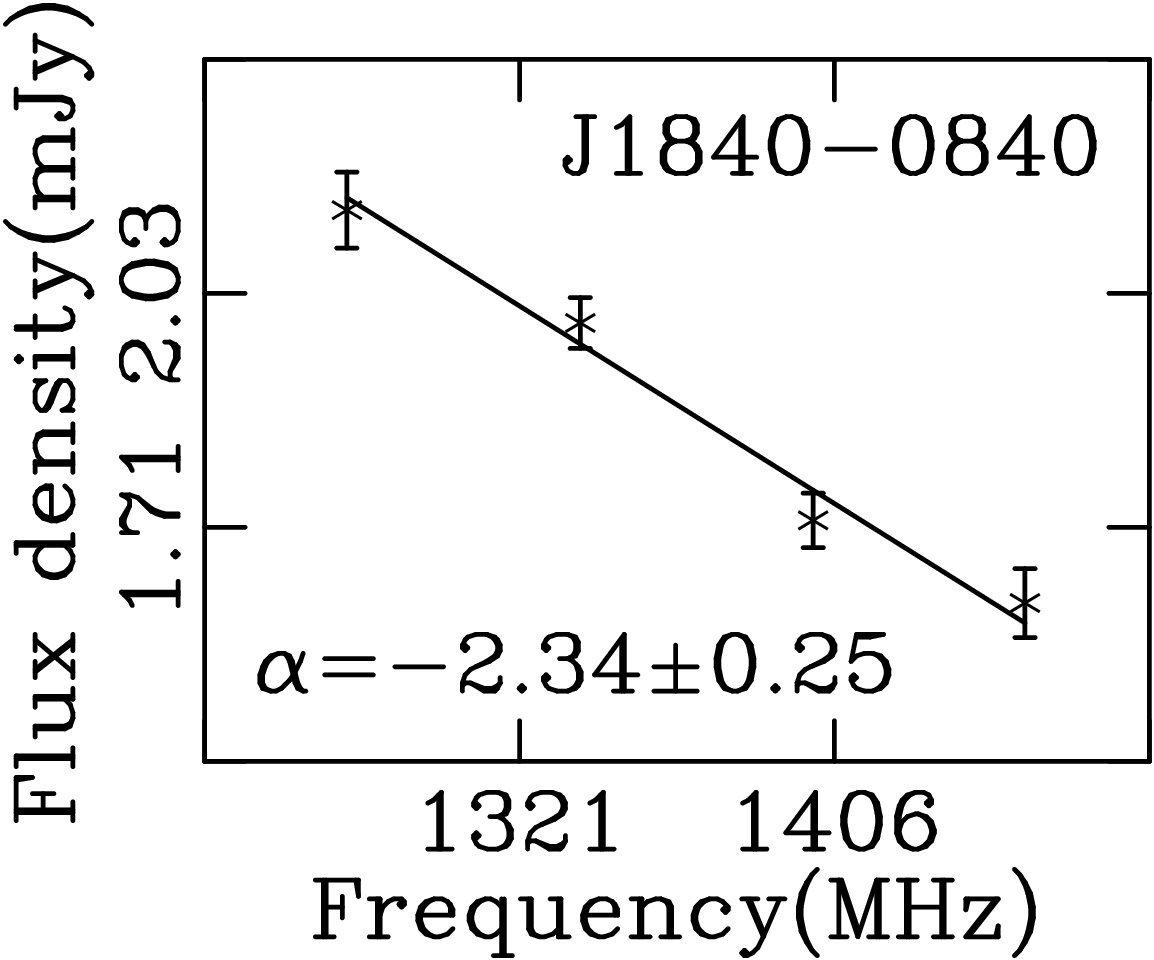}               
\includegraphics[bb = 110 120 518 685,clip,angle=-90,width=0.16\textwidth]{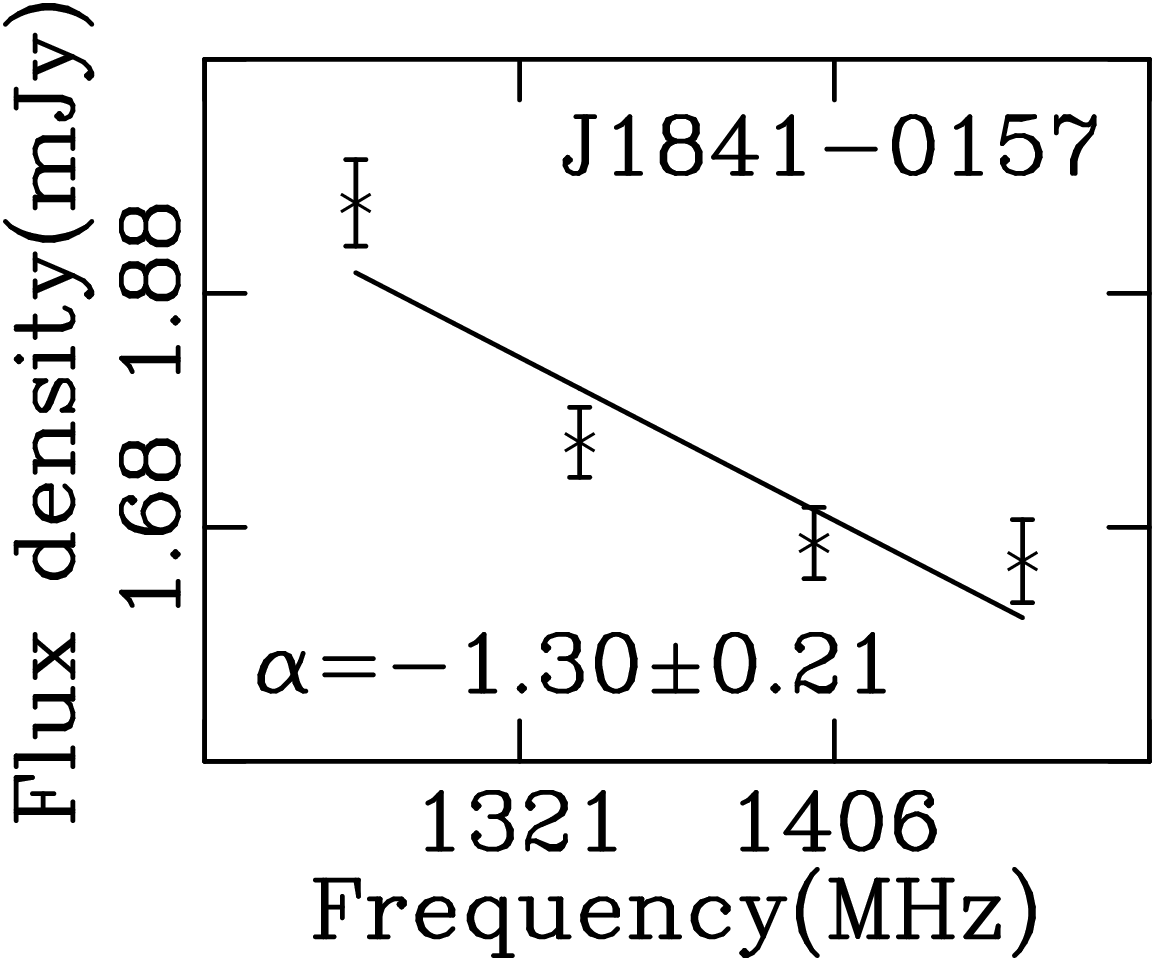}
\\[2mm]                
\includegraphics[bb = 110 72 518 685,clip,angle=-90,width=0.175\textwidth]{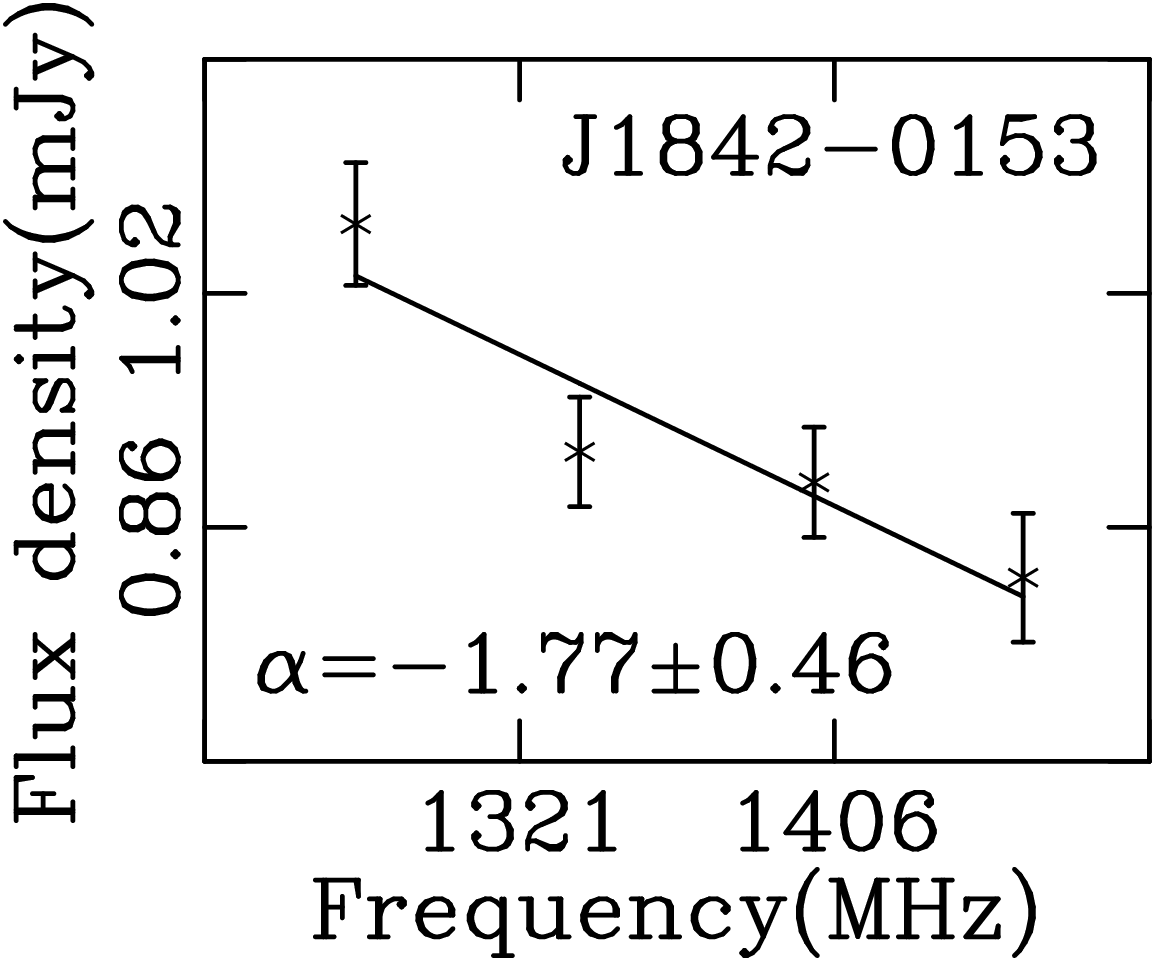}               
\includegraphics[bb = 110 120 518 685,clip,angle=-90,width=0.16\textwidth]{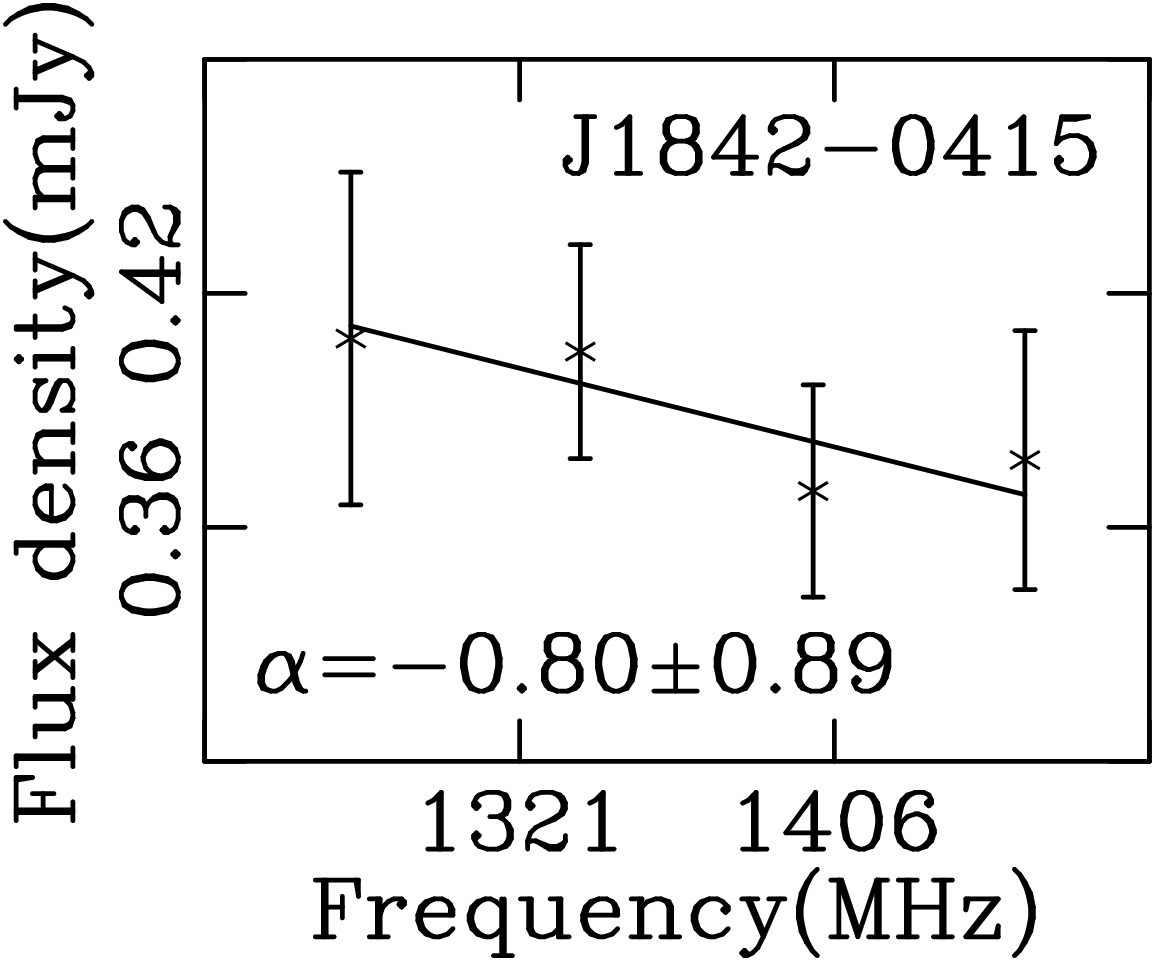}               
\includegraphics[bb = 110 120 518 685,clip,angle=-90,width=0.16\textwidth]{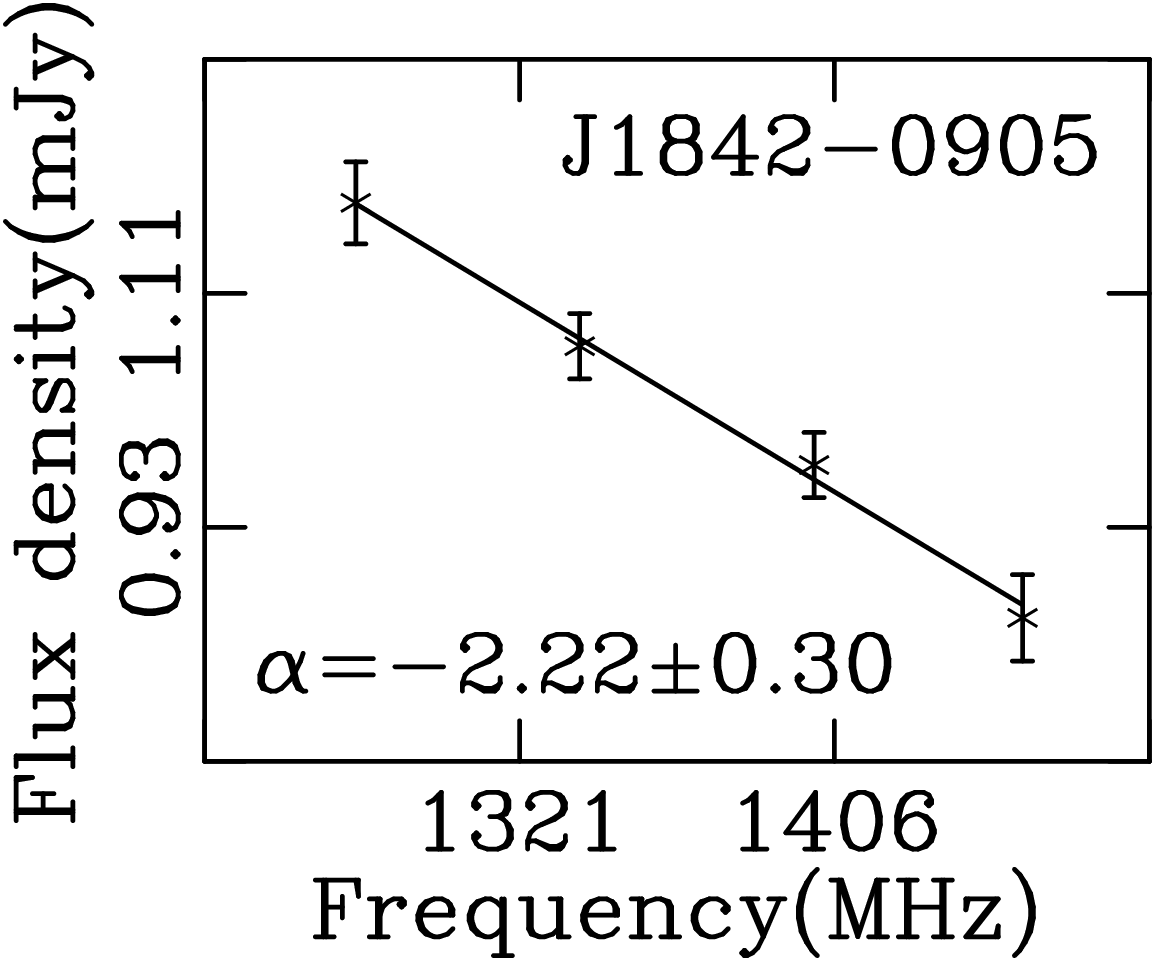}               
\includegraphics[bb = 110 120 518 685,clip,angle=-90,width=0.16\textwidth]{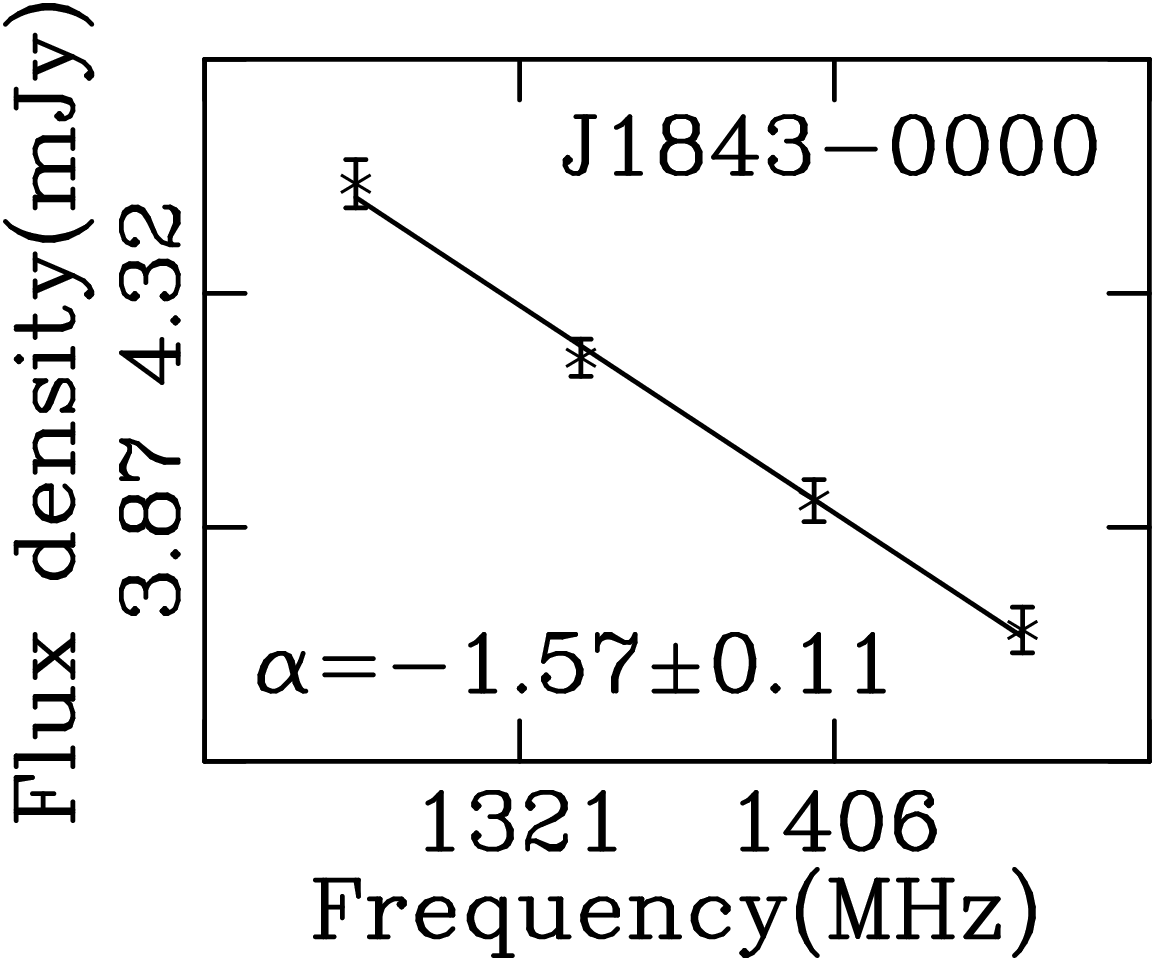}               
\includegraphics[bb = 110 120 518 685,clip,angle=-90,width=0.16\textwidth]{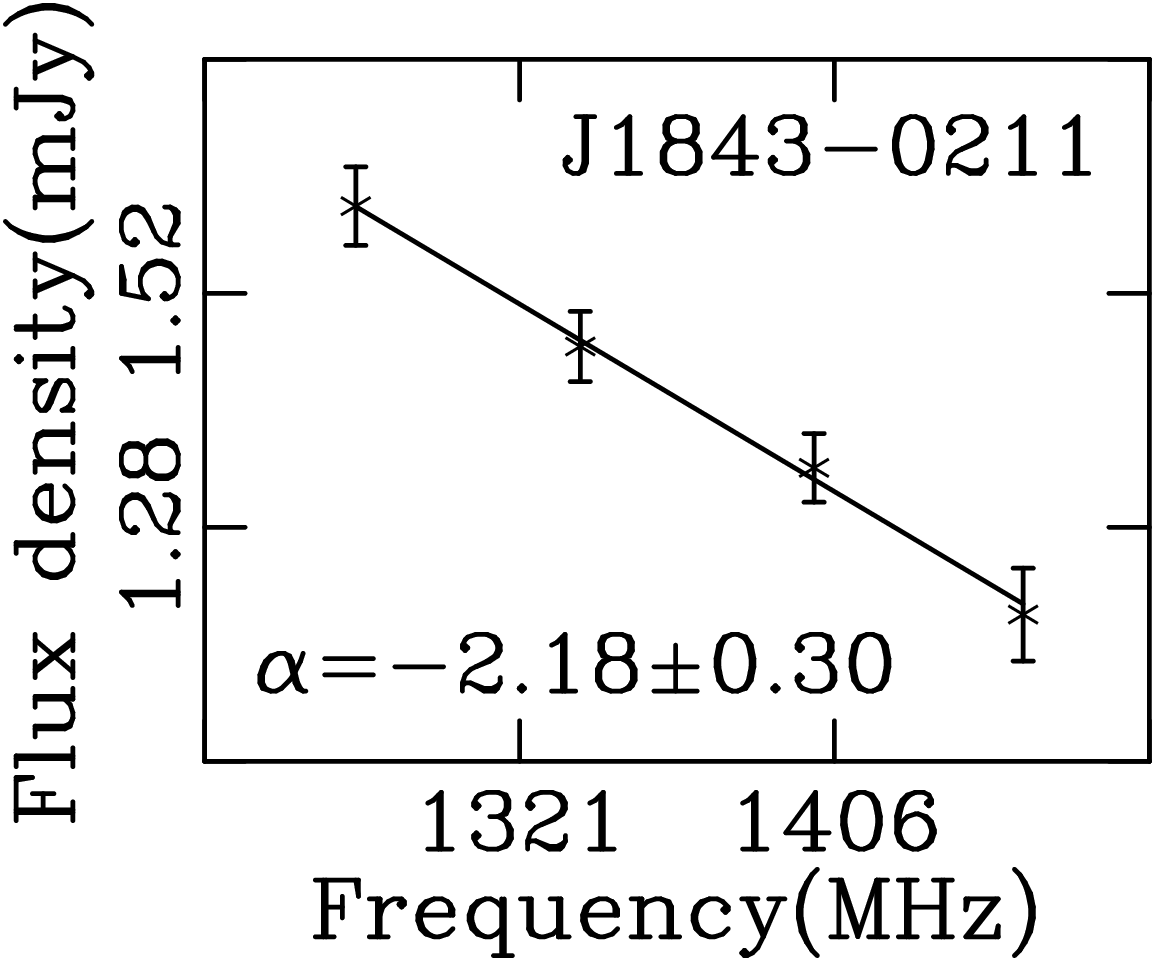}               
\includegraphics[bb = 110 120 518 685,clip,angle=-90,width=0.16\textwidth]{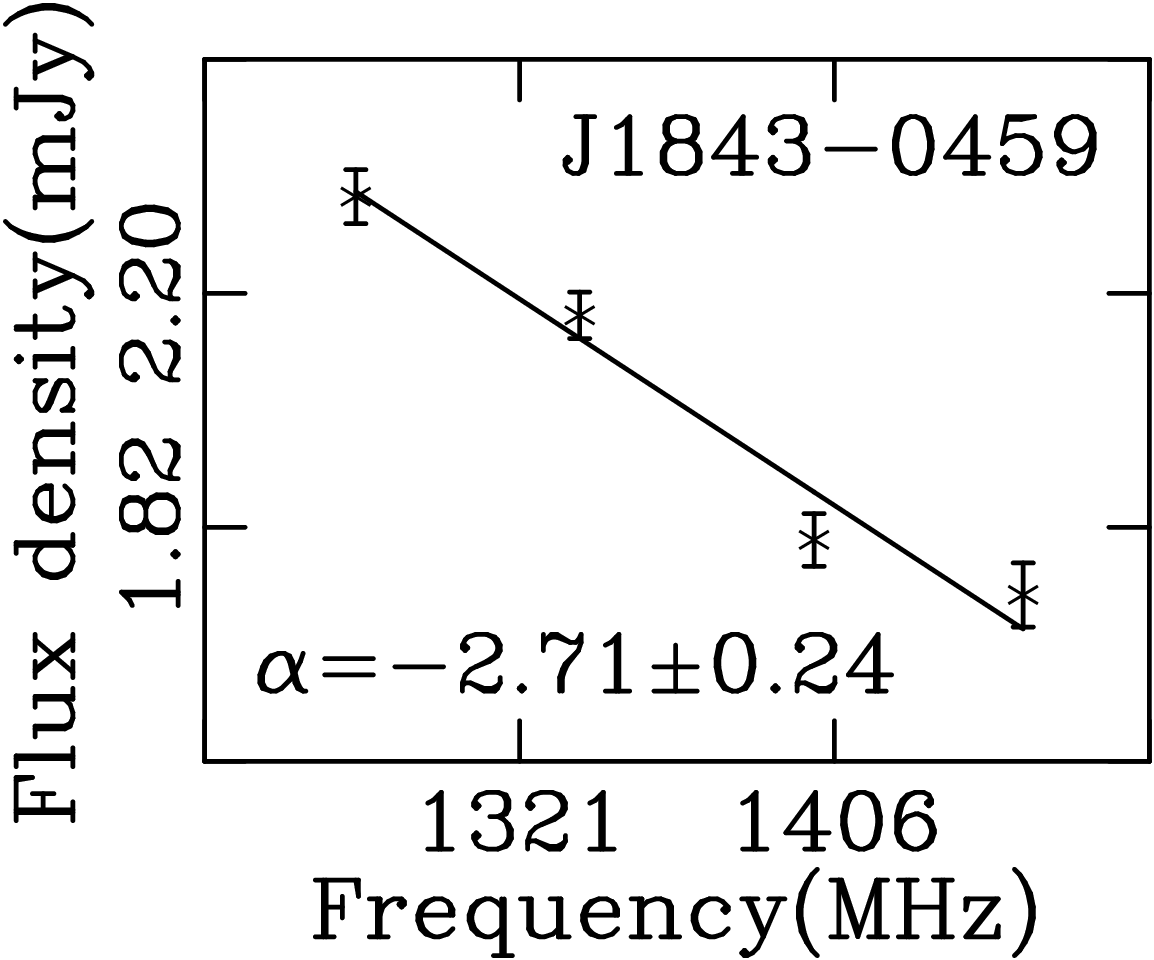}
\\[2mm]                
\includegraphics[bb = 110 72 518 685,clip,angle=-90,width=0.175\textwidth]{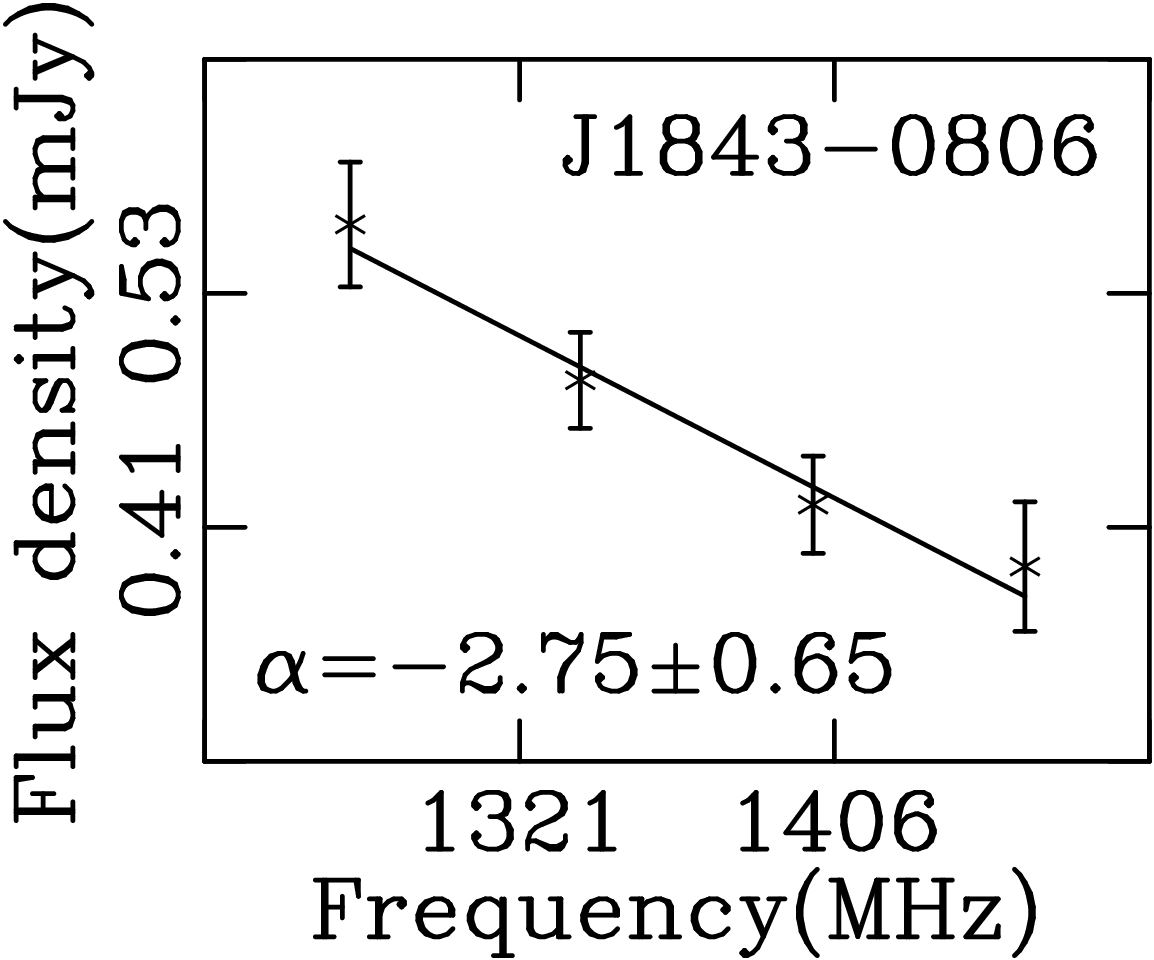}               
\includegraphics[bb = 110 120 518 685,clip,angle=-90,width=0.16\textwidth]{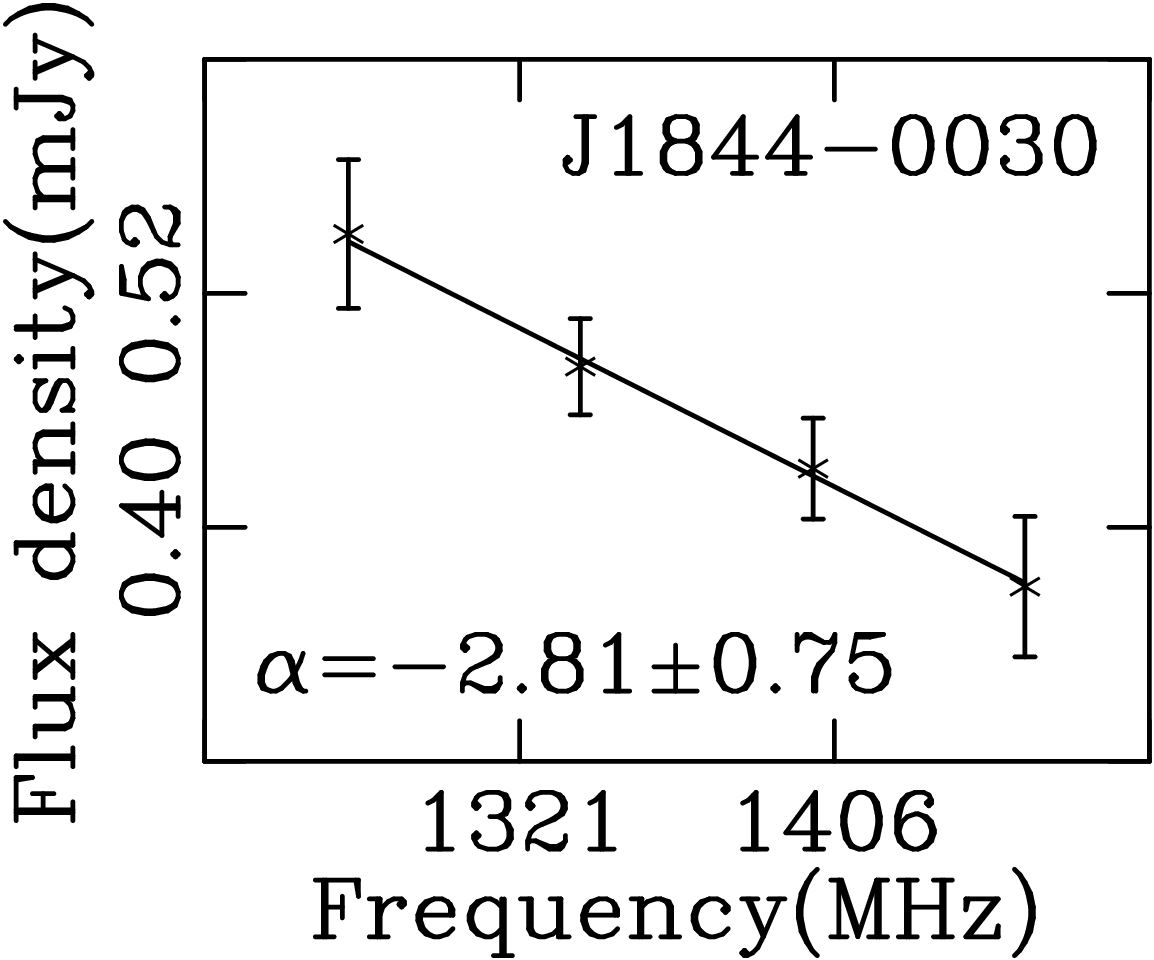}               
\includegraphics[bb = 110 120 518 685,clip,angle=-90,width=0.16\textwidth]{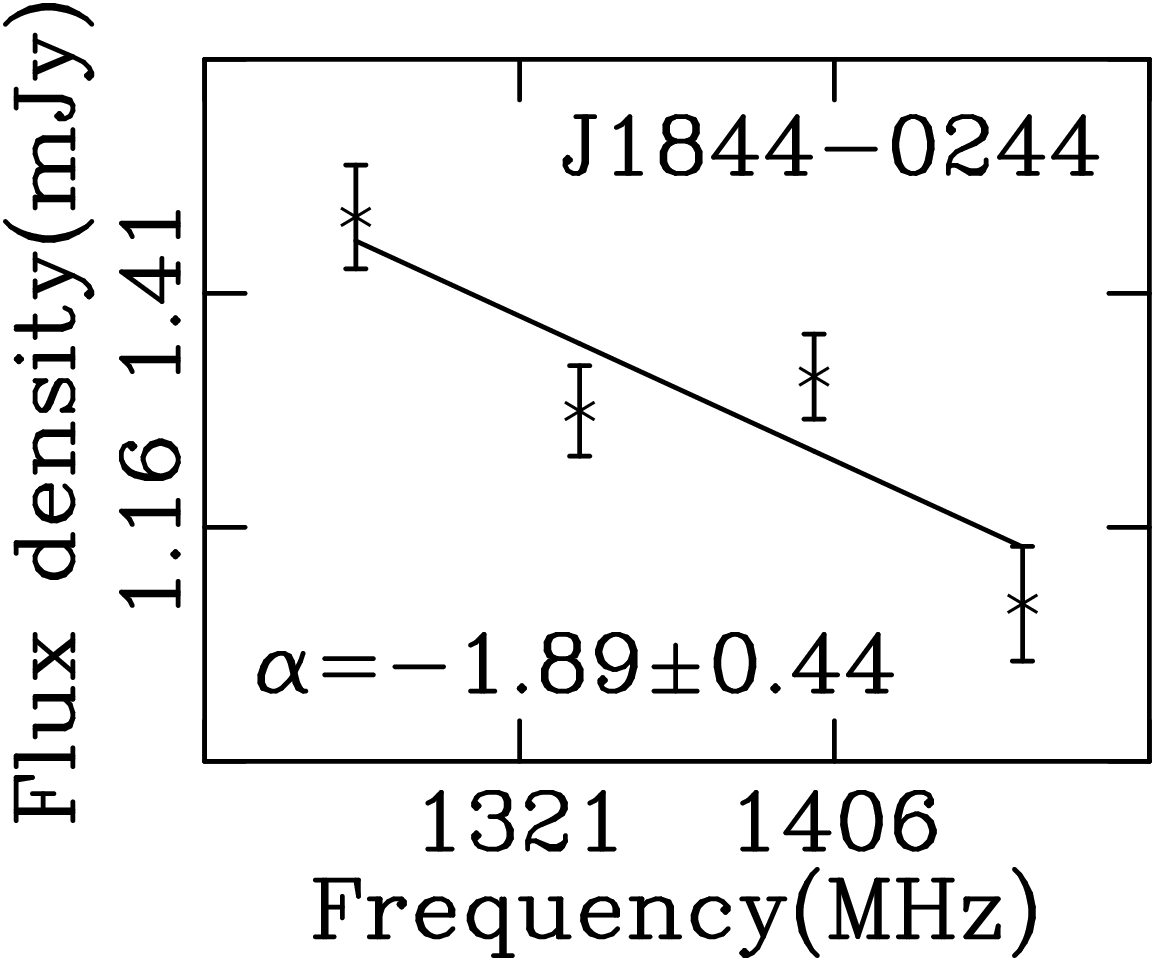}               
\includegraphics[bb = 110 120 518 685,clip,angle=-90,width=0.16\textwidth]{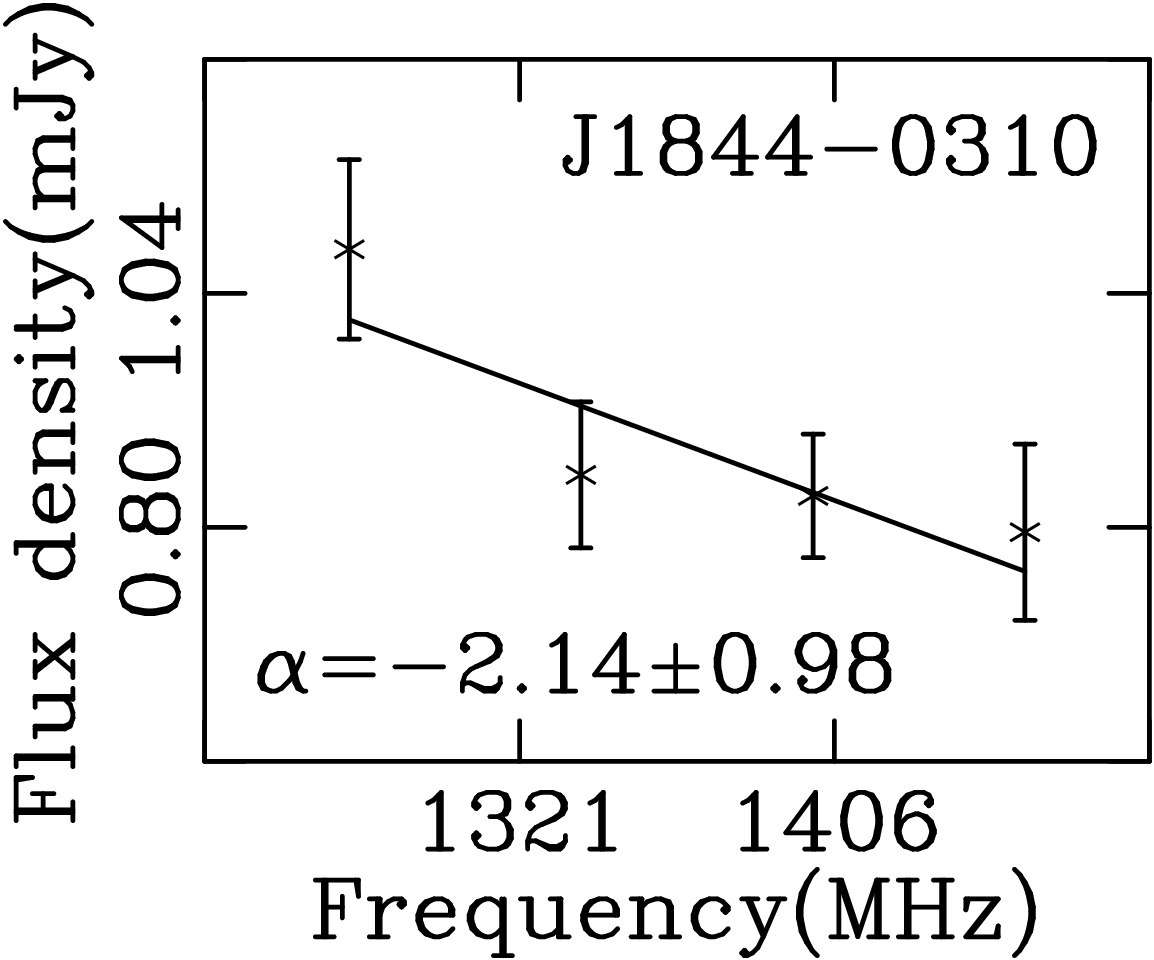}               
\includegraphics[bb = 110 120 518 685,clip,angle=-90,width=0.16\textwidth]{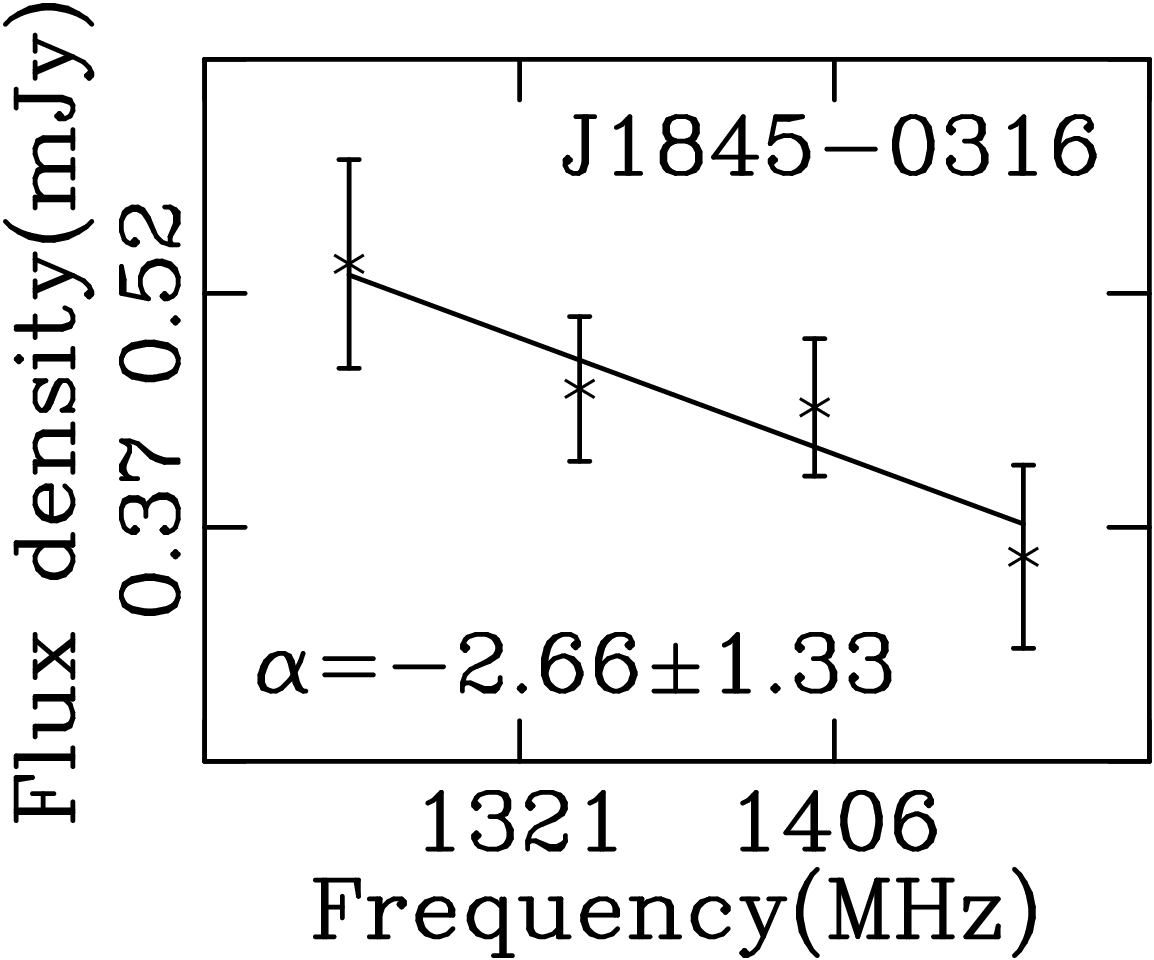}               
\includegraphics[bb = 110 120 518 685,clip,angle=-90,width=0.16\textwidth]{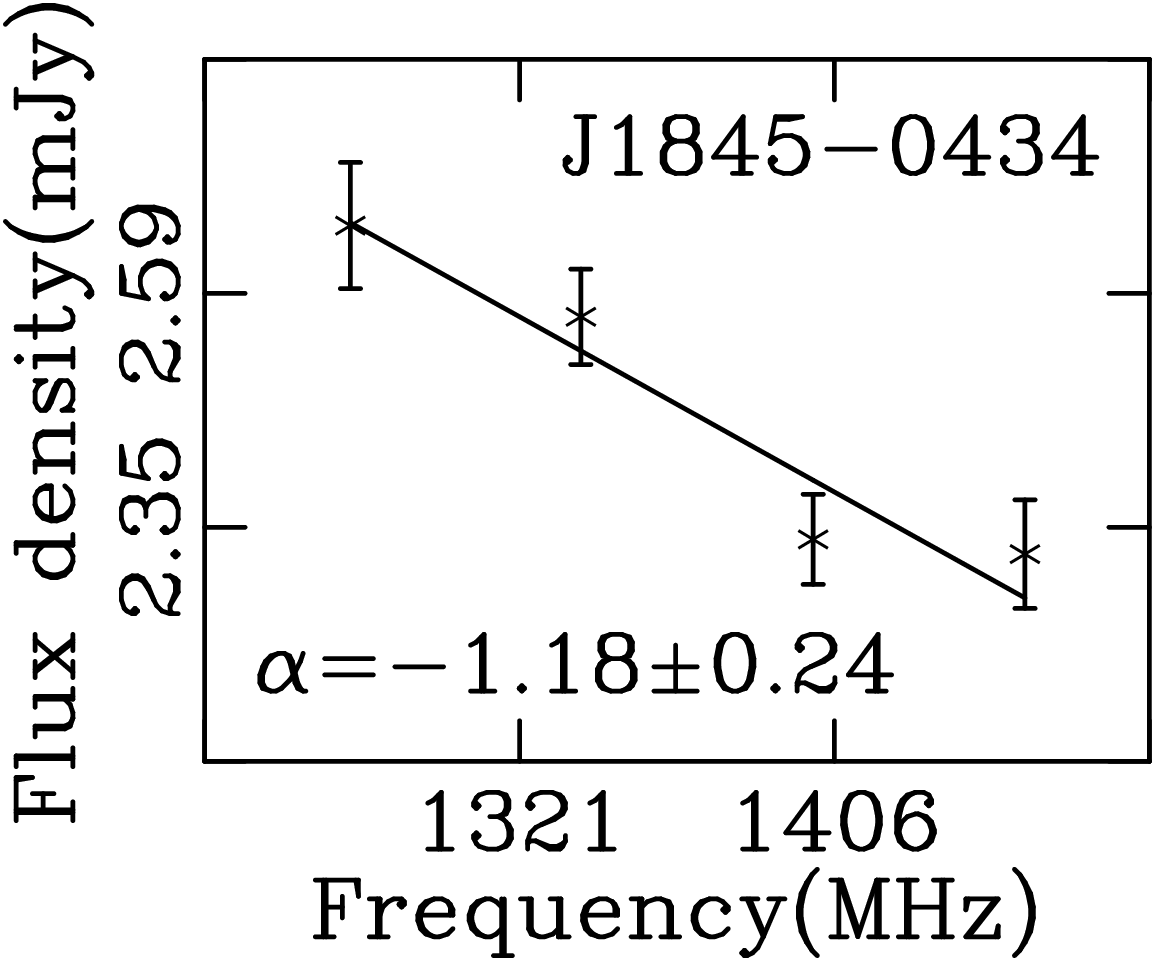}
\\[2mm]                
\includegraphics[bb = 110 72 518 685,clip,angle=-90,width=0.175\textwidth]{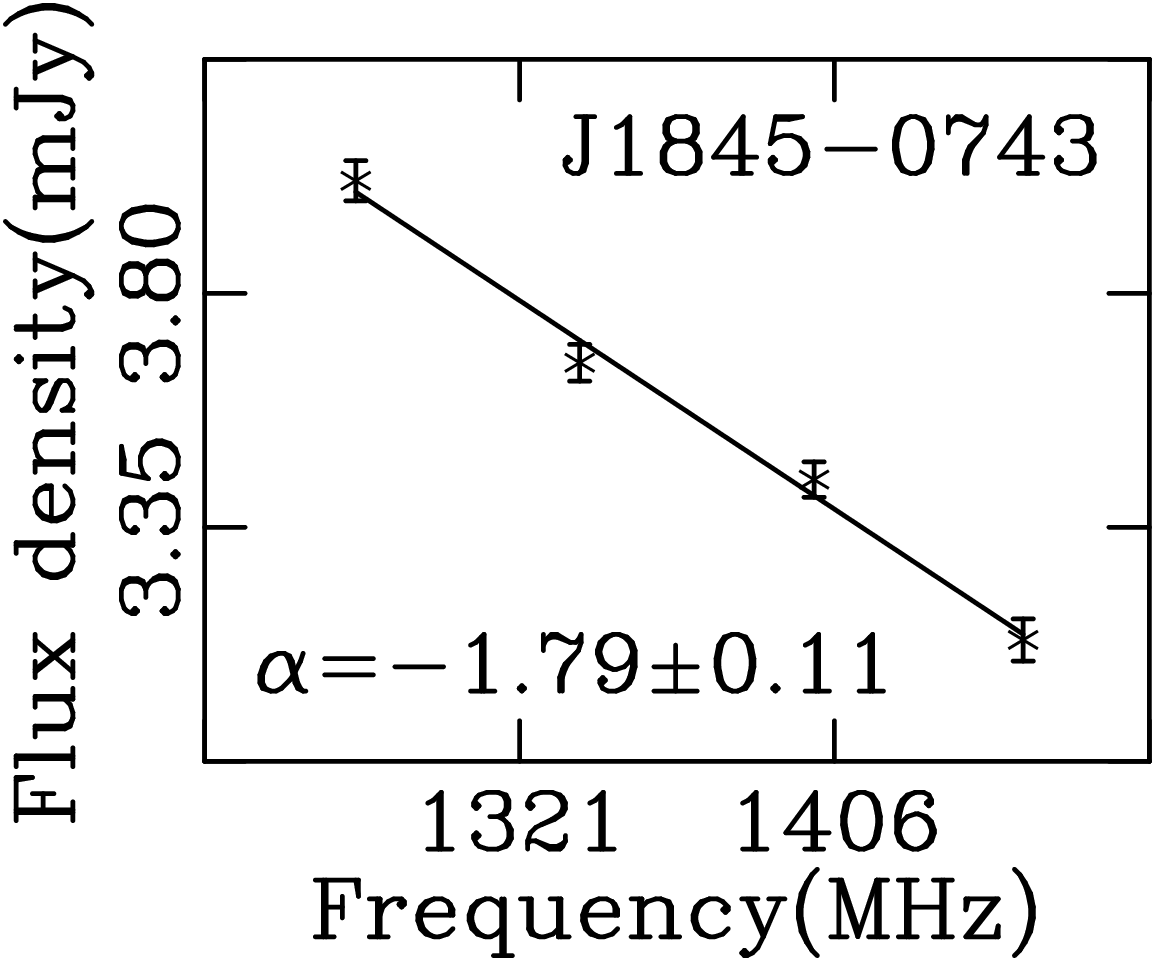}               
\includegraphics[bb = 110 120 518 685,clip,angle=-90,width=0.16\textwidth]{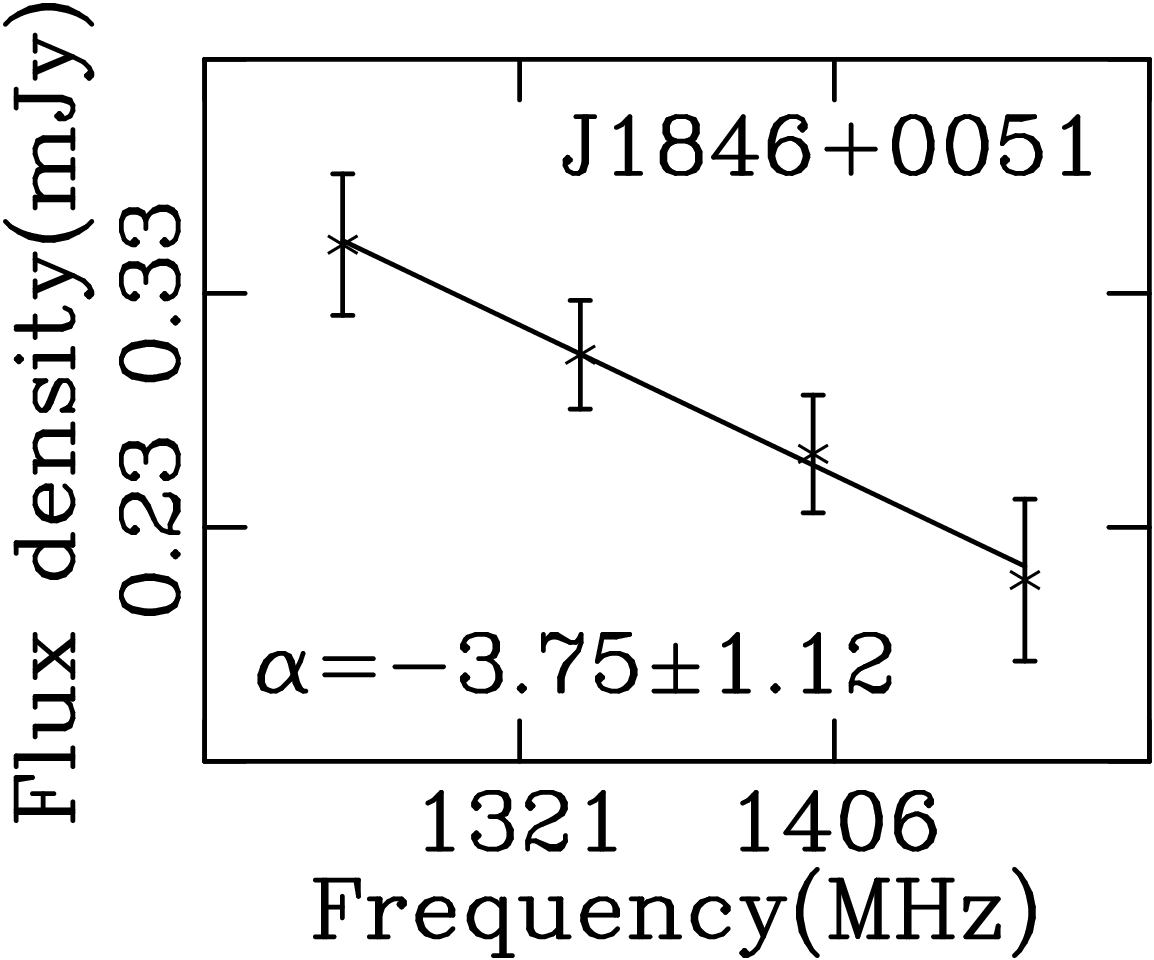}               
\includegraphics[bb = 110 120 518 685,clip,angle=-90,width=0.16\textwidth]{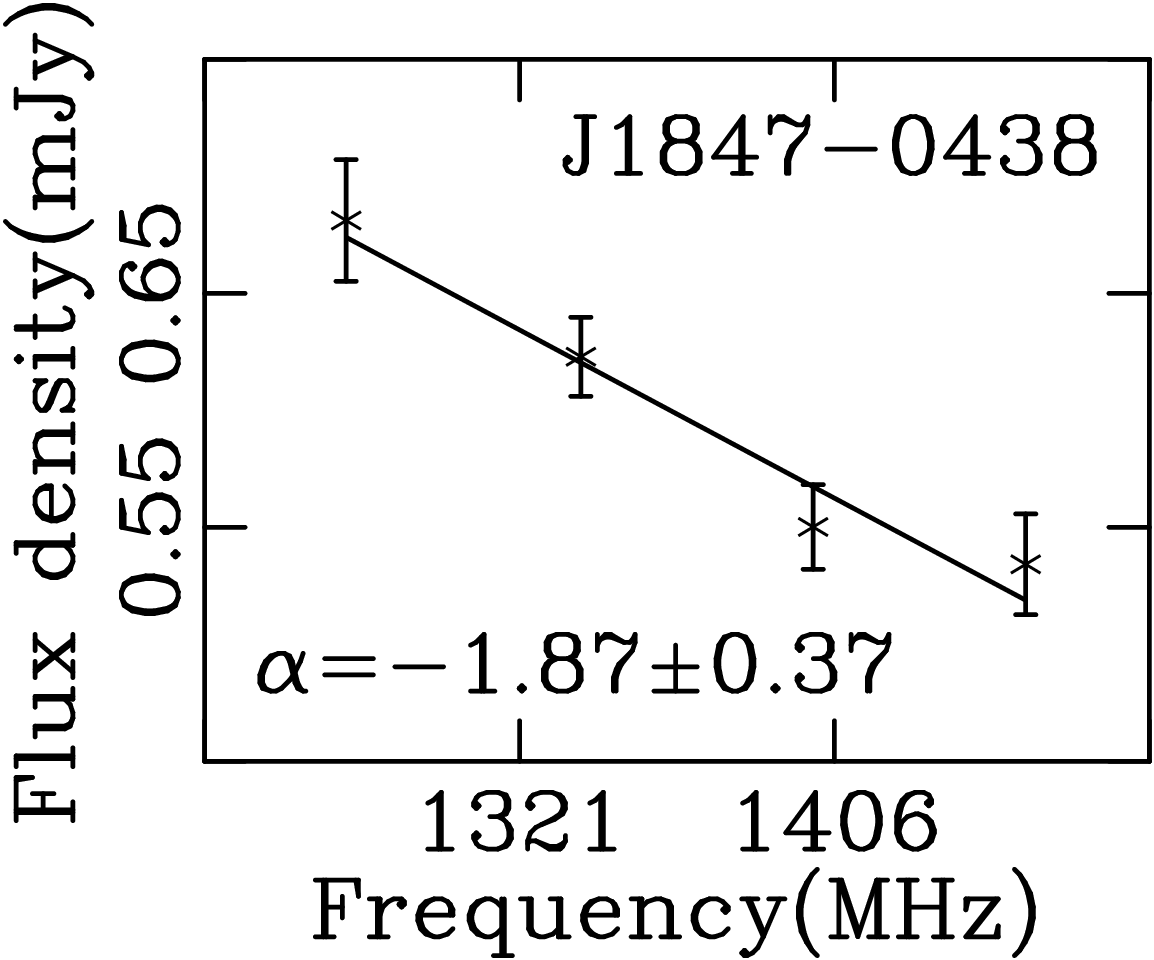}               
\includegraphics[bb = 110 120 518 685,clip,angle=-90,width=0.16\textwidth]{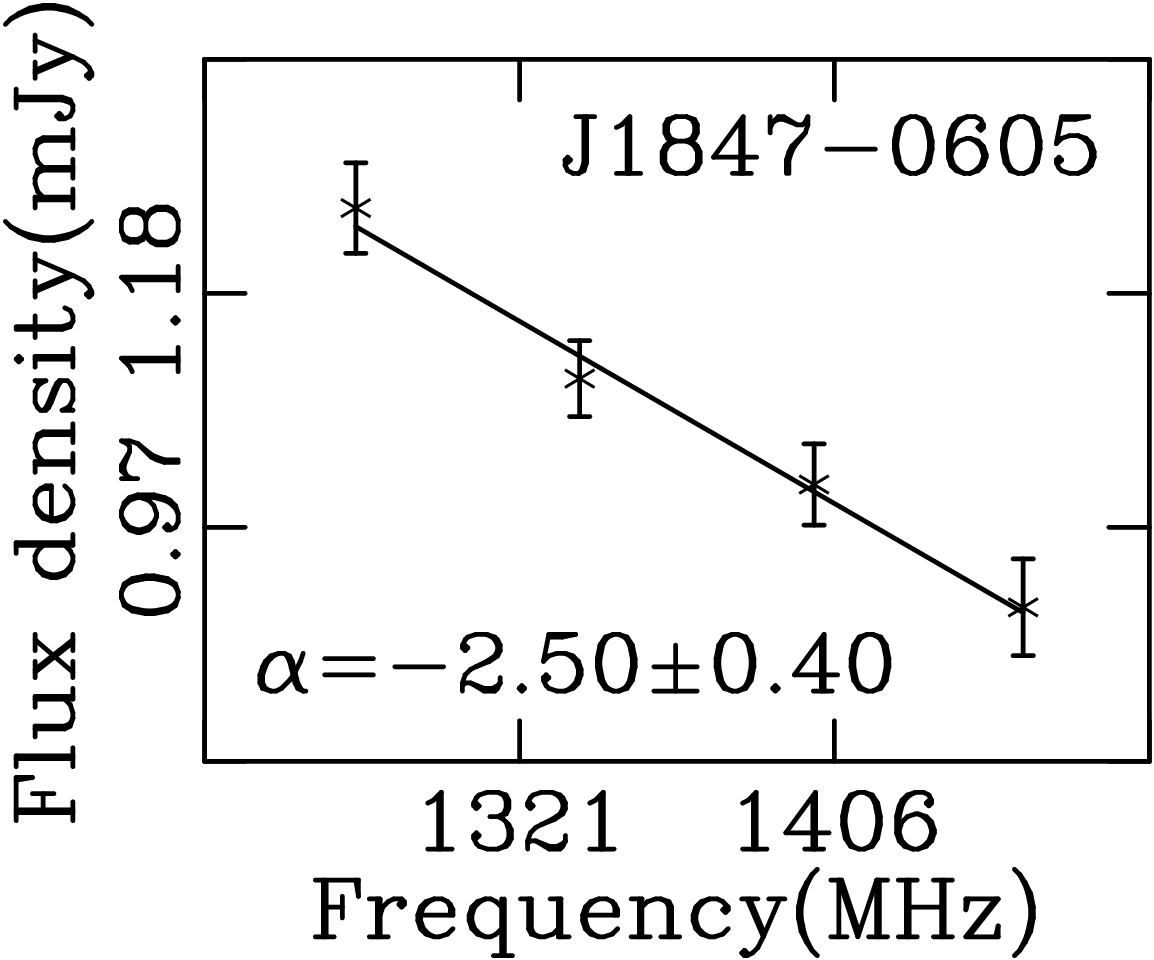}               
\includegraphics[bb = 110 120 518 685,clip,angle=-90,width=0.16\textwidth]{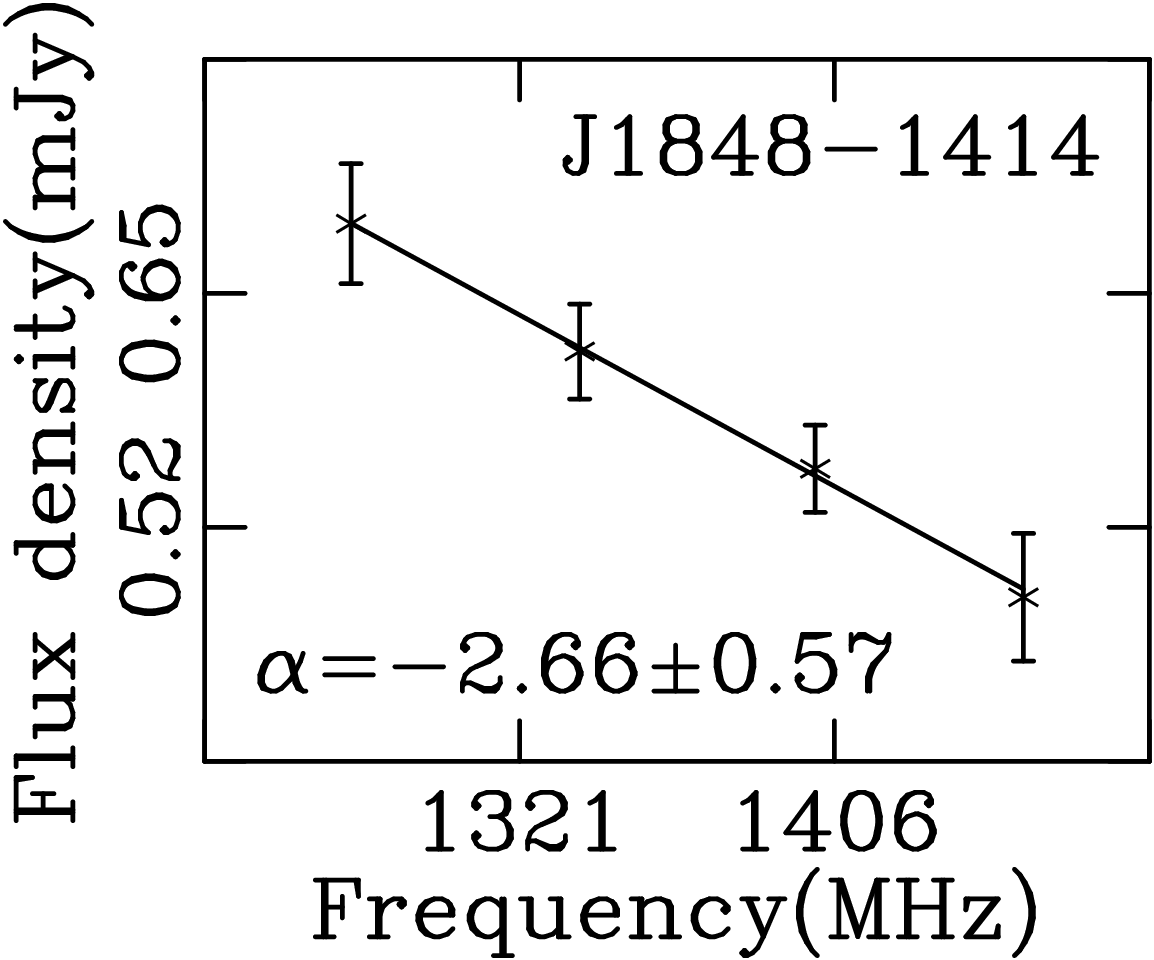}               
\includegraphics[bb = 110 120 518 685,clip,angle=-90,width=0.16\textwidth]{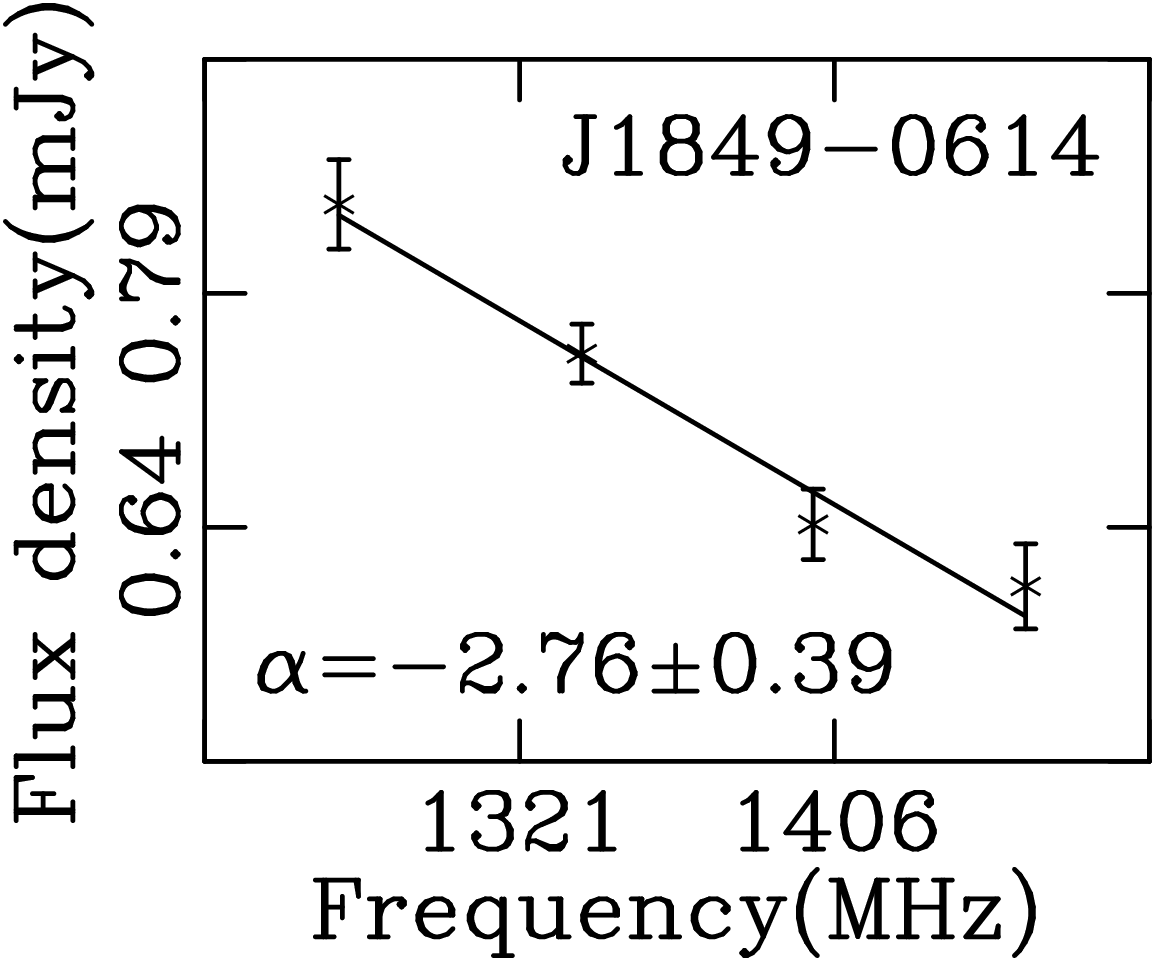}
\\[2mm]                
\includegraphics[bb = 110 72 518 685,clip,angle=-90,width=0.175\textwidth]{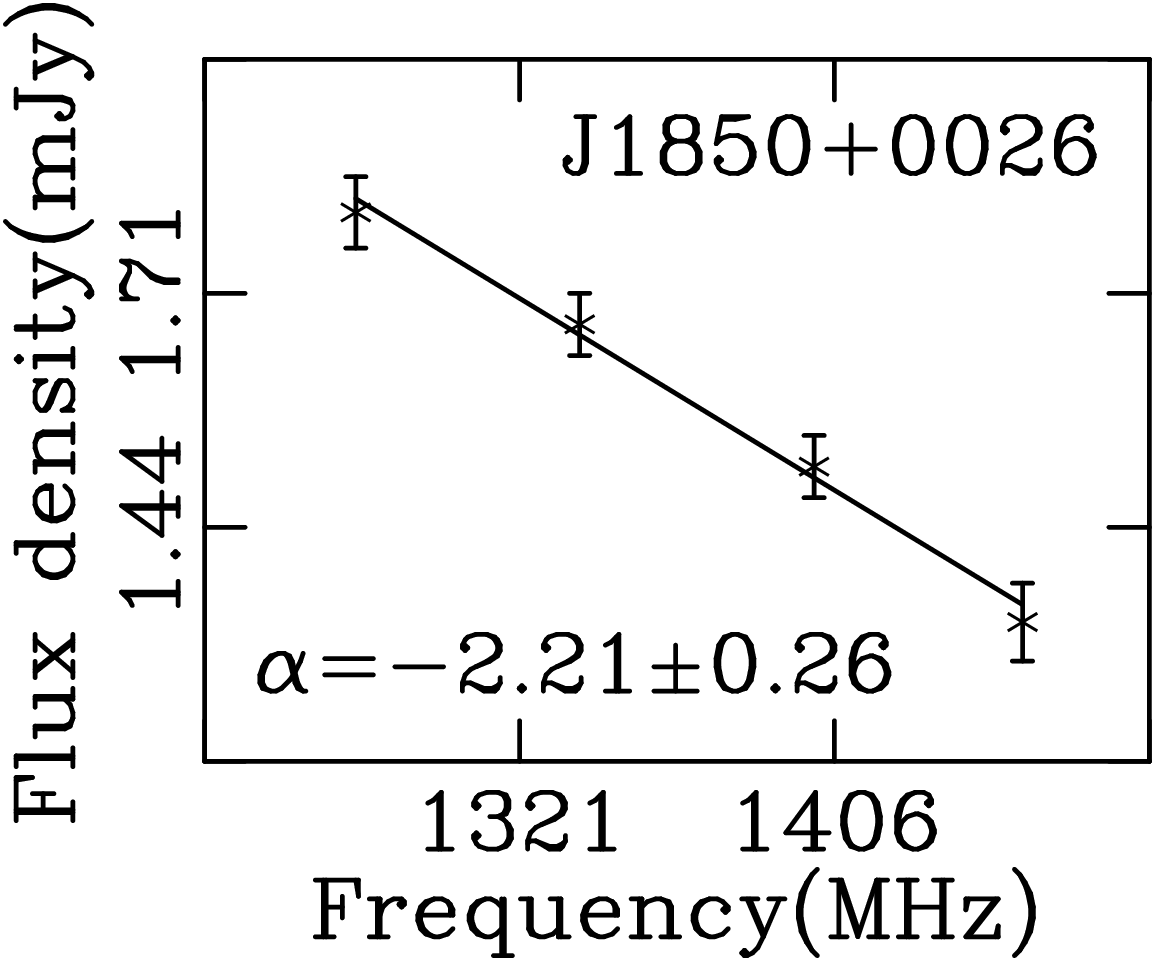}               
\includegraphics[bb = 110 120 518 685,clip,angle=-90,width=0.16\textwidth]{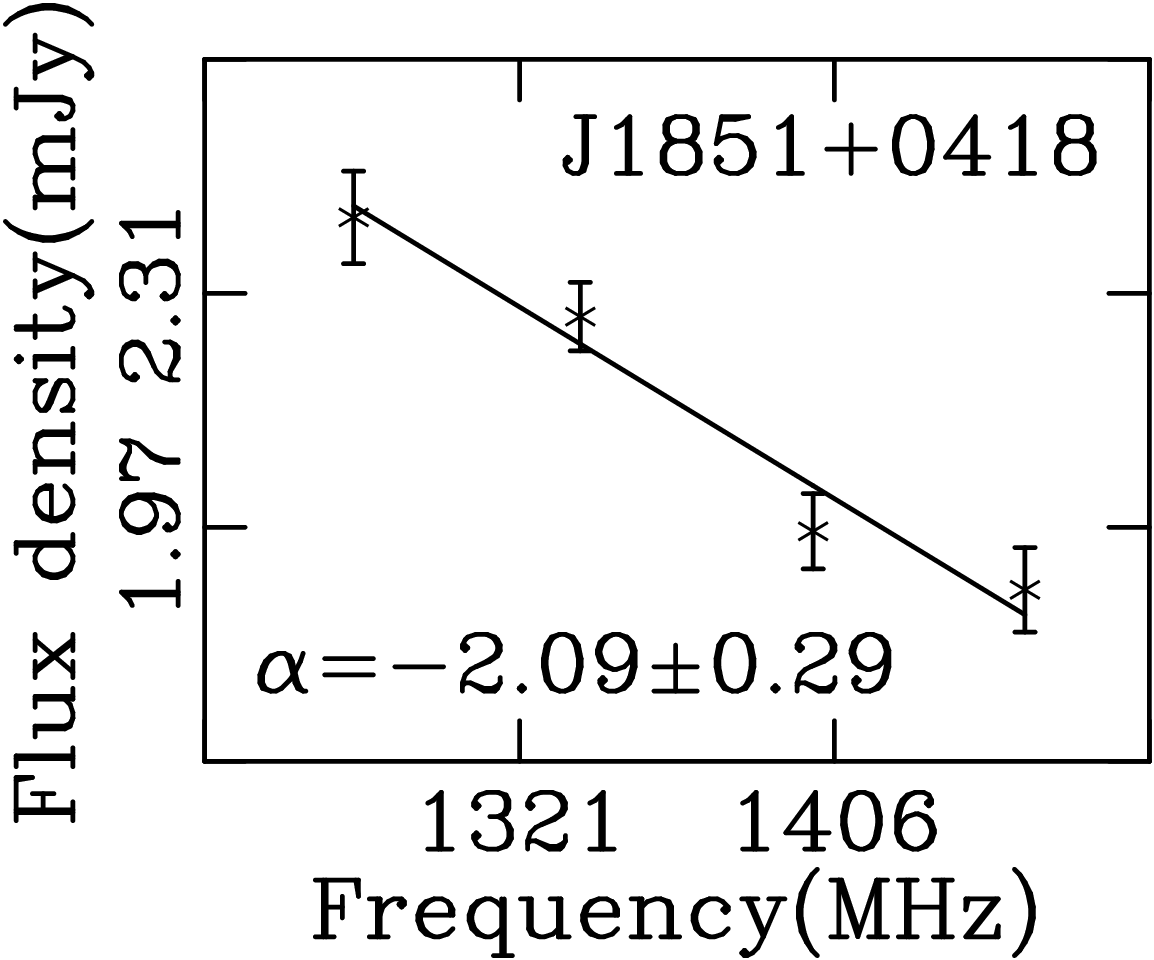}               
\includegraphics[bb = 110 120 518 685,clip,angle=-90,width=0.16\textwidth]{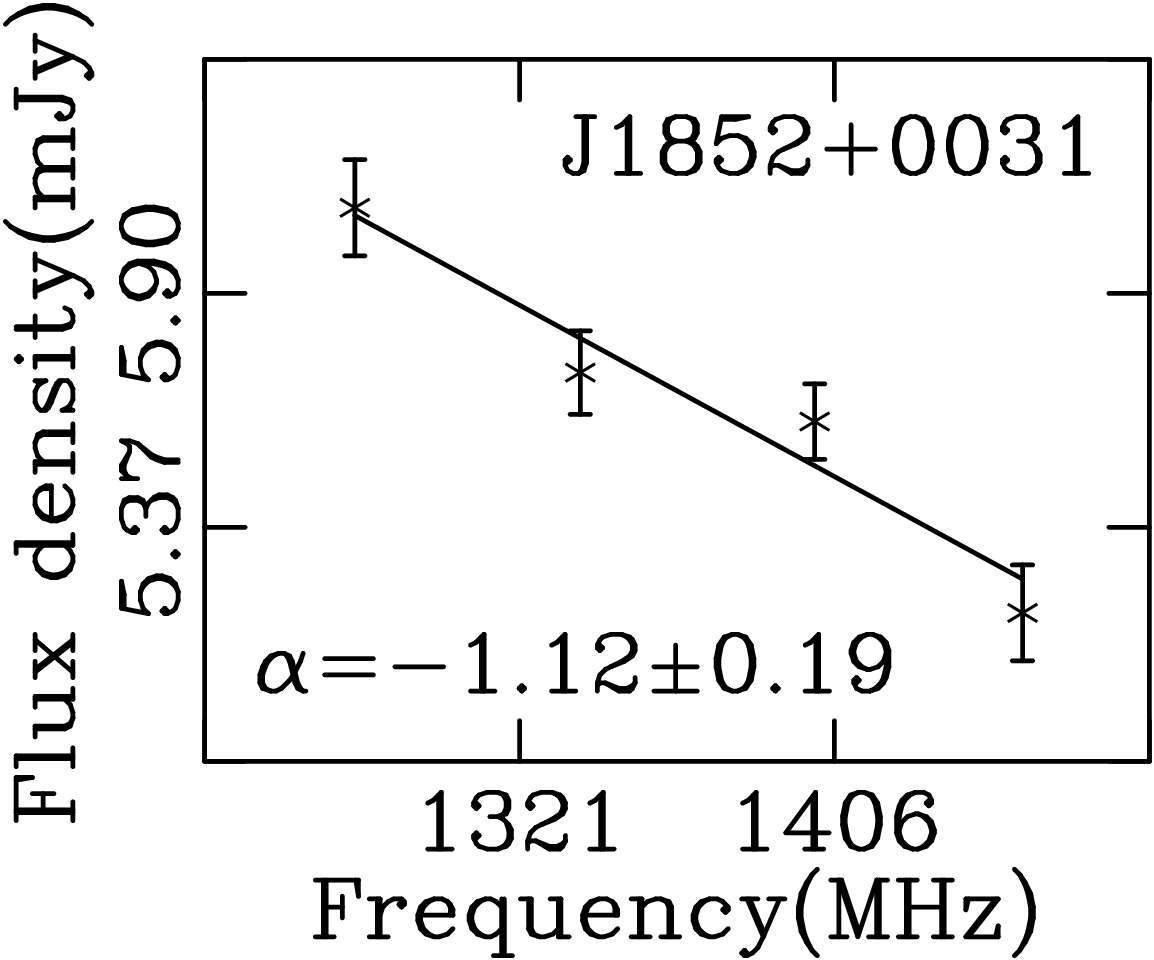}               
\includegraphics[bb = 110 120 518 685,clip,angle=-90,width=0.16\textwidth]{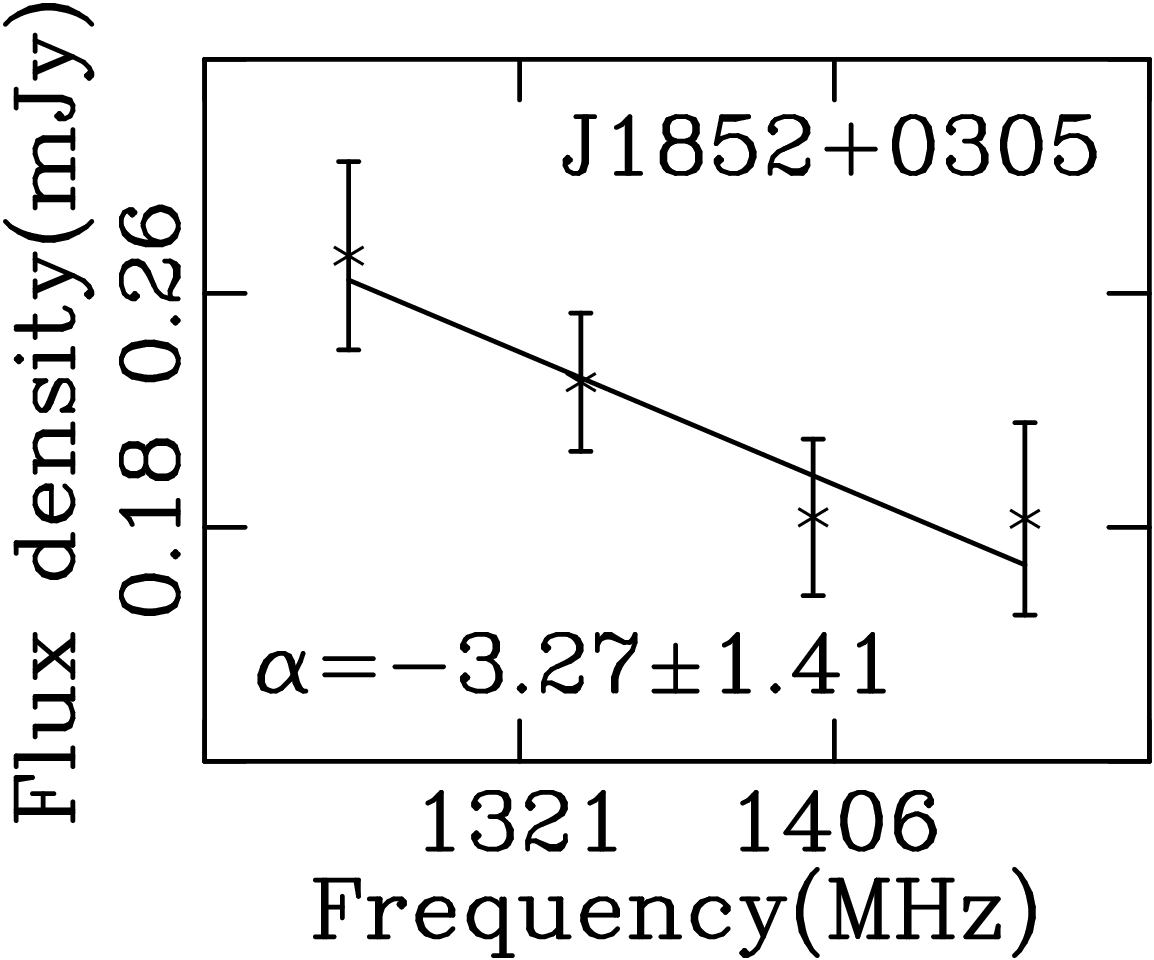}               
\includegraphics[bb = 110 120 518 685,clip,angle=-90,width=0.16\textwidth]{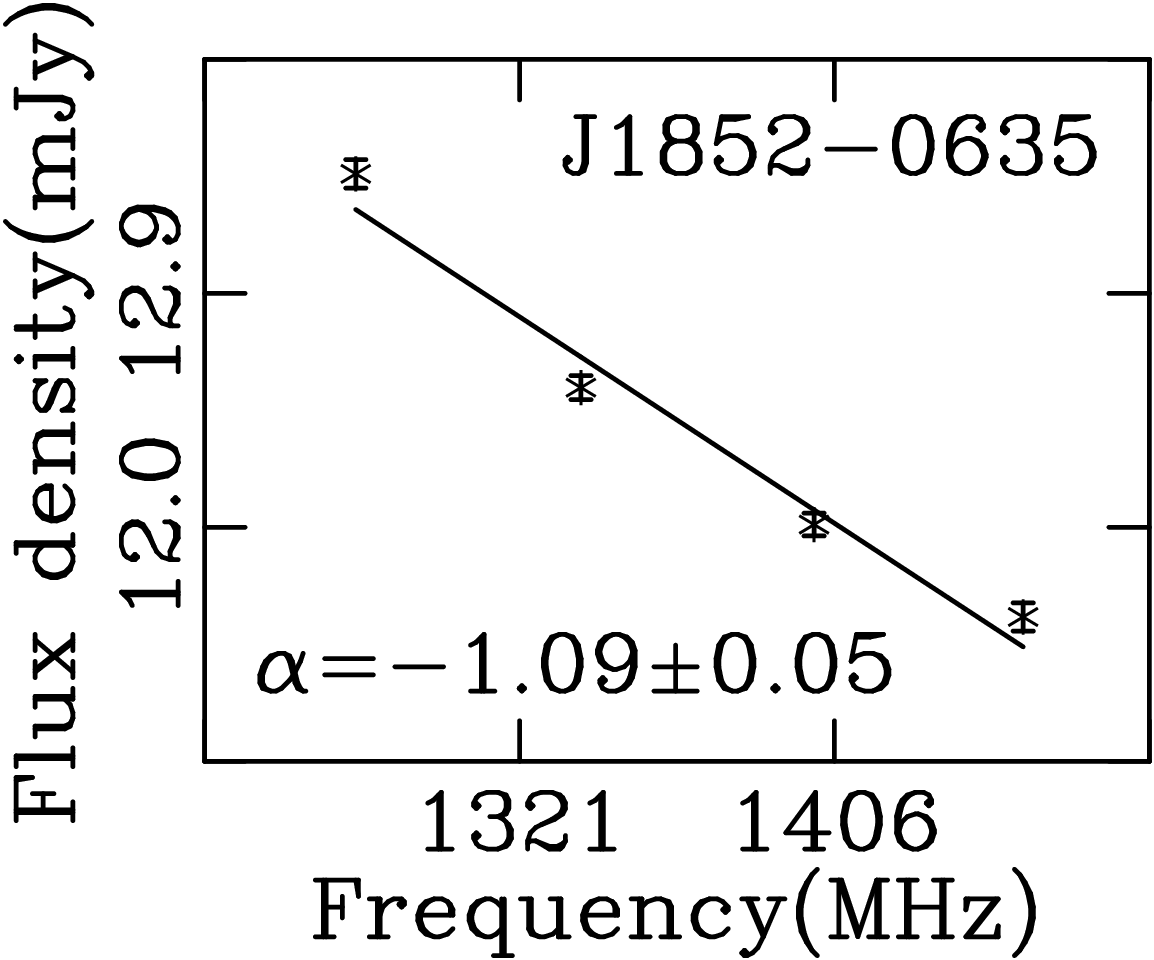}               
\includegraphics[bb = 110 120 518 685,clip,angle=-90,width=0.16\textwidth]{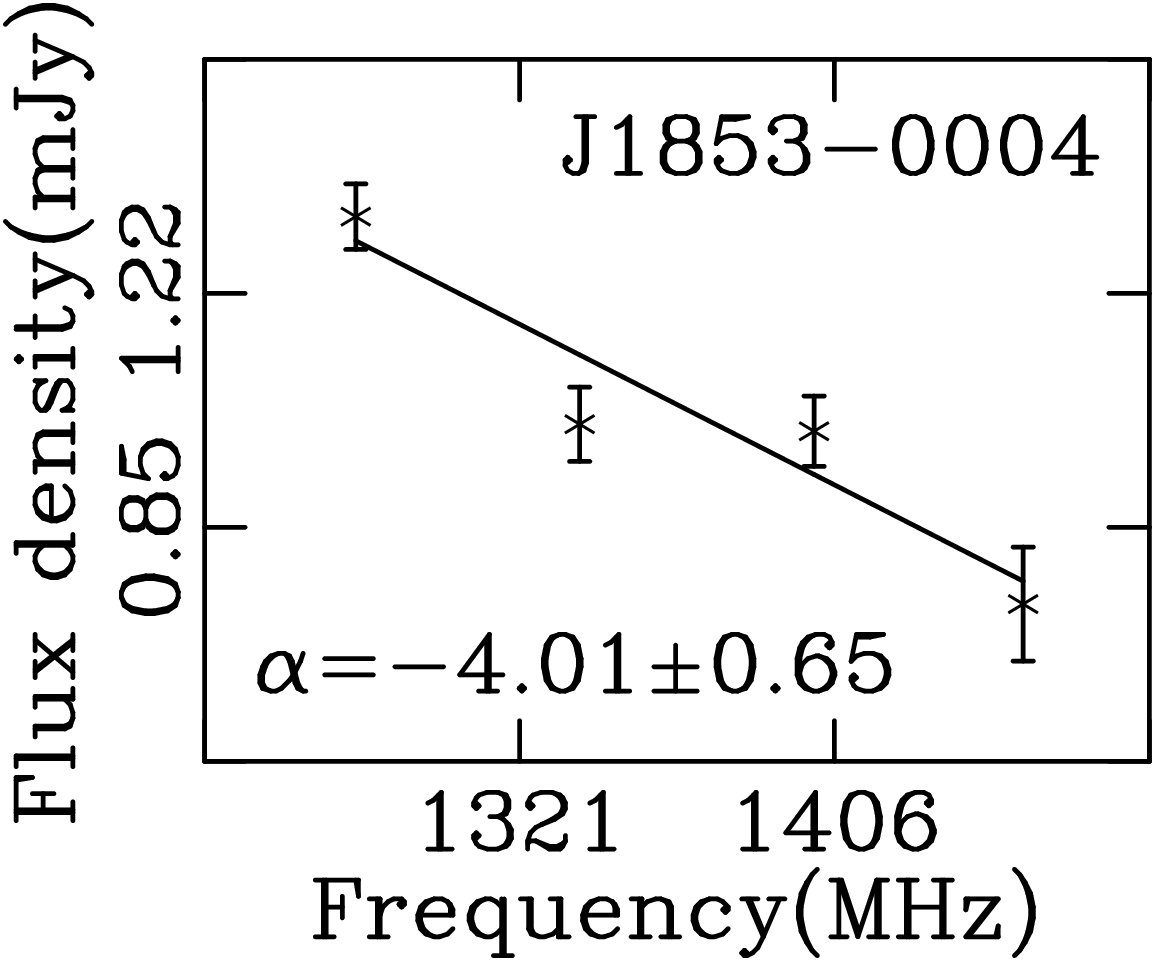}
\\[2mm]                
\includegraphics[bb = 110 72 570 685,clip,angle=-90,width=0.175\textwidth]{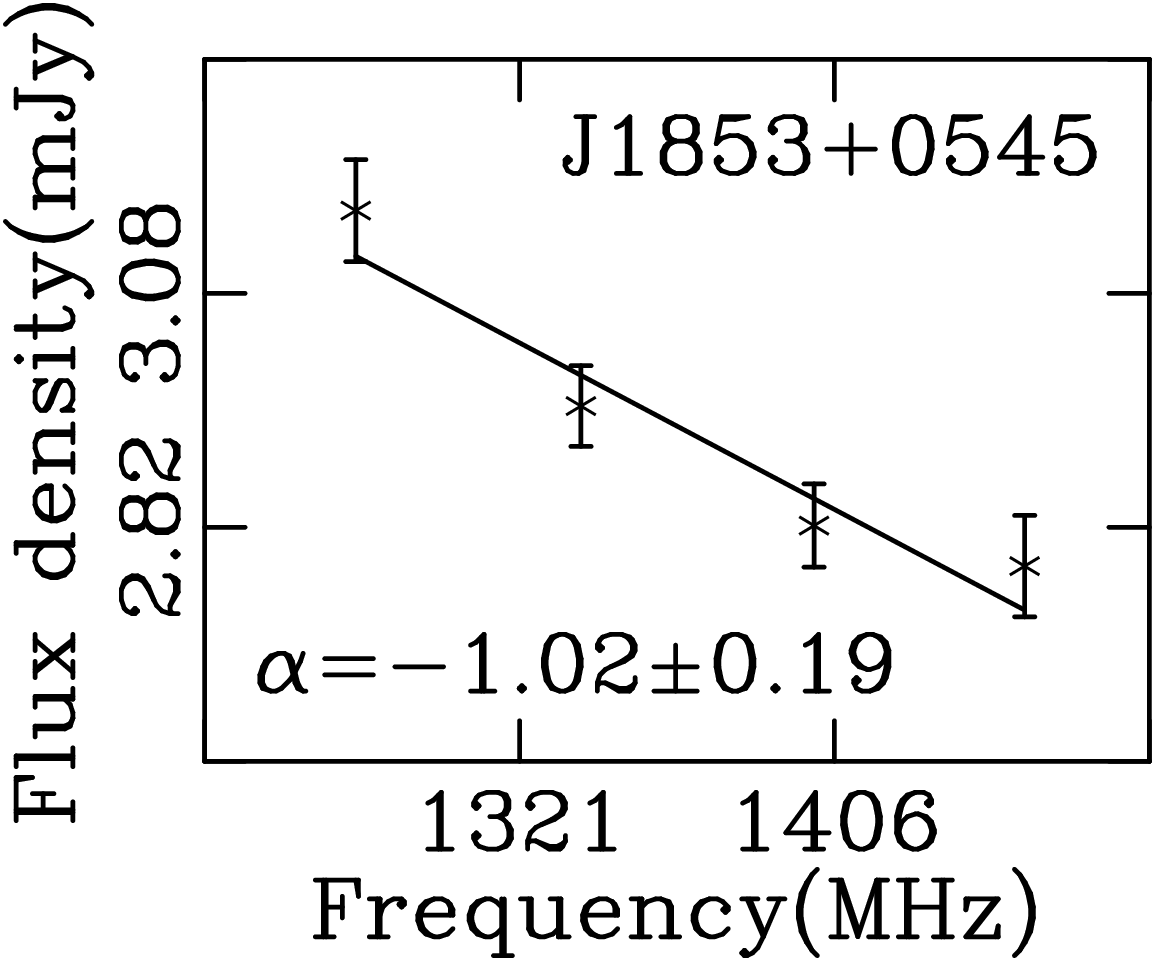}               
\includegraphics[bb = 110 120 570 685,clip,angle=-90,width=0.16\textwidth]{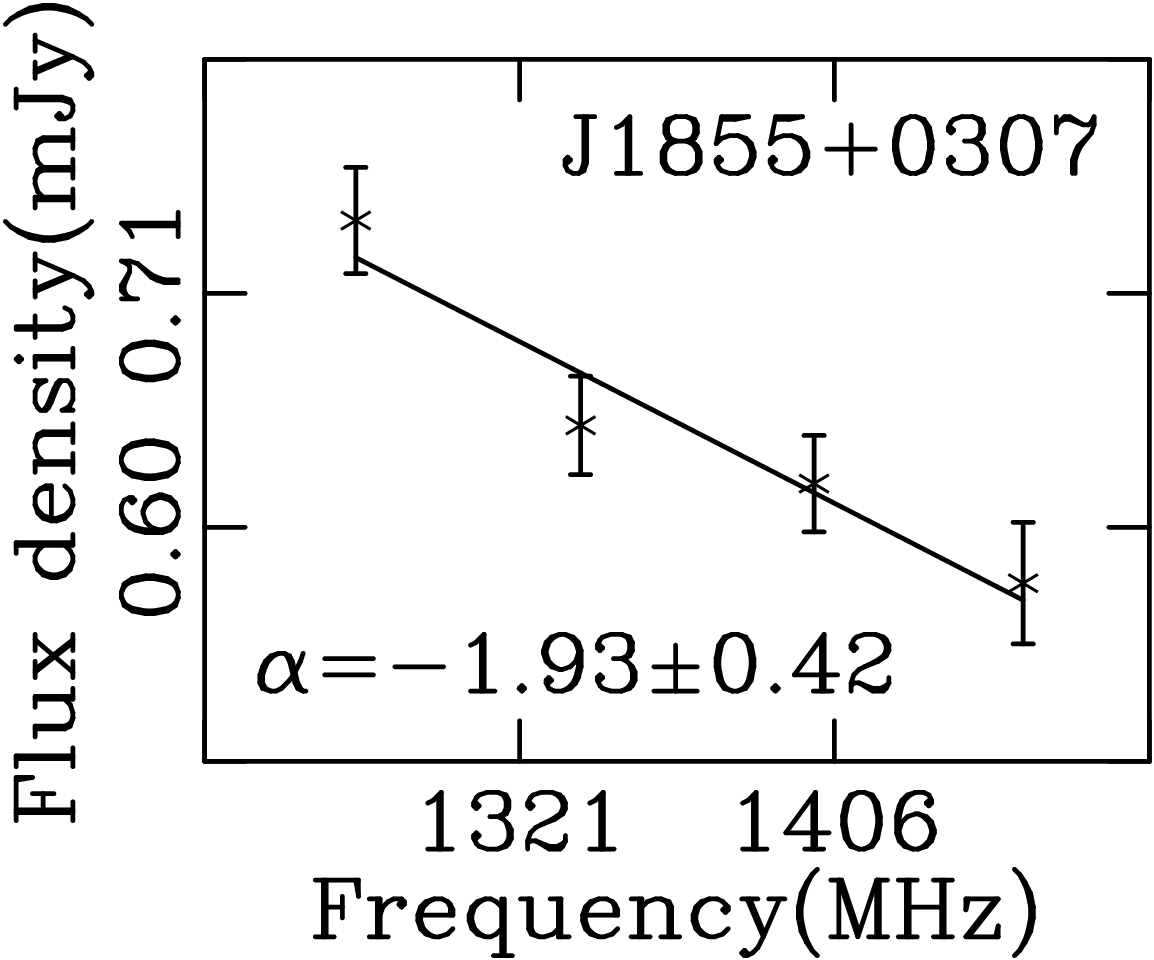}               
\includegraphics[bb = 110 120 570 685,clip,angle=-90,width=0.16\textwidth]{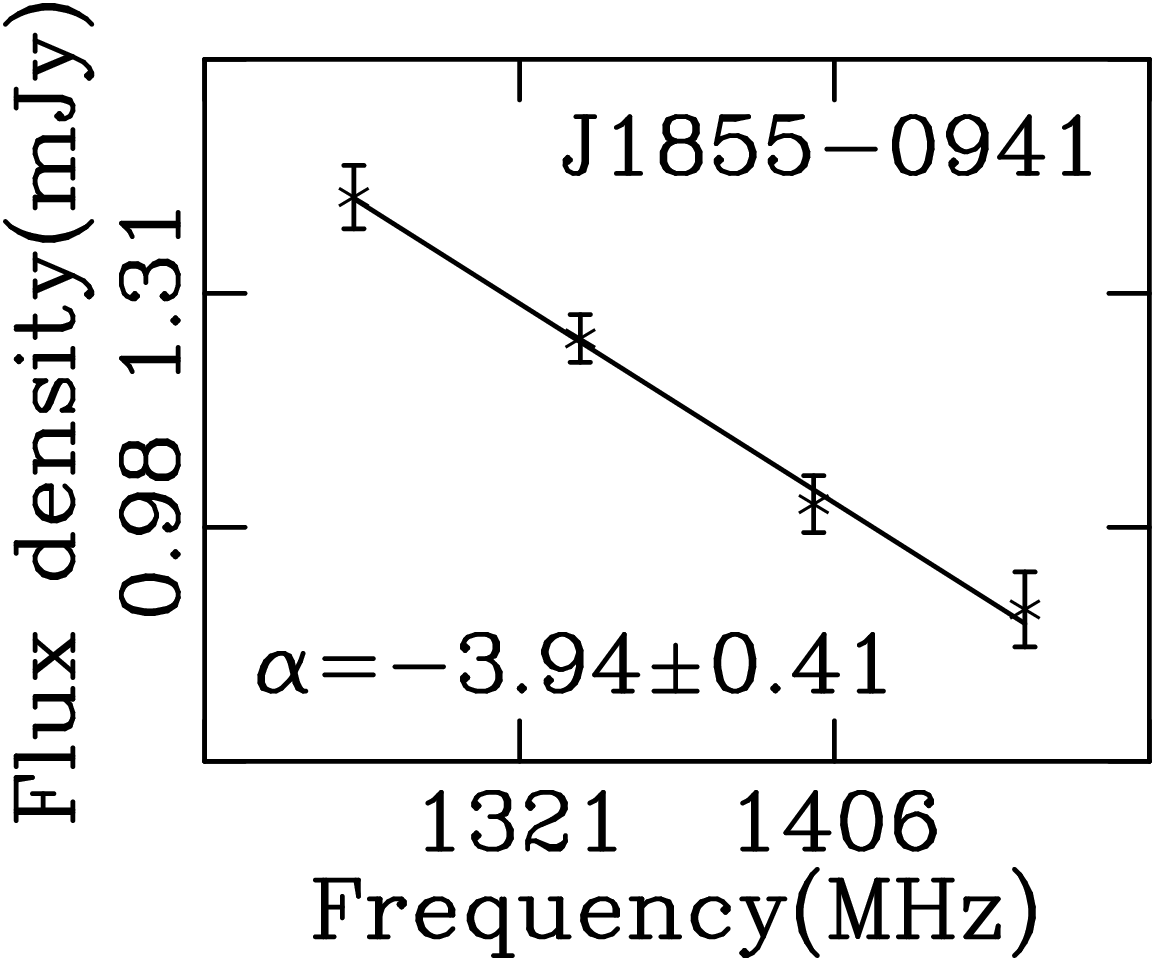}               
\includegraphics[bb = 110 120 570 685,clip,angle=-90,width=0.16\textwidth]{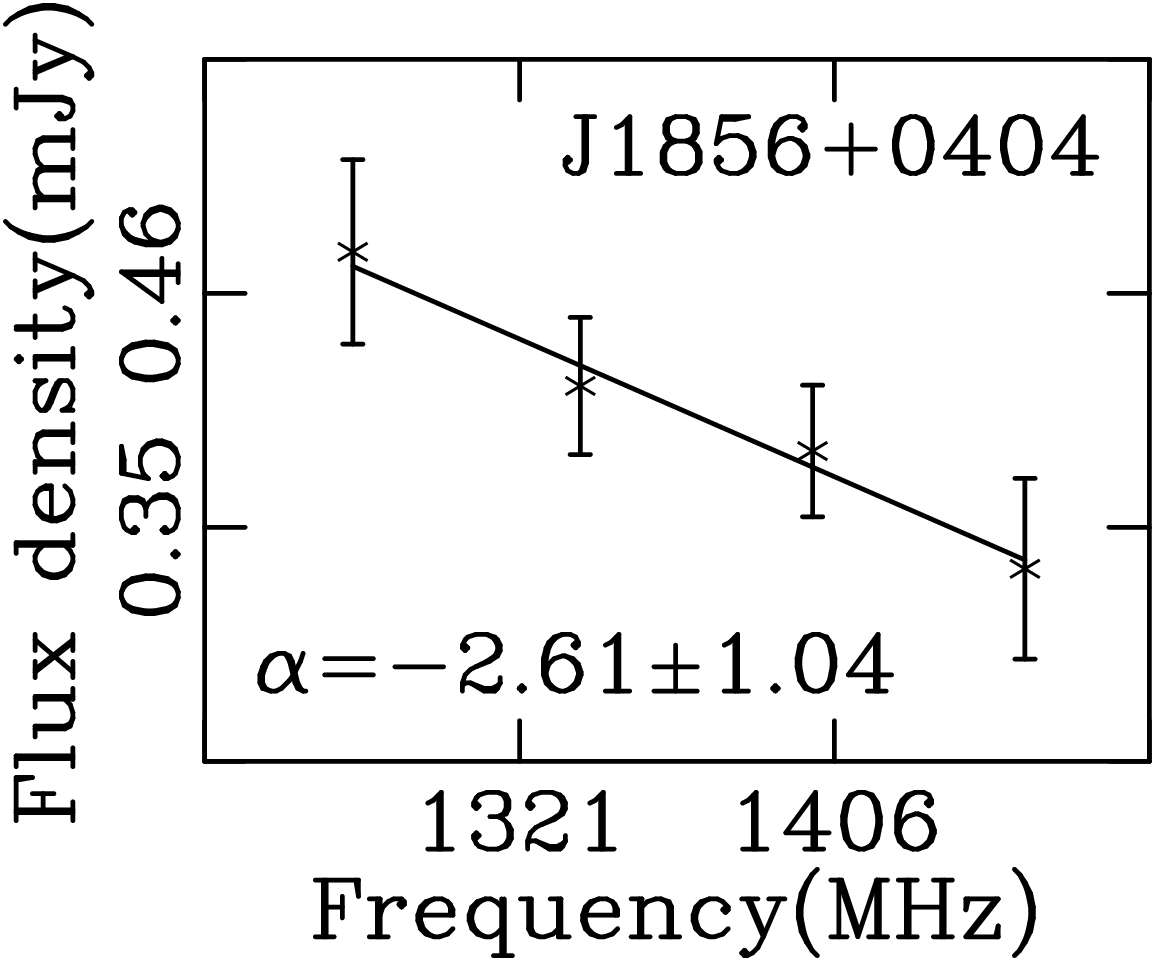}               
\includegraphics[bb = 110 120 570 685,clip,angle=-90,width=0.16\textwidth]{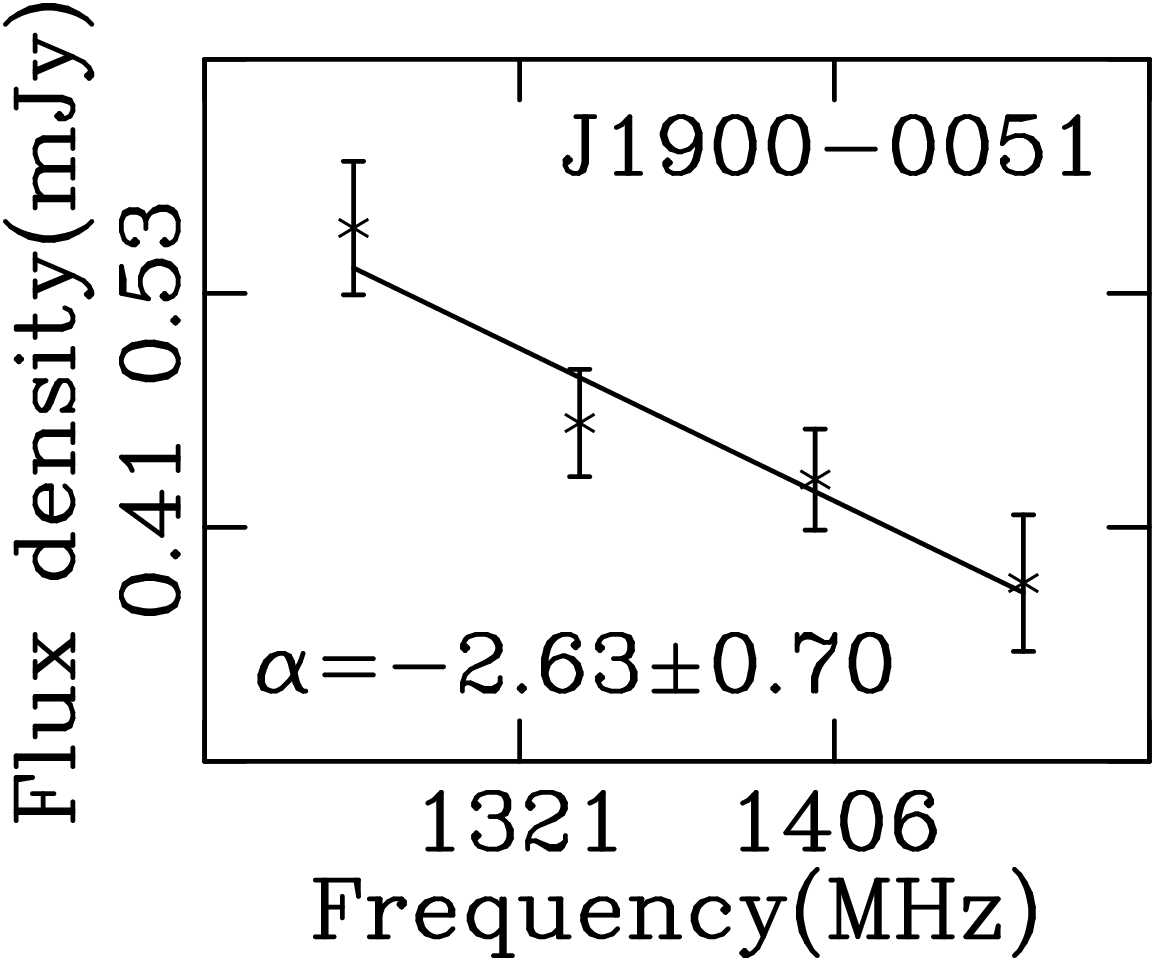}               
\includegraphics[bb = 110 120 570 685,clip,angle=-90,width=0.16\textwidth]{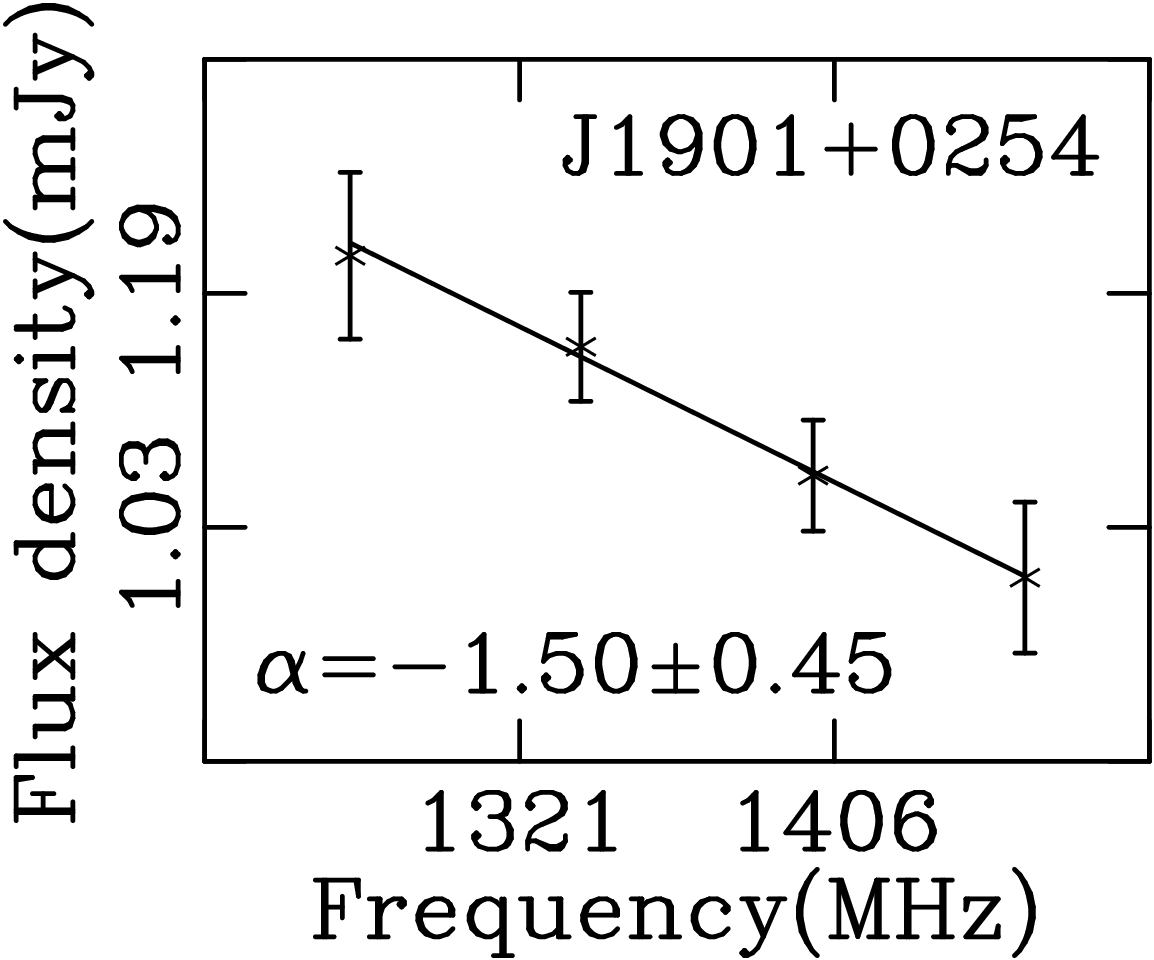}
\center
\addtocounter{figure}{-1}
\caption{--- continued.}
\end{figure}
\begin{figure}
  \center             
\includegraphics[bb = 110 72 518 685,clip,angle=-90,width=0.175\textwidth]{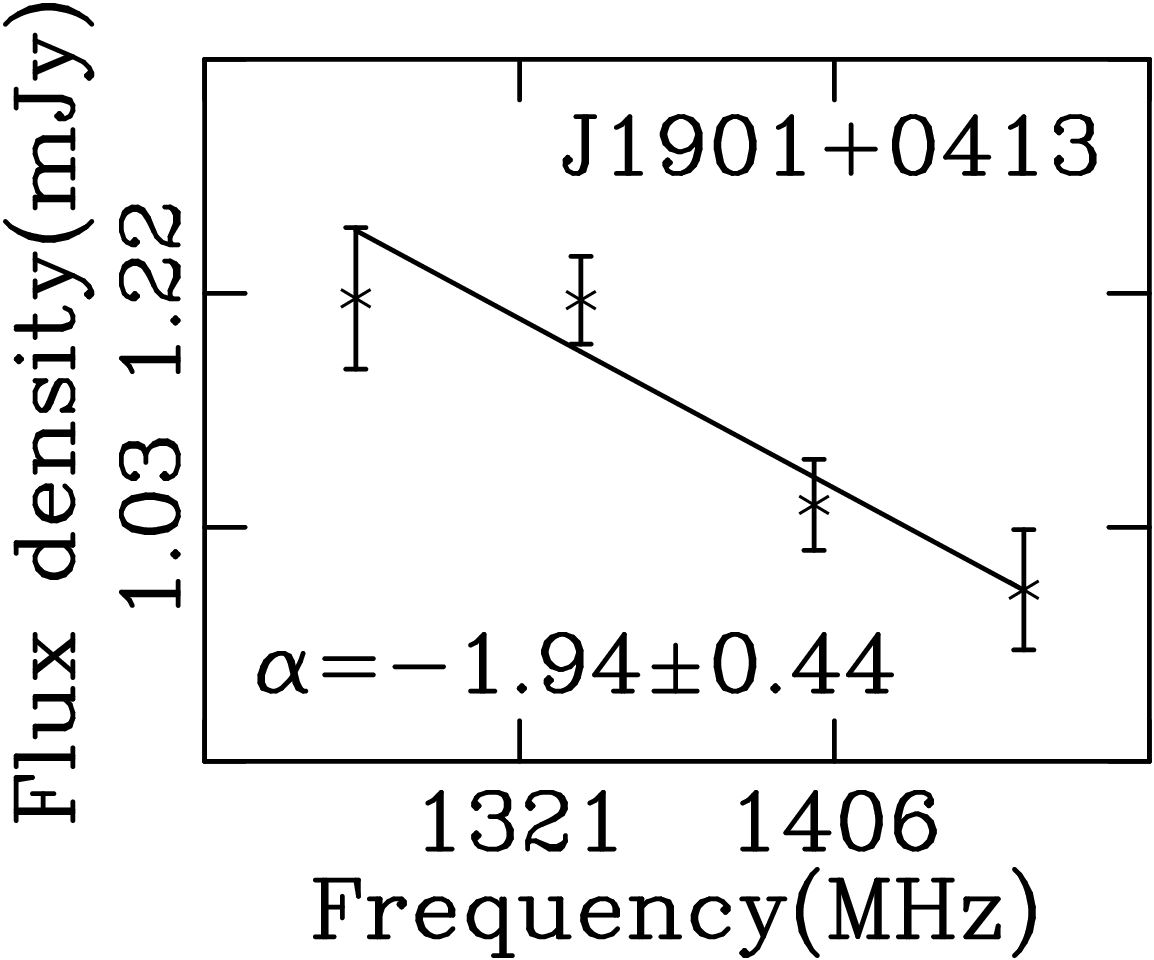}               
\includegraphics[bb = 110 120 518 685,clip,angle=-90,width=0.16\textwidth]{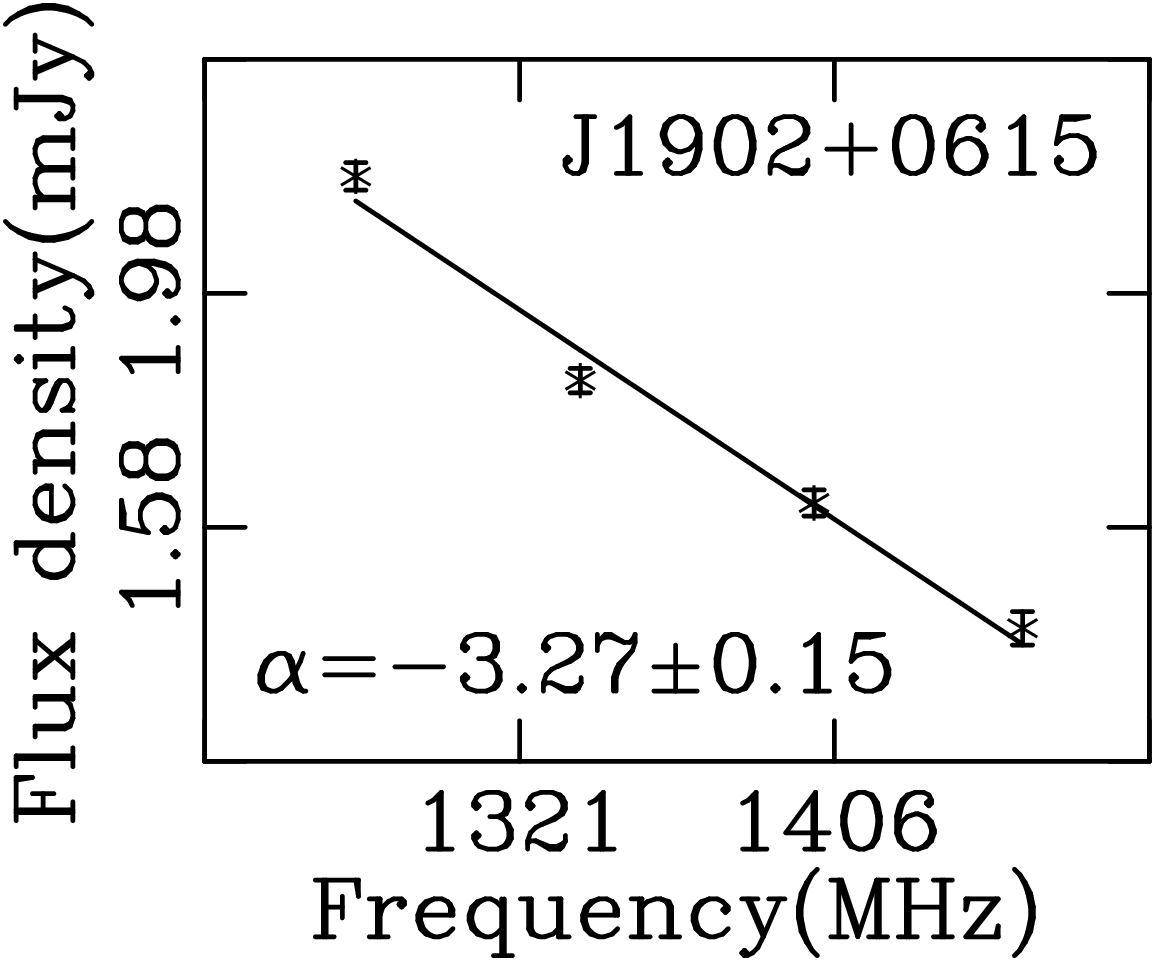}               
\includegraphics[bb = 110 120 518 685,clip,angle=-90,width=0.16\textwidth]{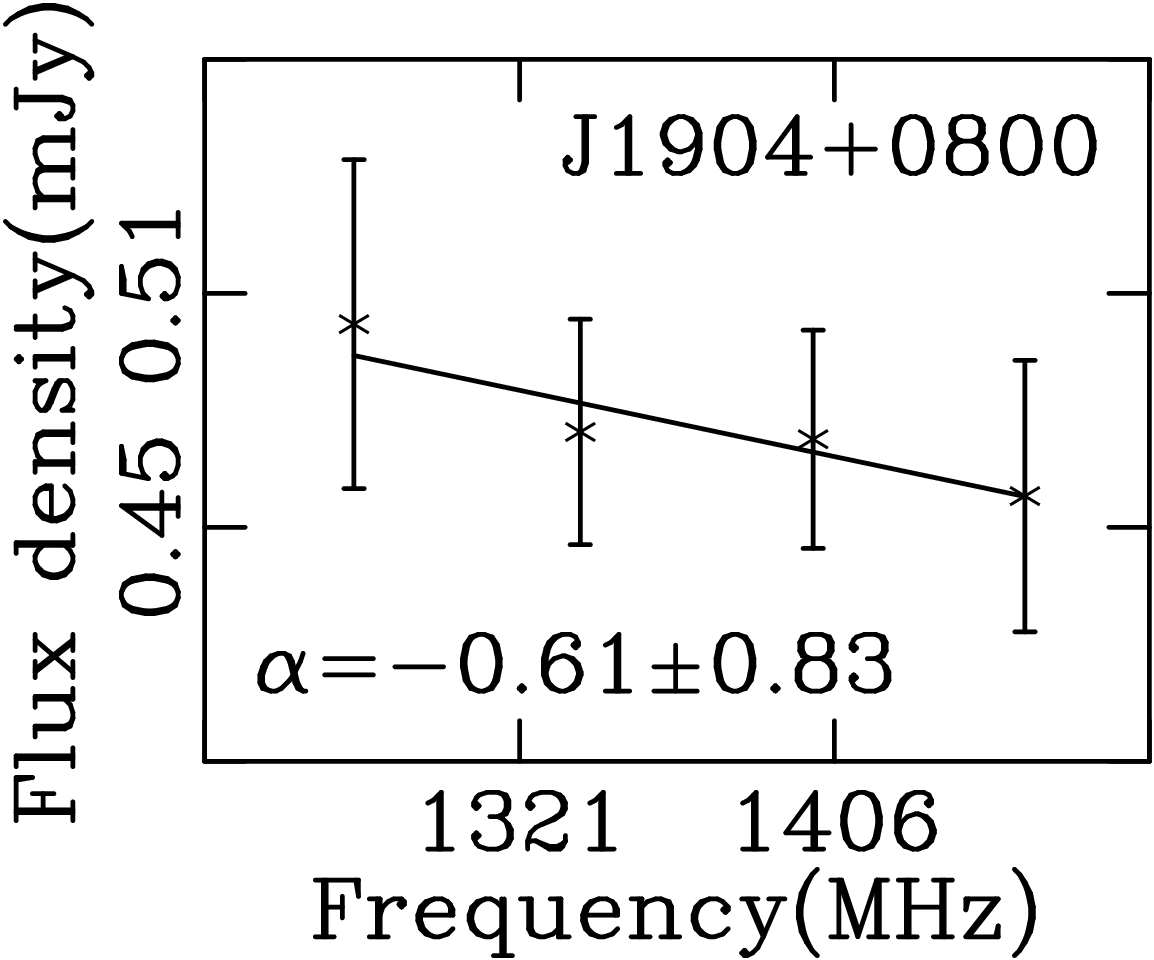}               
\includegraphics[bb = 110 120 518 685,clip,angle=-90,width=0.16\textwidth]{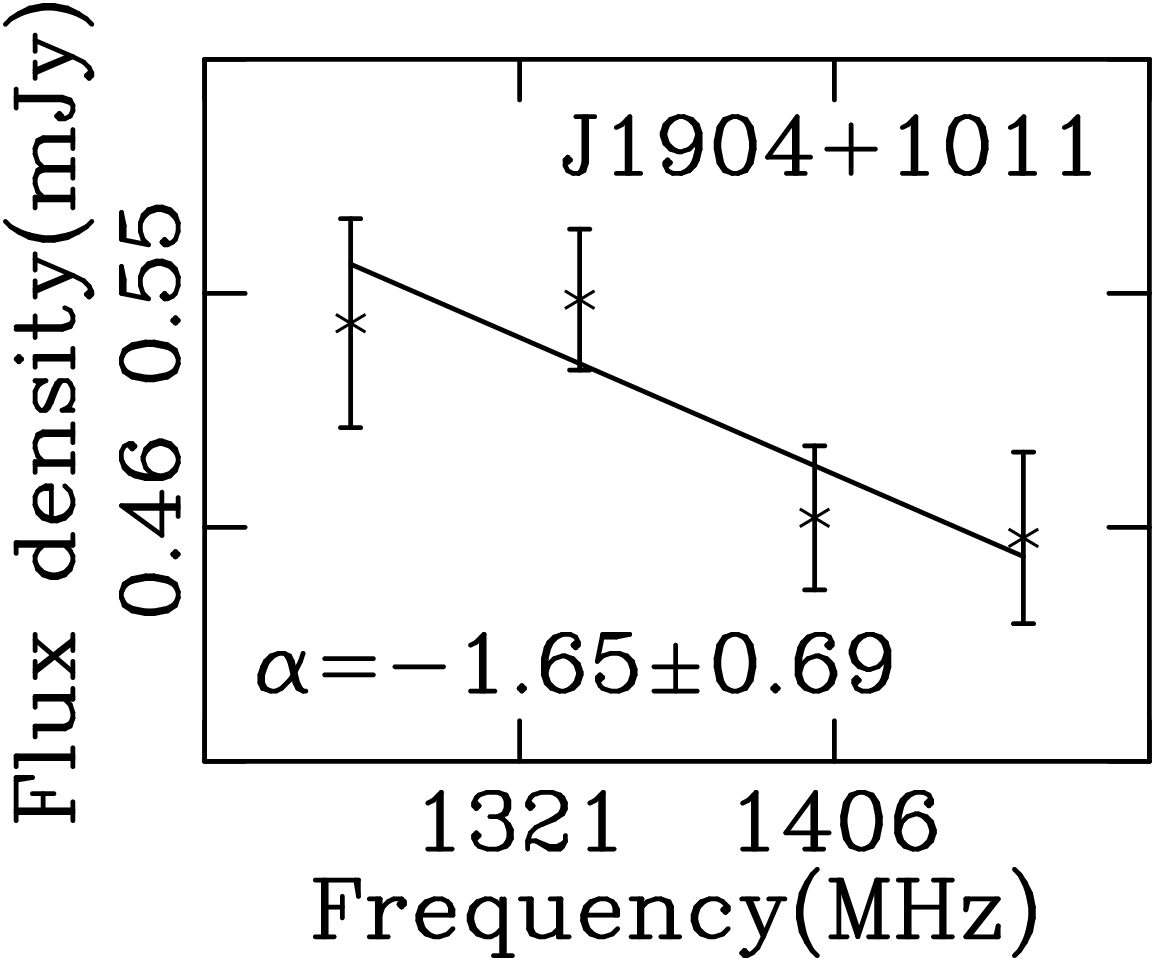}               
\includegraphics[bb = 110 120 518 685,clip,angle=-90,width=0.16\textwidth]{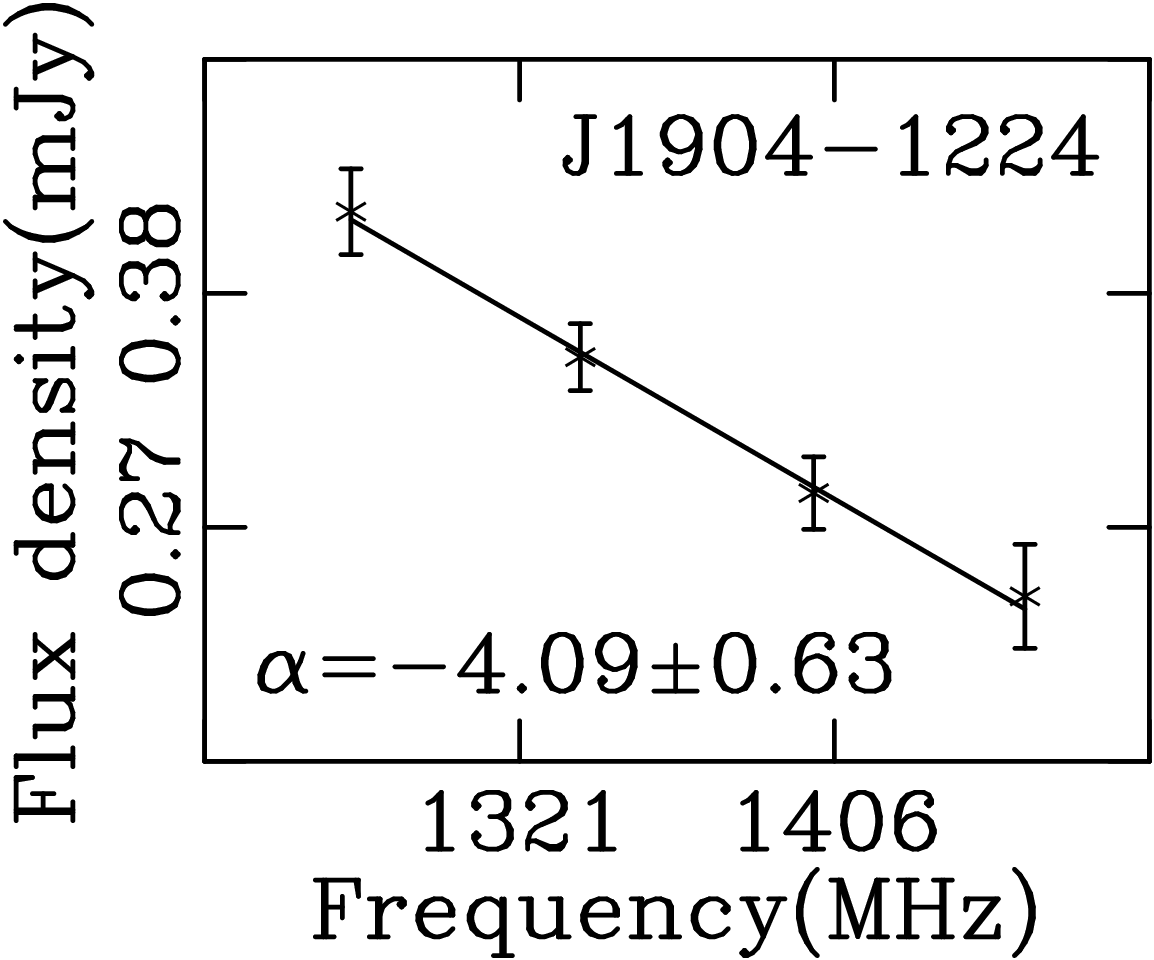}               
\includegraphics[bb = 110 120 518 685,clip,angle=-90,width=0.16\textwidth]{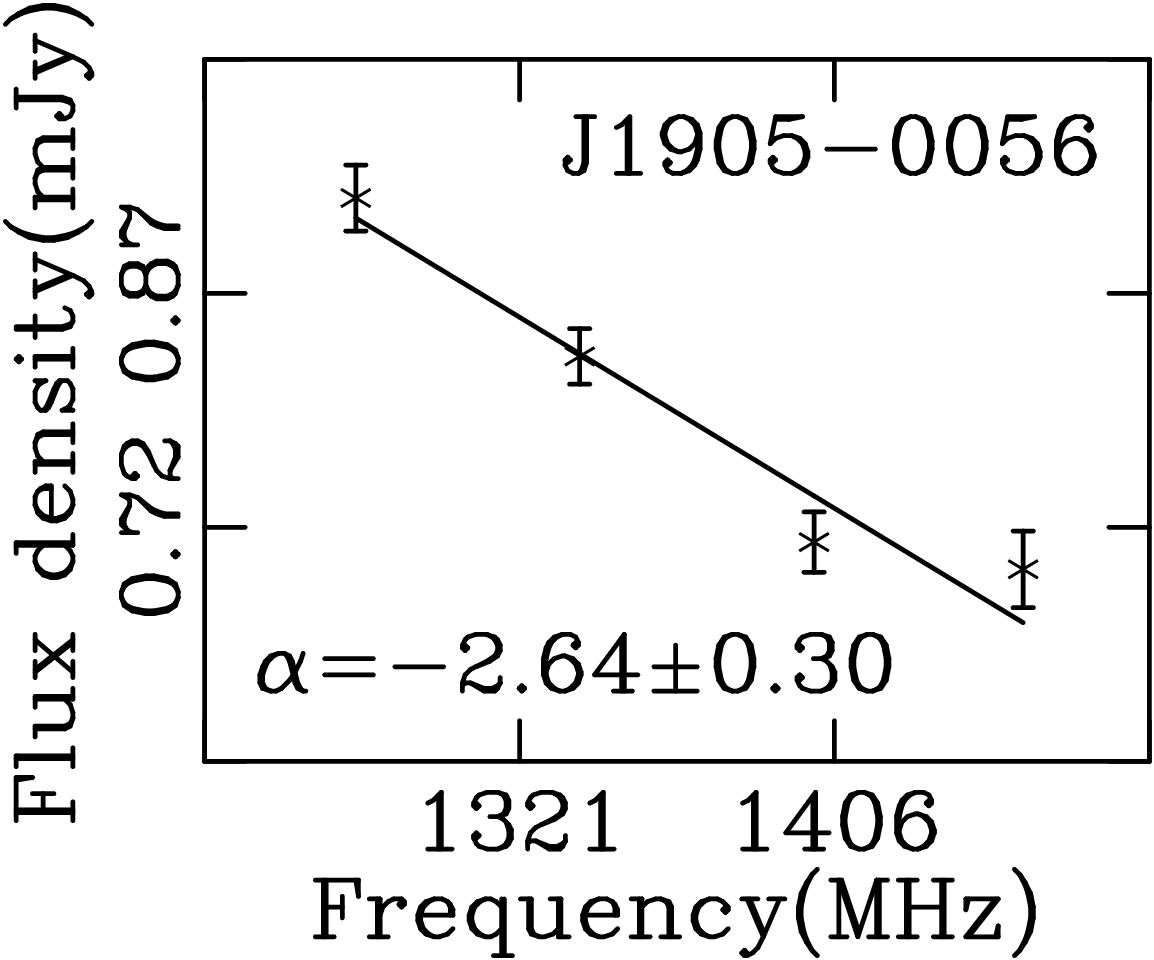}
\\[2mm]                        
\includegraphics[bb = 110 72 518 685,clip,angle=-90,width=0.175\textwidth]{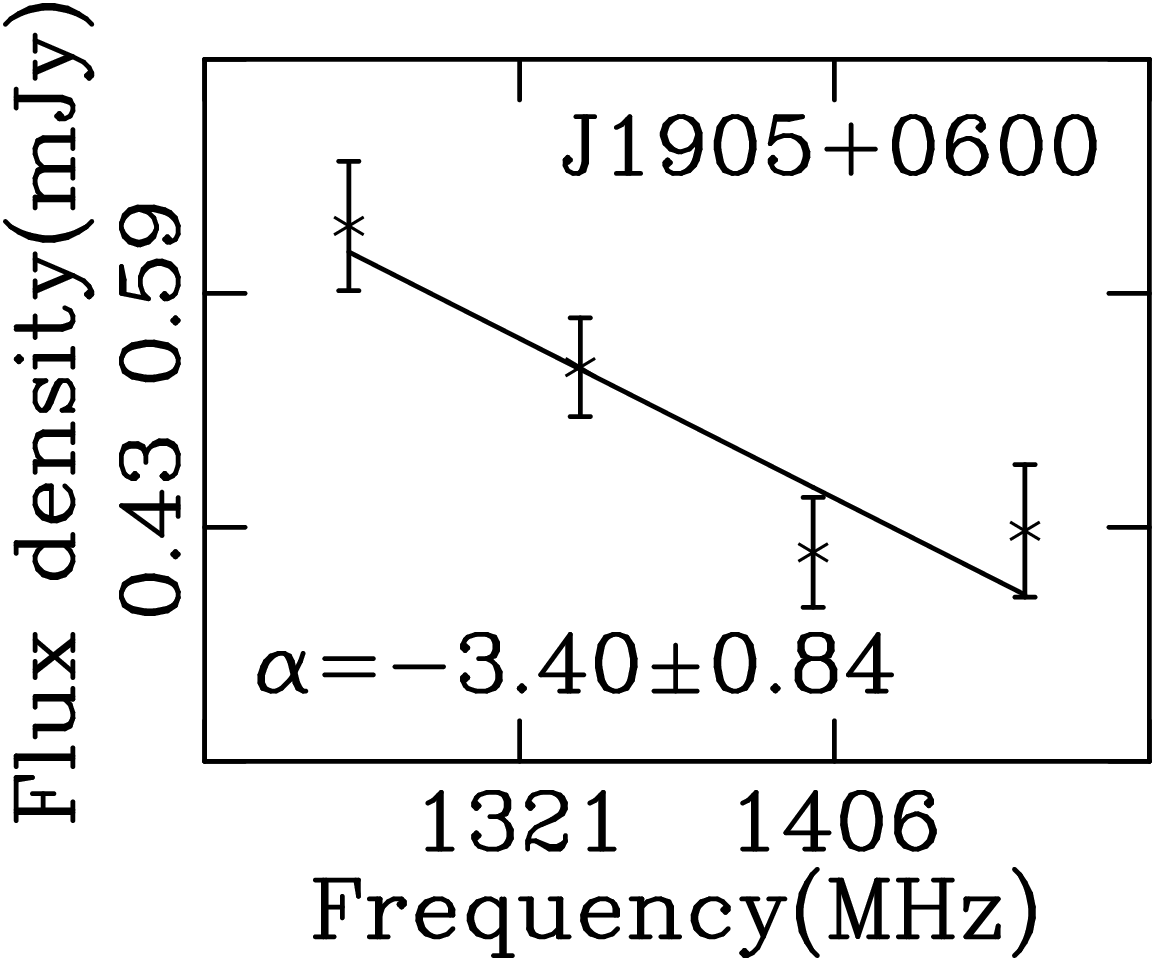}               
\includegraphics[bb = 110 120 518 685,clip,angle=-90,width=0.16\textwidth]{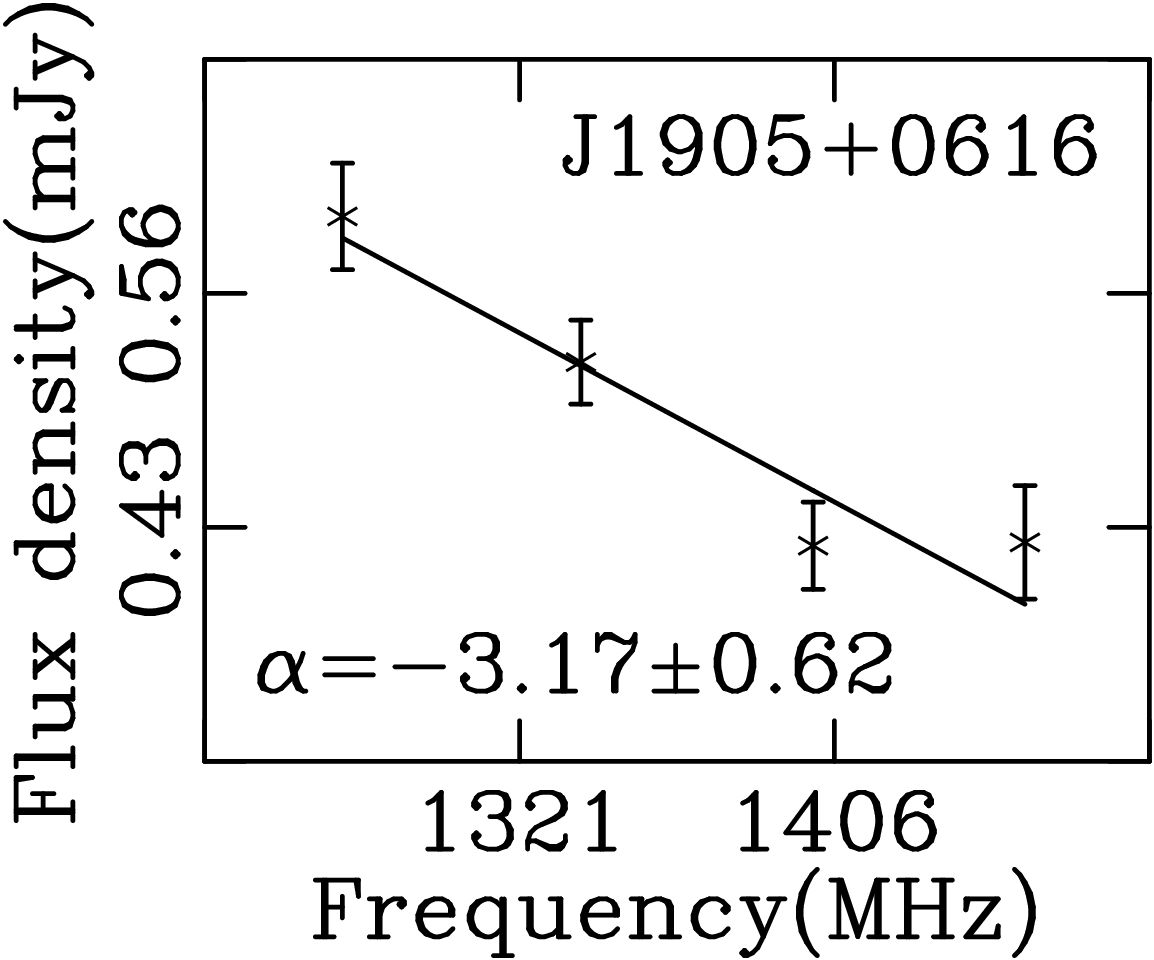}               
\includegraphics[bb = 110 120 518 685,clip,angle=-90,width=0.16\textwidth]{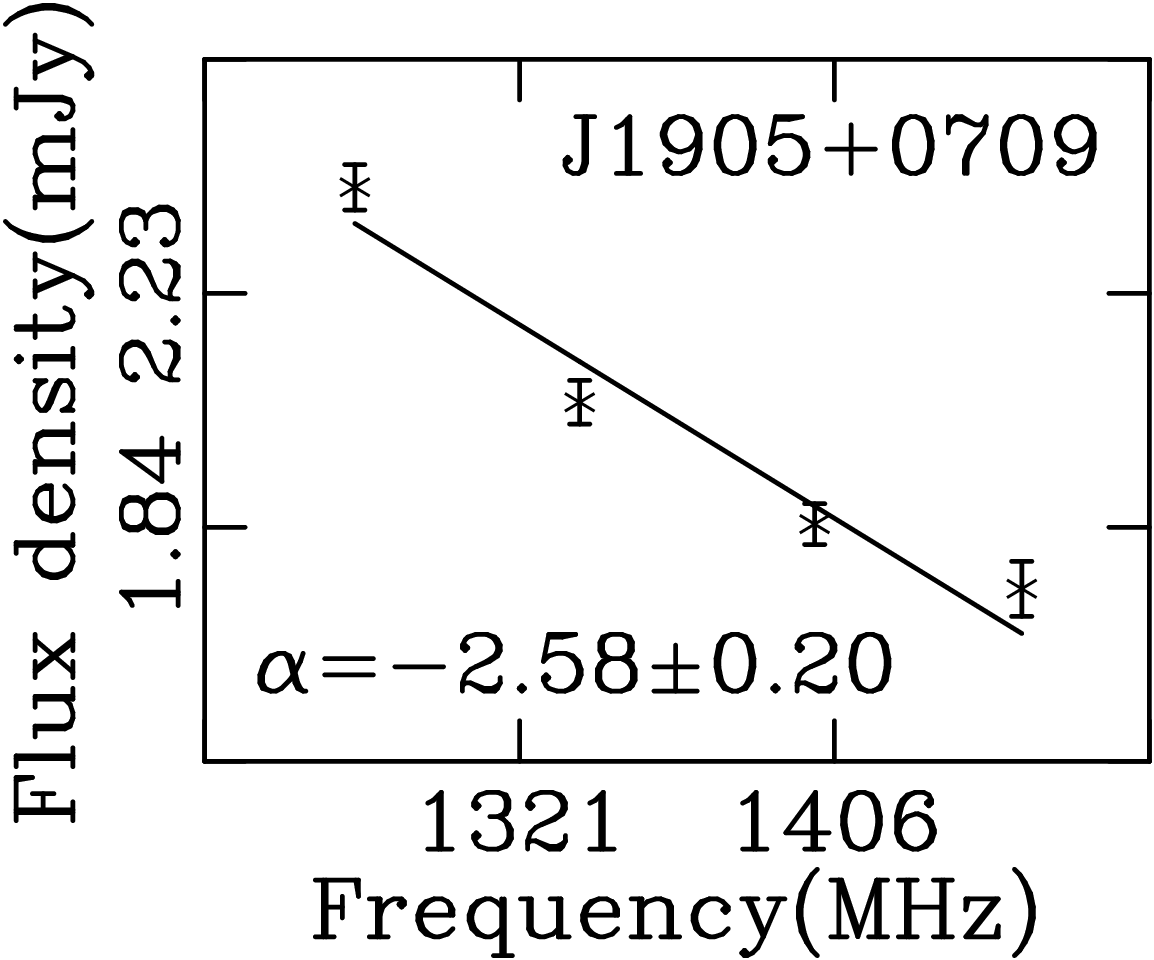}               
\includegraphics[bb = 110 120 518 685,clip,angle=-90,width=0.16\textwidth]{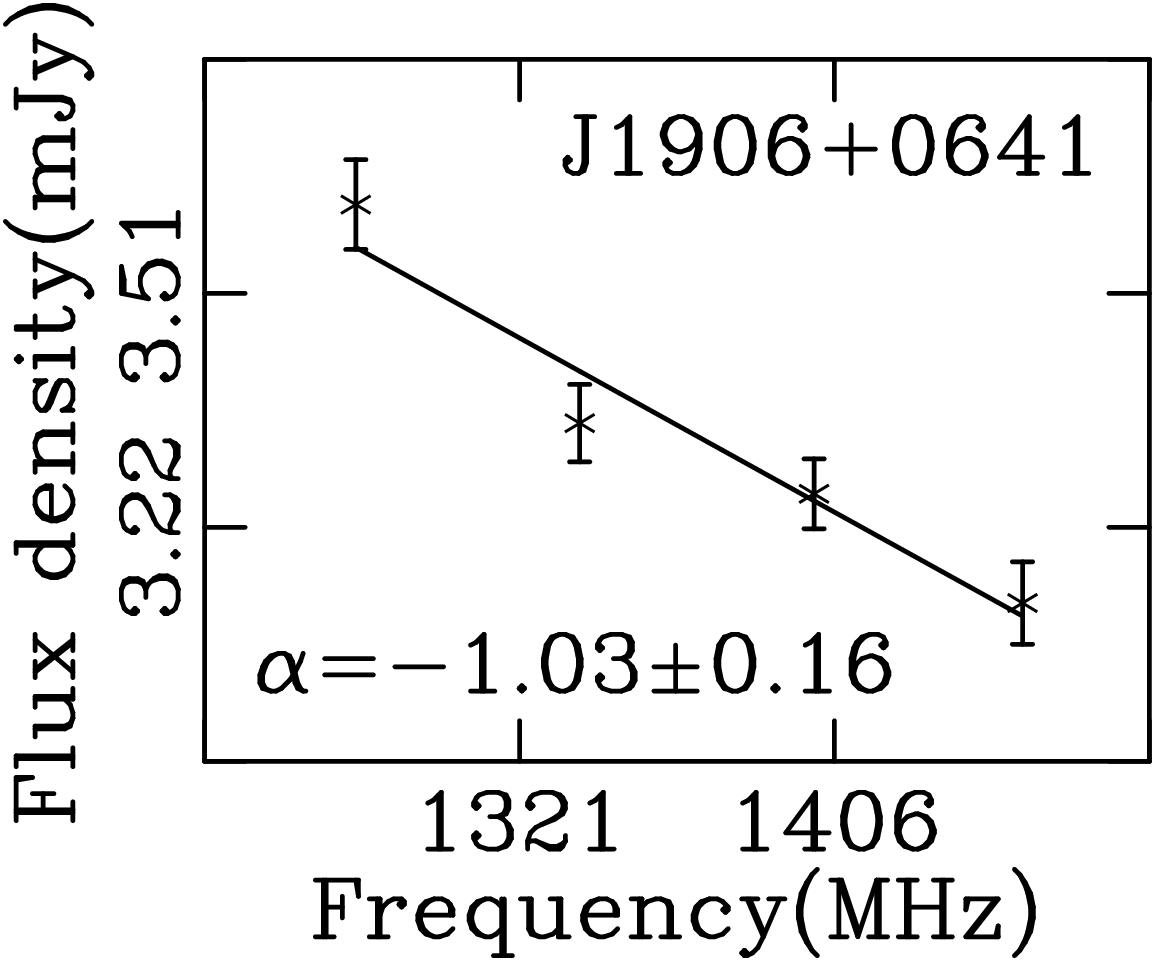}               
\includegraphics[bb = 110 120 518 685,clip,angle=-90,width=0.16\textwidth]{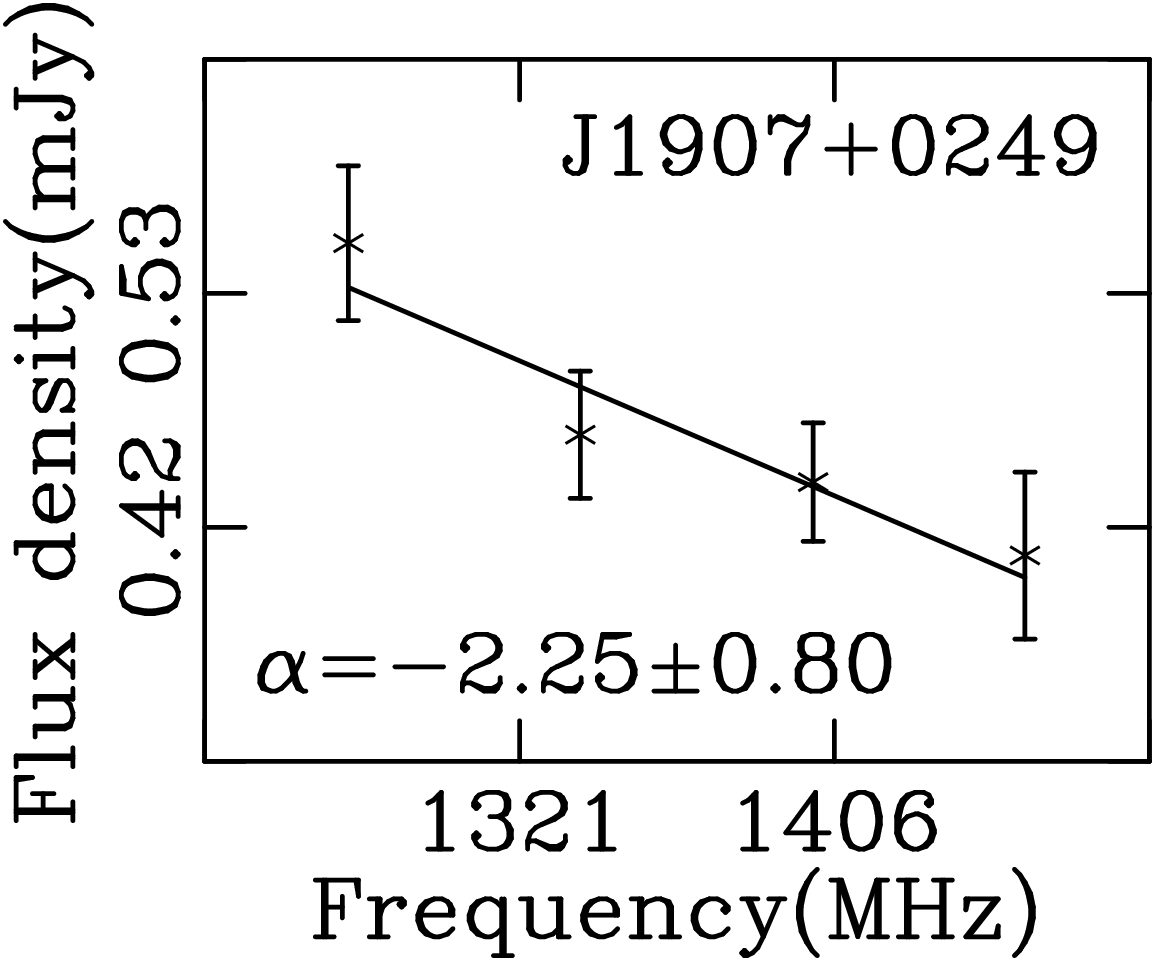}               
\includegraphics[bb = 110 120 518 685,clip,angle=-90,width=0.16\textwidth]{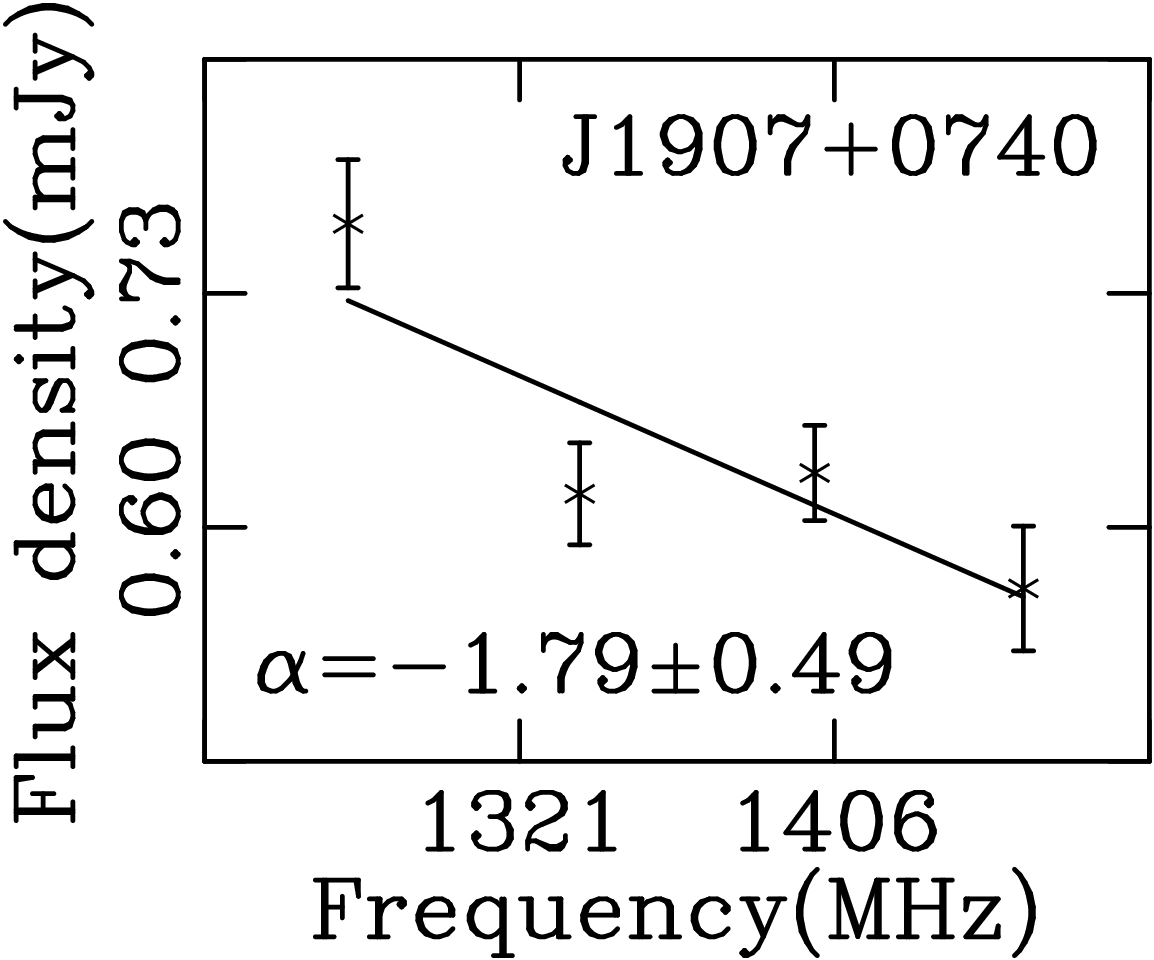}
\\[2mm]               
\includegraphics[bb = 110 72 518 685,clip,angle=-90,width=0.175\textwidth]{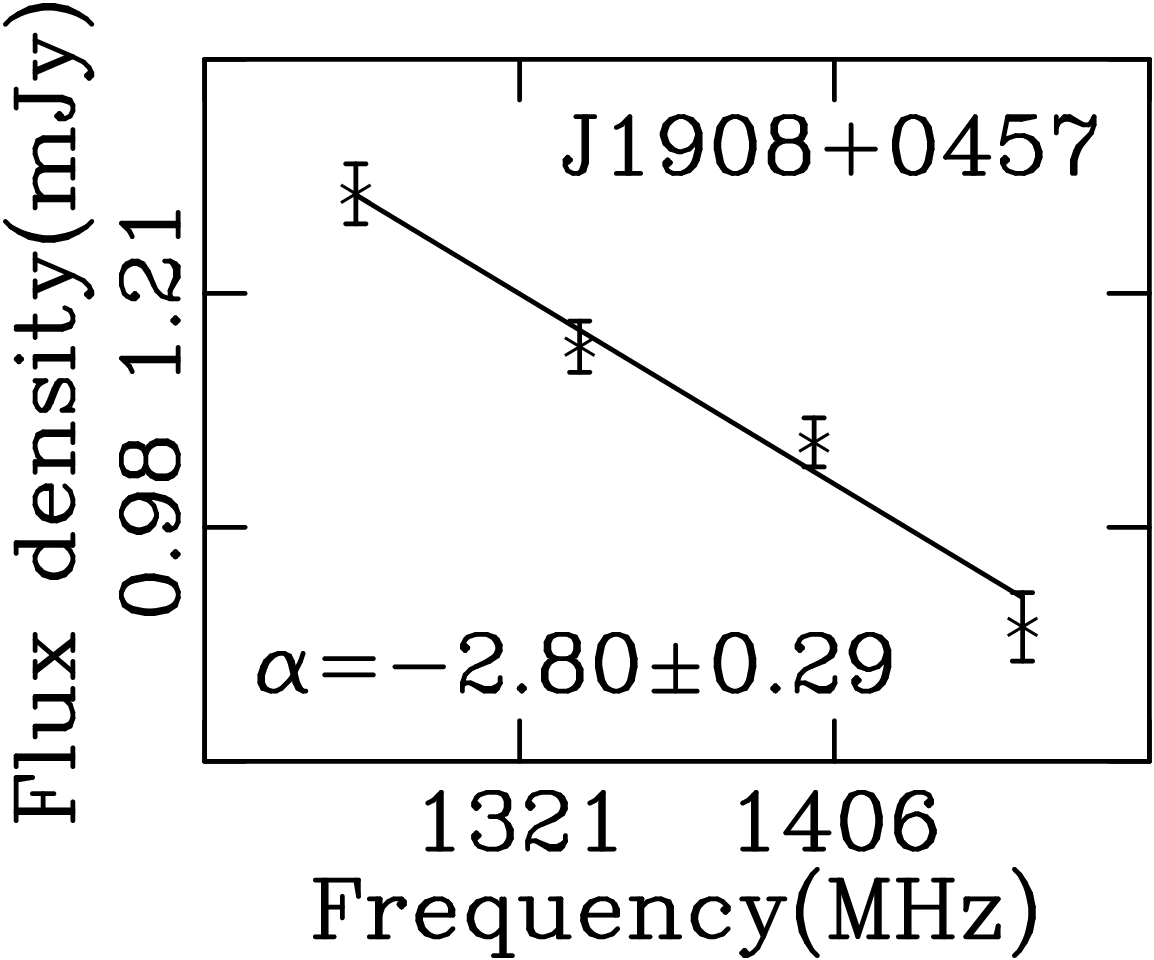}               
\includegraphics[bb = 110 120 518 685,clip,angle=-90,width=0.16\textwidth]{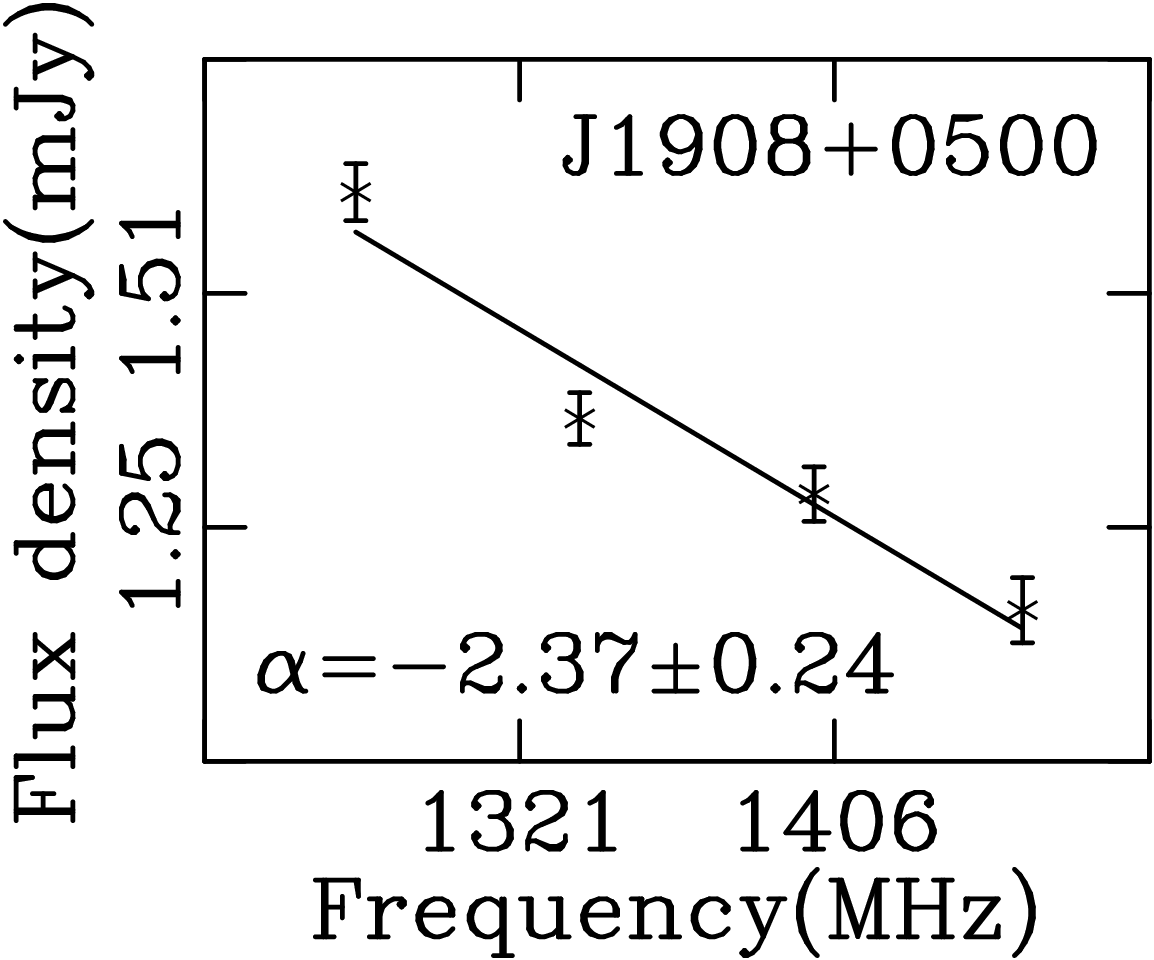}               
\includegraphics[bb = 110 120 518 685,clip,angle=-90,width=0.16\textwidth]{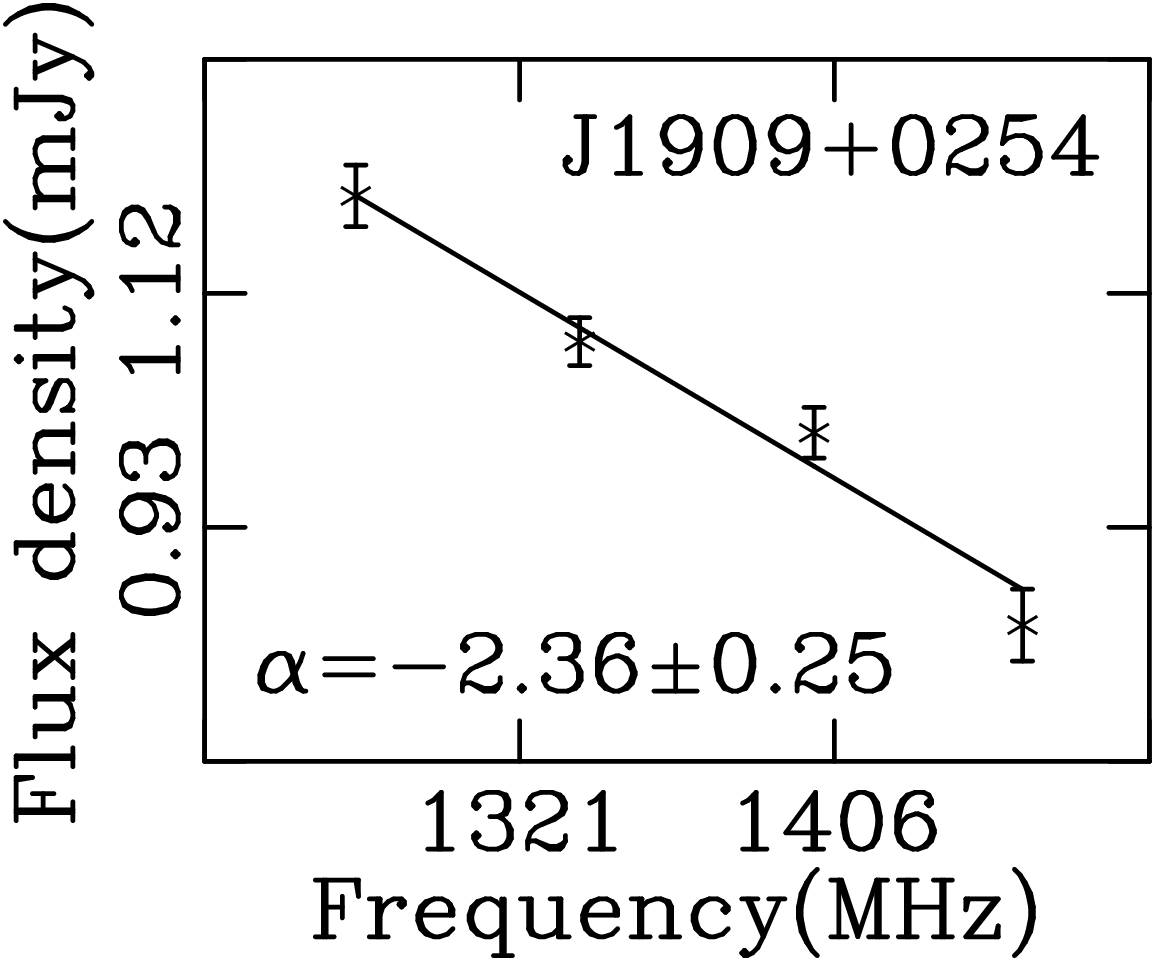}               
\includegraphics[bb = 110 120 518 685,clip,angle=-90,width=0.16\textwidth]{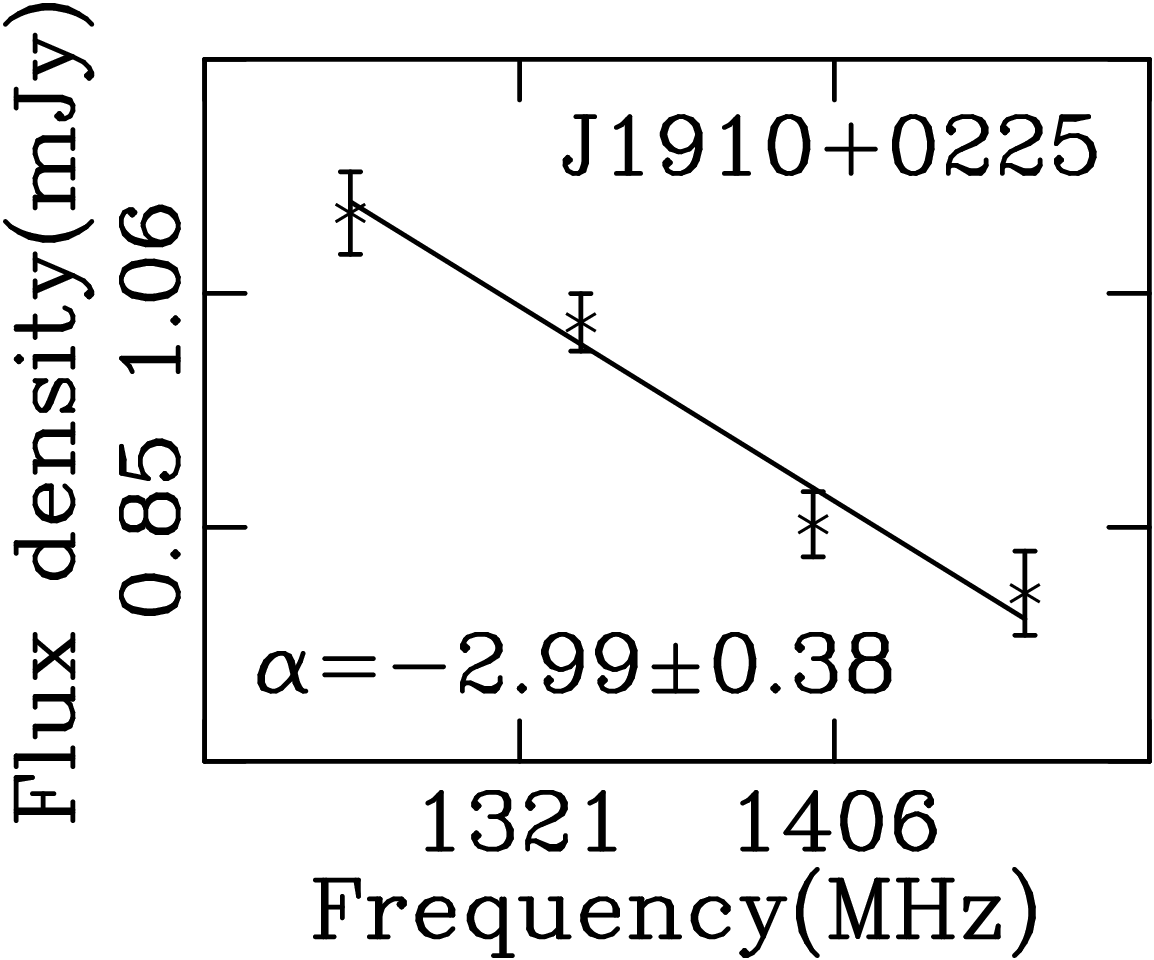}               
\includegraphics[bb = 110 120 518 685,clip,angle=-90,width=0.16\textwidth]{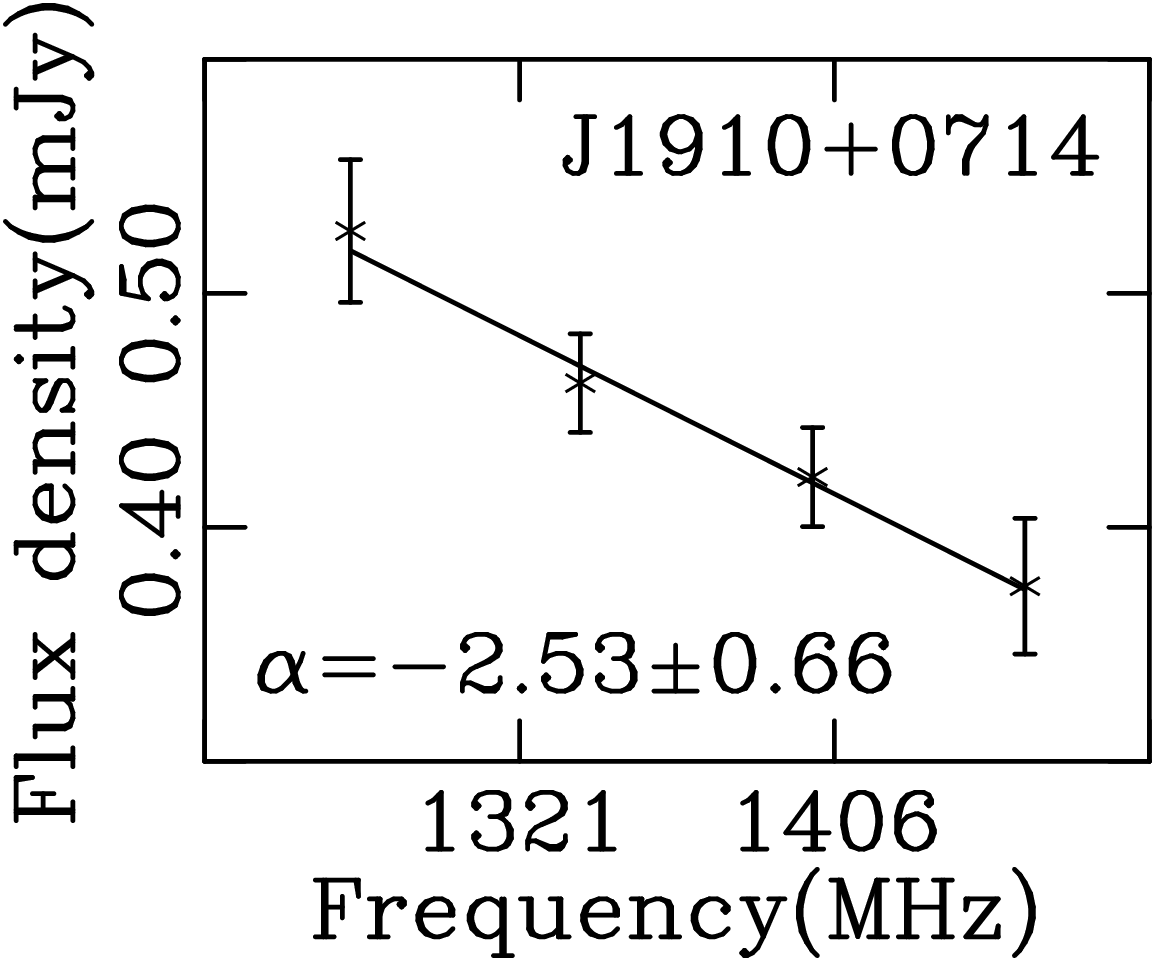}               
\includegraphics[bb = 110 120 518 685,clip,angle=-90,width=0.16\textwidth]{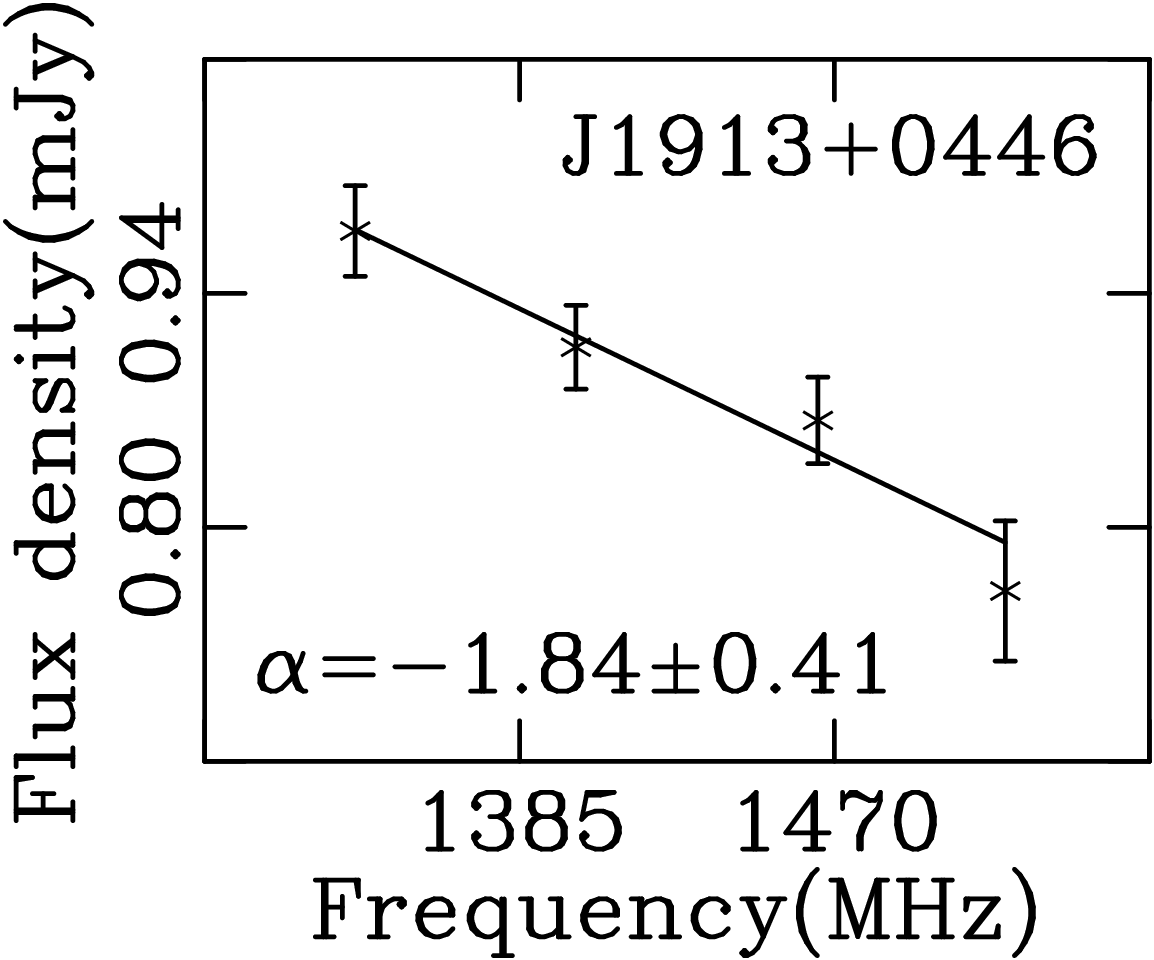}
\\[2mm]               
\includegraphics[bb = 110 72 518 685,clip,angle=-90,width=0.175\textwidth]{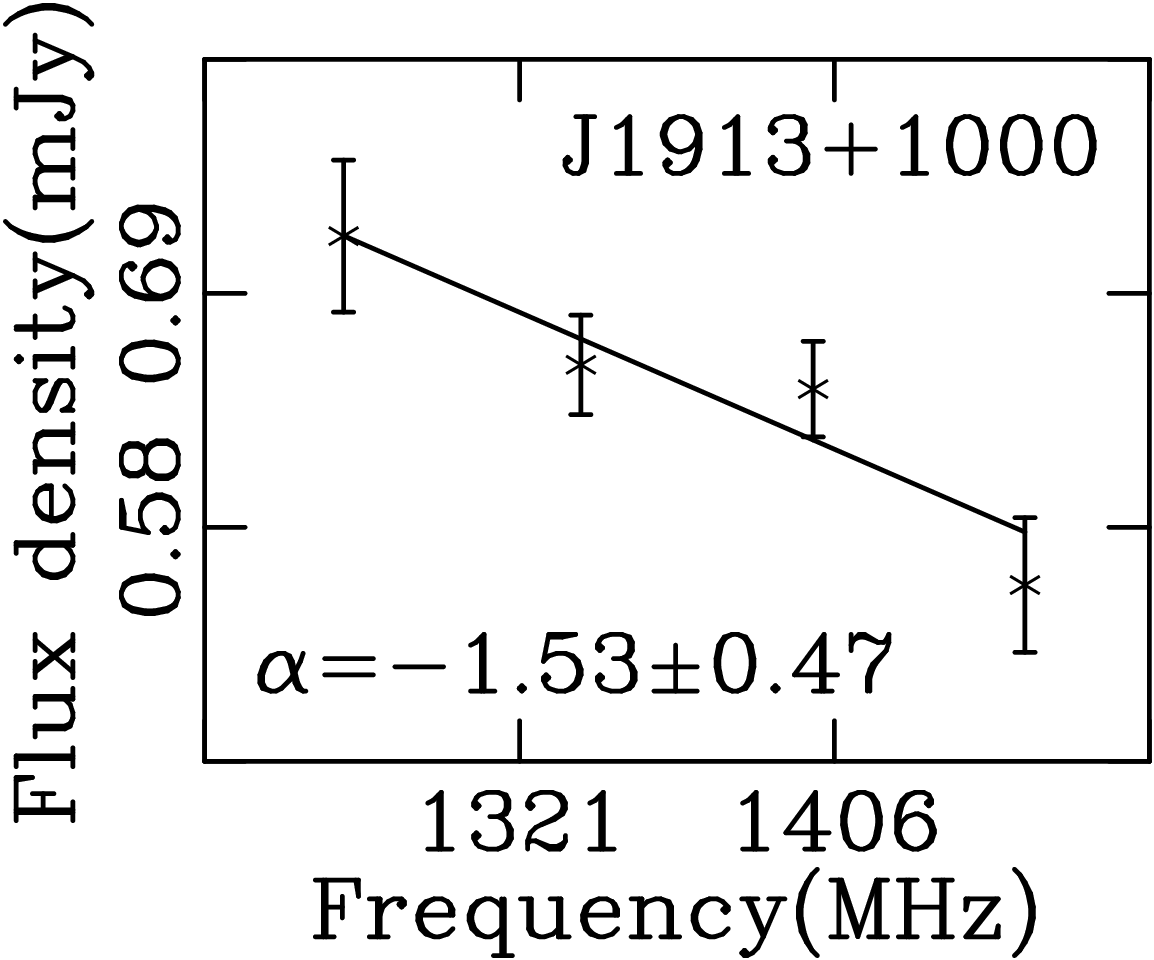}               
\includegraphics[bb = 110 120 518 685,clip,angle=-90,width=0.16\textwidth]{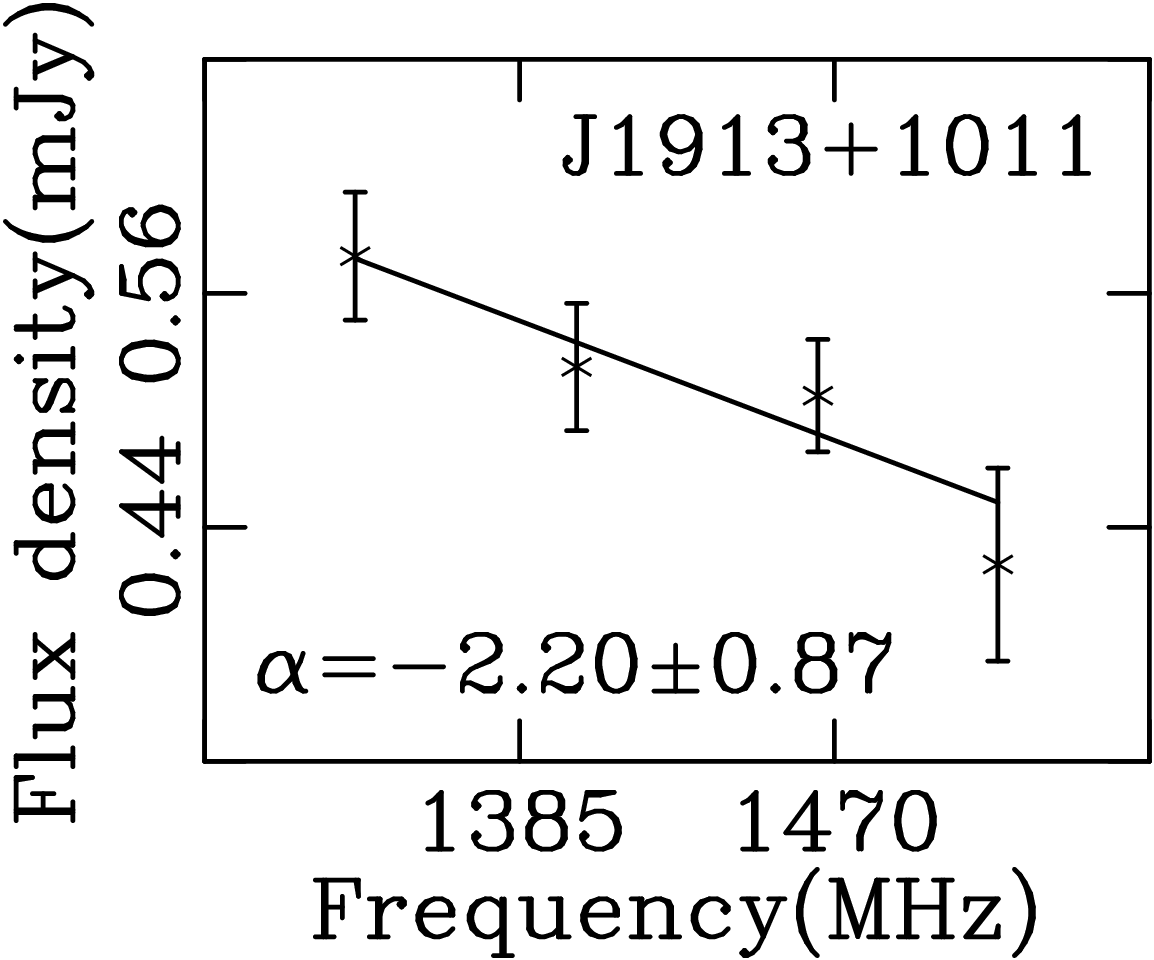}               
\includegraphics[bb = 110 120 518 685,clip,angle=-90,width=0.16\textwidth]{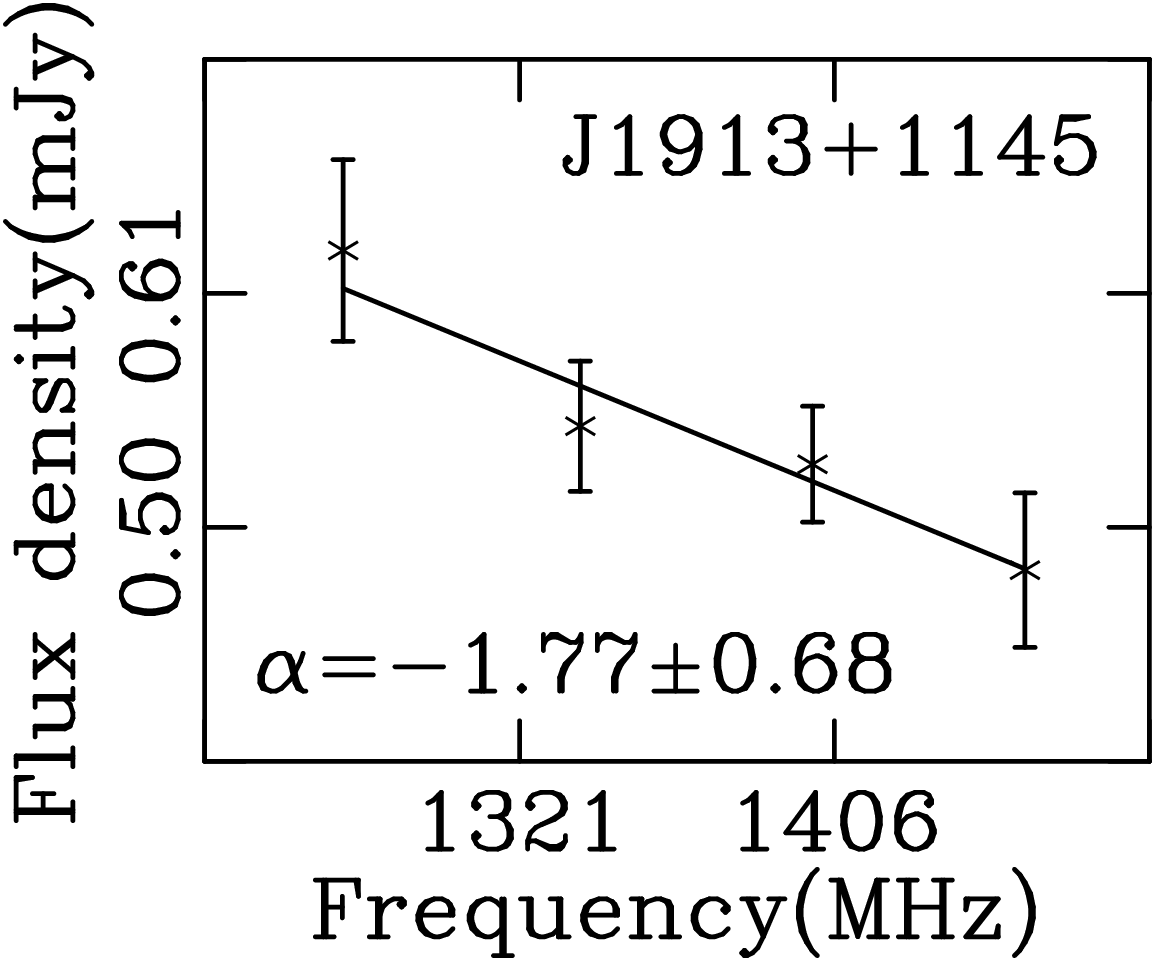}               
\includegraphics[bb = 110 120 518 685,clip,angle=-90,width=0.16\textwidth]{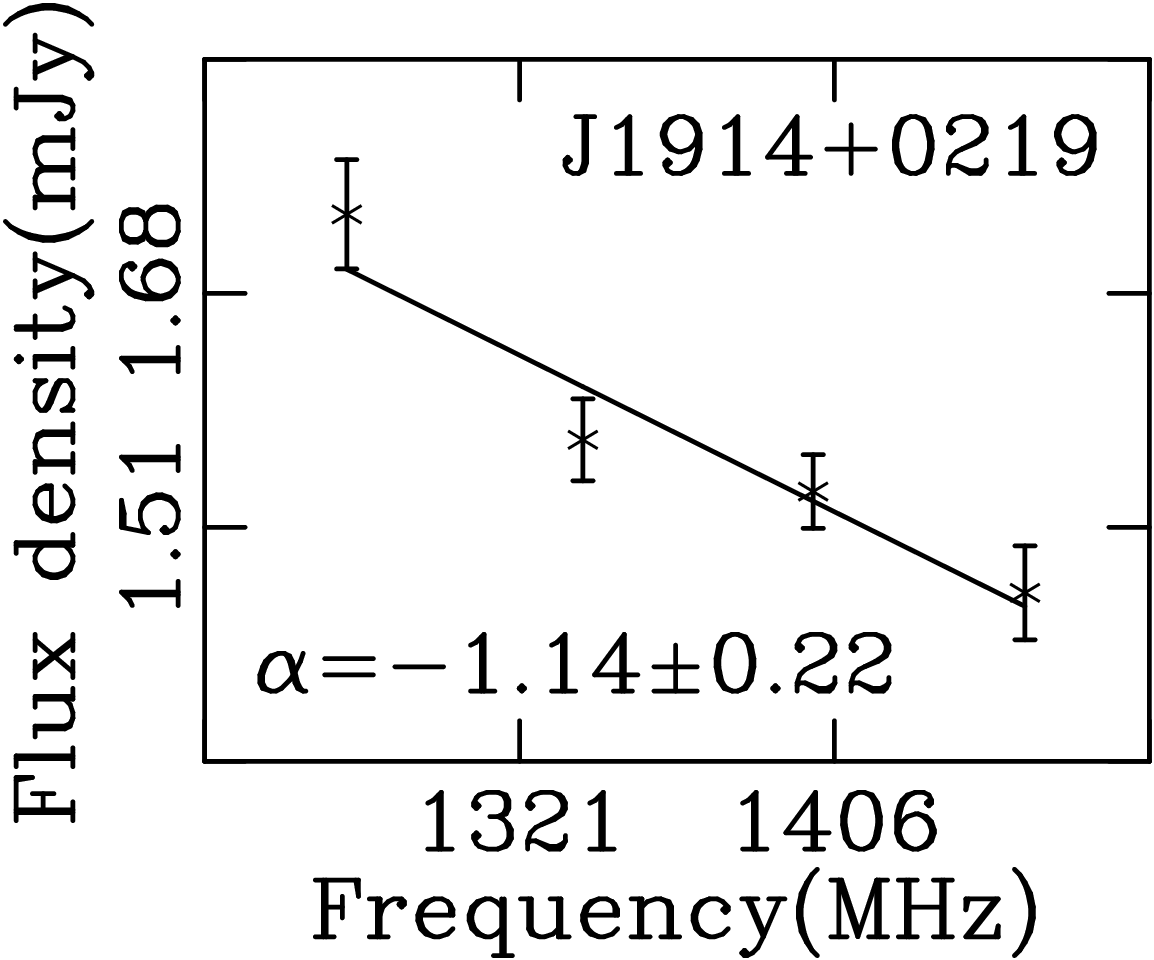}               
\includegraphics[bb = 110 120 518 685,clip,angle=-90,width=0.16\textwidth]{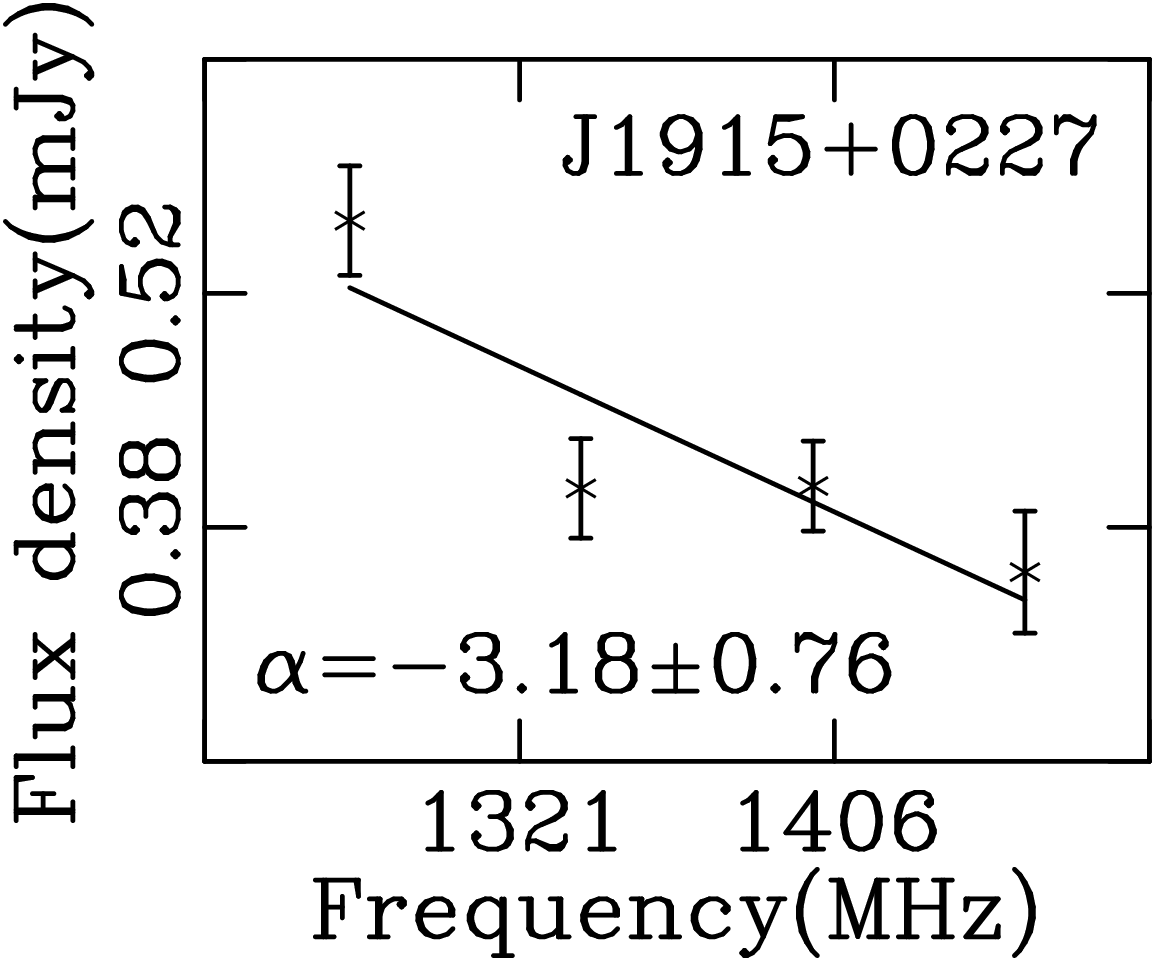}               
\includegraphics[bb = 110 120 518 685,clip,angle=-90,width=0.16\textwidth]{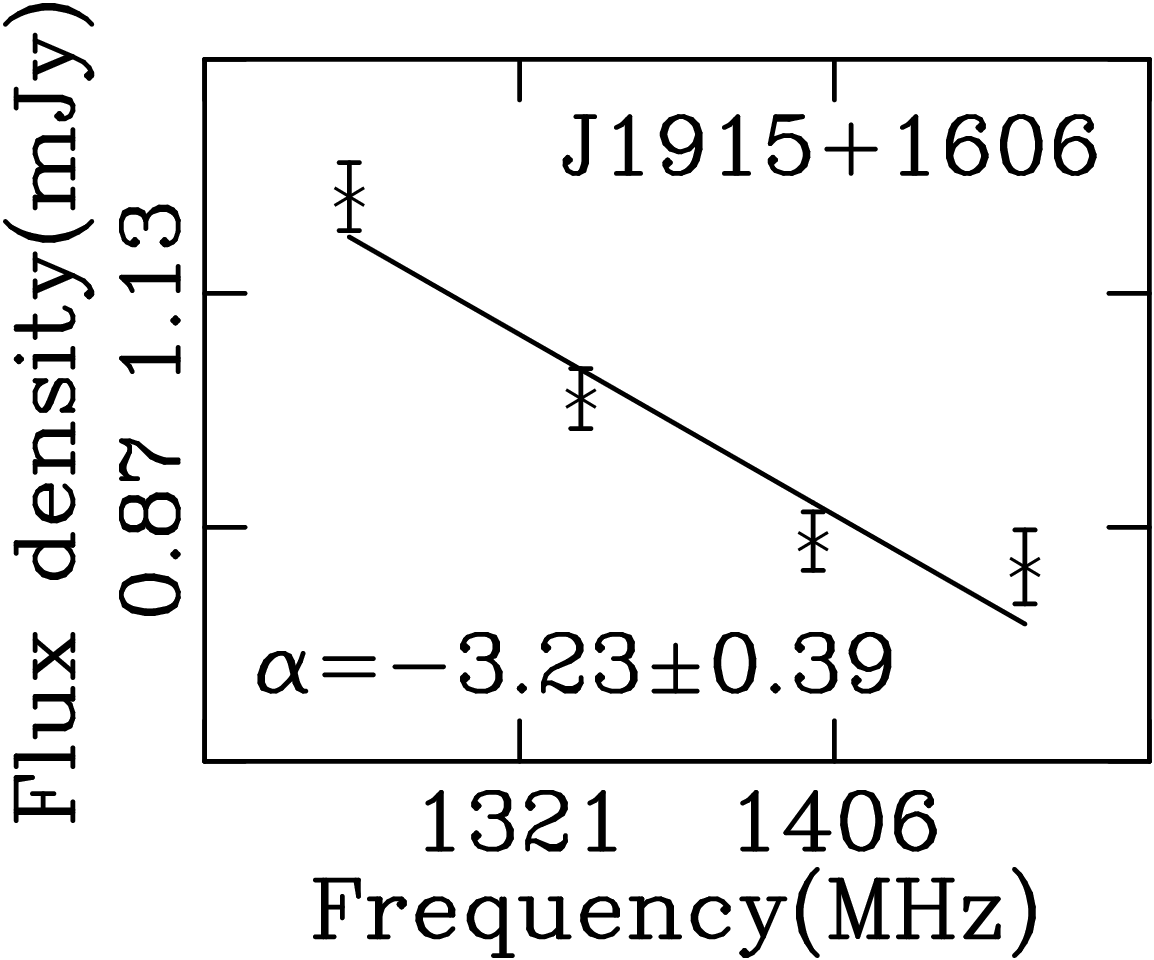}
\\[2mm]               
\includegraphics[bb = 110 72 570 685,clip,angle=-90,width=0.175\textwidth]{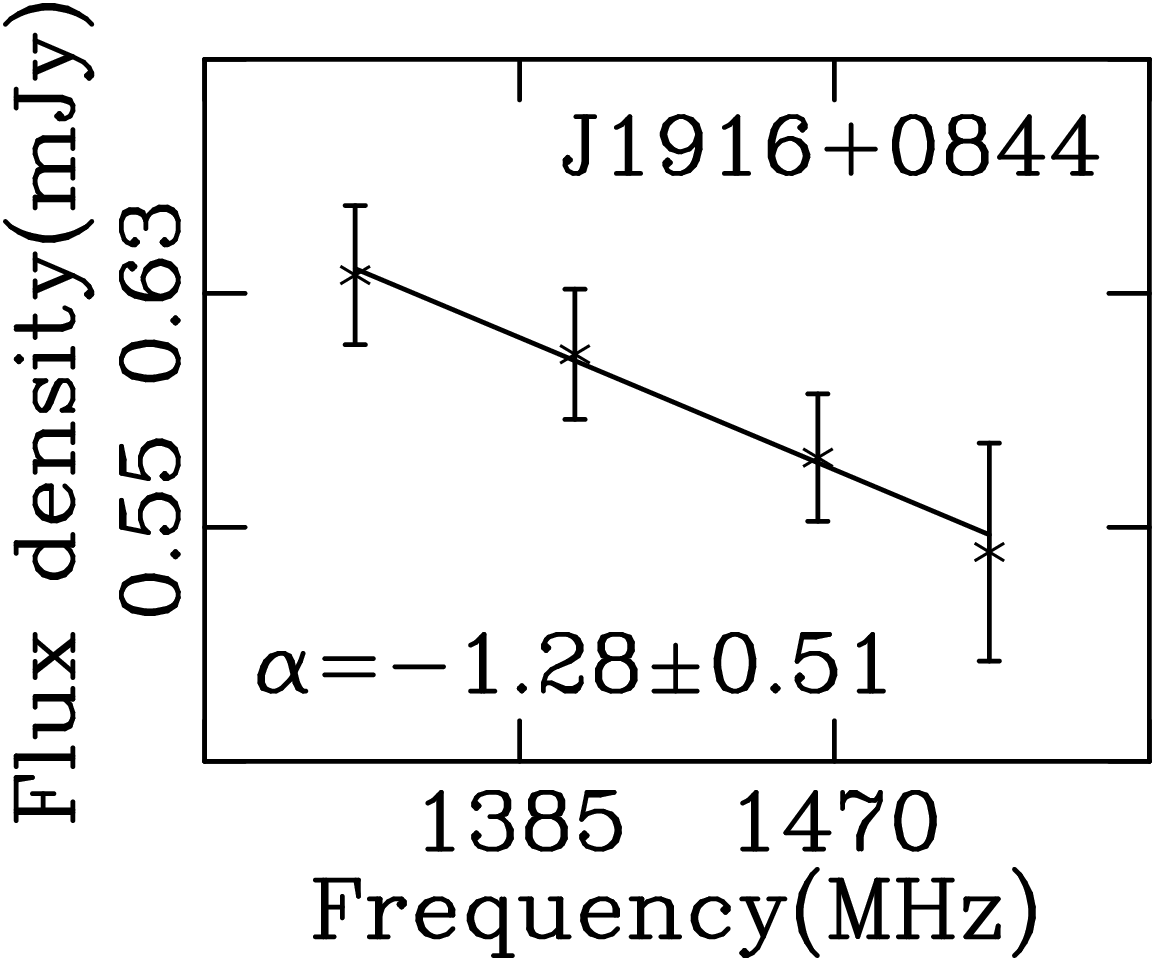}               
\includegraphics[bb = 110 120 570 685,clip,angle=-90,width=0.16\textwidth]{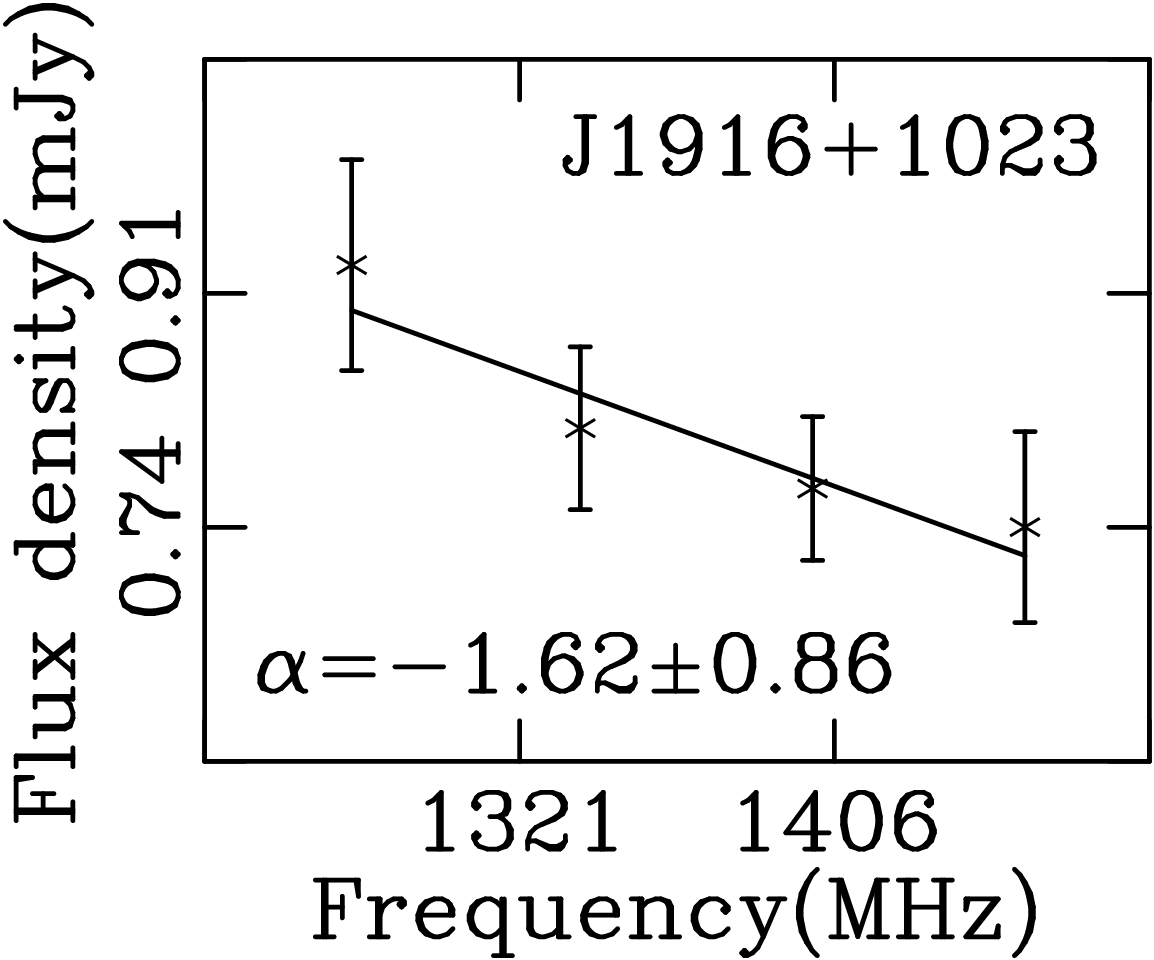}               
\includegraphics[bb = 110 120 570 685,clip,angle=-90,width=0.16\textwidth]{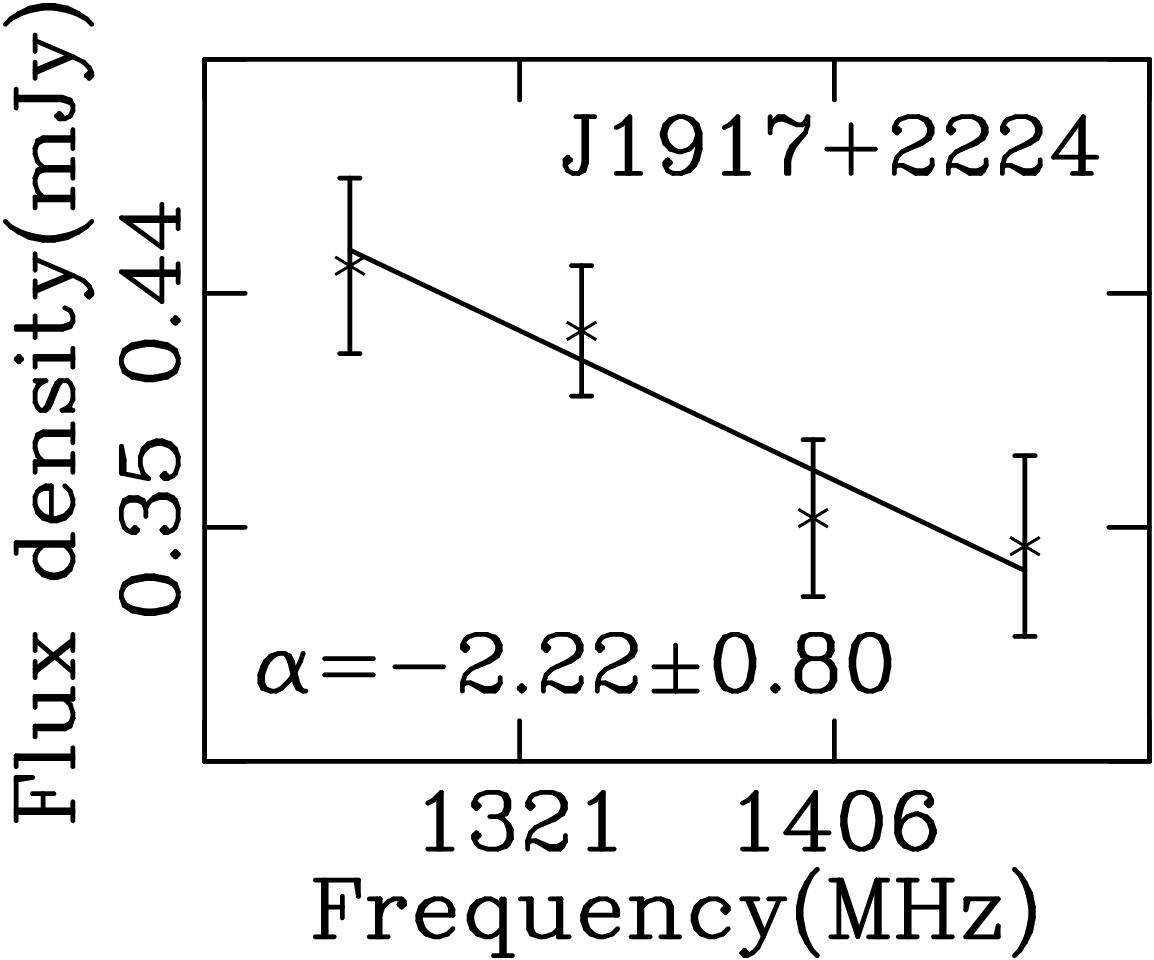}               
\includegraphics[bb = 110 120 570 685,clip,angle=-90,width=0.16\textwidth]{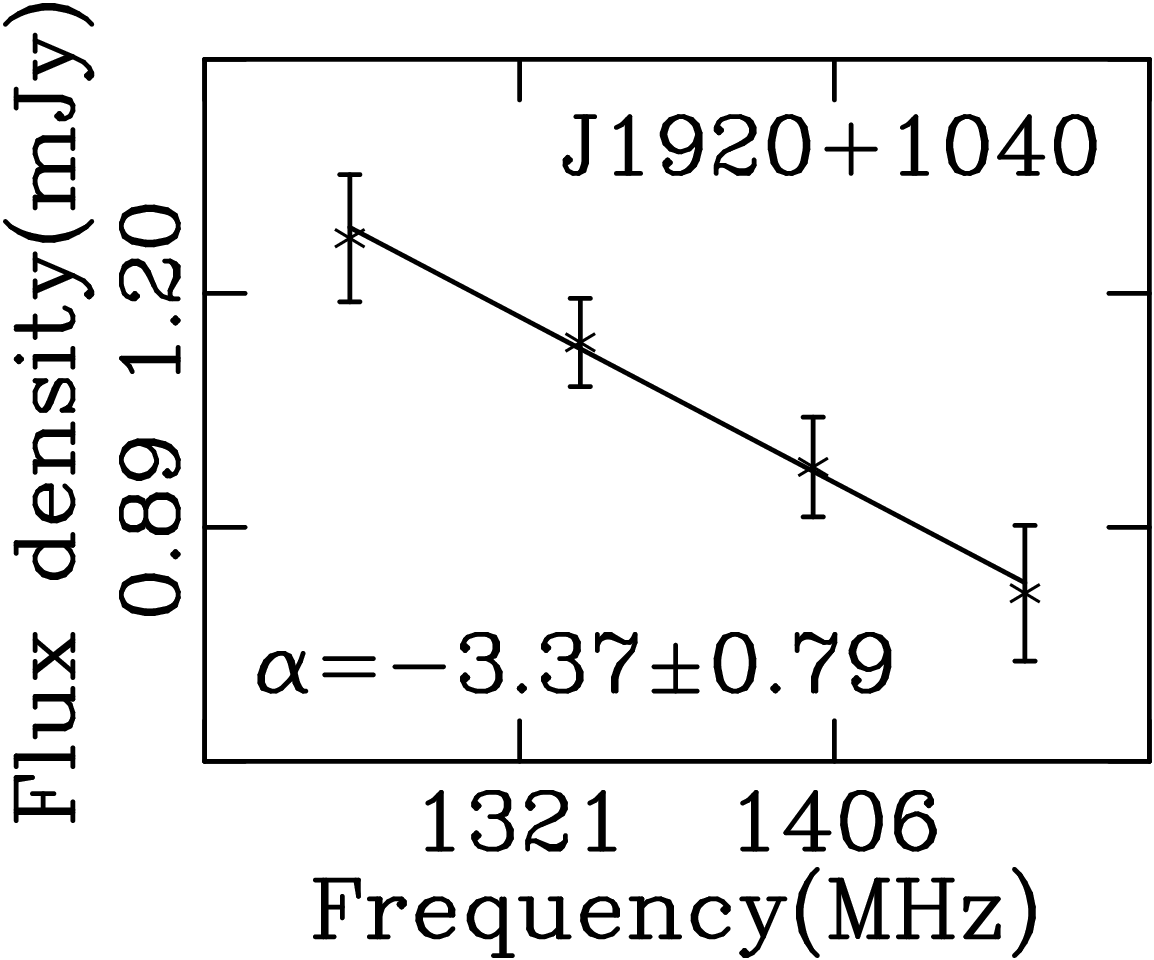}               
\includegraphics[bb = 110 120 570 685,clip,angle=-90,width=0.16\textwidth]{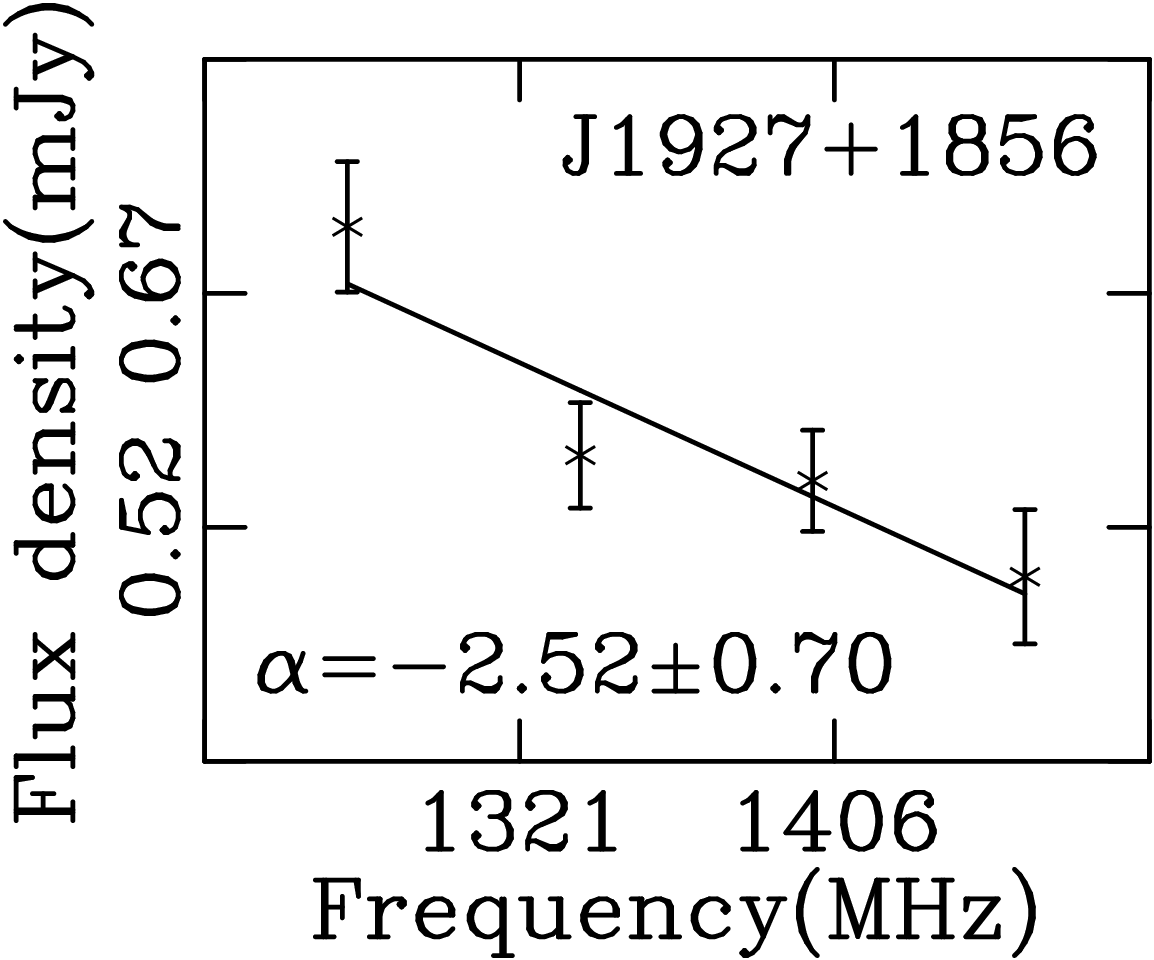}               
\includegraphics[bb = 110 120 570 685,clip,angle=-90,width=0.16\textwidth]{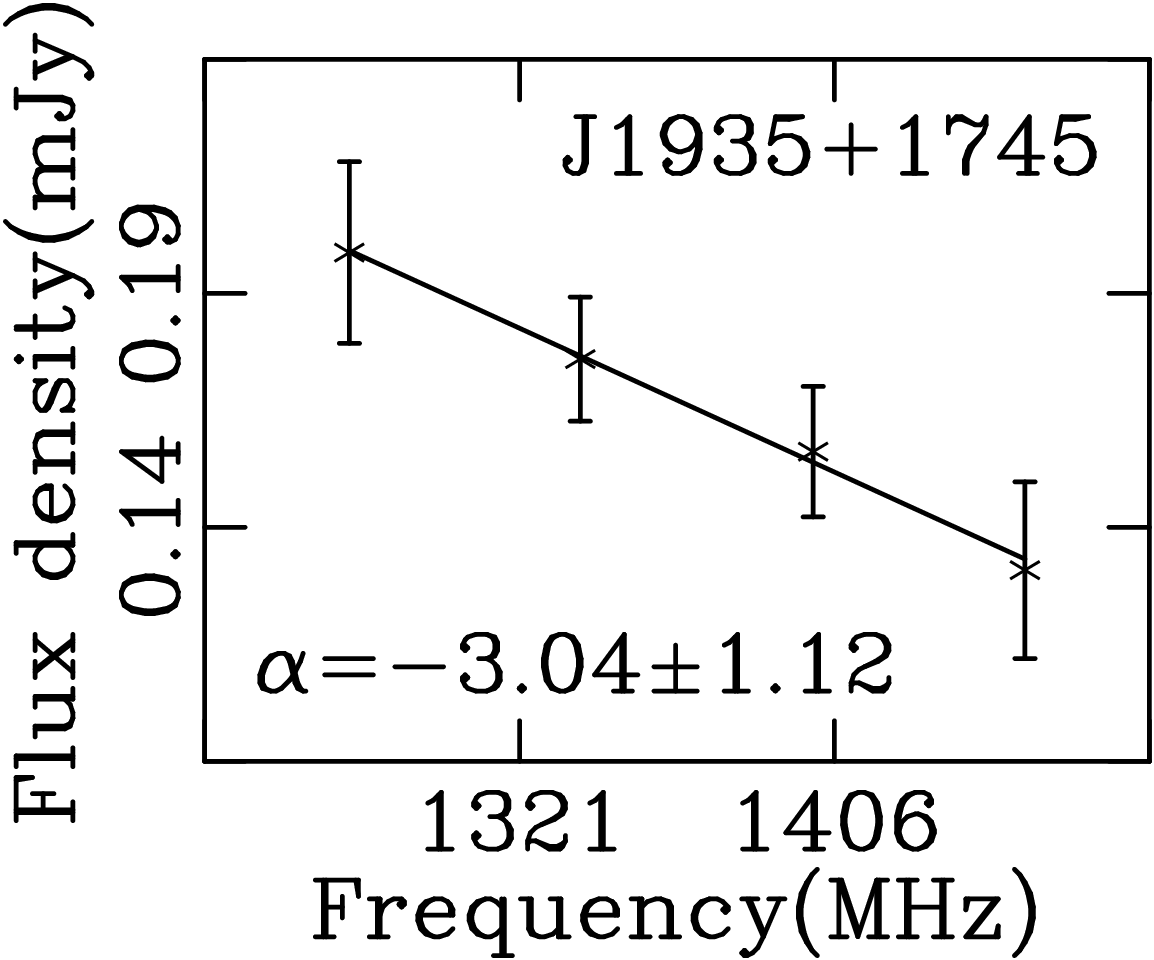}   
\center
\addtocounter{figure}{-1}
\caption{--- continued.}
\label{fig1}
\end{figure}


To get the spectral index, we applied a weighted least-squares approach by fitting
a power law to the mean flux densities at these four sub-bands, and
calculated the standard error $\sigma_{\alpha}$ of $\alpha$ by the
$\chi^{2}$ minimization technique. We found that most pulsars either
scintillate or are too weak at some or all subbands, which means that we can not
 properly fit the spectral index. Only when pulsar dispersion
measure $DM \gs $100~cm$^{-3}$pc and pulsars are strong enough, the
scintillation effect can be averaged out. The flux densities of these
pulsars at four sub-bands are fitted with a power law to get the
spectral indices as shown in Figure~\ref{fig1}.

\begin{figure}
\centering
\includegraphics[width=0.4\textwidth,angle=270]{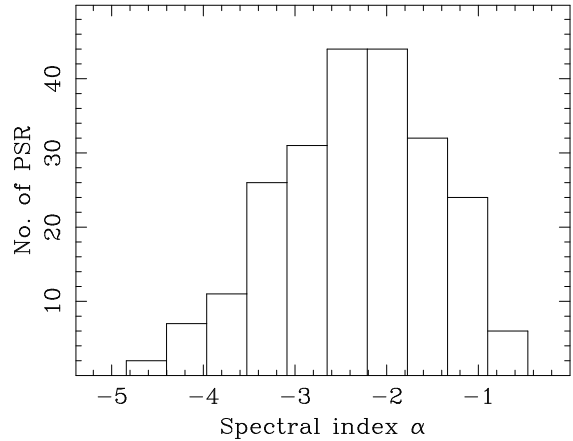}
\caption{Distribution of spectral indices for the 228 pulsars peaks
  at around $-2.2$ in the range of $-4.8$ to $-0.5$. }
\label{fig2}
\end{figure}

\section{Results and discussions}

We calculated the flux densities of 228 pulsars at four sub-bands which can be
consistently fitted with a power law, as listed in
Table~\ref{tab2}. Note that previously, pulsars were observed for
spectral indices using different telescopes by different authors,
and/or calibrated by different sources or processed with different
procedures, which may introduce different systematical
uncertainties. Our measurements presented here are carried out in a
relatively smaller frequency range, and the flux densities are
measured by one set of instruments in one wide band observation and
are consistently calibrated by one source with the same procedure,
which can avoid not only systematical uncertainties but also pulsar
flux density changes due to long-term deflective or reflective
scintillation effects. The spectral indices we obtained are
distributed in the range of $-4.8$ to $-0.5$, with a peak around
$-2.2$, as showed in Figure~\ref{fig2}.


\begin{table}[htpb]
\center
\caption{Comparison of spectral indexes of 28 pulsars we determined
  with those in literatures.}
\label{tab3}
\begin{footnotesize}
\begin{tabular}{cclll}
\hline
Pulsar Name &Our $\alpha$   & Other $\alpha$  &Frequency (MHz) &Reference            \\
\hline                                                                               
J1744-3130 &$-1.66\pm0.41$  &$ -1.1$          & 610-1170       &Dembska et al. 2014  \\
J1751-3323 &$-1.53\pm0.24$  &$ -0.32$         & 610-1170       &Dembska et al. 2014  \\
J1812-2102 &$-1.00\pm0.33$  &$ -2\pm  0.1$    & 610-4850       &Dembska et al. 2014  \\ 
J1835-1020 &$-1.22\pm0.17$  &$ -0.8$          & 610-1170       &Dembska et al. 2014  \\
J1842-0905 &$-2.22\pm0.30$  &$ -1.2$          & 610-1170       &Dembska et al. 2014  \\
J1905+0616 &$-3.17\pm0.62$  &$-0.88\pm 0.25$  & 610-4850       &Dembska et al. 2014  \\
J1644-4559 &$-2.90\pm0.00$  &$ -2.1\pm  0.1$  &1400-6500       &Bates et al. 2011    \\
J1707-4053 &$-2.99\pm0.20$  &$ -2.3\pm  0.1$  &1400-6500       &Bates et al. 2011    \\ 
J1804-0735 &$-2.35\pm0.67$  &$ -1.3\pm 0.31$  & 400-1400       &Maron et al. 2000    \\
J1816-1729 &$-1.97\pm0.26$  &$   -1\pm 0.14$  & 600-1600       &Maron et al. 2000    \\
J1818-1422 &$-2.46\pm0.14$  &$ -1.6\pm 0.22$  & 900-1600       &Maron et al. 2000    \\ 
J1822-1400 &$-2.25\pm0.43$  &$ -0.7\pm 0.22$  & 600-1400       &Maron et al. 2000    \\
J1832-1021 &$-2.93\pm0.22$  &$ -1.3\pm 0.15$  & 400-1600       &Maron et al. 2000    \\ 
J1835-0643 &$-2.01\pm0.47$  &$ -0.4\pm 0.35$  & 600-1600       &Maron et al. 2000    \\ 
J1836-1008 &$-2.98\pm0.14$  &$ -2.1\pm 0.09$  & 400-1600       &Maron et al. 2000    \\
J1836-0436 &$-1.85\pm0.18$  &$ -1.9\pm  0.3$  & 600-1600       &Maron et al. 2000    \\ 
J1837-0653 &$-1.58\pm0.16$  &$ -1.2\pm 0.24$  & 600-1600       &Maron et al. 2000    \\ 
J1845-0434 &$-1.18\pm0.24$  &$ -0.8\pm 0.29$  & 600-1400       &Maron et al. 2000    \\ 
J1852+0031 &$-1.12\pm0.19$  &$ -2.4\pm 0.12$  &1400-4900       &Maron et al. 2000    \\
J1851+0418 &$-2.09\pm0.29$  &$ -1.4$          & 600-1400       &Maron et al. 2000    \\
J1902+0615 &$-3.27\pm0.15$  &$ -2.2\pm  0.1$  & 400-4900       &Maron et al. 2000    \\
J1905-0056 &$-2.64\pm0.30$  &$ -1.9\pm 0.11$  & 400-1400       &Maron et al. 2000    \\ 
J1905+0709 &$-2.58\pm0.20$  &$ -1.3\pm  0.1$  & 600-1400       &Maron et al. 2000    \\ 
J1906+0641 &$-1.03\pm0.16$  &$ -0.7\pm 0.21$  & 400-1600       &Maron et al. 2000    \\
J1909+0254 &$-2.36\pm0.25$  &$ -2.8\pm 0.11$  & 400-1400       &Maron et al. 2000    \\ 
J1915+1606 &$-3.23\pm0.39$  &$ -1.4\pm 0.24$  & 400-1400       &Maron et al. 2000    \\
J0738-4042 &$-1.89\pm0.01$  &$ -0.69$         & 400-1400       &Slee et al. 1986     \\ 
J0952-3839 &$-4.27\pm0.54$  &$ -1.96$         & 160-400        &Slee et al. 1986     \\
\hline
\end{tabular}
\end{footnotesize}
\end{table}

\begin{figure}
  \center
  \includegraphics[width=0.3\textwidth,angle=-90]{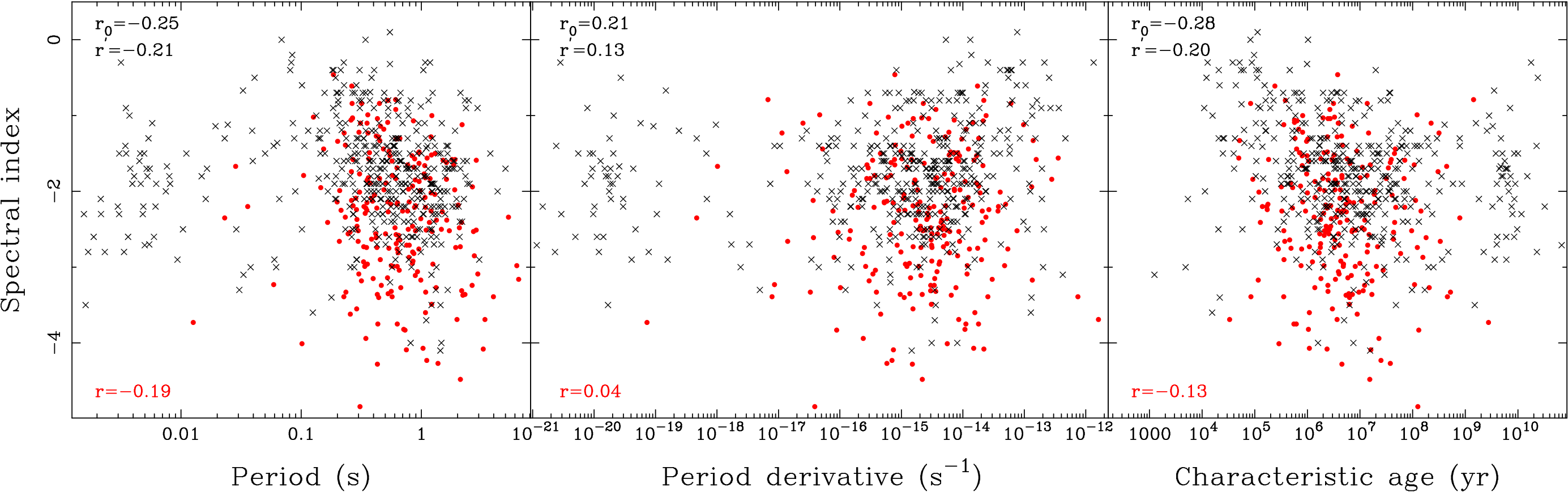}\\[0.2cm]
  \includegraphics[width=0.3\textwidth,angle=-90]{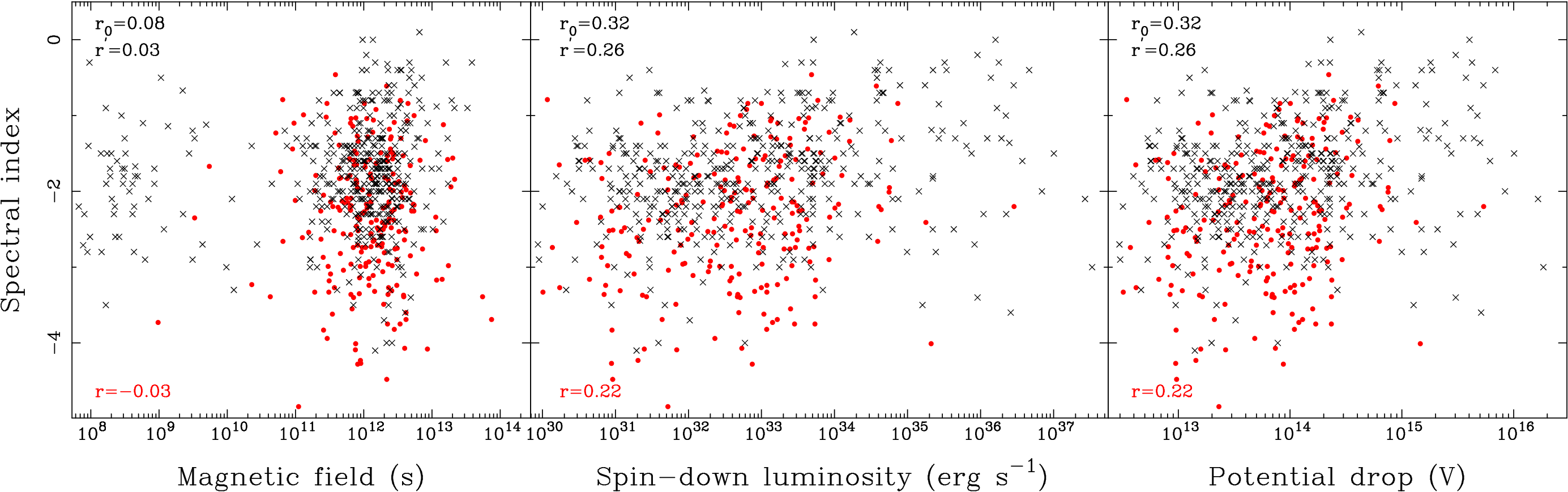}
  \center
  \caption{Spectral indices we determined (dots) and those in
    literatures (cross) are plotted against pulsar characteristic
    parameters: period $P$, period derivative $\dot{P}$, age $\tau$,
    and surface magnetic fields $B$, spin-down luminosity $\dot{E}$,
    and polar gap potential drop $\Delta \Psi$. The Spearman rank
    correlation parameters are marked for normal pulsars in
    literature ($r_0$) and in our sample ($r$) as well as the combined
    sample ($r'$).}
    \label{fig3}
\end{figure}

\begin{table}
\center
\caption{The Spearman rank correlation parameters for spectral indices
  against $P$, $\tau$, $\dot{P}$, $B$, $\dot{E}$ and $\Delta\Psi$ for
  372 normal pulsars in literature ($r_0$) and 224 normal pulsars
  (excluded 4 millisecond pulsars) in our sample ($r$) as well as 572
  pulsars in the combined sample ($r'$).  }
  \label{tab4}
  \begin{tabular}{@{}lcccccc}
      \hline
      & $\alpha$ vs. ${P}$ &$\alpha$ vs. ${\dot{P}}$  &  $\alpha$ vs. $\tau$
      &$\alpha$ vs. $B$ &$\alpha$ vs. $\dot{E}$  & $\alpha$ vs. $\bigtriangleup\Psi$  \\
      \hline
372 pulsars ($r_{0}$) & $-0.25$ &  0.21  & $-0.28$  &  0.08   & 0.32 & 0.32  \\
224 pulsars ($r$)    & $-0.19$ &  0.04  & $-0.13$  & $-0.03$ & 0.22 & 0.22  \\
572 pulsars ($r^{'}$) &$-0.21$ & 0.13  & $-0.20$  & 0.03    & 0.26 & 0.26 \\
  \hline
  \end{tabular}
\end{table}

We also compiled spectral indices of pulsars in the
literature. Previously, spectral indices have been determined for 426
pulsars
\citep{s73,kxl+98,mgj+94,lyl+95,tbm+98,kkw+98,mkk+00,srw75,ikms79,
  kll+99,c78,dkj+14,bjl+11,sab86,dcl09,ffb91,hmm+11,jml+09,lr99,
  lzb+00,lx00,mac+02,mal96,mms00,rcc+70,rcg+68}. In our sample, 28
pulsars have their spectral indices determined in the literature, and 200 weak
pulsars have newly determined spectral indices.  We compared the
spectral indices of 28 pulsars we measured with those in literature in
Table~\ref{tab3}, and found that our values are systematically
steeper. Our results seem to be reasonable because the frequency range for flux densities we
measured is at the high end of the multi-frequency wide range in
literature and because flux densities at lower frequencies always
show the turn-over which causes a global flat spectrum.

To find clues for pulsar emission process, we plot in
Figure~\ref{fig3} the spectral indices of pulsars against pulsar
rotation period $P$, the spin-down rate $\dot{P}$, and the
characteristic age $\tau=P/2\dot{P}$, the surface magnetic field
strength $B_{\mathrm{S}} \propto\sqrt{P\dot{P}}$, the spin-down energy
loss rate $\dot{E}\propto\dot{P}/P^{3}$, and also the potential drop
in the polar gap $\Delta\Psi\propto {P^{-3/2}{\dot{P}}^{1/2}}$ which
is related to particle acceleration. The spearman rank correlation
parameter ($r$) is used to describe the degree of correlation
\citep{ptvf92}, in which $r=1$ stands for strong correlation and $r=0$
for no correlation. We confirm the conclusion of previous authors
\citep[e.g.][]{lyl+95} that there are some very weak correlations
($r\sim0.3$) between spectral indices and $\dot{E}$ and $\Delta\Psi$
(see Table~\ref{tab4}), which implies that the emission properties are
physically related to particle acceleration process.

\section{Conclusions}

We obtained the flux densities at the 4 subbands for 228 pulsars
observed the Parkes 64~m telescope at 1.4~GHz, and determined their
spectral indices, among which 200 are obtained for the first time. The
indices have a broad distribution in the range of $-4.8$ to
$-0.5$. Such observations for getting spectral indices have a few advantages:
1) flux densities are measured by a single instrument and calibrated with
a consistent approach using the same flux calibrator, 2) one
observation session can avoid the short-term scintillation effect when
the DM is large enough and also avoid the long-term flux changes due
to diffractive and reflective scintillation, 3) in such a small
frequency range the profile components would not disappear
quickly. The disadvantage is that the uncertainty associated with such a determined
spectral index is large due to the limited frequency range. The best
spectral measurements in future would be made for each profile
component within one ultra wide band. The weak correlations between
spectral indexes and $\dot{E}$ and $\Delta\Psi$ imply that the
radiation properties are physically related to particle acceleration
process.

\normalem

\begin{acknowledgements}
The authors are supported by
  the National Natural Science Foundation of China (No. 11473034) and
  by the Strategic Priority Research Programme ``The Emergence of
  Cosmological Structures'' of the Chinese Academy of Sciences, Grant
  No. XDB09010200.
\end{acknowledgements}

\bibliographystyle{raa}
\bibliography{spec}

\label{lastpage}
\end{document}